
\input amssym.def
\input amssym


\def\Rrm{{\Bbb R}}
\def\Nrm{{\Bbb N}}
\def\Crm{{\Bbb C}}
\def\Zrm{{\Bbb Z}}
\def\un{{\Bbb I}}

\font\sss=msam10
\font\ssss=msam8
\font\titre=cmbx12
\font\mathgras=cmmib10
\font\mathgra=cmmib8
\font\mathgr=cmmib6


\def\psaut{\vskip 5pt plus 1pt minus 1pt}
\def\saut{\vskip 10pt plus 2pt minus 3pt}
\def\gsaut{\vskip 20pt plus 3pt minus 4pt}

\def\sqr#1#2{{\vcenter{\vbox{\hrule height.#2pt
\hbox{\vrule width.#2pt height#1pt \kern #1pt
\vrule width.#2pt}
\hrule height.#2pt}}}}
\def\carre{\hbox{\sss \char'003}}
\def\carrre{\hbox{\ssss \char'003}}

\def\fl{\rightarrow}

\def\sump_#1{\setbox0=\hbox{$\scriptstyle{#1}$}
 \setbox2=\hbox{$\displaystyle{\sum}$}
 \setbox4=\hbox{${}'\mathsurround=0pt$}
 \dimen0=.5\wd0 \advance\dimen0 by-.5\wd2
 \ifdim\dimen0>0pt
 \ifdim\dimen0>\wd4 \kern\wd4 \else\kern\dimen0\fi\fi
 \mathop{{\sum}'}_{\kern-\wd4 #1}}
\def\suma_#1^#2{\setbox0=\hbox{$\scriptstyle{#1}$}
 \setbox2=\hbox{$\displaystyle{\sum}$}
 \setbox4=\hbox{$\scriptstyle{#2}\mathsurround=0pt$}
 \dimen0=.5\wd0 \advance\dimen0 by-.5\wd2
 \ifdim\dimen0>0pt
 \ifdim\dimen0>\wd4 \kern\wd4 \else\kern\dimen0\fi\fi
 \mathop{{\sum}^{#2}}_{\kern-\wd4 #1}}
\def\psd{\hbox{$\subset\kern-10pt \times$}}
\def\ssd{\hbox{$\subset\kern-10pt +$}}
\def\sg{\hbox{\mathgras\char'033}}
\def\ssg{\hbox{\mathgra\char'033}}
\def\sssg{\hbox{\mathgr\char'033}}

\def\ds{\displaystyle}
\def\p{{\frak p}}
\def\r{{\cal P}}

\def\scr{\scriptstyle}
\def\sscr{\scriptscriptstyle}

\def\fhd{{\setbox1=\hbox{\rm \char'136}\dot{f}
    \kern-3.75pt\raise4pt\box1}\kern-1pt}
\def\fhdn{{\setbox1=\hbox{\rm \char'136}\dot{f}^{(n)}
    \kern-15.5pt\raise4pt\box1}\kern10.8pt}
\def\fhdnz{{\setbox1=\hbox{\rm \char'136}\dot{f}^{(n)}_0
    \kern-15.5pt\raise4pt\box1}\kern10.8pt}
\def\fhdnm{{\setbox1=\hbox{\rm \char'136}\dot{f}^{(n)}_\mu
    \kern-15.5pt\raise4pt\box1}\kern10.8pt}
\def\fhdz{{\setbox1=\hbox{\rm \char'136}\dot{f}_0
    \kern-7.4pt\raise4pt\box1}\kern2.5pt}
\def\fhdm{{\setbox1=\hbox{\rm \char'136}\dot{f}_\mu
    \kern-8.3pt\raise4pt\box1}\kern3.3pt}
\def\vvv{\hbox{$\vert\!\vert\!\vert$}}

\font\sc=cmcsc10
\def\refBFFLSI{[1]}
\def\refBFFLSII{[2]}
\def\refCAL{[3]}
\def\refD{[4]}
\def\refFPS{[5]}
\def\refFS{[6]}
\def\refFSYM{[7]}
\def\refFST{[8]}
\def\refFSTA{[9]}
\def\refGR{[10]}
\def\refHA{[11]}
\def\refHKG{[12]}
\def\refHKGD{[13]}
\def\refHSL{[14]}
\def\refKI{[15]}
\def\refKII{[16]}
\def\refSERG{[17]}
\def\refSI{[18]}
\def\refSTI{[19]}
\def\refST{[20]}
\def\refVW{[21]}
\def\refW{[22]}

\nopagenumbers
\magnification=1200
\centerline {\titre Asymptotic completeness, global existence and the infrared}
\saut
\centerline {\titre problem for the Maxwell-Dirac equations}
\gsaut
\saut
\centerline {\bf Mosh\'e FLATO$^1$, Jacques C. H. SIMON$^1$
and Erik TAFLIN$^{1,2}$}
\saut
\centerline {$^1$ D\'epartement de Math\'ematiques, Universit\'e de Bourgogne,}
\centerline {B.P. 138, F-21004 DIJON Cedex, France.}
\centerline {$^2$ Permanent address: Direction Scientifique,
Union des Assurances de Paris,}
\centerline {9 place Vend\^ome, F-75052 PARIS Cedex 01, France.}
\gsaut
\gsaut
\noindent{\bf Abstract:} In this monograph we prove that the nonlinear Lie
algebra representation given by the manifestly covariant Maxwell-Dirac (M-D)
equations is integrable to a global nonlinear representation $U$ of the
Poincar\'e group ${\cal P}_0$ on a differentiable manifold ${\cal U}_\infty$
of small initial conditions for the M-D equations. This solves, in
particular, the Cauchy problem for the M-D equations, namely existence of
global solutions for initial data in ${\cal U}_\infty$ at $t=0$. The
existence of modified wave operators $\Omega_+$ and $\Omega_-$ and
asymptotic completeness is proved. The asymptotic representations
$U^{(\varepsilon)}_g = \Omega^{-1}_\varepsilon \circ U_g \circ
\Omega_\varepsilon$, $\varepsilon = \pm$, $g \in {\cal P}_0$,
turn out to be nonlinear. A cohomological interpretation of the
results in the spirit of nonlinear representation theory and its connection
to the infrared tail of the electron is given.
\vskip3cm
\centerline {\it THIS MONOGRAPH IS DEDICATED TO THE MEMORY}

\centerline {\it OF THE CHIEF CREATOR OF  QUANTUM ELECTRODYNAMICS,}

\centerline {\it A GIANT IN CONTEMPORARY PHYSICS,}

 \centerline {\it A GREAT HUMAN BEING AND UNFORGETTABLE FRIEND --
 JULIAN SCHWINGER}
 \vskip1cm
 \copyright 1995 by M. Flato, J.C.H. Simon and E. Taflin

\vfill\eject

\headline={\ifodd\pageno\rightheadline \else\leftheadline\fi}
\def\rightheadline{\tenrm\hfil MAXWELL - DIRAC \hskip.2cm
EQUATIONS \hfil\folio}
\def\leftheadline{\tenrm\folio\hfil FLATO \hskip.2cm SIMON \hskip.2cm
TAFLIN \hfil}
\voffset=2\baselineskip

{\titre 1. Introduction}
\saut
It is well-known that the construction of the observables on the Fock space
of QED (Quantum Electrodynamics) requires infrared corrections to eliminate
the infrared divergencies in the perturbative expression of the quantum
scattering operator.  These corrections are introduced by hand with the
purpose to give a posteriori a finite theory. In this monograph we shall prove
rigorous results which we have obtained concerning the infrared problem for
the (classical) Maxwell-Dirac equations.  Our belief is that such results
can a priori be of interest for QED, especially for the infrared regime and
combined with the deformation-quantization approach \refBFFLSI, \refBFFLSII,
\refD. Our results show in particular that, also in the classical case one
obtains infrared divergencies, if one requires free asymptotic fields as
it is needed in QED. But before continuing the physical motivation of the
paper,
we shall describe the mathematical context.

1.1 {\sc The Mathematical Framework}

1.1.a {\it The Equations.} We shall use conventional notations:
electron charge $e=1$; Dirac matrices $\gamma^\mu$, $0 \leq \mu \leq 3$;
metric tensor
$g^{\mu \nu}$, $g^{00} = 1$, $g^{ii} = -1$ for $1 \leq i \leq 3$ and
$g^{\mu \nu} = 0$
for $\mu \neq \nu$; $\gamma^\mu  \gamma^\nu + \gamma^\nu
\gamma^\mu = 2g^{\mu \nu}$; $\partial_\mu = \partial / (\partial y^\mu)$;
$\carre = \partial_\mu \partial^\mu$ (with the Einstein  summation convention
and the index raising convention). Then the classical Maxwell-Dirac (M-D)
equations read:

$$\eqalignno{
\carre A_\mu &= \overline{\psi}\gamma_\mu  \psi,\quad 0 \leq \mu \leq 3,
& (1.1{\rm a})\cr
(i \gamma^\mu \partial_\mu + m) \psi &= A_\mu \gamma^\mu \psi,\quad m > 0,
& (1.1{\rm b})\cr \partial_\mu A^\mu &= 0,& (1.1{\rm c})\cr  }$$
where $\overline{\psi} = \psi^+   \gamma_0$, $ \psi^+$ being the Hermitian
conjugate of the Dirac spinor $\psi$. We write equations (1.1a) and (1.1b) as
an evolution equation:
$$\eqalignno{
{d \over dt}(A_\mu (t), \dot{A}_\mu (t)) &= (\dot{A}_\mu (t), \Delta
A_\mu (t)) + (0, \overline{\psi} (t) \gamma_\mu \psi (t)),&(1.2{\rm a})\cr
{d \over dt}\psi (t) &= {\cal D} \psi (t) - i A_\mu (t)
\gamma^0\gamma^\mu\psi (t),&(1.2{\rm b})\cr
}$$
where ${\cal D} = - \sum^3_{j=1}\gamma^0\gamma^j
\partial_j + i \gamma^0 m$, $\Delta = \sum^3_{j=1}\partial^2_j$,
$t \in \Rrm$ and where $A_\mu (t)$, $\dot{A}_\mu (t)\colon\Rrm^3 \fl \Rrm$,
$\psi (t)\colon\Rrm^3 \fl \Crm^4$. The Lorentz gauge condition
(1.1c) takes on initial conditions
$A_\mu (t_0)$, $\dot{A}_\mu (t_0)$, $0 \leq \mu \leq 3$, and $\psi (t_0)$
at $t=t_0$
the form (cf. \refGR, \refFST)
$$\eqalignno{
\dot{A}^0 (t_0) + \sum^3_{i=1}\partial_i A^i (t_0) &= 0,&(1.3{\rm a})\cr
\Delta A^0 (t_0) + \vert \psi (t_0) \vert^2 + \sum^3_{i=1}
\partial_i\dot{A}^i (t_0) &= 0, &(1.3{\rm b})\cr
}$$
where $A^\mu = g^{\mu \nu} A_\nu$.

1.1.b {\it Poincar\'e covariance: linear part and representation spaces.}
Since equations (1.1a), (1.1b) and (1.1c) are manifestly covariant under
the action of the universal covering group ${\cal P}_0 = \Rrm^4 \psd
SL(2, \Crm)$ of the Poincar\'e group, it is easy to complete the time
translation generator, formally defined by (1.2a) and (1.2b), to a nonlinear
representation of the whole Lie algebra $\p = \Rrm^4 \ssd {\frak{sl}}(2,\Crm)$
of
${\cal P}_0$. To do this we first introduce topological vector
spaces on which the representations will be defined.  Let $M^\rho$,
$- 1/2 <\rho < \infty$ be the completion of
$S(\Rrm^3,\Rrm^4) \oplus S(\Rrm^3,\Rrm^4)$ with respect to the norm
$$\Vert (f, \dot{f}) \Vert^{}_{M^\rho} = \big(\Vert \vert \nabla \vert^\rho f
\Vert^2_{L^2 (\Rrm^3, \Rrm^4)} + \Vert \vert \nabla \vert^{\rho-1} \dot{f}
\Vert^2_{L^2 (\Rrm^3, \Rrm^4)}\big)^{1/2}, \eqno{(1.4{\rm a})}$$
where $S(\Rrm^3, \Rrm^4)$ is the Schwartz space of test functions from
$\Rrm^3$ to $\Rrm^4$ and where $\vert\nabla\vert=(-\Delta)^{1/2}$.
Let $D = L^2 (\Rrm^3, \Crm^4)$ and let $E^\rho = M^\rho \oplus D$,
$- 1/2 < \rho < \infty$, be the Hilbert space with norm
$$\Vert (f, \dot{f}, \alpha) \Vert^{}_{E^\rho} =
\big(\Vert (f, \dot{f}) \Vert^2_{M^\rho} +
\Vert \alpha \Vert^2_D\big)^{1/2}. \eqno{(1.4{\rm b})}$$
When there is no possibility of confusion we write $E$ (resp. $M$) instead of
$E^\rho$ (resp. $M^\rho$) for $1/2 < \rho < 1$. $M^{\circ\rho}$ is the closed
subspace of elements $(f, \dot{f}) \in M^\rho$, $- 1/2 < \rho < \infty$
such that
$$\eqalignno{
\dot{f}_0 &= \sum_{1 \leq i \leq 3}\partial_i f_i,&(1.4{_rm c})\cr
f_0 &= - \sum_{1 \leq i \leq 3} \vert\nabla \vert^{-2}\partial_i\dot{f}_i,\cr
}$$
where $f = (f_0, f_1, f_2, f_3)$ and $\dot{f} = (\dot{f}_0, \dot{f}_1,
\dot{f}_2,
\dot{f}_3)$. A solution $B_\mu$ of $\carre B_\mu = 0$, $0 \leq \mu \leq 3$,
with initial conditions $(f, \dot{f}) \in M^\rho$ satisfies the gauge
condition $\partial_\mu B^\mu = 0$ if and only if $(f, \dot{f})
\in M^{\circ\rho}$. We define $E^{\circ\rho} = M^{\circ\rho} \oplus D.$

Let $\Pi = \{ P_\mu, M_{\alpha \beta} \big\vert 0 \leq \mu \leq 3,
0 \leq\alpha < \beta \leq 3 \}$ be a standard basis of the Poincar\'e Lie
algebra $\p =\Rrm^4 \ssd {\frak{sl}}(2, \Crm)$, where $P_0$ is the time
translation
generator, $P_i$, $1\leq i \leq 3$ the space translation generators, $M_{ij}$,
$1 \leq i < j \leq 3$, are  the space rotation generators and $M_{0j}$,
$1 \leq j \leq 3$, are  the boost generators. We define
$M_{\alpha \beta} = - M_{\beta \alpha}$ for $0 \leq \beta
\leq \alpha \leq 3$. There is a linear (strongly) continuous representation
$U^1$ of ${\cal P}_0$ in $E^\rho$, $- 1/2 < \rho < \infty$, (see Lemma 2.1)
with space of differentiable
vectors $E^\rho_\infty$, the differential of which is the following
linear representation $T^1$ of $\p$ in $E^\rho_\infty$:
$$\eqalignno{
 T^1_{P_0}(f, \dot{f}, \alpha) &= (\dot{f}, \Delta f, {\cal D} \alpha),
\hskip86mm&(1.5{\rm a})\cr
T^1_{P_i} (f, \dot{f}, \alpha) &= \partial_i (f, \dot{f}, \alpha),
\quad  1 \leq i \leq 3,& (1.5{\rm b})\cr
(T^1_{M_{ij}} (f, \dot{f}, \alpha)) (x) &= - (x_i\partial_j - x_j
\partial_i)(f, \dot{f}, \alpha) (x)+ (n_{ij}f, n_{ij}\dot{f}, \sigma_{ij}
\alpha) (x)&(1.5{\rm c})\cr
& \hskip 20mm 1 \leq i<j \leq 3,\sigma_{ij} = 1/2\gamma_i\gamma_j \in
{\frak{su}}(2),
n_{ij} \in {\frak {so}}(3),\cr
(T^1_{M_{0i}} (f, \dot{f}, \alpha)) (x) &= \big(x_i\dot{f}(x),
\sum^3_{j=0}\partial_j \big(x_i\partial_j f(x)\big), x_i
{\cal D} \alpha (x)\big)
+ (n_{0i}f,n_{0i}\dot{f},\sigma_{0i}
\alpha) (x),&(1.5{\rm d})\cr
& \hskip 20mm 1 \leq i \leq 3,\sigma_{0i} = 1/2
\gamma_0\gamma_i,n_{0i} \in {\frak {so}}(3,1),\cr
}$$
where $x = (x_1, x_2, x_3)$ are related to contravariant coordinates
$y^\mu$ by $x_i = y^i$ (we use this notation to avoid writing
components of vectors in $\Rrm^3$ with uppper indices that could be
confused with powers). The explicit form of $n_{\alpha \beta}$,
$0 \leq\alpha < \beta \leq 3$, defining a vector representation
of $\p$ in $\Rrm^4$, is not important here.
We remark that $U^1$ leaves $E^{\circ\rho}$ invariant.

Following closely \refST, we introduce the graded sequence of Hilbert spaces
$E^\rho_i$, \penalty-9000 $i \geq 0$, $E^\rho_0 = E^\rho$, where
$E^\rho_\infty \subset E^\rho_j \subset E^\rho_i$ for $i \leq j$. $E^\rho_i$
is the space of $C^i$-vectors of the representation $U^1$.
Suppose given an ordering $X_1 < X_2 <\cdots < X_{10}$
on $\Pi$. Then in the universal enveloping algebra $U (\p)$ of $\p$, the
subset of all products $X^{\alpha_1}_1\cdots X^{\alpha_{10}}_{10}$,
$0 \leq \alpha_i$, $1 \leq i \leq 10$, is a basis $\Pi'$ of $U (\p)$.
If $Y = X^{\alpha_1}_1 \cdots
X^{\alpha_{10}}_{10}$, then we define $\vert Y \vert = \vert \alpha \vert =
\sum_{0 \leq i \leq 10}   \alpha_i$. Let $E^\rho_i$, $i \in \Nrm$
be the completion of $E^\rho_\infty$ with respect to the norm
$$\Vert u \Vert_{E^\rho_i} = \Big(\sum_{\scr Y \in \Pi'\atop\scr \vert Y
\vert \leq i} \Vert T^1_Y (u) \Vert^2_{E^\rho}\Big)^{1/2}, \eqno{(1.6{\rm
a})}$$
where $T^1_Y$,  $Y \in U (\p)$ is defined by the canonical extension
of $T^1$ to the enveloping algebra $U (\p)$ of $\p$. Let $M^{\circ\rho}_i =
M^\rho_i \cap M^{\circ\rho}$, $E^{\circ\rho}_i = E^\rho_i \cap E^{\circ\rho}$,
$M^{\circ\rho}_\infty = M^\rho_\infty \cap M^{\circ\rho}$ and
$E^{\circ\rho}_\infty = E^\rho_\infty \cap E^{\circ\rho}$,
where $M^\rho_i$ (resp. $M^\rho_\infty$) is the image of $E^\rho_i$
(resp. $E^\rho_\infty$) in $M^\rho$ under the canonical
projection of $E^\rho$ on $M^\rho$. Let $D_i$ (resp. $D_\infty$) be the
image of
$E^\rho_i$ (resp. $E^\rho_\infty$) in $D$ under the canonical projection
of $E^\rho$ on $D$. To understand better what the elements of the spaces
$E^\rho_i$ are, we introduce the seminorms $q^{}_n$, $n \geq 0$,
on $E^\rho_\infty$, where
$$\eqalignno{
q^{}_n (u) &=(q^M_n (v)^2 + q^D_n (\alpha)^2)^{1/2},
\quad u = (v, \alpha) \in E^\rho_\infty,
v \in M^\rho_\infty,\alpha \in D_\infty,\cr
q^M_n (v) &= \Big(\sum_{\vert \mu \vert \leq \vert \nu \vert \leq n}
\Vert M_\mu \partial^\nu v \Vert^2_{M^\rho}\Big)^{1/2},\cr
q^D_n (\alpha) &= \Big(\sum_{\scr \vert \mu \vert \leq n\atop\scr \vert
\nu \vert \leq n} \Vert M_\mu   \partial^\nu   \alpha \Vert^2_D\Big)^{1/2},
\cr }$$
where $\mu = (\mu_1,\mu_2,\mu_3)$, $\nu = (\nu_1,\nu_2,\nu_3)$ are
multi-indices, $\partial^\mu = (\partial_1)^{\mu_1} (\partial_2)^{\mu_2}
(\partial_3)^{\mu_3}$ and $M_\mu (x) = x^{\mu_1}_1   x^{\mu_2}_2
x^{\mu_3}_3$. The norms $\Vert\cdot\Vert^{}_{E^\rho_n}$ and $q^{}_n$
are equivalent (see Theorem 2.9). This shows in particular that
$D_\infty = S (\Rrm^3, \Crm^4)$.
Moreover, if $1/2 < \rho < 1$ and $(f, \dot{f}) \in M^\rho_\infty$,
then (see Theorem 2.12 and Theorem 2.13)
$$\eqalignno{
(1 + \vert x \vert)^{3/2 - \rho}   \vert f(x) \vert &\leq C \Vert (f,0)
\Vert_{M^\rho_1},& (1.6{\rm b})\cr
(1+ \vert x \vert)^{5/2 - \rho + \vert \nu \vert}   (\vert \partial^\nu
  \partial_i   f(x) \vert + \vert \partial^\nu   \dot{f}
(x) \vert) & \leq C_{\vert \nu \vert}
\Vert (f, \dot{f}) \Vert_{M^\rho_{\vert \nu \vert + 2}},
&(1.6{\rm c})\cr
}$$
for $\vert \nu \vert \geq 0$, $1 \leq i \leq 3$ and
if $f\in L^q$, $q=6/(3-2\rho)$,
$x^\alpha
\partial^\beta\partial_i f \in L^p$ and $x^\alpha
\partial^\beta \dot f \in L^p$, $p = 6/(5-2 \rho)$,
$\vert \alpha \vert \leq \vert \beta \vert \leq n$, $1 \leq i \leq 3$, then
$$\Vert (f, \dot{f}) \Vert_{M^\rho_n} \leq C_n   \sum_{0 \leq \vert \alpha
\vert \leq \vert \beta \vert \leq n}\Big(\sum_{0 \leq i \leq 3}
\Vert x^\alpha   \partial^\beta \partial_i f \Vert_{L^p} +
\Vert x^\alpha \partial^\beta
\dot{f} \Vert_{L^p}\Big). \eqno{(1.6{\rm d})}$$
In particular, it follows that, if
$$\eqalignno{
\vert \partial^\nu f(x) \vert &\leq C_{\vert \nu \vert}
(1+\vert x \vert)^{\rho - 3/2 - \vert \nu \vert - \varepsilon},& (1.6{\rm
e})\cr
\vert \partial^\nu \dot{f} (x) \vert &\leq C_{\vert \nu\vert}
(1 + \vert x \vert)^{\rho - 5/2 - \vert \nu\vert - \varepsilon},\cr
}$$
for each $\vert\nu \vert \geq 0$ and some $\varepsilon > 0$,
then $(f,\dot f)\in
M^\rho_\infty$,
$1/2 < \rho < 1$. Thus $M^\rho_\infty$ contains  long-range
potentials.

1.1.c {\it Nonlinear Poincar\'e covariance.}
A nonlinear representation (see Lemma 2.17) $T$ of $\p$ in $E^\rho_\infty$,
in the sense of \refFPS, is obtained by the fact that the M-D equations are
manifestly covariant:
$$T_X = T^1_X + T^2_X,\quad X \in \p, \eqno{(1.7)}$$
where $T^1$ is given by (1.5a)--(1.5d) and, for $u = (f, \dot{f}, \alpha)
\in  E^\rho_\infty$,
$$\eqalignno{
T^2_{P_0} (u) &= (0, \overline{\alpha}   \gamma   \alpha,
-i f_\mu \gamma^0\gamma^\mu  \alpha),
\quad \gamma = (\gamma_0, \gamma_1, \gamma_2, \gamma_3), &(1.8{\rm a})\cr
T^2_{P_i} &= 0\quad \hbox{for $1 \leq i \leq 3$},\quad
T^2_{M_{ij}} = 0 \quad\hbox{for
$1 \leq i < j \leq 3$},& (1.8{\rm b})\cr
(T^2_{M_{0j}} (u)) (x) &= x_j (T^2_{P_0} (u)) (x),
\quad 1 \leq j \leq 3, x = (x_1, x_2, x_3).& (1.8{\rm c})\cr
}$$
The gauge condition (1.1c) takes on initial data $u = (f, \dot{f}, \alpha)$
the form
$$ \dot{f}_0 = \sum_{1 \leq i \leq 3} \partial_i  f_i, \hskip.5cm
\Delta f_0 = \sum_{1 \leq i \leq 3}\partial_i \dot f_i - \vert
\alpha \vert^2.
\eqno (1.1{\rm c}') $$
The subspace (topological) $V^\rho_\infty$, $1/2 <\rho < 1$, of elements
in $E^\rho_\infty$ which satisfy these gauge conditions, is diffeomorphic
to $E^{\circ\rho}_\infty$ (see Theorem 6.11). The
problem to integrate globally the nonlinear Lie algebra representation $T$
therefore consists of proving the existence of an open neighbourhood
${\cal U}_\infty$ of zero in $V^\rho_\infty$ and a group action
$U\colon{\cal P}_0 \times {\cal U}_\infty \fl {\cal U}_\infty$,
which is $C^\infty$ and such that $U_g (0) = 0$ for
$g \in {\cal P}_0$ and $d\over {dt}$ $U_{\exp (tX)} (u) \mid_{t=0}
= T_X\hskip.1cm  u, X \in \p$.

Continuing to follow \refST, we extend the linear map $X \mapsto T_X$,
from $\p$ to the vector space of all differentiable maps from $E^\rho_\infty$
to $E^\rho_\infty$,
to the enveloping algebra $U (\p)$ by defining inductively: $T_{\un} = I$,
where $\un$ is the identity element in $U (\p)$, and
$$T_{YX} = D T_Y.T_X, \quad Y \in U (\p), X \in \p.
\eqno{(1.9)}$$
Here $(DA.B) (f)$ is the Fr\'echet derivative of $A$ at the point $f$ in the
direction $B(f)$.\footnote{$^{\rm 1)}$}{We also use the notation
$(D^nA)(f;f_1,\ldots,f_n)$
for the $n$-th derivative of $A$ at $f$\penalty-10000 in the directions
$f_1,\ldots,f_n$ and the notation $D^nA.(f_1,\ldots,f_n)$ for the
function\penalty-10000 $f\mapsto (D^nA)(f;f_1,\ldots,f_n)$.}
It was proved in \refST\ that this is a linear map from $U (\p)$
to the vector space of differentiable maps from $E^\rho_\infty$ to
$E^\rho_\infty$ (see formula (1.10) of \refST\ and the sequel).
For completeness we recall the proof in \refST. The vector field $T_X$,
$X\in \p$, defines a linear differential operator $d^{}_X$ of degree at most
one operating on the space $C^\infty (E^\rho_\infty)$ by $d^{}_X F= DF.T_X$,
$F\in C^\infty (E^{}_\infty)$. The fact that $X\mapsto T^{}_X$ is a nonlinear
representation of $\p$ on $E^\rho_\infty$ implies that $X\mapsto d^{}_X$
is a linear representation of $\p$ on linear differential operators of degree
at most one. This linear continuous representation has a canonical extension
$Y\mapsto d^{}_Y$ to $U(\p)$
 on linear differential operators of arbitrary order operating on
$C^\infty (E^\rho_\infty)$. If $e^{}_Y$, $Y\in U(\p)$, is the part of
$d^{}_Y$ of degree not higher than one, then $Y\mapsto e^{}_Y$ is
a linear map of $U(\p)$ into the space of linear partial
differential operators of degree at most one. Let $Y\in U(\p)$.
We write $Y=Z+a$, where $a\in\Crm\cdot\un$ and $Z$ has no component
on $\Crm\cdot\un$ (relative to the natural graduation of $U(\p)$).
Then the previous definition of $T^{}_Y$ gives  $e^{}_Y F=DF.T_Z+aF$,
which proves that $Y\mapsto T_Y$ is a linear map on $U(\p)$.
Moreover, it was proved in \refST\ that
$$\eqalignno{
{d \over dt}T_{Y (t)} (u(t)) &= T_{P_0 Y(t)} (u(t)),
\quad Y \in U (\p,) &(1.10)\cr
\noalign{\hbox{where}}
Y(t) &= \exp (t\, {\rm ad}_{P_0})Y,\quad {\rm ad}_{P_0}Z=[P_0,Z], &(1.11)\cr
\noalign{\hbox{if}}
{d \over dt}u(t) &= T_{P_0} (u(t)). &(1.12)\cr
}$$
We note that definition (1.11) makes sense since $({\rm ad}_{P_0})^n$,
$n\geq0$, is a linear map from $U(\p)$ to $U(\p)$ leaving invariant the
subspace of elements of degree at most $l$, $l\geq0$, in $U(\p)$ and since
then $\exp (t\, {\rm ad}_{P_0})Y =\sum_{n\geq0}(n!)^{-1}(t\,
{\rm ad}_{P_0})^n Y$ converges absolutely.
For completeness we also recall the proof of (1.10). Since
${d\over dt} Y(t) ={\rm ad}_{P_0} Y(t)$, it follows from (1.9) and (1.12),
that
$$\eqalignno{
{d \over dt}T_{Y (t)} (u(t)) &= T_{{d\over dt}Y(t)}(u(t))
+(DT_{Y (t)}.T_{P_0})(u(t))\cr
&=T_{[P_0,Y(t)]}(u(t)) + T_{Y(t)P_0}(u(t))=T_{P_0Y(t)}.\cr
}$$

$T^n_Y$, $Y \in U (\p)$, denotes the $n$-homogeneous part of $T_Y$
and we shall identify $T^n_Y$ with a \hbox{$n$-linear} symmetric map and
also with a continuous linear map from  $\hat{\otimes}^n   E^\rho_\infty$
into $E^\rho_\infty$, where $\hat{\otimes}^n$ is the $n$-fold complete
tensor product endowed with the projective tensor product topology.
$T^M_Y (u)$, $T^{Mn}_Y (u)$ and $U^M_g(u)$
(resp. $T^D_Y (u)$, $T^{Dn}_Y
(u)$ and $U^D_g(u)$) is the projection of $T_Y (u)$, $T^n_Y (u)$ and
$U^{}_g(u)$ on $M^\rho$ (resp. $D$). We also define $\tilde{T}_Y$,
$Y \in U (\p)$ by  $$T_Y = T^1_Y + \tilde{T}_Y. \eqno{(1.13)}$$
Since it does not bring any contradiction, we shall also denote by
$T^{M1}$ or $T^{1M}$  (resp. $T^{D1}$ or $T^{1D}$) the linear representation
of $\p$ which is the restriction of the linear representation $T^1$ to
$M^\rho$ (resp. $D$). Similarly,
$U^{M1}_g$ or $U^{1M}_g$ (resp. $U^{D1}_g$ or $U^{1D}_g$) denotes the
restriction of $U^1_g$ to $M^\rho$ (resp. $D$).

1.2 {\sc The Infrared Problem}

On the classical level the infrared problem consists of determining to which
extent the long-range interaction created by the coupling $A^\mu  j_\mu$
between the electromagnetic potential $A_\mu$ and the current
$j_\mu = \overline{\psi} \gamma_\mu \psi$ is an obstruction for the separation,
when $\vert t \vert \fl \infty$, of the nonlinear relativistic system into
two asymptotic isolated relativistic systems, one for the electromagnetic
potential $A_\mu$ and one for the Dirac field $\psi$.
It will be proved here that there is
such an obstruction, which in particular implies that {\it asymptotic
in and out states do not transform according to a linear representation of
the Poincar\'e group}. This constitutes a serious problem for the second
quantization of the asymptotic (in and out going) fields, since the particle
interpretation usually requires free relativistic
fields, i.e. at least a linear representation of the Poincar\'e
 group ($U^1$ in our case).
Therefore we have to introduce nonlinear representations $U^{(-)}$ and
$U^{(+)}$ of ${\cal P}_0$, in the sense of \refFPS, which can give the
transformation of the asymptotic in and out states under ${\cal P}_0$ and
which can permit a particle interpretation.

There are two reasons which
permit to determine the class of asymptotic representations
$U^{(\varepsilon)}$, $\varepsilon=\pm$.
First, the classical observables, $4$-current density, $4$-momentum
and $4$-angular momemtum are invariant under gauge transformations.
Second, if the evolution equations become linear after a gauge transformation
one can use the freedom of gauge in the second quantization of the fields.
It is therefore reasonable to postulate that the asymptotic representations
$g \mapsto U^{(\varepsilon)}_g$ are linear modulo a nonlinear gauge
transformation depending on $g$ and respecting the Lorentz gauge condition.
We shall make precise the meaning of this statement at the end of this
introduction.

Moreover to determine the class of admissible modified
wave operators it is reasonable to postulate that solutions of the M-D
equations (1.1a)--(1.1c) should converge when  $t\fl \pm\infty$ in
$E^\rho$ to free solutions (i.e. solutions of equations (1.1a)--(1.1c) but
with vanishing right-hand side) modulo a gauge transformation, not even
respecting the Lorentz gauge condition; such transformations are admissible
since they leave invariant the observables.

In mathematical terms the infrared problem of the M-D equations then consists
of determining two diffeomorphisms
$\Omega_\varepsilon\colon{\cal O}^{(\varepsilon)}_\infty \fl {\cal U}_\infty$,
$\varepsilon = \pm$, the modified wave operators,
where ${\cal O}^{(\varepsilon)}_\infty$ is
an open neighbourhood of zero in $E^{\circ\rho}_\infty$ and where
${\cal U}_\infty$ is a neighbourhood of zero in $V^\rho_\infty$, satisfying
$$U^{(\varepsilon)}_g = \Omega^{-1}_\varepsilon   \circ   U_g
\circ   \Omega_\varepsilon, \quad  g \in {\cal P}_0,   \varepsilon =
\pm, \eqno{(1.14)}$$
where the asymptotic representations are $C^\infty$ functions
$U^{(\varepsilon)}\colon{\cal P}_0
\times E^{\circ \rho}_\infty \fl E^{\circ \rho}_\infty$.
In order to satisfy the two preceding postulates, we impose supplementary
conditions on $U^{(\varepsilon)}$ and $\Omega_\varepsilon$, which we shall
justify at the end of this introduction.
Let the Fourier transformation $f \mapsto \hat{f}$ be defined by
$$\hat{f} (k) = (2 \pi)^{- 3/2}   \int_{\Rrm^3}   e^{-ikx}
f(x)   dx. \eqno{(1.15)}$$
The orthogonal projections $P_\varepsilon (-i \partial)$ in $D$ on initial
data with energy sign $\varepsilon$, $\varepsilon = \pm$, for the Dirac
equation are given by:
$$(P_\varepsilon(-i\partial)\alpha)^\wedge (k)=
P_\varepsilon (k)\hat\alpha(k) = {1 \over 2} \Big(I + \varepsilon
\Big(- \sum^3_{j=1}   \gamma^0
  \gamma^j   k_j + m \gamma^0\Big) \omega(k)^{-1}\Big)\hat\alpha(k),
  \eqno{(1.16)}$$
where $\omega(k) = (m^2 + \vert k \vert^2)^{1/2}$.
We  postulate that the asymptotic representations have the following
form:
$$\eqalignno{
U^{(+)}_g (u) &= (U^{(+)M}_g (u),U^{(+)D}_g (u)),\quad U^{(+)M} = U^{1M},
&(1.17{\rm a})\cr
(U^{(+)D}_g (u))^\wedge (k) &= \sum_{\varepsilon = \pm}   e^{i \varphi_g (u, -
\varepsilon k)} P_\varepsilon (k)   (U^{1D}_g \alpha)^\wedge (k),
\quad g \in {\cal P}_0,  &(1.17{\rm b})\cr  }$$
where $u = (f, \dot{f}, \alpha) \in E^{\circ\rho}_\infty$, the function
$(g,u,k)
\mapsto  \varphi_g (u, - \varepsilon k)$ from ${\cal P}_0
\times E^{\circ\rho}_\infty \times \Rrm^3$
to $\Rrm$ is $C^\infty$ and if $(h_g (u)) (t,x)
= \varphi_g (u , mx (t^2 - \vert x \vert^2)^{-1/2})$,
$t > 0$, $\vert x \vert < t$, then $\carre h_g (u) = 0$.
Moreover growth conditions on $\varphi_g$ should be satisfied,
so that the main contribution to the phase and its derivatives
in $g$ and $k$ comes from $U^{1D}_g$, the restriction of $U^1$ to $D$.
Finally we impose the asymptotic condition
$$\eqalignno{
&\Vert U^M_{\exp (t P_0)} (\Omega_+ (u)) - U^{1M}_{\exp (t P_0)}
(f, \dot{f}) \Vert^{}_{M^\rho}& (1.17{\rm c})\cr
&\qquad{}+ \Vert U^D_{\exp (t P_0)} (\Omega_+ (u)) - \sum_{\varepsilon =
\pm}e^{is^{(+)}_\varepsilon (u,t,-i \partial)} P_\varepsilon (-i \partial)
U^{1D}_{ \exp (t P_0)} \alpha \Vert^{}_D \fl 0,\cr }$$
when $t \fl \infty$, where $u = (f, \dot{f}, \alpha) \in {\cal U}_\infty$,
$(u,t,k) \mapsto {s}^{(+)}_\varepsilon (u,t,k)$ is a $C^\infty$ function
from ${\cal U}_\infty \times \Rrm \times \Rrm^3$ to $\Rrm$,
${s}^{(+)}_\varepsilon (u,t,-i \partial)$ is the operator
defined by inverse Fourier transform of the multiplication
operator $k \mapsto {s}^{(+)}_\varepsilon (u,t,k)$. Moreover
${s}^{(+)}_\varepsilon (u,t,k)$ should satisfy growth conditions so that
its  major contribution comes from $U^{1D}_{\exp (t P_0)}(u)$.
There are similar conditions for $U^{(-)}$, $\Omega_-$ and
${s}^{(-)}_\varepsilon$, $\varepsilon = \pm$. We note that if
${s}^{(+)}_\varepsilon$ is determined, then $\Omega_{+}$ is determined and
so is $U^{(+)}.$
It was proved in \refFST, that on a set of asymptotic states
$(f, \dot{f}, \alpha)$ such that
$\hat{f}, \fhd \in C^\infty_0 (\Rrm^3- \{ 0 \})$ and $\hat{\alpha}
\in C^\infty_0 (\Rrm^3)$\footnote{$^{\rm 2)}$}{$C^k_0(X)$, $k\geq0$,
denotes the space of $k$ times continuously differentiable functions
on $X$ with compact support.},  it is possible to choose (formula (3.33a)
of \refFST) $${s}^{(+)}_\varepsilon (u,t,k) = - \vartheta \big(A^{(+)} (u),
(t,-\varepsilon tk/\omega(k))\big), \eqno{(1.18)}$$
where $A^{(+)} (u)$ is a certain approximate solution of the M-D equations
absorbing the long-range part of $A$ for a solution $(A, \psi)$,
and where
$$\vartheta (H, y) = \int_{L(y)} H_\mu (z)   d z^\mu,
\quad y \in \Rrm^4,\eqno{(1.19)}$$
where $H\colon \Rrm^4\fl\Rrm^4$ and $L(y) = \{ z \in \Rrm^4 \big\vert z = sy,
0 \leq s \leq 1 \}$.
The function $(t,k) \mapsto {s}^{(+)}_\varepsilon
(u,t,k)$ was determined by the fact, that has been proved in \refFST,
that $S_\varepsilon (u,t,k)
= \varepsilon   \omega(k) t + {s}^{(+)}_\varepsilon (u,t,k)$ has to be in a
certain  sense an approximate solution of the Hamilton-Jacobi equation for
a relativistic electron in an external electromagnetic potential:
$$\eqalignno{
&\big({\partial \over \partial t}{\cal S}_\varepsilon (u,t,k) +
A_0 (t, - \nabla_k {\cal S}_\varepsilon (u,t,k))\big)^2& (1.20{\rm a})\cr
&\qquad{}- \sum^3_{i=1}   \big(k_i + A_i (t, - \nabla_k
{\cal S}_\varepsilon (u,t,k))\big)^2  = m^2.\cr  }$$
We proved and used the fact that
$$y^\mu   \partial_\mu  \vartheta (H,y) = y^\mu  H_\mu (y),
\eqno{(1.20{\rm b})}$$
to establish that $S_\varepsilon (u,t,k)$ is an approximate solution
of the Hamilton-Jacobi equation.

1.3 {\sc Presentation of the main results}

1.3.a  {\it Some notations}.
Let $u_+ = (f, \dot{f}, \alpha) \in E^{\circ\rho}_\infty$ and let
(we do not write explicitely the multiplication sign $\times$ in formulas
extending over more than one line at cut at a multiplication)
$$\eqalignno{
J^{(+)}_\mu (t,x) &= (m/t)^3 \big(\omega(q(t,x)) / m\big)^5& (1.21)\cr
&\qquad{}\sum_{\varepsilon = \pm}  ((P_\varepsilon (-i \partial)
\alpha)^\wedge (\varepsilon  q (t,x) / m))^+ \gamma^0  \gamma_\mu
((P_\varepsilon (-i \partial)) \alpha)^\wedge (\varepsilon q (t,x) / m),\cr
}$$
for $t > 0$,   $\vert x \vert < t$, where $q (t,x) = - mx / (t^2 -
\vert x \vert^2)^{1/2}$ and, for $\vert x \vert > 0$, $0 \leq t \leq
\vert x \vert$, let $J^{(+)}_\mu (t,x) = 0$.
Since $\alpha \in S (\Rrm^3, \Crm^4)$, it follows that $J^{(+)}_\mu
\in C^\infty ((\Rrm^+ \times \Rrm^3) - \{ 0 \})$ and that its support is
contained in the forward light cone. We choose $\chi \in C^\infty (\Rrm)$,
$\chi (\tau) = 0$ for $\tau \leq 1$, $\chi (\tau) = 1$
for $\tau \geq 2$, $0 \leq \chi (\tau) \leq 1$ for $\tau \in \Rrm$,
$\chi^{}_0 \in C^\infty ([0,\infty[)$, $\chi^{}_0(\tau)=1$ for
$\tau \geq 2$, $0 \leq \chi^{}_0 (\tau) \leq 1$ for $\tau \in [0,\infty[$,
and introduce $A^{(+)} =
A^{(+)1} + A^{(+)2}$ by
$$\left.\matrix{
\hfill(A^{(+)1}(t))(x)
&= \chi^{}_0 ((t^2-\vert x\vert^2)^{1/2})(B^{(+)1}(t))(x)\hfill\cr
\hfill(A^{(+)2} (t)) (x)
&= \chi ((t^2 - \vert x \vert^2)^{1/2})(B^{(+)2} (t)) (x),\hfill\cr
}\right\} \quad {\rm for}\ t\geq \vert x\vert,$$
 and \hfill    (1.22{\rm a})
$$\left.\matrix{
\hfill(A^{(+)1}(t))(x) &= \chi^{}_0(0)(B^{(+)1}(t))(x),\hfill\cr
\hfill(A^{(+)2} (t)) (x) &= 0,\hfill\cr
}\right\} \quad {\rm for}\ t< \vert x\vert,$$
where
$$\eqalignno{
B^{(+)1}_\mu (t) &= \cos (\vert \nabla \vert t) f_\mu + \vert \nabla
\vert^{-1} \sin (\vert \nabla \vert t)   \dot{f}_\mu,\quad   t
\in \Rrm,& (1.22{\rm b})\cr
B^{(+)2}_\mu (t) &= - \int^\infty_t   \vert \nabla \vert^{-1}\sin (\vert
\nabla \vert (t-s)) J^{(+)}_\mu (s, \cdot) ds, \quad  t > 0. &(1.22{\rm c})\cr
}$$
The cut-off function $\chi$ has been introduced to exclude in a Lorentz
invariant way the points $(0,x)$, $x \in \Rrm^3$ from the support of
$A^{(+)2}$. $U^{(+)}$ is defined by formulas
(1.17a) and (1.17b) and by $\varphi_g = 0$ for $g \in SL (2, \Crm)$ and
$$\varphi_g (u, - \varepsilon k) = \vartheta^\infty \big((\chi^{}_0\circ\rho)
(B^{(+)1} (\cdot + a) - B^{(+)1} (\cdot)),
(\omega(k), - \varepsilon k)\big), \eqno{(1.23a)}$$
for $g = \exp (a^\mu   P_\mu)$, $k \in \Rrm^3$, where
$$ \vartheta^\infty (H,y) = \int^\infty_0   y^\mu   H_\mu (sy)
ds,\quad   y \in \Rrm^4, \hskip.2cm {\rm and} \hskip.2cm 
\rho(t,x) = (t^2-\vert x\vert)^{1/2}. \eqno{(1.23{\rm b})}$$ 
We have made the identification convention that
$B^{(+)1} (t,x)=(B^{(+)1} (t))(x)$. We note that the function
$y \mapsto \vartheta^\infty (H,y)$ is homogeneous of degree zero, so we could
also have taken the argument $(1, - \varepsilon k / \omega (k))$
instead of $(\omega (k), - \varepsilon k)$,
which corresponds to the choice in \refFSTA. The phase function
${s}^{(+)}_\varepsilon$ in (1.17c), which determines $\Omega_+$,
is defined by formula (1.18) and the choice of
$A^{(+)}=A^{(+)1}+A^{(+)2}$ given by (1.22a).
With the choice $\chi^{}_0 =1$ the function $h^{}_g$ introduced after (1.17b)
is given by
$(h^{}_g(u))(y)=\vartheta^\infty(B^{(+)1}(\cdot + a) -B^{(+)1}(a),y)$
and satisfies $\carre h^{}_g(u)=0$ if $(f,\dot f)\in E^{\circ\rho}$.

1.3.b {\it Statements.} We state the main results of this article for the case
where $t \fl+ \infty$. There are analog results for $t \fl - \infty$.
\saut{\it
\noindent{\bf Theorem I.} Let $1/2 < \rho < 1$. If $n \geq 4$
then $U^{(+)}\colon {\cal P}_0 \times E^{\circ\rho}_n \fl E^{\circ\rho}_n$
is a continuous nonlinear
representation of ${\cal P}_0$ in $E^{\circ\rho}_n$ and, in addition,
the function $U^{(+)}\colon
{\cal P}_0 \times E^{\circ\rho}_\infty \fl E^{\circ\rho}_\infty$ is $C^\infty$.
 Moreover $U^{(+)}$ is not equivalent by a $C^2$ map to a linear
 representation on $E^{\circ\rho}_\infty$: If $U^{(+)}_1$ and $U^{(+)}_2$
 are defined via (1.23a) by two
choices of the function $\chi^{}_0$, they are equivalent.
}
\saut
\noindent Theorem I partially sums up Theorems 3.12--3.14.
\saut{\it
\noindent{\bf Theorem II.} Let $1/2 < \rho < 1$. There exist an
open neighbourhood ${\cal U}_\infty$
(resp. ${\cal O}^{(+)}_\infty$) of zero in $V^\rho_\infty$
(resp. $E^{\circ\rho}_\infty$),
a diffeomorphism $\Omega_+ \colon{\cal O}^{(+)}_\infty \fl {\cal U}_\infty$
and a $C^\infty$ function
$U\colon {\cal P}_0 \times {\cal U}_\infty \fl {\cal U}_\infty$,
defining a nonlinear representation
of ${\cal P}_0$, such that:
$$\eqalignno{
\hbox{\rm i)}&\ {d \over dt}   U_{\exp (tX)} (u) = T_X (U_{\exp (tX)} (u)),
\quad X \in \p,
  t \in \Rrm,   u \in {\cal U}_\infty,\hskip43mm\cr
\hbox{\rm ii)}&\ \Omega_+ \circ U^{(+)}_g = U_g \circ \Omega_+\cr
\hbox{\rm iii)}&\ \lim_{t \fl \infty}   \Big(\Vert U^M_{\exp (t P_0)}
(\Omega_+ (u)) -
U^{M1}_{\exp (t P_0)}   (f, \dot{f}) \Vert^{}_{M^\rho}\cr
&\qquad\qquad{}+ \Vert U^D_{\exp (t P_0)} (\Omega_+ (u)) -
\sum_{\varepsilon = \pm} e^{i {s}^{(+)}_\varepsilon (u, t, -i \partial)}
P_\varepsilon (-i \partial) U^{D1}_{\exp (t P_0)} \alpha \Vert^{}_D\Big) = 0,
\cr }$$
for $u = (f, \dot{f}, \alpha) \in {\cal O}^{(+)}_\infty$.
}\saut
This theorem (see Theorem 6.19) solves in particular {\it the Cauchy
problem} for small initial data and proves {\it asymptotic
completeness. }
By the construction of the wave operator $\Omega_+$ in chapter 6, the solution
$(A(t,\cdot),\dot A(t,\cdot),\psi(t,\cdot))=U_{\exp(tP_0)}(u)$
of the Cauchy problem satisfies
$$\sup_{\scr x\in \Rrm^3\atop\scr t\geq 0}
\Big((1+\vert x\vert +t)^{3/2-\rho}\vert A_\mu (t,x)\vert+
(1+\vert x\vert +t)\vert \partial_\nu A_\mu (t,x)\vert+
(1+\vert x\vert +t)^{3/2}\vert\psi(t,x)\vert\Big)<\infty.$$

1.3.c  {\it Cohomology.} These results and conditions (1.14) and (1.17c)
have a natural {\it cohomological interpretation.}
We only consider the case where $t \fl \infty$, and since
the representations $U^{(+)}$ defined for different $\chi^{}_0$ via (1.23a) are
equivalent, we only consider the case where $\chi^{}_0=1$.
A necessary condition for $U^{(+)}$ and $\Omega_+$ to be a solution of equation
(1.14) is that the formal power series development of $U^{(+)}_g$,
$U_g$ and $\Omega_+$ in the initial conditions satisfy the cohomological
equations defined in \refFPS and \refFS. In particular (after trivial
transformations) the second order terms $U^{(+)2}_g$, $U^2_g$ and
$\Omega^2_+$ must satisfy  $$\delta   \Omega^2_+ =
C^2,\quad  \delta R^{(+)2} = 0, \eqno{(1.24)}$$
where $\delta$ is the coboundary operator defined by the representation
$(g,A) \mapsto U^1_g    A^2 (\otimes^2   U_{g-1})$ of the Poincar\'e group
${\cal P}_0$ on
bilinear symmetric maps $A^2$ from $E^\rho_\infty \fl E^\rho_\infty$ and where
$C^2 = R^{(+)2} - R^2$
is the cocycle defined by $R^2_g = U^2_g (U^1_{g-1} \otimes U^1_{g-1})$ and
$R^{(+)2}_g =
U^{(+)2}_g (U^1_{g-1} \otimes U^1_{g-1})$, $g \in {\cal P}_0$. Equation (1.24)
shows that the cochain $R^{(+)2}$ has to be a cocycle and then that the cocycle
$C^2$ has to be a  coboundary. This is equivalent to the existence of a
solution $U^{(+)}$, $\Omega_+$ of equation (1.14) modulo terms of order at
least three. There are similar equations for higher order  terms:
$$\delta \Omega^n_+ = C^n, \quad   \delta R^{(+)n} = 0, \eqno{(1.25)}$$
where $C^n$ and $R^{(+)n}$ are functions of $\Omega^2_+,\ldots,\Omega^{n-1}_+$
and  $U^{(+)2},\ldots,U^{(+)n}$.

In a previous article \refFST, we proved that
there exist a modified wave operator and global solutions of the M-D equations
for a set of scattering data $(f, \dot{f}, \alpha)$,
which is a subset of the spaces $E^{\circ\rho}_\infty$ introduced in the
present paper, satisfying $\hat{f}_\mu (k) = \fhdm (k) = 0$ for $k$ in a
neighbourhood of zero, i.e. $f_\mu$ and $\dot{f}_\mu$ have no low frequencies
component. It follows from that paper that the usual wave operator
(i.e. $U^{(+)} = U^1, {s}^{(+)}_\varepsilon=0$
in (1.17c)) does not exist, even in this case where there are no low
frequencies. In fact there is an obstruction to the existence of such a
solution of equation (1.14) for $n=3$  due to the self-coupling of $\psi$
with the electromagnetic potential created by the current
$\overline{\psi}   \gamma_\mu   \psi$. However, as was proved in \refFST\
(see also Lemma 3.2 of the present paper) there exists a modified wave operator
satisfying (1.25) for $n \geq 2$, with $U^{(+)} = U^1$. Moreover, as we have
already pointed out, the phase function ${s}^{(+)}_\varepsilon$ is given by
formula (1.18) (see formulas (1.5), (3.28) and (3.33) of \refFST).

Thus in the absence of low frequencies in the scattering data, for
the electromagnetic potential, we can choose $\Omega_+$ such that it
intertwines the linear representation  $U^1$ and the nonlinear representation
$U$ of ${\cal P}_0$. Now if $(f, \dot{f}) \in M^{\circ\rho}_\infty$,
$1/2 < \rho < 1$, have nontrivial low frequency part, as is the case for
the Coulomb  potential $(\hat{f}_\mu (k) \simeq \vert k \vert^{-2})$
and which is necessary to assume in order to have asymptotic completeness,
then there is a cohomological obstruction already for $n=2$ if we want to
obtain $U^{(+)2} = 0$. The essential point now is that the cocycle $R^2$ can
be split into a trivializable part $- C^2$ (a coboundary) and a
nontrivializable part $R^{(+)2}$, which defines $U^{(+)2}$ and therefore
the whole representation $U^{(+)}$. We shall call this nontrivial part
the {\it infrared cocycle}. According to the definition of $R^{(+)2}$
and formula (1.23a) it follows that $R^{(+)2}_g = 0$ for $g \in SL (2, \Crm)$,
$R^{(+)2}_g = (R^{(+)2M}_g, R^{(+)2D}_g)$, $R^{(+)2M} = 0$ and
$$\eqalignno{
&(R^{(+)2D}_g (u_1 \otimes u_2))^\wedge (k)&(1.26)\cr
&\qquad{}= {i \over 2}\sum_{\varepsilon = \pm}
\big(\vartheta^\infty (B^{(+)1}_1 (\cdot)
- B^{(+)1}_1 (\cdot - a), (\omega(k), - \varepsilon k)\big)
  P_\varepsilon (k)   \hat{\alpha}_2 (k) \cr
&\qquad\qquad{}+ \vartheta^\infty \big(B^{(+)1}_2 (\cdot)
- B^{(+)1}_2 (\cdot - a),
(\omega(k), - \varepsilon k)\big)
  P_\varepsilon (k)   \hat{\alpha}_1 (k)),\cr
}$$
for $g = \exp (a^\mu   P_\mu)$, $u_i = (f_i, \dot{f}_i, \alpha_i) \in
E^{\circ \rho}_\infty$,   $i \in \{ 1,2 \}$, and where $B^{(+)1}_{i \mu}$
is the corresponding free field given  by (1.22b).

When the limit $\vartheta(B^{(+)1}, (t, - t \varepsilon k / \omega(k))
\fl \vartheta^\infty (B^{(+)1},
(\omega(k), - \varepsilon k))$ exists for $t \fl \infty$, then the
infrared cocycle is in fact a
coboundary. In particular, this is the case when $\hat{f}_\mu,\fhdm$
are equal to zero in a neighbourhood of zero.

1.4 {\sc Physical Remarks}

1.4.a  {\it Physical motivation of the asymptotic condition.}
Next, we return to the physical justification of conditions
(1.17a)--(1.17c), and to be more specific, as this is the main case of
this paper, we choose $s^{(+)}_\varepsilon$ to be given by (1.18) and
$A^{(+)}$ to be given by (1.22a)--(1.22c) with $\chi^{}_0=1$. According to
statement iv) of Theorem 6.16 and Theorem 6.19, the asymptotic condition in
statement iii) of Theorem~II is equivalent to
$$\Vert (A(t),\dot A(t)) -(A_L(t),\dot{A}_L(t)) \Vert^{}_{M^\rho}
+\Vert\psi(t) -(e^{-i\varphi}\psi^{}_L)(t) \Vert^{}_{D} \fl 0,
\eqno{(1.27)}
$$
when $t\fl\infty$, where $(A(t),\dot A(t),\psi(t))=U_{\exp(tP_0)}
(\Omega_+(u))$, $(A_L(t),\dot{A}_L(t),\psi^{}_L(t))=$\penalty-10000
$U^1_{\exp(tP_0)}u$
and where for a fixed $u=(f,\dot f,\alpha)\in {\cal O}^{(+)}_\infty$,
$(t,x)\mapsto \varphi (t,x)$, $(t,x)\in \Rrm^+\times\Rrm^3$ is a
$C^\infty$ function given in Theorem 6.16. When needed we shall write
$(u,(t,x))\mapsto\varphi(u,(t,x))$ to stress the dependence of
$\varphi$ on $u$. It follows from Theorem A.1 that $\varphi$
in (1.27) can be replaced by $\vartheta (A^{(+)},\cdot)$, but for
technical reasons we keep $\varphi$. In fact, from a heuristic point of view,
since $\vartheta (A^{(+)},(t,x))=
-s^{}_\varepsilon(u,t,-\varepsilon x/(t^2 - \vert x\vert^2)^{1/2})$,
(1.27) with $\varphi$ replaced by $\vartheta (A^{(+)},\cdot)$ is obtained
from the leading term in the stationary phase expansion
 of $\sum_\varepsilon e^{is^{}_\varepsilon(u,t,-i\partial)}
U^{1D}_{\exp(t P_0)}\alpha$.

The energy-momentum tensor for the M--D system is, in one of its forms,
given
by
$$
t^{\mu\nu}=-F^{\mu\alpha}F^{\nu}{}_\alpha+ {1\over 4}
g^{\mu\nu}F_{\alpha\beta}F^{\alpha\beta}
+{1\over2}(\overline{\psi}\gamma^\mu (i\partial^\nu -A^\nu)\psi
+(\overline{(i\partial^\nu -A^\nu)\psi})\gamma^\mu\psi),
\eqno{(1.28)}
$$
where $0\leq \mu \leq 3$, $0\leq \nu \leq 3$ and $F_{\mu\nu}=\partial_\mu
A_\nu -\partial_\nu A_\mu$. The current density vector is given
by
$$j^\mu =\overline{\psi}\gamma^\mu \psi, \quad 0\leq \mu \leq 3.
\eqno{(1.29)}
$$
$t^{\mu\nu}$ and  $j^\mu$ are invariant under gauge transformations:
$\psi'=e^{i\lambda}\psi$, $A'_\mu=A_\mu-\partial _\mu\lambda$
(not necessarily respecting the Lorentz gauge condition
that gives here $\carre\lambda=0$).
We also introduce the energy-momentum tensor and
the current density vector for the system of free fields $A_L$
and $\psi^{}_L$ by
$$
t^{\mu\nu}_L=-F^{\mu\alpha}_LF^{\nu}_L{}_\alpha+ {1\over 4}
g^{\mu\nu}F_{L \alpha\beta}F^{\alpha\beta}_L
+{1\over2}(\overline{\psi}_L\gamma^\mu (i\partial^\nu)\psi^{}_L
+(\overline{(i\partial^\nu)\psi^{}_L})\gamma^\mu\psi^{}_L),
\eqno{(1.30)}$$
and
$$j^\mu_L =\overline{\psi}_L\gamma^\mu \psi^{}_L,
\eqno{(1.31)}
$$
where $F_{L\mu\nu}=\partial_\mu A_{L\nu} -\partial_\nu A_{L\mu}$.
Since $t^{\mu\nu}$ can be written as
$$
t^{\mu\nu}=-F^{\mu\alpha}F^{\nu}{}_\alpha+ {1\over 4}
g^{\mu\nu}F_{\alpha\beta}F^{\alpha\beta}
+{1\over2}(\overline{\psi}'\gamma^\mu (i\partial^\nu -A'^\nu)\psi'
+(\overline{(i\partial^\nu -A'^\nu)\psi})\gamma^\mu\psi'),
$$
where $\psi'=e^{i\varphi}\psi$, $A'_\mu=A_\mu-\partial _\mu\varphi$
and $F_{\mu\nu}=\partial_\mu A_\nu -\partial_\nu A_\mu$, it follows
from statement iii) of Theorem 6.16 that
$$\lim_{t\fl\infty}\int_{\Rrm^3} \vert t^{\mu\nu}(t,x) -t^{\mu\nu}_L (t,x)
\vert\, dx = \lim_{t\fl\infty}\Vert A'(t)\Vert^{}_{L^\infty}
\Vert\psi^{}_L(t)\Vert^{2}_{D}.
$$
By unitarity of $U^{1D}$, $\Vert\psi^{}_L(t)\Vert^{2}_{D}=
\Vert\alpha\Vert^{2}_{D}$.
Using statement iii) of Theorem 6.16 for $\Vert A(t)\Vert^{}_{L^\infty}$
and using statement ii) of Lemma 4.4  for $\partial_\mu\vartheta
(A^{(+)1},\cdot)$ and estimate (6.246) for
$\partial_\mu(\varphi -\vartheta(A^{(+)1},\cdot))$, it follows that
$\Vert A'(t)\Vert^{}_{L^\infty}\fl0$, when $t\fl\infty$. Therefore
$$\lim_{t\fl\infty}\int_{\Rrm^3} \vert t^{\mu\nu}
(t,x) -t^{\mu\nu}_L (t,x) \vert\, dx = 0.  \eqno{(1.32)}
$$
Similarly
$$\lim_{t\fl\infty}\int_{\Rrm^3} \vert j^{\mu}(t,x) -j^{\mu}_L (t,x)
\vert\, dx = 0. \eqno{(1.33)}  $$
Limits (1.32) and (1.33) show that, as far as one measures energy-momentum and
current, the solutions of the M-D system are asymptotically
indistinguishable from free solutions because of gauge invariance.
Condition (1.17c) is therefore natural. A similar discussion can be made
for the angular momentum tensor; however it is technically more involved,
so we omit it.

1.4.b {\it Gauge transformations.}
A gauge transformation $\psi'=e^{i\lambda}\psi$, $A'_\mu=A_\mu-\partial_\mu
\lambda$, respecting the Lorentz gauge condition, i.e. $\carre \lambda =0$,
transforms a solution $(A,\psi)$ of the M--D equations (1.1a)--(1.1c)
into a solution $(A',\psi')$ of the M--D equations.
Let $v \in {\cal U}_\infty$. $v$ is the initial data for the solution
$(A,\psi)$ of the M--D equations, and if $\lambda$
is sufficiently small and regular, the initial data $v'$ for the
solution $(A',\psi')$ is in ${\cal U}_\infty$,
since ${\cal U}_\infty$ is open.
Let $u=\Omega_+^{-1}(v)$ and $u'=\Omega_+^{-1}(v')$.
Since $\lambda(y)= \lambda(0)+\vartheta(\Lambda,y)$,
with $\Lambda_\mu = \partial_\mu \lambda$, it follows from (1.27) that
$$\Vert (A'(t),\dot{A}'(t)) -(A'_L(t),\dot{A}'_L(t)) \Vert^{}_{M^\rho}
+\Vert\psi'(t) -e^{-i\varphi(u',(t,\cdot))}\psi'^{}_L(t) \Vert^{}_{D} \fl 0,
\eqno{(1.34)}
$$
when $t\fl\infty$, where $(A'_L(t),\dot{A}'_L(t),\psi'^{}_L(t))=
U^1_{\exp(tP_0)}u'$ and $u'=(f',\dot{f}',\alpha')$,
$f'^{}_\mu(x)=f^{}_\mu(x)-(\partial_\mu\lambda)(0,x)$,
$\dot f'^{}_\mu(x)=\dot f^{}_\mu(x)-(\partial_\mu\partial_0\lambda)(0,x)$,
$\alpha'=e^{i\lambda (0)}\alpha$.
Since we can choose the constant $\lambda (0)$ arbitrarily, it follows that
the admissible gauge transformations $u \mapsto u'$ of the scattering data
 are given by
$$A'_{L\mu}=A_{L\mu}-\partial_\mu\lambda,\quad \psi'^{}_L=e^{ic}\psi^{}_L,
\eqno{(1.35)}
$$
where $c\in\Rrm$ and $\lambda$ is such that $\carre \lambda =0$, and such
that $(\Lambda (0,\cdot),\dot\Lambda (0,\cdot))\in E^{\circ\rho}_\infty$,
$\Lambda_\mu(t,x)=(\partial_\mu\lambda)(t,x)$,
$\dot\Lambda_\mu(t,x)=(\partial_0\partial_\mu\lambda)(t,x)$.

We now introduce the notion of {\it gauge-projective map}. Let
$Q\colon E^{\circ\rho}_\infty\fl E^{\circ\rho}_\infty$ be a
$C^\infty$ map leaving ${\cal O}^{(+)}_\infty$ invariant. More general
situations are possible, but to fix the ideas we make this hypothesis.
If there exists a gauge transformation $G$ of the form (1.35) such that
$$\eqalignno{
&\Vert U^M_{\exp (tP_0)}\big(\Omega_+ (Q(u))\big)
-U^{1M}_{\exp (tP_0)}(f,\dot f)\Vert^{}_{M^\rho}&(1.36)\cr
&\quad{}+ \Vert U^D_{\exp (tP_0)}\big(\Omega_+ (Q(u))\big)
-\sum_\varepsilon e^{is^{(+)}_\varepsilon(G(u),t)}
P_\varepsilon(-i\partial)U^{1D}_{\exp (tP_0)}\alpha\Vert^{}_D \fl 0,\cr
}$$
where $u=(f,\dot f,\alpha)$, then we say that $Q$ is a gauge-projective
map.

To show that the {\it asymptotic representation
$g\mapsto U^{(+)}_g$ is equal to $U^1$ modulo
a gauge-projective map} depending on $g$, we write
the asymptotic condition (1.17c), with $U^{(+)}_g(u)$ instead of $u$:
$$\eqalignno{
&\Vert U^M_{\exp (tP_0)}\big(\Omega_+ (U^{(+)}_g (u))\big) -
U^{1M}_{\exp (tP_0)}(U^{1M}_g (f,\dot f))\Vert^{}_{M^\rho}&(1.37)\cr
&\quad{}+\Vert  U^D_{\exp (tP_0)}
\big(\Omega_+(U^{(+)}_g  (u))\big)-
\sum_\varepsilon e^{is^{(+)}_\varepsilon (U^{(+)}_g(u),t,-i\partial)}
P_\varepsilon (-i\partial)U^{1D}_{\exp (tP_0)}U^{(+)D}_g(u)\Vert^{}_D \fl 0,
\cr }$$
when $t\fl\infty$.
Definition (1.23a) of the function $\varphi^{}_g$ defining $U^{(+)}_g$
by (1.17a) and (1.17b), shows that
$$
\varphi^{}_g (u,-\varepsilon k)
-\vartheta\big(B^{(+)1}(\cdot+a) -B^{(+)1}(\cdot),
(t,-\varepsilon t k/\omega(k))\big)\fl 0,
\eqno{(1.38)}$$
when $t\fl \infty$, for $g=\exp (a^\mu P_\mu)$. We recall that
$\varphi^{}_g(u)=0$
for $g\in SL(2,\Crm)$ and we note that $\carre \vartheta (H,\cdot)=0$ for
$H(y)=B^{(+)1}(y+a)-B^{(+)1}(y)$.
It follows from (1.37) and (1.38) that, if $\lambda =\vartheta (H,\cdot)$,
then $$\eqalignno{
&\Vert U^M_{\exp (tP_0)}\big((\Omega_+ (U^{(+)}_gu))\big) -
U^{1M}_{\exp (tP_0)}(U^{1M}_g(f,\dot f))\Vert^{}_{M^\rho}&(1.39)\cr
& {}+\Vert \!U^D_{\exp (tP_0)} \!\big((\Omega_+ (U^{(+)}_gu))\big)
\!\!-\!\! \sum_\varepsilon \!e^{is^{(+)}_\varepsilon
(U^{(+)}_g(u),t,-i\partial)} \!e^{-i q(t, -i\varepsilon t\partial)}\!
P_\varepsilon (-i\partial)
U^{1D}_{\exp (tP_0)}U^{1D}_g\alpha\Vert^{}_D\!\! \fl\!\! 0,\cr }$$
when $t\fl\infty$, where $q(t,k) =\lambda (t, -tk/\omega(k))$.
The definition of $A^{(+)}$ and $\varphi(u,(t,x))$ give
that $s^{(+)}_\varepsilon (U^{(+)}_g(u),t,k)=s^{(+)}_\varepsilon
(U^{1}_g(u),t,k)$. Since $\lambda =\vartheta (\Lambda,\cdot)$, where
$\Lambda_\mu =\partial_\mu\lambda$, it follows from (1.39) that
$$\eqalignno{
&\Vert U^M_{\exp (tP_0)}\big( (\Omega_+ (U^{(+)}_gu))\big)
-U^{1M}_{\exp (tP_0)}U^{1M}_g(f,\dot{f})\Vert^{}_{M^\rho}&(1.40)\cr
&\quad{}+\Vert U^{D}_{\exp (tP_0)}\big( (\Omega_+ (U^{(+)}_gu))\big)
-\sum_\varepsilon e^{is^{(+)}_\varepsilon (G(U^1_gu),t,-i\partial)}
P_\varepsilon (-i\partial)U^{1D}_{\exp (tP_0)}U^{1D}_g\alpha\Vert^{}_D\fl0,\cr
}$$
when $t\fl\infty$,
where $G$ is the gauge transformation given by $\lambda$ (as in (1.35) with
$c=0$). This shows that $Q_g = U^{(+)}_g U^1_{g^{-1}}$ is
a gauge-projective transformation according to (1.36).

1.4.c {\it Organization of the monograph.}

In chapter 2, we prove that the sequence $E^\rho_n$ of Hilbert spaces
admits a family of smoothing operators (Theorem 2.5) in the sense of \refSERG,
in order to use an implicit function theorem in the Fr\'echet space
$E^\rho_\infty$.

In chapter 3, we prove that $U^{(+)}$ is a nonlinear representation
(Theorem 3.12) which is nonlinearizable (Theorem 3.13).

In chapter 4, we construct approximate solutions $(A_n,\phi_n)$, $n\geq0$,
(see formulas (4.135a)--(4.135b))
of the M--D system. They are obtained, essentially by using stationary phase
methods and by iterating formulas (4.137b)--(4.137c). Their decrease
properties are given by Theorem 4.9 and Theorem 4.10. They are
approximate solutions in the sense that a certain remainder
term $(\Delta^M_n,\Delta^D_n)$ (see (4.140a)--(4.140b)) satisfies
Theorem 4.11. In particular the remainder term for the electromagnetic
potential
$\Delta^M_n$ is short-range (it belongs to $L^2 (\Rrm^3,\Rrm^4)$ for each
fixed time).

In chapter 5, we prove equal time weighted $L^2$--$L^2$ and $L^2$--$L^\infty$
estimates for the linear inhomogeneous Dirac equation
$(i\gamma^\mu \partial_\mu + m -\gamma^\mu G_\mu)h=g$,
with external field $G$ (Theorem 5.5 and Theorem 5.8). Combination of these
estimates with
an energy estimate (Corollary 5.2) leads to existence of $h$  and to
estimates on $h$, adapted to the nonlinear problems treated in chapter 6.

In chapter 6, we end the construction of an approximate solution $(A^*,\phi^*)$
(formulas (6.1a)--(6.2b) and (6.30)) satisfying the Lorentz gauge
condition (Proposition 6.2).
The existence of the rest term $(K,\Phi)$ (see formulas (6.30--(6.31c))
follows by the construction of a contraction mapping (formula (6.33))
in a Banach space ${\cal F}_N$ (Corollary 6.8).
In particular $K(t)\in L^2 (\Rrm^3,\Rrm^4)$.
The existence of solutions of the M--D equations (Theorem 6.10) and
the existence of a modified wave operator, denoted temporarily by
$\Omega_1$ (formula (6.172), Theorem 6.12) are then proved.
The existence of $\Omega^{-1}_1$ is shown (Theorem 6.13)
by using an implicit functions theorem in the Fr\'echet space $E^\rho_\infty$.
The disadvantage of $\Omega_1$ is that it gives a nonexplicit expression for
the the asymptotic representation because it requires the use of the
solutions of the M--D systems. To overcome this problem,  we introduce the
final modified wave operator $\Omega^{(+)}$
(formulas (6.227a), (6.227b) and (6.276)) and its extension $\Omega_+$
to a Poincar\'e invariant domain. $\Omega_+$ has the advantage of
giving a simple expression (see (1.17a) and (1.17b)) for the asymptotic
representation $U^{(+)}$.

This monograph (at least in so far as the M-D equations are concerned)
is largely self-contained. A few technical results are borrowed from
quoted references. It may be read with little or no knowledge of
nonlinear group representation theory, but the latter is crucial to a
real understanding of the methods and choice of spaces.

1.4.d {\it Final remarks.}

\noindent i) It can be natural to think of the underlying classical
theory of QED not as the M-D
equations with $c$-number spinor components but as a theory with
anticommuting spinor
components. Since the second order terms in the spinor component
of the wave operator originates
in the coupling $A_\mu   \psi$ of a $c$-number free electromagnetic potential
$A_\mu$ and a free Dirac field, we believe that the infrared cocycle
will remain the same
for a theory with anticommuting spinor components. Further for
higher order terms, the
nilpotency property $\psi_\alpha (x)^2 = 0$ can ameliorate the
infrared problems arising from
the self-coupling.
\psaut
\noindent ii) The observables, $4$-current, $4$-momentum and
$4$-angular momentum, defined for the
asymptotic representation $U^{(+)}$, which is ``gauge
projectively linear", converge when
$t \fl \infty$ to the usual free field observables.
Therefore there should be no observable
phenomena which distinguish $U^{(+)}$ from $U^1$, at least
as far as these observables are
concerned. Since the representation $U$ is equivalent to
$U^{(+)}$ by $\Omega_+$, we shall
call $U$ a ``gauge projectively linearizable" nonlinear
representation. Of course if the
Dirac field or the electromagnetic potential
 itself was an observable, then this would no longer be
true, i.e. it would then
be possible to distinguish $U^{(+)}$ and $U^1$ by the observables.
\psaut
\noindent iii) One should note that the phase factor (3.33a of \refFST),
which  looks as a very familiar
factor in abelian and non abelian gauge theories, was obtained in \refFST\
in the different context of the Hamilton-Jacobi equation associated with
the full Maxwell-Dirac equations. Would one  have taken initial data for
the $A_\mu$ field decreasing slower than $r^{- 1/2}$ and suitable initial
data for ${d \over dt}   A_\mu$, one would obtain phase factors with higher
powers of $A_\mu$.
\psaut
\noindent iv) The same methods can be used for nonabelian gauge theories
(of the Yang-Mills type) coupled with fermions. The aim here is to
separate asymptotically the linear (modulo an infrared problem that can
be a lot worse in the nonabelian case) equation for the spinors from the
pure Yang-Mills equation (the $A_\mu$ part).
The next step would then be to linearize analytically the pure Yang-Mills
equation (that is known \refFSYM\  to be formally linearizable),
and then to combine all this with the deformation-quantization approach
to deal rigorously with the corresponding quantum field theories.
\vfill\eject


\noindent{\titre 2. The nonlinear representation $T$ and spaces of
differentiable \hbox{\titre vectors}.}
\saut
In this chapter we prove properties of $T_Y$, $Y \in U(\p)$, and of
spaces $E_i$, $i \geq 0$, in order be able to use an {\it implicit
functions theorem in Fr\'echet spaces. }

To begin with, we introduce the linear representation of $\r^{}_0$, with
differential $T^1$, given by equation (1.5). We shall denote in the same way
the operators $T^1_X \in E^\rho$,  $X \in g$, and its closure. It follows
that $T^1_{P_\mu}$ are skew-adjoint, so $\exp (tT^1_{P_\mu})$,  $t \in \Rrm$,
is unitary on $E^\rho$. Let $b_\mu$ (resp. $\beta$) be the solutions of
equation $\carre  b_\mu = 0$
(resp. $(i  \gamma^\mu\partial_\mu + m)  \beta = 0$) with initial conditions
$a_\mu,  \dot{a}_\mu$ (resp. $\alpha$), where
$(a, \dot{a}, \alpha) \in E^\rho$. We define the representation
$U^1$ of $\r^{}_0$ by $U^1_g (a, \dot{a}, \alpha) = (a_g, \dot{a}_g,
\alpha_g)$,
$(a, \dot{a}, \alpha) \in E^\rho$,  $g = (n, L) \in\Rrm^4 \psd
SL(2,  \Crm)$, and
$$\eqalignno{
a_g (\vec{y}) & = \Lambda_L  b (\Lambda^{-1}_L ((0, \vec{y}) - n)),
& (2.1\hbox{a})\cr
\dot{a}_g (\vec{y}) & = \Lambda_L  \big({\partial \over \partial y^0}
b(\Lambda^{-1}_L ((y^0, \vec{y}) - n)\big)_{y^0 = 0}, & (2.1\hbox{b})\cr
\alpha_g (\vec{y}) & = \Gamma (L)  \beta (\Lambda^{-1}_L (0, \vec{y}) - n),
&(2.1\hbox{c})\cr
}$$
where $L \mapsto \Lambda_L$ is the canonical projection of $SL(2,  \Crm)$
onto $SO(3,1)$ and $\Gamma$ is the Dirac spinor representation of
$SL(2,\Crm)$. We define $\Lambda_g = \Lambda_L$.

As $T^1_{M_{0i}}$, $1 \leq i \leq 3$, is not in general skew-adjoint on
$E^\rho$, we state the following result of which we omit the proof as it
is straightforward by using Fourier decomposition:
\saut
\noindent{\bf Lemma 2.1.}
{\it
$g \mapsto U^1_g$ is a strongly continuous linear representation of
$\r^{}_0$ on $E^\rho$, with
$\rho > -{1/2},$ and $g \mapsto (\Lambda_g^{-1}  \oplus
\Lambda_g^{-1}  \oplus I) U_g^{1}$ is unitary in $E^{1/2}$.
}\saut
In order prove properties of the space of differentiable vectors
$E^\rho_\infty$ of $U^1$, it will be useful to know that the Fourier
transform of $U^1$ is a unitary representation after a multiplication by
a $C^\infty$-multiplier.
\saut
\noindent{\bf Theorem 2.2.}
{\it
Let $h_g (p_0, \vec{p}) = (p_0 / (\Lambda^{-1}_g  p)_0)^{-\rho + 1/2}$,
$(p_0, \vec{p}) \in \Rrm^4$,  $g \in \r^{}_0$ and

$u = (a, \dot{a}, \alpha) \in E^\rho$,  $\rho > -{1/2}$. Let,

$$V_g  u = \big(\Lambda^{-1}_g  h_g (\vert \nabla \vert, -i \nabla)
a_g,  \Lambda^{-1}_g  h_g (\vert \nabla \vert, -i \nabla)
\dot{a}_g,  \alpha_g\big),$$
where $(a_g, \dot{a}_g, \alpha_g) = U^1_g  u$. Then $g \mapsto V_g$ is
a unitary representation of $\r^{}_0$ on $E^\rho$.The representations
$(\Lambda^{-1} \oplus \Lambda^{-1} \oplus I) U^1$ on $E^{1/2}$ and $V$
on $E^\rho$ are unitarily equivalent. Moreover the representations
$U^1$ on $E^\rho$ and $V$ on $E^\rho$ have the same Hilbert space
of $C^n$-vectors, namely $E^\rho_n$, $n \geq 0.$}
\saut
\noindent{\it Proof.}
The map $g \mapsto H_g$ from $\r^{}_0$ to $L^\infty (\Rrm^3, GL(4, \Rrm))$,
defined by $H_g (\vec{p}) = \Lambda^{-1}_g  h_g
(\vert \vec{p} \vert, \vec{p})$  is $C^\infty$.
It follows then from the definition of $E^\rho$ that the map
$g \mapsto V_g u$
is $C^n$ if and only if this is the case for the map $g \mapsto U^1_g u$.
By construction, the space of $C^n$-vectors for $U^1$ is $E^\rho_n$.
This proves the last part of the proposition. By direct calculation one
finds that for $u \in S(\Rrm^3,  \Rrm^4)  \oplus  S(\Rrm^3,
 \Rrm^4)  \oplus  S(\Rrm^3,  \Crm^4)$
$$V_g  u = (\vert \nabla \vert^{- \rho + 1/2}  \oplus
\vert \nabla \vert^{-\rho + 1/2}  \oplus  I)  (\Lambda^{-1}_g
 \oplus  \Lambda ^{-1}_g  \oplus  I)
 U^1_g (\vert \nabla \vert^{\rho - 1/2}  \oplus
\vert \nabla \vert^{\rho - 1/2}  \oplus  I) u.$$

As $\vert \nabla \vert^{\rho - 1/2}  \oplus
\vert \nabla \vert^{\rho - 1/2} \oplus  I\colon E^\rho \fl E^{1/2}$
is an isomorphism and as the representation $(\Lambda^{-1}  \oplus
\Lambda^{-1}  \oplus I)$
$U^1$ is a unitary representation on $E^{1/2}$, it follows that $V$ is
unitary in $E^\rho$. This proves the theorem.

To be able to use the implicit functions theorem in the
Fr\'echet space $E_\infty$, we shall establish the existence of
smoothing operators (cf. \refSERG). As this can be done in a general
context of unitary representations, in the following lemma and theorem,
$V$ will be an arbitrary unitary representation of a Lie group ${\cal G}$ on a
Hilbert space $H$ and $S$ will be the corresponding representation on
$H_{\infty}$ (the space of $C^\infty$-vectors) of the Lie algebra ${\frak g}$
of ${\cal G}$ by
differentiation and $\Pi$ is a general basis of ${\frak g}$. The only fact
which is really used is that the Laplace operator for the representation
$V$, ${\bf\Delta} = \sum_{X \in \Pi} (S_X)^2$ in $H$ with domain $H_\infty$
 is essentially self-adjoint
(see \refW\ Theorem 4.4.4.3). Let $\overline{\bf\Delta}$
denote the closure of $\bf\Delta$.
\saut
\noindent{\bf Lemma 2.3.}
{\it
The domain of $(1 - \overline{\bf\Delta})^{n/2}$ is $H_n$ (the space of
$C^n$-vectors of $V$), and
the norms $\Vert\cdot\Vert^{}_{H^{}_n}$ and
$\Vert (1 - \overline{\bf\Delta})^{n/2}\cdot \Vert^{}_H$ are equivalent, i.e.
$$C^{-1}_n  \Vert f \Vert^{}_{H^{}_n} \leq \Vert
(1 - \overline{\bf\Delta})^{n/2} f \Vert^{}_H \leq C_n  \Vert f
\Vert^{}_{H^{}_n},\quad  C_n > 0, f \in H_n,  n \geq 0.$$
}\saut
\noindent{\it Proof.}
Let $\Pi=\{X_1,\ldots,X_r\}$, $r=\dim {\frak g}$. The statement is true
for $n=0$. Let $n\geq 1$ and let $f\in H_{\infty}$. Then
$$\eqalignno{
\Vert f \Vert^2_{H^{}_n} & \leq \Vert f \Vert^2_{H_{n-1}} +
\sum_{1\leq i_1,\ldots,i_n\leq r}
 \Vert S_{X_{i_1}\cdots X_{i_n}} f \Vert^2& {(2.2)}\cr
& = \Vert f \Vert^2_{H_{n-1}} + \sum_{1\leq i_1,\ldots,i_n\leq r}
(-1)^n (f, S_{X_{i_n}\cdots X_{i_1}X_{i_1}\cdots X_{i_n}} f),
\cr
}$$
as $S_X$, $X \in {\frak g}$, is skew-symmetric on $H_\infty$.
By successive commutations we obtain
$$\eqalignno{
\quad (-1)^n  \sum_{1\leq i_1,\ldots,i_n\leq r}
& X_{i_n}\cdots X_{i_1}X_{i_1}\cdots X_{i_n} & {(2.3)}  \cr
& = (-1)^n  \sum_{1\leq i_1,\ldots,i_n\leq r}(X_{i_1})^2\cdots (X_{i_n})^2
{}+ A^{}_{2n-1} +\cdots + A^{}_1 \cr
& \cr
& = (-{\bf\Delta})^n + A^{}_{2n-1} +\cdots+ A^{}_1, \cr
}$$
where $A_l$ is a (noncommutative) polynomial of degree $l$ in $X_j \in \Pi$,
$1 \leq j \leq r$. As $S_{X_j}$ is a skew-symmetric operator on
$H_\infty$, we have
$$\vert (f, S_{A_l} f) \vert \leq C \Vert f \Vert^{}_{H^{}_q}
\Vert f \Vert^{}_{H^{}_{l-q}},\quad 0 \leq q \leq l, \eqno{(2.4)}$$
where $C$ depends on $A_l$.

Equality (2.3) and inequality (2.4) give
$$\eqalign{
\sum_{1\leq i_1,\ldots,i_n\leq r}
\Vert S_{X_{i_1}\cdots X_{i_n}} f \Vert^2 & \leq
(f, (- {\bf\Delta})^n f) + C_n \Vert f \Vert^{}_{H^{}_n}  \Vert f
\Vert^{}_{H^{}_{n-1}}\cr
& \leq (f, (1 - {\bf\Delta})^n f) + C_n \Vert f \Vert^{}_{H^{}_n}
\Vert f \Vert^{}_{H^{}_{n-1}}.
}$$
It follows from the last inequality and (2.2) that
$$\eqalignno{
\Vert f \Vert^2_{H^{}_n} & \leq (f, (1 - {\bf\Delta})^n f) +
C_n \Vert f \Vert^{}_{H^{}_n}
\Vert f \Vert^{}_{H^{}_{n-1}} + \Vert f \Vert^2_{H^{}_{n-1}} & {(2.5)}\cr
& \leq (f, (1 - {\bf\Delta})^n f) + 2  C'_n \Vert f \Vert^{}_{H^{}_n}
\Vert f \Vert^{}_{H^{}_{n-1}}.\cr
}$$
Assuming that we have proved that
$$\Vert f \Vert^{}_{H^{}_l} \leq C_l \Vert (1 - \overline{\bf\Delta})^{l/2} f
\Vert^{}_H,
\quad f \in H_\infty,  0 \leq l \leq n-1,$$
which is true for $l = 0$, we get by induction using (2.5) that
$$\Vert f \Vert^2_{H^{}_n} \leq \Vert (1 - \overline{\bf\Delta})^{n/2} f
\Vert^2_H + 2 C_n\Vert f \Vert^{}_{H^{}_n}
\Vert (1 - \overline{\bf\Delta})^{{n-1 \over 2}} f \Vert^{}_H$$
for some $C_n$. This inequality implies
$$\Vert f \Vert^{}_{H^{}_n} \leq \big(\Vert (1 - \overline{\bf\Delta})^{n/2} f
\Vert^2_H
+ C^2_n \Vert (1 - \overline{\bf\Delta}\big)^{{n-1 \over 2}} f \Vert^2_H)^{1/2}
+ C_n \Vert (1 - \overline{\bf\Delta})^{{n-1 \over 2}}f \Vert^{}_H,$$
which by induction proves that (redefining $C_n$)
$$\Vert f \Vert^{}_{H^{}_n} \leq C_n \Vert (1 - \overline{\bf\Delta})^{n/2} f
\Vert^{}_H,
\quad f \in H_\infty,  n \geq 0. \eqno{(2.6)}$$
To prove the second inequality in the theorem for $f \in H_\infty$ we observe
that $(1 - {\bf\Delta})^n = S_{A'_{2n}}$, where $A'_{2n}$ is a
noncommutative polynomial of degree $2n$ in $X_j \in \Pi$, $1 \leq j \leq r$.
It follows now as in (2.4) that
$$\Vert (1 - \overline{\bf\Delta})^{n/2} f \Vert^2_H =
(f, (1 - {\bf\Delta})^n f) = (f, S_{A'_{2n}} f)\leq C^2_n \Vert f
\Vert^2_{H^{}_n},$$
for some $C_n$.

This inequality and (2.6) give  $$C^{-1}_n \Vert f \Vert^{}_{H^{}_n} \leq
\Vert (1 - \overline{\bf\Delta})^{n/2} f \Vert^{}_H \leq
C_n \Vert f \Vert^{}_{H^{}_n},\quad  f \in H_\infty,  n \geq 0. \eqno{(2.7)}$$
As $(1 - \overline{\bf\Delta})^{n/2}$ is a maximal closed operator, it follows
from (2.7) that the domain of $(1 - \overline{\bf\Delta})^{n/2}$ is $H_n$.
By continuity (2.7) is then true for $f \in H_n$, which proves the lemma.

We can now easily prove the existence of smoothing operators for the sequence
$H_i$,  $i \geq 0$, by using the spectral decomposition of
$(1 - \overline{\bf\Delta})^{1/2}$.
\saut
\noindent{\bf Theorem 2.4.}
{\it
We denote by $L_b (H, H_\infty)$ the space of linear continuous mappings from
$H$ to $H_\infty$ endowed with the topology of uniform convergence on bounded
sets. There exists a $C^\infty$ one-parameter family
$\Gamma_\tau \in L_b (H, H_\infty)$, $ \tau > 0$, such that if $f \in H_l$,
$l \geq 0$, then
$$\eqalign{
\hbox{\rm i)}\ & \Vert \Gamma_\tau f \Vert^{}_{H^{}_n}
\leq C_{n,l} \Vert f \Vert^{}_{H^{}_l}, \quad n \leq l,\cr
\hbox{\rm ii)}\ & \Vert \Gamma_\tau f \Vert^{}_{H^{}_n}
\leq C_{n,l}\  \tau^{n-l} \Vert f \Vert^{}_{H^{}_l},\quad n \geq l,\cr
\hbox{\rm iii)}\ & \Vert (1 - \Gamma_\tau) f \Vert^{}_{H^{}_n}
\leq C_{n,l}\  \tau^{n-l}\Vert f \Vert^{}_{H^{}_l},\quad  n \leq l,\cr
\hbox{\rm iv)}\ & \Vert {d \over d \tau} \Gamma_\tau f \Vert^{}_{H^{}_n}
\leq C_{n,l}\  \tau^{n-l-1}\Vert f \Vert^{}_{H^{}_l}, \quad  n \geq 0,
l \geq 0. \hskip56.92mm\cr
}$$
}\saut
\noindent{\it Proof.}
Let $(1 - \overline{\bf\Delta})^{1/2}$ denote the positive self-adjoint
square root of the positive self-adjoint  operator
$(1 - \overline{\bf\Delta}) \geq 1$. Since, according to Lemma 2.3, the norms
$\Vert \cdot\Vert^{}_{H^{}_n}$ and $\Vert (1 -
\overline{\bf\Delta})^{n/2}\cdot\Vert^{}_H$
are equivalent, it is enough to prove i)--iv)
with $\Vert \cdot \Vert^{}_{H^{}_n}$ replaced by
$\Vert (1 - \overline{\bf\Delta})^{n/2}\cdot\Vert^{}_H.$

Let
$$(1 - \overline{\bf\Delta})^{1/2} = \int^\infty_1  \lambda\,
de_\lambda,\eqno{(2.8)}$$
where $\lambda \mapsto e_\lambda$  is
the spectral resolution of $(1 - \overline{\bf\Delta})^{1/2}$ and let
$\varphi \in C^\infty_0 (\Rrm)$, where $\varphi (s) = 1$ for
$\vert s \vert \leq 1$, $\varphi (s) = 0$ for
$\vert s \vert \geq 2$ and $0 \leq \varphi (s) \leq 1$ for $s \in \Rrm$.
We define $\Gamma_\tau$, $\tau > 0$, by
$$\Gamma_\tau f = \int^\infty_1  \varphi (\lambda / \tau)  d(e_\lambda f),
\quad  f \in H,  \tau > 0. \eqno{(2.9)}$$
Since the map $\tau \mapsto \varphi_\tau$,  $\varphi_\tau (\lambda) = \varphi
(\lambda / \tau)$, is $C^\infty$ from $]0, \infty[$ to $L^\infty (\Rrm)$, it
follows using (2.9) that the map $\tau \mapsto \Gamma_\tau$ is $C^\infty$
from $]0, \infty[$ to $L_b (H, H_\infty).$

We have for $f$ belonging to the domain of $(1 - \overline{\bf\Delta})^{l/2}$,
$$\eqalignno{
\Vert (1 - \overline{\bf\Delta})^{n/2}  \Gamma_\tau f \Vert^2_H  & =
\int^\infty_1 \lambda^{2n}  (\varphi (\lambda / \tau))^2  d \Vert e_\lambda f
\Vert^2_H & {(2.10)}\cr
& = \int^\infty_1  \lambda^{2(n-l)} (\varphi ({\lambda / \tau}))^2
 \lambda^{2l}  d \Vert e_\lambda f \Vert^2_H\cr
& \leq \sup_{\lambda \geq 1}  \big(\lambda^{2 (n-l)}
(\varphi (\lambda / \tau))^2\big)
\Vert (1 - \overline{\bf\Delta})^{l/2} f \Vert^{}_H,\quad  n,l \geq 0.\cr
}$$
If $n \leq l$, then
$$\sup_{\lambda \geq 1} \big(\lambda^{2 (n-l)}
(\varphi (\lambda / \tau))^2\big) \leq 1, \quad \tau > 0,$$
which together with (2.10) proves statement i). If $n \geq l$, then
$$\eqalign{
\sup_{\lambda \geq 1}  \big(\lambda^{2 (n-l)} (\varphi (\lambda / \tau))^2\big)
& = \tau^{2 (n-l)}  \sup_{s \geq \tau^{-1}}  s^{2 (n-l)} (\varphi (s))^2\cr
& \leq \tau^{2 (n-l)}  \sup_{s \in \Rrm}  s^{2 (n-l)} (\varphi (s))^2\cr
& \leq 2^{2 (n-l)}  \tau^{2 (n-l)},\quad   \tau > 0,
}$$
which together with (2.10) proves statement ii). We have for $f$ belonging to
the domain of $(1 - \overline{\bf\Delta})^{l/2}$ and $n \leq l$,
$$\eqalign{
\Vert (1 - \overline{\bf\Delta})^{n/2}  (1 - \Gamma_\tau) f \Vert^2_H
& = \int^\infty_1\lambda^{2n} (1 - \varphi (\lambda / \tau))^2
d \Vert e_\lambda f \Vert^2_H\cr
& = \tau^{2 (n-l)}  \int^\infty_1  ({\lambda/\tau})^{2 (n-l)}
 (1 - \varphi (\lambda/\tau))^2 \lambda^{2l} d\Vert e_\lambda f \Vert^2_H\cr
& \leq \tau^{2 (n-l)}\sup_{s \in \Rrm} s^{2 (n-l)} (1 - \varphi (s))^2
\Vert (1 - \overline{\Delta})^{l/2} f \Vert^2_H.\cr
}$$
This inequality gives using that
$\sup_{s \in \Rrm}  s^{2 (n-l)} (1 - \varphi (s))^2 \leq 1$, $n \leq l$,
$$\Vert (1 - \overline{\bf\Delta})^{n/2} (1 - \Gamma_\tau) f \Vert^2_H
\leq \tau^{2 (n-l)} \Vert (1 - \overline{\bf\Delta}^{l / 2} f \Vert^{}_H,
\quad n \leq l,  \tau > 0.  \eqno{(2.11)}$$
Inequality (2.11) proves statement iii).

Finally we have, since $\tau \mapsto \varphi_\tau \in L^\infty (\Rrm)$
is differentiable,
$$(1 - \overline{\bf\Delta})^{n/2}  {d \over d \tau}  \Gamma_\tau f =
- \int^\infty_1  \lambda^n  {\lambda / \tau^2}
\varphi' ({\lambda/ \tau})  d (e_{\lambda} f),$$
where $\varphi'$ is the derivative of $\varphi$. If $n \geq 0$,  $l \geq 0$
and $f$ belongs to the domain of $(1 - \overline{\bf\Delta})^{l / 2}$,
it then follows that
$$\eqalign{
\Vert (1 - \overline{\bf\Delta})^{n/2}  {d \over d \tau}  \Gamma_\tau
f \Vert^2_H & = \tau^{-2}  \int^\infty_1  \lambda^{2 (n-l)}
({\lambda / \tau})^2 (\varphi' (\lambda / \tau))^2  \lambda^{2l}
 d \Vert e_\lambda f \Vert^2_H \cr
& \leq \tau^{2 (n-l-1)} \sup_{s \in \Rrm}  \big(s^{2 (n-l+1)}
(\varphi' (s))^2\big)  \Vert (1 - \overline{\bf\Delta})^{l/2} f
\Vert^2_H.\cr
}$$
By the definition of $\varphi$ we have $\varphi' (s) = 0$ for
$\vert s \vert \leq 1$ and $\vert s \vert \geq 2$.
The last inequality then gives
$$\Vert (1 - \overline{\bf\Delta})^{n/2}
{d \over d \tau}  \Gamma_\tau f \Vert^2_H
\leq C_{n,l}\,\tau^{2 (n-l-1)} \Vert (1 - \overline{\bf\Delta})^{l/2} f
\Vert^2_H, \quad n,l \geq 0,$$
where $C_{n,l} = \sup_{s \in \Rrm}
\big(s^{2 (n-l+1)} (\varphi' (s))^2\big) < \infty$.
This proves the statement iv).

We are now prepared to prove the existence of a smoothing operator for
the sequence of Hilbert spaces defined by (1.6).
\saut
\noindent{\bf Theorem 2.5.}
{\it
There exists a $C^\infty$ one-parameter family
$\Gamma_\tau \in L_b (E, E_\infty)$, $\tau > 0$, such that if
$f \in E_l$,  $l \geq 0$, then statements
i)--iv) of Theorem 2.4 hold with $H_n$ and $H_l$ replaced by $E_n$
and $E_l$ respectively.
}\saut
\noindent{\it Proof.}
According to Theorem 2.2, $E_n$, $ n \geq 0$, is the Hilbert space of
$n$-differentiable vectors of the unitary representation
$g \mapsto V_g$ of ${\cal P}_0$ in $E$. It follows
then from Theorem 2.4 that there exists a $C^\infty$ one-parameter family
with the announced properties.

The existence of a smoothing operator guarantees that the norms $\Vert \cdot
\Vert^{}_{E^{}_n}$ satisfy a convexity property, which we make explicit now:
\saut
\noindent{\bf Corollary 2.6.}
{\it
Let $0 \leq n_2 \leq n \leq n_1$,  $n_1 \neq n_2$. Then
$$\Vert f \Vert^{}_{E^{}_n} \leq C_{n_1, n_2}  \Vert f \Vert^{{n-n_2 \over
n_1 - n_2}}_{E^{}_{n_1}}  \Vert f
\Vert^{{n_1 - n \over n_1 - n_2}}_{E^{}_{n_2}}, \quad f \in E_{n_1}.$$
Moreover if $N_0 \leq n_i \leq N$,  $i = 1,2$, and $n_1 + n_2 \leq N_0 + N$,
then
$$\Vert f \Vert^{}_{E^{}_{n_1}}  \Vert g \Vert^{}_{E^{}_{n_2}}
\leq C_N \big(\Vert f \Vert^{}_{E^{}_{N_0}}
\Vert g \Vert^{}_{E^{}_N} + \Vert f \Vert^{}_{E^{}_N}  \Vert g
\Vert^{}_{E^{}_{N_0}}\big),
\quad  f,g \in E_N.$$
}\saut
\noindent{\it Proof.}
The first statement follows from Theorem 2.2.2 of \refSERG\ and
statements ii) and iii) of Theorem 2.4.
To prove the second statement let $n_1 + n_2 = N' + N$, where
$N' \leq N_0$ and let $n_1 = a  N + (1 - a) N'$, where
$0 \leq a \leq 1$. Then $n_2 = (1 - a)N + a  N'$. It follows from the
first statement of the corollary that
$$\Vert f \Vert^{}_{E^{}_{n_1}} \leq C'_N  \Vert f \Vert^a_{E^{}_N}
\Vert f \Vert^{1-a}_{E^{}_{N'}},\quad  \Vert g \Vert^{}_{E^{}_{n_2}} \leq C'_N
\Vert g \Vert^{1-a}_{E^{}_N}  \Vert g \Vert^a_{E^{}_{N'}}.$$
This gives
$$\eqalign{
\Vert f \Vert^{}_{E^{}_{n_1}}  \Vert g \Vert^{}_{E^{}_{n_2}} & \leq C''_N
(\Vert f \Vert^{}_{E^{}_N} \Vert g \Vert^{}_{E^{}_{N'}})^a
(\Vert f \Vert^{}_{E^{}_{N'}}  \Vert g \Vert_{E^{}_N})^{1-a}\cr
 & \leq C''_N \big(a \Vert f \Vert^{}_{E^{}_N}  \Vert g \Vert^{}_{E^{}_{N'}} +
 (1-a) \Vert f \Vert^{}_{E^{}_{N'}}  \Vert g \Vert^{}_{E^{}_N}\big).\cr
}$$
Since $\Vert f \Vert^{}_{E^{}_{N'}} \leq \Vert f \Vert^{}_{E^{}_{N_0}}$,
$ \Vert g \Vert^{}_{E^{}_{N'}}\leq \Vert g \Vert^{}_{E^{}_{N_0}}$,
this proves the corollary.

To establish estimates it will be useful to have more explicit
expressions for the norms $\Vert \cdot \Vert^{}_{E^{}_n}$,
than those given in (1.6).

For $Y = X_1  X_2\cdots X_L$, $L \geq 1$, and $X_1,\ldots,X_L \in \Pi$,
let $\vert Y \vert = L$ and ${\cal L} (Y)$ be the number of factors equal to
$M_{\mu \nu}$, $0 \leq \mu < \nu \leq 3$, in $Y$. Let ${\cal L} (Y) = 0$
and $\vert Y \vert = 0$, for $Y = \un$.
\saut
\noindent{\bf Lemma 2.7.}
{\it
There exist linear maps $Y \mapsto Q^{}_i (Y)$, $Y \mapsto R^{}_i (Y)$,
$i = 1,2$, and $Y \mapsto \Gamma (Y)$ from $U(\p)$ into the space of
differential operators on $\Rrm^3$ with polynomial coefficients such that
$$T^1_Y (u) = \big(Q^{}_1 (Y) f + R^{}_1 (Y) \dot{f},  R^{}_2 (Y) f +
Q^{}_2 (Y) \dot{f},  \Gamma (Y) \alpha\big), \eqno{(2.12)}$$
for $u = (f, \dot{f}, \alpha) \in E_\infty$,  $Y \in U(\p)$.
If $Y = \un$, then $Q^{}_i (Y) = I$, $R^{}_i (Y) = 0$, $i = 1,2$, and
$\Gamma (Y) = I$, where $I$ is the identity mapping. If $Y = X_1\cdots X_L$,
$X_i \in \Pi$, $1 \leq i \leq L$, $L \geq 1$, then the polynomials
$Q^{}_i (Y, x, \xi)$, $R^{}_i (Y, x, \xi)$ and $\Gamma (Y, x, \xi)$,
$x, \xi \in \Rrm^3$, associated to
$Q^{}_i (Y, x, -i \nabla)$, $R^{}_i (Y, x, -i \nabla)$ and
$\Gamma (Y, x, -i\nabla)$ respectively, satisfy for $a \neq 0$:
$$\eqalignno{
Q^{}_i (Y, a^{-1} x, a \xi) & = a^{\vert Y \vert - {\cal L} (Y)} Q^{}_i
(Y,x, \xi), \quad  \quad\hskip1pt\deg  Q^{}_i (Y) \leq \vert Y \vert +
{\cal L} (Y), & {(2.13{\rm a})}\cr R^{}_1 (Y, a^{-1} x, a \xi) & =
 a^{\vert Y \vert - {\cal L} (Y)-1}  R^{}_1 (Y, x, \xi), \quad
 \deg  R^{}_1 (Y) \leq \vert Y \vert + {\cal L} (Y) - 1,
& {(2.13{\rm b})}\cr
R^{}_2 (Y, a^{-1} x, a \xi) & = a^{\vert Y \vert - {\cal L} (Y) + 1}
R^{}_2 (Y, x, \xi), \quad \deg  R^{}_2 (Y) \leq \vert Y \vert +
{\cal L} (Y) + 1, & {(2.13{\rm c})}\cr
}$$
The degree of a polynomial, denoted by $\deg$, is the total degree in
$(x, \xi)$. If $\deg_x$ (resp. $\deg_\xi$) denotes the degree relative
to the variable $x$ (resp. $\xi$), then
$$\deg_\xi  \Gamma (Y, x, \xi) \leq \vert Y \vert, \quad \deg_x \Gamma
(Y, x, \xi) \leq {\cal L} (Y). \eqno{(2.14)}$$
}\saut
\noindent{\it Proof.}
Since $T^1_{\un}$ is the identity operator in $E$, we have $Q^{}_i
({\un}) = I$, $R^{}_i ({\un}) = 0$, $i=1,2$, and $\Gamma ({\un}) = I$. For
$Y = X_1\cdots X_L$, $X_i \in \Pi$, $L \geq 1$, we prove the lemma
by induction in $\vert Y \vert = L$. For $\vert Y \vert = 1$, it follows from
definition (1.5) of $T^1_X$, $X \in \Pi$, that formula (2.12), properties
(2.13) and (2.14) are satisfied. Suppose (2.12), (2.13) and (2.14) are true
for $1 \leq \vert Y \vert = L$. If $Y' = YX$, $X \in \Pi$, then $T^1_{Y'} =
T^1_Y  T^1_X$, $\vert Y' \vert = \vert Y \vert + 1$ and ${\cal L} (Y') =
{\cal L} (Y) + {\cal L} (X)$.

It follows from the induction hypothesis and (2.12) that
$$\eqalignno{
Q^{}_1 (Y', x, -i \nabla) &= Q^{}_1 (Y, x, -i \nabla)  Q^{}_1 (X, x, -i \nabla)
+ R^{}_1 (Y, x, -i \nabla)   R^{}_2 (X, x, -i \nabla),\qquad &
{(2.15\hbox{a})}\cr
Q^{}_2 (Y', x, -i \nabla)& = R^{}_2 (Y, x, -i \nabla)  R^{}_1 (X, x, -i \nabla)
+ Q^{}_2 (Y, x, -i \nabla)  Q^{}_2 (X, x, -i \nabla), & {(2.15\hbox{b})}\cr
R^{}_1 (Y', x, -i \nabla)& = Q^{}_1 (Y, x, -i \nabla)  R^{}_1 (X, x, -i \nabla)
+ R^{}_1 (Y, x, -i \nabla)  Q^{}_2 (X, x, -i \nabla), & {(2.15\hbox{c})}\cr
R^{}_2 (Y', x, -i \nabla)& = R^{}_2 (Y, x, -i \nabla)  Q^{}_1 (X, x, -i \nabla)
+ Q^{}_2 (Y, x, -i \nabla)  R^{}_2 (X, x, -i \nabla), & {(2.15\hbox{d})}\cr
\Gamma (Y', x, -i \nabla)& = \Gamma (Y, x, -i \nabla)  \
\Gamma (X, x, -i \nabla). & {(2.15\hbox{e})}\cr
}$$
Let $D_1 (x, \xi)$, $D_2 (x, \xi)$ be two polynomials in $x, \xi \in \Rrm^3$
with $\deg  D_i = d_i$, $i = 1,2$, and let
$$D_i (a^{-1} x, a \xi) = a^{\nu_i} D_i (x, \xi),\quad  a \neq 0,
\nu_i \in {\Zrm}.$$
It then follows using the Leibniz rule that there is a polynomial $D$
in $x$ and $\xi$
such that $D (x, -i \nabla) = D_1 (x, -i \nabla)  D_2 (x, -i \nabla)$
and
$$\deg  D = d_1 + d_2,  D(a^{-1} x, a \xi) = a^{\nu_1 + \nu_2}
D (x, \xi). \eqno{(2.16)}$$
We apply (2.16) to $Q^{}_1 (Y')$ in (2.15a), which gives according to (2.13a)
$$\eqalign{
\deg  Q^{}_1 (Y') & \leq \max \big(\deg  Q^{}_1 (Y) + \deg  Q^{}_1 (X),
\deg  R^{}_1 (Y) + \deg  R^{}_2 (X)\big)\cr
&\leq \max  \big(\vert Y \vert + {\cal L} (Y) + \vert X \vert + {\cal L} (X),
 \vert Y \vert + {\cal L} (Y) - 1 + \vert X \vert + {\cal L} (X) + 1\big)\cr
& =\vert Y \vert + \vert X \vert + {\cal L} (Y) + {\cal L} (X)
= \vert Y' \vert +{\cal L} (Y')\cr
}$$
and
$$\eqalign{
Q^{}_1 (Y', a^{-1} x, a \xi)
& = a^{\vert Y \vert - {\cal L} (Y) + \vert X \vert -
{\cal L} (X)} Q^{}_1 (Y', x, \xi)\cr
& = a^{\vert Y' \vert - {\cal L} (Y')} Q^{}_1 (Y', x, \xi).\cr
}$$
This proves that (2.13a) is true for $L+1$ and $i=1$. Application of (2.16)
to $Q^{}_2 (Y')$, $R^{}_1 (Y')$, $R^{}_2 (Y')$,
$\Gamma (Y')$ in (2.15b)--(2.15e) proves  similarly that (2.13a)--(2.13c)
are true for $L+1$.

Let $M_i (x, \xi)$, $i=1,2$, be two polynomials in $x, \xi \in \Rrm^3$. Using
Liebniz rule it follows that there is a polynomial $M (x, \xi)$ such
that $M (x, -i \nabla) = M_1 (x, -i \nabla)  M_2 (x, -i \nabla)$ and that
$$\deg_\xi  M = \deg_\xi M_1 + \deg_\xi  M_2,\quad
\deg_x  M = \deg_x  M_1 + \deg_x  M_2. \eqno{(2.17)}$$
Formula (2.15e) and relation (2.17) prove (2.14) for $L+1$. This proves the
lemma.
\saut
\noindent{\bf Lemma 2.8.}
{\it
If $u = (f, \dot{f}, \alpha) \in E_1^{\rho}$, ${1 \over 2} < \rho < 1$, then
$$\Big(\sum_{0 \leq \vert \beta \vert \leq \vert \delta \vert \leq 1}
\Vert M_\beta \partial^\delta (f, \dot{f}) \Vert^2_{M^\rho}
+ \sum_{\scr \vert \beta \vert \leq 1\atop\scr \vert \delta \vert \leq 1}
\Vert M_\beta  \partial^\delta  \alpha\Vert^2_{L^2}\Big)^{1 \over 2}
\leq C \Vert u \Vert^{}_{E^{}_1},$$
for some constant $C$ depending only on $\rho$.
Here $M_\beta (x) = x^\beta = x_1^{\beta_1}x_2^{\beta_2}x_3^{\beta_3}$.
}\saut
\noindent{\it Proof.}
Let $g \mapsto V_g$ be the unitary representation of $\r^{}_0$ in $E^\rho$
defined in Theorem 2.2. Let $X \mapsto \xi^{}_X$, be the corresponding Lie
algebra representation of $\p$ in the space of differential vectors of $V$,
which according to Theorem 2.2 is $E_\infty$. We write
$$\xi^{}_X  u = (\xi^M_X (f, \dot{f}), \xi^D_X \alpha),
\quad u = (f, \dot{f}, \alpha) \in E_\infty. \eqno{(2.18)}$$
Since, by Theorem 2.2, the Hilbert space of $C^1$-vectors of $V$ is $E_1$,
we have
$$\Big(\Vert u \Vert^2_E + \sum_{X \in \Pi}
\Vert \xi^{}_X u \Vert^2_E\Big)^{{1 \over 2}}
\leq C \Vert u \Vert^{}_{E^{}_1},\quad  u \in E_\infty, \eqno{(2.19)}$$
for some constant $C$.

We shall prove the lemma by giving an upper bound for the Laplacian
${\bf\Delta}_E$ of the representation $V$:
$${\bf\Delta}_E u = \sum_{X \in \Pi}
\xi^2_X  u = ({\bf\Delta}_M (f, \dot{f}),{\bf\Delta}_D \alpha),
\eqno{(2.20\hbox{a})}$$
where
$${\bf\Delta}_M = \sum_{X \in \Pi} (\xi^M_X)^2,
\quad {\bf\Delta}_D = \sum_{X \in \Pi}(\xi^D_X)^2. \eqno{(2.20\hbox{b})}$$
According to Theorem 2.2, the ``Dirac part" of $T^1$ is identical to $\xi^D$.
Expressions (1.5a)--(1.5d) give after rearrangement of the terms
$$\eqalign{
{\bf\Delta}_D
& = {\cal D}^2 + \sum_{1 \leq i \leq 3}
\partial^2_i
+\sum_{1 \leq i \leq j \leq 3}
(x_i \partial_j - x_j  \partial_i + \sigma_{ij})^2
+ \sum_{1 \leq i \leq 3}
(x_i  {\cal D} + \sigma_{0i})^2\cr
& = 2 \Delta - m^2 + \sum_{i<j}
\big((x_i  \partial_j - x_j  \partial_i)^2
+ 2 (x_i  \partial_j - x_j  \partial_i) \sigma_{ij} + \sigma^2_{ij}\big)\cr
&\quad {}+ \sum_i
\big(x_i  {\cal D}^2 x_i + x_i  {\cal D} [x_i, {\cal D}]
+ x_i  {\cal D}  \sigma_{0i} + x_i  \sigma_{0i}
{\cal D} + \sigma^2_{0i}\big).\cr
}$$
Using the fact that
$$[x_i, {\cal D}] = - \gamma_0  \gamma_i = - 2  \sigma_{0i},\quad
 (\sigma_{ij})^2 = - {1 \over 4},\quad  (\sigma_{0i})^2 = {1 \over 4},$$
we obtain
$$\eqalign{
{\bf\Delta}_D & =  2 \Delta - m^2 + \sum_{i<j} \big((x_i  \partial_j - x_j
\partial_i)^2 + 2 (x_i  \partial_j - x_j  \partial_i) \sigma_{ij} -
{1 \over 4}\big)\cr
&\quad {}+ \sum_i \big(x_i (\Delta - m^2) x_i + x_i [\sigma_{0i}, {\cal D}]
+ {1 \over 4}\big).\cr
}$$
A direct calculation gives
$$\eqalign{
[\sigma_{0j}, {\cal D}] & = im  \gamma_j + \sum_l  {1 \over 2}
(\gamma_j  \gamma_l - \gamma_l  \gamma_j) \partial_l\cr
& = im  \gamma_j - 2 \sum_{l \neq j}  \sigma_{jl}
\partial_l,\cr
}$$
which shows that
$$\sum_j  x_j [\sigma_{0j}, {\cal D}] = im  \sum_j
x_j  \gamma_j - 2 \sum_{j<l}  \sigma_{jl}
(x_j  \partial_l - x_l  \partial_j).$$
Therefore
$${\bf\Delta}_D = 2 \Delta - m^2 + \sum_{i<j} (x_i  \partial_j - x_j
\partial_i)^2 + \sum_i  x_i (\Delta - m^2) x_i
+ im  \sum_l  x_l  \gamma_l. \eqno{(2.21)}$$
Using the fact that
$$\eqalignno{
\sum_{i<j} (x_i  \partial_j - x_j  \partial_i)^2
& = \sum_{i,j}(\partial_j  x_i  x_i  \partial_j - \partial_i
x_i  x_j  \partial_j),& {(2.22\hbox{a})}\cr
\noalign{\hbox{and that}}
\sum_i  x_i  \Delta  x_i
& = \sum_{i,j}
\partial_j  x_i  x_i  \partial_j, & {(2.22\hbox{b})}\cr
}$$
we obtain
$${\bf\Delta}_D = - m^2 + 2 \Delta - m^2 \vert x \vert^2 + 3 \sum_{i,j}
\partial_j  x_i  x_i  \partial_j
- \sum_{i,j}  \partial_i  x_i  x_j  \partial_j +
im  \sum_l  x_l  \gamma_l. \eqno{(2.23)}$$
It follows from (2.23) that
$$\eqalignno{
(\alpha, - {\bf\Delta}_D  \alpha) &=  m^2 \Vert \alpha \Vert^2_{L^2}
+ 2 \sum_i \Vert \partial_i \alpha \Vert^2_{L^2}
+ m^2 \sum_i \Vert x_i \alpha \Vert^2_{L^2}& (2.24)\cr
& \quad {}+ 3 \sum_{i,j} \Vert x_i  \partial_j  \alpha \Vert^2_{L^2}
-\Vert \sum_{i}x_i\partial_i \alpha \Vert^2_{L^2}
- m  \sum_j (\alpha, i \gamma_j  x_j \alpha)\cr
& \geq m^2 \Big(\Vert \alpha \Vert^2_{L^2}
+ \sum_i \Vert x_i  \alpha \Vert^2_{L^2}\Big)
+2 \sum_i \Vert \partial_i  \alpha \Vert^2_{L^2}\cr
& \quad {}+ 2 \sum_{i,j} \Vert x_i  \partial_j  \alpha \Vert^2_{L^2}
- m \sum_j \Vert x_j  \alpha \Vert^{}_{L^2}  \Vert \alpha \Vert^{}_{L^2}.
\cr
}$$
As pointed out in the proof of Theorem 2.2, the unitary representation $V$
in $E^\rho$ is equivalent to a unitary representation $V'$ in
$E^{{1/2}}$ by the isomorphism
$\vert \nabla \vert^{\rho - {1 \over 2}}
\oplus I \colon M^\rho \oplus D \fl M^{{1/2}}\oplus D$.
$V$ is a direct sum: $V = V^{M^\rho} \oplus V^D$ and $V' = V^{M^{1/2}}
\oplus V^D$. The generators of $V^{M^{1/2}}$ are given by
$$\eqalignno{
\xi^{M^{1/2}}_{P_0} (f, \dot{f}) & = (\dot{f}, \Delta f),& (2.25)\cr
\xi^{M^{1/2}}_{P_i} & = \partial_i,\quad  1 \leq i \leq 3,\cr
\xi^{M^{1/2}}_{M_{ij}} & = m_{ij},\quad 1 \leq i \leq j \leq 3,\cr
m_{ij} & = - x_i  \partial_j + x_j  \partial_i,\cr
\xi^{M^{1/2}}_{M_{0i}} (f, \dot{f}) & = \Big(x_i  \dot{f},
\sum^3_{j=1}  \partial_j  x_i  \partial_j  f\Big),
\quad  1 \leq i \leq 3.\cr
}$$
The Laplacian ${\bf\Delta}_{M^{1/2}}$ is diagonal.
Let ${\bf\Delta}_{M^{1/2}} (f, \dot{f}) =
({\bf\Delta}^{(1)}_{M^{1/2}} f,  {\bf\Delta}^{(2)}_{M^{1/2}} \dot{f}).$
Using (2.25) a direct calculation gives:
$$\eqalignno{
{\bf\Delta}^{(1)}_{M^{1/2}} &= 2 \Delta + \sum_{i,j} (2 \partial_j  x_i
 x_i \partial_j - \partial_i  x_i  x_j
\partial_j) - \sum_i x_i\partial_i, & {(2.26\hbox{a})}\cr
{\bf\Delta}^{(2)}_{M^{1/2}} &= 2 \Delta + \sum_{i,j} (2 \partial_j  x_i
 x_i  \partial_j - \partial_i  x_i  x_j
 \partial_j) + \sum_i x_i\partial_i. & {(2.26\hbox{b})}\cr
 }$$
The equivalence of $V^{M^\rho}$ on $M^\rho$ and $V^{M^{1/2}}$ on $M^{1/2}$
shows that ${\bf\Delta}_{M^\rho} = ({\bf\Delta}^{(1)}_{M^\rho},
{\bf\Delta}^{(2)}_{M^\rho}) =
\vert \nabla \vert^{{1 \over 2} - \rho}
{\bf\Delta}_{M^{1/2}} \vert \nabla \vert^{\rho - {1 \over 2}}.$
This fact shows that if $v = (f, \dot{f}) \in M^\rho_\infty$:
$$(v, {\bf\Delta}^{}_{M^\rho} v)^{}_{M^\rho}
= (\vert \nabla \vert^{\rho + {1 \over 2}} f,
{\bf\Delta}^{(1)}_{M^{1/2}} \vert \nabla \vert^{\rho - {1 \over 2}} f)
+ (\vert \nabla \vert^{\rho - {3 \over 2}} \dot{f},
{\bf\Delta}^{(2)}_{M^{1/2}} \vert \nabla \vert^{\rho - {1 \over 2}} \dot{f})
\eqno{(2.27)}$$
Since $M^\rho$ is a real vector space and $[x_i  \partial_j,
\vert \nabla \vert^a] = a \vert \nabla \vert^{a - 2}\partial_i  \partial_j$,
it follows from (2.26) and (2.27) that
$$\eqalignno{
(v, - {\bf\Delta}_{M^\rho} v)_{M^\rho} = {}
& 2 \Vert \nabla v \Vert^2_{M^\rho} + 2 \sum_{i,j}
\Vert x_i  \partial_j  v \Vert^2_{M^\rho}
- \Vert \sum_i x_i\partial_i v \Vert^2_{M^\rho} & {(2.28)}\cr
&\qquad {}+ C_1 (\rho) \Vert \vert \nabla \vert^\rho f \Vert^2_{L^2}
+ C_2 (\rho) \Vert \vert \nabla \vert^{\rho - 1}\dot{f} \Vert^2_{L^2},\cr
}$$
where $C_1 (\rho)$ and $C_2 (\rho)$ are two real numbers.
This gives the inequality
$$(v, - {\bf\Delta}^{}_{M^\rho} v)^{}_{M^\rho}
\geq 2 \Vert \nabla v \Vert^2_{M^\rho} +
\sum_{i,j} \Vert x_i  \partial_j  v \Vert^2_{M^\rho}
+ C(\rho) \Vert v \Vert^2_{M^\rho},\eqno{(2.29)}$$
where $C(\rho) = \min (C_1 (\rho), C_2 (\rho))$.
We define the norm $q^{}_1$ by
$$q^{}_1 (u) = \Big(\sum_{\vert \beta \vert \leq \vert \delta \vert \leq 1}
\Vert M_\beta \partial^\delta (f, \dot{f}) \Vert^2_{M^\rho}
+ \sum_{\scr\vert \beta \vert \leq 1\atop\scr\vert \delta \vert \leq 1}
\Vert M_\beta  \partial^\delta \alpha \Vert^2_{L^2}\Big)^{{1 \over 2}}
\eqno{(2.30)}$$
for $u = (f, \dot{f}, \alpha) \in E^\rho_\infty$.

It follows from inequalities (2.24) and (2.29) that
$$(u, - {\bf\Delta}^{}_E u)_{E^\rho} \geq \mu (\rho)^2 (q^{}_1 (u))^2 -
2 \nu_1 (\rho) q^{}_1 (u) \Vert u \Vert^{}_E - \nu_2 (\rho)^2
\Vert u \Vert^2_E $$
where $\mu (\rho) > 0$ and $\nu_i (\rho) \geq 0$,  $\nu_2 (\rho) \geq 0$.
This inequality gives that
$$q^{}_1 (u) \leq {1 \over \mu} \big((u, - {\bf\Delta}^{}_{E} u)
+ (\nu^2_1 + \nu^2_2) \Vert u \Vert^2_E\big)
+ \nu_1 \Vert u \Vert^{}_E, \quad \mu > 0,$$
where we have omitted to indicate the dependence on $\rho$ of the constants.
$\xi^{}_X$, $X \in \p$,
is skew-symmetric with domain $E_\infty$, so $\sum_{X \in \Pi}
\Vert \xi^{}_X u \Vert^2_E =
(u, - {\bf\Delta}^{}_{E} u)$,  $u \in E_\infty$.
This fact and the last inequality
show that
$$q^{}_1 (u) \leq C \Big(\sum_{X \in \Pi} \Vert \xi^{}_X u \Vert^2_E
+ \Vert u \Vert^2_E\Big)^{{1 \over 2}},
\quad  u \in E_\infty, \eqno{(2.31)}$$
for some constant $C$. It follows from inequalities (2.19) and (2.31) that
(with
a new constant $C$)
$$q^{}_1 (u) \leq C \Vert u \Vert^{}_{E^{}_1},\quad  u \in E_\infty.$$
By continuity this inequality is true for $u \in E_1$, which proves
the lemma.

To state the next theorem we introduce the following {\it seminorms} $q^{}_n$,
$n \geq 0$, on $E_\infty$:
$$(q^{}_n (u))^2 = (q^M_n (v))^2 + (q^D_n (\alpha))^2, \quad u = (v,
\alpha) \in E_\infty,  v \in M_\infty,  \alpha \in D_\infty,
\eqno{(2.32\hbox{a})}$$
$$\eqalignno{
q^M_n (v) & = \Big(\sum_{\vert \mu \vert \leq \vert \nu \vert \leq n}
\Vert M^\mu
\partial^\nu v \Vert^2_M\Big)^{{1 \over 2}}\cr
\noalign{\hbox{and}}
q^D_n (\alpha) & = \Big(\sum_{\scr \vert \mu \vert \leq n\atop\scr
\vert \nu \vert \leq n}
\Vert M^\mu  \partial^\nu \alpha \Vert^2_D\Big)^{{1 \over 2}},&{(2.32\hbox{b})}
 }$$
where $\mu$ and $\nu$ are multiindices and $M^{\mu}$ is defined as in
Lemma 2.8. As will be proved, $q^{}_n$ has a continuous extension to $E_n$.
\saut
\noindent{\bf Theorem 2.9.}
{\it
There exist constants $C_n > 0$ such that
$$C^{-1}_n \Vert u \Vert^{}_{E^{}_n} \leq q^{}_n (u) \leq C_n \Vert u
\Vert^{}_{E^{}_n},
\quad n \geq 0.$$
Moreover the linear space
$$E_{c}=M_{c}\oplus D_{c}
= \{(f,\dot{f}, \alpha)\in E_{\infty}\big\vert \hat{f}, \fhd\in
C^{\infty}_{0}(\Rrm^3 - \{0\}, \Crm^4),
\hat{\alpha}\in C^{\infty}_{0}(\Rrm^3, \Crm^4)  \}$$
is dense in $E_{\infty}$.
}\saut
\noindent{\it Proof.}
According to definition (1.6a) of $\Vert\cdot  \Vert^{}_{E^{}_n}$ and
expression
(2.12) for $T^1_Y u$, $Y \in \Pi'$,
$u = (f, \dot{f}, \alpha)\in E_\infty$ of Lemma 2.7 we have
$$\eqalignno{
\Vert u \Vert^2_{E^{}_n} =& \sum_{\scr Y \in \Pi'\atop\scr\vert Y \vert \leq n}
\Big(\Vert \Gamma (Y)\alpha \Vert^2_{L^2}
+\Vert \vert \nabla \vert^\rho
(Q^{}_1 (Y) f + R^{}_1 (Y) \dot{f}) \Vert^2_{L^2}  & {(2.33)} \cr
& \qquad {}+\Vert \vert \nabla \vert^{\rho - 1}
(R^{}_2 (Y) f + Q^{}_2 (Y) \dot{f}) \Vert^2_{L^2}\Big).\cr
}$$
We shall estimate the terms on the right-hand side of (2.33). It follows at
once
from inequality (2.14) in Lemma 2.7 that
$$\Vert \Gamma (Y) \alpha \Vert^{}_{L^2} \leq C_n  q^D_n (\alpha),
\quad Y \in \Pi',  \vert Y \vert \leq n, \eqno{(2.34)}$$
and from (2.13a) that
$$\big(\Vert \vert \nabla \vert^\rho  Q^{}_1 (Y) f \Vert^2_{L^2} + \Vert \vert
\nabla \vert^{\rho - 1}  Q^{}_2 (Y) \dot{f} \Vert^2_{L^2}\big)^{{1 \over 2}}
\leq C_n  q^M_n (f, \dot{f}),
\quad Y \in \Pi',  \vert Y \vert \leq n.\eqno{(2.35)}$$
We observe that
$$\Vert \vert \nabla \vert^\rho  R^{}_1 (Y) \dot{f} \Vert^2_{L^2}
= \sum_{1 \leq j \leq 3}
\Vert \vert \nabla \vert^{\rho - 1} \partial_j  R^{}_1 (Y) \dot{f}
\Vert^2_{L^2}.$$
We have $\partial_j  R^{}_1 (Y) \dot{f}
= R^{}_{1j} (Y) \dot{f} + P_{1j} (Y) \dot{f}$, where
$$R^{}_{1j} (Y, x, \xi) = {\partial \over \partial x_j} R^{}_1 (Y, x, \xi),
P_{1j} (Y, x, \xi) = R^{}_1 (Y, x, \xi) i \xi^{}_j.$$
Since $R^{}_1$ satisfies (2.13b), the polynomials $R^{}_{1i}$ and $P_{1i}$
satisfy condition (2.13a) of $Q^{}_i$. This shows that
$$\Vert \vert \nabla \vert^\rho  R^{}_1 (Y) \dot{f} \Vert^2_{L^2}
\leq C_n \sum_{\vert \mu \vert \leq \vert \nu \vert \leq n}
\Vert \vert \nabla\vert^{\rho - 1} M_\mu  \partial^\nu \dot{f} \Vert^2_{L^2},
\quad Y \in \Pi',  \vert Y \vert \leq n. \eqno{(2.36)}$$
It follows from (2.13c) that $R^{}_2
(Y, x, \xi) = \sum_{i \leq j \leq 3}  r^{(1)}_j (Y, x, \xi) i \xi^{}_j$, where
$r^{(1)}_j$ satisfies condition (2.13a) of $Q^{}_i$.
Then $R^{}_2 (Y) f = \sum_{1 \leq i \leq 3}
 r^{(1)}_j (Y)  \partial_j f = \sum_{1 \leq i \leq 3}
\partial_j  r^{(1)}_j  (Y) f + R^{(1)}_2 (Y) f$,
where $R^{(1)}_2 (Y, x, \xi), r^{(1)}_j (Y, x, \xi)$, once more satisfies
the homogeneity condition in (2.13c) and
$\deg  R^{(1)}_2 \leq \vert Y \vert + {\cal L} (Y) - 1$, with $R^{(1)}_2 = 0$
if this number is strictly smaller than one.
We now repeat this argument with
$R^{(l+1)}_2$ instead of $R^{(l)}_2,  R^{(0)}_2 = R^{}_2$ until we have
$R^{(L)}_2 = 0$ for some $L \leq {\cal L} (Y) + 1$.
The corresponding sequence $r^{(1)},\ldots,r^{(L)}$, gives
$$R^{}_2 (Y) f = \sum_{1 \leq j \leq 3}  \partial_j  r^{}_j (Y)f,
 r^{}_j (Y, x, \xi) = \sum^L_{l=1}  r^{(l)}_j (Y, x, \xi), \eqno{(2.37)}$$
where $r^{}_j$ satisfies (2.13a). Equalities (2.37) and (2.13a) show that
$$\eqalignno{
\Vert \vert \nabla \vert^{\rho - 1} R^{}_2 (Y) f \Vert^{}_{L^2}
& \leq \sum_j \Vert \vert\nabla \vert^{\rho - 1}
\partial_j  r^{}_j (Y) f \Vert^{}_{L^2} & {(2.38)}\cr
&\leq \sum_j \Vert \vert \nabla \vert^\rho r^{}_j (Y) f \Vert^{}_{L^2}\cr
&\leq C_n \Big(\sum_{\vert \mu \vert \leq \vert \nu \vert \leq n}
\Vert \vert \nabla \vert^\rho M_\mu
\partial^\nu f \Vert^{2}_{L^2}\Big)^{{1 \over 2}},
\quad Y \in \Pi', \vert Y\vert\leq n.\cr
}$$
It follows from expression (2.33)
of $\Vert u \Vert^2_{E_n}$ and inequalities (2.34), (2.35),
(2.36) and (2.38) that
$\Vert u \Vert^{}_{E^{}_n} \leq C_n\, q^{}_n (u)$, $u \in E_\infty$, $n \geq
0$,
for some constants $C_n$. This proves the first inequality of the theorem.

To prove the second inequality of the theorem, by induction we note that
$q^{}_0 (u) = \Vert u \Vert^{}_{E^{}_0} = \Vert u \Vert^{}_E$ by definition and
suppose that $q^{}_i (u) \leq C_i \Vert u \Vert^{}_{E^{}_i}$
for $0 \leq i \leq n$.
It follows from definition (2.32a) of $q^M_n$ that
$$(q^M_{n+1} (v))^2 = (q^M_n (v))^2
+ \sum_{\scr \vert \mu \vert \leq n\atop \scr \vert \nu \vert = n+1}
\Vert M_\mu  \partial^\nu  v \Vert^2_M +
\sum_{\scr \vert \mu \vert = n+1\atop \scr \vert \nu \vert = n+1}
\Vert M_\mu \partial^\nu  v \Vert^2_M, \eqno{(2.39)}$$
where $v \in M_\infty$. We estimate the two last two terms on the
right-hand side of this equality. According to the induction hypothesis
$$\eqalignno{
\sum_{\scr \vert \mu \vert \leq n\atop\scr \vert \nu \vert = n+1}
\Vert M_\mu \partial^\nu  v \Vert^2_M
& = C''_{n}\sum_{1 \leq i \leq 3}
\sum_{\scr \vert \mu \vert \leq n\atop\scr\vert \nu \vert = n}
\Vert M_\mu  \partial^\nu \partial_i  v \Vert^2_M & (2.40)\cr
&\leq C''_{n}\sum_{1 \leq i \leq 3}
(q^M_n (\partial_i  v))^2 \cr
&\leq C''_{n} C^2_n  \sum_{1 \leq i \leq 3}
\Vert T^{1M}_{P_i} v \Vert^2_{M^{}_n} \cr
& \leq C'_n  \Vert v \Vert^2_{M^{}_{n+1}},\cr
}$$
for some constants $C'_n$ and $C''_{n}$. For the term
with $\vert \mu \vert = n+1$, and $\vert \nu \vert = n+1$
it follows, using the same argument which led to (2.37), that
$$M_\mu  \partial^\nu = \sum_{l,k}  x_l  \partial_k
 r^{\mu \nu}_{lk},\quad  r^{\mu \nu}_{lk} (x, \xi) = r^{\mu \nu}_{lk}
(a^{-1} x, a \xi), \quad  \deg  r^{}_{lk} \leq 2n. \eqno{(2.41)}$$
This and Lemma 2.8, with $\alpha = 0$, show that, for
$\vert \mu \vert = \vert \nu \vert = n+1$,
$$\Vert M_\mu  \partial^\nu  v \Vert^{}_M \leq \sum_{l,k} \Vert x_l
 \partial_k  r^{\mu \nu}_{lk}  v \Vert^{}_M \leq C
 \sum_{l,k} \Vert r^{\mu \nu}_{lk}  v \Vert^{}_{M^{}_1}, \eqno{(2.42)}$$
$\vert \mu \vert = \vert \nu \vert = n+1$, where $r^{\mu \nu}_{lk} =
r^{\mu \nu}_{lk} (x, -i \nabla).$
It follows from property (2.41) of $r^{\mu \nu}_{lk}$, the
definition of $q^M$ and (2.42) that for
$\vert \mu \vert = \vert \nu \vert = n+1$, and suitable constants
$C$, $C_n$ and $C'$ that
$$\eqalign{
\Vert M_\mu  \partial^\nu  v \Vert^{}_M
& \leq C  \sum_{l,k}
\Big(\sum_{X \in \Pi}  \Vert T^{1M}_X  r^{\mu \nu}_{lk}
v \Vert^2_M + \Vert v \Vert^2_M\Big)^{{1 \over 2}}\cr
& \leq C'  \sum_{l,k}  \Big(\sum_{X \in \Pi}  \big(\Vert r^{\mu \nu}_{lk}
 T^{1M}_X  v \Vert^{}_M + \Vert [T^{1M}_X,  r^{\mu \nu}_{lk}]v
 \Vert^{}_M\big) +  \Vert v \Vert^{}_M\Big)\cr
&\leq C_n  \sum_{X \in \Pi}  q^{}_n (T^{1M}_X  v)
+ C' \sum_{l,k}  \Vert [T^{1M}_X,  r^{\mu \nu}_{lk}] v \Vert^{}_{M},\cr
}$$
where $r^{\mu \nu}_{lk} = r^{\mu \nu}_{lk} (x, -i \nabla)$.

According to the induction hypothesis this gives with new constants
$$\eqalignno{
\Vert M_\mu  \partial^\nu  v \Vert^{}_M
& \leq C'_n \Big(\sum_{X \in \Pi} \Big(\Vert T^{1M}_X  v \Vert^{}_{M^{}_{n}}
+ C'  \sum_{l,k}\Vert [T^{1M}_X,  r^{\mu \nu}_{lk}] v
\Vert^{}_M\Big)\Big) & {(2.43)}\cr
& \leq C_n  \Vert v \Vert^{}_{M^{}_{n+1}} + C  \sum_{X \in \Pi}
\sum_{l,k}  \Vert [T^{1M}_X,  r^{\mu \nu}_{lk}] v \Vert^{}_M,\cr
}$$
$\vert \mu \vert = \vert \nu \vert = n+1$,
$r^{\mu \nu}_{lk} = r^{\mu \nu}_{lk}(x, -i \nabla)$.
Let $v = (f, \dot{f}) \in M_\infty$ and let $r$ be one of the partial
differential operators $r^{\mu \nu}_{lk}$,
$\vert \mu \vert = \vert \nu \vert =n+1$. Definition (1.5) of $T^{1}$
restricted to $M_\infty$ give for $v = (f, \dot{f})$
$$\eqalignno{
&\sum_{X \in \Pi}  \Vert [T^{1M}_X, r] v \Vert^{}_M &(2.44)\cr
& \quad{}\leq   C  \Big(\sum_i \Vert [\partial_i, r] v \Vert^{}_M
+ \sum_{i<j}  \Vert [x_i  \partial_j - x_j  \partial_i, r]
v \Vert^{}_M + \Vert \vert \nabla \vert^{\rho - 1} [\Delta, r] f
\Vert^{}_{L^2}\cr
&\qquad {} +  \sum_i  \Vert \vert \nabla \vert^\rho  [x_i, r]
\dot{f} \Vert^{}_{L^2} + \sum_i  \Vert \vert \nabla
\vert^{\rho - 1} [\sum_l  \partial_l  x_i  \partial_l, r] f
\Vert^{}_{L^2}\Big),\quad r = r (x, -i \nabla).
}$$
By the same argument which led to (2.37) it follows that there exist
polynomials $p (X, x, \xi)$, for $X = P_\mu$ and for $X = M_{ij}$,
$1 \leq i < j \leq 3$, and polynomials $p^{(i)} (X, x, \xi)$, for
$X = M_{0j}$, $i=1,2$, and $1 \leq j \leq 3$, such that
$$\eqalignno{
[\partial_i,  r (x, -i \nabla)] & = p(P_i, x, -i \nabla), & (2.45) \cr
[\Delta, r(x, -i \nabla)] & = p (P_0, x, -i \nabla),\cr
[x_i  \partial_j - x_j  \partial_i,  r(x, -i \nabla)] & =
p (M_{ij}, x, -i \nabla),\quad  1 \leq i < j \leq 3,\cr
[x_i,  r (x, -i \nabla)] & = p^{(1)} (M_{0i}, x, -i \nabla), \cr
[\sum_l \partial_l  x_i  \partial_l,  r(x, -i \nabla)] & = p^{(2)}
(M_{0i}, x, -i \nabla).\cr
}$$
Since, according to (2.41) $r$ satisfies $r(a^{-1} x, a \xi) = r(x, \xi)$
and $\deg r \leq 2n$, it follows from (2.45) that
$$\eqalignno{
p (P_i, a^{-1} x, a \xi) & = a  p (P_i, x, \xi),
\hskip20.5mm\deg  p(P_i) \leq 2 n-1, &{(2.46a)}\cr
p (P_0, a^{-1} x, a \xi) & = a^2  p(P_0, x, \xi),
\hskip18.5mm\deg p(P_0) \leq 2n,  &{(2.46b)}\cr
p (M_{ij}, a^{-1} x, a \xi) & = p (M_{ij}, x, \xi),
\hskip20mm\deg  p(M_{ij}) \leq 2n, &{(2.46c)}\cr
p^{(1)} (M_{0i}, a^{-1} x, a \xi) & = a^{-1}  p^{(1)} (M_{0i}, x, \xi),
\hskip10.5mm\deg  p^{(1)} (M_{0i}) \leq 2 n-1, &{(2.46d)}\cr
p^{(2)} (M_{0i}, a^{-1} x, a \xi) & = a  p^{(2)} (M_{0i}, x, \xi),
\hskip14.2mm\deg  p^{(2)} (M_{0i}) \leq 2 n+1. &{(2.46e)}\cr
}$$
It follows from (2.45), (2.46a) and (2.46c) that
$$\sum_i  \Vert [\partial_i, r] v \Vert^{}_M + \sum_{i<j}
\Vert [x_i \partial_j - x_j  \partial_i, r] v \Vert \leq C_n q^M_n (v).
\eqno{(2.47)}$$
Due to (2.46b) and (2.46e) we can write:
$$\eqalignno{
\hbox{\rm a)}\ &p (P_0, x, -i \nabla) = \sum_j   \partial_j
p_j (P_0, x, -i \nabla),\quad \deg  p_j (P_0) \leq 2 n-1,\hskip42mm&{(2.48)}\cr
& p_j (P_0, a^{-1} x, a \xi) = a  p_j (P_0, x, \xi),\cr
 & \cr
\hbox{\rm b)}\ & p^{(2)} (M_{0j}, x, -i \nabla) = \sum_l  \partial_l
 p^{(2)}_l (M_{0j}, x, -i \nabla),
\quad \deg  p^{(2)}_l (M_{0j}) \leq 2n, & {(2.49)}\cr
& p^{(2)}_l (M_{0j}, a^{-1} x, a \xi) = p^{(2)}_l (M_{0j}, x, \xi).\cr
}$$
It follows from (2.45), (2.48) and (2.49) that,
(leaving out the arguments $x, -i \nabla$)
$$\eqalignno{
&\Vert \vert \nabla \vert^{\rho - 1}  [\Delta, r] f \Vert^{}_{L^2} + \sum_j
\Vert \vert \nabla \vert^{\rho - 1} [\sum_k  \partial_k  x_j
 \partial_k, r] f \Vert^{}_{L^2} & {(2.50)}\cr
&\qquad{} \leq \sum_j  \Vert \vert \nabla \vert^{\rho - 1} \partial_j
p_j (P_0) f \Vert^{}_{L^2} + \sum_{j,l}  \Vert \vert \nabla \vert^{\rho - 1}
 \partial_l  p^{(2)}_l (M_{0j}) f \Vert^{}_{L^2}\cr
&\qquad{}\leq \sum_j  \Vert \vert \nabla \vert^\rho  p_j (P_0) f \Vert^{}_{L^2}
+
\sum_{j,l} \Vert \vert \nabla \vert^\rho  p^{(2)}_l (M_{0j}) f
\Vert^{}_{L^2}\cr
&\qquad{}\leq C_n  q^M_n  ((f, 0)),\cr
}$$
for some constant $C_n$. Due to (2.46d) we can write
$$\partial_k [x_l, r] = \partial_k  p^{(1)} (M_{0l}, x, -i \nabla)
= p^{(1)}_k (M_{0l}, x, -i \nabla), \eqno{(2.51)}$$
$$\deg  p^{(1)}_k (M_{0l}) \leq 2n,\quad  p^{(1)}_k (M_{0l}, a^{-1} x,
a \xi) = p^{(1)}_k (M_{0l}, x, \xi).$$
It follows from (2.51) that
$$\eqalignno{
\sum_i  \Vert \vert \nabla \vert^\rho [X_i, r] \dot{f} \Vert^{}_{L^2}
& \leq
C  \sum_{i,j} \Vert \vert \nabla \vert^{\rho - 1} \partial_j [X_i, r] \dot{f}
\Vert^{}_{L^2}
& {(2.52)}\cr
& = C  \sum_{i,j}  \Vert \vert \nabla \vert^{\rho - 1}
p^{(1)}_j (M_{0i}) \dot{f} \Vert^{}_{L^2}\cr
&\leq C_n  q^M_n  ((0, \dot{f})).\cr
}$$
We obtain from inequalities (2.44), (2.47), (2.50) and (2.52) that
$$\sum_{X \in \Pi}  \Vert [T^{1M}_X, r] v \Vert^{}_M \leq C_n  q^M_n
(v), \eqno{(2.53)}$$
for some constant $C_n$.

It follows from (2.43) and (2.53) that (with new $C_n$)
$$\Vert M_\mu  \partial^\nu  v \Vert^{}_M \leq C_n (\Vert v
\Vert^{}_{M^{}_{n+1}} +
q^M_n (v)),\quad  \vert \mu \vert = \vert \nu \vert = n+1. \eqno{(2.54)}$$
It now follows from inequalities (2.39), (2.40) and (2.54) that
$$q^M_{n+1} (v) \leq C'_n \big(q^M_n (v) + \Vert v \Vert^{}_{M^{}_{n+1}}\big),
\eqno{(2.55)}$$
and then by the induction hypothesis that
$$q^M_{n+1} (v) \leq C_n  \Vert v \Vert^{}_{M^{}_{n+1}}, \eqno{(2.56)}$$
for some constant $C_n.$

The inequality $q^D_{n+1} (\alpha) \leq C_n  \Vert \alpha
\Vert^{}_{D^{}_{n+1}}$
is proved in a similar way, but one has only to consider the degree of
the corresponding polynomials $r (x, \xi)$ and not their homogeneity
properties. Since the proof of this part is very similar, we omit it.
Together with (2.56) we then have
$$q^{}_{n+1} (u) \leq C_n       \Vert u \Vert^{}_{E^{}_{n+1}},$$
which according to the induction hypothesis proves the second inequality
of the theorem.

To prove that $E_{c}$ is dense in $E_{\infty}$ we first observe that
$C^{\infty}_{0}(\Rrm^3,\Crm)$ is dense in $S(\Rrm^3,\Crm)$ which after
inverse Fourier transform shows that $D_{c}$ is dense in $D_{\infty}$.
We therefore only have to prove that $M_{c}$ is dense in $M_{\infty}$.
Let $\varphi \in C^{\infty}_{0}(\Rrm^3)$, $0\leq \varphi(k)\leq 1$ for
$k\in\Rrm^3$, $\varphi(k)=1$ for $\vert k\vert\leq 1$,
$\varphi(k)=0$ for $\vert k\vert\geq 2$. Let $\varphi_{n}(k)
=\varphi(n^{-1}k)(1-\varphi(nk))$ for $k\in\Rrm^3$, $n\in\Nrm$, $n\geq 2$,
and let $\psi_{n}=1-\varphi_{n}$. Then $0\leq \psi_{n}(k)\leq 1$  for
$k\in\Rrm^3$ and $\psi_{n}(k)=0$ for $2/n \leq \vert k\vert \leq n$.
Moreover if
$\vert \alpha\vert \geq 1$ then $ \vert k\vert^{\vert\alpha\vert}
\vert \partial^{\alpha}\psi_{n}(k)\vert \leq 2^{\vert\alpha\vert}
C_{\vert\alpha\vert}$ for $\vert k\vert \leq 2/n$ and
$\vert \partial^{\alpha}\psi_{n}(k)\vert \leq n^{-\vert\alpha\vert}
C_{\vert\alpha\vert}$ for $\vert k\vert \geq n$, where the $C_{l}$ are
constants
independent of $n$ and $k$. For $(f,\dot{f})\in M_{\infty}$ we define
$\hat{f}^{(n)}(k)=\varphi_{n}(k)\hat{f}(k)$ and
$\fhdn(k)=\varphi_{n}(k)\fhd(k)$. Then
$(f^{(n)},\dot{f}^{(n)})\in M_{c}$ and
$\hat{f}^{(n)}- \hat{f}= \psi_{n}\hat{f}$,
$\fhdn- \fhd= \psi_{n}\fhd$.
The first inequality of the theorem and the Plancherel theorem now show that
$$\eqalign{
& \Vert (\hat{f}^{(n)}- \hat{f},\fhdn- \fhd)
\Vert^{2}_{M^{}_N}\cr
&\qquad{} \leq C''_{N}\sum_{\vert \mu\vert\leq\vert\nu\vert\leq N}
\int_{\Rrm^3-B_n}
\left( \vert k\vert^{2\rho}\vert \partial^{\mu}(k^{\nu}
\psi_{n}(k)\hat{f}(k))
\vert^{2}+\vert k\vert^{2\rho-2}
\vert \partial^{\mu}(k^{\nu} \psi_{n}(k)\fhd(k))
\vert^{2}\right) dk,\cr
}$$
$n\geq 2$, where $B_n =\{k\in\Rrm^3\big\vert 2/n \leq \vert k\vert\leq n\}$.
The estimates
of $\psi_n$ and commutation of $\partial_{\mu}$ and $k^{\nu}$ give that
$$\eqalign{
&\Vert (\hat{f}^{(n)}- \hat{f},\fhdn- \fhd)
\Vert^{2}_{M^{}_N}\cr
&\qquad{}\leq C''_{N}\sum_{\vert \mu\vert\leq\vert\nu\vert\leq N}
\int_{\Rrm^3-B_n}
\left( \vert k\vert^{2\rho}\vert \partial^{\mu}(k^{\nu}\hat{f}(k))
\vert^{2}+\vert k\vert^{2\rho-2}
\vert \partial^{\mu}(k^{\nu} \fhd(k))
\vert^{2}\right) dk.\cr
}$$
The last inequality converges to zero when $n\fl \infty$ since
$q^{}_{N}((f,\dot{f}))$ is finite. This proves the theorem.

The elements of the space have important asymptotic decrease properties:
\saut
\noindent{\bf Lemma 2.10.}
{\it
Let $1 \leq p < \infty$ and let $f \in L^p (\Rrm^3)$ be such that $M_\alpha
\partial^\beta  f \in L^p (\Rrm^3),  M_\alpha (x) =x^\alpha$, for
$0 \leq \vert \alpha \vert \leq \vert \beta \vert \leq n$. If $\nu$ is a
multi-index
and $(n - \vert \nu \vert)  p > 3$, then (after a change on a set of measure
zero)
$$(1 + \vert x\vert)^{3/p + \vert \nu \vert} \vert \partial^\nu  f(x) \vert
\leq C_{\nu, p}  \sum_{\vert \alpha \vert \leq \vert \beta \vert \leq n}
 \Vert M_\alpha  \partial^\beta  f \Vert^{}_{L^p}.
\eqno{(2.57)}$$
}\saut
\noindent{\it Proof.}
A Sobolev embedding gives at once with $l p > 3$ that
$$\Vert \partial^\nu  f \Vert^{}_{L^\infty}
\leq C  \sum_{\vert \mu \vert \leq l}
 \Vert \partial^{\mu+\nu} f \Vert^{}_{L^p} \leq C  \sum_{\vert \beta
\vert \leq n}  \Vert \partial^\beta  f \Vert^{}_{L^p}, \eqno{(2.58)}$$
and that $\partial^\nu  f$ is a continuous function vanishing at $\infty.$

Let $\varphi \in C^\infty_0 (\Rrm^3)$, $\varphi (x) = 1$ for $1/2
\leq \vert x \vert \leq 2$, $\varphi (x) = 0$ for $\vert x \vert \leq
1/3$,
$\varphi (x) = 0$ for $\vert x \vert > 3$ and $\varphi (x) \geq 0$,
$ x \in \Rrm^3$. We define $g_a (x) = f (ax)$. Since $a^{\vert \nu \vert}
(\partial^\nu f) (ax) = \partial^\nu  g_a (x)$ and $\partial^\nu
g_a (x) = \partial^\nu (g_a  \varphi) (x)$ for $1/2 \leq \vert x \vert
\leq 2$, we obtain from (2.58)
$$\eqalignno{
\sup_{{1/2} \leq \vert x \vert \leq 2}  \vert a^{\vert \nu \vert}
(\partial^\nu  f) (ax) \vert & \leq \Vert \partial^\nu  g_a
\varphi \Vert^{}_{L^\infty} & (2.59)\cr
& \leq \sum_{\vert \beta \vert \leq n}  \Vert \partial^\beta (g_a
 \varphi) \Vert^{}_{L^p} \cr
 &\leq C_n  \sum_{\vert \beta_1 \vert +
 \vert \beta_2 \vert\leq n}
 \Vert (\partial^{\beta_1} g_a) (\partial^{\beta_2} \varphi) \Vert^{}_{L^p},
\cr
 }$$
where $(n - \vert \nu \vert)  p > 3$ and where we have used the Leibniz
rule in the last inequalities.

A change of coordinates gives
$$\Vert (\partial^{\beta_1} g_a) (\partial^{\beta_2} \varphi) \Vert^{}_{L^p} =
a^{\vert \beta_1 \vert
- 3/p} \Big(\int \vert (\partial^{\beta_1} f) (x)  (\partial^{\beta_2} \varphi)
({x/a}) \vert^p dx\Big)^{1/p}.$$
Since the integrand vanishes outside the set
$\{ x \big\vert {a/3} \leq \vert x \vert \leq 3a \}$, it follows that
$$\eqalign{
a^{3/p} \Vert (\partial^{\beta_1} g_a) (\partial^{\beta_2} \varphi)
\Vert^{}_{L^p} & \leq C_{n,p}  \Vert \partial^{\beta_2} \varphi
\Vert^{}_{L^\infty} \Vert \vert x \vert^{\vert \beta_1 \vert}
\partial^{\beta_1} f \Vert^{}_{L^p}\cr
& \leq C'_{n,p}  \sum_{\vert \alpha \vert \leq \vert \beta_1 \vert}
 \Vert M_\alpha  \partial^{\beta_1} f \Vert^{}_{L^p}\cr
& \leq C'_{n,p}  \sum_{\vert \alpha \vert \leq \vert \beta \vert\leq n}
\Vert M_\alpha  \partial^{\beta} f \Vert^{}_{L^p}.\cr
}$$
This inequality and (2.59) show that for ${1/2} \leq \vert y \vert \leq 2$
$$a^{3/p + \vert \nu \vert} \vert (\partial^\nu f) (ay) \vert \leq C_{\nu,p}
 \sum_{\vert \alpha \vert \leq \vert \beta \vert \leq n}
\Vert M_\alpha  \partial^\beta  f \Vert^{}_{L^p},$$
which with $a = \vert x \vert > 0,  y = {x/\vert x \vert}$, together
with (2.58) gives estimate (2.57). This proves the lemma.
\saut
\noindent{\bf Remark 2.11.}
Lemma 2.10 is still true if $p = \infty$, $n = \vert \nu \vert$. In this
case the proof is trivial.
\saut
\noindent{\bf Theorem 2.12.}
{\it
If $(f, 0) \in M^\rho_1,  {1/2} < \rho < 1$, then
$$(1 + \vert x\vert)^{3/2 - \rho} \vert f(x) \vert \leq C  \Vert (f,0)
\Vert^{}_{M^\rho_1} \eqno{(2.60{\rm a})}$$
and if $(f, \dot{f}) \in M^\rho_{\vert \nu \vert + 2},  1/2 < \rho < 1$,
then
$$(1 + \vert x \vert)^{5/2 + \vert \nu \vert - \rho} \big(\vert \partial^\nu
\partial_i  f(x) \vert + \vert \partial^\nu  \dot{f} (x) \vert\big)
\leq C_{\vert \nu \vert}  \Vert (f, \dot{f})
\Vert^{}_{M^{}_{\vert \nu \vert + 2}}.\eqno{(2.60\hbox{b})}$$
}\saut
\noindent{\it Proof.}
Let $p = 6 (3 - 2 \rho)^{-1}$. Then $3 < p < 6$ for $1/2 < \rho < 1$. Suppose
for the moment that $f \in C^\infty_0 (\Rrm^3)$. The inequality
(cf. Theorem 4.5.3. of \refHA)
$$\Vert f \Vert^{}_{L^p (\Rrm^3)} \leq C_p  \Vert \vert \nabla \vert^\rho
 f \Vert^{}_{L^2 (\Rrm^3)}, \quad  p = 6 (3 - 2 \rho)^{-1}, \eqno{(2.61)}$$
where the constant $C_p$ is independent of the support of $f$, and inequality
(2.57) with $n - \vert \mu \vert = 1 > {3/p}$ of lemma 2.10 then give
$$\eqalignno{
(1 + \vert x \vert)^{3/2 - \rho+\vert \mu\vert}
\vert \partial^\mu  f(x) \vert
& \leq C_{\mu, \rho}  \sum_{\vert \alpha \vert \leq \vert \beta \vert \leq
\vert \mu \vert + 1}  \Vert \vert \nabla \vert^\rho  M_\alpha
 \partial^\beta  f \Vert^{}_{L^2}  & (2.62\hbox{a})\cr
& \leq C'_{\mu, \rho}  q^M_{\vert \mu \vert + 1}  (f,0),\cr
}$$
where $q^M$ was defined in (2.32a). It follows from this inequality and Theorem
2.9 that
$$(1 + \vert x \vert)^{3/2 - \rho + \vert \mu \vert}  \vert \partial^\mu
 f(x) \vert \leq C_{\mu, \rho}  \Vert (f,0)
 \Vert^{}_{M^{}_{1+ \vert \mu \vert}},
\eqno{(2.62\hbox{b})}$$
for $(f, 0) \in M_{1+ \vert \mu \vert}$. As a matter of fact,
$C^\infty_0 (\Rrm^3,\Rrm^4) \oplus C^\infty_0 (\Rrm^3, \Rrm^4)$ is dense
in $S (\Rrm^3, \Rrm^4) \oplus S(\Rrm^3, \Rrm^4)$, which by definition is
dense in $M^\rho_n$. In particular, with $\mu = 0$, this proves (2.60a).

According to (2.62a) we have
$$(1 + \vert x \vert)^{3/2 - \rho + \vert \mu \vert}  \vert \partial^\mu
 x_i  \dot{f} (x) \vert \leq C_{\mu, \rho}  q^M_{\vert \mu \vert + 1}
 (Q^{}_i  \dot{f}, 0), \eqno{(2.63)}$$
where $(Q^{}_i  \dot{f}) = x_i  \dot{f}(x)$. By the definition of $q^M$
$$\eqalignno{
q^M_{\vert \mu \vert + 1}  (Q^{}_i  \dot{f}, 0)^2 &
= \sum_{\vert \alpha \vert\leq \vert \beta \vert \leq \vert \mu
\vert + 1}
\Vert \vert \nabla \vert^\rho M_\alpha  \partial^\beta  Q^{}_i  \dot{f}
\Vert^2_{L^2}& {(2.64)}\cr
&= \sum_{\scr \vert \alpha \vert \leq \vert \beta \vert \leq
\vert \mu \vert + 1\atop \scr 1 \leq j \leq 3}
\Vert \vert \nabla \vert^{\rho - 1}  \partial_j
M_\alpha  \partial^\beta  Q^{}_j  \dot{f} \Vert^2_{L^2}.\cr
}$$
But $\partial_j  M_\alpha  \partial^\beta  Q^{}_l =
Q^\beta_{j \alpha l} (x, -i \nabla)$, where $Q^\beta_{j \alpha l} (x, \xi)$
is a polynomial of degree
$\vert \alpha \vert + \vert \beta \vert + 2 \leq 2  \vert \mu \vert + 4$
and $Q^\beta_{j \alpha l} (a^{-1} x, a \xi) = a^{\vert \beta \vert -
\vert \alpha \vert} Q^\beta_{j \alpha l} (x, \xi)$. Therefore
$$q^M_{\vert \mu \vert + 1} (Q^{}_i  \dot{f}, 0) \leq C_{\vert \mu \vert}
 q^M_{\vert \mu \vert + 2} (0, \dot{f}),$$
which together with (2.63) and (2.64) shows that
$$(1 + \vert x \vert)^{3/2 - \rho + \vert \mu \vert} \vert \partial^\mu
x_i  \dot{f} \vert \leq C_{\rho, \mu}  q^M_{\vert \mu \vert + 2}
 (0, \dot{f}) \leq C'_{\rho, \mu} \Vert (0, \dot{f})
 \Vert^{}_{M^\rho_{\vert \mu \vert + 2}},
\eqno{(2.65)}$$
where the last inequality follows from Theorem 2.9.

In a similar way it follows from (2.62a) that
$$\vert \partial^\mu  \dot{f} (x) \vert \leq C_{\mu, \rho}
\Vert (0, \dot{f}) \Vert^{}_{M^\rho_{\vert \mu \vert + 2}}. \eqno{(2.66)}$$
Inequalities (2.65) and (2.66) with $\mu = 0$ prove (2.60b) with $\nu = 0$ and
$f = 0$. Since $[\partial^\mu, x_i]$ is a monomial of degree
$\vert \mu \vert - 1$ in $\nabla$, it follows from (2.65) and (2.66)
by induction that (2.60b) is true for $\vert \nu \vert \geq 0$, with $f = 0.$

Let $\partial^\mu = \partial^\nu  \partial_i$ in (2.62b). Then (2.60b),
with $\dot{f} = 0$, follows from (2.62b). Since (2.60b) is true for the
particular cases $(f, 0) \in M^\rho_{\vert \nu \vert + 2}$ and $(0, \dot{f})
\in M^\rho_{\vert \nu \vert + 2}$, it is also true for $(f, \dot{f}) \in
M^\rho_{\vert \nu \vert + 2}$, because
$\Vert (f,0) \Vert^{}_{M_n} + \Vert (0, \dot{f}) \Vert^{}_{M_n} \leq \sqrt 2
\Vert (f, \dot{f} \Vert^{}_{M_n}$. This proves the theorem.

We shall prove that $M^\rho$ {\it contains the long range potentials}
which it is expected to contain. It follows from the next theorem that
$(f, \dot{f}) \in  E^\rho_\infty$
if $(1 + \vert x \vert)^{a + \vert \nu \vert} \vert \partial^\nu  f(x) \vert$
and $(1 + \vert x \vert)^{a + 1 + \vert \nu \vert} \vert \partial^\nu
\dot{f} (x) \vert$ are uniformly bounded in $x$ for some $a > {3/2} - \rho$.
\psaut
\noindent{\bf Theorem 2.13.}
{\it
Let $1/2 < \rho < 1,  p = 6(5 - 2 \rho)^{-1},  q = 6(3 - 2 \rho)^{-1},$
$f \in L^q$ and $\dot{f} \in L^p$. If $M_\alpha  \partial^\beta
\partial_i  f \in L^p (\Rrm^3, \Rrm^4)$ and $M_\alpha  \partial^\beta
 \dot{f} \in L^p (\Rrm^3, \Rrm^4)$ for $0 \leq \vert \alpha \vert \leq
\vert \beta \vert,  1 \leq i \leq 3$, then $(f, \dot{f}) \in M_n$ and
$$\Vert (f, \dot{f}) \Vert^{}_{M^{}_n}
\leq C_n \Big(\sum_{\scr 0 \leq \vert \alpha \vert \leq
\vert \beta \vert \leq n\atop\scr
1 \leq i \leq 3} \Vert M_\alpha  \partial^\beta
 \partial_i  f \Vert^{}_{L^p} + \sum_{0 \leq \vert \alpha \vert \leq
\vert \beta \vert \leq n} \Vert M_\alpha  \partial^\beta  \dot{f}
\Vert^{}_{L^p}\Big),\quad  n \geq 0.$$
}
\noindent{\it Proof.}
Suppose for the moment that $\dot{f} \in C^\infty_0 (\Rrm^3)$.
The inequality (cf. Theorem 4.5.3. of \refHA)
$$\Vert \vert \nabla \vert^{\rho - 1} \dot{f} \Vert^{}_{L^2 (\Rrm^3)} \leq C_p
\Vert \dot{f} \Vert^{}_{L^p (\Rrm^3)}, \quad p = 6(5 - 2 \rho)^{-1},
\eqno{(2.67)}$$
where $C_p$ is independent of the support of $\dot{f}$, gives $\Vert (0,
\dot{f}) \Vert^{}_{M^\rho} \leq C_p \Vert f \Vert^{}_{L^p},
f \in C^\infty_0$. Since $C^\infty_0$
is dense $L^p$, for our given $p$, and dense in $S (\Rrm^3)$,
it follows by continuity and by the definition of the space $M^\rho$ that
$$\Vert (0,\dot{f}) \Vert^{}_{M^\rho} \leq C_p \Vert \dot{f} \Vert^{}_{L^p},
\quad p = 6(5 - 2 \rho)^{-1},  \dot{f} \in L^p, \eqno{(2.68)}$$
and that $(0, \dot{f}) \in M^\rho$ for $\dot{f} \in L^p.$

According to inequality (2.68), Theorem 2.9 and definition (2.32a) of $q^M_n$
we have
$$\Vert (0, \dot{f}) \Vert^{}_{M^\rho_n} \leq C_{\rho,n}  q^M_n (0, \dot{f})
\leq C'_{\rho,n}  \sum_{0 \leq \vert \alpha \vert \leq \vert \beta \vert \leq
n}
\Vert M_\alpha  \partial^\beta  \dot{f} \Vert^{}_{L^p}, \eqno{(2.69)}$$
$p = 6(5 - 2 \rho)^{-1}$, $n \geq 0$, if $M_\alpha  \partial^\beta
 \dot{f} \in L^p$ for $0 \leq \vert \alpha \vert \leq \vert\beta\vert
 \leq n$.
Since $\Vert (f,0) \Vert^{}_{M^\rho_n} \leq C_{\rho, n}  q^M_n (f,0)$
according to Theorem 2.9 and
$q^M_n  (f,0) \leq C_n \sum_i  q^M_n (0, \partial_i  f)$
using definition (2.32a) of
$q^M_n$ it follows from (2.69) and the triangle inequality that the
inequality of the theorem is true.
\saut
\noindent{\bf Corollary 2.14.}
{\it
Let $n \geq 0$, $1/2 < \rho < 1$, $f \in C^{n+1} (\Rrm^3, \Rrm^4)$,
$\dot{f} \in C^n (\Rrm^3, \Rrm^4)$ and let
$$
\Gamma_{n,a} (f, \dot{f})
= \sum_{\vert \nu \vert \leq n+1}  \sup_x
\big((1 + \vert x \vert)^{a + \vert \nu \vert} \vert \partial^\nu
f(x) \vert\big) + \sum_{\vert \nu \vert \leq n} \sup_x
\big((1 + \vert x\vert)^{a + 1 + \vert \nu \vert}
\vert \partial^\nu  \dot{f} (x) \vert\big).$$
If $\Gamma_{n,a}  (f, \dot{f}) < \infty$ for some $a > 3/2 - \rho$, then
$(f, \dot{f}) \in M^\rho_n$ and $\Vert (f, \dot{f})
\Vert^{}_{M^{\rho}_n} \leq C_n  \Gamma_{n,a}  (f, \dot{f}).$
}\saut
\noindent{\it Proof.}
Let $a$ be such that $\Gamma_{n,a}  (f,\dot{f})$ is finite. Then the
hypotheses of Theorem 2.13 are satisfied and the right-hand side of the
inequality in that theorem is bounded by $C_n  \Gamma_{n, a}$,
after redefining the constants. This proves the corollary.

Later we shall need estimates of weighted supremum norms of solution of the
homogeneous wave equation. These estimates follow directly from
Kirchhoff's formula and Theorem 2.12.
\saut
\noindent{\bf Proposition 2.15.}
{\it
If $n \geq 1$, $(f, \dot{f}) \in M^\rho_{n+2}$, $1/2 < \rho < 1$,
then the solution $u$ of the wave equation $\carre  u = 0$, with initial
conditions $u(0, x) = f(x)$,  ${\partial \over \partial t} u (t,x)
\vert_{t=0} = \dot{f}(x)$,
satisfies
$$\eqalign{
&(1 + \vert x \vert + \vert t \vert)^{3/2 - \rho}
\vert u(t,x) \vert +(1 + \vert x \vert + \vert t \vert)\cr
&\qquad\sum_{1 \leq \vert \nu \vert + l \leq n}
(1 + \big\vert\vert t \vert - \vert x \vert\big\vert)^{1/2 - \rho +
\vert \nu \vert + l}
\vert \partial^\nu\big({\partial \over \partial t}\big)^l  u(t,x) \vert\cr
&\qquad\qquad{}\leq C_{n,\rho}  \Vert (f, \dot{f}) \Vert^{}_{M^{\rho}_{n+2}}
}$$
}\saut
\noindent{\it Proof.}
Give first $f, \dot{f} \in S (\Rrm^3,\Rrm^4)$. Then
$\partial^\alpha (\partial / \partial t)^m u (t,x)
= u_{\alpha,m} (t,x),  \alpha = (\alpha_1, \alpha_2,\alpha_3)$,
is the solution of the wave equation with initial conditions $f_{\alpha,m},
 \dot{f}_{\alpha,m}$, where $(f_{\alpha,m}, \dot{f}_{\alpha,m}) =
 T^{M1}_{P^\beta}(f, \dot{f})$, where
 $\beta = (m, \alpha_1, \alpha_2, \alpha_3),  P^\beta
 =P^{\beta_0}_0  P^{\beta_1}_1  P^{\beta_2}_2  P^{\beta_3}_3$
is an element in the enveloping algebra $U(\p)$ and where $T^{M1}_X$
denotes the restriction to the space $M^{\rho}$ of the linear Lie algebra
representation $T^1_X$ as
defined by (1.5). As the initial data are in $S(\Rrm^3,\Rrm^4)$ it is sure
that the solution is given by Kirchhoff's formula (cf. \refHA):
$$u_\beta (t,x) = (4 \pi)^{-1}  \int_{\vert \omega \vert = 1}
\Big(f_\beta (x + t \omega) + t \sum_{i}\omega^{i}\partial_{i}
f_\beta (x + t \omega) +t \dot{f}_\beta (x + t \omega)\Big)  d \omega.
\eqno{(2.70)}$$
Let $\Gamma_{n,a}$ be as in Corollary 2.14 and let
$$j_a (t, x) = (4 \pi)^{-1}  \int_{\vert \omega \vert = 1} (1 + \vert x +
\omega t \vert^2)^{- a / 2} d \omega,\quad  a \in \Rrm. \eqno{(2.71)}$$
It follows from (2.70) and (2.71) that
$$\vert u_\beta (t,x) \vert \leq \Gamma_{0,a}  (f_\beta, \dot{f}_\beta)
 \big(j_a (t,x) + \vert t \vert  j_{a+1} (t,x)\big). \eqno{(2.72)}$$
The easy explicit evaluation of the integral in (2.71) in polar coordinates
leads to
$$\eqalignno{
j_a (t,x) & \leq C_a (1 + \vert x \vert + \vert t \vert)^{-a}  \quad
\hbox{for}\ 0 < a < 2, & {(2.73a)}\cr
j_a (t,x) & \leq C_a (1 + \vert x \vert + \vert t \vert)^{-2}
(1 + \big\vert\vert t \vert - \vert x \vert\big\vert)^{2-a} \quad
\hbox{ for } a > 2. & {(2.73\hbox{b})}\cr
}$$
We choose $a = 3/2 - \rho + \vert \beta \vert$ in (2.72), where
$1/2 < \rho < 1.$ Then $1/2 + \vert \beta \vert < a < 1 + \vert \beta \vert$
and it follows from (2.73) that
$$\eqalignno{
j_a (t,x) + \vert t \vert  j_{a+1} (t,x) & \leq C_{\rho, \vert \beta \vert}
(1 + \vert x \vert + \vert t \vert)^{- (3/2 - \rho)},\quad
\vert \beta \vert = 0, & {(2.74\hbox{a})}\cr
j_a (t,x) + \vert t \vert  j_{a+1} (t,x)
&\leq C_{\rho, \vert \beta \vert} (1 + \vert x \vert + \vert t \vert)^{-1} (1 +
\big\vert \vert t \vert -
\vert x \vert\big\vert)^{- (1/2 - \rho + \vert \beta \vert)},
\quad \vert \beta \vert \geq 1,\qquad\quad & {(2.74\hbox{b})}\cr
}$$
where $a = 3/2 - \rho + \vert \beta \vert.$

Estimates (2.72) and (2.74) show that, if
$$\eqalign{
Q^{}_n (u)&  = \sup_{t,x}       \big((1 + \vert x \vert +
\vert t \vert)^{3/2 - \rho}  \vert u (t,x) \vert\big)\cr
&\qquad {}+ \sum_{1 \leq \vert \beta \vert \leq n}  \sup_{t,x}  \big((1 +
\vert x \vert + \vert t \vert)  (1 + \big\vert \vert t \vert - \vert x
\vert\big\vert)^{1/2 - \rho + \vert \beta \vert}
 \vert u_\beta (t,x) \vert\big),\cr
 }$$
then
$$Q^{}_n (u) \leq \sum_{0 \leq \vert \beta \vert \leq n}
C_{\rho, \vert \beta \vert}
 \Gamma_{0, 3/2 - \rho + \vert \beta \vert} (f_\beta, \dot{f}_\beta).
\eqno{(2.75)}$$
It follows from the definition of $\Gamma_{n,a}$ and from
$T^1_{P^\beta} = \partial^\alpha \left({\scr 0\ I\atop\scr \Delta\
0}\right)^t$,
$\beta = (t,\alpha)$, that
$$\Gamma_{0, 3/2 - \rho + \vert \beta \vert} (f_\beta, \dot{f}_\beta) \leq
\Gamma_{\vert \beta \vert, 3/2 - \rho}  (f, \dot{f}),$$
which inserted into (2.75) gives (with new constant)
$$Q^{}_n (u) \leq C_{\rho,n}  \Gamma_{n, 3/2 - \rho}  (f, \dot{f}).
\eqno{(2.76)}$$
Inserting the result $\Gamma_{n, 3/2 - \rho} (f, \dot{f}) \leq C_{n, \rho}
\Vert (f, \dot{f}) \Vert^{}_{M^{}_{n+2}}$ from Theorem 2.12 into
inequality (2.76) we obtain for some constant $C_{\rho,n}$
$$Q^{}_n (u) \leq C_{\rho,n}  \Vert (f, \dot{f}) \Vert^{}_{M^{}_{n+2}}.
\eqno{(2.77)}$$
Since $S (\Rrm^3,\Rrm^4) \oplus S (\Rrm^3,\Rrm^4)$ is dense in the space
$M_{n+2}$ by construction,
it follows that (2.77) is true for $(f, \dot{f}) \in M_{n+2}$.
This proves the proposition.

It follows directly from definitions (1.5), (1.7) and (1.8) of $T_X,
X \in \p$, that $[T_X, T_Y] \equiv DT_X. T_Y - DT_Y.T_X = T_{[X,Y]}$
on the space of $C^\infty$ functions. The next  lemmas and corollary
prove in particular that $T_X$, $X \in \p$, is a continuous polynomial
from $E_\infty$ to $E_\infty$, which assures that $X \mapsto T_X$ is a
{\it nonlinear representation} of $\p$ in $E_\infty.$
\saut
\noindent{\bf Lemma 2.16.}
{\it
Let $N \geq 0$, $u_i = (f_i, \dot{f}_i, \alpha_i) \in E_\infty$,
$i = 1,2$, and let $p = 6(5 - 2 \rho)^{-1}$. Then
$$\eqalignno{
& \Vert T^2_{P_0} (u_1 \otimes u_2) \Vert^{}_{E^{}_N}
\leq C_N
\Big(\sum_{\vert \mu \vert \leq\vert \nu_1 \vert + \vert \nu_2 \vert \leq N}
\Vert M_\mu \vert \partial^{\nu_1}
\alpha_1 \vert\vert  \partial^{\nu_2} \alpha_2 \vert \Vert^{}_{L^p}\cr
&\qquad {}+ \sum_{\scr \vert \mu \vert \leq N\atop\scr
\vert \nu_1 \vert + \vert \nu_2 \vert \leq N}
\big(\Vert M_\mu  \vert \partial^{\nu_1} f_1 \vert\vert \partial^{\nu_2}
\alpha_2 \vert \Vert^{}_{L^2} + \Vert M_\mu  \vert \partial^{\nu_1} f_2
\vert\vert \partial^{\nu_2} \alpha_1 \vert  \Vert^{}_{L^2}\big)\Big)\cr
\noalign{\hbox{\it and}}
& \Vert T^2_{M_{0j}} (u_1 \otimes u_2) \Vert^{}_{E^{}_N}
\leq C_N \Big(\sum_{\vert \mu \vert
\leq \vert \nu_1 \vert + \vert \nu_2 \vert + 1 \leq N+1}
\Vert M_\mu \vert \partial^{\nu_1} \alpha_1 \vert\vert
\partial^{\nu_2} \alpha_2 \vert  \Vert^{}_{L^p}\cr
&\qquad {}+ \sum_{\scr \vert \mu \vert \leq N+1\atop\scr
\vert \nu_1 \vert + \vert \nu_2 \vert \leq N}
\big(\Vert M_\mu \vert \partial^{\nu_1} f_1 \vert\vert \partial^{\nu_2}
\alpha_2 \vert \Vert^{}_{L^2} + \Vert M_\mu \vert \partial^{\nu_1} f_2
\vert\vert \partial^{\nu_2} \alpha_1 \vert
\Vert^{}_{L^2}\big)\Big).\cr
}$$
}\saut
\noindent{\it Proof.}
It follows from (1.8a) that
$$\eqalignno{
& \Vert T^2_{P_0} (u_1 \otimes u_2) \Vert^2_{E^{}_N}& {(2.78)}\cr
&\qquad{}= \Vert (0, {1 \over 2} (\overline{\alpha}_1
 \gamma  \alpha_2 + \overline{\alpha}_2  \gamma
\alpha_1)) \Vert^2_{M^{}_n}
+ \Vert {1 \over 2} \Big(\sum^3_{\mu = 0} (f_{1 \mu}  \gamma^0
\gamma^\mu  \alpha_2 + f_{2 \mu}  \gamma^0  \gamma^\mu
 \alpha_1)\Big) \Vert^2_{D^{}_n}.\cr
}$$
It follows from Theorem 2.13 and Liebniz rule that
$$\Vert (0, \overline{\alpha}_i  \gamma  \alpha_j) \Vert^{}_{M^{}_n} \leq
C_n  \sum_{\vert \beta \vert \leq \vert \nu_1 \vert + \vert \nu_2 \vert
\leq n} \Vert M_\beta \vert \partial^{\nu_1} \alpha_i \vert \vert
\partial^{\nu_2} \alpha_j \vert
 \Vert^{}_{L^p}, \eqno{(2.79)}$$
where $p = 6 (5 - 2 \rho)^{-1}$. By definition (2.32b) of $q^D_n$ and by
the first part of Theorem 2.9:
$$\eqalignno{
\sum^3_{\mu = 0}  \Vert f_{i \mu}  \gamma^0
\gamma^\mu  \alpha_j \Vert^{}_{D^{}_n} & \leq C'_n  \sum^3_{\mu = 0}
 q^D_n (f_{i \mu}  \gamma^0  \gamma^\mu
\alpha_j) & {(2.80)}\cr
& \leq C''_n  \sum_{\scr\vert \beta \vert \leq n\atop\scr \vert \nu_1
\vert + \vert \nu_2 \vert
\leq n} \Vert M_\beta \vert \partial^{\nu_1} f_i \Vert \partial^{\nu_2}
\alpha_j \vert \Vert^{}_{L^2}.\cr
}$$
The triangle inequality, inequalities (2.78), (2.79) and (2.80) prove
 the first inequality in the lemma. The proof of the second inequality is
 so similar that we omit it.
\saut
\noindent{\bf Lemma 2.17.}
{\it
If $u_i \in E_\infty$, $i = 1,2$, and $X \in \Pi$, then
$$\eqalign{
\hbox{\rm i)}\ & \Vert T^2_X (u_1 \otimes u_2) \Vert^{}_{E^{}_N}
\leq C_{N} \big(\Vert u_1 \Vert^{}_{E^{}_{N+1}}
 \Vert u_2 \Vert^{}_{E^{}_0} + \Vert u_1 \Vert^{}_{E^{}_0}
 \Vert u_2 \Vert^{}_{E^{}_{N+1}}\big),
\quad N \geq 0,\cr
\hbox{\rm ii)}\ & \Vert T^2_X (u_1 \otimes u_2) \Vert^{}_{E^{}_N}
\leq C^{}_N \big(\Vert u_1 \Vert^{}_{E^{}_N}
 \Vert u_2 \Vert^{}_{E^{}_1} + \Vert u_1 \Vert^{}_{E^{}_1}
 \Vert u_2 \Vert^{}_{E^{}_N}\big)^{3/2 - \rho}\cr
&\quad \big(\Vert u_1 \Vert^{}_{E^{}_{N+1}}
\Vert u_2 \Vert^{}_{E^{}_0} + \Vert u_1 \Vert^{}_{E^{}_0}
 \Vert u_2 \Vert^{}_{E^{}_{N+1}}\big)^{\rho - 1/2}, \quad N \geq 0,\cr
\hbox{\rm iii)}\ &\Vert T^2_X (u_1 \otimes u_2) \Vert^{}_E
\leq C  \min \big(\Vert u_1 \Vert^{}_{E^{}_1}
\Vert u_2 \Vert^{\rho - 1/2}_{E^{}_1}
\Vert u_2 \Vert^{3/2 - \rho}_{E^{}_0},
\Vert u_1 \Vert^{\rho - 1/2}_{E^{}_1}
\Vert u_1 \Vert^{3/2 - \rho}_{E^{}_0}
\Vert u_2 \Vert^{}_{E^{}_1}\big).\hskip3.66mm\cr
}$$
}\saut
\noindent{\it Proof.}
Since $T^2_X = 0$ if $X \in \Pi$ and $X \neq P_0$,  $X \neq M_{0j}$,
$j = 1,2,3$, we have according to Lemma 2.16 for $X \in \Pi$,  $u_i = (f_i,
 \dot{f}_i, \alpha_i) \in E_\infty$, $i = 1,2$,
$$\eqalignno{
\Vert T^2_X (u_1 \otimes u_2) \Vert^{}_E
& \leq C \Big(\sum_{\vert \mu \vert \leq 1}
\Vert M_\mu \vert \alpha_1 \vert  \vert \alpha_2 \vert \Vert^{}_{L^p}
&{(2.81)}\cr
&\qquad{}+ \sum_{\vert \mu \vert \leq 1} \big(\Vert M_\mu \vert f_1 \vert
\vert
\alpha_2 \vert \Vert^{}_{L^2} + \Vert M_\mu \vert f_2 \vert  \vert \alpha_1
\vert \Vert^{}_{L^2}\big)\Big),\quad  p = 6(5 - 2 \rho)^{-1}.\cr
}$$
Since $\Vert M_\mu \vert \alpha_1 \vert  \vert \alpha_2 \vert \Vert^{}_{L^p}
\leq \Vert \alpha_1 \Vert^{}_{L^2}  \Vert M_\mu  \alpha_2 \Vert^{}_{L^q}$,
$p = 6(5 - 2 \rho)^{-1}$, $q = 3(1 - \rho)^{-1}$ and
$\Vert M_\mu \alpha_2 \Vert^{}_{L^q}$ $ \leq C  \sum_{\vert \nu \vert \leq 1}
 \Vert \partial^\nu M_\mu  \alpha_2 \Vert^{}_{L^2}$, we obtain
$$\eqalignno{
\sum_{\vert \mu \vert \leq 1} \Vert M_\mu \vert \alpha_1 \vert  \vert
\alpha_2 \vert \Vert^{}_{L^p} & \leq C  \Vert \alpha_1 \Vert^{}_{L^2}
\sum_{\scr\vert \mu \vert \leq 1 \atop\scr \vert \nu \vert \leq 1}
\Vert M_\mu  \partial^\nu
 \alpha_2 \Vert^{}_{L^2}&{(2.82)}\cr
& \leq C'  \Vert \alpha_1 \Vert^{}_D  \Vert \alpha_2 \Vert^{}_{D^{}_1},
\quad p = 6(5 - 2 \rho)^{-1},  1/2 < \rho < 1.
}$$
The inequalities $\Vert M_\mu \vert f_1 \vert  \vert \alpha_2 \vert
\Vert^{}_{L^2}
\leq \Vert f_1 \Vert^{}_{L^q}  \Vert M_\mu  \alpha_2 \Vert^{}_{L^{3 / \rho}}$,
$q = 6(3 - 2 \rho)^{-1}$, $\Vert f_1 \Vert^{}_{L^q} \leq C $
$ \Vert (f_1,0) \Vert^{}_M$
(cf. (2.61)) and $\Vert M_\mu  \alpha_2 \Vert^{}_{L^{3 / \rho}} \leq C
\sum_{\vert \nu \vert \leq 1} \Vert \partial^\nu  M_\mu  \alpha_2
\Vert^{}_{L^2}$
give
$$\sum_{\vert \mu \vert \leq 1} \Vert M_\mu \vert f_1 \vert
\vert \alpha_2 \vert \Vert^{}_{L^2} \leq C \Vert (f_1,0) \Vert^{}_M
\Vert \alpha_2 \Vert^{}_{D^{}_1}.
\eqno{(2.83)}$$
Since $\Vert (f_i,0) \Vert^{}_{M^{}_n} \leq \Vert u_i \Vert^{}_{E^{}_n},
\Vert \alpha_i \Vert^{}_{D^{}_n}
\leq \Vert u_i \Vert^{}_{E^{}_n}$ it follows from (2.81), (2.82) and
(2.83) that
$$T^2_X (u_1 \otimes u_2) \Vert^{}_E \leq C \big(\Vert u_1 \Vert^{}_E
\Vert u_2 \Vert^{}_{E^{}_1} +
\Vert u_1 \Vert^{}_{E^{}_1}  \Vert u_2 \Vert^{}_E\big),$$
which proves the first statement of the lemma in the case where $N = 0.$

Let $u = (f, \dot{f}, \alpha) \in E_\infty$ and $\vert \mu \vert \leq 1$.
It follows from (2.60a) of Theorem 2.12 and from the inequality
$$\eqalignno{
\Vert (1 + \vert x \vert)^{\rho - 1/2} \alpha \Vert^{}_{L^2}
&\leq \Vert (1 + \vert x \vert)\alpha \Vert^{\rho - 1/2}_{L^2}
\Vert \alpha \Vert^{3/2 - \rho}_{L^2} \leq \Vert \alpha
\Vert^{\rho - 1/2}_{D^{}_1}  \Vert \alpha \Vert^{3/2 - \rho}_D,\cr
\noalign{\hbox{that}}
\Vert M_\mu \vert f\vert  \vert \alpha \vert \Vert^{}_{L^2}
&\leq \sup_x \big((1 + \vert x \vert)^{3/2 - \rho} \vert f(x) \vert\big)
\Vert(1 + \vert x \vert)^{\rho - 1/2} \alpha \Vert^{}_{L^2}&{(2.84)}\cr
&\leq C  \Vert (f,0) \Vert^{}_{M^{}_1}  \Vert \alpha
\Vert^{\rho - 1/2}_{D^{}_1}  \Vert \alpha \Vert^{3/2 - \rho}_D,
\vert \mu \vert \leq 1.\cr
}$$
It follows from Lemma 2.16 that, for $X \in \Pi$,
$\Vert T^2_X (u_1 \otimes u_2) \Vert^{}_{E^{}_N}$ is bounded by a sum
 of terms of the form
$$C_N  \Vert M_\mu \vert M_{\mu_1}  \partial^{\nu_1}
\alpha_1 \vert \vert M_{\mu_2}  \partial^{\nu_2}  \alpha_2 \vert
 \Vert^{}_{L^p}
= C_N  I (\mu, \mu_1, \mu_2, \nu_1, \nu_2),\quad p = 6(5 - 2 \rho)^{-1},
\eqno{(2.85)}$$
where $\vert \mu \vert \leq 1,  \vert \nu_1 \vert + \vert \nu_2 \vert \leq
N,  \vert \mu_1 \vert \leq \vert \nu_1 \vert,  \vert \mu_2 \vert
\leq \vert \nu_2 \vert$ and of terms of the form
$$\eqalignno{
&C_N  J (\mu, \mu_1, \mu_2, \nu_1, \nu_2)&(2.86)\cr
&\qquad {}=C_N (\Vert M_\mu \vert M_{\mu_1}  \partial^{\nu_1} f_1 \vert
\vert M_{\mu_2}\partial^{\nu_2} \alpha_2 \vert  \Vert^{}_{L^2} +
\Vert M_\mu \vert M_{\mu_1}  \partial^{\nu_1} f_2
\vert \vert M_{\mu_2}  \partial^{\nu_2}\alpha_1 \vert  \Vert^{}_{L^2}),\cr
}$$
where $\vert \mu \vert \leq 1,  \vert \nu_1 \vert + \vert \nu_1 \vert \leq
N,  \vert \mu_1 \vert + \vert \mu_2 \vert \leq N.$

Let first $\vert \nu_1 \vert \geq \vert \nu_2 \vert$ in (2.85). Then (2.82)
gives
$$\eqalignno{
\Vert M_\mu \vert M_{\mu_1}  \partial^{\nu_1} \alpha_1 \vert \vert M_{\mu_2}
 \partial^{\nu_2} \alpha_2 \vert  \Vert^{}_{L^p} & \leq C_{\vert \nu_1 \vert}
\Vert M_{\mu_1}  \partial^{\nu_1} \alpha_1 \Vert^{}_D  \Vert M_{\mu_2}
 \partial^{\nu_2} \alpha_2 \Vert^{}_{D^{}_1}&{(2.87)}\cr
&\leq C_{\vert \nu_1 \vert}  \Vert \alpha_1 \Vert^{}_{D^{}_{\vert \nu_1 \vert}}
\Vert \alpha_2 \Vert^{}_{D^{}_{\vert \nu_2 \vert + 1}} \cr
& \leq C_{\vert \nu_1 \vert}
\Vert u_1 \Vert^{}_{E^{}_{\vert \nu_1 \vert}}
\Vert u_2 \Vert^{}_{E^{}_{\vert \nu_2 \vert + 1}}\cr
& = I'  (\vert \nu_1 \vert,  \vert \nu_2 \vert),\cr
}$$
where we have used Theorem 2.9. Since $\vert \nu_1 \vert +
\vert \nu_2 \vert + 1\leq N + 1$, $\vert \nu_1 \vert \leq N$,
$\vert \nu_2 \vert + 1 \leq [{N \over 2}]
+ 1 \leq N$ for $N \geq 1$, Corollary 2.6 gives with $N_0 = 1$
$$I' (\vert \nu_1 \vert,  \vert \nu_2 \vert) \leq C'_N \big(\Vert u_1
\Vert^{}_{E^{}_1}
 \Vert u_2 \Vert^{}_{E^{}_N} + \Vert u_1 \Vert^{}_{E^{}_N}  \Vert u_2
 \Vert^{}_{E^{}_1}\big).
\eqno{(2.88)}$$
If we now take $\vert \nu_1 \vert < \vert \nu_2 \vert$, we obtain
the same estimate (2.88), so (2.88) is true for $\vert \nu_1 \vert +
\vert \nu_2 \vert \leq N, N \geq 1$. From (2.85) and (2.88) it follows that
$$I (\mu, \mu_1, \mu_2, \nu_1, \nu_2) \leq C'_N \big(\Vert u_1
\Vert^{}_{E^{}_1} \Vert u_2 \Vert^{}_{E^{}_N} + \Vert u_1
\Vert^{}_{E^{}_N}  \Vert u_2  \Vert^{}_{E^{}_1}\big),
\eqno{(2.89)}$$
with the range of multi-indices defined in (2.85).

Let $\vert \nu_1 \vert > \vert \nu_2 \vert$ in (2.86). We can choose
$\mu_1$ and $\mu_2$, without changing the value of $J (\mu, \mu_1,
\mu_2, \nu_1, \nu_2),$ such that $\vert \mu_1 \vert \leq \vert \nu_1 \vert,
\vert \mu_2 \vert \leq N - \vert \nu_1 \vert$. Inequality (2.83) then gives
$$\eqalignno{
\Vert M_\mu \vert M_{\mu_1}  \partial^{\nu_1} f_1 \vert\vert M_{\mu_2}
 \partial^{\nu_2} \alpha_2 \vert  \Vert^{}_{L^2}
 &\leq C_{\vert M_1 \vert}
\Vert (M_{\mu_1}  \partial^{\nu_1} f_1, 0) \Vert^{}_M  \Vert M_{\mu_2}
 \partial^{\nu_2} \alpha_2 \Vert^{}_{D^{}_1} &{(2.90)}\cr
&\leq C'_{\vert \nu_1 \vert} \Vert (f_1,0) \Vert^{}_{M^{}_{\vert \nu_1 \vert}}
\Vert \alpha_2 \Vert^{}_{D^{}_{1+N - \vert \nu_1 \vert}}\cr
&\leq C'_{\vert \nu_1 \vert} \Vert u_1 \Vert^{}_{E^{}_{\vert \nu_1 \vert}}
\Vert u_2 \Vert^{}_{E^{}_{1+N - \vert \nu_1 \vert}},\cr
}$$
where we have used Theorem 2.9. An estimate for the second term in (2.86)
is obtained by permuting $u_1$ and $u_2$ in (2.90). This gives
$$J (\mu, \mu_1, \mu_2, \nu_1, \nu_2)
\leq C_N \big(\Vert u_1 \Vert^{}_{E^{}_{\vert \nu_1 \vert}}
\Vert u_2 \Vert^{}_{E^{}_{N+1 - \vert \nu_1 \vert}}
+ \Vert u_1 \Vert^{}_{E^{}_{N+1 - \vert \nu_1 \vert}} \Vert u_2
\Vert^{}_{E^{}_{\vert \nu_1 \vert}}\big),\quad \vert \nu_1 \vert > \vert
\nu_2 \vert.
$$
Since $\vert \nu_1 \vert + \vert \nu_2 \vert
\leq N$ and $N+1 - \vert\nu_1\vert \leq N$ in this inequality,
Corollary 2.6 gives
$$J (\mu, \mu_1, \mu_2, \nu_1, \nu_2)
\leq C_N \big(\Vert u_1 \Vert^{}_{E^{}_1}  \Vert u_2 \Vert^{}_{E^{}_N} +
\Vert u_1 \Vert^{}_{E^{}_N}
 \Vert u_2 \Vert^{}_{E^{}_1}\big),\quad \vert \nu_1 \vert > \vert \nu_2 \vert.
\eqno{(2.91)} $$
Let $\vert \nu_1 \vert \leq \vert \nu_2 \vert$ in (2.86).
As above we can choose $\vert \mu_1 \vert \leq \vert \nu_1 \vert$
and $\vert \mu_2 \vert \leq N - \vert \nu_1 \vert$.
Application of the inequality (2.84) to the two terms in (2.86) gives
$$\eqalign{
J (\mu, \mu_1, \mu_2, \nu_1, \nu_2)
& \leq C_N  \Big(\Vert (f_1,0)
\Vert^{}_{M^{}_{\vert \nu_1 \vert}}  \Vert \alpha_2
\Vert^{\rho - 1/2}_{D^{}_{1+N - \vert \nu_1 \vert}}
 \Vert \alpha_2 \Vert^{3/2 - \rho}_{D^{}_{N - \vert \nu_1 \vert}}\cr
&\qquad{}+ \Vert (f_2,0) \Vert^{}_{M^{}_{\vert \nu_1 \vert}}  \Vert \alpha_1
\Vert^{\rho - 1/2}_{D_{1+N - \vert \nu_1 \vert}}
\Vert \alpha_1 \Vert^{3/2 - \rho}_{D^{}_{N -\vert \nu_1 \vert}}\Big),
\quad  \vert \nu_1 \vert \leq \vert \nu_2 \vert.\cr
}$$
It now follows from the definition of the norms that
$$\eqalignno{
J (\mu, \mu_1, \mu_2, \nu_1, \nu_2)
& \leq C_n  \Big(\Vert u_1
\Vert^{}_{E^{}_{\vert \nu_1 \vert}}
 \Vert u_2 \Vert^{\rho - 1/2}_{E^{}_{N+1 - \vert \nu_1 \vert}}
\Vert u_2 \Vert^{3/2 - \rho}_{E^{}_{N - \vert \nu_1 \vert}}&{(2.92)}\cr
&\qquad{}+ \Vert u_2 \Vert^{}_{E^{}_{\vert \nu_1 \vert}}
\Vert u_1 \Vert^{\rho - 1/2}_{E^{}_{N+1 - \vert \nu_1 \vert}}
\Vert u_1 \Vert^{3/2 - \rho}_{E^{}_{N - \vert \nu_1 \vert}}\Big),
\quad \vert \nu_1 \vert\leq\vert \nu_2 \vert. \cr
}$$
Application of Corollary 2.6 to the terms
$(\Vert u_1 \Vert^{}_{E^{}_{\vert \nu_1 \vert}}
\Vert u_2 \Vert^{}_{E^{}_{N+1 - \vert \nu_1 \vert}})^{\rho - 1/2}$
and\penalty-10000
$(\Vert u_1 \Vert^{}_{E^{}_{\vert \nu_1 \vert}}
\Vert u_2 \Vert^{}_{E^{}_{N - \vert \nu_2 \vert}})^{3/2 - \rho}$
and the corresponding terms with $u_1$ and $u_2$ permuted in (2.92), gives
$$\eqalignno{
J (\mu, \mu_1, \mu_2, \nu_1, \nu_2)
& \leq C_N  \big(\Vert u_1 \Vert^{}_E
\Vert u_2 \Vert^{}_{E^{}_{N+1}} + \Vert u_1 \Vert^{}_{E^{}_{N+1}}  \Vert u_2
\Vert^{}_E\big)^{\rho - 1/2}&{(2.93)} \cr
&\qquad \big(\Vert u_1 \Vert^{}_{E^{}_1}  \Vert u_2 \Vert^{}_{E^{}_N} +
\Vert u_1
\Vert^{}_{E^{}_N}
 \Vert u_2 \Vert^{}_{E^{}_1}\big)^{3/2 - \rho}, \quad \vert \nu_1 \vert \leq
\vert \nu_2 \vert.\cr
}$$
Factorization of the right-hand side of (2.91) into a factor with
exponent $\rho - 1/2$
and a factor with exponent $3/2 - \rho$ and the application of Corollary 2.6
to the terms in the first factor show that (2.93) is also true for
$\vert \nu_1 \vert >\vert \nu_2 \vert$.
This proves the second statement of the lemma for $N>1$. The case
$N=0$ is the same as in the first statement of the lemma.

The first statement of the lemma for $N \geq 1$
follows from the second by application
of Corollary 2.6 to the factor with exponent $3/2 - \rho$.

Finally statement iii) follows from (2.82), application of estimate
(2.83) to one of the terms
$\Vert M_\mu \vert f_1 \vert  \vert \alpha_2 \vert  \Vert^{}_{L^2}$ or
$\Vert M_\mu \vert f_2 \vert  \vert \alpha_1 \vert
\Vert^{}_{L^2}$,
and application to the other terms of estimate (2.84). This proves the lemma.
\saut
\noindent{\bf Corollary 2.18.}
{\it
If $u \in E_\infty$ and $X \in \Pi$, then
$$\eqalignno{
\Vert T^2_X (u) \Vert^{}_{E^{}_N} &
\leq C_N (\Vert u \Vert^{}_{E^{}_1}  \Vert u \Vert^{}_{E^{}_N})^{3/2 - \rho}
(\Vert u \Vert^{}_{E^{}_0}  \Vert u \Vert^{}_{E^{}_{N+1}})^{\rho - 1/2},
\quad  N \geq 1,\cr
\noalign{\hbox{\it and}}
\Vert T^2_X (u) \Vert^{}_{E^{}_N} & \leq C  \Vert u \Vert^{}_{E^{}_0}
\Vert u \Vert^{}_{E^{}_{N+1}}, \quad N \geq 0.\cr
}$$
}\saut
We shall need an analogy of Lemma 2.17 and Corollary 2.18 for $T_Y = T^1_Y
+ \tilde{T}_Y,$
where $Y \in \Pi'$, the basis for the enveloping algebra $U(\p).$
\saut
\noindent{\bf Lemma 2.19.}
{\it
If $u_1,\ldots,u_n \in E_\infty$ and $Y \in \Pi'$, then
$$\eqalignno{
\hbox{\rm i)}\ &\Vert T^n_Y  (u_1  \otimes\cdots\otimes u_n) \Vert^{}_{E^{}_N}
\leq C \ds{\sum_i} \ds{\prod_{1 \leq l \leq n-1}}  \Vert u_{i_l} \Vert^{}_E
\Vert u_{i_n} \Vert^{}_{E^{}_{N+ \vert Y \vert}},\cr
\noalign{\hbox{\it for $n \geq 1$, $N \geq 0$ and}}
\hbox{\rm ii)}\ & \Vert T^n_Y (u_1  \otimes\cdots\otimes  u_n)
\Vert^{}_{E^{}_N}\cr
&\hskip-10pt\leq C \Big(\sum_i
\Vert u_{i_1} \Vert^{}_{E^{}_{N+ \vert Y \vert - 1}}
\Vert u_{i_2} \Vert^{}_{E^{}_1}  \prod^n_{l=3}  \Vert u_{i_l}
\Vert^{}_E\Big)^{3/2 - \rho}
\Big(\sum_i  \Vert u_{i_1} \Vert^{}_{E^{}_{N+ \vert Y \vert}}  \Vert
u_{i_2} \Vert^{}_{E^{}_1}  \prod^n_{l=3}
\Vert u_{i_l} \Vert^{}_E\Big)^{\rho - 1/2},\cr
}$$
for $n \geq 2$ and $\vert Y \vert + N \geq 1$. Here the summation is over all
permutation $i$ of $(1,\ldots,n)$ and the constant $C$ depends on
$\vert Y \vert,  n,  N,\rho.$
}\saut
\noindent{\it Proof.}
We prove the first statement by induction. It is true for $Y = {\un}$,
because $T_{{\un}} (u) = u$. It follows from Lemma 2.17 that it
is also true for $Y = X \in \Pi.$
Suppose it is true for $\vert Y \vert \leq L$. If $Y' = YX,  \vert Y \vert
\leq L,  X \in \Pi$, then it follows from definition (1.9) of $T_{YX}$
that with $I_q = \otimes^q I$,  $I=$ identity in $E$,
$$
T^n_{YX} = \sum_{0 \leq q \leq n-1}  T^n_Y  (I_q
\otimes  T^1_X  \otimes  I_{n - q - 1}) \tau^{}_n
+ \sum_{0 \leq q \leq n - 2}  T^{n-1}_Y  (I_q
\otimes  T^2_X  \otimes  I_{n - q - 2}) \tau^{}_n,\eqno{(2.94)}$$
where $\tau^{}_n$ is the normalized symmetrization operator on
$\hat{\otimes}^n E$ ($=E\hat{\otimes}E\hat{\otimes}\cdots\hat{\otimes}E$, $n$
times).
By the induction hypothesis we have, after reindexation for $n \geq 2$,
$$\eqalign{
&\Vert T^{n-p+1}_Y  (I_q  \otimes  T^p_X \otimes  I_{n - q - p})
(\tau^{}_n  \otimes^n_{j=1}
 u_j) \Vert^{}_{E^{}_N}\cr
&\qquad{}\leq C  \sum_i  \Vert u_{i_1} \Vert^{}_E\cdots\Vert u_{i_{n-2}}
\Vert^{}_E\cr
&\qquad\qquad{}\big(\Vert u_{i_{n-1}} \Vert^{}_E  \Vert T^1_X  u_{i_{n}}
\Vert^{}_{E^{}_{\vert Y \vert + N}} + \Vert T^1_X  u_{i_{n-1}}
\Vert^{}_E
\Vert u_{i_{n}} \Vert^{}_{E^{}_{\vert Y \vert + N}}\big),
\quad p = 1,2,\, n-p \geq 0.\cr
}$$
Since, according to the definition of $\Vert\cdot  \Vert^{}_{E^{}_l}$
and Corollary 2.6
$$\eqalign{
&\Vert u_{i_{n-1}} \Vert^{}_E  \Vert T^1_X  u_{i_{n}}
\Vert^{}_{E^{}_{\vert Y \vert + N}} + \Vert T^1_X  u_{i_{n-1}}
\Vert^{}_E
\Vert u_{i_{n}} \Vert^{}_{E^{}_{\vert Y \vert + N}}\cr
&\qquad{}\leq C'  \big(\Vert u_{i_{n-1}} \Vert^{}_E  \Vert u_{i_n}
\Vert^{}_{E^{}_{\vert Y \vert +1+N}}
+ \Vert u_{i_{n-1}} \Vert^{}_{E^{}_{\vert Y \vert +1+N}}  \Vert u_{i_n}
\Vert^{}_E\big),\cr
}$$
and since the case $n = 1$ is trivial, we obtain
$$\eqalignno{
&\Vert T^{n-p+1}_Y  (I_q  \otimes  T^p_X
\otimes  I_{n-q-p})  (\tau^{}_n  \otimes^n_{j=1}
u_j) \Vert^{}_{E^{}_N}&{(2.95)}\cr
&\qquad{}\leq C''  \sum_i  \prod^{n-1}_{j=1}  \Vert u_{i_j} \Vert^{}_E
 \Vert u_{i_n} \Vert^{}_{E^{}_{N+1+ \vert Y \vert}},
 \quad n \geq 1, p = 1,2, n-p \geq 0.
}$$
 Formula (2.94) and inequality (2.95) prove, after changing the constant
$C$, that the first statement of the lemma is true for $\vert Y' \vert = L+1.$
So, by induction it is true for all $\vert Y \vert \geq 0.$

According to Theorem 2.4 of \refST\ we have, for $X \in \Pi$ and $Z \in \Pi'$,
$$T^n_{XZ} = \sum_{n_1 + n_2 = n}  \sump_{Z,2} T^2_X (T^{n_1}_{Z_1}
 \otimes  T^{n_2}_{Z_2})  \tau^{}_n + T^1_X
T^n_Z, \quad n \geq 2, \eqno{(2.96)}$$
where $\sump_{Z,2}$ is a sum over a subset of couples $(Z_1, Z_2)$ with
$Z_i \in \Pi'$ and $0 \leq \vert Z_i \vert \leq \vert Z \vert$,
$\vert Z_1 \vert + \vert Z_2 \vert = \vert Z \vert$ and where $\tau^{}_n$
is the normalized symmetrization
operator on $\hat{\otimes}^n  E$. Let $Y = XZ,  X \in \Pi,
 Z \in \Pi'$, let $f_l \in E_\infty,  1 \leq l \leq n_1,
g_j \in E_\infty,  1 \leq j \leq n_2$, and let $f = f_1 \otimes
\cdots\otimes  f_{n_1},  g = g_1  \otimes\cdots\otimes
 g_{n_2}$. According to statement ii) of Lemma 2.17, we obtain
$$\eqalignno{
& \Vert T^2_X  (T^{n_1}_{Z_1}  (f)  \otimes
T^{n_2}_{Z_2}  (g)) \Vert^{}_{E^{}_N}& (2.97)\cr
&\qquad{}\leq C_N  \big(\Vert T^{n_1}_{Z_1}  (f) \Vert^{}_{E^{}_N}
\Vert T^{n_2}_{Z_2}  (g) \Vert^{}_{E^{}_1} + \Vert T^{n_1}_{Z_1}  (f)
\Vert^{}_{E^{}_1}  \Vert T^{n_2}_{Z_2}  (g)
\Vert^{}_{E^{}_N}\big)^{3/2 - \rho}\cr
&\qquad\qquad{} \big(\Vert T^{n_1}_{Z_1}  (f) \Vert^{}_{E^{}_{N+1}}
\Vert T^{n_2}_{Z_2}
 (g) \Vert^{}_E + \Vert T^{n_1}_{Z_1}  (f) \Vert^{}_E
\Vert T^{n_2}_{Z_2}(g) \Vert^{}_{E^{}_{N+1}}\big)^{\rho - 1/2},
\quad N \geq 0.\cr
}$$
Let $N \geq 1$. Then it follows from the first inequality of Lemma 2.19
which
is already proved and from the second inequality of Corollary 2.6, that
$$\eqalignno{
& \Vert T^{n_1}_{Z_1}  (f) \Vert^{}_{E^{}_N}  \Vert T^{n_2}_{Z_2}
 (g) \Vert^{}_{E^{}_1} +  \Vert T^{n_1}_{Z_1}  (f) \Vert^{}_{E^{}_1}
\Vert T^{n_2}_{Z_2}  (g) \Vert^{}_{E^{}_N} & {(2.98)}\cr
&\qquad{} \leq C_{N,n,\vert Z \vert}  \sum_{l,j}  \Vert f_{l_2} \Vert^{}_E
\cdots
\Vert f_{l_{n_1}} \Vert^{}_E  \Vert g_{j_2} \Vert^{}_E\cdots\Vert g_{j_{n_2}}
\Vert^{}_E\cr
&\qquad\qquad{}  \big(\Vert f_{l_1} \Vert^{}_{E^{}_{N+\vert Z_1 \vert}}
\Vert g_{j_1}
\Vert^{}_{E^{}_{1+\vert Z_2 \vert}}
+ \Vert f_{l_1} \Vert^{}_{E^{}_{1+\vert Z_1 \vert}}  \Vert g_{j_1}
\Vert^{}_{E^{}_{N+\vert Z_2 \vert}}\big)\cr
&\qquad{} \leq C'_{N,n,\vert Z \vert}  \sum_{l,j}  \Vert f_{l_2}
\Vert^{}_E\cdots
\Vert f_{l_{n_1}} \Vert^{}_E  \Vert g_{j_2} \Vert^{}_E\cdots
\Vert g_{j_{n_2}}
\Vert^{}_E\cr
&\qquad\qquad{} \big(\Vert f_{l_1} \Vert^{}_{E^{}_{N+\vert Z \vert}}
\Vert g_{j_1}
\Vert^{}_{E^{}_1} +
\Vert f_{l_1} \Vert^{}_{E^{}_1}  \Vert g_{j_1}
\Vert^{}_{E^{}_{N+\vert Z \vert}}\big),
\quad N \geq 1.\cr
}$$
For the last inequality we have also used that
$\vert Z_1 \vert + \vert Z_2 \vert= \vert Z \vert$.
We obtain in the same way for $N \geq 0$,
$$\eqalignno{
&\Vert T^{n_1}_{Z_1}  (f) \Vert^{}_{E^{}_{N+1}}  \Vert T^{n_2}_{Z_2}
(g) \Vert^{}_E + \Vert T^{n_1}_{Z_1}  (f) \Vert^{}_E
\Vert T^{n_2}_{Z_2}  (g) \Vert^{}_{E^{}_{N+1}}&{(2.99)}\cr
&\qquad{}\leq C_{N,n,\vert Z \vert}  \sum_{l,j}  \Vert f_{l_2}
\Vert^{}_E\cdots
\Vert f_{l_{n_1}} \Vert^{}_E  \Vert g_{j_2} \Vert^{}_E\cdots
\Vert g_{n_2} \Vert^{}_E\cr
&\qquad\qquad{}\big(\Vert f_{l_1} \Vert^{}_{E^{}_{N+\vert Z \vert + 1}}
\Vert g_{j_1} \Vert^{}_E
+ \Vert f_{l_1} \Vert^{}_E  \Vert g_{j_1}
\Vert^{}_{E^{}_{N+\vert Z \vert +1}}\big),
\quad  N \geq 0.\cr
}$$
Let $u_1,\ldots,u_n \in E_\infty$. Then it follows
from inequalities (2.97), (2.98)
and (2.99) that
$$\eqalignno{
&\Vert T^2_X (T^{n_1}_{Z_1}  \otimes  T^{n_2}_{Z_2})
 \tau^{}_n (\otimes^n_{l=1}  u_l) \Vert^{}_{E^{}_N} &{(2.100)}\cr
&\qquad{}\leq C_{N,n,\vert Z \vert}  \Big(\sum_i  \Vert u_{i_1}
\Vert^{}_{E^{}_{N+\vert Z \vert}} \Vert u_{i_2} \Vert^{}_{E^{}_1}
\Vert u_{i_3} \Vert^{}_E\cdots\Vert u_{i_n} \Vert^{}_E\Big)^{3/2 - \rho}\cr
&\qquad\qquad{}\Big(\sum_i      \Vert u_{i_1}
\Vert^{}_{E^{}_{N+\vert Z \vert + 1}}
\Vert u_{i_2} \Vert^{}_E\cdots \Vert u_{i_n} \Vert^{}_E\Big)^{\rho - 1/2}\cr
&\qquad{}\leq C_{N,n,\vert Z \vert}
\Big(\sum_i  \Vert u_{i_1} \Vert^{}_{E^{}_{N+\vert Z \vert}}
 \Vert u_{i_2} \Vert^{}_{E^{}_1}  \Vert u_{i_3}
 \Vert^{}_E\cdots\Vert u_{i_n}
 \Vert^{}_E\Big)^{3/2 - \rho}\cr
&\qquad\qquad{} \Big(\sum_i  \Vert u_{i_1}
\Vert^{}_{E^{}_{N+\vert Z \vert + 1}}
\Vert u_{i_2} \Vert^{}_{E^{}_1}  \Vert u_{i_3}
\Vert^{}_E\cdots \Vert u_{i_n}
\Vert^{}_E\Big)^{\rho - 1/2},\quad      N \geq 1.\cr
}$$
Let $f_1,\ldots,f_{n_1}$, $g_1,\ldots,g_{n_2}$, $f,g$ be defined as
previously. Statement iii) of Lemma 2.17 gives
$$\eqalignno{
 \Vert T^2_X (T^{n_1}_{Z_1}  (f)  \otimes  T^{n_2}_{Z_2}
 (g)) \Vert^{}_E
&\leq C  \min \Big(\Vert T^{n_1}_{Z_1}  (f) \Vert^{}_{E^{}_1}
\Vert T^{n_2}_{Z_2}  (g) \Vert^{\rho - 1/2}_{E^{}_1}  \Vert T^{n_2}_{Z_2}
 (g) \Vert^{3/2 - \rho}_E\quad & (2.101)\cr
&\qquad{}\Vert T^{n_1}_{Z_1}  (f)
\Vert^{\rho - 1/2}_{E^{}_1}  \Vert T^{n_1}_{Z_1}
 (f) \Vert^{3/2 - \rho}_E
 \Vert T^{n_2}_{Z_2}  (g) \Vert^{}_{E^{}_1}\Big).\cr
 }$$
It follows from the first inequality of Lemma 2.19 that
$$\eqalignno{
&\Vert T^{n_1}_{Z_1}  (f) \Vert^{}_{E^{}_1}  \Vert T^{n_2}_{Z_2}
(g) \Vert^{\rho - 1/2}_{E^{}_1}  \Vert T^{n_2}_{Z_2}  (g)
\Vert^{3/2 - \rho}_{E^{}_0}&{(2.102)}\cr
&\qquad{} \leq C_{\vert Z \vert,n}  \Big(\sum_{l,j}
\Vert f_{l_1} \Vert^{}_{E^{}_{1+\vert Z_1 \vert}}
 \Vert f_{l_2} \Vert^{}_E\cdots\Vert f_{l_{n_1}} \Vert^{}_E
 \Vert g_{j_1} \Vert^{}_{E^{}_{1+\vert Z_2 \vert}}
 \Vert g_{j_2} \Vert^{}_E\cdots\Vert g_{j_{n_2}}
 \Vert^{}_E\Big)^{\rho - 1/2}\cr
&\qquad\qquad{} \Big(\sum_{l,j}  \Vert f_{l_1}
\Vert^{}_{E^{}_{1+\vert Z_1 \vert}}
\Vert f_{l_2} \Vert^{}_E\cdots\Vert f_{l_{n_1}} \Vert^{}_E
\Vert g_{j_1} \Vert^{}_{E^{}_{\vert Z_2 \vert}}
 \Vert g_{j_2} \Vert^{}_E\cdots\Vert g_{j_{n_2}}
 \Vert^{}_E\Big)^{3/2 - \rho}.\cr
 }$$
Inequalities (2.101) and (2.102) give
$$
\Vert T^2_X  (T^{n_1}_{Z_1}  \otimes  T^{n_2}_{Z_2})
 \tau^{}_n (\otimes^n_{j=1}  u_j) \Vert^{}_E
\leq C_{n, \vert Z \vert}  \min (Q (\vert Z_1 \vert, \vert Z_2 \vert),
 Q (\vert Z_2 \vert, \vert Z_1 \vert)), \quad  n \geq 2,\eqno{(2.103)}$$
where
$$\eqalignno{
Q (\vert Z_1 \vert, \vert Z_2 \vert) & = \Big(\sum_i
\Vert u_{i_1} \Vert^{}_{E^{}_{1+\vert Z_1 \vert}}
 \Vert u_{i_2} \Vert^{}_{E^{}_{1+\vert Z_2 \vert}}
 \Vert u_{i_3} \Vert^{}_E\cdots\Vert u_{i_n} \Vert^{}_E\Big)^{\rho - 1/2}
 &{(2.104)}\cr
&\qquad{} \Big(\sum_i  \Vert u_{i_1} \Vert^{}_{E^{}_{1+\vert Z_1 \vert}}  \Vert
u_{i_2} \Vert^{}_{E^{}_{\vert Z_2 \vert}}  \Vert u_{i_3} \Vert^{}_E
\cdots\Vert u_{i_n} \Vert^{}_E\Big)^{3/2 - \rho}.\cr
}$$
If $\vert Z_1 \vert = \vert Z_2 \vert = 0$, then
$$\eqalignno{
 Q (0,0) & \leq \Big(\sum_i  \Vert u_{i_1} \Vert^{}_{E^{}_1}  \Vert u_{i_2}
\Vert^{}_{E^{}_1}  \Vert u_{i_3} \Vert^{}_E \cdots
 \Vert u_{i_n} \Vert^{}_E\Big)^{\rho - 1/2}
&{(2.105\hbox{a})}\cr
&\qquad{}\Big(\sum_i  \Vert u_{i_1} \Vert^{}_E  \Vert u_{i_2} \Vert^{}_{E^{}_1}
 \Vert u_{i_3} \Vert^{}_E\cdots\Vert u_{i_n} \Vert^{}_E\Big)^{3/2 - \rho}.\cr
 }$$
If $\vert Z_1 \vert = \vert Z_2 \vert \geq 1$, then $1 +
\vert Z_1 \vert \leq \vert
Z \vert$, so it follows from the second inequality of Corollary~2.6 that
$$\eqalignno{
Q (\vert Z_1 \vert, \vert Z_2 \vert)
&\leq C \Big(\sum_i  \Vert u_{i_1}
\Vert^{}_{E^{}_{1 + \vert Z \vert}}  \Vert u_{i_2} \Vert^{}_{E^{}_1}
\Vert u_{i_3}\Vert^{}_E\cdots\Vert u_{i_n} \Vert^{}_E\Big)^{\rho - 1/2}
&{(2.105\hbox{b})}\cr
&\qquad{}\Big(\sum_i  \Vert u_{i_1}
\Vert^{}_{E^{}_{\vert Z \vert}}  \Vert u_{i_2}
\Vert^{}_{E^{}_1}  \Vert u_{i_3} \Vert^{}_E\cdots
\Vert u_{i_n} \Vert^{}_E\Big)^{3/2 - \rho},
\quad \vert Z_1 \vert = \vert Z_2 \vert \geq 1.\cr
}$$
If $\vert Z_1 \vert < \vert Z_2 \vert$, then
$\vert Z_1 \vert + 1 \leq \vert Z_2 \vert,$
so it follows from Corollary 2.6 that
$$\eqalignno{
Q (\vert Z_1 \vert, \vert Z_2 \vert)
&\leq C \Big(\sum_i  \Vert u_{i_1} \Vert^{}_{1+\vert Z \vert}
 \Vert u_{i_2} \Vert^{}_{E^{}_1}  \Vert u_{i_3} \Vert^{}_E\cdots\Vert
u_{i_n} \Vert^{}_E\Big)^{\rho - 1/2}&{(2.105\hbox{c})}\cr
&\qquad{}\Big(\sum_i
\Vert u_{i_1} \Vert^{}_{E^{}_{\vert Z \vert}}  \Vert u_{i_2}
\Vert^{}_{E^{}_1}  \Vert u_{i_3} \Vert^{}_E \cdots
\Vert u_{i_n} \Vert^{}_E\Big)^{3/2 - \rho},
\quad \vert Z_1 \vert < \vert Z_2 \vert.\cr
}$$
It follows from inequalities (2.103) and (2.105) that
$$\eqalignno{
&\Vert T^2_X (T^{n_1}_{Z_1}  \otimes  T^{n_2}_{Z_2})
\tau^{}_n       (\otimes^n_{j=1}  u_j) \Vert^{}_E &(2.106)\cr
&\qquad\leq C_{n, \vert Z \vert}
\Big(\sum_i  \Vert u_{i_1} \Vert^{}_{E^{}_{1+\vert Z \vert}}
 \Vert u_{i_2} \Vert^{}_{E^{}_1}  \Vert u_{i_3} \Vert^{}_E \cdots\Vert
u_{i_n} \Vert^{}_E\Big)^{\rho - 1/2}\cr
&\qquad\qquad{}\Big(\sum_i      \Vert u_{i_1} \Vert^{}_{E^{}_{\vert Z \vert}}
\Vert u_{i_2}\Vert^{}_{E^{}_1}  \Vert u_{i_3} \Vert^{}_E\cdots
\Vert u_{i_n} \Vert^{}_E\Big)^{3/2 - \rho},\cr
}$$
where $\vert Z_1 \vert + \vert Z_2 \vert = \vert Z \vert.$
Inequalities (2.100) and (2.106) show that
$$\eqalignno{
&\Vert T^2_X (T^{n_1}_{Z_1}  \otimes  T^{n_2}_{Z_2})
\tau^{}_n       (\otimes^n_{j=1}  u_j) \Vert^{}_{E^{}_N}&{(2.107)}\cr
&\qquad \leq C_{N,n,\vert Z \vert} \Big(\sum_i  \Vert u_{i_1}
\Vert^{}_{E^{}_{N+\vert Z \vert}}
 \Vert u_{i_2} \Vert^{}_{E^{}_1}  \Vert u_{i_3} \Vert^{}_E \cdots
 \Vert u_{i_n} \Vert^{}_{E}\Big)^{3/2 - \rho}\cr
&\qquad\qquad\Big(\sum_i  \Vert u_{i_1} \Vert^{}_{E^{}_{N+\vert Z \vert + 1}}
\Vert u_{i_2} \Vert^{}_{E^{}_1}
\Vert u_{i_3} \Vert^{}_E\cdots\Vert u_{i_n} \Vert^{}_{E}\Big)^{\rho - 1/2},
\quad N \geq 0,\cr
}$$
where $\vert Z_1 \vert + \vert Z_2 \vert = \vert Z \vert,  \vert Z_j \vert
\geq 0,  n \geq 2.$

We now prove the second statement of the lemma by induction in
$\vert Y \vert$, $Y = XZ$, using formula (2.96). If $\vert Y \vert = 0$
then the statement is true since $T^n_{\bf 1} = 0$ for $n \geq 2$.
Let $\vert Y \vert = L+1$. Then, supposing
inequality ii) of the lemma true for all $\vert Y \vert \leq L$, we get
$$\eqalignno{
\Vert T^1_X  T^n_Z  (\otimes^n_{j=1}  u_j) \Vert^{}_{E^{}_N}
&\leq \Vert T^n_Z  (\otimes^n_{j=1}  u_j) \Vert^{}_{E^{}_{N+1}}&{(2.108)}\cr
&\leq C_{\vert Z \vert, N,n}  \Big(\sum_i
\Vert u_i\Vert^{}_{E^{}_{N+\vert Z \vert}}
 \Vert u_{i_2} \Vert^{}_{E^{}_1}  \prod^n_{l=3}  \Vert
u_{i_l} \Vert^{}_E\Big)^{3/2 - \rho}\cr
&\qquad\Big(\sum_i\Vert u_{i_1} \Vert^{}_{E^{}_{N+1+\vert Z \vert}}  \Vert
u_{i_2} \Vert^{}_{E^{}_1}  \prod^n_{l=3}  \Vert u_{i_l} \Vert^{}_E
\Big)^{\rho - 1/2}.\cr
}$$
It now follows from equality (2.96) and inequalities (2.107) and (2.108) that
inequality ii) of the lemma is true. This proves the lemma.

According to Lemma 2.19 $T^n_Y,  Y \in \Pi'$, has continuous extensions
to spaces larger than $E_\infty$.
These extensions are denoted by the same symbol
$T^n_Y.$
\saut
\noindent{\bf Remark 2.20.}
Let $Y \in \Pi'$ and $N \geq 0$. Then $T^n_Y$ is a continuous linear map from
$\hat{\otimes}^n  E_{N+\vert Y \vert}$ to $E_N.$
\saut
\noindent Lemma 2.19  immediately gives estimates for the polynomial $T_Y.$
\saut
\noindent{\bf Corollary 2.21.}
{\it
$T_Y,  Y \in \Pi'$ is a continuous polynomial from $E_{N+\vert Y \vert}$
to $E_N$ satisfying:
\psaut
\noindent\hbox{\rm \phantom{ii}i)}
$\Vert T_Y (u) \Vert^{}_{E^{}_N} \leq C_{N,\vert Y \vert}
(\Vert u\Vert^{}_E)
\Vert u \Vert^{}_{E^{}_{N+\vert Y \vert}}$,
\psaut
\noindent\hbox{\rm \phantom{i}ii)}
$\Vert \tilde{T}_Y (u) \Vert^{}_{E^{}_N} \leq C_{N,\vert Y \vert}
(\Vert u \Vert^{}_E)
 \Vert u \Vert^{}_E  \Vert u \Vert^{}_{E^{}_{N+\vert Y \vert}}$,
\psaut
\noindent\hbox{\rm iii)}
$\Vert \tilde{T}_Y (u) \Vert^{}_{E^{}_N} \leq C_{N,\vert Y \vert}
(\Vert u \Vert^{}_E) \Vert u \Vert^{}_{E^{}_1}
(\Vert u \Vert^{}_{E^{}_{N+\vert Y \vert - 1}})^{3/2 - \rho}
\Vert u \Vert^{\rho - 1/2}_{E^{}_{N+\vert Y \vert}}$,
\psaut
\noindent where $C_{N,\vert Y\vert }$ is an increasing continuous
function and $\tilde{T}_Y =T_Y - T^1_Y$.
}\saut
Let $Y \mapsto a^{}_Y$ be a linear function from $U(\p)$ to a Banach space
$B$.
We introduce (cf.~(2.38) of \refST)
$$\wp^{B}_N (a) = \Big(\sum_{\scr Y \in \Pi'\atop\scr
\vert Y \vert \leq N}  \Vert
a_Y \Vert^2_B\Big)^{1/2}, \quad N \geq 0. \eqno{(2.109)}$$
When $B=E$ we write $\wp^{}_{N}$ instead of $\wp^{E}_{N}$.
According to definition (1.6a) of $\Vert\cdot\Vert^{}_{E^{}_N}$ we have
$$\wp^{}_N (T^1  (u)) = \Vert u \Vert^{}_{E^{}_N}, \quad N \geq 0.
\eqno{(2.110)}$$
We can now prove that the linear representation $T^1$ of $U(\p)$ is
bounded by the nonlinear representation $T$, and vice-versa,
on a $E$-neighbourhood of zero in $E_\infty$.
\saut
\noindent{\bf Theorem 2.22.}
{\it
There is a neighbourhood ${\cal O}$ of zero in $E$ such that for
$u \in E_N \cap {\cal O}$:
$$\eqalign{
\hbox{\rm i)}\ &\wp^{}_N (T (u)) \leq C_N  \Vert u \Vert^{}_{E^{}_N},\quad
N \geq 0,\cr
\hbox{\rm ii)}\ &\Vert u \Vert^{}_{E^{}_N} \leq C_N  \wp^{}_N  (T(u)),
\quad 0 \leq N \leq 1,\cr
\hbox{\rm iii)}\ &\Vert u \Vert^{}_{E^{}_N} \leq F_N  (\wp^{}_1 (T(u)))
\wp^{}_N
(T(u)),\quad  N \geq 0.\hskip63.95mm\cr
}$$
Here $C_N$ is a constant and $F_N$ an increasing continuous function,
only depending on the neighbourhood ${\cal O}.$
}\saut
\noindent{\it Proof.}
For $K > 0$ we define ${\cal O} = \{ u \in E \big\vert
\Vert u \Vert^{}_E < K \}.$
Let $Y \in \Pi'$, $ \vert Y \vert \leq N$ and $N \geq 0.$
It follows from i) of Corollary 2.21 that for $u \in E_N \cap {\cal O}$,
$$\Vert T_Y (u) \Vert^{}_{E} \leq C_{0, \vert Y \vert} (\Vert u \Vert^{}_E)
  \Vert u \Vert^{}_{E^{}_{\vert Y \vert}}
\leq C_{0, \vert Y \vert} (K)  \Vert u \Vert^{}_{E^{}_N}.$$
Summation over $Y \in \Pi'$,  $\vert Y \vert \leq N$ now gives inequality
i) of the theorem for $N \geq 0.$

To prove the second inequality we note that it is true for
$N = 0$ and that, according to ii) of Corollary 2.21,
$$\eqalign{
\wp^{}_1 (\tilde{T} (u))
& \leq \Big(\sum_{\scr Y \in \Pi'\atop \scr \vert Y \vert \leq 1}
 (C_{0, \vert Y \vert}  \big(\Vert u \Vert^{}_E)  \Vert u \Vert^{}_E
 \Vert u \Vert^{}_{E^{}_{\vert Y \vert}}\big)^2\Big)^{1/2}\cr
&\leq C (K) K  \Vert u \Vert^{}_{E^{}_1}, \quad u \in E_1 \cap {\cal O},\cr
}$$
where $C(K)$ is continuous and increasing in $K \geq 0$. For a given
$\chi$, $0 < \chi < 1$, we choose $K$ such that $C(K)  K \leq \chi.$
Then
$$\wp^{}_1 (\tilde{T} (u)) \leq \chi  \Vert u \Vert^{}_{E^{}_1},
\quad  u \in E_1 \cap {\cal O}.$$
Since $\Vert u \Vert^{}_{E^{}_1} = \wp^{}_1 (T^1 u) \leq \wp^{}_1 (T(u))
+ \wp^{}_1 (\tilde{T} (u))$ we obtain $\Vert u \Vert^{}_{E^{}_1} \leq \chi
\Vert u \Vert^{}_{E^{}_1} + \wp^{}_1 (T(u)),$ which gives
$$\Vert u \Vert^{}_{E^{}_1} \leq (1 - \chi)^{-1}  \wp^{}_1 (T(u)),
\quad  u \in E_1\cap {\cal O}, \eqno{(2.111)}$$
where $0 < \chi < 1$. This proves inequality ii).

We prove the third inequality by induction. It follows from inequality ii)
that it is true for $N = 0$ and $N = 1$. Suppose it is true up to $N - 1,
N \geq 1$. Then inequality iii) of Corollary 2.21 gives for
$\vert Y \vert \leq N, Y \in \Pi'$,
$$\eqalign{
\Vert \tilde{T}_Y (u) \Vert^{}_E
&\leq C_{0, \vert Y \vert} (K)  \Vert u \Vert^{}_{E^{}_1}
 \Vert u \Vert^{3/2 - \rho}_{E^{}_{N - 1}}
 \Vert u \Vert^{\rho - 1/2}_{E^{}_N}\cr
&\leq C_{0, \vert Y \vert} (K)  C_1  \wp^{}_1 (T(u))
\big(F_{N-1} (\wp^{}_1 (T(u)))  \wp^{}_{N-1} (T(u))\big)^{3/2 - \rho}
\Vert u \Vert^{\rho - 1/2}_{E^{}_N},\cr
}$$
where inequality ii) and the induction hypothesis were used for the
second inequality.
Since $\Vert u \Vert^{}_{E^{}_N} \leq \wp^{}_N (T(u)) +
\wp^{}_N (\tilde{T} (u))$ we get after
summation over $Y$
$$\Vert u \Vert^{}_{E^{}_N} \leq \wp^{}_N (T(u))
+ H_N \big(\wp^{}_1 (T(u))\big)  \wp^{}_{N-1} (T(u))^{3/2 - \rho}
 \Vert u \Vert^{\rho - 1/2}_{E^{}_N}, \eqno{(2.112)}$$
where $H_N$ is an increasing continuous function depending only on $K.$
Since $\wp^{}_1 (T(u)) = \wp^{}_{N-1} (T(u)) = 0$ if
$\wp^{}_N (T(u)) = 0,  N \geq 1$,
it follows from (2.112) that inequality~iii) of the theorem is true for
$\wp^{}_N (T(u)) = 0.$
Let $\wp^{}_N (T(u)) > 0$ and let
$x = \Vert u \Vert^{}_{E^{}_N} / \wp^{}_N (T(u))$. Since $\wp^{}_{N-1}
(T(u)) \leq \wp^{}_N (T(u))$, inequality (2.112) gives
$$x \leq 1 + A^{}_N  x^{\rho - 1/2},\quad       x \geq 0, \eqno{(2.113)}$$
where $A_N = H_N (\wp^{}_1 (T(u)))$. If $x \leq 1$ it then follows from the
definition of $x$ that inequality iii) of the theorem is true.
If $x > 1$, it then follows
from (2.113) that $x \leq 1 + A^{}_N  x^{1/2}$, since $0 < \rho < 1/2$,
which shows that $x^{1/2} \leq A^{}_N/2 +
((A_N/2)^2 + 1)^{1/2} \leq 2 ((A_N/2)^2 + 1)^{1/2}.$
So $x \leq A^2_N + 4$ and $\Vert u \Vert^{}_{E^{}_N} \leq (A^2_N + 4)
\wp^{}_N (T(u))$,
which proves that inequality iii) of the theorem is also true for $x > 1$.
This proves the theorem.
\vfill\eject

\noindent{\titre 3. The asymptotic nonlinear representation.}
\saut
If we suppose that the limit (1.17c) exists and that
$t \mapsto u(t) = \Omega_+(U^{(+)}_{\exp(t  P_0)} (v))$ is
a solution of the evolution equation (1.2), then we can easily find the
explicit expressions (1.17a) and (1.17b) of $U^{(+)}_g$, $g \in {\cal P}_0$,
since the Maxwell-Dirac equations are manifestly covariant.
Let $T^{(+)}$ be the Lie algebra representation of
$\p$ which is the differential of $U^{(+)}$. In this chapter we
shall prove that $U^{(+)}_g$ maps $E^{0 \rho}_N$ into $E^{0 \rho}_N,$ for $N$
sufficiently large, that $T^{(+)}_X$ maps $E^{}_\infty$ into $E^{}_\infty$ and
deduce properties of the $C^n$-vectors of $U^{(+)}$.
We give a direct proof (theorem 3.14) of the fact that
$X \mapsto T^{(+)}_X$ is a representation of $\p$. We return to the
question of the existence of the limit (1.17c) in chapter 6.

To begin with we study {\it the phase function} in (1.23a) of the
definition of $U^{(+)}$.
\saut
\noindent{\bf Lemma 3.1.}
{\it
Let $f$ be a continuous function on $D^{}_0 = \{ (t,x) \in \Rrm \times \Rrm^3
\big\vert t > \vert x \vert$ and $t > 0 \}$, which is absolutely
Lebesgue integrable
on every half-line in $D^{}_0$ starting at the origin.
$f$ is defined to be zero outside $D^{}_0$. Let
$$\hat{g} (k) = \int^\infty_0  f(\tau  \omega(k)/m,
\tau  k/m)  d \tau, \quad k \in \Rrm^3, w(k) = (m^2 + k^2)^{1/2}.$$
\psaut
\noindent\hbox{\rm \phantom{i}i)} if $f(t,x)=0$ for $0\leq t \leq 1$ and
$f\in L^\infty(\Rrm^+,L^2(\Rrm^3))$, then,

$\Vert (\omega(-i \partial))^{-2} g \Vert^{}_{L^2} \leq C
\ds{\sup_{t \geq 1}} \Vert f(t) \Vert^{}_{L^2}$,
\psaut
\psaut
\noindent\hbox{\rm ii)} if the function $(t,x)\mapsto
(1+t)(1+t-\vert x \vert)^\varepsilon f(t,x)$ belongs to
$L^\infty(D^{}_{0})$, where $\omega(k) = (m^2+k^2)^{1/2}$
and $\omega(-i\partial)=(m^2+\vert {-}i\partial\vert^2)^{1/2}=
(m^2 - \Delta)^{1/2}$, then,

$\vert \hat{g}(k)\vert \leq \ds{m \over\omega(k)} (\ln (1+2 ({\omega(k)
/ m})^2) + \varepsilon^{-1})
\ds{\sup_{(t,x) \in D^{}_0}}  ((1+t)(1+t-\vert x \vert)^\varepsilon
\vert f(t,x) \vert)$, $ \varepsilon > 0$.
\psaut
}\saut
\noindent{\it Proof.}
To prove the first statement of the lemma, let $D^{}_1 =
\{ (t,x) \in \Rrm \times \Rrm^3 \big\vert t
\geq (1+\vert x \vert^2)^{1/2} \}$
and let
$$\hat{g}_0 (k) = \int^1_0  f(\tau \omega(k)/m,  \tau k/m)
d \tau,\quad  \hat{g}_1 (k) = \int^\infty_1  f(\tau \omega(k)/m,
\tau k/m)  d \tau. \eqno{(3.1)}$$
For $\hat{g}_1$, Schwarz inequality gives
$$\eqalign{
\vert \hat{g}_1 (k) \vert & \leq \Big(\int^\infty_1  \tau^{- 2 \delta}
 d \tau\Big)^{1/2}  \Big(\int^\infty_1  \tau^{2 \delta}
\vert f(\tau \omega(k)/m,  \tau  k/m) \vert^2  d \tau\Big)^{1/2}\cr
&\leq \Big(C_\delta \int^\infty_1      \tau^{2 \delta}
\vert f(\tau  \omega(k)/m,  \tau k/m) \vert^2  d \tau\Big)^{1/2},
\quad  \delta > 1/2.\cr
 }$$
Plancherel theorem now gives
$$\eqalign{
\Vert \omega(-i \partial)^{-2}  g_1 \Vert^2_{L^2} & = \Vert\omega^{-2}
\hat{g} \Vert^2_{L^2}\cr
&\leq C_\delta  \int_{\Rrm^3}  dk  \int^\infty_1
 d \tau  \tau^{2 \delta}  \vert f(\tau \omega(k)/m,
 \tau k/m) \vert^2  \omega(k)^{-4}.\cr
 }$$
Making the variable transformation $t = \tau  \omega(k)/m$,
$x = \tau  k/m$, with Jacobian equal to $m^2  \tau^3/\omega(k)$,
in the right-hand side of the last inequality and using the fact that
$\omega(k) = m  t / \tau$, we obtain
$$\Vert \omega(-i \partial)^{-2}  g_1 \Vert^2_{L^2} \leq C_\delta
\int_{D^{}_1}  \vert f(t,x) \vert^2  (1-(x/t)^2)^{\delta}
 t^{2 \delta - 3}  m^{-5}  dt  dx.$$
Since $t \geq 1$ in $D^{}_1$ we obtain, choosing $1/2 < \delta < 1$,
$$\Vert \omega(-i \partial)^{-2}  g_1 \Vert^2_{L^2} \leq C'_\delta
m^{-5}  \sup_{t \geq 1}  \Vert f(t) \Vert^2_{L^2}. \eqno{(3.2)}$$
For $\hat{g}_0$, we have by Schwarz inequality
$$\vert \hat{g}_0 (k) \vert^2 \leq \int^1_0  \vert f(\tau
\omega(k)/m,  \tau  k/m) \vert^2  d \tau.$$
Multiplication by $\omega(k)^{-4}$, integration on $k$ and
changing variables as in the case for $\hat{g}_1$ give, since
$D^{}_1 \subset D^{}_0$,
$$\Vert\omega (-i \partial)^{-2}  g_0 \Vert^2_{L^2}
\leq \int_{D^{}_0 - D^{}_1}
 \vert f(t,x) \vert^2  t^{-3}  m^{-5}  dt
 dx.$$
Since $t \geq 1$ in the support of $f$, we obtain
$$\Vert\omega (-i \partial)^{-2}  g_0 \Vert^2_{L^2} \leq C      m^{-5}
 \sup_{t \geq 1}  \Vert f(t) \Vert^2_{L^2}. \eqno{(3.3)}$$
Definition (3.1), inequalities (3.2) and (3.3) prove statement i)
of the lemma.

To prove the second statement, let
$$C = \sup_{(t,x) \in D^{}_0}  ((1+t)  (1+t-\vert x \vert)^{\varepsilon}
 \vert f(t,x) \vert).$$
It follows from the definition of $\hat{g}$ that
$$\vert \hat{g} (k) \vert \leq C  \int^\infty_0  (1+\tau
\omega(k)/m)^{-1}  (1+ \tau\omega(k)/m - \tau  \vert k \vert/m)^{- \varepsilon}
 d \tau .$$
Since $\omega(k) - \vert k \vert \geq m^2 / (2\omega (k))$ we obtain,
with $t = \tau\omega(k)/m$,
$$\eqalign{
\vert \hat{g} (k) \vert & \leq C  {m \over \omega(k)}  \int^\infty_0
(1+t)^{-1}  (1+t m^2/ (2 \omega(k)^2))^{- \varepsilon}  dt\cr
&\leq C  {m \over \omega(k)}  \Big(\int^{2(\omega(k)/m)^2}_0
(1+t)^{-1}  dt + \int^\infty_{2(\omega(k)/m)^2}  t^{-(1+\varepsilon)}
 dt\  (2 (\omega(k)/m)^2)^\varepsilon\Big)\cr
&= C  {m \over \omega(k)}  (\ln (1+2 (\omega(k)/m)^2) + \varepsilon^{-1}),\cr
}$$
which proves the last statement of the lemma.

In the case where $f$ in Lemma 3.1 is a solution of the wave equation and
if the spatial Fourier transform of $f$ vanishes in a neighbourhood of zero,
then $\hat g$ exists and its $L^{\infty}$-norm can be estimated directly in
terms of weighted $L^{2}$-norms of the Fourier transform of the  initial data.
We recall that $M^{\rho}_c$, $\rho\in]-1/2,\infty[$,
was defined in Theorem~2.9, and since there is no possibility of
confusion we keep the previous notation
$\rho(t,x)=(t^2 - \vert x\vert^2)^{1/2}$ for $\vert t\vert \geq \vert x\vert$.
\saut
\noindent{\bf Lemma 3.2.}
{\it
Let $(h, \dot{h}) \in M^\rho_c$, let
$$f (t,x) = \chi^{}_{0} (\rho (t,x))  \big(\cos ((- \Delta)^{1/2} t)  h
+ (- \Delta)^{- 1/2}  \sin  ((- \Delta)^{1/2} t)  \dot{h}\big)
(x),$$
where $\chi^{}_{0} \in C^\infty ([0,\infty[)$, and let
$\hat{g}$ be  defined as in Lemma~3.1.
\psaut
\noindent\hbox{\rm \phantom{i}i)}
If $- 1/2 < \rho$, $a,b\in \Rrm$, $a-\rho>-1/2$, $a-\rho-b<-1/2$ and
if $0\leq \chi^{}_{0} (\tau) \leq 1$ for $\tau \in [0,\infty[$,
$\chi^{}_{0} (\tau) = 1$ for $\tau \geq 2$, then
$$\Vert\hat{g} \Vert^{}_{L^\infty} \leq C_{\rho,a,b}
\Vert \vert \nabla \vert^{-a}  (1+\vert\nabla\vert)^b(h, \dot{h})
\Vert^{}_{M^\rho}.$$
\noindent\hbox{\rm ii)}
If $1/2 < \rho<1$, and if $\chi^{}_{0} (\tau) = 0$ for $\tau \geq 2$, then
$$
\Vert\omega^{3/2-\rho}\hat g\Vert^{}_{L^\infty}\leq
C_\rho\Vert(h, \dot{h}) \Vert^{}_{M^\rho}.
$$}

Before proving the lemma we remark that the definition of $\hat{g}$ makes
sense.
In fact if $(h, \dot{h})$ is as in the lemma, then $f$ is absolutely integrable
on half-lines starting at the origin according to Proposition 2.15.
Since  $M^{\rho}_{c}$ is dense in $M^{\rho}_{\infty}$ according to Theorem~2.9,
it follows from the inequality in Lemma~3.2 that the linear map
$(h, \dot{h}) \mapsto\hat{g}$ has a unique continuous extension to the Hilbert
space $\vert\nabla\vert^{a}(1+\vert\nabla\vert)^{-b}M^{\rho}$.
\saut
\noindent{\it Proof.}
To prove statement i), let $0<r<R$ be such that the supports of
$\hat h$ and $\hat{\dot h}$ are
contained in $\{p\in \Rrm \big\vert r\leq \vert p\vert$ $\leq R\}$.
We extend the domain of definition of $\chi^{}_{0}$ to $\Rrm$ by defining
$\chi^{}_{0}(\tau)=0$
for $\tau<0$. Let $\varphi \in S(\Rrm)$, $0\leq\varphi(\tau)\leq1$,
$\varphi(\tau)=1$ for $\vert\tau\vert\leq1$ and
$\varphi(\tau)=0$ for $\vert\tau\vert\geq2$.
Let $\chi^{}_{1}=(1-\varphi)\chi^{}_{0}$ and
let $\theta \in S' (\Rrm)$ (resp. $\theta_{1}\in S'(\Rrm)$)
be the inverse Fourier transform of $\sqrt{2 \pi}
\chi^{}_{0}$ (resp. $\sqrt{2 \pi}\chi^{}_{1}$).
Since the derivative $\chi'^{}_{1} \in C^\infty_0$ it follows that
$s \mapsto s  \theta_{1} (s)$ is a function in $S (\Rrm)$, so $\theta_{1}$
restricted to $\Rrm - \{0\}$ is a $C^\infty$ function satisfying
$\theta_1(s)\leq A_n (1+\vert s\vert)^{-n}\vert s\vert^{-1}$, $n\geq 0$,
$s\neq0$
for some constants $A_n$. Since $\chi^{}_0 -\chi^{}_1 \in C_0^\infty
(\Rrm-\{0\})$ and
since $\chi^{}_0 -\chi^{}_1$ has bounded left and right
derivatives of all orders at zero, it
follows that $\theta -\theta_1$ is an entire function satisfying
$\vert\theta(s)-\theta_1(s)\vert\leq A
(1+\vert s\vert)^{-1}$, $s\in \Rrm$, for
some $A>0$. Therefore
$$\vert \theta (s) \vert \leq A \vert s \vert^{-1}, \quad s \neq 0,
\eqno{(3.4\hbox{a})}$$
for some constant $A$.

Since $h \in S ( \Rrm^3)$, we get by Fourier transformation
$$\eqalign{
&(e^{ i\varepsilon \tau  \omega(k) \vert \nabla \vert/m}  h)
(\tau k/m)\cr
&\qquad {} = (2 \pi)^{-3/2}  \int  e^{i \tau  (\omega(k)
\vert p \vert \varepsilon + k \cdot p)/m}  \hat{h} (p)
 dp,\quad   \tau \in \Rrm,  k \in \Rrm^3.\cr
}$$
This gives
$$\eqalign{
&\int^\infty_{- \infty}  (e^{i \varepsilon\tau   \omega(k) \vert
\nabla \vert/m}  h)  (\tau k/m)  \chi^{}_0 (\tau) d \tau\cr
&\qquad {}= (2 \pi)^{-3/2}  \int  \chi^{}_0 (\tau)  e^{i \tau (\omega(k)
 \vert p \vert \varepsilon + k\cdot p)/m}
\hat{h} (p)  dp  d \tau.\cr
}$$
Since ${1 \over m}  (\omega(k)  \vert p \vert + \varepsilon k\cdot p)
\geq (2  \omega(k))^{-1}  m \vert p \vert \geq (2  \omega(k))^{-1}
 mr\ $ in the support of $\hat{h}$, we obtain
$$\eqalignno{
&\int^\infty_{- \infty}  (e^{i \varepsilon \tau  \omega(k)
\vert \nabla \vert/m} h)  (\tau k/m)  \chi^{}_0 (\tau)
d \tau&(3.4\hbox{b})\cr
&\qquad{}= (2 \pi)^{-2}  \int   \theta (m^{-1}  (\varepsilon
 \omega(k)  \vert p \vert + k.p))  \hat{h} (p) dp,\cr
}$$
where according to (3.4a)
$$\vert \theta (m^{-1}  (\varepsilon  \omega(k)  \vert p
\vert + k\cdot p)) \vert
\leq A  m \vert \omega(k)  \vert p \vert + \varepsilon
k\cdot p \vert^{-1}\leq 2A\omega(k)/(m\vert p\vert),
\quad \vert p \vert > 0.$$
Let $d,b\in\Rrm$, $d> -1/2$ and $d-b<-1/2$.
Schwarz inequality and Plancherel theorem give
$$\eqalignno{
&\big\vert \int^\infty_{- \infty}       (e^{i \varepsilon \tau  \omega(k)
 \vert \nabla \vert/m} h)  (\tau k/m)
\chi^{}_0 (\tau)d \tau \big\vert &{(3.5)}\cr
&\quad{}\leq \Big(\int_{r\leq\vert p \vert \leq R}
\vert \theta (m^{-1} (\varepsilon
 \omega (k)     \vert p \vert + k\cdot p)) \vert^2  \vert p \vert^{2d}
(1+\vert p\vert)^{-2b} dp\Big)^{1/2}
\Vert \vert \nabla \vert^{-d} (1+\vert \nabla\vert)^{b} h \Vert^{}_{L^2}.\cr
}$$
Let $F_\varepsilon (s) = \varepsilon  \int_s^{\varepsilon\infty}
\vert \theta (\xi) \vert^2  d \xi$. Then $0\leq F_\varepsilon (s)
\leq B \vert s \vert^{-1}$, for $\varepsilon s > 0,  \varepsilon = \pm$ and
some $B>0$. We have
$$\eqalign{
I (k) & = \int_{r\leq\vert p \vert \leq R}  \vert \theta (m^{-1} (\varepsilon
\omega (k)  \vert p \vert + k\cdot p)) \vert^2 \vert p \vert^{2d}
(1+\vert p\vert)^{-2b} dp\cr
&{}= 2 \pi  \int_{r}^R  d \vert p \vert
\vert p \vert^{2+2d}(1+\vert p\vert)^{-2b}
\int^\pi_0  d {\nu}  \sin {\nu}  \big\vert \theta (m^{-1}
(\varepsilon \omega (k)
\vert p \vert + \cos  {\nu}  \vert k \vert  \vert p \vert))
\big\vert^2 \cr
&{}= 2 \pi  \int_{\vert p \vert \geq r} d \vert p \vert
\vert p \vert^{1+2d}(1+\vert p\vert)^{-2b}
m \vert k \vert^{-1}  \big(F_\varepsilon
(m^{-1}  \varepsilon (\omega(k) - \vert k \vert)  \vert p \vert)\cr
&\qquad {}- F_\varepsilon  (m^{-1}  \varepsilon (\omega(k) + \vert k \vert)
 \vert p \vert)\big).\cr
}$$
This gives $I(k) \leq C_{d,b}\  \omega(k)$ for
$\vert k \vert \geq m$. It follows directly from
the definition of $I(k)$ and from inequality $\vert \theta (m^{-1}
(\varepsilon \omega (k)  \vert p \vert + k\cdot p)) \vert \leq C
\vert p \vert^{-1}$
for $\vert k \vert \leq m$, $\vert p \vert >0$, that $I(k) \leq C'_{d,b}$
for $\vert k \vert \leq m$. Inequality (3.5) now gives
$$\eqalignno{
&\big\vert \int^\infty_{- \infty}(e^{i \varepsilon \tau\omega(k)
\vert \nabla \vert/m} h)  (\tau k/m)  \chi^{}_0 (\tau) d \tau \big\vert
&(3.6)\cr
&\quad{}\leq C_{d,b}  \Vert \vert \nabla \vert^{-d}(1+\vert\nabla\vert)^b
 h \Vert^{}_{L^2},
\quad  d>-1/2,d-b<-1/2.\cr
}$$
Since
$$\hat{g} (k) = \int^\infty_{- \infty}
f(\tau\omega (k)/m,  \tau k/m)  d \tau$$
and
$$f (\tau\omega(k)/m,  \tau  k/m) =
{\chi^{}_0(\tau)\over 2}  \sum_{\varepsilon = \pm}  (e^{\varepsilon i \tau
 \omega(k)  \vert \nabla \vert/m} h - i e^{\varepsilon i \tau
\omega(k)  \vert \nabla \vert/m}  \vert \nabla \vert^{-1}
\dot{h}),$$
the inequality in the lemma follows from (3.6) with $h$ and (3.6)
with $\vert \nabla \vert^{-1}  \dot{h}$ instead of $h$ and by
defining $a=d+\rho$.
This proves statement
i) of the lemma.

To prove statement ii) of the lemma, we observe that in this case, instead of
inequality (3.4a), we have
$$\vert\theta (s)\vert\leq A(1+\vert s\vert)^{-1},\quad s\in \Rrm.$$
In the same way as we obtained (3.5), it now follows that
$$\eqalignno{
&\big\vert \int^\infty_{- \infty}  (e^{i \varepsilon\tau   \omega(k) \vert
\nabla \vert/m}  h)  (\tau k/m)  \chi^{}_0 (\tau) d \tau\big\vert\cr
&\qquad {}\leq \Big( \int \vert\theta(m^{-1} ( \varepsilon \omega(k)
 \vert p \vert + k\cdot p))\vert^2 \vert p \vert^{-2\rho}dp\Big)^{1/2}
\Vert \vert\nabla\vert^{\rho} h \Vert^{}_{L^2},\quad 1/2<\rho<1.\cr
}$$
The last two  inequalities give that
$$\eqalignno{
&\big\vert \int^\infty_{- \infty}  (e^{i \varepsilon\tau   \omega(k) \vert
\nabla \vert/m}  h)  (\tau k/m)  \chi^{}_0 (\tau) d \tau\big\vert\cr
&\qquad {}\leq C_\rho (\omega(k))^{-3/2+\rho}
\Vert \vert\nabla\vert^{\rho} h \Vert^{}_{L^2},\quad 1/2<\rho<1.\cr
}$$
The inequality of statement ii) now follows as in the proof of statement i).
This proves the lemma.

We are now ready to study {\it the phase function} in the definitions
(1.17a) and (1.17b) of $U^{(+)}$.
For $(f, \dot{f}) \in M^\rho_\infty$, $1/2 < \rho < 1$,
we introduce, if $X=a^\nu P^{}_\nu$,
$$(\Phi_{\varepsilon,X}  (f, \dot{f})) (k) = \int^\infty_0
a^\nu l^\mu_\varepsilon  \chi^{}_{0} (m \tau)  B^{(+) 1}_{\nu \mu}
(\tau  l_\varepsilon)  d \tau,\quad     0 \leq \nu \leq 3,
 \varepsilon = \pm, \eqno{(3.7)}$$
where $l^0_\varepsilon = \omega (k)$,  $l^j_\varepsilon = - \varepsilon
k_j$,  $1 \leq j \leq 3,$
$$\eqalignno{
B^{(+)1}_\mu (t) & = \cos  ((- \Delta)^{1/2}  t)  f_\mu +
(- \Delta)^{-1/2}  \sin ((- \Delta)^{1/2}  t)  \dot{f}_\mu, &
{(3.8\hbox{a})}\cr
B^{(+)1}_{0 \mu} (t) &= {d \over dt}  B^{(+)1}_\mu (t),&(3.8\hbox{b}) \cr
B^{(+)1}_{j \mu} (t) &= \partial_j  B^{(+)1}_\mu (t),\quad  1 \leq j \leq 3,\cr
}$$
and where $\chi^{}_0 \in C^\infty ([0,\infty[)$, $\chi^{}_0 (s) = 1$ for
$s \geq 2$ and
$0 \leq \chi^{}_0 (s) \leq 1$. We use Einstein's
summation convention over $\mu$ and $\nu$ in (3.7). It follows directly from
Proposition 2.15 and statement ii) of Lemma 3.1 that
$k \mapsto (\Phi_{\varepsilon, X}
(f, \dot{f})) (k)$ is a $C^\infty$ function for
$(f, \dot{f}) \in M^\rho_\infty.$

We denote by $\Phi_\epsilon(f,\dot f)$ the linear map $X\mapsto
\Phi_{\epsilon,X}(f,\dot f)$ from $\Rrm^4$ to $C^\infty(\Rrm^3)$.

To state the next proposition we note that
$$\eqalignno{
\eta^{(\varepsilon)}_{M^{}_{ij}}& = - k_i  {\partial \over \partial k_j} +
k_j  {\partial \over \partial k_i},\quad  1 \leq i < j \leq 3,
&{(3.9a)}\cr
\eta^{(\varepsilon)}_{M^{}_{0i}}& = - \varepsilon
{\partial \over \partial k_j}
\omega (k),\quad  1 \leq i \leq 3,  \varepsilon = \pm, &{(3.9b)}\cr
}$$
are the generators for a representation $\eta^{(\varepsilon)}$ of
$so (3,1)$ on
$C^\infty (\Rrm^3, \Crm)$
and that the matrices $n_{\mu \nu}$,  $0 \leq \mu < \nu \leq 3$, in
(1.5) define a representation which we denote $X \mapsto n^{}_X$ of
$so (3,1)$ in $\Crm^4$.
\saut
\noindent{\bf Proposition 3.3.}
{\it
Let $N \geq 0$ be an integer and $\Pi''$ be the restriction of
the standard basis $\Pi'$ of $U(\p)$ to the enveloping
algebra $U (so (3,1))$. Then
$$\eqalignno{
\hbox{\rm i)}\  &\sum_{\scr  Y \in \Pi''\atop\scr\vert Y \vert \leq N}
\Vert \omega^{-5/2 -a} \eta^{(\varepsilon)}_Y  \Phi_\varepsilon  (f, \dot{f})
\Vert^{}_{L^2} \leq C_{\rho, N,a,b}  \Vert (1-\Delta)^{b/2} (f, \dot{f})
\Vert^{}_{M^\rho_N},\hskip33mm\cr
\noalign{\hbox{if $(1-\Delta)^{b/2}(f, \dot{f}) \in M^\rho_N$,
$- 1/2 < \rho \leq 3/2$,
$a>0$, $b>3/2-\rho$,}}
\hbox{\rm ii)}\ & \sum_{\scr Y \in \Pi''\atop\scr\vert Y \vert \leq N }
\Vert \omega^{-1}\eta^{(\varepsilon)}_Y  \Phi_\varepsilon  (f, \dot{f})
\Vert^{}_{L^\infty} \leq C_{\rho, N}  \Vert(1-\Delta)^{b/2} (f, \dot{f})
\Vert^{}_{M^\rho_N},\cr
\noalign{\hbox{if $(1-\Delta)^{b/2}(f, \dot{f}) \in M^\rho_N$,
$-1/2 < \rho < 3/2$, $b>3/2-\rho$,}}
\hbox{\rm iii)}\ & \sum_{\scr  Y \in \Pi''\atop\scr\vert Y \vert \leq N}
\vert (\eta^{(\varepsilon)}_Y  \Phi_\varepsilon  (f, \dot{f})(k)\vert
\leq C_{\rho,N} (1+ \ln(1+\omega (k)/m))
\Vert (f, \dot{f}) \Vert^{}_{M^\rho_{N+3}},\cr
\noalign{\hbox{if $(f, \dot{f}) \in M^\rho_{N+3}$, $1/2 < \rho < 1$.}}
}$$
}\saut
\noindent{\it Proof.}
Let $(f, \dot{f}) \in M^{}_c$, where $M^{}_c$ is defined in Theorem 2.9.
The initial conditions $f^{(\nu)}_{\mu}, \dot{f}^{(\nu)}_{\mu}$ of the solution
$B_{\nu\mu}$, given by (3.8b), of the wave equation satisfy
$$\Vert \vert\nabla\vert^{-1}(1-\Delta)^{b/2}(f^{(\nu)}_{}, \dot{f}^{(\nu)}_{})
\Vert^{}_{M^\rho_0}
\leq\Vert(1-\Delta)^{b/2}(f, \dot{f})\Vert^{}_{M^\rho_0},\quad b\in\Rrm.$$
It follows from definition (3.7) of $\Phi_{\varepsilon,X}$, $X\in\p$, and from
Lemma 3.2 (with $a=1$) that
$$\Vert \omega^{-1} \Phi_{\varepsilon, P_\nu} (f, \dot{f})\Vert^{}_{L^\infty}
\leq C_{\rho,b}\Vert(1-\Delta)^{b/2}(f, \dot{f})\Vert^{}_{M^\rho_0},
\eqno{(3.10)}$$
where $\varepsilon = \pm$, $0\leq\nu\leq3$, $-1/2<\rho<3/2$, $b>3/2-\rho$.
We define a group representation $V^{(\varepsilon)}$, $\varepsilon=\pm$,
of $SL(2,\Crm)$
on the space of distributions $r\in S'(\Rrm^3,\Crm^4)$ by
$$(V_A^{(\varepsilon)}r)(k)=\Lambda(A)F^{(\varepsilon)}
(\Lambda(A)^{-1}(\omega(k),
-\varepsilon k)),$$
where $A\mapsto\Lambda(A)$ is the canonical projection of $SL(2,\Crm)$ onto
$SO(3,1)$ and where the function $F^{(\varepsilon)}\colon
\{p\in\Rrm^4\big\vert p_0>0,
p^\mu p_\mu =m^2\}\fl \Crm^4$ is given by $F^{(\varepsilon)}(\omega(k),
-\varepsilon k)=r(k)$. Let $r^{(\varepsilon)}_\nu(f,\dot f) =
\Phi_{\varepsilon, P_\nu} (f, \dot{f})$.
It follows from definition (3.7) of $\Phi_{\varepsilon,X}$ that
$V_A^{(\varepsilon)}r^{(\varepsilon)}(f,\dot f)=r^{(\varepsilon)}
(U^{M1}_{(0,A)}(f,\dot f))$.
Since the differential of $V^{(\varepsilon)}$ is the Lie algebra
representation $\eta^{(\varepsilon)} + n$ and since $(f, \dot{f})
\in M^{}_c\subset M^{\rho}_\infty$, it follows that
$$\eta^{(\varepsilon)}_X  \Phi_{\varepsilon, P_\nu}  (f, \dot{f})
+(n^{}_{X})^{\mu}{}_{\nu}\Phi_{\varepsilon, P_\mu}  (f, \dot{f})
=\Phi_{\varepsilon, P_\nu}  (T^{M1}_X (f, \dot{f})),\quad X\in\p,
\eqno{(3.11)}$$
so
$$\sum_\nu  \vert n^{(\varepsilon)}_X  \Phi_{\varepsilon, P_\nu}
(f, \dot{f})) \leq C  \sum_\nu  (\vert \Phi_{\varepsilon, P_\nu}
(T^{M1}_X (f, \dot{f})) \vert
+ \vert \Phi_{\varepsilon, P_\nu} (f, \dot{f}) \vert),\quad  X \in \Pi
\cap so (3,1),$$
which extends to the enveloping algebra:
$$\sum_\nu  \vert \eta^{(\varepsilon)}_Y  \Phi_{\varepsilon, P_\nu}
 (f, \dot{f}) \vert \leq C_{\vert Y \vert}  \sum_\nu
\sum_{\scr Z \in \Pi''\atop\scr \vert Z \vert \leq \vert Y \vert }
\vert \Phi_{\varepsilon, P_\nu}
 (T^{M1}_Z  (f, \dot{f})) \vert,\quad Y \in \Pi''. \eqno{(3.12)}$$

The inequality in statement ii) of the proposition, for
$(f, \dot{f}) \in M^{}_c$,
now follows from inequality (3.10) and
inequality (3.11), since
$$\Vert (1-\Delta)^{b/2}T^{M1}_Y (f, \dot{f}) \Vert^{}_{M^\rho}
\leq C_{\vert Y\vert,b}\Vert(1-\Delta)^{b/2}(f, \dot{f}) \Vert^{}_{M^\rho_N},
\quad Y\in\Pi'', \vert Y\vert\leq N.$$
We note that $M^{}_c$ is dense in $M^{}_\infty$ according to Theorem 2.19.
So $M^{}_c$ is dense in $M^{\rho}_N$ which proves statement ii).
Statement i) follows trivially from statement ii).

To prove statement iii) we suppose that $(f, \dot{f}) \in M^{}_c$.
It follows from statement ii) of Lemma 3.1 that
$$
\eqalignno{
\vert \Phi_\varepsilon (f, \dot{f}) \vert(k)
&\leq C \big(\ln (1+2 ({\omega(k)/ m})^2) +
({3/2 } - \rho)^{-1}\big)& (3.13)\cr
&\qquad \sum_{\mu, \nu}  \sup_{\scr  t \geq 0\atop\scr  t \geq \vert x \vert}
 ((1+t)  (1+t - \vert x \vert)^{3/2-\rho}  \vert
B^{(+)1}_{\nu, \mu}  (t,x) \vert),\quad \rho < 3/2.\cr
}$$
According to definition (3.8) of $B^{(+)1}_{\nu \mu}$, we obtain using
Proposition 2.15 that statement~iii) is true for $N=0$ and
$(f, \dot{f}) \in M^{}_c$. Inequality (3.12) then proves
the inequality in statement~iii)
for $(f, \dot{f}) \in M^{}_c$. Since $M^{}_c$ is dense
in $M^\rho_{3+N}$, this proves the proposition.

We now define rigourously $T^{(+)}$ by $T^{(+)} = T^1 + T^{(+)2}$,
where $T^{(+)2}_{M^{}_{\mu \nu}} = 0$
for

$0 \leq \mu < \nu \leq 3$ and, if $u = (f, \dot{f}, \alpha) \in
E^{}_\infty, 0\leq \mu \leq 3,$

$$\eqalignno{
&(T^{(+)2}_{P_\mu} (u))^{\wedge}  (k) = i (0,0,  \sum_{\varepsilon = \pm}
(\Phi_{\varepsilon, P_\mu} (f, \dot{f})) (k)  P_\varepsilon  (k)
\hat{\alpha} (k)), \quad &(3.14)\cr
}$$

Proposition 3.3 and the algebraic properties of $T^{(+)}$ will permit
us to prove that $T^{(+)}$ is a nonlinear representation of
$\p$ on $E^{}_\infty$.

\saut

\noindent{\bf Theorem 3.4.}
{\it
$T^{(+)}_X$,  $X \in \p$ is a continuous polynomial of degree two from
$E^\rho_\infty$ to $E^\rho_\infty$ and $T^{(+)}$ is a nonlinear
representation of $\p$ on $E^\rho_\infty$, $1/2 < \rho < 1$.
}\saut
\noindent{\it Proof.}
$T^1_X$, $X \in \p$ is a continuous linear map from $E^{}_\infty$
to $E^{}_\infty$ by the definition of $E^{}_\infty$. It follows
from definition (3.9b) of $\eta^{(\varepsilon)}_{M^{}_{0i}}$,
$1 \leq i \leq 3$, and statement~iii) of Proposition~3.3 that
$\Phi_{\varepsilon, P_\nu}$ is a continuous linear map from
$M^\rho_\infty$ into the space of $C^\infty$ functions on $\Rrm^3$
uniformly bounded together with their derivatives by a constant times the
function $k\mapsto (\omega(k))$. This proves that the bilinear
map $M^\rho_\infty \times D^{}_\infty
\fl S (\Rrm^3,\Crm^4) = D^{}_\infty$ defined by
$((f,\dot{f}), \alpha) \mapsto \Phi_{\varepsilon, P_\nu}  (f,\dot f)
P_\varepsilon  \hat{\alpha}$ is continuous, so $T^{(+)}_X\colon
E^\rho_\infty \fl E^\rho_\infty$ is continuous.

To prove that $T^{(+)}$ is a representation we have to show that
$$T^{(+)}_{XY} - T^{(+)}_{YX} = T^{(+)}_{[X,Y]}, \quad  X, Y \in \p.
\eqno{(3.15)}$$
Since $T^{(+)}_X = T^1_X$ for $X \in sl (2,  \Crm)$, (3.15) is true for
$X, Y \in sl (2,  \Crm)$. It follows from definition (1.5) of $T^1$ and
definition (3.9) of $\eta^{(\varepsilon)}$ that
$$(T^{D1}_X  \alpha)^{\wedge} = \sum_{\varepsilon = \pm}
(\eta^{(\varepsilon)}_X + \sigma^{}_X)  P_\varepsilon  \hat{\alpha},
\quad X\in {\frak{sl}}(2,  \Crm), \eqno{(3.16)}$$
where $X\mapsto \sigma^{}_X$ is the representation of ${\frak{sl}}(2,\Crm)$ on
$\Crm^4$
defined by the matrices $\sigma^{}_{\mu \nu}$,
\hbox{$0 \leq \mu < \nu \leq 3$,}
and where $P_\varepsilon$ is given by (1.16). A direct calculation shows that
$$(- n^{}_X  \Phi_\varepsilon)_\mu = \Phi_{\varepsilon, [X, P_\mu]},
\quad X \in {\frak{sl}}(2,  \Crm),  0 \leq \mu \leq 3. \eqno{(3.17)}$$
Formulas (3.11), (3.16) and (3.17) give (since $\eta^{(\varepsilon)}_X$
is a derivation
and since $T^{D1}_X$ commutes with the projector $P_\varepsilon (-i \partial))$
with $X \in {\frak{sl}}(2,\Crm)$,  $Y \in \Rrm^4$,
$$\eqalignno{
(T^{D2}_{XY}    (f, \dot{f}, \alpha))^{\wedge}
&= (T^{D1}_X T^{D2}_Y(f, \dot{f}, \alpha))^{\wedge}& (3.18)\cr
&= i\sum_{\varepsilon = \pm}  (\eta^{(\varepsilon)}_X +
\sigma^{}_X)  \Phi_{\varepsilon, Y}     (f, \dot{f})  (P_\varepsilon
 (-i \partial) \alpha)^{\wedge}\cr
&= i  \sum_{\varepsilon = \pm}  ((\eta^{(\varepsilon)}_X
 \Phi_{\varepsilon, Y}  (f, \dot{f}))  P_\varepsilon
 \hat{\alpha} + \Phi_{\varepsilon, Y}  (f, \dot{f})
(T^{D1}_X  P_\varepsilon  (-i \partial)) \alpha)^{\wedge}\cr
&= i \sum_{\varepsilon = \pm}  (\Phi_{\varepsilon, Y}
 (T^{M1}_X (f, \dot{f})) +  \Phi_{\varepsilon, [X,Y]}
(f, \dot{f}))  (P_\varepsilon  (-i \partial) \alpha)^{\wedge}\cr
&\qquad {}+i  \sum_{\varepsilon = \pm}  \Phi_{\varepsilon, Y}
(f, \dot{f}))  (P_\varepsilon  (-i \partial) T^{D1}_X
\alpha)^{\wedge}. \cr
}$$
On the other hand we have
$$\eqalignno{
(T^{D2}_{YX}    (f, \dot{f}, \alpha))^{\wedge}
&= ((DT^{D2}_Y.T^1_X) (f, \dot{f}, \alpha))^{\wedge}& (3.19)\cr
&= i    \sum_{\varepsilon = \pm}  \Phi_{\varepsilon,Y}
(T^{M1}_X  (f, \dot{f}))  (P_\varepsilon  (-i \partial)
\alpha)^{\wedge}
+i  \sum_{\varepsilon = \pm}    \Phi_{\varepsilon, Y}
(f, \dot{f}))  (P_\varepsilon  (-i \partial)  T^{D1}_X
 \alpha)^{\wedge}.
}$$
Equalities (3.18) and (3.19) give
$$\eqalignno{
((T^{(+) D2}_{XY} - T^{(+) D2}_{YX})    (f, \dot{f}, \alpha))^{\wedge}
&= i    \sum_{\varepsilon = \pm}  \Phi_{\varepsilon, [X,Y]}
 (f, \dot{f})  (P_\varepsilon  (- i\partial) \alpha)^{\wedge}& (3.20)\cr
&=(T^{(+) D2}_{[X,Y]}  (f, \dot{f}, \alpha))^\wedge,
\quad X \in {\frak {sl}}(2,  \Crm),  Y \in \Rrm^4,\cr
}$$
where the last step follows from definition (3.14) and from the
fact that $[X,Y]$
is in the subalgebra $\Rrm^4$ of $\p$. This proves equality (3.15) for $X \in
sl (2,  \Crm)$, $Y \in \Rrm^4$, since $[T^1_X, T^1_Y] = T^1_{[X,Y]}$.
Now let $X, Y \in \Rrm^4$;
then for $u = (f, \dot{f}, \alpha) \in E^\rho_\infty$
$$\eqalignno{
(T^{(+)D}_{XY} (u))^\wedge & = ((DT^{(+)D}_X. T^{(+)}_Y) (u))^\wedge&(3.21)\cr
&= (T^{(+)D1}_X  T^{(+)D}_Y (u))^\wedge + ((DT^{(+)D2}_X.T^{(+)}_Y)
(u)^\wedge\cr
&= (T^{(+)D1}_X  T^{(+)D}_Y (u))^\wedge + i  \sum_{\varepsilon = \pm}
 \Phi_{\varepsilon, X}  (T^{M1}_Y  (f, \dot{f}))
 (P_\varepsilon  (-i \partial) \alpha)^\wedge\cr
&\qquad {}+i  \sum_{\varepsilon = \pm}  \Phi_{\varepsilon,X}
(f, \dot{f})  (P_\varepsilon  (-i \partial)  (T^{(+)D}_Y(u))^\wedge\cr
&= (T^{D1}_{XY}  \alpha)^\wedge  +i  \sum_{\varepsilon = \pm}
 \Phi_{\varepsilon, Y}  (f, \dot{f})  (P_\varepsilon
(-i \partial)  T^{D1}_X  \alpha)^\wedge\cr
&\qquad {}+i  \sum_{\varepsilon,X}  \Phi_{\varepsilon,X}
(T^{M1}_Y  (f, \dot{f}))  (P_\varepsilon (-i \partial) \alpha)^\wedge\cr
&\qquad {}+i  \sum_{\varepsilon = \pm}  \Phi_{\varepsilon,X}
(f, \dot{f})  (P_\varepsilon (-i \partial)  T^{D1}_Y \alpha)^\wedge\cr
&\qquad {}- \sum_{\varepsilon, \varepsilon' = \pm}  \Phi_{\varepsilon,X}
 (f, \dot{f})  P_\varepsilon  \Phi_{\varepsilon',Y}
 (f, \dot{f})  P_{\varepsilon'}  \hat{\alpha}.\cr
}$$
It follows from (3.21) that for $X,Y \in \Rrm^4$
$$\eqalignno{
&((T^{(+)D}_{XY} - T^{(+)D}_{YX}) (u))^\wedge &(3.22)\cr
&\qquad{}= i    \sum_\varepsilon  (\Phi_{\varepsilon,X}
(T^{M1}_Y  (f, \dot{f})) - \Phi_{\varepsilon,Y}  (T^{M1}_X
 (f, \dot{f})))  (P_\varepsilon (-i \partial) \alpha)^\wedge.\cr
}$$
It follows from definitions (3.7) of $\Phi$ and (3.8) of $B$ that
$$\Phi_{\varepsilon,X}  (T^{M1}_Y  (f, \dot{f})) - \Phi_{\varepsilon,Y}
 (T^{M1}_X  (f, \dot{f})) = 0,\quad X,Y \in \Rrm^4,
\eqno{(3.23)}$$
which according to (3.22) shows that (3.15) is true for $X,Y \in \Rrm^4$. This
proves the theorem.
\saut
\noindent{\bf Lemma 3.5.}
{\it
If $(f, \dot{f}, \alpha) \in E^\rho_\infty$,  ${1\over2} < \rho < 1$,
$0 \leq \mu \leq 3$,  $\varepsilon = \pm$, $0 < a \leq 1$ and
${3\over2}-\rho < b \leq 1$, then
$$\eqalign{
\hbox{\rm i)}\ \Vert \Phi_{\varepsilon,P_\mu}  (f, \dot{f})\hat{\alpha}
\Vert^{}_{L^2}
& \leq C_{\rho,a}\Vert (f, \dot{f} \Vert^{}_{M^\rho_3}
\Vert \omega(-i \partial)^a\alpha \Vert^{}_{D^{}_0}\cr
&\leq C'_{\rho,a}  \Vert (f, \dot{f}) \Vert^{}_{M^\rho_3}  \Vert
\alpha \Vert^{1-a}_{D^{}_0}  \Vert \omega(-i \partial) \alpha
\Vert^{a}_{D^{}_0},
\cr
\hbox{\rm ii)}\ \Vert \Phi_{\varepsilon,P_\mu}  (f, \dot{f}) \hat{\alpha}
\Vert^{}_{L^2}
&\leq C_{\rho,b}  \Vert(1-\Delta)^{b/2} (f, \dot{f}) \Vert^{}_{M^\rho_0}
\Vert\omega(-i\partial) \alpha\Vert^{}_{D^{}_0}\cr
&\leq C'_{\rho,b}\Vert (f, \dot{f}) \Vert^{1-b}_{M^\rho_0}
\Vert(1-\Delta)^{1/2} (f, \dot{f}) \Vert^{b}_{M^\rho_0}
\Vert\omega(-i\partial) \alpha\Vert^{}_{D_0}. \cr
}$$}\saut
\noindent{\it Proof.}
It follows from statement iii) of Proposition 3.3 that
$$\eqalign{
\Vert \Phi_{\varepsilon, P_\nu}  (f, \dot{f})  \hat{\alpha}\Vert^{}_{L^2}
&\leq \Vert \omega^{-a}  \Phi_{\varepsilon, P_\nu}  (f, \dot{f})
\Vert^{}_{L^\infty}  \Vert \omega^a     \hat{\alpha} \Vert^{}_{L^2}\cr
&\leq C_{\rho, a}  \Vert (f, \dot{f}) \Vert^{}_{M^{}_3}  \Vert \omega^a
 \hat{\alpha} \Vert^{}_{L^2},\quad  a > 0.\cr
}$$
Since $\Vert \omega^a  \hat{\alpha} \Vert^{}_{L^2} \leq \Vert
\omega (-i \partial)
 \alpha \Vert^a_{L^2}  \Vert \alpha \Vert^{1-a}_{L^2}$,
$0 < a \leq 1$, we obtain the inequality in statement~i).

We get using statement ii) of Proposition 3.3
$$\eqalign{
\Vert \Phi_{\varepsilon,P_\nu}(f, \dot{f}) \hat{\alpha}\Vert^{}_{L^2}
& \leq\Vert \omega^{-1}  \Phi_{\varepsilon, P_\nu}(f, \dot{f})
\Vert^{}_{L^\infty}\Vert \omega  \hat{\alpha} \Vert^{}_{L^2}\cr
&\leq C_{\rho,b}  \Vert(1-\Delta)^{b/2} (f, \dot{f}) \Vert^{}_{M^\rho_0}
\Vert\omega(-i\partial){\alpha} \Vert^{}_{L^2}, \quad b>3/2-\rho.\cr
}$$
Since $\Vert \omega^{5/2 - \rho}  \hat{\alpha} \Vert^{}_{L^2} \leq \Vert
\omega(-i \partial)^3\alpha \Vert^{}_{L^2} \leq C
\Vert \alpha \Vert^{}_{D^{}_3}$
and by Sobolev embedding $\Vert \omega^3  \hat{\alpha} \Vert^{}_{L^\infty}
\leq C \Vert (1-\Delta)\omega^3  \hat{\alpha} \Vert^{}_{L^2} \leq C' \Vert
\alpha \Vert^{}_{D^{}_3}$
and $\Vert (f^{(i)}, \dot{f}^{(i)}) \Vert^{}_{M^\rho}
\leq C \Vert (f, \dot{f}) \Vert^{}_{M^\rho}$. This proves the first
inequality in statement ii) of the lemma. The second follows from the
first, since
$$\Vert(1-\Delta)^{b/2} (f, \dot{f}) \Vert^{}_{M^\rho_0}
\leq\Vert(f, \dot{f}) \Vert^{1-b}_{M^\rho_0}
\Vert(1-\Delta)^{1/2} (f, \dot{f}) \Vert^{b}_{M^\rho_0}\quad
\hbox{\rm for}\ b\leq 1.$$
This proves the lemma.
\saut
\noindent{\bf Lemma 3.6.}
{\it
If $u_1, u_2 \in E^{}_\infty$,  $X \in \Pi$,  $N \geq 0$ and
${3\over2} - \rho < a \leq 1$, then
$$\eqalign{
\hbox{\rm i)}\
\Vert T^{(+)2}_X  (u_1 \otimes u_2) \Vert^{}_{E^{}_N}
&\leq C_N  (\Vert u_1 \Vert^{}_{E^{}_{1}}  \Vert u_2 \Vert^{}_{E^{}_{N+1}} +
\Vert u_1 \Vert^{}_{E^{}_{N+1}}  \Vert u_2 \Vert^{}_{E^{}_{1}}),\hskip100mm\cr
\hbox{\rm ii)}\ \Vert T^{(+)2}_X  (u_1 \otimes u_2) \Vert^{}_{E^{}_N}
&\leq C_{N,a}   (\Vert u_1 \Vert^{}_{E^{}_{3}}  \Vert u_2 \Vert^{}_{E^{}_N} +
\Vert u_1 \Vert^{}_{E^{}_N}  \Vert u_2 \Vert^{}_{E^{}_{3}})^{1-a}\cr
&\qquad (\Vert u_1 \Vert^{}_{E^{}_{3}}  \Vert u_2 \Vert^{}_{E^{}_{N+1}}
+ \Vert u_1 \Vert^{}_{E^{}_{N+1}} \Vert u_2 \Vert^{}_{E^{}_{3}})^a,\cr
}$$
$$\eqalign{
\hbox{\rm iii)}\ \Vert T^{(+)2}_X(u_1 \otimes u_2) \Vert^{}_E
&\leq C  \min   \big(\Vert u_1 \Vert^{}_{E^{}_3}  \Vert u_2
\Vert^{1-a}_{E^{}_0}
 \Vert u_2 \Vert^{a}_{E^{}_1},  \Vert u_2 \Vert^{}_{E^{}_3}
\Vert u_1 \Vert^{1-a}_{E^{}_0}  \Vert u_1
\Vert^{a}_{E^{}_1}\big).\hskip100mm\cr
}$$
}\saut
\noindent{\it Proof.}
Let ${\cal O}_i$,  $i = 1,2$, be two orderings on the basis $\Pi$ of $\p$
and let $\Pi'_i$ be the canonical basis of $U(\p)$ corresponding to
the ordering ${\cal O}_i$. The norms on the space of $C^N$-vectors given
by (1.6a) corresponding to ${\cal O}_1$ and ${\cal O}_2$ respectively,
are equivalent. Let ${\cal O}_1$ be an ordering such that
$P_\alpha < M^{}_{\mu \nu}$ and ${\cal O}_2$ an ordering such that
$M^{}_{\mu \nu} < P_\alpha$ for $0 \leq \mu < \nu \leq 3$ and
$0 \leq \alpha \leq 3.$

If $Y \in U ({\frak {sl}}(2, \Crm))$, then $T^{(+)}_Y = T^1_Y$ so it follows
from (1.9) that $T^1_Y  T^{(+)2}_X = T^{(+)2}_{YX}$ for $X \in \p$. If moreover
$X \in \Pi \cap \Rrm^4$ and $Y \in \Pi'_2$, then $YX \in \Pi'_2$. Since
$YX$ can be written, as it is seen by commutation, as a linear combination
of elements $Z'Z
\in \Pi'_1$, where $Z' \in \Rrm^4$, $Z \in \Pi'_1 \cap U({\frak {sl}}(2,
\Crm))$
and $\vert Z \vert \leq \vert Y \vert$, we get, remembering that
$T^{(+)2}$ vanishes on ${\frak {sl}}(2,\Crm)$
$$\sum_{\vert Y \vert \leq N}  \sum_{X \in \Pi}  \Vert T^1_Y
 T^{(+)2}_X  (u_1 \otimes u_2) \Vert^2_E
\leq C_N  \sum_{\vert Z \vert \leq N}  \sum_{X \in \Pi}
 \Vert T^{(+)2}_{XZ}  (u_1 \otimes u_2) \Vert^2_E,\eqno{(3.24\hbox{a})}$$
where the summation is taken over
$$Y \in U({\frak {sl}}(2,  \Crm)) \cap \Pi'_2,\quad  Z \in
U({\frak {sl}}(2, \Crm)) \cap \Pi'_1, \eqno{(3.24\hbox{b})}$$
and $u_1, u_2 \in E^{}_\infty.$

It follows by definition (3.14) of $T^{(+)2}$ that
$$\eqalignno{
&T^1_Y  T^{(+)2}_X  (u_1 \otimes u_2)& (3.25)\cr
&\qquad{}= T^{(+)2}_X   \big((f_1, \dot{f}_1,  T^{D1}_Y  \alpha_1)
 \otimes  (f_2, \dot{f}_2,  T^{D1}_Y  \alpha_2)\big),
\quad Y \in U (\Rrm^4),\cr
}$$
where $(f_i, \dot{f}_i, \alpha_i) = u_i \in E^{}_\infty$ for $i = 1,2$.
Inequality (3.24a) with domain of summation (3.24b) and equality
(3.25) give, using that different orderings on $\Pi$ define equivalent norms,
$$\eqalignno{
&\sum_{X \in \Pi}  \Vert T^{(+)2}_X  (u_1 \otimes u_2) \Vert^2_{E^{}_N}
&{(3.26\hbox{a})}\cr
&\qquad \leq C_N  \sum_{\vert Y \vert + \vert Z \vert \leq N}  \sum_{X \in \Pi}
 \Vert T^{(+)2}_{XZ}  \big((f_1, \dot{f}_1,  T^{D1}_Y
 \alpha_1)  \otimes  (f_2, \dot{f}_2, T^{D1}_Y  \alpha_2)\big) \Vert^2_E,\cr
}$$
with summation over
$$Y \in U (\Rrm^4) \cap \Pi'_2,\quad  Z \in U ({\frak {sl}}(2,  \Crm)) \cap
\Pi'_2. \eqno{(3.26\hbox{b})}$$
According to Theorem 2.4 of \refST\ (cf. (2.96)) and  since $T^{(+)}_Z =
T^1_Z$ for $Z \in U({\frak {sl}}(2,  \Crm))$, it is enough to estimate
$$ I_X  (Z_1, Z_2, Y)
=\Vert T^{(+)2}_X  (T^1_{Z_1}   u^Y_1  \otimes T^1_{Z_2}  u^Y_2 \Vert^{}_E,
\eqno{(3.27)}$$
$X \in \Pi$,  $Z_1, Z_2 \in U({\frak {sl}}(2,  \Crm)) \cap \Pi'_2$,
$Y \in U(\Rrm^4) \cap \Pi'_2$,  $\vert Z_1 \vert + \vert Z_2 \vert +
\vert Y \vert \leq N$, $u^{Y}_i=(f_i,\dot f_i,T^{D1}_Y \alpha_i)$,
in order to establish a bound on the right-hand side of (3.26a).
It follows from definition (3.14) of $T^{(+)}$ and Lemma 3.5 that,
for $0 < a \leq 1$ and $3/2 -\rho < b \leq 1$,
$$\eqalignno{
\Vert T^{(+)2}_X  (u_1 \otimes u_2) \Vert^{}_E
&\leq C_a  \Big(\min  \big(\Vert (f_1, \dot{f}_1) \Vert^{}_{M^{}_3}
\Vert \alpha_2 \Vert^{1-a}_{D^{}_0}  \Vert \omega(-i \partial)
\alpha_2 \Vert^a_{D^{}_0},
 &{(3.28)}\cr
&\qquad\qquad  \Vert (f_1, \dot{f}_1) \Vert^{1-b}_{M^{}_0}
\Vert(1-\Delta)^{1/2} (f_1, \dot{f}_1) \Vert^{b}_{M^{}_0}
  \Vert \omega(-i\partial)\alpha_2 \Vert^{}_{D^{}_0}\big)\cr
&\qquad {}+ \min  \big(\Vert (f_2, \dot{f}_2) \Vert^{}_{M^{}_3}
\Vert \alpha_1 \Vert^{1-a}_{D^{}_0}  \Vert \omega(-i \partial)
\alpha_1 \Vert^a_{D^{}_0},\cr
&\qquad\qquad \Vert (f_2, \dot{f}_2) \Vert^{1-b}_{M^{}_0}
\Vert(1-\Delta)^{1/2} (f_2, \dot{f}_2) \Vert^{b}_{M^{}_0}
  \Vert \omega(-i\partial)\alpha_1 \Vert^{}_{D^{}_0}\big)\Big),\cr
}$$
where $u_i = (f_i, \dot{f}_i, \alpha_i) \in E^{}_\infty$.
Definition (3.27) and inequality
(3.28) prove that
$$\eqalignno{
I_X  (Z_1, Z_2, Y)
&\leq C_a  \Big(\min  \big(\Vert (f_1, \dot{f}_1)
\Vert^{}_{M^{}_{\vert Z_1 \vert+3}}
 \Vert \alpha_2 \Vert^{1-a}_{D^{}_{\vert Z_2 \vert + \vert Y \vert}}
\Vert \alpha_2 \Vert^a_{D^{}_{\vert Z_2 \vert + \vert Y \vert + 1}},
&{(3.29)}\cr
&\qquad\qquad \Vert (f_1, \dot{f}_1) \Vert^{1-b}_{M^{}_{\vert Z_1 \vert}}
\Vert (f_1, \dot{f}_1) \Vert^{b}_{M^{}_{\vert Z_1\vert +1}}
\Vert \alpha_2\Vert^{}_{D^{}_{\vert Z_2 \vert + \vert Y \vert + 1}}\big)\cr
&\qquad{}+ \min  \big(\Vert (f_1, \dot{f}_2)
\Vert^{}_{M^{}_{\vert Z_2 \vert + 3}}
\Vert \alpha_1 \Vert^{1-a}_{D^{}_{\vert Z_1 \vert + \vert Y \vert}}
\Vert\alpha_1 \vert^a_{D^{}_{\vert Z_1 \vert + \vert Y \vert + 1}},\cr
&\qquad\qquad \Vert (f_2, \dot{f}_2) \Vert^{1-b}_{M^{}_{\vert Z_2 \vert}}
\Vert \alpha_1
\Vert (f_2, \dot{f}_2) \Vert^{b}_{M^{}_{\vert Z_1+1}}
\Vert \alpha_1\Vert^{}_{D^{}_{\vert Z_1 \vert + \vert Y \vert + 1}}
\big)\Big).\cr
}$$
It follows from (3.26a), (3.27) and (3.29) that for $3/2 -\rho < a \leq 1$ and
$a=b$
$$\eqalignno{
&\Big(\sum_{X \in \Pi}\Vert T^{(+)2}_X(u_1
\otimes u_2)\Vert^2_{E^{}_N}\Big)^{1/2}& (3.30)\cr
&\qquad{}\leq C_{N,a}  \sum_{N_1+N_2+n \leq N}
\Big(\min \big(\Vert u_1 \Vert^{}_{E^{}_{N_1+3}}
\Vert u_2 \Vert^{1-a}_{E^{}_{N_2+n}}
 \Vert u_2 \Vert^a_{E^{}_{N_2+n+1}},\cr
&\qquad\qquad\qquad{}  \Vert u_1 \Vert^{1-a}_{E^{}_{N_1}}
 \Vert u_1 \Vert^a_{E^{}_{N_1+1}}
\Vert u_2 \Vert^{}_{E^{}_{N_2+n+1}}\big)\cr
&\qquad\qquad{}+ \min \big(\Vert u_2 \Vert^{}_{E^{}_{N_2+3}}
\Vert u_1 \Vert^{1-a}_{E^{}_{N_1+n}}
\Vert u_1 \Vert^a_{E^{}_{N_1+n+1}},\cr
&\qquad\qquad\qquad{}\Vert u_2 \Vert^{}_{E^{}_{N_2}}
\Vert u_2 \Vert^{a}_{E^{}_{N_2+1}}
\Vert u_1 \Vert^{}_{E^{}_{N_1+n+1}}\big)\Big)\cr
&\qquad{}\leq  C'_{N,a}  \sum_{N_1+N_2 \leq N}
\Big(\min \big(\Vert u_1 \Vert^{}_{E^{}_{N_1+3}}
\Vert u_2 \Vert^{1-a}_{E^{}_{N_2}}
\Vert u_2 \Vert^a_{E^{}_{N_2+1}},\cr
&\qquad\qquad\qquad{}\Vert u_1 \Vert^{1-a}_{E^{}_{N_1}}
\Vert u_1 \Vert^{a}_{E^{}_{N_1+1}}
\Vert u_2 \Vert^{}_{E^{}_{N_2+1}}\big)\cr
&\qquad\qquad{}+ \min \big(\Vert u_2 \Vert^{}_{E^{}_{N_2+3}}
\Vert u_1 \Vert^{1-a}_{E^{}_{N_1}}
\Vert u_1 \Vert^a_{E^{}_{N_1+1}},\cr
&\qquad\qquad\qquad{}\Vert u_2 \Vert^{}_{E^{}_{N_2}}
\Vert u_2 \Vert^{a}_{E^{}_{N_2+1}}
\Vert u_1 \Vert^{}_{E^{}_{N_1+1}}\big)\Big).\cr
}$$
To prove statement iii) we observe that for $N=2$,
the right hand side of (3.30) is smaller than
$$\eqalign{
&C'_{0,a}
\Big(\min \big(\Vert u_1 \Vert^{}_{E^{}_{3}}
\Vert u_2 \Vert^{1-a}_{E^{}_{0}}
\Vert u_2 \Vert^a_{E^{}_{1}},
\Vert u_1 \Vert^{1-a}_{E^{}_{0}}
\Vert u_1 \Vert^{a}_{E^{}_{1}}
\Vert u_2 \Vert^{}_{E^{}_{1}}\big)\cr
&\qquad\qquad{}+ \min \big(\Vert u_2 \Vert^{}_{E^{}_{3}}
\Vert u_1 \Vert^{1-a}_{E^{}_{0}}
\Vert u_1 \Vert^a_{E^{}_{1}},
\Vert u_2 \Vert^{1-a}_{E^{}_{0}}
\Vert u_2 \Vert^{a}_{E^{}_{1}}
\Vert u_1 \Vert^{}_{E^{}_{1}}\big)\Big)\cr
&\qquad\leq 2C'_{0,a}
\min \big(\Vert u_1 \Vert^{}_{E^{}_{3}}
\Vert u_2 \Vert^{1-a}_{E^{}_{0}}
\Vert u_2 \Vert^a_{E^{}_{1}},
\Vert u_2 \Vert^{}_{E^{}_{3}}
\Vert u_1 \Vert^{1-a}_{E^{}_{0}}
\Vert u_1 \Vert^{a}_{E^{}_{1}}\big).\cr
}$$

To prove statement i) we choose on the right-hand side of (3.30)
the second term in both  minima.
Since $\Vert u_i \Vert^{1-a}_{E^{}_{N_i}}
\Vert u_i \Vert^a_{E^{}_{N_i+1}} \leq
\Vert u_i \Vert^{}_{E^{}_{N_i+1}}$ we obtain the majorization
$$2C'_{N,a}  \sum_{N_1+N_2 \leq N}
\Vert u_1 \Vert^{}_{E^{}_{N_1+1}}  \Vert u_2 \Vert^{}_{E^{}_{N_2+1}}.$$
In this expression no seminorm has a higher index than $N+1$.
Corollary 2.6 now
gives, since $N_1+N_2 \leq N$, that the last expression is smaller than
$$C''_{N,a}  (\Vert u_1 \Vert^{}_{E^{}_1}  \Vert u_2 \Vert^{}_{E^{}_{N+1}}
+ \Vert u_1 \Vert^{}_{E^{}_{N+1}}  \Vert u_2 \Vert^{}_{E^{}_1}),
\quad N \geq 0,$$
which proves statement i).

To prove statement ii) we choose on the right-hand side of (3.30)
the second term in the first minimum for $N_2 \leq N-1$ and the first term for
$N_2=N$.
A similar choice for the second minimum gives the following majorization
of the last member of inequality (3.30):

$$\eqalign{
C'_{N,a}  \Big(&\sum_{\scr N_1+N_2 \leq N\atop\scr  N_2 \leq N-1}
\Vert u_1 \Vert^{1-a}_{E^{}_{N_1}}  \Vert u_2 \Vert^{a}_{E^{}_{N_1+1}}
\Vert u_2 \Vert^{}_{E^{}_{N_2+1}}
+ \Vert u_1 \Vert^{}_{E^{}_{3}}
\Vert u_2 \Vert^{1-a}_{E^{}_N}
\Vert u_2 \Vert^{a}_{E^{}_{N+1}}\cr
{}+ &\sum_{\scr N_1+N_2 \leq N\atop\scr  N_1 \leq N-1}
\Vert u_2 \Vert^{1-a}_{E^{}_{N_2}}
\Vert u_2 \Vert^{a}_{E^{}_{N_2+1}}
\Vert u_1 \Vert^{}_{E^{}_{N_1+1}}
+ \Vert u_2 \Vert^{}_{E^{}_{3}}
\Vert u_1 \Vert^{1-a}_{E^{}_N}
\Vert u_1 \Vert^{a}_{E^{}_{N+1}}\Big).\cr
}$$
It follows, using the second inequality of Corollary 3.6, that this expression
is bounded by
$$\eqalign{
&C''_{N,a}  (\Vert u_1 \Vert^{}_{E^{}_3}  \Vert u_2 \Vert^{}_{E^{}_N} +
\Vert u_1 \Vert^{}_{E^{}_N}  \Vert u_2 \Vert^{}_{E^{}_3})^{1-a}\cr
&\qquad\qquad(\Vert u_1 \Vert^{}_{E^{}_3}       \Vert u_2 \Vert^{}_{E^{}_{N+1}}
+ \Vert u_1 \Vert^{}_{E^{}_{N+1}}
 \Vert u_2 \Vert^{}_{E^{}_3})^a,  \quad 3/2-\rho < a \leq 1,  N \geq 0,\cr
}$$
which proves  statement ii). This proves the lemma.
\saut
\noindent{\bf Corollary 3.7.}
{\it
If $N\geq 0$, $u \in E^{}_\infty$ and $X \in \Pi$, then
$$\eqalign{
\hbox{\rm i)}&\
\Vert T^{(+)2}_X  (u) \Vert^{}_{E^{}_N}
\leq C_{N,a}  \Vert u \Vert^{}_{E^{}_3}
\Vert u \Vert^{1-a}_{E^{}_N}
\Vert u \Vert^{a}_{E^{}_{N+1}},\quad 3/2-\rho<a\leq 1,\cr
\hbox{\rm ii)}&\
\Vert T^{(+)2}_X  (u) \Vert^{}_{E^{}_N}
\leq C_N  \Vert u \Vert^{}_{E^{}_1}
\Vert u \Vert^{}_{E^{}_{N+1}}.\hskip100mm\cr
}$$
}\saut
The properties in Corollary 2.6 of the spaces $E^{}_N$ and the estimates
in Lemma
3.6 of $T^{(+)2}_X$,  $X \in \p$, lead to estimates of $T^{(+)}_Y$,
$Y \in \Pi'$. The proof of this is so similar to that of Lemma 2.19
that we only state the result.
\saut
\noindent{\bf Lemma 3.8.}
{\it
If $u_1,\ldots,u_n \in E^{}_\infty$ and $Y \in \Pi'$, then
$$\eqalign{
\hbox{\rm i)}&\
\Vert T^{(+)n}_Y  (u_1 \otimes\cdots\otimes u_n) \Vert^{}_{E^{}_N}
\leq C  \sum_i  \prod_{1 \leq l \leq n-1}
\Vert u_{i_l} \Vert^{}_{E^{}_3}
\Vert u_{i_n} \Vert^{}_{E^{}_{N+\vert Y \vert}},\
\hbox{for $n \geq 1$, $N \geq 0$,}\hskip100mm\cr
\hbox{\rm ii)}&\
\Vert T^{(+)n}_Y  (u_1 \otimes\cdots \otimes u_n) \Vert^{}_{E^{}_N}
\leq C_a  \Big(\sum_i  \Vert u_{i_1} \Vert^{}_{E^{}_{N+\vert Y \vert-1}}
 \prod^n_{l=2}  \Vert u_{i_l} \Vert^{}_{E^{}_3}\Big)^{1-a}\cr
&\hskip50mm\Big(\sum_i
\Vert u_{i_1} \Vert^{}_{E^{}_{N+\vert Y \vert}}  \prod^n_{l=2}
 \Vert u_{i_l} \Vert_{E^{}_3}\Big)^a, \quad 3/2 -\rho  < a\leq 1,\cr
}$$
for $n \geq 2$ and $\vert Y \vert + N \geq 1$.
Here the summation is taken over all permutations $i$ of $(1,\ldots,n)$
and the constants $C$, $C_a$ depend on $\vert Y \vert$, $n$,  $N$, $\rho$.
}\saut
To prove that {\it the nonlinear representation $T^{(+)}$ majorizes the linear
representation} $T^1$ in $E^{}_N$ we first need to prove this for
$E^{}_3$. This can be done easily using the fact that
$\alpha \mapsto T^{(+)D}_Y  ((f, \dot{f}, \alpha))$
is a linear function for $Y \in U (\Rrm^4).$
\saut
\noindent{\bf Lemma 3.9.}
{\it
Let $u = (f, \dot{f}, \alpha) \in E^{}_\infty$, $g = (f,\dot f,0)$ and
$v = (0,0, \alpha)$. If $Y \in U(\Rrm^4)$ then
$$T^{(+)Dn}_Y  (u \otimes\cdots\otimes u) = n  T^{(+)Dn}_Y
 (g \otimes\cdots\otimes g \otimes v),\quad  n \geq 1.$$
}\saut
\noindent{\it Proof.}
The statement is trivial for $Y = {\un}$. For $Y \in \Pi \cap \Rrm^4$ it
follows
from definition (3.14) of $T^{(+)}$. Suppose it is true for $\vert Y \vert =
L$,
$Y \in \Pi' \cap U (\Rrm^4)$ and let $X \in \Pi \cap \Rrm^4$. Then it
follows from definition (1.9) and the induction hypothesis that:
$$\eqalign{
&T^{(+)Dn}_{YX}  (u \otimes\cdots\otimes u) = n  T^{(+)Dn}_Y
 \Big(\sum_{0 \leq q \leq n-1}  I_q \otimes T^1_X \otimes I_{n-q-1}
 (g \otimes\cdots\otimes g \otimes v)\Big)\cr
&\qquad{}+ n  {2 (n-1) \over n}  T^{(+)D(n-1)}_Y  (g \otimes \cdots
\otimes g \otimes T^{(+)2}_X  (g \otimes v)) = n  T^{(+)Dn}_{YX}
 (g \otimes\cdots\otimes g \otimes v),\cr
}$$
where we have used that $T^{(+)n}$ is symmetric. This proves the lemma.

Lemma 3.8 and Lemma 3.9 have the following immediate corollary
which we state without proof:
\saut
\noindent{\bf Corollary 3.10.}
{\it
$T^{(+)}_Y$, $Y \in \Pi'$, is a $C^\infty$ polynomial
from $E^{}_{N+\vert Y \vert}$
to $E^{}_N$, for $N + \vert Y \vert \geq 3$. If $u = (f, \dot{f}, \alpha) \in
E^{}_{N + \vert Y \vert}\cap E^{}_{3}$, $g = (f, \dot{f})$, then
$$\eqalignno{
\hbox{\rm i)}&\ \Vert T^{(+)}_Y  (u) \Vert^{}_{E^{}_N}
\leq \Vert u \Vert^{}_{E^{}_{N + \vert Y \vert}}
+ \Vert \tilde{T}^{(+)}_Y  (u) \Vert^{}_{E^{}_{N}},\cr
\noalign{\hbox{where $N + \vert Y \vert \geq 0$,}}
\hbox{\rm ii)}&\ \Vert \tilde{T}^{(+)}_Y  (u) \Vert^{}_{E^{}_N}
\leq C_{N, \vert Y \vert}
(\Vert g \Vert^{}_{M^{}_3})  (\Vert g \Vert^{}_{M^{}_3}
\Vert \alpha \Vert^{}_{D^{}_{N + \vert Y \vert}}
+ \Vert g \Vert^{}_{M^{}_{N + \vert Y \vert}}
\Vert \alpha \Vert^{}_{D_3}),\cr
\noalign{\hbox{where $N + \vert Y \vert \geq 0$, $C_{0,0}=0$,}}
\hbox{\rm iii)}&\ \Vert \tilde{T}^{(+)}_Y  (u) \Vert^{}_{E^{}_N}
\leq C_{N, \vert Y \vert,a}
 (\Vert g \Vert^{}_{M^{}_3})
(\Vert g \Vert^{}_{M^{}_3}  \Vert \alpha
\Vert^{}_{D^{}_{N + \vert Y \vert - 1}} +
\Vert g \Vert^{}_{M^{}_{N + \vert Y \vert - 1}}
\Vert \alpha \Vert^{}_{D^{}_3})^{1-a}\cr
&\hskip60mm (\Vert g \Vert^{}_{M^{}_3}
\Vert \alpha \Vert^{}_{D^{}_{N + \vert Y \vert}}
+ \Vert g \Vert^{}_{M^{}_{N + \vert Y \vert}}
\Vert \alpha \Vert^{}_{D^{}_3})^a,
\hskip12mm\cr
}$$
where $3/2 -\rho < a \leq 1$, $N + \vert Y \vert \geq 1$.
$C_{N, \vert Y \vert}$
and $C_{N, \vert Y \vert,a}$ are increasing continuous functions from $\Rrm^+$
to $\Rrm^+$, $\tilde{T}^{(+)} = T^{(+)} - T^1$ and $\tilde{T}^{(+)}_\un =0$.
}\saut
We can now prove that the linear representation $T^1$ is bounded by the
nonlinear representation $T^{(+)}.$
\saut
\noindent{\bf Theorem 3.11.}
{\it
$$\eqalignno{
\hbox{\rm i)}&\ \wp^{}_N  (T^{(+)}  (u)) \leq C_N  (\Vert u \Vert^{}_{E^{}_3})
 \Vert u \Vert^{}_{E^{}_N},\quad  N \geq 0,\cr
\hbox{\rm ii)}&\ \hbox{If $K > 0$ is sufficiently small,
$u = (f, \dot{f}, \alpha) \in E$,
$g = (f, \dot{f}) \in M^{}_3$ and $\Vert g \Vert^{}_{M^{}_3} \leq K$ then}\cr
&\hskip25mm \Vert u \Vert^{}_{E^{}_N} \leq F_N  (\wp^{}_3 (T^{(+)}  (u)))
\wp^{}_N  (T^{(+)}  (u)),\quad  N \geq 3.\cr
\hbox{\rm iii)}&\ \hbox{If $u\in E^{}_3$ then}\cr
&\hskip25mm\Vert u \Vert^{}_{E^{}_N} \leq C_N(\Vert u \Vert^{}_{E^{}_3})
(\wp^{}_N (T^{(+)}  (u))
+\Vert u \Vert^{}_{E^{}_{3}}),\quad  N \geq 3.\cr
&\hskip-12pt\hbox{$C_N$ and $F_N$ are increasing continuous
functions from $\Rrm^+$ to $\Rrm^+.$}\cr
}$$
}\saut
\noindent{\it Proof.}
$u \mapsto T^{(+)}_Y  (u)$, $Y\in\Pi'$, is a polynomial.
It follows from statements i) and ii) of Corollary 3.10 that
$$\Vert T^{(+)}_Y  (u) \Vert^{}_E \leq
\Vert u \Vert^{}_{E^{}_Y} + C_{\vert Y \vert}
 (\Vert u \Vert^{}_{E^{}_3})  \Vert u \Vert^{}_{E^{}_3}  \Vert
u \Vert^{}_{E^{}_{\vert Y \vert}}$$
for $Y \in \Pi'$ and $\vert Y \vert \geq 0$, where $C_{\vert Y \vert}$ is a
continuous increasing function. According to definition (2.109)
of $\wp^{}_N$ we then have
$$\wp^{}_N (T^{(+)}     (u)) \leq C'_N  (\Vert u \Vert^{}_{E^{}_3})
\Vert u \Vert^{}_{E^{}_N},\quad  N \geq 0,$$
which proves statement i) of the theorem.

Statement ii) of Corollary 3.10 shows that, if $u = (f, \dot{f}, \alpha) \in
E$,
$g = (f, \dot{f})$, then
$$\eqalign{
\Vert T^1_Y  u \Vert^{}_E & \leq \Vert T^{(+)}_Y  (u) \Vert^{}_E +
\Vert \tilde{T}^{(+)}_Y  (u) \Vert^{}_E\cr
& \leq \Vert T^{(+)}_Y  (u) \Vert^{}_E + C_{\vert Y \vert}
(\Vert g \Vert^{}_{M^{}_3})  (\Vert g \Vert^{}_{M^{}_3}
\Vert \alpha \Vert^{}_{D^{}_{\vert Y \vert}} +
\Vert g \Vert^{}_{M^{}_{\vert Y \vert}}
\Vert \alpha \Vert^{}_{D^{}_3}),\cr
}$$
for $Y \in \Pi'$ and $\vert Y \vert \geq 0$. This gives
$$\Vert u \Vert^{}_{E^{}_N} \leq \wp^{}_N (T^{(+)}  (u))
+ C'_N  (\Vert g \Vert^{}_{M^{}_3})  (\Vert g \Vert^{}_{M^{}_3}
\Vert \alpha \Vert^{}_{D^{}_N} + \Vert g \Vert^{}_{M^{}_N}
\Vert \alpha \Vert^{}_{D^{}_3}), \quad N \geq 0,\eqno{(3.31)}$$
where $C'_N$ is a continuous increasing function.

Inequality (3.31) and $\Vert \alpha \Vert^{}_{D^{}_3} \leq
\Vert u \Vert^{}_{E^{}_3}$ give
$$\Vert u \Vert^{}_{E^{}_3} \leq \wp^{}_3  (T^{(+)}     (u)) + 2
C'_3  (\Vert g \Vert^{}_{M^{}_3})  \Vert g \Vert^{}_{M^{}_3}
\Vert u \Vert^{}_{E^{}_3}.$$
Since $C'_3$ is continuous we can choose $K > 0$ such that $2  C'_3 (K)
 K < {1/ 2}$, which gives
$$\Vert u \Vert^{}_{E^{}_3} \leq 2  \wp^{}_3  (T^{(+)}  (u)).
\eqno{(3.32\hbox{a})}$$
This proves statement ii) for $N=3$. Suppose for the moment that
statement iii) of the theorem is true. It then follows from inequality
(3.32) and by defining $F_N=3C_N$, $N\geq4$, where $C_N$ is given by
statement iii), that statement ii) of the theorem is true.

We need to prove statement iii) of the theorem.
It follows from statement iii) of Corollary 3.10 that
$$
\Vert u \Vert^{}_{E^{}_N} \leq \wp^{}_N  (T^{(+)}(u))
+ C_{N,a}  (\Vert g \Vert^{}_{M^{}_3})
\Vert u \Vert^{}_{E^{}_3}
\Vert u \Vert^{1-a}_{E^{}_{N-1}}
\Vert u \Vert^{a}_{E^{}_N},
\eqno{(3.32\hbox{b})}$$
$  3/2 -\rho < a \leq 1$,  $N \geq 1$. Let $u\neq0$, since the case
$u=0$ is trivial. Then
$\wp^{}_N(T^{(+)}(u))>0$. Let
$$x^{}_N=\Vert u\Vert^{}_{E^{}_N}/ \wp^{}_N(T^{(+)}(u)).$$
Inequality (3.32b), with $a<1$, shows that
$$x^{}_N\leq 1+A_Nx^{1-a}_{N-1}x^{a}_{N},\quad
3/2 -\rho < a \leq 1,  N \geq 4,$$
where
$$A_N= C_{N,a}  (\Vert g \Vert^{}_{M^{}_3}) \Vert u \Vert^{}_{E^{}_3}
\big(\wp^{}_{N-1}(T^{(+)}(u))/\wp^{}_N(T^{(+)}(u))\big)^{1-a}.$$
As we shall show at the end of this proof, there exists $q^{}_{a}\in
\Rrm^+$, independent of $N$, $A_N$, $x^{}_{N-1}$, $x^{}_{N}$,
such that $x^{}_N\leq q^{}_{a} (1+A^{1/(1-a)}_N x^{}_{N-1}).$
This gives that
$$\Vert u \Vert^{}_{E^{}_N} \leq q^{}_{a}
\big(\wp^{}_N (T^{(+)}(u)) +
(C_{N,a} (\Vert g \Vert^{}_{M^{}_3})
\Vert u \Vert^{}_{E^{}_3})^{1/(1-a)}
\Vert u \Vert^{}_{E^{}_{N-1}}\big), \quad N\geq 4.$$
Induction in $N$ now gives the inequality of statement iii).

Let $x,y,c\in [0,\infty[$, let $b\in ]0,1[$ and let $x\leq 1+ cy^{1-b}
x^b$. If $x\geq 1$, then it follows that $x\leq x^b + cy^{1-b}x^b$,
which gives that $x\leq (1+ cy^{1-b})^{1/(1-b)}$.
Since the function $z\mapsto z^{1/(1-b)}$, $z\geq0$, is convex, it follows
that $x\leq 2^{b/(1-b)}(1+c^{1/(1-b)}y)$. The right-hand side of this
inequality is larger than $1$, so it follows that
$x\leq 2^{b/(1-b)}(1+c^{1/(1-b)}y)$ for $x\geq 0$.
This completes the proof.

We next prove that $U^{(+)}$ extended to $E$, defining $\varphi_g, g \in
{\cal P}_0$, by (1.23a) for $u \in E$, is a {\it continuous representation} in
 $E^{}_N$, $N\geq0$, and that it has the same $C^\infty$-vectors as $U^1$
 for $N \geq 3$ (i.e. the map $g \mapsto U^{(+)}_g (u)$ from ${\cal P}_0
 \hskip.1cm  t_0\hskip.1cm E_N$ is $C^\infty$ if and only if $u \in E_\infty$).
We also prove {\it analyticity} properties of $ u\mapsto U_g^{(+)}(u)$.
\saut
\noindent{\bf Theorem 3.12.}
{\it

\noindent\hbox{\rm \phantom{ii}i)}
Let $N \geq 0$. Then $(g,u)\mapsto U_g^{(+)}(u)$ is a continuous function
from ${\cal P}_0\times E^{}_N$ to $E^{}_N$ and $E^{\circ\rho}_N$
is invariant under $U^{(+)}$.
\psaut
\noindent\hbox{\rm \phantom{i}ii)} If $N\geq 3$, then $E^{}_\infty$
is the set of
$C^\infty$ vectors for $U^{(+)}$ in $E^{}_N$, and
$U^{(+)}\colon {\cal P}_0\times E^{}_\infty \fl E^{}_\infty$
is a $C^\infty$ function and $T^{(+)}_X (u) = (d/dt)\hskip.1cm
U^+ _{exp (tX)}(u)
\mid_{t=0}, u \in E_\infty.$
\psaut
\noindent\hbox{\rm iii)} If $b>0$, $N\geq3$ and if ${\cal J}u=
(f,\dot f, (1-\Delta)^{-1/2}\alpha)$,
$u=(f,\dot f,\alpha)$, then $u\mapsto {\cal J}^bU_g^{(+)}(u)$ is
a real analytic map from $E^{}_N$ to $E^{}_N$ for $g\in{\cal P}_0$.
Moreover if $\sum_{n\geq0} K^n_{g,g_0}(u_0;u-u_0)$ is the Taylor
development of $U^1_{g^{-1}}U^{(+)}_g(u) - U^1_{g_0^{-1}}
U^{(+)}_{g_0}(u_0) - (u-u_0)$ at $u_0$, then there
exists $R>0$ such that
$$\lim_{g\fl g_0}\ \sup_{\Vert u-u_0\Vert^{}_{E^{}_N}\leq R}\
\sum_{n\geq1}\Vert K^n_{g,g_0}(u_0;u-u_0)\Vert^{}_{E^{}_N}=0.$$
}\saut
\noindent{\it Proof.}
Let $u=(f,\dot f,\alpha)\in E^{}_c$, where $E^{}_c$
was defined in Theorem 2.9, $g=(f,\dot f)$ and let
$h(a)=\exp(a^\nu P_\nu)$, $a_0$, $a_1$, $a_2$, $a_3\in\Rrm$.
It follows from definitions (1.17a) and (1.17b) of $U^{(+)}$ and
from definition (1.23a) of the phase function $\varphi$, that
$$\eqalignno{
(U^{(+)D}_{h(a)}(u))^\wedge
&=e^{}_{h(a)}  (g)(U^{1D}_{h(a)}\alpha)^\wedge,&{(3.33\hbox{a})}\cr
\noalign{\hbox{where}}
((e^{}_{h(a)}(g)) \alpha)^\wedge  (k)
&= \sum_{\varepsilon=\pm}\exp(i(q^{}_{\varepsilon}(g,a))(k))
P^{}_{\varepsilon}(k)\hat\alpha(k),&(3.33\hbox{b})\cr
(q^{}_{\varepsilon}(g,a))(k)
&=\varphi_{\exp (a^\nu P_\nu)}(u,-\varepsilon k).&(3.33\hbox{c})\cr
}$$
Let $F\colon\Rrm^4\fl\Rrm^4$ be the solution of the wave equation
with initial data $(U^{1M}_{h(a)}-I)g$. According to
(1.23a) and (3.33c),
$$(q^{}_{\varepsilon}(g,a))(k)=\int_0^\infty \chi^{} _0(\tau m)
l^\mu_\varepsilon(k) F_\mu (\tau l_\varepsilon(k))d\tau,\eqno{(3.34)}$$
where $l^0_\varepsilon(k)=\omega(k)$ and
$l^j_\varepsilon(k)=-\varepsilon k_j$ for $1\leq j\leq 3$.
Since $(U^{1M}_{h(a)}-I)g\in M^{}_c$, it follows from Lemma 3.2
that
$$\vert q^{}_{\varepsilon}(g,a)\vert(k)\leq C_{\rho,b}\ \omega(k)
\Vert \vert \nabla\vert^{-1/2} (1+\vert \nabla\vert)^{b}
(U^{1M}_{h(a)}-I)g\Vert^{}_{M^\rho_0}, \eqno{(3.35\hbox{a})}$$
where $1/2<\rho<1$ and $ 1-\rho<b\leq 1/2$. Since
$\vartheta^\infty$, defined in (1.23b), satisfies
$\vartheta^\infty(H,\Lambda y)=\vartheta^\infty(\Lambda^{-1} H,y)$ for
$\Lambda\in SO(3,1)$, it follows by differentiation with respect to
$\Lambda$ in this expression and from (3.34) and (3.35a), that
$k\mapsto (q^{}_{\varepsilon}(g,a))(k)$ is a $C^\infty$ function
and that
$$\vert \eta^{(\varepsilon)}_Zq^{}_{\varepsilon}(g,a)\vert(k)
\leq C_{\rho,b}\ \omega(k)
\Vert \vert \nabla\vert^{-1/2} (1+\vert \nabla\vert)^{b}
T^{1M}_{Z}
(U^{1M}_{h(a)}-I)g\Vert^{}_{M^\rho_0}, \eqno{(3.35\hbox{b})}$$
for $Z\in U({\frak {sl}}(2,\Crm))$, where the representation
$\eta^{(\varepsilon)}$  is defined in (3.9b). Using that
$$T^{1M}_{X}U^{1M}_{h(a)}=U^{1M}_{h(a)}T^{1M}_{X-[a^\mu P_{\mu}, X]}$$
for $X\in {\frak {sl}}(2,\Crm)$,
that $[P_\mu,X]\in\Rrm^4$ and that $U^{1M}$ is strongly
continuous on $M^\rho_0$, it follows from (3.35b) that
$$\eqalignno{
&\sum_{\scr  Z \in \Pi'\cap U({\frak {sl}}(2,\Crm))\atop\scr\vert Z\vert \leq
N}
\vert \eta^{(\varepsilon)}_Zq^{}_{\varepsilon}(g,a)\vert(k)&(3.35\hbox{c})\cr
 &\qquad{}\leq C_{\rho,N,b}\ \omega(k)
\Big(\sum_{\scr Z \in \Pi'\cap U({\frak {sl}}(2,\Crm))\atop\scr \vert Z\vert
\leq N} \Vert \vert \nabla\vert^{-1/2} (1+\vert \nabla\vert)^{b}
(U^{1M}_{h(a)}-I)T^{1M}_{Z}g\Vert^{}_{M^\rho_0}\cr
&\qquad\qquad{}+\sum_{\scr Y\in\Pi'\atop\scr \vert Y\vert \leq N-1}
\Vert \vert \nabla\vert^{1/2} (1+\vert \nabla\vert)^{b}
T^{1M}_{Y}g\Vert^{}_{M^\rho_0}\Big),\cr
}$$
for $N\geq0$, $\vert a\vert =\sum_\nu \vert a^\nu\vert \leq 1$.
Using elementary properties of the exponential function, it follows that
$$\Vert \vert \nabla\vert^{-1/2} (1+\vert \nabla\vert)^{b}
(U^{1M}_{h(a)}-I)T^{1M}_{Z}g\Vert^{}_{M^\rho_0}
\leq C_b \vert a\vert^{1/2-b}\Vert T^{1M}_{Z}g\Vert^{}_{M^\rho_0},
\quad \vert a\vert\leq 1.$$
The norm $\Vert \vert \nabla\vert^{1/2} (1+\vert \nabla\vert)^{b}
\cdot \Vert^{}_{M^\rho_0}$ is weaker than the norm
$\Vert \vert \nabla\vert^{1/2}\cdot \Vert^{}_{M^\rho_0}$. This shows,
together with inequality (3.35c), that $k\mapsto(q^{}_{\varepsilon}(g,a))(k)$
is a $C^N$ function and that
$$ \sum_{\scr Z \in \Pi'\cap U({\frak {sl}}(2,\Crm)) \atop\scr\vert Z\vert
\leq N} \vert \eta^{(\varepsilon)}_Zq^{}_{\varepsilon}(g,a)\vert(k)
\leq \vert a\vert^{1/2-b}C_{\rho,N,b}\ \omega(k)
\Vert g\Vert^{}_{M^\rho_N},\eqno{(3.36)}$$
where $N\geq0$, $1/2<\rho<1$, $1-\rho<b\leq1/2$, $\vert a\vert\leq 1$
and where $g \in M^\rho_N$,  since $M^\rho_\infty$ is dense in $M^\rho_N$
and $M_c$ is dense in $M^\rho_\infty$, according to Theorem 2.9.

Definition (3.34) of $q^{}_{\varepsilon}(g,a)$  shows that
${\partial\over\partial a^\mu}q^{}_{\varepsilon}(g,a)
= \Phi^{}_{\varepsilon, P_\mu}(U^{1M}_{h(a)}g)$.
This gives that
$$q^{}_{\varepsilon}(g,a)=\int_0^1
a^\mu \Phi^{}_{\varepsilon, P_\mu}(U^{1M}_{h(as)}g)ds.\eqno{(3.37)}$$
This equality, statement iii) of Proposition 3.3 and the strong continuity
of $U^{1M}$ on $M^\rho_N$ show that
$$ \sum_{\scr  Z \in \Pi'\cap U({\frak{sl}}(2,\Crm))\atop\scr\vert Z\vert \leq
N}
\vert \eta^{(\varepsilon)}_Zq^{}_{\varepsilon}(g,a)\vert(k)
\leq C_{\rho,N}\vert a\vert
(1+\ln(1+\omega(k)/m))\Vert g\Vert^{}_{M^\rho_{N+3}},\eqno{(3.38)}$$
where $N\geq0$, $1/2<\rho<1$, $\vert a\vert\leq 1$
and  $g \in M^\rho_{N+3}$.

Let $N\geq0$, $g \in M^\rho_{N}$, $a\in\Rrm^4$, $\vert a\vert\leq 1$,
$\alpha \in D^{}_{N}$, and let $Y \in \Pi'\cap U(\Rrm^4)$ and
$Z \in \Pi'\cap U({\frak{sl}}(2,\Crm))$,
$\vert Y\vert +\vert Z\vert\leq N$. We obtain using definition (3.9) of
$\eta^{(\varepsilon)}_{}$
$$\eqalignno{
&(T^{D1}_{YZ}(e^{}_{h(a)}(g)-I) P^{}_{\varepsilon}(-i\partial)
\alpha)^\wedge&(3.39)\cr
&\qquad{}=\sum_{Z',Z''\in D(Z)}
C(Z,Z',Z'')(\eta^{(\varepsilon)}_{Z'}(e^{iq^{}_{\varepsilon}(g,a)}-1))
P^{}_{\varepsilon}(-i\partial)T^{D1}_{YZ''}
\alpha)^\wedge,\cr
}$$
where $C(Z,Z',Z'')\in \Rrm$ and $D(Z)$ is the set of $Z',Z''\in
\Pi'\cap U({\frak{sl}}(2,\Crm))$ such that $\vert Z\vert=\vert Z'\vert+\vert
Z''\vert$.
We have used here that $T^{D1}_{X} e^{}_{h(a)}(g)\alpha
=e^{}_{h(a)}(g)T^{D1}_{X}\alpha$, $X\in \Rrm^4$, according
to definition (3.33b) of $e^{}_{h(a)}(g)$. It follows from equation
(3.39) that
$$\eqalignno{
&\Vert (e^{}_{h(a)}(g)-I)\alpha\Vert^{}_{D^{}_N}&(3.40)\cr
&\qquad \leq C_N
\sum_{\scr Y\in \Pi', Z \in \Pi'\cap U({\frak{sl}}(2,\Crm))
\atop\scr\vert Z\vert+\vert Y\vert \leq N, \varepsilon=\pm}
\Vert (\eta^{(\varepsilon)}_{Z}
(e^{iq^{}_{\varepsilon}(g,a)}-1))
P^{}_{\varepsilon}(-i\partial)T^{D1}_{Y}
\alpha)^\wedge\Vert^{}_{L^2}.\cr
}$$

If $Z_1,\ldots,Z_n\in\Pi'\cap U({\frak{sl}}(2,\Crm))$, $n\geq1$,
$L=\vert Z_1\vert+\cdots+ \vert Z_n\vert$, then it follows from inequality
(3.36) and Corollary 2.6 that
$$
\vert(\eta^{(\varepsilon)}_{Z_1}q^{}_{\varepsilon}(g,a))(k)\cdots
(\eta^{(\varepsilon)}_{Z_n}q^{}_{\varepsilon}(g,a))(k)\vert
\leq \vert a\vert^{n(1/2-b)}\omega(k)^n C_{\rho,b,N,L}
\Vert g\Vert^{n-1}_{M^\rho_0}\Vert g\Vert^{}_{M^\rho_L},\eqno{(3.41)}$$
for  $\vert a\vert\leq 1$, $1/2<\rho<1$, $1-\rho<b\leq1/2$.
Inequality (3.40), Leibniz rule and inequality (3.41)
with $L=\vert Z\vert$ and $\vert Z_i\vert\geq 1$ for
$1\leq i\leq n$, give that
$$\eqalignno{
&\Vert (e^{}_{h(a)}(g)-I)\alpha\Vert^{}_{D^{}_N}&(3.42)\cr
&\qquad\leq C_N \sum_{\scr  Y\in \Pi' \atop\scr\vert Y\vert=N}
\Vert (e^{}_{h(a)}(g)-I)T^{D1}_{Y}\alpha\Vert^{}_{D^{}_0}
+C_N\vert a\vert^{1/2-b}
\sum_{\scr N_1+N_2\leq N \atop\scr N_1 \geq 1 }
\Vert g\Vert^{}_{M^\rho_{N_1}}\Vert \alpha\Vert^{}_{D^{}_{N_1+N_2}},\cr
}$$
where $C_N$ depends only on $\rho$, $b$ and
$\Vert g\Vert^{}_{M^\rho_{0}}$. Given
a bounded subset $K\subset \Rrm^3$ and a bounded subset $B\subset M^\rho_0$,
according to inequality (3.36) for
$\vert a\vert\leq 1$, $\exp(iq^{}_{\varepsilon}(g,a))(k))-1$
converges to zero, uniformly for $(k,g)\in K\times B$,
(resp. $(k,a)\in K\times \{a\in\Rrm^4\big\vert \vert a\vert\leq 1 \}$)
when $a\fl 0$ (resp. $g\fl0$ in $M^\rho_{0}$).

Moreover, since
$\vert\exp(iq^{}_{\varepsilon}(g,a))(k))-1\vert\leq 2$, it follows
from inequality (3.40), with $N=0$, and from the Lebesgue dominated
convergence theorem, that for $Y\in\Pi'$, $\vert Y\vert\leq N$,
$N\geq 0$, $\alpha\in D^{}_N$,
$$\Vert (e^{}_{h(a)}(g)-I)T^{D1}_{Y}\alpha\Vert^{}_{D^{}_0}\fl 0$$
uniformly on $\{a\big\vert\vert a\vert\leq 1\}$ (resp. bounded subsets of
$M^\rho_0$) when $g\fl 0$ in $M^\rho_0$ (resp. $a\fl 0$).
Inequality (3.42) now shows that, if $B$ is a bounded subset of
$M^\rho_N$ and $\alpha\in D^{}_{N}$, then
$(e^{}_{h(a)}(g)-I)\alpha\fl 0$ in $D^{}_{N}$ uniformly
on $\{ \vert a\vert\leq 1\}$ (resp. $B$) when $g\fl 0$
in $M^\rho_N$ (resp. $a\fl 0$). Moreover it follows from inequality
(3.42) that
$\Vert e^{}_{h(a)}(g)\alpha\Vert^{}_{D^{}_N}\leq C_B \Vert \alpha
\Vert^{}_{D^{}_N}$ for $\vert a\vert\leq 1$, $g\in B$. Since
$$e^{}_{h(a)}(g_1)\alpha_1-\alpha=e^{}_{h(a)}(g_1)(\alpha_1-\alpha) +
(e^{}_{h(a)}(g_1)-I)\alpha,$$
it then follows that $\Vert e^{}_{h(a)}(g_1)\alpha_1 -\alpha\Vert^{}_{D^{}_N}
\fl 0$, when $(a,g_1,\alpha_1)\fl(0,(g,\alpha))$ in $\Rrm^4\times
E^\rho_N$. Equality (3.33a) and the fact that $U^{1D}$ is a strongly
continuous linear representation of ${\cal P}_0$ in $D^{}_N$
give that the function $(a,u)\mapsto U^{(+)D}_{h(a)} (u)$ from
$\{\vert a\vert\leq 1\}\times E^\rho_N$ to $D^{}_N$ is continuous at $(0,u)$
and therefore that the function $(a,u)\mapsto U^{(+)}_{h(a)} (u)$ from
$\{\vert a\vert\leq 1\}\times E^\rho_N$ to $E^\rho_N$
is continuous at $(0,u)$, since $U^{(+)} = U^{1M}\oplus U^{(+)D}$
and $U^{1M}$ is a strongly continuous linear representation of
${\cal P}_0$ in $E^{\rho}_N$.
The fact that the set of elements $U^{(+)D}_{h(a)}$, $a\in\Rrm^4$,
is a connected group
now shows that $(a,u)\mapsto U^{(+)}_{h(a)} (u)$ is a continuous  map
from $\Rrm^4\times E^\rho_N$ to $E^\rho_N$. Finally, since
$U^{(+)}_A= U^1_A$ for $A\in SL(2,\Crm)$ and since $U^1$ is a strongly
continuous linear representation, it follows that
$(j,u)\mapsto U^{(+)}_{j} (u)$ is a continuous  map
from ${\cal P}_0\times E^\rho_N$ to $E^\rho_N$, $N\geq 0$.
$M^{\circ\rho}_N$ is invariant under $U^{1M}$, so by the definition of
$U^{(+)}$, the space $M^{\circ\rho}_N$ is invariant under $U^{(+)}$.

To prove statement iii), let $e^{n-1}_{h(a)}(g)$, $n\geq 1$,
$g\in M^\rho_\infty$, $a \in\Rrm^4$, be defined by
$$(e^{n-1}_{h(a)}(g)\alpha)^\wedge (k) =
\sum_{\varepsilon =\pm}{1\over (n-1)!}(i (q^{}_{\varepsilon}(g,a))(k))^{n-1}
P^{}_{\varepsilon}(k) \hat\alpha(k),\eqno{(3.43)}$$
where $\alpha\in D^{}_\infty$ and $k\in \Rrm^3$. Similarly
as we obtained inequality (3.40), it follows from (3.43) that
$$\eqalignno{
&\Vert e^{n-1}_{h(a)}(g)\alpha \Vert^{}_{D^{}_N}&(3.44)\cr
&\qquad\leq C_N ((n-1)!)^{-1}
\sum_{\scr Y\in \Pi', Z \in \Pi'\cap U({\frak{sl}}(2,\Crm))
\atop\scr\vert Z\vert+\vert Y\vert \leq N, \varepsilon=\pm }
\Vert (\eta^{(\varepsilon)}_{Z}
(q^{}_{\varepsilon}(g,a)^{n-1}))
(P^{}_{\varepsilon}(-i\partial)T^{D1}_{Y}
\alpha)^\wedge\Vert^{}_{L^2},\cr
}$$
where $N\geq 0$, $(g,\alpha)\in E^\rho_\infty$, $a \in\Rrm^4$ and $n\geq 1$.

Let $N\geq3$, $n\geq2$, $Z$, $Z_1,\ldots,Z_{n-1}\in\Pi'\cap U(
{\frak{sl}}(2,\Crm))$,
$\vert Z_1\vert+\ldots+\vert Z_{n-1}\vert =\vert Z\vert$ and let
$0\leq \vert Z\vert \leq N$. If $L$ is the number of elements $Z_i$
such that $\vert Z_i\vert +3>N$, then $L(N-2)\leq \vert Z\vert$.
This gives that $L=0$ for $\vert Z\vert\leq N-3$, $L\leq 1$ for
$\vert Z\vert=N-2$, $L\leq 2$ for $\vert Z\vert=N-1$ and
$L\leq 2$ for $\vert Z\vert=N$. Therefore if $L\geq1$, then
$N-\vert Z\vert+L\leq 3$. To estimate the product
$$\vert(\eta^{(\varepsilon)}_{Z_1}q^{}_{\varepsilon}(g,a))\cdots
(\eta^{(\varepsilon)}_{Z_{n-1}}q^{}_{\varepsilon}(g,a))\vert(k)$$
we can suppose that $\vert Z_i\vert\geq\vert Z_j\vert$ for $i<j$. If
$L=0$ then we use inequality (3.38) for all the factors
and if $L\geq1$ then we use inequality (3.36), with
$1-\rho<b\leq1/2$, for the factors
$\eta^{(\varepsilon)}_{Z_i}q^{}_{\varepsilon}(g,a)$, $1\leq i\leq L$,
and inequality (3.38) for the other factors. This gives
$$\eqalign{
&\vert(\eta^{(\varepsilon)}_{Z_1}q^{}_{\varepsilon}(g,a))\cdots
(\eta^{(\varepsilon)}_{Z_{n-1}}q^{}_{\varepsilon}(g,a))\vert(k)\cr
&\qquad \leq\Big(
C_{\rho,b,N}\vert a\vert^{1/2-b} \omega(k))^L
\Vert g\Vert^{}_{M^{}_{\vert Z_1\vert}}\cdots
\Vert g\Vert^{}_{M^{}_{\vert Z_L\vert}}\cr
&\qquad\qquad\prod_{i=L+1}^{n-1}
C_{\rho,\vert Z_i\vert}\Vert g\Vert^{}_{M^{}_{\vert Z_i\vert+3}}
\vert a\vert
(1+\ln(1+\omega(k)/m))\Big),\quad
\vert a\vert\leq 1.\cr
}$$
Using Corollary 2.6 and redefining the constants, we obtain that
$$\eqalign{
&\vert(\eta^{(\varepsilon)}_{Z_1}q^{}_{\varepsilon}(g,a))\cdots
(\eta^{(\varepsilon)}_{Z_{n-1}}q^{}_{\varepsilon}(g,a))\vert(k)\cr
&\qquad \leq C_N C^{n-1}_0
\Vert g\Vert^{n-2}_{M^{}_{3}}\Vert g\Vert^{}_{M^{}_{N}}
\vert a\vert^{(n-1)(1/2-b)}(1+\ln(1+\omega(k)/m))^{n-1}
\omega(k)^L.\cr
}$$
Since $N-\vert Z\vert +L\leq 3$, when $L\geq 1$ and since $0\leq L\leq 3$,
we now obtain from (3.44), by using Leibniz rule,
$$\eqalign{
\Vert e^{n-1}_{h(a)}(g)\alpha \Vert^{}_{D^{}_N}
& \leq C_N C^{n-1}_0((n-1)!)^{-1}\vert a\vert^{(n-1)(1/2-b)}\cr
&\qquad\quad\Vert g\Vert^{n-2}_{M^{}_{3}}\Vert g\Vert^{}_{M^{}_{N}}
\sum_{\scr Y\in \Pi'
\atop\scr\vert X\vert+\vert Y\vert \leq N}
\Vert (1+\ln(1+\omega(\cdot)/m))^{n-1}(T^{D1}_{Y}\alpha)^\wedge
(\cdot)\Vert^{}_{L^2}.\cr
}$$
If $0<s<1$, then
$$\Vert (1+\ln(1+\omega(\cdot)/m))^{n-1}
(1+\omega(\cdot)/m)^{-s}\Vert^{}_{L^\infty}
\leq \big({n-1\over s}\big)^{n-1}\exp(1-(n-1)/s),
$$
so using the Stirling formula for $(n-1)!$ we obtain, with new constants,
$$\eqalignno{
&\Vert e^{n-1}_{h(a)}(g)\alpha \Vert^{}_{D^{}_N}&(3.45) \cr
&\quad{} \leq C_N C^{n-1}_0(se^{(1-s)/s})^{-(n-1)}
\vert a\vert^{(n-1)(1/2-b)}
\Vert g\Vert^{n-2}_{M^{}_{3}}\Vert g\Vert^{}_{M^{}_{N}}
\Vert (1-\Delta)^{s/2}\alpha \Vert^{}_{D^{}_N},\cr
}$$
for $N\geq3$, $1-\rho<b\leq1/2$, $n\geq2$, $0<s<1$, where we have
used that the norms $\Vert (1-\Delta)^{s/2}\cdot \Vert^{}_{D^{}_N}$
and $\Vert(1+\omega(-i\partial)/m)^{s}\cdot\Vert^{}_{D^{}_N}$
are uniformly equivalent for $0\leq s\leq 1$.
For given $\rho$, $b$, $s$ such that $1/2<\rho<1$, $1-\rho<b\leq1$
and $0<s<1$, there exists, according to inequality (3.45), $r>0$
and $C^{(r)}_N\geq0$, $N\geq3$, such that
$$\sum_{n\geq 2}\Vert e^{(n-1)}_{h(a)}(g)
(1-\Delta)^{-s/2}\alpha \Vert^{}_{D^{}_N}
\leq C^{(r)}_N\vert a\vert^{1/2-b} \Vert g\Vert^{}_{M^{}_{N}}
\Vert\alpha \Vert^{}_{D^{}_N},
\eqno{(3.46)}$$
for $N\geq3$, $(g,\alpha)\in E^{}_{N}$, $ \Vert g\Vert^{}_{M^{}_{3}}\leq r$,
$\vert a \vert\leq1$.

Let $u=(g,\alpha)\in E^{}_{N}$, $u_0=(g_0,\alpha_0)\in E^{}_{N}$
and  $N\geq3$. According to definition (3.33b) of $e^{}_{h(a)}$,
it follows that $e^{}_{h(a)}(g)=e^{}_{h(a)}(g-g_0)e^{}_{h(a)}(g_0)$.
The equality
$$\eqalign{
e^{}_{h(a)}(g)\alpha-  e^{}_{h(a)}(g_0)\alpha_0
&=e^{}_{h(a)}(g_0)(\alpha-\alpha_0)\cr
&\qquad{}+(e^{}_{h(a)}(g-g_0)-I)e^{}_{h(a)}(g_0)(\alpha-\alpha_0)\cr
&\qquad{}+(e^{}_{h(a)}(g-g_0)-I)e^{}_{h(a)}(g_0)(\alpha_0),\cr
}$$
the definitions
$$\eqalign{
F^{1}_{h(a)}(u_0;u-u_0)
&=(e^{}_{h(a)}(g_0)-I)(\alpha-\alpha_0)\cr
&\qquad{}+e^{(1)}_{h(a)}(g-g_0)e^{}_{h(a)}(g_0)\alpha_0,\cr
F^{n}_{h(a)}(u_0;u-u_0)
&=e^{(n-1)}_{h(a)}(g-g_0)e^{}_{h(a)}(g_0)(\alpha-\alpha_0)\cr
&\qquad{}+e^{(n)}_{h(a)}(g-g_0)e^{}_{h(a)}(g_0)\alpha_0,\quad n\geq2,\cr
}$$
the fact that
$\Vert e^{}_{h(a)}(g)\beta \Vert^{}_{D^{}_N}
\leq C'_N (g_0) \Vert\beta \Vert^{}_{D^{}_N}$, which has been
established in the proof of statement i), and inequality (3.46), show
that if $r$ and $C^{(r)}_N$ are as in (3.46), then the series
$$\sum_{n\geq1} \Vert F^{n}_{h(a)}
({\cal J}^s u_0;{\cal J}^s(u-u_0))\Vert^{}_{D^{}_N}$$
converges uniformly for $\Vert g-g_0\Vert^{}_{M^{}_{3}}\leq r$,
$$\eqalign{
&\sum_{n\geq 1}F^{n}_{h(a)}({\cal J}^s u_0;{\cal J}^s(u-u_0))\cr
&\qquad{}=e^{}_{h(a)}(g)(1-\Delta)^{-s/2}\alpha
-e^{}_{h(a)}(g_0)(1-\Delta)^{-s/2}\alpha_0
-(1-\Delta)^{-s/2}(\alpha-\alpha_0)\cr
}$$
and that
$$\eqalignno{
&\sum_{n\geq1} \Vert F^{n}_{h(a)}
({\cal J}^s u_0;{\cal J}^s(u-u_0))\Vert^{}_{D^{}_N}&(3.47)\cr
&\qquad{}\leq\Vert (e^{}_{h(a)}(g_0)-I)
(1-\Delta)^{-s/2}(\alpha-\alpha_0)\Vert^{}_{D^{}_N}\cr
&\qquad\qquad{}
+C^{(r)}_N\vert a\vert^{1/2-b}C'_N(g_0)\Vert g-g_0\Vert^{}_{M^{}_{N}}
(\Vert\alpha-\alpha_0\Vert^{}_{D^{}_N}
+\Vert\alpha_0\Vert^{}_{D^{}_N}),\quad \vert a\vert\leq 1, N\geq3.\cr
}$$
It follows from (3.38), with $N=0$, and from (3.42), with $N\geq3$ and
$1-\rho<b\leq1/2$, that
$$\Vert (e^{}_{h(a)}(g_0)-I)
(1-\Delta)^{-s/2}(\alpha-\alpha_0)\Vert^{}_{D^{}_N}\fl 0$$
uniformly on bounded subsets of elements $\alpha-\alpha_0\in D^{}_N$,
when $a\fl0$. Choosing $1-\rho<b\leq1/2$ in (3.47), we obtain that
$$\lim_{a\fl0}\
\sup_{\scr\Vert u-u_0\Vert^{}_{E^{}_{N}}\leq R
\atop\scr\Vert g-g_0\Vert^{}_{M^{}_{3}}\leq r}\
\sum_{n\geq1} \Vert F^{n}_{h(a)}
({\cal J}^s u_0;{\cal J}^s(u-u_0))\Vert^{}_{D^{}_N}=0,
\eqno{(3.48)}$$
for all $N\geq3$, $R>0$. By definition (3.33b) of $e^{}_{h(a)}$,
it follows that
$$S^{}_{h(a)}(u)=U^{1}_{h(a)^{-1}}U^{(+)}_{h(a)}(u),$$
then $S^{}_{h(a)}(u)=(g, e^{}_{h(a)}(g)\alpha)$, for $u=(g,\alpha)$.
Therefore we have proved that
$$S^{}_{h(a)}\circ{\cal J}^s\colon
E^{}_N\fl E^{}_N,\quad  N\geq3, \vert a\vert\leq 1,$$
is a real analytic
function and, if
$$S^{}_{h(a)}(u)=\sum_{n\geq0}S^{n}_{h(a)}(u_0;u-u_0)$$
is the power series development of $S^{}_{h(a)}(u)$
at $u_0$, choosing $0<R\leq r$
in (3.48),  we have:
$$\eqalignno{
&\lim_{a\fl0}\Vert S^{}_{h(a)}({\cal J}^s u)-S^{}_{h(a)}({\cal J}^s u_0)
- {\cal J}^s (u-u_0)\Vert^{}_{E^{}_N}& (3.49)\cr
&\qquad{}\leq\lim_{a\fl0}\ \sup_{\Vert u\Vert^{}_{E^{}_{N}}\leq R}\
\Big(\sum_{n\geq2}
\Vert S^{n}_{h(a)}({\cal J}^su_0;{\cal J}^s(u-u_0))\Vert^{}_{E^{}_{N}}\cr
&\qquad\qquad{}+
\Vert S^{1}_{h(a)}({\cal J}^su_0;{\cal J}^s(u-u_0))
-{\cal J}^s(u-u_0)\Vert^{}_{E^{}_{N}}\Big) = 0.\cr
}$$
Since $U^{(+)}$ is a group representation and since ${\cal J}^s$ commutes
with $U^{(+)}_{h(a)}$, we can conclude that
$S^{}_{h(a)}\circ{\cal J}^s\colon E^{}_N\fl E^{}_N$
is real analytic for $a \in\Rrm^4$ and that inequality (3.49) is true with
$a\fl a_0$, $a_0 \in\Rrm^4$, instead of $a\fl 0$. Since $U^{(+)}$ is linear on
$SL(2,\Crm)$, this proves statement iii) of the theorem.

To prove statement ii), let $u\in E^{}_N$ be a $C^\infty$-vector
for $U^{(+)}$ in $E^{}_N$, $N\geq3$. Then $\wp^{}_L (T^{(+)}(u))<\infty$,
$L\geq 0$, according to definition (2.109) of $\wp^{}_L$ and that of
$C^\infty$-vectors. It follows from statement iii) of Theorem 3.11 that
$$\Vert u\Vert^{}_{E^{}_L}\leq C_L (\Vert u\Vert^{}_{E^{}_3})
(\wp^{}_L (T^{(+)}(u)) + \Vert u\Vert^{}_{E^{}_3}).$$
This proves that $u\in E^{}_\infty$.
Now, let $u$ be a $C^\infty$-vector of $U^1$ in $E^{}_N$, $N\geq3$,
i.e. $u\in E^{}_\infty$. Then it follows from statement i) of Theorem 3.11
that $\wp^{}_L (T^{(+)}(u))<\infty$ for $L\geq 0$, which proves that
$u$ is a differential vector for $U^{(+)}$.

According to Corollary 3.10 $T^{(+)}_Y \colon E^{}_\infty\fl E^{}_\infty$,
$Y\in\Pi'$, is a $C^\infty$ function. Statement~iii) of the theorem shows
in particular that $(g,u)\mapsto (D^nU^{(+)}_g)(u;u_1,\ldots,u_n)$ is a
continuous function from ${\cal P}_0\times E^{}_\infty$
into $E^{}_\infty$ for $u_i\in E^{}_\infty$. Therefore
$(D^n(T^{(+)}_Y\circ U^{(+)}_g))(u;u_1,\ldots,u_n)$ is continuous from
${\cal P}_0\times E^{}_\infty$ into $E^{}_\infty$. It then follows
from the definition of $T^{(+)}$ that the function
$(g,u)\mapsto (U^{(+)}_g)(u)$ from ${\cal P}_0\times E^{}_\infty$
to $E^{}_\infty$ is $C^\infty$. This proves the theorem.

The representation $U^{(+)}$ of ${\cal P}_0$ on $E^{\rho}_\infty$
or on $E^{\circ\rho}_\infty$ is {\it not linearizable} by a $C^2$ map. This is
a particular case of next theorem:
\saut
\noindent{\bf Theorem 3.13.}
{\it
Let $1/2<\rho<1$. There does not exist a neighbourhood ${\cal O}$ of
zero in $E^{\circ\rho}_\infty$ and a $C^2$ map $F\colon{\cal O}\fl
E^{\rho}_0$ such that $F(0)=0$, $(DF)(0;u)=u$ and such that
$F(U^{(+)}_g(u))=U^{1}_gF(u)$ for all $u$ in a neighbourhood of
zero in $E^{\circ\rho}_\infty$ and all $g$ in a neighbourhood of the
identity in ${\cal P}_0$.
}\saut
\noindent{\it Proof.}
Suppose on the contrary that ${\cal O}$ and $F\colon{\cal O}\fl
E^{\rho}_0$ exist and let $F^1u=u$, $F^2(u_1\otimes u_2)=
1/2 (D^2F)(0;u_1,u_2)$ for $u$, $u_1$, $u_2 \in E^{\circ\rho}_\infty$.
Let $L(B_1,B_2)$ be the linear continuous operators from $B_1$ to
$B_2$, where $B_1$ and $B_2$ are topological vector spaces.
$F^2\in L(E^{\circ\rho}_\infty\hat\otimes E^{\circ\rho}_\infty,E^{\rho}_0)$
since $F$ is a $C^2$ map and $F^2(u_1\otimes u_2)=F^2(u_2\otimes u_1)$.
Let $R^{(+)2}_g=U^{(+)}_g(U^{1}_{g^{-1}}\otimes U^{1}_{g^{-1}})$,
$g\in {\cal P}_0$. According to statements i) and ii) of Theorem~3.12,
$g\mapsto R^{(+)2}_g(u_1\otimes u_2)$ is $C^\infty$ from ${\cal P}_0$ to
$E^{\circ\rho}_\infty$ for $u_1$, $u_2 \in E^{\circ\rho}_\infty$ and
according to statement iii) of Theorem~3.12 $R^{(+)2}_g\in
L(E^{\circ\rho}_\infty\hat\otimes E^{\circ\rho}_\infty,E^{\rho}_0)$
for  $g\in {\cal P}_0$. Differentiating twice with respect to $u$ at $u=0$,
the equality
$F(U^{(+)}_g(u))=U^{1}_gF(u)$, which by hypothesis is true
for all $g$ in a neighbourhood of the identity in ${\cal P}_0$ and
for all $u$ in a neighbourhood of zero in $E^{\circ\rho}_\infty$,
we obtain that
$U^{1}_gF^2(U^{1}_{g^{-1}}\otimes U^{1}_{g^{-1}})-F^2=-R^{(+)2}_g$
for $g$ in a neighbourhood of the identity in ${\cal P}_0$.
Using the fact that $R^{(+)2}$ is a $1$-cocycle for the ${\cal P}_0$-module
$L(E^{\circ\rho}_\infty\hat\otimes E^{\circ\rho}_\infty,E^{\rho}_0)$
defined by $(g,Q)\mapsto U^{1}_g Q (U^{1}_{g^{-1}}\otimes U^{1}_{g^{-1}})$,
i.e. $R^{(+)2}_{gh}=U^{1}_gR^{(+)2}_{h} (U^{1}_{g^{-1}}\otimes U^{1}_{g^{-1}})
+ R^{(+)2}_{g}$, it then follows that
$$U^{1}_gF^2(U^{1}_{g^{-1}}\otimes U^{1}_{g^{-1}})-F^2=-R^{(+)2}_g,\quad
g\in{\cal P}_0.\eqno{(3.50)}$$
Since  $g\mapsto F^2(U^{1}_g u_1 \otimes U^{1}_g u_2)$ and
$g\mapsto R^{(+)2}_g(U^{1}_g u_1 \otimes U^{1}_g u_2)$,
$u_1$, $u_2 \in E^{\circ\rho}_\infty$, are $C^\infty$ from ${\cal P}_0$ to
$E^{\rho}_0$, it follows that $F^2(u_1\otimes u_2)\in
E^{\rho}_\infty$ for $u_1$, $u_2 \in E^{\circ\rho}_\infty$ and that
$F^2\in L(E^{\circ\rho}_\infty\hat\otimes E^{\circ\rho}_\infty,
E^{\rho}_\infty)$.

The definition (1.17a) and (1.17b) of $U^{(+)}$ shows that $R^{(+)2}_g=
(0,R^{(+)D2}_g)$, where $R^{(+)D2}_g$ $\in L(E^{\circ\rho}_\infty\hat\otimes
E^{\circ\rho}_\infty,E^{\rho}_\infty)$ is given by
$$\eqalignno{
&(R^{(+)D2}_g(u_1\otimes u_2))^\wedge(k)&(3.51)\cr
&\qquad{}={i\over 2}\sum_{\varepsilon=\pm}
(\varphi_g(U^{1}_{g^{-1}} u_1,-\varepsilon k)P^{}_{\varepsilon}(k)
\hat\alpha_2(k)
+\varphi_g(U^{1}_{g^{-1}} u_2,-\varepsilon k)P^{}_{\varepsilon}(k)
\hat\alpha_1(k))\cr
}$$
and where $\varphi$ is given by (1.23a) and (1.23b).
Let $F^2=(F^{M2},F^{D2})$, where $F^{M2}\in
L(E^{\circ\rho}_\infty\hat\otimes E^{\circ\rho}_\infty,M^{\rho}_\infty)$
and $F^{D2}\in
L(E^{\circ\rho}_\infty\hat\otimes E^{\circ\rho}_\infty,D^{\rho}_\infty)$,
let $G^2((f,\dot f)\otimes\alpha)=F^{D2}((f,\dot f,0)\otimes(0,0,\alpha))$
and let $r^2_g((f,\dot f)\otimes\alpha)=R^{(+)D2}_g
((f,\dot f,0)\otimes(0,0,\alpha))$. Then $G^2\in
L(M^{\circ\rho}_\infty\hat\otimes D^{}_\infty,D^{}_\infty)$
satisfies, according to (3.50), the equality
$$U^{1D}_gG^2(U^{1M}_{g^{-1}}\otimes U^{1D}_{g^{-1}})-G^2=-r^{2}_g,\quad
g\in{\cal P}_0.\eqno{(3.52)}$$
$G^2$ is the unique element in
$L(M^{\circ\rho}_\infty\hat\otimes D^{}_\infty,D^{}_0)$
satisfying (3.52). As a matter of fact, if $H\in
L(M^{\circ\rho}_\infty\hat\otimes D^{}_\infty,D^{}_0)$
satisfies the equation
$$U^{1D}_gH((f,\dot f)\otimes\alpha)=H(U^{1M}_g(f,\dot f)
\otimes U^{1D}_g\alpha)$$
for $g\in {\cal P}_0$, $(f,\dot f)\in M^{\circ\rho}_\infty$ and $\alpha\in
D^{}_\infty$, it follows that $H((f,\dot f)\otimes\alpha)\in
D^{}_\infty$ and that $T^{1D}_XH=HS^{}_X$, $X\in \p$, where
$S^{}_X=T^{1M}_X\otimes I + I\otimes T^{1D}_X$.
This equality extends to the enveloping algebra, i.e.
$T^{1D}_YH=HS^{}_Y$ for $Y\in U(\p)$. Using
$\sum_{0\leq \mu\leq3}T^{1D}_{P_\mu P^\mu}= -m^2$ and
$\sum_{0\leq \mu\leq3}T^{1M}_{P_\mu P^\mu}= 0$, we obtain that
$$ H\Big(\sum_{0\leq \mu\leq3}T^{1M}_{P_\mu}\otimes T^{1D}_{P^\mu}\Big)=0
\eqno{(3.53)}$$
Since $\vert p_1\vert \omega(p_2) - p_1\cdot p_2\geq
 \vert p_1\vert m^2 /(2\omega(p_2))$ for $p_1,p_2\in \Rrm^3$,
it follows by studying the expression
$$\sum_{0\leq \mu\leq3}((T^{1M}_{P_\mu}(f,\dot f))^\wedge
\otimes (T^{1D}_{P^\mu}\alpha)^\wedge(p_1,p_2)$$
that the map
$$\sum_{0\leq \mu\leq3}T^{1M}_{P_\mu}\otimes T^{1D}_{P^\mu}
\colon M^{\circ\rho}_c\hat\otimes D^{}_\infty\fl
M^{\circ\rho}_c\hat\otimes D^{}_\infty,$$
where $M^{}_c$ was defined in Theorem 2.9, is surjective.
Therefore, since $M^{\circ\rho}_c$ is dense in $M^{\circ\rho}_\infty$,
$H$ vanishes on a dense subset of
$M^{\circ\rho}_\infty\hat\otimes D^{}_\infty$,
which by continuity proves that $H=0$. Hence $G^2$ is the unique element in
$L(M^{\circ\rho}_\infty\hat\otimes D^{}_\infty,D^{}_0)$ satisfying (3.52).

Let $(f,\dot f)\in M^{\circ\rho}_c$ and let
$$(\Theta^{}_\varepsilon((f,\dot f)))(k)= \vartheta^{\infty}(\chi^{}_0\circ
\rho B^{(+)1}, (\omega(k),-\varepsilon k)),\quad \varepsilon=\pm,
\eqno{(3.54)}$$
where $\vartheta^{\infty}$, $\chi^{}_0$, $\rho$ and $B^{(+)1}$ are as
in (1.22b)--(1.23b). It follows from statement i) of Lemma 3.2 that the
function
$k\mapsto\omega(k)^{-1}(\Theta^{}_\varepsilon((f,\dot f)))(k)$ is an element
of $L^\infty(\Rrm^3)$, so $\Theta^{}_\varepsilon((f,\dot f))\hat
\alpha\in L^2(\Rrm^3,\Crm^4)$ for $\alpha\in D^{}_\infty$. Therefore
$G^2((f,\dot f)\otimes\alpha)$ defined by
$$ (G^2((f,\dot f)\otimes\alpha))^\wedge
={-i\over 2}\sum_{\varepsilon=\pm}
(\Theta^{}_\varepsilon((f,\dot f)))(k)P^{}_{\varepsilon}(k)\hat\alpha(k),
\eqno{(3.55)}$$
is an element of $D^{}_0$ and a direct calculation shows that
$$(U^{1D}_gG^2(U^{1M}_{g^{-1}}\otimes U^{1D}_{g^{-1}})-G^2+r^{2}_g)
((f,\dot f)\otimes\alpha)=0,$$
for $g \in {\cal P}_0$, $(f,\dot f)\in M^{\circ\rho}_c$ and
$\alpha\in D^{}_\infty$. Since $M^{\circ\rho}_c$ is dense
in $M^{\circ\rho}_\infty$, it follows that the unique solution
$G^2\in L(M^{\circ\rho}_\infty\hat\otimes D^{}_\infty,D^{}_0)$ of equation
(3.52) is given by continuous extension of $G^2$ defined in (3.55)
for $(f,\dot f)\in M^{\circ\rho}_c$. Let
$(f,\dot f)\in M^{\circ\rho}_\infty$ and let
$(f^{(n)},\dot f^{(n)})\in M^{\circ\rho}_c$, $n\geq0$, be a sequence which
converges in $M^{\circ\rho}_\infty$ to $(f,\dot f)$.
Since $G^2\in L(M^{\circ\rho}_\infty\hat\otimes D^{}_\infty,D^{}_\infty)$,
it follows that $G^2((f^{(n)},\dot f^{(n)})\otimes\alpha)$ converges
in $D^{}_\infty$ to $G^2((f,\dot f)\otimes\alpha)$ for $\alpha\in D^{}_\infty$,
which in particular shows that
$(G^2((f^{(n)},\dot f^{(n)})\otimes\alpha))^\wedge(0)$ converges in
$\Crm^4$ to $(G^2((f,\dot f)\otimes\alpha))^\wedge(0)$.
For given $\varepsilon=\pm$, we choose $\alpha\in D^{}_\infty$ such
that $\hat\alpha(0)=P^{}_{\varepsilon}(0)\hat\alpha(0)$ and
$\vert\hat\alpha(0)\vert =1$. It then follows from (3.55) that
$$(\Theta^{}_\varepsilon((f^{(n)},\dot f^{(n)})))(0)
=2i(\hat\alpha(0))^+(G^2((f^{(n)},\dot f^{(n)})\otimes\alpha))^\wedge(0),$$
which proves that the sequence
$\Theta^{}_\varepsilon((f^{(n)},\dot f^{(n)}))(0)$, $n\geq0$, converges
in $\Rrm$, when $n\fl\infty$. By definition (3.54)
of $\Theta^{}_\varepsilon$ and definition (1.23b) of $\vartheta^{\infty}$,
it follows that
$$(\Theta^{}_\varepsilon((f^{(n)},\dot f^{(n)})))(0)
=\int_0^\infty \chi^{}_0(\tau) Q^{(n)}(\tau,0)d\tau,\eqno{(3.56)}$$
where $Q^{(n)}=Q^{(n)}_1 +Q^{(n)}_2$ and
$$\eqalign{
Q^{(n)}_1(\tau,0)&=(\cos(\tau\vert \nabla\vert)f^{(n)}_0)(0),\cr
Q^{(n)}_2(\tau,0)&=\vert \nabla\vert^{-1}
\sin(\tau\vert \nabla\vert)\dot f^{(n)}_0)(0).\cr
}$$
Let $\theta\in S'(\Rrm)$ be the inverse Fourier transform of
$\chi^{}_0\colon \Rrm\fl\Rrm$, where $\chi^{}_0(\tau)=0$
for $\tau<0$. Let $\lambda(\tau) =1$ for $\tau\geq0$
and $\lambda(\tau) =0$ for $\tau<0$ and let $\theta^{}_0$
be the inverse Fourier transform of $\sqrt{2\pi}\lambda$.
Then $\theta^{}_0(s)=i(s+i0)^{-1}$, $\theta-\theta^{}_0$ is an entire
analytic function since $\chi^{}_0-\lambda$ has compact support and
$\vert(\theta-\theta^{}_0)(s)\vert\leq (1+\vert s\vert)^{-1}$.
Equality (3.4b) gives, noting that $0\not\in {\rm supp} \hat f^{(n)}_0$,
$$\eqalignno{
\int_0^\infty \chi^{}_0(\tau) Q^{(n)}_1(\tau,0)d\tau
&=(2(2\pi)^2)^{-1}\int((\theta-\theta^{}_0)(\vert p\vert)
+ (\theta-\theta^{}_0)(-\vert p\vert)\hat f^{(n)}_0(p) dp
&(3.57\hbox{a})\cr
&\cr
\noalign{\hbox{and}}
&\cr
\int_0^\infty \chi^{}_0(\tau) Q^{(n)}_2(\tau,0)d\tau
&=(2\pi)^{-2}\int\vert p\vert^{-2}\fhdnz(p) dp&(3.57\hbox{b})\cr
&\qquad{}+(2\pi)^{-2}\int(2i\vert p\vert)^{-1}((\theta-\theta^{}_0)
(\vert p\vert)
- (\theta-\theta^{}_0)(-\vert p\vert))\fhdnz(p) dp.\cr
}$$
The properties of $\theta-\theta^{}_0$ show that
$$\eqalign{
&(2(2\pi)^2)^{-1}
\int(\vert(\theta-\theta^{}_0)(\vert p\vert)\vert
+\vert (\theta-\theta^{}_0)(-\vert p\vert)\vert)
(\vert\hat f^{(n)}_0(p)\vert+ \vert p\vert^{-1}
\vert\fhdnz(p)\vert )dp\cr
&\hskip50mm{}\leq C_\rho \Vert(f^{(n)},\dot f^{(n)}) \Vert^{}_{M^\rho},
\quad 1/2<\rho<1.\cr
}$$
This inequality and inequalities (3.56), (3.57a) and (3.57b) give that
$$\eqalignno{
&(\Theta^{}_\varepsilon((f^{(n)},\dot f^{(n)})))(0)
\leq C_\rho \Vert(f^{(n)},\dot f^{(n)}) \Vert^{}_{M^\rho}
+(2\pi)^{-1}\int\vert p\vert^{-2}\fhdnz(p)dp,&(3.58)\cr
&\big\vert (\Theta^{}_\varepsilon((f^{(n)},\dot f^{(n)})))(0)
-(2\pi)^{-2}\int\vert p  \vert^{-2}\fhdnz(p)dp\big\vert\cr
&\hskip40mm{}\leq C_\rho \Vert(f^{(n)},\dot f^{(n)}) \Vert^{}_{M^\rho},
\quad \varepsilon=\pm,1/2<\rho<1.\cr
}$$
Let $1<b<1/2+\rho$, $\psi\in C^\infty_0(\Rrm^3)$, $0\leq \psi(p)\leq 1$
for $p\in \Rrm^3$, $\psi(p)=\psi(-p)$ for $p\in \Rrm^3$,
$\psi(p)=1$ for $\vert p  \vert\leq1$ and $\psi(p)=0$
for $\vert p  \vert\geq2$. Let $f_0=0$, $\dot f_j =0$ for $1\leq j\leq3$,
$\fhdz(p)= \psi(p)\vert p  \vert^{-a}$ and $\hat f_j(p) =
-i(p_j/\vert p  \vert^{2})\fhdz(p)$ for $1\leq j\leq3$.
Let $\hat f^{(n)}_\mu(p) = (1- \psi(np))\hat f_\mu(p)$
and $\fhdnm(p) = (1- \psi(np))\fhdm(p)$,
for $0\leq \mu\leq3$ and $n\geq0$. It follows using the equivalence
of the norms $\Vert\cdot\Vert^{}_{E^{}_l}$ and $q^{}_l$ given by Theorem
2.9 that $(f,\dot f)\in M^{\rho}_\infty$ and
$\Vert(f^{(n)},\dot f^{(n)})-(f,\dot f)\Vert^{}_{M^{\rho}_\infty}\fl 0$
when $n\fl\infty$. Obviously $(f^{(n)},\dot f^{(n)})\in M^{\rho}_c$.
Moreover conditions (1.4c) are satisfied for $(f,\dot f)$ (resp.
$(f^{(n)},\dot f^{(n)})$, $n\geq 0$), so
$(f,\dot f)\in M^{\circ\rho}_\infty$
(resp. $(f^{(n)},\dot f^{(n)})\in M^{\circ\rho}_c$, $n\geq 0$).
The right-hand side of inequality (3.58) is uniformly bounded in $n$
for the sequence $(f^{(n)},\dot f^{(n)})$ so constructed, but the integral
on the left-hand side goes to infinity when $n$ goes to infinity.
This proves that the sequence
$(\Theta^{}_\varepsilon((f^{(n)},\dot f^{(n)})))(0)$, $n\geq 0$,
does not converges in $\Rrm$ when $n\fl \infty$. This is in contradiction
with the fact that it follows, as we proved  from the hypothesis that this
sequence converges in $\Rrm$. This shows that the map $F\colon{\cal O}\fl
E^{\rho}_0$ does not exist, which proves the theorem.

We shall prove that asymptotic reprensentations $U^{(+)}$ defined by
formulas (1.17a), (1.17b) and (1.23a) for different choices of the function
$\chi^{}_0$ are equivalent. For $i\in\{1,2\}$ let $\chi^{}_i\in C^\infty([0,
\infty[)$,
$0\leq \chi^{}_i(\tau)\leq 1$ for $\tau\geq0$, $\chi^{}_i(\tau)=1$ for
$\tau\geq2$, let $U^{(+)}_i$ be the corresponding asymptotic representation
given by formulas (1.17a), (1.17b) and (1.23a) with $\chi^{}_i$ replacing
$\chi^{}_0$, and let $A^{(+)1}_i$ be defined correspondingly by formula (1.22a)
with $\chi^{}_i$ replacing $\chi^{}_0$. We introduce the notation
$\Theta_\varepsilon((f,\dot f))$ for $\varepsilon=\pm$, $(f,\dot f)\in M^\rho$
by
$$\big(\Theta_\varepsilon((f,\dot f))\big)(k)=\vartheta^\infty
(A^{(+)1}_1 -A^{(+)1}_2, (\omega(k),-\varepsilon k)).
\eqno{(3.59\hbox{a})}$$
For $u=(f,\dot f,\alpha)\in E^\rho$, let $F(u)=(f,\dot f,F^D(u))$, where
$$
(F^D(u))^{\wedge}(k)=\sum_{\varepsilon=\pm}
\exp\big(-i(\Theta_\varepsilon((f,\dot f)))(k)\big)P_\varepsilon(k)
\hat\alpha(k).
\eqno{(3.59\hbox{b})}$$
\saut
\noindent{\bf Theorem 3.14.}
{\it
Let $1/2<\rho<1$. If $n,N\in\Nrm$, then $F,F^{-1}\in
C^n (E_{N+n},E_{N})$, $F,F^{-1}\in C^\infty(E_\infty,E_\infty)$
and $U^{(+)}_{2g}=F\circ U^{(+)}_{1g}\circ F^{-1}$ for $g\in {\cal P}_0$.
}
\saut
\noindent{\it Proof.}
To prove that $F\in \hskip-.8pt C^n (E_{N+n},E_{N})$
it is sufficient to prove that $F^D\in \hskip-.8pt C^n (E_{N+n},D_{N})$.
Let $\chi^{}_0=\chi^{}_1-\chi^{}_2$. Then $\chi^{}_0\in C^\infty([0,\infty[)$
and $\chi^{}_0(\tau)=0$ for $\tau\geq2$. Since $\vartheta^\infty (H,\Lambda y)
=\vartheta^\infty (\Lambda^{-1}H, y)$,
$\Lambda\in SO(3,1)$, the function $k\mapsto
\big(\Theta_\varepsilon((f,\dot f))\big)(k)$
is $C^\infty$ on $\Rrm^3$.
It follows from (3.59a) and statement ii) of Lemma 3.2 that
$$
\vert\big(\Theta_\varepsilon((f,\dot f))\big)(k)\vert\leq
C_\rho \omega(k)^{\rho-1/2}\Vert (f,\dot f)\Vert^{}_{M^\rho_0},\quad
(f,\dot f)\in M^\rho_c.
$$
The fact that $M^\rho_c$ is dense is $M^\rho_\infty$, according to Theorem 2.9,
then shows
$$
\vert\big(\eta^{(\varepsilon)}_Z \Theta_\varepsilon((f,\dot
f))\big)(k)\vert\leq
C_\rho \omega(k)^{\rho-1/2}\Vert T^{1M}_Z(f,\dot f)\Vert^{}_{M^\rho_0},
\eqno{(3.60)}$$
for $Z\in \Pi'\cap U({\frak{sl}}(2,\Crm))$, $(f,\dot f)\in M^\rho_{\vert
Z\vert}$.
Let $G_\varepsilon(u)=\exp\big(-i\Theta_\varepsilon((f,\dot f))\big)\hat\alpha$
for $u=(f,\dot f,\alpha)\in M^\rho_0$. Inequality (3.60), with $Z=\un$,
Lebesgue's dominated convergence theorem and Plancherel theorem show that
$\omega^lG_\varepsilon\in C^0(E^\rho_l,L^2(\Rrm^3,\Crm^4))$ for $l\in\Nrm$.
Let $Z\in \Pi'\cap U({\frak{sl}}(2,\Crm))$. Then
$\eta^{(\varepsilon)}_ZG_\varepsilon(u)$
is a linear combination of expressions:
$$
(\eta^{(\varepsilon)}_{Z_1})\Theta_\varepsilon((f,\dot f))\cdots
(\eta^{(\varepsilon)}_{Z_j})\Theta_\varepsilon((f,\dot f))
G_\varepsilon((f,\dot f,\beta^{(\varepsilon)}_{j+1})),
$$
where $(\beta^{(\varepsilon)}_{j+1})^\wedge =\eta^{(\varepsilon)}_{Z_{j+1}}
\hat\alpha$,
$Z_i \in \Pi'\cap U({\frak{sl}}(2,\Crm))$ for $1\leq i\leq j+1$,
$\vert Z_{1}\vert +\cdots
+\vert Z_{j+1}\vert=\vert Z\vert$, $\vert Z_{i}\vert\geq 1$
for $1\leq i\leq j$ and when $j\geq0$ and the expression is reduced to
$G_\varepsilon((f,\dot f,\beta^{(\varepsilon)}_{1}))$ if $j=0$.
Since $\omega^l G_\varepsilon\in C^0(E^\rho_l,L^2)$ and $\rho-1/2\leq1$, it
follows from inequality (3.60) and from the equivalence of norms
in Theorem 2.9 that
$\omega^l\eta^{(\varepsilon)}_Z G_\varepsilon(u)\in
C^0(E^\rho_{\vert Z\vert +l},L^2(\Rrm^3,\Crm^4))$. Using that
$
(F^D(u))^\wedge =
\sum_{\varepsilon}G_\varepsilon((f,\dot f,P_\varepsilon(-i\partial)
\alpha))$
and the equivalence of norms in Theorem 2.9, we obtain
$$
T^{1D}_Y F^D \in C^0(E^\rho_{\vert Y\vert}, D_0),\quad Y\in\Pi'.
\eqno{(3.61)}$$

Derivation of $F^D$ shows that
$P_\varepsilon((D^nF^D)(u;u_1,\ldots,u_n))^\wedge$,
where $u=(f,\dot f,\alpha)$ and $u_i=(f_i,\dot f_i,\alpha_i)$ is a linear
combination of terms:
$$\eqalignno{
&\Theta_\varepsilon((f_{i_1},\dot f_{i_1}))\cdots
\Theta_\varepsilon((f_{i_n},\dot f_{i_n}))
P_\varepsilon(F^D((f,\dot f,\alpha)))^\wedge&(3.62\hbox{a})\cr
\noalign{\noindent and}
&\Theta_\varepsilon((f_{i_1},\dot f_{i_1}))\cdots
\Theta_\varepsilon((f_{i_{n-1}},\dot f_{i_{n-1}}))
P_\varepsilon(F^D((f,\dot f,\alpha_{i_n})))^\wedge,&(3.62\hbox{b})\cr
}$$
where $(i_1,\ldots,i_n)$ is a permutation of $(1,\ldots,n)$. This proves,
together with (3.60) and (3.61), that the function
$u\mapsto  T^{1D}_Y (D^nF^D)(u;u_1,\ldots,u_n)$ is an element of
$C^0(E_{\vert Y\vert+n}, D_0)$, for $Y\in\hskip-.6pt\Pi'$ and
$(u_1,\ldots,u_n)\in E_{\vert Y\vert+n}$. This shows that the map
$u\mapsto\hskip-.7pt  (D^nF^D)(u;u_1,\ldots,u_n)$ is an element of
$C^0(E_{N+n}, D_N)$, for $n,N\in\Nrm$, which then shows that
$F^D\in C^n(E_{N+n}, D_N)$ and $F^D\in C^\infty(E_\infty,E_\infty)$.
Since the inverse of $F$ is given by $(f,\dot f,\alpha)\mapsto
F((-f,-\dot f,\alpha))$, it also follows that $F^{-1}$ has these properties.
A short calculation shows that $F$ intertwines $U^{(+)}_1$ and $U^{(+)}_2$.
This proves the theorem.
\vfill\eject

\noindent{\titre 4. Construction of the approximate solution.}
\saut
The basic technical tool in the proof of the existence of modified wave
operators for the M-D equation is here the same as in \refFST, namely the
{\it construction of approximate solutions absorbing the slowly decaying part
of the solution of the M-D equation}. The existence of the remainder (the
difference between the exact and the approximate solution) can then be
proved using only standard Sobolev estimates. In order to be able to use
the implicit functions theorem in Fr\'echet spaces for the existence of the
inverse of the modified wave operator, we shall establish precise estimates
giving the loss of order of seminorms. This is done by the study of the
equations for the enveloping algebra after a transformation compensating
the long range Fourier phase.

We shall need a lemma which is an analog of Theorem 3.5 in \refFST\
adapted to our spaces $E^{}_N$. Let first $t \mapsto f_j (t), j = 1,2$, be
continuous functions from $\Rrm^+$ to $D^{}_N$ and let
$$\eqalignno{ &(G_{\varepsilon, \mu} (f_1, f_2)) (t)&(4.1\hbox{a})\cr
&\qquad{}= -
\int^\infty_t {\sin ((- \Delta)^{1/2} (t-s)) \over (- \Delta)^{1/2}} (e^{i
\varepsilon \omega(-i \partial)s} f_1 (s))^+ \gamma^{}_0
\gamma_\mu  (e^{i \varepsilon  \omega(-i \partial)s}
f_2 (s))  ds,\quad t \geq 0,\cr
}$$
and
$$\eqalignno{ &(\dot{G}_{\varepsilon, \mu}(f_1, f_2)) (t)&(4.1\hbox{b})\cr
&\qquad{}=- \int^\infty_t
\cos ((- \Delta)^{1/2}  (t-s))  (e^{i \varepsilon
\omega(- i\partial)s}  f_1 (s))^+ \gamma^{}_0  \gamma_\mu
 (e^{i \varepsilon  \omega(-i \partial)s}  f_2 (s))ds,\quad t \geq 0.\cr
}$$
\saut
\noindent{\bf Lemma 4.1.}
{\it
Let $n \geq 0$, $N \geq 0$, $q \geq 0$ be integers, $\alpha = (
\alpha_1, \alpha_2, \alpha_3)$ a multi-index and let $f_j \colon \Rrm^+ \fl
D^{}_N,
j = 1,2$, be $C^q$ functions. Let $G$ and $\dot{G}$ be given by (4.1a) and
(4.1b).  If $N$ is chosen sufficiently large, depending on $\vert \alpha
\vert$ and $n$, then
$$\eqalignno{
&(1+t)^{\vert \alpha \vert + q - \vert \beta \vert + \chi + \rho - 1/2}
\Vert x^\beta  \big({d \over dt}\big)^q \partial^\alpha  (G
(f_1, f_2) (t),  \dot{G} (f_1, f_2) (t)) \Vert^{}_{M^\rho_n}& (4.2)\cr
&\qquad\leq C_{n, \vert \alpha \vert, \vert \beta \vert, q, \rho}
\sum_{q^{}_1+q^{}_2=q}
 \sup_{s \geq t}  \big((1+s)^{q+\chi}  \Vert \big({d \over ds}\big)^{q^{}_1}
 f_1 (s) \Vert^{}_{D^{}_N}  \Vert \big({d \over ds}\big)^{q^{}_2}
f_2 (s) \Vert^{}_{D^{}_N}\big),\quad  t \geq 0,
}$$
where $0 \leq \rho \leq 1$, $\chi + q + \vert \alpha \vert - \vert \beta
\vert +
\rho - 1/2 > 0.$ Moreover there is $N'$ depending on $\vert \alpha \vert$ such
that
$$\eqalignno{
&\big\vert \big({\partial \over \partial t}\big)^q \partial^\alpha
G(f_1, f_2) (t,x) \big\vert + (1+t+\vert x \vert) \big\vert \big({\partial
\over \partial t}\big)^q \partial^\alpha \dot{G} (f_1, f_2) (t,x)
\big\vert&(4.3)\cr
&\quad{}\leq C_{\vert \alpha \vert, q, \chi,
\varepsilon} (1+\vert x \vert + t)^{- (1+\vert \alpha \vert +q+\chi -
\varepsilon)} (1+t)^{- \varepsilon}\cr
&\qquad
\sum_{q^{}_1 + q^{}_2 = q}
 \sup_{s \geq t} \big((1+s)^{q+\chi}
 \Vert\big({d \over ds}\big)^{q^{}_1} f_1 (s)
 \Vert^{}_{D^{}_{N'}} \Vert \big({d \over ds}\big)^{q^{}_2} f_2 (s)
 \Vert^{}_{D^{}_{N'}}\big),\quad t \geq 0, \chi \geq 0, \varepsilon > 0.  }$$
 }\saut
\noindent{\it Proof.}
After the change of variable $s - t \fl s$, we have
$$\eqalignno{
(G_{\varepsilon, \mu}  (f_1, f_2)) (t) & = \int^\infty_0
{\sin  ((- \Delta)^{1/2} s) \over (- \Delta)^{1/2}}  J_\mu
 (s+t)  ds & (4.4\hbox{a})\cr
\noalign{\hbox{and}}
(\dot{G}_{\varepsilon, \mu}  (f_1, f_2)) (t) & = - \int^\infty_0
\cos  ((- \Delta)^{1/2} s)  J_\mu  (s+t)
ds,& (4.4\hbox{b})\cr
}$$
where
$$J_\mu  (s+t) = (e^{i \varepsilon  \omega(-i \partial) (s+t)}
 f_1 (s+t))^+  \gamma^{}_0      \gamma_\mu  (e^{i
\varepsilon  \omega (-i \partial) (s+t)}  f_2 (s+t)). \eqno{(4.4\hbox{c})}$$
The function $k \mapsto K(k) = \vert k \vert^{-1}  \sin (\vert k \vert)$
is an entire analytic function on $\Crm^3$.
As it is seen by considering its Taylor development for small $k$,
$$\vert K_\gamma (k) \vert \leq C_\gamma  (1+\vert k \vert)^{-1},
\quad k \in \Rrm^3, \eqno{(4.5)}$$
where $K_\gamma (k) = {\partial^\gamma \over \partial k^\gamma}  K (k)$.

Therefore, if $\beta = (\beta_1, \beta_2, \beta_3)$ is a multi-index,
$$x^\beta {\sin ((- \Delta)^{1/2} s) \over (- \Delta)^{1/2} s} s =
\sum_{\gamma_1 + \gamma_2 = \beta} C_{\gamma_1, \gamma_2} K_{\gamma_2} (-i
s \partial) x^{\gamma_1} s^{1+\vert \gamma_2 \vert},
\eqno{(4.6)}$$
where $C_{\gamma_1, \gamma_2}$ are constants given by Leibniz formula.

Similarly if $L (k) = \cos  (\vert k \vert)$, then the entire analytic
function $L$ satisfies
$$\vert L_\gamma (k) \vert \leq C_\gamma,\quad  k \in \Rrm^3,
L_\gamma (k) = {\partial^\gamma \over \partial k^\gamma}  L(k) \eqno{(4.7)}$$
and we obtain
$$x^\beta  \cos ((- \Delta)^{1/2} s) = \sum_{\gamma_1 + \gamma_2 = \beta}
 C_{\gamma_1, \gamma_2}  L_{\gamma_2}  (-i s \partial)
 x^{\gamma_1}  s^{\vert \gamma_2 \vert}. \eqno{(4.8)}$$
We next estimate the $L^p$-norm, $1 \leq p \leq \infty$,  of $x^\gamma
\partial^\alpha  J_\mu (t)$,  $\vert \gamma \vert \leq \vert \alpha \vert$,
$t \geq 0$. To do this, it is sufficient to estimate the $L^p$-norm of
$$\Gamma_{\gamma, \alpha} (t) = x^\gamma  \partial^\alpha
(e^{-i \varepsilon  \omega(-i \partial) t}  h_1 (t))
(e^{i \varepsilon  \omega(-i \partial) t}  h_2 (t)),
\quad t \geq 0, \eqno{(4.9)}$$
where $h_j (t) \in D_\infty$,   $j = 1,2$. According to Theorem~A.1 of
the appendix there are, for given $n\geq 0$, functions $r^{}_j (t)
\in D^{}_\infty$
with $\hbox{\rm supp}\  r^{}_j (t) \subset \{ x \big\vert\vert x \vert
\leq t \}$,
such that
$$\eqalignno{
&\Vert x^\gamma  \partial^\alpha  (e^{i \nu(j) \varepsilon
 \omega(-i \partial) t}  h_j (t) - e^{i \nu (j) \varepsilon
m  \rho (t)}  r^{}_j (t)) \Vert^{}_{L^2}&(4.10)\cr
&\qquad{}\leq C_{n, \vert \alpha \vert}  \Vert h_j (t)
\Vert^{}_{D^{}_{N+2 \vert \alpha \vert}}
 t^{- (n+1- \vert \gamma \vert)},\quad  t > 0,\quad
}$$
where $\rho(t)$ is the function $x\mapsto(t^2 -\vert x\vert^2)^{1/2}$
for $\vert x\vert\leq t$ and $\rho(t)(x)=0$ otherwise, and
$$\Vert \rho (t)^{-l}  \partial^\alpha  r^{}_j (t) \Vert^{}_{L^2} \leq
C_{n, \vert \alpha \vert, l}  \Vert h_j (t)
\Vert^{}_{D^{}_{N+2 \vert \alpha \vert +l}}
 t^{- \vert \alpha \vert - l},\quad     t > 0,  l \geq 0,\eqno{(4.11)}$$
where $\rho (t)^{-l}  \partial^\alpha  r^{}_j (t)$
is defined as being equal to zero for
$\vert x\vert\geq t$, and  where $\nu (1) = - 1$,
$\nu (2) = 1$ and where $N$,
depending on $n$, is sufficiently large. Again,
according to Theorem~A.1 the functions
$r^{}_j$, $j = 1,2$, satisfy
$$\eqalignno{
&\Vert x^\gamma  \partial^\alpha  (e^{i \nu (j) \varepsilon
 \omega(-i \partial) t}  h_j (t) - e^{i \nu (j) \varepsilon m \rho(t)}
 r^{}_j (t)) \Vert^{}_{L^\infty}&(4.12)\cr
&\qquad{}\leq C_{n, \vert \alpha \vert}
\Vert h_j (t) \Vert^{}_{D^{}_{N+2 \vert \alpha \vert}}
t^{- (n + 5/2 - \vert \gamma \vert)},\quad t > 0,\cr
}$$
and
$$\eqalignno{
&\Vert \rho(t)^{-l}  \partial^\alpha    r^{}_j (t)
\Vert^{}_{L^\infty}&(4.13)\cr
&\qquad{}\leq C_{n, \vert \alpha \vert, l}
\Vert h_j (t) \Vert^{}_{D^{}_{N+2 \vert \alpha \vert +l}}
t^{- 3/2 - \vert \alpha \vert - l},  \quad t > 0,  l \geq 0.\cr
}$$
If
$$\delta_j (t) = e^{i \nu (j) \varepsilon  \omega(-i \partial) t}
h_j (t) - e^{i \nu (j) \varepsilon m \rho (t)}  r^{}_j (t),$$
then it follows from (4.9) that
$$\eqalign{
&\Gamma_{\gamma, \alpha} (t) = x^\gamma  \partial^\alpha
\big(r^{}_1 (t)  r^{}_2 (t) + e^{ - i m \rho (t)}  r^{}_1 (t)
\delta_2 (t)\cr
&\qquad{}+ \delta_1 (t)  e^{i m \rho (t)}  r^{}_2 (t) + \delta_1 (t)
 \delta_2 (t)\big).\cr
}$$
Leibniz rule and the fact that $\hbox{\rm supp}\
r^{}_j (t) \subset \{ x \big\vert
\vert x \vert \leq t \}$ give
$$\eqalignno{
&\Vert \Gamma_{\gamma, \alpha} (t) - x^\gamma   \partial^\alpha
(r^{}_1 (t)  r^{}_2 (t)) \Vert^{}_{L^p}&(4.14)\cr
&\qquad{}\leq C_{\vert \alpha \vert}
\sum_{\alpha_1 + \alpha_2 = \alpha}
\Big(t^{\vert \gamma \vert}  \Vert (\partial^{\alpha_1}  e^{-im
\rho (t)}  r^{}_1 (t))  (\partial^{\alpha_2}  \delta_2 (t))
\Vert^{}_{L^p}\cr
&\qquad\qquad{}+ t^{\vert \gamma \vert}  \Vert (\partial^{\alpha_1}
\delta_1 (t))
 (\partial^{\alpha_2}  e^{i m \rho (t)}  r^{}_2 (t))\Vert^{}_{L^p}\cr
&\qquad\qquad{}+ \Vert x^\gamma  (\partial^{\alpha_1}  \delta_1 (t))
(\partial^{\alpha_2}  \delta_2 (t)) \Vert^{}_{L^p}\Big),\quad  t > 0,
1 \leq p \leq \infty. \cr
}$$

Since $\vert \gamma \vert \leq \vert \alpha \vert$ we can write
(for given $\alpha_1, \alpha_2$ with $\alpha_1 + \alpha_2 = \alpha$)
$\gamma = \gamma_1 + \gamma_2$
with $\vert \gamma_j \vert \leq \vert \alpha_j \vert,  j = 1,2.$ Doing
this we obtain from (4.10) and (4.12) that
$$\eqalignno{
&\Vert x^\gamma  (\partial^{\alpha_1}   \delta_1 (t))
(\partial^{\alpha_2}  \delta_2 (t)) \Vert^{}_{L^p}&(4.15)\cr
&\qquad{}\leq \Vert x^{\gamma_1} \partial^{\alpha_1}  \delta_1 (t)
\Vert^{1/p}_{L^2}
\Vert x^{\gamma_1}  \partial^{\alpha_1}  \delta_1 (t)
\Vert^{(p-1)/p}_{L^\infty}
\Vert x^{\gamma_2}  \partial^{\alpha_2}  \delta_2 (t)
\Vert^{1/p}_{L^2}
 \Vert x^{\gamma_2}  \partial^{\alpha_2}  \delta_2 (t)
\Vert^{(p-1)/p}_{L^\infty} \cr
&\qquad{}\leq C_{n, \vert \alpha \vert}  t^{- (2n+2+3(p-1)/p -
\vert\gamma\vert}
\Vert h_1 (t) \Vert^{}_{D^{}_{N+ \vert \alpha \vert}}
\Vert h_2 (t) \Vert^{}_{D^{}_{N+ \vert \alpha \vert}},
\quad t > 0,  1 \leq p \leq \infty.\cr
}$$
We have used here the fact that
$$\Vert fg \Vert^{}_{L^p} \leq \Vert f \Vert^{1/p}_{L^2}
\Vert f \Vert^{(p-1)/p}_{L^\infty}
\Vert g \Vert^{1/p}_{L^2}  \Vert g \Vert^{(p-1)/p}_{L^\infty},
\quad 1 \leq p \leq \infty.$$
It follows directly by induction that there are polynomials in $(t,x)$,
$F^{(l)}_\gamma$ of degree not higher than $\vert \gamma \vert$, such that, if
$\vert x\vert < t$,
$$\partial^\gamma  e^{i m \rho} = e^{i m \rho}  \sum^{\vert \gamma \vert - 1}_
{l = 0}  {1 \over \rho^{\vert \gamma \vert + l}}  F^{(l)}_\gamma,
\quad \vert \gamma \vert \geq 1. \eqno{(4.16)}$$
Equality (4.16) gives
$$\vert \partial^\gamma  e^{i m \rho (t,x)} \vert \leq C_{\vert \gamma \vert}
 (1+t)^{\vert \gamma \vert}  \sum^{\vert \gamma \vert}_{l=0}
 \rho (t,x)^{- (\vert \gamma \vert + l)}, \quad
\vert x \vert \leq t,  t > 0,\vert \gamma \vert \geq 0. \eqno{(4.17)}$$

Since the support of $r^{}_j (t)$ is contained in $\{ x \big\vert  \vert x
\vert
\leq t \}$,  $t > 0$, it follows from (4.17) and Schwarz inequality that
$$\eqalignno{
\hskip-3.5pt&\sum_{\alpha_1 + \alpha_2 = \alpha}  \Vert (\partial^{\alpha_1}
e^{- im \rho (t)}  r^{}_1 (t))  (\partial^{\alpha_2}
\delta_2 (t)) \Vert^{}_{L^p}\cr
&\qquad{}\leq C_{\vert \alpha \vert}  \sum_{\alpha_1 + \alpha_2 +
\alpha_3 = \alpha}
 \Vert (\partial^{\alpha_1}  e^{-i m \rho (t)})
(\partial^{\alpha_2}  r^{}_1 (t))  (\partial^{\alpha_3}
\delta_2 (t)) \Vert^{}_{L^p}\cr
&\qquad{}\leq C'_{\vert \alpha \vert}  \sum_{\alpha_1 + \alpha_2 +
\alpha_3 = \alpha}
 (1+t)^{\vert \alpha_1 \vert}  \sum^{\vert \alpha_1 \vert}_{l=0}
 \Vert \rho (t)^{- (\vert \alpha_1 \vert + l)}  (\partial^{\alpha_2}
 r^{}_1 (t))  (\partial^{\alpha_3}  \delta_2 (t)) \Vert^{}_{L^p}\cr
&\qquad{}\leq C'_{\vert \alpha \vert}  \sum_{\alpha_1 + \alpha_2 +
\alpha_3 = \alpha}
 \sum^{\vert \alpha_1 \vert}_{l=0}  (1+t)^{\vert \alpha_1 \vert}
 \Vert \rho (t)^{- (\vert \alpha_1 \vert + l)}  \partial^{\alpha_2}
 r^{}_1 (t) \Vert^{1/p}_{L^2}\cr
&\qquad\qquad{}\Vert \rho (t)^{- (\vert \alpha_1 \vert + l)}
\partial^{\alpha_2}
 r^{}_1 (t) \Vert^{(p-1)/p}_{L^\infty}  \Vert \partial^{\alpha_3}
 \delta_2 (t) \Vert^{1/p}_{L^2}  \Vert \partial^{\alpha_3}
 \delta_2 (t) \Vert^{(p-1)/p}_{L^\infty},\quad t > 0,  1 \leq p \leq
 \infty.\cr
}$$

It now follows from inequalities (4.10), (4.11), (4.12) and (4.13) that
$$\eqalign{
&\sum_{\alpha_1 + \alpha_2 = \alpha}  \Vert (\partial^{\alpha_1}
e^{-i m \rho (t)}  r^{}_1 (t))  (\partial^{\alpha_2}
\delta_2 (t)) \Vert^{}_{L^p}\cr
&\qquad{}\leq C_{n, \vert \alpha \vert}  \Vert h_1 (t) \Vert^{}_{D^{}_{N'}}
\Vert h_2 (t) \Vert^{}_{D^{}_{N'}}\cr
&\qquad\qquad{}\sum_{\alpha_1 + \alpha_2 + \alpha_3 = \alpha}
\sum^{\vert \alpha _1 \vert}_{l=0}
(1+t)^{\vert \alpha_1 \vert}  t^{- (n+1+3 (p-1)/p + \vert \alpha_1 \vert +
\vert \alpha_2 \vert + l)},\quad t > 0,  1 \leq p \leq \infty,\cr
}$$
where $N'$ depends only on $n$ and $\vert \alpha \vert$.

For $t \geq 1$, we get
$$\eqalignno{
&\sum_{\alpha_1 + \alpha_2 = \alpha}  \Vert (\partial^{\alpha_1}
e^{- i m \rho (t)}  r^{}_1 (t))  (\partial^{\alpha_2}
\delta_2 (t)) \Vert^{}_{L^p}&(4.18)\cr
&\qquad{}\leq C_{n, \vert \alpha \vert}  t^{- (n+1+3(p-1)/p)}   \Vert
h_1 (t) \Vert^{}_{D^{}_{N'}}  \Vert h_2 (t) \Vert^{}_{D^{}_{N'}},
\quad  t \geq 1, 1 \leq p \leq \infty,
}$$
for some constant $C_{n, \vert \alpha \vert}$.

Inequalities (4.11) and (4.13) give similarly
$$\eqalignno{
&\sum_{\alpha_1 + \alpha_2 = \alpha}  \Vert (\partial^{\alpha_1}
r^{}_1 (t))  (\partial^{\alpha_2}  r^{}_2 (t)) \Vert^{}_{L^p}&(4.19)\cr
&\qquad{}\leq C_{n, \vert \alpha \vert}
\Vert h_1 (t) \Vert^{}_{D^{}_{N+2 \vert \alpha \vert}}
\Vert h_2 (t) \Vert^{}_{D^{}_{N+2 \vert \alpha \vert}}
t^{- (\vert\alpha \vert + 3 (p-1)/p)},\quad t > 0,  1 \leq p \leq \infty.\cr
}$$

Inequalities (4.14), (4.15) and (4.18) give for $t \geq 1$,
$$\eqalignno{
&\Vert \Gamma_{\gamma, \alpha} (t) - x^\gamma   \partial^\alpha (r^{}_1 (t)
r^{}_2 (t)) \Vert^{}_{L^p}&(4.20)\cr
&\qquad{}\leq C_{n, \vert \alpha \vert}  \Vert h_1 (t) \Vert^{}_{D^{}_{N'}}
\Vert h_2 (t) \Vert^{}_{D^{}_{N'}}  (t^{-2(n+1)} + t^{- (n+1)})
t^{\vert \gamma \vert - 3(p-1)/p}\cr
&\qquad{}\leq C_{n, \vert \alpha \vert}  \Vert h_1 (t) \Vert^{}_{D^{}_{N'}}
 \Vert h_2 (t) \Vert^{}_{D^{}_{N'}}  t^{- (n+1 - \vert \gamma \vert
+ 3 (p-1)/p)},\quad t \geq 1,  1 \leq p \leq \infty.\cr
}$$

Choosing $n = \vert \alpha \vert - 1$ for $\vert \alpha \vert \geq 1$ and $n=0$
for $\vert \alpha \vert = 0$ in (4.20), and using (4.19), we obtain
$$\eqalignno{
&\Vert \Gamma_{\gamma, \alpha} (t) \Vert^{}_{L^p}&(4.21)\cr
&\qquad{}\leq C_{\vert \alpha \vert}
\Vert h_1 (t) \Vert^{}_{D^{}_N}  \Vert h_2 (t) \Vert^{}_{D^{}_N}  t^{- (\vert
\alpha \vert - \vert \gamma \vert + 3(p-1)/p)},\quad t \geq 1,
\vert \gamma \vert \leq \vert \alpha \vert, 1\leq p \leq \infty,\cr
}$$
where $N$ is redefined and depends only on $\vert \alpha \vert$.
Since
$$\eqalignno{
\Vert x^\gamma  \partial^\alpha  e^{- i \omega (- \partial) t}
 h_1 (t) \Vert^{}_{L^2}
&\leq C_{\vert \alpha \vert}  \Vert h_1 (t)
\Vert^{}_{D^{}_{\vert \alpha \vert}},
\quad 0 \leq t \leq 1,\vert \gamma \vert \leq \vert \alpha \vert&(4.22)\cr
\noalign{\hbox{and}}
\Vert x^\gamma  \partial^\alpha  e^{i \omega(-i \partial) t}
 h_2 (t) \Vert^{}_{L^\infty}
&\leq C_{\vert \alpha \vert}
\Vert h_2 (t)\Vert^{}_{D^{}_{\vert \alpha \vert + 2}},
\quad  0 \leq t \leq 1,\vert \gamma \vert \leq \vert \alpha \vert,\cr
}$$
which is obtained by using
$\Vert f\Vert^{}_{L^\infty} \leq C \Vert (1-\Delta) f \Vert^{}_{L^2}$,
we obtain from (4.9) and Schwarz inequality that
$$\eqalignno{
&\Vert \Gamma_{\gamma, \alpha} (t) \Vert^{}_{L^p}&(4.23)\cr
&\qquad{}\leq C'_{\vert \alpha \vert}
\Vert h_1 (t) \Vert^{}_{D^{}_{\vert \alpha \vert + 2}}
\Vert h_2 (t) \Vert^{}_{D^{}_{\vert \alpha \vert +2}},
\quad 0 \leq t \leq 1,\vert \gamma \vert \leq \vert \alpha \vert,
1 \leq p \leq \infty.\cr
}$$

Inequalities (4.21) and (4.23) give (with new $C_{\vert \alpha \vert}$)
$$\eqalignno{
&\Vert \Gamma_{\gamma, \alpha} (t) \Vert^{}_{L^p}&(4.24)\cr
&\qquad{}\leq C_{\vert \alpha \vert}
\Vert h_1 (t) \Vert^{}_{D^{}_N}  \Vert h_2 (t) \Vert^{}_{D^{}_N}
(1+t)^{- (\vert \alpha \vert - \vert \gamma \vert + 3(p-1)/p)},\quad  t \geq 0,
 1 \leq p \leq \infty,\cr
}$$
for some $N$ depending only on $\vert \alpha \vert$.

It follows from the expression (4.4a) of $G_{\varepsilon, \mu}$ and equality
(4.6) that
$$\eqalignno{
&x^\beta  \partial^\alpha  (G_{\varepsilon, \mu} (f_1, f_2)) (t)& (4.25)\cr
&\qquad{}= \sum_{\gamma_1 + \gamma_2 = \beta}   C_{\gamma_1, \gamma_2}
\int^\infty_0  K_{\gamma_2}  (-is \partial )  x^{\gamma_1}
s^{1+\vert \gamma_2 \vert}  \partial^\alpha  J_\mu
(s+t)  ds, \quad t \geq 0.\cr
}$$
According to definition (4.4c) of $J_\mu$ and estimate (4.24)
for $\Gamma_{\gamma, \alpha}$ defined by (4.9), we have
$$\eqalignno{
&\Vert x^\gamma  \partial^\alpha  J_\mu (t) \Vert^{}_{L^p}&(4.26)\cr
&\qquad{}\leq C_{\vert \alpha \vert}  \Vert f_1 (t) \Vert^{}_{D^{}_N}
\Vert f_2 (t) \Vert^{}_{D^{}_N}  (1+t)^{- (\vert \alpha \vert -
\vert \gamma \vert +
3 (p-1)/p)},\quad  t \geq 0, \vert \alpha \vert \geq \vert \gamma \vert,
1 \leq p \leq \infty.\cr
}$$

Since $\vert p \vert^\rho  \vert K_\gamma (p) \vert \leq
C_{\vert \gamma \vert}
 (1+\vert p \vert)^{-(1-\rho)}$ according to (4.5) we have
$$\eqalignno{
t^{\rho}\Vert \vert \nabla \vert^\rho  K_\gamma (-i t\partial ) f
\Vert^{}_{L^2}
&=\Vert \vert -it \partial  \vert^\rho  K_\gamma
(-it \partial) f \Vert^{}_{L^2} &(4.27)\cr
&\leq C_{\vert \gamma \vert} \Vert f \Vert^{}_{L^2},
\quad t \geq 0,  0 \leq \rho \leq 1.\cr
}$$
Inequalities (4.25), (4.26) and (4.27)
give for $\vert \beta \vert \leq \vert \alpha \vert$
(with a new $C_{\vert \alpha \vert}$)
$$\eqalignno{
&\Vert \vert \nabla \vert^\rho  x^\beta  \partial^\alpha
(G_{\varepsilon, \mu}  (f_1, f_2)) (t) \Vert^{}_{L^2}&(4.28)\cr
&\qquad{}\leq C_{\vert \alpha \vert}
 \sum_{\gamma_1 + \gamma_2 = \beta}
\int^\infty_0  s^{1 - \rho}  s^{\vert \gamma_2 \vert}
\Vert f_1 (s+t) \Vert^{}_{D^{}_N}  \Vert f_2 (s+t) \Vert^{}_{D^{}_N}
(1+s+t)^{- (3/2+ \vert \alpha \vert - \vert \gamma_1 \vert)}  ds\cr
&\qquad{}\leq C'_{\rho, \vert \alpha \vert}  \sup_{s \geq t}  ((1+s)^q
 \Vert f_1 (s) \Vert^{}_{D^{}_N}  \Vert f_2 (s) \Vert^{}_{D^{}_N})
(1+t)^{- (\rho - 1/2 + \vert \alpha \vert - \vert \beta \vert + q)},\cr
}$$
$t \geq 0$, $\vert \beta \vert \leq \vert \alpha \vert$,
$q + \vert \alpha \vert - \vert
\beta \vert + \rho - 1/2 > 0$.

It follows from (4.4b), (4.7) and (4.8) that
$$\eqalignno{
&\Vert \vert \nabla \vert^{\rho -1}  x^\beta  \partial^\alpha
 (\dot{G}_{\varepsilon, \mu}  (f_1, f_2)) (t) \Vert^{}_{L^2}
&(4.29)\cr
&\qquad{}\leq C_{\vert \alpha \vert}  \sum_{\gamma_1 + \gamma_2 = \beta}
\int^\infty_0  s^{\vert \gamma_2 \vert}  \Vert \vert \nabla \vert^{\rho - 1}
x^{\gamma_1}  \partial^\alpha   J_\mu (s+t) \Vert^{}_{L^2}
ds,\quad \vert \beta \vert \leq \vert \alpha \vert.\cr
}$$
According to (2.67) we obtain, since
$C^\infty_0 (\Rrm^3)$ is dense in $L^p$, $1 \leq p < \infty$, that
$$\Vert \vert \nabla \vert^{\rho - 1}  x^\gamma  \partial^\alpha
 J_\mu (t) \Vert^{}_{L^2} \leq C \Vert x^\gamma  \partial^\alpha
 J_\mu (t) \Vert^{}_{L^p},\quad  t \geq 0,$$
where $p= 6(5-2 \rho)^{-1}$, $0 \leq \rho \leq 1$. We obtain from (4.26)
that
$$\eqalignno{
&\Vert \vert \nabla \vert^{\rho - 1}  x^\gamma  \partial^\alpha
 J_\mu (t) \Vert^{}_{L^2} &(4.30)\cr
&\qquad{}\leq C_{\vert \alpha \vert}  \Vert f_1 (t) \Vert^{}_{D^{}_N}
 \Vert f_2 (t) \Vert^{}_{D^{}_N}
(1+t)^{- (\vert \alpha \vert - \vert \gamma \vert + 1/2 + \rho)}, \quad
t \geq 0,  0 \leq \rho \leq 1,\vert \gamma \vert \leq \vert \alpha \vert.\cr
}$$

Inequalities (4.29) and (4.30) show that (with a new $C_{\vert \alpha \vert}$)
$$\eqalignno{
&\Vert \vert \nabla \vert^{\rho - 1}  x^\beta   \partial^\alpha
 (\dot{G}_{\varepsilon, \mu}  (f_1, f_2)) (t) \Vert^{}_{L^2}&(4.31)\cr
&\qquad{}\leq C_{\vert \alpha \vert}\sup_{s \geq 0}
(\Vert f_1 (s) \Vert^{}_{D^{}_N} \Vert f_2 (s) \Vert^{}_{D^{}_N})\cr
&\qquad\qquad{}\sum_{\gamma_1 + \gamma_2 = \beta}
\int^\infty_0  (1+s+t)^{\vert \gamma_2 \vert}
(1+t+s)^{- (\vert \alpha \vert - \vert \gamma_1 \vert + 1/2 + \rho)}ds\cr
&\qquad{}\leq C'_{\vert \alpha \vert}  \sup_{s \geq t}  \big((1+s)^q
\Vert f_1 (s) \Vert^{}_{D^{}_N}  \Vert f_2 (s) \Vert^{}_{D^{}_N}\big)
(1+t)^{- (\vert \alpha \vert - \vert \gamma \vert - 1/2 + \rho + q)},
\quad t \geq 0,\cr
}$$
where $0 \leq \rho \leq 1$, $\vert \beta \vert \leq \vert \alpha \vert$
and $\vert \alpha \vert - \vert \gamma \vert + \rho - 1/2 + q > 0$.
Inequalities
(4.28) and (4.31) prove inequality (4.2) of the lemma in the case where $q=0.$
The case $q \geq 1$ follows by differentiation of expressions (4.4a) and (4.4b)
with respect to  $t$, and then by using inequalities (4.28) and (4.31) for
$q \geq 1$. This proves
inequality (4.2).

To prove inequality (4.3), introduce
$$I_{\gamma, \alpha, \mu} (t) = x^\gamma  \partial^\alpha
J_\mu (t),\quad  t \geq 0, \eqno{(4.32)}$$
and let $v_j (t)$ be the stationary phase development of
$e^{i \varepsilon \omega(-i \partial)t} f_j (t)$ up to order $n$.
We observe that for $t > 0,$ Plancherel theorem gives
$$\eqalignno{
\Vert (I_{\gamma, \alpha, \mu} (t))^\wedge \Vert^{}_{L^1}
&\leq \Big(\int (1+\vert k \vert^2  t^2)^{-2}  dk\Big)^{1/2}
\Vert (1 -t^2 \Delta )  I_{\gamma, \alpha, \mu} (t) \Vert^{}_{L^2}&(4.33)\cr
&= C  t^{- 3/2}  \Vert (1 - t^2\Delta  )
I_{\gamma, \alpha, \mu} (t) \Vert^{}_{L^2},\quad  t > 0.\cr
}$$
Commutation of $\Delta$ and $x^\gamma$ gives
$$\eqalignno{
&\Vert (1 - t^2\Delta)  I_{\gamma, \alpha, \mu} (t) \Vert^{}_{L^2}& (4.34)\cr
&\qquad{}\leq C_{\vert \gamma \vert, \vert \alpha \vert}
\Big(\Vert I_{\gamma, \alpha, \mu}(t) \Vert^{}_{L^2}
+ t^2  \sum_{\vert \delta \vert = \vert \gamma \vert - 2}
\Vert I_{\delta, \alpha, \mu} (t) \Vert^{}_{L^2}\cr
&\qquad\qquad{}+ t^2\sum_{\scr \vert \delta \vert=
\vert \gamma \vert - 1\atop\scr
\vert \lambda \vert = \vert \alpha \vert + 1}
\Vert I_{\delta, \lambda, \mu} (t) \Vert^{}_{L^2}
+ t^2  \sum_{\vert \lambda \vert = \vert \alpha \vert + 2}
\Vert I_{\gamma, \lambda, \mu} (t) \Vert^{}_{L^2}\Big),\cr
}$$
where sum over $\delta$ is absent if the upper bound of
$\vert \delta \vert$
is strictly negative.

According to (4.26) and (4.32) we now get
$$\eqalign{
&\Vert (1 - t^2\Delta )  I_{\gamma, \alpha, \mu} (t) \Vert^{}_{L^2}\cr
&\qquad{}\leq C_{\vert \alpha \vert}
\Vert f_1 (t) \Vert^{}_{D^{}_N}  \Vert f_2 (t) \Vert^{}_{D^{}_N}
(1+t)^{- (\vert \alpha \vert - \vert \gamma \vert + 3/2)}, \quad t \geq 0,
\vert \gamma \vert \leq \vert \alpha \vert,\cr
}$$
which together with (4.33), shows that (with a new $C_{\vert \alpha \vert}$)
$$\eqalignno{
&\Vert (I_{\gamma, \alpha, \mu} (t))^\wedge \Vert^{}_{L^1}&(4.35\hbox{a})\cr
&\qquad{}\leq C_{\vert \alpha \vert}
 \Vert f_1 (t) \Vert^{}_{D^{}_N}  \Vert f_2 (t) \Vert^{}_{D^{}_N}
t^{- (\vert \alpha \vert - \gamma \vert + 3)},\quad  t \geq 1,
\vert \gamma \vert \leq \vert \alpha \vert,\cr
}$$
where $N$ depends only on $\vert \alpha \vert$. Similarly, we obtain using
(4.20)
$$\Vert \big(I_{\gamma, \alpha, \mu} (t) - x^\gamma  \partial^\alpha
(v^+_1 (t)  \gamma^{}_0  \gamma_\mu  v_2 (t))\big)^\wedge
\Vert^{}_{L^1}
\leq C_{\vert \alpha \vert, n}  \Vert f_1 (t) \Vert^{}_{D^{}_N}
\Vert f_2 (t) \Vert^{}_{D^{}_N}  t^{- (n+2- \vert \gamma \vert)}.
\eqno{(4.35\hbox{b})}$$
On the other hand, we also have
$$\Vert (I_{\gamma, \alpha, \mu} (t))^\wedge \Vert^{}_{L^1} \leq C
\Vert (1-\Delta)
 I_{\gamma, \alpha, \mu} (t) \Vert^{}_{L^2},\quad  t \geq 0, \eqno{(4.36)}$$
which, by the method used from (4.33) to (4.35), gives
$$\Vert (I_{\gamma, \alpha, \mu} (t))^\wedge \Vert^{}_{L^1} \leq
C_{\vert \alpha \vert}
 \Vert f_1 (t) \Vert^{}_{D^{}_N}  \Vert f_2 (t) \Vert^{}_{D^{}_N},
\quad t \geq 0,  \vert \gamma \vert \leq \vert \alpha \vert. \eqno{(4.37)}$$
Inequalities (4.35) and (4.37) give (with a new $C_{\vert \alpha \vert}$)
$$\Vert (I_{\gamma, \alpha, \mu} (t))^\wedge \Vert^{}_{L^1} \leq
C_{\vert \alpha \vert}
 \Vert f_1 (t) \Vert^{}_{D^{}_N}  \Vert f_2 (t) \Vert^{}_{D^{}_N}
(1+t)^{- (\vert \alpha \vert - \vert \gamma \vert + 3)},\quad  t \geq 0,
 \vert \gamma \vert \leq \vert \alpha \vert,\eqno{(4.38)}$$
where $N$ depends only on $\vert \alpha \vert$.

Since $\Vert f \Vert^{}_{L^\infty} \leq \Vert \hat{f} \Vert^{}_{L^1}$
it follows from inequality (4.5), equality (4.25) and definition (4.32), that
$$\eqalign{
&\Vert x^\beta  \partial^\alpha  (G_{\varepsilon, \mu}
(f_1, f_2)) (t) \Vert^{}_{L^\infty}\cr
&\qquad{}\leq C_{\vert \alpha \vert}  \sum_{\gamma_1 +
\gamma_2 = \beta}
\int^\infty_0  s^{1+\vert \gamma_2 \vert}  \Vert (I_{\gamma_1,
\alpha, \mu}  (t+s))^\wedge \Vert^{}_{L^1}  ds,\quad  t \geq 0,
\vert \beta \vert \leq \vert \alpha \vert.\cr
}$$
It now follows from (4.38) that (with a new $C_{\vert \alpha \vert}$)
$$\eqalignno{
&\Vert x^\beta  \partial^\alpha  (G_{\varepsilon, \mu}
(f_1, f_2)) (t) \Vert^{}_{L^\infty}&(4.39)\cr
&\qquad{}\leq C_{\vert \alpha \vert}  \sum_{\gamma_1 +
\gamma_2 = \beta}
\int^\infty_0  s^{1+\vert \gamma_2 \vert}  (1+s+t)^{- (\vert
\alpha \vert - \vert \gamma_1 \vert + 3)}  \Vert f_1 (t+s) \Vert^{}_{D^{}_N}
\Vert f_2 (t+s) \Vert^{}_{D^{}_N}  ds\cr
&\qquad{}\leq C'_{\vert \alpha \vert}  \sup_{s \geq 0}
(\Vert f_1 (s) \Vert^{}_{D^{}_N}
 \Vert f_2 (s) \Vert^{}_{D^{}_N})
(1+t)^{-( \vert \alpha \vert - \vert \beta \vert + 1)}, \quad t \geq 0,
 \vert \beta \vert \leq \vert \alpha \vert, \cr
}$$
where $N$ depends only on $\vert \alpha \vert$.

The proof of
$$\eqalignno{
&\Vert x^\beta  \partial^\alpha  (\dot{G}_{\varepsilon, \mu}
 (f_1, f_2)) (t) \Vert^{}_{L^\infty}&(4.40)\cr
&\qquad{}\leq C'_{\vert \alpha \vert}
\sup_{s \geq 0}  (\Vert f_1 (s) \Vert^{}_{D^{}_N}
\Vert f_2 (s) \Vert^{}_{D^{}_N})
(1+t)^{- (\vert \alpha \vert - \vert \beta \vert + 2)},\quad  t \geq 0,
\vert \beta \vert \leq \vert \alpha \vert, \cr
}$$
is so similar to the proof of (4.39) that we omit it.

Let $\vert t \vert \leq \vert x \vert$, then it follows
from inequalities (2.60a)
and (2.60b) of Theorem (2.12), with $1/2 < \rho < 1$, that
$$\eqalignno{
&(1+\vert x \vert+t)^{3/2 - \rho + \vert \alpha \vert}
(\vert \partial^\alpha
 G(f_1, f_2) (t,x) \vert + (1+\vert x \vert +
 \vert t \vert) \vert \partial^\alpha
 \dot{G} (f_1, f_2) (t,x) \vert)&(4.41)\cr
&\qquad{}\leq C_{\vert \alpha \vert}
(1+\vert x \vert)^{3/2 - \rho + \vert \alpha \vert}
(\vert \partial^\alpha  G (f_1, f_2) (t,x) \vert + (1+\vert x \vert)
\vert \partial^\alpha  \dot{G} (f_1, f_2) (t,x) \vert)\cr
&\qquad{}\leq C'_{\vert \alpha \vert}  \Vert (G (f_1, f_2) (t),  \dot{G}
(f_1, f_2) (t) \Vert^{}_{M^\rho_{\vert \alpha \vert + 2}},
\quad t \geq 0,\vert x \vert \leq t.\cr
}$$
Inequality (4.2) now shows that
$$\eqalignno{
&(1+\vert x \vert + t)^{3/2 - \rho + \vert \alpha \vert}  (\vert
\partial^\alpha
 G (f_1, f_2) (t,x) \vert + (1+\vert x \vert + t) \vert \partial^\alpha
 \dot{G} (f_1, f_2) (t,x) \vert)&(4.42)\cr
&\qquad{}\leq C_{\vert \alpha \vert}  \sup_{s \geq 0}
(\Vert f_1 (s) \Vert^{}_{D^{}_{N'}}
 \Vert f_2 (s) \Vert^{}_{D^{}_{N'}})  (1+t)^{- \rho + 1/2},
\quad \vert x \vert \leq t,  t \geq 0,\cr
}$$
where $N'$ depends only on $\vert \alpha \vert$. Inequalities (4.39), (4.40)
and
(4.42) with $\varepsilon = \rho - 1/2$, prove the last statement of the lemma
in
the case where $q = 0$ and $\chi = 0$. The cases $q > 0$ or $\chi > 0$ are
obtained
by derivation of expressions (4.4a) and (4.4b) and by using the formulas (4.5)
to (4.8). This proves the lemma since $D^{}_\infty$ a dense in $D^{}_N$.

The proof of Lemma 4.1 has a useful corollary giving the decay rate
of the rest term of $G_{\varepsilon, \mu}$ after substraction of the
main contribution given
by stationary phase development of $e^{i \varepsilon \omega(-i \partial) t}
f_j (t),  j =1,2$, in (4.1). To state the result let, $s$
being a fixed parameter,
$$\sum_{0 \leq l \leq n}  g^{}_{j, \varepsilon, l}  (s,t,x)
 e^{i \varepsilon \rho (t,x) m},\quad  t > 0,$$
be the stationary phase development up to order $n$ of
$e^{i \varepsilon \omega(-i \partial) t}f_j (s)$, $j = 1,2$,
$f_j (s) \in D^{}_\infty$,
$s \geq 0$, defined by (A.1a) and (A.1b). We define
$$r^{}_{j, \varepsilon, n}  (t,x) = \sum_{0 \leq l \leq n}
g^{}_{j, \varepsilon,l}
 (t,t,x), \quad t > 0, \eqno{(4.43)}$$
and
$$\eqalignno{
&(R_{\varepsilon, \mu,n} (f_1, f_2)) (t)& (4.44\hbox{a})\cr
&\qquad{}= - \int^\infty_t  {\sin
((- \Delta)^{1/2}  (t-s)) \over (- \Delta)^{1/2}}  (r^{}_{1, \varepsilon,n}
(s))^+  \gamma^{}_0  \gamma_\mu  r^{}_{2, \varepsilon, n} (s)
 ds, \quad  t > 0,\cr
&(\dot{R}_{\varepsilon, \mu,n} (f_1, f_2)) (t)&(4.44\hbox{b})\cr
&\qquad{}= - \int^\infty_t
\cos((- \Delta)^{1/2}  (t-s))  (r^{}_{1, \varepsilon,n} (s))^+
\gamma^{}_0  \gamma_\mu  r^{}_{2, \varepsilon,n} (s)
ds, \quad t > 0. \cr
}$$
\saut
\noindent{\bf Corollary 4.2.}
{\it
Let $(G, \dot{G})$ and $(R, \dot{R})$ be defined by (4.1) and (4.44)
respectively and let $f_j \colon \Rrm^+ \fl D^{}_N$,  $j=1,2$, be $C^0$
functions.
Let $n \geq 0$ be an integer and let $\alpha, \beta$ be multi-indices.
If $N$, depending on $n$, $\vert \alpha \vert$ and $\vert \beta \vert,$
is chosen sufficiently large, then
$$\eqalignno{
&\Vert x^\beta  \partial^\alpha  (G_{\varepsilon, \mu} (f_1,
f_2) - R_{\varepsilon, \mu,n} (f_1, f_2)) (t) \Vert^{}_{L^2} &(4.45{\rm a})\cr
&\qquad{}+ t \Vert x^\beta  \partial^\alpha  (\dot{G}_{\varepsilon, \mu}
(f_1, f_2) - \dot{R}_{\varepsilon, \mu,n} (f_1, f_2)) (t) \Vert^{}_{L^2}\cr
&\qquad\qquad{}\leq C_{n, \vert \alpha \vert}  \sup_{s \geq 0}
(\Vert f_1 (s) \Vert^{}_{D^{}_N}
 \Vert f_2 (s) \Vert^{}_{D^{}_N})  t^{- (n - \vert \beta \vert + 1/2)},
\quad t \geq 1,  n \geq \vert \beta \vert,\cr
}$$
$$\eqalignno{
&\Vert \partial^\alpha  (G_{\varepsilon, \mu} (f_1, f_2) - R_{\varepsilon,
\mu,n}
(f_1, f_2)) (t) \Vert^{}_{L^\infty}&(4.45{\rm b})\cr
&\qquad{}+ t \Vert \partial^\alpha  (\dot{G}_{\varepsilon, \mu} (f_1, f_2) -
\dot{R}_{\varepsilon, \mu,n} (f_1, f_2)) (t) \Vert^{}_{L^\infty}\cr
&\qquad\qquad{}\leq C_{n, \vert \alpha \vert}  \sup_{s \geq 0}
(\Vert f_1 (s) \Vert^{}_{D^{}_N}
\Vert f_2 (s) \Vert^{}_{D^{}_N})  t^{- (n+2+\vert \alpha \vert)},
\quad  t \geq 1,n\geq 0,\cr
}$$
$$\eqalignno{
&t^{\vert \alpha \vert - \vert \beta \vert + \rho - 1/2}
\Vert x^\beta \partial^\alpha  (R_{\varepsilon, n} (f_1, f_2) (t),
\dot{R}_{\varepsilon, n} (f_1, f_2) (t)) \Vert^{}_{M^\rho} &(4.46)\cr
&\qquad{}\leq C_{n, \vert \alpha \vert}  \sup_{s \geq 0}
(\Vert f_1 (s) \Vert^{}_{D^{}_{N}}
 \Vert f_2 (s) \Vert^{}_{D^{}_N}), \quad t \geq 1,  \vert \alpha \vert
\geq \vert \beta \vert,1/2 < \rho < 1,\cr
}$$
and
$$\eqalignno{
&(1+\vert x \vert + \vert t \vert)^{3/2 - \rho + \vert \alpha \vert}
(\vert \partial^\alpha  R_{\varepsilon, \mu,n}  (f_1, f_2) (t,x) \vert \cr
&\qquad{}+(1+\vert x  \vert + \vert t \vert)  \vert \partial^\alpha
\dot{R}_{\varepsilon, \mu,n}(f_1, f_2) (t,x) \vert)&(4.47)\cr
&\qquad\qquad{}\leq C_{n, \vert \alpha \vert}  \sup_{s \geq 0}
(\Vert f_1 (s) \Vert^{}_{D^{}_{N}}
\Vert f_2 (s) \Vert^{}_{D^{}_N})  t^{- \rho + 1/2},
\quad t \geq 1.\cr
}$$
}\saut
\noindent{\it Proof.}
Inequality (4.45a) follows from the estimate (4.20):
$$\eqalign{
&\Vert \Gamma_{\gamma, \alpha} (t) - x^\gamma   \partial^\alpha
(r^{}_1 (t)  r^{}_2 (t)) \Vert^{}_{L^2}\cr
&\qquad{}\leq C_{n, \vert \alpha \vert}
\Vert h_1 (t) \Vert^{}_{D^{}_{N'}}  \Vert h_2 (t) \Vert^{}_{D^{}_{N'}}
t^{- (1+n)}  t^{\vert \gamma \vert - 3/2}, \quad t > 0,\cr
}$$
and by observing that
$$\eqalign{
&\big\Vert \int^\infty_0  {\sin ((- \Delta)^{1/2}  s) \over (- \Delta)^{1/2}}
 \big(\Gamma_{\gamma, \alpha} (t+s) - x^\gamma  \partial^\alpha
 (r^{}_1 (s+t)  r^{}_2 (s+t))\big)  ds \big\Vert^{}_{L^2}\cr
&\qquad{}+ t  \big\Vert \int^\infty_0  \cos ((- \Delta)^{1/2} s)
\big(\Gamma_{\gamma, \alpha} (t+s) - x^\gamma  \partial^\alpha
(r^{}_1 (s+t)  r^{}_2 (s+t))\big)  ds \big\Vert^{}_{L^2}\cr
&\qquad\qquad{}\leq C'_{n, \vert \alpha \vert}  \sup_{s \geq 0}
(\Vert h_1 (s) \Vert^{}_{D^{}_{N'}}
\Vert h_2 (s) \Vert^{}_{D^{}_{N'}})  t^{- (n-\vert \gamma \vert + 1/2)},
\quad t \geq 1.\cr
}$$
The proofs of (4.46) and (4.47) are contained in the proof of (4.2) and (4.3)
because of the bound (4.14). Inequality (4.45b) follows from (4.35) with $n$
replaced by $n+1 + \vert \alpha \vert$ and by estimating all terms in $v^+_1
(t)
 \gamma^{}_0  \gamma_\mu  v_2 (t)$ of order higher than
$n$. This proves the corollary.

We can now prove an analog of Theorem 3.5 of \refFST\ adapted to the spaces
$E^{}_N$.
Let $t \mapsto f_j (t)$,  $j=1,2$, be $C^2$ functions from $\Rrm^+$ to
$D^{}_N$,
let $m_1, m_2 > 0$,  $\varepsilon_1, \varepsilon_2 = \pm 1$,
$M = \vert \varepsilon_1  m_1 - \varepsilon_2  m_2 \vert > 0$,
let $\varepsilon$ be the sign of $- \varepsilon_1  m_1 + \varepsilon_2
 m_2$ and let $\omega^{}_M (k) = (M^2 + \vert k \vert^2)^{1/2}$,  $k \in
\Rrm^3$.
Introduce
$$\eqalignno{
(H_\mu (f_1, f_2)) (t) &= - \int^\infty_t  {\sin  ((- \Delta)^{1/2} (t-s))
\over (- \Delta)^{1/2}}  (I_\mu (f_1, f_2)) (s)  ds,
\quad t \geq 0, &(4.48\hbox{a})\cr
(\dot{H}_\mu (f_1, f_2)) (t) &= - \int^\infty_t  \cos  ((- \Delta)^{1/2} (t-s))
 (I_\mu (f_1, f_2)) (s)  ds,\quad  t \geq 0, &(4.48\hbox{b})\cr
\noalign{\hbox{where}}
(I_\mu (f_1, f_2)) (t) &= (e^{i \varepsilon_1   \omega_{m_1} (-i \partial) t}
 f_1 (t))^+  \gamma^{}_0  \gamma_\mu  (e^{i
\varepsilon_2  \omega_{m_2} (-i \partial) t}  f_2 (t)).&(4.48\hbox{c})\cr
}$$
If $t \mapsto f_j (t)$, $j=1,2$, are bounded and $C^\infty$ from $\Rrm^+$
to $D^{}_\infty$, then it turns out that the asymptotic behaviour of
$H_\mu$ is that of $e^{i \varepsilon  \omega^{}_M (-i \partial) t}
h(t)$, where $k \mapsto
(h(t))^\wedge  (k)$ is regular outside $k=0$. This fact, which follows
from the proof of the next lemma, was already proved in \refFST\
in a different setting.
\saut
\noindent{\bf Lemma 4.3.}
{\it
Let $(H, \dot{H})$ be defined by (4.48) and let $f_j \colon \Rrm^+
\fl D^{}_N$,
$j = 1,2$, be $C^3$ functions. Let $n \geq 0$ be an integer. If $N$
is chosen sufficiently large, then
$$\eqalign{
\hbox{\rm i)}&\ \sum_{\vert \alpha \vert \leq n}  \Vert \partial^\alpha
((H (f_1, f_2)) (t),  (\dot{H} (f_1, f_2)) (t)) \Vert^{}_{M^\rho}\cr
&\qquad{}\leq C_{n, \rho}  b_N  (f_1, f_2) (t)  (1+t)^{- (\rho + 1/2)},
\quad  - 1/2 < \rho \leq 1,  t \geq 0,\cr
&\cr
\hbox{\rm ii)}&\ \sum_{\vert \alpha \vert \leq n}  \Vert \partial^\alpha
\vert \nabla \vert^{\rho - 1}  ((H (f_1, f_2)) (t),  (\dot{H}
(f_1, f_2)) (t)) \Vert^{}_{L^2}\cr
&\qquad{}\leq C_{n, \rho}  b_N (f_1, f_2) (t)   (1+t)^{- (\rho + 1/2)},
\quad  1/2 < \rho \leq 1,  t \geq 0,\cr
&\cr\hbox{\rm iii)}&\ \vert ((H (f_1, f_2)) (t)(x)\vert
+ \vert ((\dot{H} (f_1, f_2)) (t)) (x) \vert\cr
&\qquad{}\leq C  b_N (f_1, f_2) (t)  (1+t+\vert x \vert)^{-3},
\quad t \geq 0,\hskip62.494mm\cr
}$$
where
$$b_N (f_1, f_2) (t) = \prod^2_{j=1}  \sum^3_{l=0}  \sup_{s \geq t}
 \big((1+s)^l  \Vert {d^l \over ds^l}  f_j (s) \Vert^{}_{D^{}_N}\big),$$
and where $N$ depends on $n$.
}\saut
\noindent{\it Proof.}
Let $h_j \in S(\Rrm^3,\Crm)$,   $j = 1,2$, and let
$$\Gamma (t) = (e^{-i \varepsilon_1  \omega_{m_1} (-i \partial)t}
h_1)  (e^{i \varepsilon_2  \omega_{m_2} (-i \partial)t}
h_2),\quad  t \geq 0,\eqno{(4.49)}$$
and let $r_j(t) = \sum_{q = 0}^n r_j^{(q)}(t)$
be the stationary phase development of
$e^{-i \varepsilon_1  \omega (-i \partial)t} h_j, j = 1,2$
up to order $n$ in Theorem A.1.
In complete analogy with the derivation of inequality
(4.20), by inequalities (4.14), (4.15), (4.18) and (4.19)
we obtain for $N$ sufficiently
large, depending on $\vert \alpha \vert$ and $n,$ and for
$\vert \beta \vert \leq n$,
$$\eqalignno{
&\Vert x^\beta  \partial^\alpha  (\Gamma (t) - e^{i \varepsilon M \rho (t)}
 r^{}_1 (t)  r^{}_2 (t)) \Vert^{}_{L^p}&(4.50)\cr
&\qquad{}\leq C_n  \Vert h_1 \Vert^{}_{D^{}_N}  \Vert h_2 \Vert^{}_{D^{}_N}
t^{- (n+1+3(p-1)/p-\vert \beta \vert)},\quad  t \geq 1,
1 \leq p \leq \infty,\cr
}$$
where $\hbox{\rm supp}\  r^{}_j (t) \subset \{ \vert x \vert \leq t \}$,

$$\Vert \rho (t)^{-l}  \partial^\alpha  r^{(q)}_j (t) \Vert^{}_{L^2}
\leq C_{n, \vert \alpha \vert,l}  \Vert h_j
\Vert^{}_{D^{}_{N+2\vert \alpha \vert+l}}
 t^{- \vert \alpha \vert - l - q},  \eqno{(4.51\hbox{a})}$$
and
$$\Vert \rho (t)^{-l}  \partial^\alpha  r^{(q)}_j (t)
\Vert^{}_{L^\infty}
\leq C_{n, \vert \alpha \vert,l}  \Vert h_j
\Vert^{}_{D^{}_{N+2\vert \alpha \vert+l}}
 t^{- \vert \alpha \vert - l - 3/2 - q}, \eqno{(4.51\hbox{b})}$$
$t > 0$, $0\leq q\leq n$, $l \geq 0$,  $\vert \alpha \vert \geq 0$. The
function
$(t,x) \mapsto r^{(q)}_j  (t,x)$ is homogeneous of degree $- 3/2 - q$.

The functions $r^{}_{(q)}$,  $0 \leq q \leq n$ defined by
$$r^{}_{(q)} (t,x) = t^{3/2 + q}  \sum_{q^{}_1+q^{}_2 = q}  r^{(q^{}_1)}_1
 (t,x)  r^{(q^{}_2)}_2  (t,x),\quad  t > 0, \eqno{(4.52)}$$
vanish outside the forward light cone and are homogeneous of degree $- 3/2$.
We can therefore apply Theorem A.2 which gives,
using (4.51a), (4.51b) and (4.52) for $\vert\beta \vert \leq L$,
$$\eqalignno{
&\Vert x^\beta  \partial^\alpha  (e^{i \varepsilon M \rho(t)}
 r^{}_{(q)} (t) - \sum_{0 \leq l \leq L}  t^{-l}  e^{i \varepsilon
 \omega^{}_M (-i \partial) t}  g^{}_{q,l}) \Vert^{}_{L^2}&(4.53\hbox{a})\cr
&\qquad{}\leq C_{L, \vert \alpha \vert}  \Vert h_1 \Vert^{}_{D^{}_{N'}}
\Vert h_2 \Vert^{}_{D^{}_{N'}}  t^{- (L+1 - \vert \beta \vert)},
\quad t > 0, \vert\beta\vert\leq L,\cr
&\Vert x^\beta  \partial^\alpha  (e^{i\varepsilon M \rho(t)}
 r^{}_{(q)} (t) - \sum_{0 \leq l \leq L}  t^{-l} e^{i \varepsilon
 \omega^{}_M (-i \partial)t}  g^{}_{q,l}) \Vert^{}_{L^\infty}&(4.53\hbox{b})\cr
&\qquad{}\leq C_{L, \vert \alpha \vert}  \Vert h_1 \Vert^{}_{D^{}_{N'}}
\Vert h_2 \Vert^{}_{D^{}_{N'}}  t^{- (L+5/2-\vert \beta \vert)},
\quad t > 0,\vert\beta\vert\leq L,\cr
}$$
and
$$\Vert g^{}_{q,l} \Vert^{}_{D^{}_j} \leq C_{L,j}
\Vert h_1 \Vert^{}_{D^{}_{N'+2j}}
 \Vert h_2 \Vert^{}_{D^{}_{N'+2j}}, \eqno{(4.54)}$$
where $N'$ depends on $L$ and $\vert \alpha \vert$ and where $g_{q, l}$
stands for the fonctions $f_l$ of theorem A.2. Defining
$$g' (t) = \sum_{q, l \geq 0}  t^{- (3/2 + q)}  g^{}_{q,l},
\quad  t > 0, \eqno{(4.55)}$$
using definition (4.52), interpolating between the
$L^2$- and $L^\infty$-estimate
in (4.53) and using (4.54), we obtain for $2 \leq p \leq \infty$,
$$\eqalignno{
&\Vert x^\beta  \partial^\alpha  (e^{i \varepsilon M \rho (t)}
 r^{}_1 (t)  r^{}_2 (t) - e^{i \varepsilon  \omega^{}_M (-i \partial)t}
 g' (t)) \Vert^{}_{L^p}&(4.56)\cr
&\qquad{}\leq C_{L, \vert \alpha \vert}  \Vert h_1 \Vert^{}_{D^{}_{N'}}
\Vert h_2 \Vert^{}_{D^{}_{N'}}  t^{- (L+5/2-\vert \beta \vert + 3(p-2)/2p)},
\quad  t \geq 1,\cr
}$$
$$\Vert \big({d \over dt}\big)^l  g' (t) \Vert^{}_{D^{}_j} \leq C_{L,j}
\Vert h_1 \Vert^{}_{D^{}_{N'+2j}}  \Vert h_2 \Vert^{}_{D^{}_{N'+2j}}
t^{- 3/2 - l},\quad t \geq 1, \eqno{(4.57)}$$
where $l \geq 0$,  $\vert \beta \vert \leq L$,  $L \geq 0$ and
where $N'$ depends on $L$ and $\vert \alpha \vert$.

Since $\Vert f \Vert^{}_{L^1} \leq C
\Vert (1+\vert x \vert)^2  f \Vert^{}_{L^2}$,
we obtain, with a new $L$ and a new $C_{L, \vert \alpha \vert}$,
$$\eqalignno{
&\Vert x^\beta  \partial^\alpha  (e^{i \varepsilon M \rho (t)}
 r^{}_1 (t)  r^{}_2 (t) - e^{i \varepsilon  \omega^{}_M (-i \partial)t}
 g' (t) \Vert^{}_{L^p}&(4.58)\cr
&\qquad{}\leq C_{L, \vert \alpha \vert}  \Vert h_1 \Vert^{}_{D^{}_{N'}}
\Vert h_2 \Vert^{}_{D^{}_{N'}}  t^{- (L+5/2-\vert \beta \vert)},
\quad t \geq 1,1 \leq p \leq \infty,    \vert \beta \vert \leq L - 2,\cr
}$$
$N'$ depends on $\vert \alpha \vert$ and $L$.

According to (4.50) and (4.58) we have,
choosing $L$ and $N$ sufficiently large,
$$\eqalignno{
&\Vert x^\beta  \partial^\alpha  (\Gamma (t) - e^{i \varepsilon
 \omega^{}_M (-i \partial)t}  g' (t)) \Vert^{}_{L^p}& (4.59)\cr
&\qquad{}\leq C_n  \Vert h_1 \Vert^{}_{D^{}_N}  \Vert h_2 \Vert^{}_{D^{}_N}
 t^{- (n+1+3(p-1)/p-\vert \beta \vert)},\quad  t \geq 1,
1 \leq p \leq \infty.\cr
}$$
We define
$$g (t) = \chi (t)  g' (t),\quad  t \in \Rrm, \eqno{(4.60)}$$
where $\chi \in C^\infty (\Rrm)$,  $0 \leq \chi (t) \leq 1$,
$\chi (t) = 0$ for $t \leq 1$,  $\chi (t) = 1$ for $t \geq 2$. It then
follows from (4.57) and (4.59) that
$$\eqalignno{
&\Vert x^\beta  \partial^\alpha  (\Gamma (t) - e^{i \varepsilon
 \omega^{}_M (-i \partial) t}  g(t)) \Vert^{}_{L^p}&(4.61)\cr
&\qquad{}\leq C_n  \Vert h_1 \Vert^{}_{D^{}_N}  \Vert h_2 \Vert^{}_{D^{}_N}
(1+t)^{- (n+1+3(p-1)/p-\vert \beta \vert)},\quad  t \geq 0,     1 \leq
p \leq \infty,
}$$
and
$$\Vert \big({d \over dt}\big)^l  g(t) \Vert^{}_{D^{}_j} \leq C_{n,j,l}
\Vert h_1 \Vert^{}_{D^{}_{N+2j}}
\Vert h_2 \Vert^{}_{D^{}_{N+2j}}  (1+t)^{- 3/2-l},
\quad t \geq 0,  j \geq 0,  l \geq 0,\eqno{(4.62)}$$
where $N$ depends on $n \geq 0$. In
fact $\Vert \Gamma (t) \Vert^{}_{L^p}
\leq C \Vert h_1 \Vert^{}_{D^{}_N}  \Vert
h_2 \Vert^{}_{D^{}_N}$ for $0 \leq t \leq 2$,
if $N$ is sufficiently large.

Let $f_j \colon \Rrm^+ \fl D^{}_\infty$ be $C^k$ and let
$$B_\mu (t,s) = (e^{i \varepsilon_1  \omega_{m_1} (-i \partial)t}
f_1 (s))^+  \gamma^{}_0  \gamma_\mu  (e^{i \varepsilon_2
\omega_{m_2} (-i \partial)t}  f_2 (s)),\quad t,s \geq 0.$$
According to (4.48c) we have
$$(I_\mu  (f_1, f_2)) (t) = B_\mu (t,t),\quad  t \geq 0.$$
Application of inequalities (4.61) and (4.62) for
$B_\mu (t,s)$, with $s$ fixed, proves that there is a $C^k$ function
$g \colon \Rrm^+ \fl D^{}_\infty$ such that
$$\eqalignno{
&\Vert x^\beta  \partial^\alpha  ((I_\mu (f_1, f_2)) (t) -
e^{i \varepsilon  \omega^{}_M (-i \partial)t}
g^{}_\mu (t)) \Vert^{}_{L^p}&(4.63)\cr
&\qquad{}\leq C_{n, \alpha}  \Vert f_1 (t) \Vert^{}_{D^{}_N}
\Vert f_2 (t) \Vert^{}_{D^{}_N}
(1+t)^{- (n+1+3(p-1)/p-\vert \beta \vert)},\quad  t \geq 0,
1 \leq p \leq \infty,
}$$
and
$$\eqalignno{
&\Vert \big({d \over dt}\big)^l  g(t) \Vert^{}_{D^{}_j}&(4.64) \cr
&\qquad{}\leq C_{n,j,l}
\sum_{0 \leq q \leq l}  \Vert (1+t)^q  \big({d \over dt}\big)^q
f_1 (t) \Vert^{}_{D^{}_N}
\Vert (1+t)^{l-q}  \big({d \over dt}\big)^{l-q}  f_2 (t) \Vert^{}_{D^{}_N}
 (1+t)^{- 3/2 - l},\cr
}$$
$ t \geq 0$, $n \geq 0$,
$j \geq 0$, $q \geq l \geq 0$,
where $N$ depends on $n$, $j$, $\vert \alpha \vert$ and $\vert \beta \vert
\leq n-2$.

Let
$$\eqalignno{
\Delta^{(n)}_{1, \mu} (t) &= - \int^\infty_t  {\sin  ((- \Delta)^{1/2}
 (t-s)) \over (- \Delta)^{1/2}}  \big((I_\mu (f_1, f_2)) (s) -
e^{i \varepsilon  \omega^{}_M (-i \partial)s}  g^{}_\mu (s)\big)
ds,\hskip3mm&(4.65\hbox{a})\cr
\dot{\Delta}^{(n)}_{1, \mu} (t) &= - \int^\infty_t  \cos ((- \Delta)^{1/2}
 (t-s))  \big((I_\mu (f_1, f_2)) (s) - e^{i \varepsilon
\omega^{}_M (-i \partial)s}  g^{}_\mu (s)\big)  ds,&(4.65\hbox{b})
}$$
$t \geq 0$, where $g$ satisfies (4.63) and (4.64). It follows from (4.6a) and
(4.6b) that, for $- 1/2 < \rho \leq 1$,
$$\eqalign{
&\sum_{\vert \alpha \vert \leq q}  \Vert x^\beta  \partial^\alpha
 (\Delta^{(n)}_1 (t),  \dot{\Delta}^{(n)}_1 (t)) \Vert^{}_{M^\rho}\cr
&\qquad{}\leq \sum_{0 \leq \mu \leq 3}  \sum_{\vert \alpha \vert \leq q}
\int^\infty_t  \Vert \vert \nabla \vert^{\rho - 1}  (s^{\vert \beta \vert}
+ \vert x \vert^{\vert \beta \vert})  \partial^\alpha  \big((I_\mu
(f_1, f_2)) (s) - e^{i \varepsilon
\omega^{}_M (-i \partial)s}  g^{}_\mu (s)\big)
\Vert^{}_{L^2}  ds.\cr
}$$
Since $\Vert \vert \nabla \vert^{\rho - 1}  f \Vert^{}_{L^2} \leq C_p
\Vert f \Vert^{}_{L^p}$,  $p = 6(5 - 2 \rho)^{-1}$,  $1 < p \leq 2$,
i.e. $- 1/2 < \rho \leq 1$, inequality (4.63) gives
$$\eqalignno{
&\sum_{\vert \alpha \vert \leq q}  \Vert x^\beta  \partial^\alpha
 (\Delta^{(n)}_1 (t),  \dot{\Delta}^{(n)}_1 (t)) \Vert^{}_{M^\rho}&(4.66)\cr
&\qquad{}\leq C_{n,q,\rho}  \sup_{s \geq 0}  (\Vert f_1 (s) \Vert^{}_{D^{}_N}
 \Vert f_2 (s) \Vert^{}_{D^{}_N})  (1+t)^{-n-\vert \beta \vert},
\quad  n \geq 1+\vert \beta \vert,  t \geq 0,  q \geq 0,\cr
}$$
where $N$ depends on $n$ and $k$.

For $n \geq 1$, let
$$\eqalignno{
H^{(n)}_\mu (t) &= - \int^\infty_t  {\sin ((- \Delta)^{1/2} (t-s)) \over
(- \Delta)^{1/2}}  e^{i \varepsilon  \omega^{}_M (-i \partial)s}
 g^{}_\mu (s)  ds,&(4.67\hbox{a})\cr
\dot{H}^{(n)}_\mu (t) &= - \int^\infty_t  \cos ((- \Delta)^{1/2}
(t-s))  e^{i \varepsilon  \omega^{}_M (-i \partial)s}  g^{}_\mu
(s)  ds,\quad  t \geq 0,&(4.67\hbox{b})\cr
}$$
where $g$ satisfies (4.63) and (4.64). We observe that
$$\eqalignno{
{\sin(\vert k \vert (t-s)) \over \vert k \vert}  e^{i \varepsilon
 \omega^{}_M (k)s} &= {\partial^l \over \partial s^l}  K_l (t,s,k),
 \quad l \geq 0, &(4.68\hbox{a})\cr
\noalign{\hbox{and}}
\cos  (\vert k \vert (t-s))  e^{i \varepsilon   \omega^{}_M (k)s} &=
{\partial^l \over \partial s^l}  \dot{K}_l (t,s,k),\quad  l \geq
0,&(4.68\hbox{b})\cr
}$$
where
$$\eqalignno{
K_l (t,s,k) &= \Big({(\varepsilon\omega^{}_M (k) + \vert k \vert)^l
- (\varepsilon\omega^{}_M (k) - \vert k \vert)^l \over i^l  2i \vert k \vert
M^{2l}}  \cos  (\vert k \vert (t-s))&(4.69a)\cr
&\qquad{}+ {(\varepsilon  \omega^{}_M (k) + \vert k \vert)^l + (\varepsilon
\omega^{}_M (k) - \vert k \vert)^l \over i^l  2  M^{2l}}
{\sin  (\vert k \vert (t-s)) \over \vert k \vert}\Big)  e^{i \varepsilon
\omega^{}_M (k) s},\cr
}$$
and
$$\dot{K}_l (t,s,k) = {\partial \over \partial t}  K_l (t,s,k).
\eqno{(4.69\hbox{b})}$$
The functions $k \mapsto K_l (t,s,k)$ and $k \mapsto \dot K_l (t,s,k)$
are real analytic functions on $\Rrm^3$.

Partial integration three times in (4.67) gives, using (4.68),
$$\eqalignno{
H^{(n)}_\mu (t) &= K_1 (t,t, - i \partial)
g^{}_\mu (t) - K_2 (t,t, -i \partial)
 {d \over dt}  g^{}_\mu (t) + \Delta^{(n)}_{2, \mu} (t),&(4.70\hbox{a})\cr
\noalign{\hbox{and}}
\dot{H}^{(n)}_\mu (t) &= \dot{K}_1 (t,t, -i \partial)   g^{}_\mu (t) -
\dot{K}_2 (t,t, - i \partial)  {d \over dt}  g^{}_\mu (t) +
\dot{\Delta}^{(n)}_{2, \mu} (t),
&(4.70\hbox{b})\cr
}$$
where
$$\eqalignno{
\Delta^{(n)}_{2, \mu} (t) &= K_3 (t,t, -i \partial)  {d^2 \over dt^2}
 g^{}_\mu (t) + \int^\infty_t  K_3 (t,s, -i \partial)
{d^3 \over ds^3}  g^{}_\mu (s)  ds, &(4.71\hbox{a})\cr
\noalign{\hbox{and}}
\dot{\Delta}^{(n)}_{2, \mu} (t) &= \dot{K}_3 (t,t, -i \partial)
{d^2 \over dt^2}
 g^{}_\mu (t) + \int^\infty_t  \dot{K}_3 (t,s, -i \partial)
{d^3 \over ds^3}  g^{}_\mu (s)  ds.&(4.71\hbox{b})\cr
}$$
According to definition (4.69) of $K$, we have
$$\eqalignno{
\vert K_l(t,s,k)\vert &\leq C_l(\omega^{}_M (k))^l
\vert k \vert^{-1},\quad l \geq 0,\cr
\vert \dot{K}_l (t,s,k) \vert & \leq C_l (\omega^{}_M (k))^l,
\quad  l \geq 0,\cr
\noalign{\hbox{and}}
\vert K_l (t,t,k) \vert & \leq l  C_l (\omega^{}_M (k))^{l-1},
\quad l \geq 0,\cr
\vert \dot{K}_l (t,t,k) \vert & \leq C_l (\omega^{}_M (k))^l,
\quad l \geq 0,\cr
}$$
$t,s \in \Rrm^+,  k \in \Rrm^3$.

These inequalities for $K$, similar inequalities for the
derivatives in the third argument of $K$ and
(4.71) give, for $k \in \Nrm$,
$$\eqalignno{
&\sum_{\vert \alpha \vert \leq k}  \Vert x^\beta  \partial^\alpha
(\Delta^{(n)}_2 (t),\dot{\Delta}^{(n)}_2
(t)) \Vert^{}_{M^\rho}\cr
&\quad{}\leq C_k  \sum_{\vert \alpha \vert \leq k + 3}
\Vert \vert x \vert^{\vert \beta \vert}  \partial^\alpha  \vert
\nabla \vert^{\rho - 1}  {d^2 \over dt^2}  g(t) \Vert^{}_{L^2}\cr
&\qquad{}+ C_k  \sum_{\vert \alpha \vert \leq k + 3}  \int^\infty_t
\Vert \vert x \vert^{\vert \beta \vert}  \partial^\alpha
\vert \nabla \vert^{\rho - 1}  {d^3 \over ds^3}  g(s) \Vert^{}_{L^2},
\quad\vert \beta \vert \leq 3.\cr
}$$
Since $\Vert \vert \nabla \vert^{\rho - 1}  f \Vert^{}_{L^2} \leq C_p
\Vert f \Vert^{}_{L^p}$, $p = 6(5-2 \rho)^{-1}$, $1 < p \leq 2$, and
since\penalty-10000
$\Vert f \Vert^{}_{L^p} \leq C' \Vert (1+\vert x \vert)^2
f \Vert^{}_{L^2}$
for $1 \leq p \leq 2$, inequality (4.64) now gives for
$- 1/2 < \rho \leq 1$ and
$\vert \beta \vert \leq 3$
$$\eqalignno{
&\sum_{\vert \alpha \vert \leq k}  \Vert x^\beta  \partial^\alpha
 (\Delta^{(n)}_2  (t),  \dot{\Delta}^{(n)}_2
(t)) \Vert^{}_{M^\rho}&(4.72)\cr
&\qquad{}\leq C_{n,k,\rho}  \prod^2_{j=1}  \sum^3_{l=0}
\sup_{s \geq t}  \big(\Vert (1+s)^l  {d^l \over ds^l}
f_j (s) \Vert^{}_{D^{}_N}\big)  (1+t)^{- 7/2 + \vert \beta \vert},
\quad t \geq 0,\cr
}$$
where $N$ depends on $n$ and $k$.

Let
$$\eqalignno{
\Delta^{(n)}_\mu &= \Delta^{(n)}_{1, \mu} + \Delta^{(n)}_{2, \mu},&(4.73)\cr
\dot{\Delta}^{(n)}_\mu &=\dot{\Delta}^{(n)}_{1, \mu} +
\dot{\Delta}^{(n)}_{2, \mu}.\cr
}$$
For $n \geq 4$ it follows from inequalities (4.66) and (4.72) that
$$\eqalignno{
&\sum_{\vert \alpha \vert \leq k}  \Vert x^\beta  \partial^\alpha
 (\Delta^{(n)}  (t),  \dot{\Delta}^{(n)}  (t)) \Vert^{}_{M^\rho}
&(4.74)\cr
&\qquad{}\leq C_{n,k,\rho}  \prod^2_{j=1}  \sum^3_{l=0}
\sup_{s \geq t}  \big((1+s)^l  \Vert {d^l \over ds^l}
f_j (s) \Vert^{}_{D^{}_N}\big)  (1+t)^{- 7/2 + \vert \beta \vert},\cr
}$$
where $t \geq 0$,  $- 1/2 < \rho \leq 1$, $n \geq 4$ and $N$
depends on $n$ and $k$.

It follows from definition (4.48) of $(H, \dot{H})$, (4.65) of
$(\Delta^{(n)}_1,
 \dot{\Delta}^{(n)}_1)$, (4.67) of $(H^{(n)},\dot{H}^{(n)})$,
(4.71) of $(\Delta^{(n)}_2,  \dot{\Delta}^{(n)}_2)$, (4.73) of $(\Delta^{(n)},
 \dot{\Delta}^{(n)})$ and from expression (4.70) that
$$\eqalignno{
H_\mu (t) &= \Delta^{(n)}_\mu   (t) + K_1 (t,t, -i \partial)
g^{}_\mu (t) - K_2 (t,t, -i \partial)  {d \over dt}  g^{}_\mu (t),\cr
\noalign{\hbox{and}}
\dot{H}_\mu (t) &= \dot{\Delta}^{(n)}_\mu  (t) + \dot{K}_1 (t,t, -i
\partial)  g^{}_\mu (t) - \dot{K}_2 (t,t, -i \partial)
{d \over dt}  g^{}_\mu (t).\cr
}$$
It now follows from the explicit expression (4.69) of $K$ and $\dot{K}$, that
$$\eqalignno{
\hskip-5mmH_\mu (t)&= \Delta^{(n)}_\mu(t)
- M^{-2}  e^{i \varepsilon  \omega^{}_M (-i \partial)t}
g^{}_\mu (t) - 2 i \varepsilon  \omega^{}_M (-i \partial)M^{-4}
e^{i \varepsilon  \omega^{}_M (-i \partial) t}{d \over dt}
g^{}_\mu (t),\hskip10mm&(4.75\hbox{a})\cr
\dot{H}_\mu (t)&= \dot{\Delta}^{(n)}_\mu(t) - i\varepsilon
\omega^{}_M (-i \partial)  M^{-2}  e^{i \varepsilon
 \omega^{}_M (-i \partial)t}  g^{}_\mu (t)& (4.75\hbox{b})\cr
&\qquad{}- (2(\omega^{}_M (-i \partial))^2 -
M^2)  M^{-4}  e^{i \varepsilon  \omega^{}_M (-i \partial)t}
 {d \over dt}  g^{}_\mu (t).\cr
}$$
Inequality (4.64) and equality (4.75) give,
$$\eqalignno{
&\sum_{\Vert \alpha \vert \leq k}  \Vert \partial^\alpha  (H
(t) - \Delta^{(n)}  (t),  0) \Vert^{}_{M^\rho} &(4.76)\cr
&\qquad{}\leq C_{n,k,\rho}  \prod^2_{j=1}  \sum^1_{l=0}
\sup_{s \geq 0}  \big(\Vert {d^l \over ds^l}  f_j (s) \Vert^{}_{D^{}_N}\big)
 (1+t)^{- 3/2}, \quad  t \geq 0,  k \geq 0,\cr
}$$
where $0 \leq \rho \leq 1$,  $n \geq 4$, and $N$ depends on $k$ and $n$.

We get from (4.75) and $\Vert \vert \nabla \vert^{\rho-1}  f
\Vert^{}_{L^2} \leq
C_p  \Vert f \Vert^{}_{L^p}$,  $p = 6(5-2 \rho)^{-1}$,
$1 < p \leq 2$,
$$\eqalignno{
&\sum_{\vert \alpha \vert \leq k}  \vert \partial^\alpha
\vert \nabla \vert^{\rho - 1}  (H (t) - \Delta^{(n)} (t),
\dot{H} (t) - \dot{\Delta}^{(n)} (t)) \Vert^{}_{L^2}\cr
&\qquad{}\leq C_\rho  \sum_{\vert \alpha \vert \leq k+2}
\big(\Vert \partial^\alpha
 e^{i \varepsilon  \omega^{}_M (-i \partial)t}g(t) \Vert^{}_{L^p} +
\Vert \partial^\alpha  e^{i \varepsilon  \omega^{}_M (-i \partial)t}
 {d \over dt}  g(t) \Vert^{}_{L^p}\big),\quad - 1/2 < \rho \leq 1.\cr
}$$
It follows using for example Corollary 2.2 of \refHKG, that
$$\Vert e^{i \varepsilon  \omega^{}_M (-i \partial)t}f \Vert^{}_{L^p} \leq
C_p  \Vert f \Vert^{}_{D^{}_J}  (1+\vert t \vert)^{3/p - 3/2},$$
for
$1 \leq p \leq 2$ for some $J > 0$. Since $3/p - 3/2 = 1 - \rho$ we obtain,
using (4.64),
$$\eqalignno{
&\sum_{\vert \alpha \vert \leq k}  \Vert \partial^\alpha
\vert \nabla \vert^{\rho - 1}  (H(t) - \Delta^{(n)} (t),  \dot{H} (t) -
\dot{\Delta}^{(n)} (t)) \Vert^{}_{L^2}&(4.77)\cr
&\qquad{}\leq C_{n,k,\rho}  \prod^2_{j=1}  \sum^1_{l=0}
\sup_{s \geq 0}  (\Vert {d^l \over ds^l}  f_j (s) \Vert^{}_{D^{}_N})
 (1+t)^{- (\rho + 1/2)},\cr
}$$
for $t \geq 0$, $- 1/2 < \rho \leq 1$,  $n \geq 4$, $k \geq 0$, where $N$
depends on $k$ and $n$.

Using that $\vert (e^{i \varepsilon  \omega^{}_M (-i \partial)t}  f)
(t,x) \vert
\leq C (1+t+\vert x \vert)^{- 3/2}  \Vert f \Vert^{}_{D^{}_J}$,
for some $J \geq 0$,
we obtain from inequality (4.64) and equality (4.75) that
$$\eqalignno{
&\vert (H(t) - \Delta^{(n)} (t),  \dot{H} (t) - \dot{\Delta}^{(n)} (t))
(x) \vert &(4.78)\cr
&\qquad{}\leq C_n  \prod^2_{j=1}  \sum^1_{l=0}  \sup_{s \geq t}
 \big(\Vert {d^l \over ds^l}  f_j (s) \Vert_{D^{}_{N_0}}\big)
(1+t+\vert x \vert)^{-3},\quad  t \geq 0,  n \geq 4\cr
}$$
for some $N_0$.

When $f_j \colon \Rrm^+ \fl D^{}_\infty$ is a $C^3$ function,
statement ii) of the lemma
now follows from inequalities (4.74) and (4.77) with $n = 4$. In fact,
for $1/2 < \rho \leq 1$:
$$\eqalignno{
\Vert \partial^\alpha  \vert \nabla \vert^{\rho - 1}(H(t),
\dot{H} (t)) \Vert^{}_{L^2}
&\leq \Vert \partial^\alpha(\Delta^{(4)} (t), 0) \Vert^{}_{M^{\rho-1}} +
\Vert \partial^\alpha  (0, \dot{\Delta}^{(4)} (t)) \Vert^{}_{M^\rho}\cr
&\qquad{}+ \Vert \partial^\alpha  \vert \nabla \vert^{\rho - 1}  (H(t) -
\Delta^{(4)} (t),  \dot{H} (t) - \dot{\Delta}^{(4)} (t)) \Vert^{}_{L^2},\cr
}$$
where the first two terms can be estimated by (4.74) since
$- 1/2 < \rho -1 \leq 1$,
and the last term by (4.77). Statement i) follows from
inequalities (4.74), (4.76)
and (4.77) for $0 \leq \rho \leq 1$. For $- 1/2 < \rho < 0$,
it follows from (4.74)
and (4.77), since in this case,
$$\eqalignno{
\Vert \partial^\alpha  (H(t),  \dot{H} (t)) \Vert^{}_{M^\rho}
&\leq
\Vert \partial^\alpha  (\Delta^{(4)} (t),  \dot{\Delta}^{(4)} (t))
\Vert^{}_{M^\rho}
+ \Vert \partial^\alpha  \vert \nabla \vert^{\rho}
(H(t) - \Delta^{(4)} (t)) \Vert^{}_{L^2}\cr
&\qquad{}+ \Vert \partial^\alpha  \vert \nabla \vert^{\rho-1}
 (\dot{H} (t) - \dot{\Delta}^{(4)} (t)) \Vert^{}_{L^2}\cr
}$$
and, since $- 1/2 < \rho < 0$ and $0 < \rho+1 < 1$, (4.77)
can be applied to the last two terms. Statement iii) of the
lemma follows from (4.74) and (4.78), since
$$\eqalignno{
&\Vert \nu (t)^3  (H(t),  \dot{H} (t)) \Vert^{}_{L^\infty}\cr
&\qquad{}\leq
\Vert \nu (t)^3  (H(t) - \Delta^{(4)} (t),  \dot{H} (t) -
\dot{\Delta}^{(4)} (t)) \Vert^{}_{L^\infty}\cr
&\qquad\qquad{}+ C  \sum_{\vert \alpha \vert \leq 2}
\Vert \partial^\alpha
 \nu (t)^3  (\Delta^{(4)} (t),  \dot{\Delta}^{(4)}
(t)) \Vert^{}_{L^2},\quad
(\nu(t))(x)=(1+t+\vert x\vert^2)^{1/2},\cr
}$$
where we have used $\Vert f \Vert^{}_{L^\infty} \leq C
\Vert f \Vert^{}_{W^{2,2}}$.
Finally it follows by continuous extension that statements
i), ii) and iii) are
valid for $C^3$ functions $f_j \colon \Rrm^+ \fl D^{}_N$
having finite $b_N$.
This ends the proof of the lemma.

We remark that statement ii) of the lemma is still true
for $- 1/2 < \rho \leq 1$,
which follows by a slight change in the proof. However
this result will not be used in this article.

To prove {\it the existence of a modified wave operator} for
$t \fl + \infty$, we first
use an asymptotic condition slightly different
from that defined by
$s^{(+)}_\varepsilon$
in (1.17c), where $s^{(+)}_\varepsilon$ is given by (1.18). We  introduce
$$\Theta' (t)  u = \Big(f, \dot{f},  \sum_{\varepsilon = \pm}
(\varepsilon\omega(-i\partial)t + s'_\varepsilon(t,-i\partial))
P_\varepsilon (-i \partial) \alpha\Big),
\quad u = (f, \dot{f}, \alpha) \in E^{}_N, \eqno{(4.79)}$$
for some $N$ sufficiently large, where
$$s'_\varepsilon (t,k) = - {\vartheta} (A', t, - {\varepsilon k t/\omega(k)} ),
\quad t \geq 0, \eqno{(4.80)}$$
$\vartheta$ is given by (1.19) and $A'$ being an electromagnetic potential
which we shall determine later.

To estimate $\vartheta$, we introduce the following representation of the
Poincar\'e Lie algebra $\p$ on functions of $(t,x)$:
$$\eqalignno{
\xi^{}_{P_0} &= {\partial \over \partial t}, \quad  \xi^{}_{P_i} = {\partial
\over
\partial x_i},  \quad 1 \leq i \leq 3,& (4.81\hbox{a})\cr
\xi^{}_{M_{0i}} &= x_i  {\partial \over \partial t} + t  {\partial \over
\partial x_i},  \quad 1 \leq i \leq 3,&(4.81\hbox{b})\cr
\xi^{}_{M_{ij}} &= -x_i  {\partial \over \partial x_j} + x_j  {\partial
\over \partial x_i},  \quad 1 \leq i < j \leq 3.& (4.81\hbox{c})\cr
}$$
We also define the representation $\xi^D$ (resp. $\xi^M$) on the space of Dirac
fields (resp. vector fields) by
$$\eqalignno{
\xi^D_{P_\mu}&= \xi^{}_{P_\mu},\hskip17.8mm  \xi^M_{P_\mu} = \xi^{}_{P_\mu},
\quad 0 \leq \mu \leq 3 & (4.81\hbox{d})\cr
\xi^D_{M_{\mu \nu}} &= \xi^{}_{M_{\mu \nu}} + \sigma_{\mu \nu},\quad
\xi^M_{M_{\mu \nu}} =
\xi^{}_{M_{\mu \nu}} + n_{\mu \nu}, \quad  0 \leq \mu < \nu \leq 3. &
(4.81\hbox{e})\cr
}$$
\saut
\noindent{\bf Lemma 4.4.}
{\it
Let $A' \colon \Rrm^+ \times \Rrm^3 \fl \Rrm^4$ be a $C^k$ function, $k \geq
0$,
and let $1/2 < \rho \leq 3/2$, then
$$\eqalignno{
& \cr
\hbox{\rm i)}&\ \vert \vartheta (A',t,x)
\vert \leq ((1 + t +\vert x \vert)^{\rho - 1/2} - 1)
C(\rho - 1/2)^{-1}  \sup_{s,y}  ((1+s+\vert y \vert)^{3/2 - \rho}
\vert A' (s,y) \vert)\cr
&\cr
\hbox{\rm ii)}&\ \vert {\partial \over \partial t}\vartheta(A',t,x)
\vert + \sum^3_{i=1}
 \vert {\partial \over \partial x_i}  \vartheta (A',t,x) \vert\cr
&\qquad{}\leq C (1+t+\vert x \vert)^{\rho - 3/2}  (\rho - 1/2)^{-1}\cr
&\qquad\qquad{}\sum_{0 \leq \nu \leq 3}  \sup_{s,y}
\Big((1+s+\vert y \vert)^{3/2 - \rho}
 \big(\vert A'_\nu (s,y) \vert + \vert A''_\nu (s,y)  \vert
 + \sum_{0 \leq \mu \leq 3}
 \vert \xi^{}_{M_{\mu \nu}} A'_\mu (s,y) \vert\big)\Big),\cr
\noalign{\hbox{where}}
&\qquad{}A''_0 (t,x) = \Big({\partial \over \partial t}
A'_0 (t,x) - \sum_{1 \leq i \leq 3}
\partial_i  A'_i (t,x)\Big) t,\cr
&\qquad{}A''_j (t,x) = \Big({\partial
\over \partial t} A'_0 (t,x) - \sum_{1 \leq i \leq 3}  \partial_i
A'_i (t,x)\Big) (- x_j),\quad \hbox{$1 \leq j \leq 3$ and $k \geq 1$,}\cr
& \cr
\hbox{\rm iii)}&\  \sum_{\vert \alpha \vert + l = n + n'} \vert {\partial^l
\over
\partial t^l}  \partial^\alpha  \vartheta (A',t,x) \vert\cr
&\qquad{}\leq C (\rho - 1/2)^{-1}  (1+t+ \vert x \vert)^{-1}  (1+ \big\vert
t - \vert x \vert \big\vert)^{-n + \rho + 1/2}\cr
&\qquad\qquad{}\sum_{0 \leq \nu \leq 3}
\sum_{\vert \beta \vert + r = n - 1 + n'}
\sup_{s,y} \Big((1+s+ \vert y \vert)  (1+\big\vert s -
\vert y \vert \big\vert)^{n-\rho-1/2}\cr
&\qquad\qquad{}\big(\vert {\partial^r \over \partial s^r}
\partial^\beta  A'_\nu
(s,y) \vert + \vert {\partial^r \over \partial s^r}  \partial^\beta
A''_\nu (s,y) \vert + \sum_{0 \leq \mu \leq 3}  \vert
{\partial^r \over \partial s^r}
 \partial^\beta  \xi^{}_{M_{\mu \nu}}  A'_\mu (s,y) \vert\big)\Big),\cr
\noalign{\hbox{if $k \geq n \geq 2$,  $n' \geq 0$.}}
\noalign{\hbox{\indent In the three cases the supremum is taken over $(s,y)
\in \Rrm^+ \times \Rrm^3$.}}
}$$
}\saut
\noindent{\it Proof.}
Statement i) follows directly from
$$\eqalignno{
\vert \vartheta (A',t,x) \vert
&\leq \int^1_0 \Big(\vert t \vert  \vert A_0
(s(t,x)) \vert + \vert x \vert  \sum^3_{i=1}  \vert A_i (s(t,x))
\vert\Big)ds\cr
&\leq \sum_{0 \leq \mu \leq 3}  \sup_{(s,y) \in \Rrm^+ \times \Rrm^3}
\Big((1+s+\vert y \vert)^{3/2 - \rho}  \vert A_\mu (s,y) \vert
\int^1_0  {t + \vert x \vert \over (1+s(t+\vert x \vert))^{3/2-\rho}}
 ds\Big),\cr
}$$
where $1/2 < \rho \leq 3/2$. To prove statement ii) we use covariant
notation and summation convention over
repeated contravariant  and covariant indices.
Let $y^0 = t$,  $y^i = x_i$,\penalty-10000
$1 \leq i \leq 3$. We note that using the gauge condition we obtain
$$\xi^{}_{P_\nu}  y_\mu  A^{' \mu} (sy)
=  (\xi^{}_{M_{\mu \nu}}  A^{' \mu}) (sy) + A'_\nu (sy) + A''_\nu (sy),
\eqno{(4.82)}$$
where $A''_\nu (z) = z_\nu  {\partial \over \partial z^\mu}
A^{' \mu} (z)$. This shows that (not using summation conventions)
$$\eqalignno{
&\vert {\partial \over \partial t}  {\vartheta} (A',t,x) \vert + \sum^3_{i=1}
 \vert {\partial \over \partial x_i} \vartheta (A',t,x) \vert\cr
&\quad{}\leq \sum_{0 \leq \nu \leq 3}  \sup_{y \in \Rrm^+ \times \Rrm^3}
 \Big((1+ \vert \vec{y} \vert + y^0)^{3/2 - \rho}  \big(\vert A'_\nu
(y) \vert + \vert A''_\nu (sy) \vert + \sum_{0 \leq \mu \leq 3}  \vert
\xi^{}_{M_{\mu \nu}}  A'_\mu (y) \vert\big)\Big)\cr
&\qquad{}\int^1_0  (1+s(t+\vert x \vert)^{\rho - 3/2}   ds,
\quad y = (y^0,  \vec{y}).\cr
}$$
Since the last integral is bounded by $C(\rho - 1/2)^{-1}
(1+t+\vert x \vert)^{\rho-3/2}$,
$\rho > 1/2$, this proves statement ii).

To prove statement iii) let $\vert \alpha \vert + l = n + n' = N$,  $n \geq 2$.
It follows from equality (4.82) and definition (4.81a) of $\xi^{}_{P_\mu}$ that
$$\eqalignno{
&\xi^{}_{P_{\nu^{}_1}}\cdots\xi^{}_{P_{\nu^{}_N}}
y_\mu  A^{' \mu} (sy)&(4.83)\cr
&\qquad{}= s^{N-1} \big(\xi^{}_{P_{\nu^{}_1}}\cdots\xi^{}_{P_{\nu^{}_{N-1}}}
\xi^{}_{M_{\mu \nu^{}_N}}
A^{' \mu} + \xi^{}_{P_{\nu^{}_1}}  A'_{\nu^{}_N}
+ \xi^{}_{P_{\nu^{}_1}}\cdots\xi^{}_{P_{\nu^{}_{N-1}}}
A''_{\nu^{}_N}\big)(sy),\cr
}$$
which gives
$$\sum_{\vert \alpha \vert + l = N}  \vert {\partial^l \over \partial t^l}
 \partial^\alpha \vartheta (A',t,x) \vert \leq Q_{n,N} (A')I_n(t,x),
\eqno{(4.84)}$$
where
$$I_n(t,x)= \int^1_0  s^{n-1}  (1+s(t+\vert x \vert))^{-1}
(1+s \big\vert t - \vert x \vert \big\vert)^{\rho - n + 1/2}  ds$$
and
$$\eqalignno{
Q_{n,N} (A') &= \sum_{0 \leq \nu \leq 3}  \sup_{y \in \Rrm^+ \times \Rrm^3}
\Big((1+y^0+\vert \vec{y} \vert)
(1+ \big\vert y^0 - \vert \vec{y} \vert\big\vert)^{n-1/2-\rho}\cr
&\qquad{}\big(\vert \nabla^{N-1}
A'_\nu (y) \vert + \vert \nabla^{N-1}
A''_\nu (y) \vert + \sum_{0 \leq \mu \leq 3}
 \vert \nabla^{N-1} (\xi^{}_{M_{\mu \nu}} A'_\mu) (y) \vert\big)\Big),\cr
}$$
$$\nabla = \Big({\partial \over \partial y^0},  {\partial \over \partial y^1},
 {\partial \over \partial y^2},  {\partial \over \partial y^3}\Big),
 \quad y = (y^0,  \vec{y}).$$
To estimate the integral $I_n (t,x)$, $n \geq 2$,
$(t,x) \in \Rrm^+ \times \Rrm^3$,
we observe that
$$I_n (t,x) \leq 2^{n+1/2-\rho}  \int^1_0  s^{n-1}
(1+s(1+t+\vert x \vert))^{-1}  (1+s(1+\big\vert t -
\vert x \vert \big\vert))^{\rho-n+1/2}
 ds.$$
This gives, for $n \geq 2$ and $1/2 < \rho \leq 3/2$,
$$\eqalignno{
I_n (t,x) &\leq 2^{n+1/2-\rho}  (1+t+\vert x \vert)^{-1}
(1+\big\vert t - \vert x \vert \big\vert)^{\rho+1/2-n}\cr
&\qquad{}\int^1_0  {s(1+t+\vert x \vert) \over 1+s(1+t+\vert x \vert)}
\Big({s(1+\big\vert t - \vert x \vert \big\vert)
\over 1+s(1+\big\vert t -
\vert x \vert \big\vert)}\Big)^{n-1/2-\rho}s^{\rho-3/2}  ds\cr
&\leq (\rho - 1/2)^{-1}  2^{n+1/2-\rho}  (1+t+\vert x \vert)^{-1}
 (1+\big\vert t - \vert x \vert\big \vert)^{\rho+1/2-n},\cr
}$$
which together with (4.84) proves statement iii) of the lemma.

There is an analog of Lemma 4.4 with $L^2$-estimates:
\saut
\noindent{\bf Lemma 4.5.}
{\it
Let $\xi^M_Y  A' \in C^0 (\Rrm^+, M^\rho)$ for $Y \in \Pi'$, let $A''$
be defined as in Lemma 4.4, let $B_\mu = - \partial_\mu  \vartheta (A')$,
$ 0 \leq \mu \leq 3$, let $Z \in \Pi'
\cap U(\Rrm^4)$, let $L \in U({\frak{sl}}(2, \Crm))$
and let $F_\mu (y) = \int^1_0  A_\mu (sy)  ds$, $y
\in \Rrm^+ \times \Rrm^3$.
\vskip10pt
\noindent \hbox{\rm i)} If $1/2 < \rho \leq 1$, then
$$\eqalignno{
&\Vert (\xi^M_{ZL} B, \xi^M_{P_0 ZL} B) (t) \Vert^{}_{M^\rho}\cr
&\qquad{}\leq C (1+t)^{-a}
\sup_{0 \leq s \leq t}  \Big((1+s)^b  (\Vert (\xi^M_{ZL} A',
\xi^M_{P_0 ZL} A') (s) \Vert^{}_{M^\rho}\cr
&\quad\qquad{}+ \Vert (\xi^M_{ZL} A'',
\xi^M_{P_0 ZL} A'')(s) \Vert^{}_{M^\rho}
+ \sum_{X \in \Pi \cap {\frak{sl}}(2, \Crm)}  \Vert (\xi^M_{ZXL} A',
\xi^M_{P_0  ZXL} A') (s) \Vert^{}_{M^\rho}\Big),\cr
}$$
where $a = \vert Z \vert + \rho - 1/2$ if $b > \vert Z \vert + \rho - 1/2$,
$a = b$ if $b < \vert Z \vert + \rho - 1/2$ and $b \in \Rrm$. The constant $C$
depends only on $\rho$, $a$, $b$.
\vskip 10pt
\noindent\hbox{\rm ii)} If $0 \leq \rho < 1$, then
$$\eqalignno{
&\Vert (1+t+\vert \cdot \vert)^{-1}
(\xi^M_{ZY} F) (t) \Vert^{}_{L^2 (\Rrm^3, \Rrm^4)}
+ \Vert (\xi^M_{P_\mu ZY} F) (t) \Vert^{}_{L^2 (\Rrm^3, \Rrm^4)}\cr
&\quad{}\leq C \Big((1+t)^{- a_1}  \sup_{0 \leq s \leq t}  \big((1+s)^{b_1}
 \Vert \vert \nabla \vert^\rho  (\xi^M_{ZY} A') (s) \Vert^{}_{L^2}\big)\cr
&\quad\quad{}+ (1+t)^{- a_2}  \sup_{0 \leq s \leq t}  \big((1+s)^{b_2}
\Vert (\xi^M_{ZY} A', \xi^M_{P_0  ZY} A') (s) \Vert^{}_{M^1}\big)\Big),
\quad Y \in \Pi',  0 \leq \mu \leq 3,  t \geq 0,\cr
}$$
for each $\tau \in ] 0,1 [$, such that $\vert Z \vert + 1/2 - (1 - \rho)
\tau > 0$, where
$$\tau a_1 + (1 - \tau) a_2 = \vert Z \vert + 1/2,$$
if $\tau b_1 + (1 - \tau) b_2 >\vert Z \vert + 1/2 - (1 - \rho) \tau$,
$$\tau a_1 + (1 - \tau) a_2 = (1 - \rho) \tau + \tau b_1 + (1 - \tau) b_2,$$
if $\tau b_1 + (1 - \tau) b_2 < \vert Z \vert + 1/2 - (1 - \rho) \tau$,
and where $a_2 = \vert Z \vert + 1/2$ if $b_2 >
\vert Z \vert + 1/2$ and $a_2 = b_2$ if
$b_2 < \vert Z \vert + 1/2$. The constant $C$
depends only on $a_1$, $a_2$, $b_1$, $b_2$,
$\rho$.
}\saut
\noindent{\it Proof.}
Since $(\xi^M_L  B)_\mu = - \partial_\mu  \vartheta (\xi^M_L A')$,
$0 \leq \mu \leq 3$, for $L \in U({\frak{sl}}(2, \Crm))$, and since
$$\vert(\xi^M_{Z'Y}  F) (y) \vert \leq \int^1_0  s^{\vert Z' \vert}
\vert (\xi^M_{Z'Y}  A') (sy) \vert  ds,\quad
Z' \in U(\Rrm^4),  y \in \Rrm^+ \times \Rrm^3,$$
it is enough to consider the case where $L = Y = {\un}$,
the identity element in $U(\p)$.

It follows from equality (4.83) with $N = \vert Z \vert + 1$,
since $\xi^M_Y = \xi^{}_Y$
for $Y \in U(\Rrm^4)$, that
$$\eqalignno{
&\big\vert ( \vert \nabla \vert^\rho  (\xi^M_Z B)_\mu (t))
(x\big)\vert&(4.85\hbox{a}) \cr
&\qquad{}\leq \int^1_0  s^{\rho + \vert Z \vert}
\Big(\sum_{0 \leq \alpha \leq 3}
 \big\vert ( \vert \nabla \vert^\rho  \xi^{}_{Z M_{\alpha \mu}}
A'_\alpha) (st, sx) \big\vert\cr
&\qquad\qquad{}+ \big\vert ( \vert \nabla \vert^\rho  \xi^{}_Z
A'_\mu) (st,sx)\big \vert +
\big\vert ( \vert \nabla \vert^\rho  \xi^{}_Z   A''_\mu) (st,sx)
\big\vert\Big)ds\cr
\noalign{\hbox{and}}
&\big\vert \big( \vert \nabla \vert^{\rho-1}
(\xi^M_{P_0 Z}  B)_\mu (t)\big) (x)\big\vert
& (4.85\hbox{b})\cr
&\qquad{}\leq \int^1_0  s^{\rho + \vert Z \vert}
\Big(\sum_{0 \leq \alpha \leq 3}
\big\vert ( \vert \nabla \vert^{\rho-1}  \xi^{}_{P_0ZM_{\alpha \mu}}
A'_\alpha) (st, sx) \big\vert\cr
&\qquad\qquad{}+ \big\vert ( \vert \nabla \vert^{\rho-1}
\xi^{}_{P_0}  A'_\mu) (st,sx) \big\vert +
\big\vert ( \vert \nabla \vert^{\rho-1}  \xi^{}_{P_0Z}  A''_\mu)
(st,sx)\big\vert\Big)ds.\cr
}$$
We prove next the following result. If $1 \leq p \leq \infty$,
$f \in C(\Rrm^+, L^p (\Rrm^3))$
$$(g^{}_t (s)) (x) = s^{\varepsilon} (f(st))  (sx),  \quad\varepsilon
> 3/p - 1, \eqno{(4.86\hbox{a})}$$
where $t,  s \in \Rrm^+$, $x \in \Rrm^3$, then
$$\big\Vert \int^1_0  g^{}_t (s)
ds \big\Vert^{}_{L^p} \leq C (1+t)^{3/p - \varepsilon-1}
 \sup_{0 \leq s \leq t}  (1+s)^\lambda  \Vert f(s) \Vert^{}_{L^p},
\eqno{(4.86\hbox{b})}$$
for $\lambda > \varepsilon - 3/p + 1$, and
$$\big\Vert \int^1_0  g^{}_t (s)
ds \big\Vert^{}_{L^p} \leq C (1+t)^{- \lambda}
 \sup_{0 \leq s \leq t}  (1+s)^\lambda  \Vert f(s) \Vert^{}_{L^p},
\eqno{(4.86\hbox{c})}$$
for $\lambda < \varepsilon - 3/p + 1$.
$C$ is a constant depending on $p, \lambda$
and $\varepsilon.$ In fact, since
$$\big\Vert \int^1_0  g^{}_t (s)  ds \big\Vert^{}_{L^p} \leq \int^1_0
 \Vert g^{}_t (s) \Vert^{}_{L^p}  ds$$
and
$$\Vert g^{}_t (s) \Vert^{}_{L^p} = s^{\varepsilon - 3/p}
\Vert f(st) \Vert^{}_{L^p},$$
it follows that
$$\big\Vert \int^1_0  g^{}_t (s)  ds \big\Vert^{}_{L^p}
\leq \int^1_0  s^{\varepsilon - 3/p}(1+st)^{- \lambda}ds
\sup_{0 \leq s' \leq t}  (1+s')^\lambda \Vert f (s') \Vert^{}_{L^p}.$$
Since $\varepsilon - 3/p > -1$, the integral in this expression exists and is
bounded by $C(1+t)^{- \varepsilon + 3/p - 1}$ if
$\varepsilon - 3/p - \lambda < -1$
and by $C (1+t)^{- \lambda}$ if $\varepsilon - 3/p - \lambda > -1,$
which is seen by making the substitution $s' = st$. This proves
estimates (4.86b) and (4.86c).

The inequality in statement i) of the lemma follows by applying
(4.86a) and (4.86b), with $p=2$ and $\varepsilon = \rho + \vert Z \vert$,
to the integrands in (4.85a)
and (4.85b), since $\varepsilon > - 1/2$.

To prove statement ii), we note that
$$\eqalign{
&\vert (1+t+\vert x \vert)^{-1}  (\xi^{}_Z  F) (t,x) \vert +
\vert (\xi^{}_{P_\mu Z}  F) (t,x) \vert\cr
&\qquad{}\leq (1+t+\vert x \vert)^{-1}  (v(t)) (x) + (u(t)) (x),
\quad t \in \Rrm^+,x \in \Rrm^3,\cr
}$$
where
$$(v(t)) (x) = \int^1_0  s^{\vert Z \vert}
\vert (\xi^{}_Z A') (st,sx) \vert  ds$$
and
$$(u(t)) (x) =
\sum_{\scr   Z' \in \Pi' \cap U(\Rrm^4)\atop\scr\vert Z' \vert
= \vert Z \vert + 1}
\int^1_0  s^{\vert Z \vert + 1}  \vert (\xi^{}_{Z'}A') (st,sx) \vert  ds.$$
Let $0 < \tau \leq 1$ and let $q^{-1} = \tau p^{-1} + (1 - \tau) / 6$, where
$p = 6 / (3-2 \rho)$, $0 \leq \rho < 1$. Then $p \leq q < 6$ and $\Vert
(1+t+\vert\cdot \vert)^{-1} \Vert^{}_{L^{q'}} \leq C_q  (1+t)^{3 / q' - 1}$,
where $q' = 2q / (q-2) > 3$. H\"older inequality gives
$$\Vert (1+t+\vert\cdot \vert)^{-1}  v(t) \Vert^{}_{L^2} \leq C_q
(1+t)^{1/2 - 3/q}  \Vert v(t) \Vert^{}_{L^q}.$$
Estimating $\Vert v(t) \Vert^{}_{L^q}$ by inequalities (4.86b) and (4.86c),
we obtain
$$\Vert (1+t+\vert\cdot\vert)^{-1}  v(t) \Vert^{}_{L^2} \leq
C (1+t)^{1/2-3/q - \lambda'}
 \sup_{0 \leq s \leq t}  (1+s)^\lambda  \Vert f(s) \Vert^{}_{L^q},$$
where $q$ is chosen such that $\vert Z \vert > 3/q - 1$
and where $\lambda' = \lambda$
if $\lambda < \vert Z \vert - 3/q + 1$ and $\lambda' =
\vert Z \vert - 3/q + 1$
for $\lambda > \vert Z \vert - 3/q + 1.$ It follows,
by the choice of $q$ and $\tau$
that $\Vert f(s) \Vert^{}_{L^q} \leq
\Vert f(s) \Vert^\tau_{L^p}  \Vert f(s)
\Vert^{1 - \tau}_{L^6},$ which together with the Sobolev
inequality (2.61) give
$$\eqalign{
&\Vert (1+t+\vert\cdot\vert)^{-1}v(t)\Vert^{}_{L^2}\cr
&\qquad{}\leq C' (1+t)^{1/2 - 3/q - \lambda'}
\sup_{0 \leq s \leq t}  \big((1+s)^\lambda  \Vert \vert \nabla \vert^\rho
 (\xi^{}_Z  A') (s) \Vert^\tau_{L^2}  \Vert \vert \nabla \vert
 (\xi^{}_Z  A') (s) \Vert^{1 - \tau}_{L^2}\big).\cr
}$$
Introduce $a_1$, $a_2$, $b_1$, $b_2 \in \Rrm$ such that
$\lambda' + 3/q - 1/2 = \tau
 a_1 + (1 - \tau) a_2$ and $\lambda = \tau  b_1 + (1 - \tau)
b_2$. Substitution of $1/q = \tau / p + (1 - \tau) / 6$,
$1/p = (3 - 2 \rho) / 6$
and the relation between $\lambda$ and $\lambda'$, give that
$$\tau a_1 + (1 - \tau) a_2 = \vert Z \vert + 1/2,$$
if $\tau b_1 + (1 - \tau) b_2 >\vert Z \vert + 1/2 - (1 - \rho) \tau$
, and
$$\tau a_1 + (1 - \tau) a_2 = (1 - \rho) \tau +
\tau b_1 + (1 - \tau) b_2,$$
if $\tau b_1 + (1 - \tau) b_2 < \vert Z \vert +
1/2 - (1 - \rho) \tau$. Moreover
$\tau \in ]0,1]$ and $\vert Z \vert + 1/2 - (1 -\rho) \tau > 0$.
By the definition
of $a_1$, $a_2$, $b_1$, $b_2$ we obtain that
$$\eqalign{
\Vert (1+t+\vert\cdot\vert)^{-1} v(t) \Vert^{}_{L^2}
&\leq C' \big((1+t)^{-a_1}
\sup_{0 \leq s \leq t}  (1+s)^{b_1}  \Vert \vert \nabla \vert^\rho
 (\xi^{}_Z  A') (s) \Vert^{}_{L^2}\big)^\tau\cr
&\qquad{}\big((1+t)^{-a_2}  \sup_{0 \leq s \leq t}  (1+s)^{b_2}
\Vert \vert \nabla \vert  (\xi^{}_Z  A') (s)
\Vert^{}_{L^2}\big)^{1 - \tau}.\cr
}$$
Using that $r^\tau_1  r^{1 - \tau}_2 \leq \tau r^{}_1 + (1-\tau) r^{}_2$,
where $r^{}_1, r^{}_2 \geq 0$ we obtain that
$\Vert (1+t+\vert\cdot\vert)^{-1} v(t) \Vert^{}_{L^2}$
is bounded by the right-hand side of the inequality in statement
ii) of the lemma. This is also the case for $\Vert u(t) \Vert^{}_{L^2}$,
which is seen by estimating
the terms in the sum defining $u(t)$ by inequalities (4.86b) and (4.86c), with
$\varepsilon = \vert Z \vert + 1$,  $\lambda = b_2$ and $\lambda' = a_2$.
This is possible since $\tau > 0$, so $a_1$ and $b_1$ are determined
by the above conditions when $\tau$ is fixed. We note that the condition
$\varepsilon > 3/p - 1$
is satisfied for $p=2$ and that $a_2 = b_2$ if $b_2 < \vert Z \vert + 1/2$ and
that $a_2 = \vert Z \vert + 1/2$ if $b_2 > \vert Z \vert + 1/2$. This proves
the
lemma.

In the next corollary, we define
$$\eqalignno{
A'_\mu (t) &= \cos ((- \Delta)^{1/2} t)  f_\mu + (- \Delta)^{- 1/2}
 \sin ((- \nabla)^{1/2} t)  \dot{f}_\mu& (4.87\hbox{a})\cr
&\qquad{}+ \sum_{\varepsilon = \pm}  (G_{\varepsilon, \mu}
(\beta^{(\varepsilon)}_1,
 \beta^{(\varepsilon)}_2)) (t),\quad 0 \leq \mu \leq 3,\cr
}$$
where $(f, \dot{f}) \in M^\rho_N$ and $\beta^{(\varepsilon)}_j
\in C^0(\Rrm^+, D^{}_N)$,
$\varepsilon = \pm$,  $j = 1,2$, for some $N$ sufficiently large
and where $G_{\varepsilon, \mu}$ is defined by (4.1a).
We suppose that the free field part of $A'_\mu$ satisfies the Lorentz
gauge condition, i.e.
$$\dot{f}_0 - \sum_{1 \leq i \leq 3}  \partial_i  f_i = 0,\quad
 \Delta  f_0 - \sum_{1 \leq i \leq 3}  \partial_i
 \dot{f}_i = 0. \eqno{(4.87\hbox{b})}$$
\saut
\noindent{\bf Corollary 4.6.}
{\it
Let $1/2 < \rho < 1$, let $(f, \dot{f}) \in M^{\circ\rho}_N$ and let
$\beta^{(\varepsilon)}_j \in C^\infty (\Rrm^+, D^{}_N)$. If $A'$
is defined by (4.87a), $B$ defined as in Lemma 4.5 and $N$ is sufficiently
large, then
$$\eqalignno{
\hbox{\rm i)}&\ \Vert (B(t), \dot{B} (t)) \Vert^{}_{M^\rho}\cr
&\qquad{}\leq C
\Vert (f, \dot{f}) \Vert^{}_{M^\rho_1}
+ C_\rho  \sum_{\varepsilon = \pm}  \sup_{0 \leq s}
\big(\Vert \beta^{(\varepsilon)}_1 (s) \Vert_{D^{}_N}
\Vert \beta^{(\varepsilon)}_2
(s) \Vert^{}_{D^{}_N}\big),\cr
\noalign{\hbox{$t \geq 0$,}}
\hbox{\rm ii)}&\ \vert \vartheta (A',t,x) \vert\cr
&\qquad{}\leq ((1+t+\vert x \vert)^{\rho - 1/2} - 1)
C_\rho \Big(\Vert (f, \dot{f}) \Vert^{}_{M^\rho_2}
+ \sum_{\varepsilon = \pm} \sup_{0 \leq s}
 \big(\Vert \beta^{(\varepsilon)}_1 (s) \Vert^{}_{D^{}_N}
 \Vert \beta^{(\varepsilon)}_2
(s) \Vert^{}_{D^{}_N}\big)\Big),\hskip3mm \cr
\noalign{\hbox{$t \geq 0$,}}
\hbox{\rm iii)}&\ \vert {\partial \over \partial t}\vartheta
(A',t,x) \vert + \vert {\partial
\over \partial x_i}\vartheta (A',t,x) \vert\cr
&\qquad{}\leq (1+t+\vert x \vert)^{\rho - 3/2}
C_\rho \Big(\Vert (f, \dot{f}) \Vert^{}_{M^\rho_3}
+ \sum_{\varepsilon = \pm} \sup_{0 \leq s}
 \big(\Vert \beta^{(\varepsilon)}_1 (s) \Vert^{}_{D^{}_N}
 \Vert \beta^{(\varepsilon)}_2
(s) \Vert^{}_{D^{}_N}\big)\Big),\cr
\noalign{\hbox{$t \geq 0$,}}
\hbox{\rm iv)}&\ \sum_{\vert \alpha \vert + l = n+n'}
\vert \big({\partial \over
\partial t}\big)^l  \partial^\alpha \vartheta  (A',t,x) \vert\cr
&\qquad{}\leq (1+t+\vert x \vert)^{-1}
(1+\big\vert t - \vert x \vert \big\vert)^{- n + \rho + 1/2}
C_{\rho,n} \Big(\Vert (f, \dot{f}) \Vert^{}_{M^\rho_{n+1+n'}}\cr
&\qquad\qquad{} +\sum_{\sscr\varepsilon = \pm\atop\sscr {\sscr l_1 + l_2
= l - 1\atop\sscr \vert \alpha_1 \vert
+ \vert \alpha_2 \vert \leq n'}}
\sup_{0 \leq s} (1+s)^{l_1-1}  \big(\Vert\big({d \over ds}\big)^{l_1}
\partial^{\alpha_1} \beta^{(\varepsilon)}_1 (s) \Vert^{}_{D^{}_N}
\Vert \big({d \over ds}\big)^{l_2}  \partial^{\alpha_2}
\beta^{(\varepsilon)}_2 (s) \Vert^{}_{D^{}_N}\big)\Big),\cr
\noalign{\hbox{$t \geq 0$, $n \geq 2$, $n' \geq 0$,
where $N$ depends only on $n$,}}
\hbox{\rm v)}&\ \hbox{if moreover $(f, \dot{f}) = 0$
and $\chi \geq 0$, then}\cr
&\ \sum_{\vert \alpha \vert + l \leq n+n'}  \vert \big({\partial
\over \partial t}\big)^l
 \partial^\alpha \vartheta  (A',t,x) \vert\cr
&\qquad{}\leq S_{n, \chi}(t,x)
\sum_{\varepsilon = \pm}  \sum_{\scr l_1 + l_2 + \vert \alpha_1 \vert +
\vert \alpha_2 \vert \leq n\atop\scr  l'_1 + l'_2 +
\vert \alpha'_1 \vert + \vert
\alpha'_2 \vert \leq n'}\cr
&\qquad\qquad{}\sup_{0 \leq s}  \big((1+s)^{l_1+l_2+\chi}
\Vert \big({d \over ds}\big)^{l_1 + l'_1} \partial^{\alpha_1}
\beta^{(\varepsilon)}_1 (s)
\Vert^{}_{D^{}_N}
\Vert \big({d \over ds}\big)^{l_2+l'_2} \partial^{\alpha_2}
\beta^{(\varepsilon)}_2 (s) \Vert^{}_{D^{}_N}\big)\cr
}$$
where $t \geq 0$,  $x \in \Rrm^3$,  $n \geq 1$,  $n' \geq 0$,
and where
$$\eqalignno{
S_{n, \chi} (t,x) &\leq C_\chi (1+t+\vert x \vert)^{-n}
\quad \hbox{for}\ \chi > 0\cr
\noalign{\noindent and}
S_{n,0} (t,x) &\leq C'_\delta (1+t+\vert x \vert)^{- n + \delta}
\quad\hbox{for}\
\chi = 0,  \delta > 0.\cr
}$$
The constants $C_\chi,  \chi > 0$ and $C'_\delta,  \delta > 0$
depend on $n$ and $n'$, and $N$ depends only on $n$.
}\saut
\noindent{\it Proof.}
It follows from Lemma 4.1 that the hypotheses of Lemma 4.5 are satisfied for
$N$ sufficiently large. We shall bound the terms on the right-hand side of the
inequality in statement i) of
Lemma 4.5. It follows from Lemma 4.1 and definitions (4.87a) and
(4.87b) of $A'$ that, for $N$ sufficiently large,
$$\eqalignno{
&\Vert (A' (t), {d \over dt} A' (t)) \Vert^{}_{M^\rho}&(4.88\hbox{a})\cr
&\qquad{}\leq \Vert (f, \dot{f}) \Vert^{}_{M^\rho}
+ (1+t)^{1/2 - \rho}    C_\rho  \sum_{\varepsilon = \pm}
\sup_{0 \leq s}  \big(\Vert \beta^{(\varepsilon)}_1 (s) \Vert^{}_{D^{}_N}
\Vert \beta^{(\varepsilon)}_2 (s) \Vert^{}_{D^{}_N}\big),\cr
&\Vert (A'' (t), {d \over dt}  A'' (t)) \Vert^{}_{M^\rho} &(4.88\hbox{b})\cr
&\qquad{}\leq (1+t)^{1/2 - \rho}  C_\rho  \sum_{\varepsilon = \pm}
 \sup_{0 \leq s}  \big(\Vert \beta^{(\varepsilon)}_1 (s) \Vert^{}_{D^{}_N}
 \Vert \beta^{(\varepsilon)}_2 (s) \Vert^{}_{D^{}_N}\big),\cr
}$$
where $A''$ is defined in Lemma 4.4, and
$$\eqalignno{
&\sum_X  \Vert (\xi^{}_X A') (t),  {d \over dt}  \xi^{}_X
A' (t)) \Vert^{}_{M^\rho}& (4.88\hbox{c})\cr
&\qquad{}\leq C \Vert (f, \dot{f}) \Vert^{}_{M^\rho_1} +
(1+t)^{1/2 - \rho}  C_\rho
 \sum_{\varepsilon = \pm}  \sup_{0 \leq s}
 \big(\Vert \beta^{(\varepsilon)}_1
(s) \Vert^{}_{D^{}_N}  \Vert \beta^{(\varepsilon)}_2 (s)
\Vert^{}_{D^{}_N}\big),\cr
}$$
where the sum is taken over $X \in {\frak{sl}}(2, \Crm) \cap \Pi$.
Statement i) now follows
from Lemma 4.5 and inequalities (4.88).

It follows directly from (4.87a), Proposition 2.15 with
$n=0$ and inequality (4.3)
of Lemma 4.1 with $\vert \alpha \vert = 0$, that
$$\eqalignno{
&(1+t+\vert x \vert)^{3/2 - \rho}  \vert A'_\mu (t,x) \vert &(4.89)\cr
&\qquad{}\leq C_\rho
\Big(\Vert (f, \dot{f}) \Vert^{}_{M^\rho_2}
+ (1+t)^{1/2 - \rho}\sum_{\varepsilon = \pm}  \sup_{0 \leq s}
\big(\Vert \beta^{(\varepsilon)}_1 (s)
\Vert^{}_{D^{}_N}\Vert \beta^{(\varepsilon)}_2
(s) \Vert^{}_{D^{}_N}\big)\Big),\quad  t \geq 0.\cr
}$$
Statement ii) of the corollary follows from (4.89) and statement i) of Lemma
4.4. The proof of statements iii) and iv) is so similar to that of ii) that
we omit it.

To prove statement v) we note that
$$\eqalign{
{d \over dt} (G_{\varepsilon, \mu} (f_1, f_2)) (t) &= G_{\varepsilon, \mu}
(i \varepsilon \omega(-i \partial)  f_1 + \dot{f}_1, f_2)
+ G_{\varepsilon, \mu}  (f_1, i \varepsilon \omega(-i \partial) f_2 +
\dot{f}_2),\cr
\dot{f}_j (t) &= {d \over dt}  f_j (t),\quad  j = 1,2,\cr
}$$
and that
$$\partial_i  G_{\varepsilon, \mu} (f_1, f_2) = G_{\varepsilon, \mu}
 (\partial_i  f_1, f_2) + G_{\varepsilon, \mu}  (f_1,
\partial_i  f_2),\quad  1 \leq i \leq 3,$$
where $\varepsilon = \pm$,  $0 \leq \mu \leq 3$. These two equalities,
inequality (4.3) of Lemma 4.1 and equality (4.83) give, with
$l + \vert \alpha \vert =
n+n',  n \geq 1,  n' \geq 0$,
$$\eqalign{
&\vert \big({\partial \over \partial t}\big)^l  \partial^\alpha
\vartheta (A',t,x) \vert\cr
&\quad{}\leq \sum_{\varepsilon = \pm}  \sum_{\scr l_1+l_2+
\vert \alpha_1 \vert + \vert \alpha_2 \vert = n\atop\scr
l'_1+l'_2+ \vert \alpha'_1
\vert + \vert \alpha'_2 \vert \leq n'}
\int^1_0  s^{n+n'-1}  (1+st+\vert sx \vert)^{- (n+\chi-\delta)}
 (1+st)^{- \delta}  ds\cr
&\qquad{}C_{n,n',\chi,\delta}  \sup_{s' \geq 0}  \Big((1+s')^{l_1+l_2+\chi}
 \Vert \big({d\over ds}\big)^{l_1+l'_1}  \partial^{\alpha'_1}
 \beta^{(\varepsilon)}_1 (s') \Vert^{}_{D^{}_N}  \Vert \big({d
\over ds}\big)^{l_2+l'_2}  \partial^{\alpha'_2}  \beta^{(\varepsilon)}_2
(s') \Vert^{}_{D^{}_{N'}}\Big),\cr
}$$
where $\chi \geq 0$,  $\delta > 0$ and where $N$ depends on $n$. If
$\chi > 0$ then we choose $\delta = \chi / 2$.
The integral in the last estimate is then bounded by
$$\int^1_0  s^{n-1}  (1+s(t+\vert x \vert))^{-(n+\chi/2)}
 ds \leq C_\chi (1+t+\vert x \vert)^{-n},$$
which proves statement v) of the corollary for $\chi > 0$. If $\chi = 0$
then the integral is bounded by
$$\int^1_0  s^{n-1}  (1+s(t+\vert x \vert))^{-n+\delta}
ds \leq C'_\delta (1+t+\vert x \vert)^{-n+\delta}, \quad 0 < \delta < 1,$$
which proves the statement for $\chi = 0$. This proves the corollary.

In order to {\it construct approximate solutions of the M-D equations,}
we introduce the Banach space
${\cal D}^{d, \chi}_{n,j}$, $n \geq 0$, $j \geq 0$,
$d \geq 0$, $\chi \geq 0$ as the subspace of elements
$(\beta^{(+)}, \beta^{(-)}) \in C^n (\Rrm^+, D^{}_j \oplus D^{}_j)$
with finite norm
$$\eqalignno{
\Vert (\beta^{(+)}, \beta^{(-)}) \Vert^{}_{{\cal D}^{d, \chi}_{n,j}}
&= \sum_{\varepsilon = \pm}  \sum_{0 \leq l \leq n}  \Big(\sup_{t \geq 0}
\big((1+t)^{\chi + l}  \Vert \big({d \over dt}\big)^l  P_\varepsilon
(-i \partial)  \beta^{(\varepsilon)} (t) \Vert^{}_{D^{}_j}\big)&(4.90)\cr
&\qquad{}+ \sup_{t \geq 0}  \big((1+t)^{d + \chi + l}  \Vert \big({d
\over dt}\big)^l
 P_{- \varepsilon} (-i \partial)  \beta^{(\varepsilon)} (t)
 \Vert^{}_{D^{}_j}\big)\Big).\cr
}$$
We also introduce the Fr\'echet spaces ${\cal D}^{d, \chi}_{\infty, \infty} =
\displaystyle{\cap_{n \geq 0,  j \geq 0}}  {\cal D}^{d, \chi}_{n,j}$.

For $(f, \dot{f}) \in M^{\circ\rho}_\infty$, for $\beta_j =
(\beta^{(+)}_j, \beta^{(-)}_j)
\in {\cal D}^{0, \chi^{}_j}_{\infty, \infty}$,  $j=1,2$, and for $\beta_3 =
(\beta^{(+)}_3, \beta^{(-)}_3) \in {\cal D}^{d, \chi^{}_3}_{\infty, \infty}$
we introduce the polynomial
$((f, \dot{f}), \beta_1, \beta_2, \beta_3) \mapsto \lambda^{(\varepsilon)} (t)
((f, \dot{f}), \beta_1,
\beta_2, \beta_3)$ by the following expression, where we have omitted
the arguments
$(f, \dot{f}), \beta_1, \beta_2,\beta_3$:
$$\eqalignno{
&\lambda^{(\varepsilon)} (t)& (4.91)\cr
&\quad{}= i  e^{-i \varepsilon \omega(-i \partial) t}
\int^\infty_t  e^{i(t-s) {\cal D}}  (A'_\mu (s) + B_\mu (s))
 \gamma^0  \gamma^\mu  e^{i \varepsilon \omega(-i \partial)s}
 \beta^{(\varepsilon)}_3 (s)  ds,\quad t \geq 0,\varepsilon = \pm,\cr
}$$
where $A'$, defined in (4.87a), is a function of $(f, \dot{f}),
(\beta^{(+)}_1,
\beta^{(-)}_1)$ and $(\beta^{(+)}_2, \beta^{(-)}_2)$, and where
$B_\mu$ is defined
as in Lemma 4.5, i.e.
$$(B_0 (t)) (x) = - {\partial \over \partial t} \vartheta (A',t,x),
\quad (B_i (t)) (x) = - {\partial \over \partial x_i} \vartheta (A',t,x),
\quad 1 \leq i \leq 3.$$
The integral in (4.91) has to be interpreted in the sense of the strong
improper Riemann integral in $L^2$. It follows from the next proposition that
$\lambda^{(\varepsilon)}$ exists for certain choices of
$\chi^{}_1$, $\chi^{}_2$, $\chi^{}_3$
and $d$:
\saut
\noindent{\bf Proposition 4.7.}
{\it
Let $1/2 < \rho < 1$, $0 \leq d \leq 1$, $q\in \Nrm$ and let
$\chi = \chi^{}_1 + \chi^{}_2 + \chi^{}_3$, $\chi^{}_j
\geq 0$. Let $(f, \dot{f}) \in
M^{\circ\rho}_\infty$, $(\beta^{(+)}_j, \beta^{(-)}_j)
\in {\cal D}^{0, \chi^{}_j}_{\infty, \infty}$
for $j=1,2$, and let $(\beta^{(+)}_3, \beta^{(-)}_3)
\in {\cal D}^{d, \chi^{}_3}_{\infty, \infty}$.
Then

\noindent\hbox{\rm \ i)} If $(\beta^{(+)}_j, \beta^{(-)}_j) = 0$
for $j=1$ or $j=2$ and
$d - \chi^{}_3 > \rho - 1/2$ then
$$\eqalignno{
&\sum_{\scr Y\in \Pi'\cap U(\Rrm^4) \atop\scr \vert Y\vert \leq q}
\Vert (\xi^{}_Y (\lambda^{(+)}, \lambda^{(-)})
\Vert^{}_{{\cal D}^{d', \chi'}_{n,l}}\cr
&\qquad{}\leq C
\sum_{\scr Y_1,Y_2\in \Pi'\cap U(\Rrm^4)\atop\scr\vert Y_1\vert
+\vert Y_2\vert \leq q }
\Vert T^{M1}_{Y_1}(f, \dot{f}) \Vert^{}_{M^\rho_N}
\Vert \xi^{}_{Y_2} (\beta^{(+)}_3, \beta^{(-)}_3)
\Vert^{}_{{\cal D}^{d, \chi^{}_3}_{L,N}},
\quad n,l \geq 0,\cr
}$$
where $d' = 1 - d$,  $\chi' = 1/2 - \rho + \chi^{}_3 + d$ and $L = n + l + 1$.
Here $C$ depends on $n$, $l$, $d$, $q$ and $\chi^{}_3$, while
$N$ depends on $n$ and $l$.

\noindent\hbox{\rm ii)} If $(f, \dot{f}) = 0$ and
$\chi^{}_3 + d > 0$, then
$$\eqalign{
&\sum_{\scr Y\in \Pi'\cap U(\Rrm^4)\atop\scr\vert Y\vert \leq q}
\Vert \xi^{}_Y (\lambda^{(+)}, \lambda^{(-)})
\Vert^{}_{{\cal D}^{d', \chi'}_{n,l}}\cr
&\ \leq C
\sum_{\scr Y_i\in \Pi'\cap U(\Rrm^4)
\atop\scr\vert Y_1\vert+\vert Y_2\vert+\vert Y_3\vert \leq q }
\Vert\xi^{}_{Y_1} (\beta^{(+)}_1, \beta^{(-)}_1)
\Vert^{}_{{\cal D}^{0, \chi^{}_1}_{L,N}}
\Vert \xi^{}_{Y_2}(\beta^{(+)}_2, \beta^{(-)}_2)
\Vert^{}_{{\cal D}^{0, \chi^{}_2}_{L,N}}
\Vert \xi^{}_{Y_3}(\beta^{(+)}_3, \beta^{(-)}_3)
\Vert^{}_{{\cal D}^{0, \chi^{}_3}_{L,N}},\cr
}$$
where  $l,n\geq 0$, $d' = 1 - d$,  $\chi' = \chi + d - \delta$
and $L = n+l+1$. Here $C$
depends on $n$, $l$, $d$, $q$ and $\chi$, while $N$ depends on $n$ and $l$.
Moreover $\delta = 0$
if $\chi^{}_1 + \chi^{}_2 > 0$ and $\delta > 0$ if $\chi^{}_1 + \chi^{}_2 = 0$.
}\saut
\noindent{\it Proof.}
Since the case $q>0$ is so similar to the case $q=0$, we only consider $q=0$.
Let $W = \Pi \cup \{D\}$ be a basis of the $11$-dimensional
Weyl Lie algebra ${\frak w}$ satisfying
$$[D, P_\mu] = - P_\mu,\quad   [D, M_{\alpha \beta}] = 0, \quad  0 \leq \mu
\leq 3,  0 \leq \alpha < \beta \leq 3,$$
let $W_1$ be the corresponding basis of the $7$-dimensional subalgebra
${\frak w}_1 = {\frak{sl}}(2, \Crm)\oplus \Rrm$ and let
$$(\xi^{}_D f) (t,x) = t {\partial \over \partial t}
f(t,x) + \sum_{0 \leq i \leq 3}
 x_i {\partial \over \partial x_i}  f(t,x). \eqno{(4.93)}$$
Let $H \in C^\infty (\Rrm^+ \times \Rrm^3)$ and let $F_s (t,x) = H(s,s x/t)$,
$t > 0$. Using definition (4.81) of $\xi^{}_X$, $X \in \Pi$, and
definition (4.93) of $\xi^{}_D$, we obtain
$$\eqalignno{
(\xi^{}_{P_i} F_S) (t,x) &= {s \over t} (\xi^{}_{P_i} H)
(s, s x/t),\quad  1 \leq
i \leq 3,&(4.94\hbox{a})\cr
(\xi^{}_{P_0} F_s) (t,x) &= {s \over t} (\xi^{}_{P_0} H)
(s, s x/t) - {1 \over t}
(\xi^{}_D H) (s, s x/t),\cr
(\xi^{}_{M_{0i}} F_s) (t,x) &= (\xi^{}_{M_{0i}} H) (s, s x/t) - {x_i \over t}
(\xi^{}_D H) (s, s x/t),\quad  0 \leq i \leq 3,&(4.94\hbox{b})\cr
(\xi^{}_{M_{ij}} F_s) (t,x) &= (\xi^{}_{M_{ij}} H) (s, s x/t),\quad
1 \leq i < j \leq 3,&(4.94\hbox{c})\cr
\xi^{}_D  F_s &= 0.&(4.94\hbox{d})\cr
}$$
Repeated use of (4.94b) and (4.94c) shows that if
$Y \in \Pi' \cap U({\frak{sl}}(2, \Crm)$
then there are elements $Z \in W'$, the standard basis of the
enveloping algebra of ${\frak w}_1$ build on $W_1$, and polynomials
$q^{}_Z$ such that
$$(\xi^{}_Y F_s) (t,x) = \sum_{\scr  Z \in W'_1\atop\scr\vert Z \vert
\leq \vert Y \vert}
 q^{}_Z (x/t) (\xi^{}_Z H) (s, s x/t),\quad  s \geq 0,  t \geq 0.$$
Since
$$\big({d \over ds} F_s\big) (t,x) = s^{-1} (\xi^{}_D H) (s,s x/t),
\eqno{(4.94\hbox{e})}$$
we obtain
$$\Vert \big(\xi^{}_Y{d^q \over ds^q} F_s\big) (1, \cdot)
\Vert^{}_{L^\infty (B)}
\leq C_{\vert Y \vert, q}\ s^{-q} \sum_{\scr Z\in W'
\atop\scr \vert Z \vert \leq \vert Y \vert + q}
 \sup_{\vert x \vert \leq 1}  \vert (\xi^{}_Z H) (s, sx) \vert,
\eqno{(4.95\hbox{a})}$$
where $Y \in \Pi' \cap U({\frak{sl}}(2, \Crm))$ and $B$ is the unit
ball in $\Rrm^3$.

It follows by induction in $\vert X \vert + q \geq 0$, $q \geq 0$,
$X \in \Pi' \cap U(\Rrm^4)$, from (4.94a) and (4.94e), that
$$\eqalignno{
&\big(\xi^{}_X {d^q \over ds^q} F_s\big) (t,x)&(4.95\hbox{b})\cr
&\quad{}= \sum_{\scr\vert Y \vert + l \leq \vert X \vert
+ q\atop\scr  \vert Y \vert \leq \vert X \vert} (\xi^{}_Y\xi^l_D H)
(s,xs/t)t^{- \vert X \vert}  s^{\vert Y \vert - q}  C_{X,Y,l,q},
\quad s > 0,  t > 0,  Y \in \Pi' \cap U(\Rrm^4),\cr
}$$
for some constants $C_{X,Y,l,q}$. This gives, for $0 < t/2 \leq s \leq 2t$,
$$\Vert \big(\xi^{}_Y {d^q \over ds^q} F_s\big)
(t, \cdot) \Vert^{}_{L^\infty (\Rrm^3)} \leq
C_{\vert Y \vert,q}\ s^{-q}  \sum_{\scr  Z \in W'\atop\scr
\vert Z \vert \leq \vert Y \vert + q} \Vert (\xi^{}_Z H)
(s, \cdot) \Vert^{}_{L^\infty (\Rrm^3)},
\eqno{(4.96)}$$
where $Y \in \Pi' \cap U(\Rrm^4)$.

It follows from definition (1.19) of $\vartheta$ and by partial
integration, that
$$\vartheta (\xi^{}_D A', y) = y_{\mu}  A'^\mu (y) - \vartheta (A',y),$$
which, together with (1.20b), give
$$\eqalignno{
\xi^{}_D {\partial \over \partial y^\mu} \vartheta (A',y) &= - {\partial
\over \partial y^\mu}  \vartheta (A',y) + {\partial \over \partial y^\mu}
(y^\nu A'_\nu (y))&(4.97)\cr
&= {\partial \over \partial y^\mu} \vartheta  (\xi^{}_D A',y),\cr
}$$
$0 \leq \mu \leq 3$. If $\xi^M_{M_{\mu \nu}} =
\xi^{}_{M_{\mu \nu}} + n_{\mu \nu}$,
$0 \leq \mu < \nu \leq 3$, where $n_{\mu \nu}$ is as in (1.5) and if
$\xi^M_D = \xi^{}_D$, then we obtain as in (4.97) that
$$(\xi^M_X  B(y))_\mu = - {\partial \over \partial y^\mu}
\vartheta (\xi^M_X A',y),\quad  X \in {\frak w}_1,  0 \leq \mu \leq 3,$$
which extends to the enveloping algebra:
$$((\xi^M_Y B) (y))_\mu = - {\partial \over \partial y^\mu}\vartheta
(\xi^M_Y A',y),\quad  Y \in U({\frak w}_1),  0 \leq \mu \leq 3.
\eqno{(4.98)}$$
Let $(\beta^{(+)}_i, \beta^{(-)}_i) = 0$ for $i = 1$ or $i = 2$.
Then it follows from definition (4.87a) of $A'$, that $A'$ is a free
field with initial conditions
$(f, \dot{f}) \in M^{\circ\rho}_\infty$. Then $\xi^M_X  A'$  is also a free
field with initial conditions $T^{M1}_X (f, \dot{f}) \in M^{\circ\rho}_\infty,$
where $T^{M1}_X$ is defined by (1.5) if $X \in \Pi$ and $(T^{M1}_D (f,
\dot{f}))
(x) = (f, \dot{f}) (x) + \sum_{0 \leq i \leq 3} x_i
\partial_i (f, \dot{f}) (x)$. Let $H_\mu (f, \dot{f}) = A'_\mu + B_\mu.$ Then
it
follows from (4.98) and the fact that the initial condition for $\xi^M_Y
A'$,  $Y \in W'$, is $T^{M1}_Y (f, \dot{f}) \in M^\rho_\infty,$ that $\xi^M_Y
 H (f, \dot{f}) = H (T^{M1}_Y (f, \dot{f}))$.
This gives according to Proposition 2.15 and statement~iii) of Corollary 4.6
$$\vert (\xi^M_Y  H (f, \dot{f})) (x,t) \vert \leq C_\rho (1+\vert x \vert +
\vert t \vert)^{\rho - 3/2}  \Vert (f,
\dot{f}) \Vert^{}_{M^\rho_{3+\vert Y \vert}},
\quad Y \in W'.$$
By the definition of $\xi^M$, we obtain that
$$\sum_{\scr Z \in W'\atop\scr  \vert Z \vert \leq i}
\vert (\xi^{}_Z (A' + B)) (t,x) \vert
\leq C_\rho (1 + \vert x \vert + \vert t \vert)^{\rho - 3/2}  \Vert (f,
\dot{f}) \Vert^{}_{M^\rho_{3+i}},\quad i \geq 0,\eqno{(4.99)}$$
when $(\beta^{(+)}_j, \beta^{(-)}_j) = 0$ for $j=1$ or $j=2$.

When $(f, \dot{f}) = 0$ in definition (4.87a) of $A',$ then we estimate
$(\xi^{}_Z (A' + B)) (t,x)$ by inequality (4.3) of Lemma 4.1 and by
statement v) of Corollary 4.6 for $\vert x \vert \leq 2t.$ This gives,
with $P^\alpha = P^{\alpha_0}_0 P^{\alpha_1}_1  P^{\alpha_2}_2
P^{\alpha_3}_3$,
$$\eqalignno{
&\sum_{\scr Z \in W'\atop\scr  \vert Z \vert \leq i}
\vert (\xi^{}_Z (A' + B)) (t,x) \vert&(4.100)\cr
&\quad{}\leq C_i  \sum_{\vert \alpha \vert \leq i}
\vert t \vert^{\vert \alpha \vert}
 \vert (\xi^{}_{P^\alpha} (A' + B)) (t,x) \vert \cr
&\quad{}\leq C'_i (1+t)  \sum_{\scr \varepsilon = \pm\atop\scr  l_1+l_2 \leq i}
\sup_{0 \leq s \leq t}  \Big((1+s)^{l_1+l_2+\chi^{}_1+\chi^{}_2}
\Vert {d^{l_1} \over ds^{l_1}}
 \beta^{(\varepsilon)}_1 (s) \Vert^{}_{D^{}_N}  \Vert {d^{l_2} \over ds^{l_2}}
 \beta^{(\varepsilon)}_2 (s) \Vert^{}_{D^{}_N}\Big),\cr
}$$
$\vert x \vert \leq 2t$,  $t \geq 0$,
for some $N$ depending on $i \geq 0$, where $\delta = 0$ if
$\chi^{}_1 + \chi^{}_2 > 0$
and $\delta > 0$ if $\chi^{}_1 + \chi^{}_2 = 0$. In the last case
$C'$ depends on $\delta$.

Let $H = A' + B$. It follows from inequalities (4.99) and (4.100) that
$$\eqalignno{
&\sum_{\scr Z \in W'\atop\scr  \vert Z \vert \leq i} \vert (\xi^{}_Z H)
(t,x) \vert&(4.101)\cr
&\qquad{}\leq C_i \Big((1+t)^{\rho - 3/2}  \Vert (f, \dot{f})
\Vert^{}_{M^\rho_{3+i}}
+ (1+t)^{-1+\delta}\cr
&\qquad\qquad{}\sum_{\scr \varepsilon = \pm\atop\scr l_1 + l_2 \leq i}
\sup_{0 \leq s \leq t}  \big((1+s)^{l_1+l_2+\chi^{}_1+\chi^{}_2} \Vert
{d^{l_1} \over ds^{l_1}}  \beta^{(\varepsilon)}_1 (s) \Vert^{}_{D^{}_N} \Vert
{d^{l_2} \over ds^{l_2}}  \beta^{(\varepsilon)}_2 (s)
 \Vert^{}_{D^{}_N}\big)\Big),\cr
}$$
$\vert x \vert \leq 2t$,  $t \geq 0$,
for some $N$ depending on $i \geq 0$ and $C_i$ depending on $\rho$,
$1/2<\rho< 1$.

The function $\lambda^{(\varepsilon)}$,  $\varepsilon = \pm$ can now
be written as
$$\lambda^{(\varepsilon)} (t) = ie^{- i \varepsilon \omega (-i \partial) t}
\int^\infty_t  e^{i(t-s) {\cal D}} \sum_{0 \leq \mu \leq 3} F_{s, \mu}
(s)  \gamma^0  \gamma^\mu  e^{i \varepsilon \omega (-i \partial)s}
\beta^{(\varepsilon)}_3 (s)  ds, \eqno{(4.102)}$$
where $(F_{s, \mu} (t)) (x) = F_{s, \mu} (t,x) = H_\mu (s,s  x/t)$, $t > 0$.
According to Theorem A.3 there are bilinear forms
$g^{(\varepsilon)}_{s, \mu,j}$
of $F_{s, \mu}$ and $\beta^{(\varepsilon)}_3 (s)$ such that
$$\eqalignno{
&\big\Vert \big({d \over dt}\big)^p  \big({d \over ds}\big)^q
\big(e^{-i \varepsilon
\omega (-i \partial)t}  F_{s, \mu} (t)  e^{i \varepsilon \omega
(-i \partial)t} \beta^{(\varepsilon)}_3 (s) - \sum_{0 \leq j \leq n} t^{-j}
g^{(\varepsilon)}_{s, \mu, j}\big) \big\Vert^{}_{D^{}_L}&(4.103)\cr
&\qquad{}\leq C_{N,q}t^{-n-p-1}
\sum_{q^{}_1+q^{}_2=q} \Big(\sum_{\scr \vert \alpha
\vert \leq L\atop\scr a \leq p}
\Vert \big({d \over dt}\big)^a \big({d \over ds}\big)^{q^{}_1}
\partial^\alpha  F_{s, \mu} (t) \Vert^{}_{L^\infty (\Rrm^3)} \cr
&\qquad\qquad{}+ \sum_{Y \in D(N)}
\Vert \big(\xi^{}_Y \big({d \over ds}\big)^{q^{}_1}F_{s, \mu}
\big)(1, \cdot) \Vert^{}_{L^\infty (B)}\Big)
\Vert \big({d \over ds}\big)^{q^{}_2}
\beta^{(\varepsilon)}_3 (s) \Vert^{}_{D^{}_N},\cr
}$$
$t \geq 1$, $n \geq 1$,  $p \geq 0$,  $q \geq 0$,
where $D(N) = \{ Y \in \Pi' \cap U({\frak{sl}}(2, \Crm))\big\vert
\vert Y \vert \leq N \}$
and where $N$ depends on $n$, $p$ and $L$. Here we have used
the Leibniz rule for $(d/ds)^q$ and the fact that
$g^{(\varepsilon)}_{s, \mu,j}$ is a bilinear form
of $F_{s,\mu}$ and $\beta^{(\varepsilon)}_3 (s)$.
According to (A.36) and (A.37)
we have
$$\widehat{g^{(\varepsilon)}}_{s, \mu,0} (k) = F_{s,\mu} (1, -
{\varepsilon k/ \omega(k)})
(\beta^{(\varepsilon)}_3 (s))^\wedge (k) \eqno{(4.104)}$$
and
$$\eqalignno{
&\Vert \big({d \over ds}\big)^q  g^{(\varepsilon)}_{s, \mu,l}
\Vert^{}_{D^{}_j}&(4.105)\cr
&\qquad{}\leq C _{j,l,q} \sum_{\scr Y \in D(M)\atop\scr  q^{}_1+q^{}_2=q}
\Vert \big(\xi^{}_Y \big({d \over ds}\big)^{q^{}_1}
 F_{s, \mu}\big) (1, \cdot) \Vert^{}_{L^\infty (B)}
\Vert \big({d \over ds}\big)^{q^{}_2}
 \beta^{(\varepsilon)}_3 (s) \Vert^{}_{D^{}_M},\cr
}$$
where $M$ depends on $j$ and $l$.

We introduce for $s > 0$,  $q \geq 0$,  $N \geq 0$,
$1/2 < \rho < 3/2$, $\delta \geq0$, $\chi^{}_1 \geq 0$ and $\chi^{}_2 \geq 0$:
$$\eqalignno{
&\Gamma^{\rho, \chi^{}_1, \chi^{}_2,\delta}_{i,q,N} (s)
= \Big((1+s)^{\rho - 3/2} \Vert (f, \dot{f})
\Vert^{}_{M^\rho_{3+i+q}}+ (1+s)^{-1+\delta}&(4.106)\cr
&\qquad{}\sum_{\scr l_1+l_2 \leq i+q\atop \scr\varepsilon=\pm}
\sup_{0 \leq \tau \leq s}  \big((1+ \tau)^{l_1+l_2+\chi^{}_1+\chi^{}_2}
\Vert {d^{l_1} \over d \tau^{l_1}}  \beta^{(\varepsilon)}_1 (\tau)
\Vert^{}_{D^{}_N}  \Vert {d^{l_2} \over d \tau^{l_2}}  \beta^{(\varepsilon)}_1
(\tau) \Vert^{}_{D^{}_N}\big)\Big).\cr
}$$
It follows from (4.95) and (4.101) that
$$\Vert \big(\xi^{}_Y{d^q \over ds^q}  F_{s, \mu}\big)
(1, \cdot) \Vert^{}_{L^\infty (B)}
\leq C_{\vert Y \vert, q}  s^{-q}  \Gamma^{\rho,\chi^{}_1,\chi^{}_2,\delta}_
{\vert Y \vert,q,N} (s),\quad  s > 0, \eqno{(4.107)}$$
for $Y \in \Pi' \cap U ({\frak{sl}}(2, \Crm))$, and it follows from (4.96)
and (4.101) that
$$\Vert \big(\xi^{}_Y{d^q \over ds^q}  F_{s, \mu}\big) (t, \cdot)
\Vert^{}_{L^\infty
(\Rrm^3)} \leq C_{\vert Y \vert,q}  s^{-q}  \Gamma^{\rho,\chi^{}_1,
\chi^{}_2,\delta}_
{\vert Y \vert,q,N} (s), \eqno{(4.108)}$$
where $Y \in \Pi' \cap U (\Rrm^4)$ and $0 < t/2 \leq s \leq 2t$. Inequalities
(4.103), (4.107) and (4.108) give, since $\Gamma$ is increasing in $q$,
$$\eqalignno{
&\Vert {d^p \over dt^p}  {d^q \over ds^q} \Big(e^{-i \varepsilon
\omega (-i \partial)t}
F_{s, \mu} (t)  e^{i\varepsilon \omega (-i \partial)t}
\beta^{(\varepsilon)}_3 (s) - \sum_{0 \leq j \leq n}  t^{-j}
g^{(\varepsilon)}_{s, \mu,j}\Big) \Vert^{}_{D^{}_L}&(4.109)\cr
&\qquad{}\leq C_{n,L,p,q}  s^{-n-p-q-1}
\Gamma^{\rho,\chi^{}_1,\chi^{}_2,\delta}_{N,q,N} (s)
\sum_{q' \leq q}  (1+s)^{q'}  \Vert {d^{q'} \over ds^{q'}}
\beta^{(\varepsilon)}_3 (s) \Vert^{}_{D^{}_N},\cr
}$$
where $t \geq 1$, $ t/2 \leq s \leq 2t$,  $p \geq 0$,
$q \geq 0$,  $n \geq 1$, $ L \geq 0$ and where $N$, depending on
$n$, $p$, $q$ and $L$, is sufficiently large. Inequalities
(4.105) and (4.107) give
$$\Vert {d^q \over ds^q}  g^{(\varepsilon)}_{s, \mu,j} \Vert^{}_{D^{}_l} \leq
C_{l,j,q}  s^{-q}  \Gamma^{\rho,\chi^{}_1,\chi^{}_2,\delta}_{M,q,M} (s)
\sum_{q' \leq q}  (1+s)^{q'}  \Vert {d^{q'} \over ds^{q'}}
\beta^{(\varepsilon)}_3 (s) \Vert^{}_{D^{}_M},
\eqno{(4.110)}$$
where $s > 0$,  $q \geq 0$,  $j \geq 0$,  $l \geq 0$ and
where $M$, depending on $l$, $j$, $q$, is sufficiently large.

It follows from (4.109) from the definition of $F_{\mu,s} (t)$
(cf. (4.102)) and
from Leibniz rule that
$$\eqalignno{
&\Vert {d^p \over ds^p}  \Big(e^{-i \varepsilon \omega (-i \partial) s}
 H_\mu (s)  e^{i \varepsilon \omega (-i \partial) s}
\beta^{(\varepsilon)}_3 (s) - \sum_{0 \leq j \leq n}  s^{-j}
g^{(\varepsilon)}_{s,\mu,j}\Big) \Vert^{}_{D^{}_L}&(4.111)\cr
&\qquad{}\leq C_{n,L,p}  s^{-n-p-1}
\Gamma^{\rho,\chi^{}_1,\chi^{}_2,\delta}_{N,P,N}(s)
\sum_{q \leq p}  (1+s)^q  \Vert {d^q \over ds^q}
\beta^{(\varepsilon)}_3 (s) \Vert^{}_{D^{}_N},\cr
}$$
where $s \geq 1$,  $p \geq 0$,  $n \geq 1$, $L \geq 0$
and where $N$, depending on $n$, $p$ and $L$, is sufficiently large.

Since $H_\mu = A'_\mu + B_\mu$, we have according to (1.20b) and (4.92) that
$t H_0 (t,x) + \sum_{1 \leq i \leq 3}$ $x_i  H_i (t,x) = 0$.
The formula (see statement (i) of Lemma 3.13 in \refFST)
$$\sum_{0 \leq \mu \leq 3}  P_\varepsilon (k)  \gamma^0
 \gamma^\mu  f_\mu  P_\varepsilon (k) = \Big(f_0 +
\sum_{1 \leq i \leq 3}  {- \varepsilon k_i \over \omega (k)}
f_i\Big)  P_\varepsilon  (k),
\quad f_\mu \in  \Crm,  0 \leq \mu \leq 3, \eqno{(4.112)}$$
then gives that
$$\sum_{0 \leq \mu \leq 3}  P_\varepsilon  (k)
\gamma^0  \gamma^\mu  H_\mu (t, - {\varepsilon kt /\omega (k)})
P_\varepsilon  (k) = 0,\quad  t \geq 0,
 k \in \Rrm^3. \eqno{(4.113)}$$
It follows from (4.104), the definition of $F_{s,\mu}$ and (4.113) that
$$\sum_{0 \leq \mu \leq 3}  P_\varepsilon  (-i \partial)
\gamma^0  \gamma^\mu g^{(\varepsilon)}_{s,\mu,0} = \sum_{0 \leq \mu \leq 3}
 P_\varepsilon  (-i \partial)  \gamma^0
\gamma^\mu  h^{(\varepsilon)}_{s,\mu,0}, \eqno{(4.114)}$$
where
$$(h^{(\varepsilon)}_{s,\mu,0})^\wedge  (k) = F_{s,\mu} (1, - {\varepsilon k
/ \omega (k)})  (P_{- \varepsilon} (-i \partial)  \beta^{(\varepsilon)}_3
(s))^\wedge  (k). \eqno{(4.115)}$$
Since $h^{(\varepsilon)}_{s,\mu,0}$ is the value of the bilinear
form $g^{(\varepsilon)}_{s,\mu,0}$
applied to $F_{s,\mu}$ and $P_{- \varepsilon}  (-i \partial)
\beta^{(\varepsilon)}_3 (s),$ inequality (4.110) gives
$$\eqalignno{
&\Vert {d^p \over ds^p}  h^{(\varepsilon)}_{s,\mu,0} \Vert^{}_{D^{}_l}\cr
&\qquad{}\leq C_{l,p}  s^{-p}  \Gamma^{\rho,\chi^{}_1,\chi^{}_2,
\delta}_{M,p,M} (s)
\sum_{q \leq p}  (1+s)^q  \Vert {d^q \over ds^q}
P_{- \varepsilon}  (-i \partial)  \beta^{(\varepsilon)}_3 (s)
\Vert^{}_{D^{}_M},\quad  s > 0,\cr
}$$
where $M$ depends on $l \leq 0$ and $p \geq 0$. By definitions (4.90)
of the norm in ${\cal D}^{d,\chi}_{n,j}$, we obtain from this inequality
and inequality (4.114)
that
$$\eqalignno{
&\Vert {d^p \over ds^p}  P_\varepsilon  (-i \partial)
\sum_{0 \leq \mu \leq 3}  \gamma^0  \gamma^\mu
g^{(\varepsilon)}_{s,\mu,0} \Vert^{}_{D^{}_l}& (4.116)\cr
&\quad{}\leq C_{l,p} \Big(s^{\rho - 3/2 - d-p-\chi^{}_3}
\Vert (f, \dot{f}) \Vert^{}_{M^\rho_N}
 \Vert (\beta^{(+)}_3, \beta^{(-)}_3) \Vert^{}_{{\cal D}^{d,
 \chi^{}_3}_{p,N}}\cr
&\qquad{}+ s^{-1-d-p+\delta-\chi^{}_3} \Vert (\beta^{(+)}_1,
\beta^{(-)}_1)
\Vert^{}_{{\cal D}^{0,\chi^{}_1}_{p,N}}
 \Vert (\beta^{(+)}_2, \beta^{(-)}_2)
 \Vert^{}_{{\cal D}^{0,\chi^{}_2}_{p,N}}
 \Vert (\beta^{(+)}_3, \beta^{(-)}_3)
 \Vert^{}_{{\cal D}^{d,\chi^{}_3}_{p,N}}\Big),\quad s>0,\cr
}$$
where $N$ depends on $p \geq 0$ and $l \geq 0$.

It follows from (4.110) that
$$\eqalignno{
\Vert {d^p \over ds^p}  g^{(\varepsilon)}_{s,\mu,j} \Vert^{}_{D^{}_l}
&\leq C_{l,j,p} \Big(s^{\rho-3/2-p-\chi^{}_3}  \Vert (f, \dot{f})
\Vert^{}_{M^\rho_N}  \Vert (\beta^{(+)}_3, \beta^{(-)}_3)
\Vert^{}_{{\cal D}^{0,\chi^{}_3}_{p,N}}&(4.117)\cr
&\qquad{}+ s^{-1-p+\delta-\chi^{}_3}  \prod_{1 \leq i \leq 3}
\Vert (\beta^{(+)}_i,
\beta^{(-)}_i) \Vert^{}_{{\cal D}^{0,\chi^{}_i}_{p,N}}\Big),\quad  s > 0,\cr
}$$
where $N$ depends on $p \geq 0$,  $j \geq 0$ and $l \geq 0$. Proposition
2.15, (4.3) of Lemma 4.1 and statement iv) of Corollary 4.6 show that
$$\eqalignno{
&\Vert {d^p \over ds^p}  e^{-i \varepsilon \omega (-i \partial)s}
 H_\mu (s)  e^{i \varepsilon \omega (-i \partial)s} \Vert^{}_{D^{}_l}
&(4.118)\cr
&\qquad{}\leq C_{l,p}   \Big(\Vert (f, \dot{f}) \Vert^{}_{M^\rho_N}  \Vert
\beta^{(+)}_3, \beta^{(-)}_3) \Vert^{}_{{\cal D}^{0,0}_{p,N}}
+ \prod_{1 \leq i \leq 3}  \Vert (\beta^{(+)}_i, \beta^{(-)}_i)
\Vert^{}_{{\cal D}^{0,0}_{p,N}}\Big),
\quad 0 \leq s \leq 2,\cr
}$$
where $N$ depends on $p \geq 0$ and $l \geq 0$.

It follows from inequalities (4.106), (4.111) with $n=1$, (4.116), (4.117) with
$j=1$ and (4.118), that
$$\eqalignno{
&\Vert {d^p \over ds^p}  P_\varepsilon (-i \partial)  \sum_{0 \leq \mu \leq 3}
 \gamma^0  \gamma^\mu  e^{-i \varepsilon \omega (-i \partial)s}
 H_\mu (s)  e^{i \varepsilon \omega (-i \partial)s}
\beta^{(\varepsilon)}_3 (s) \Vert^{}_{D^{}_l}&(4.119)\cr
&\qquad{}\leq C_{l,p} \Big((1+s)^{\rho -3/2-d-p-\chi^{}_3}
\Vert (f, \dot{f}) \Vert^{}_{M^\rho_N}
 \Vert (\beta^{(+)}_3, \beta^{(-)}_3)
 \Vert^{}_{{\cal D}^{d,\chi^{}_3}_{p,N}}\cr
&\qquad\quad{}+ (1+s)^{-d-1-p-\chi^{}_3+\delta}  \Vert (\beta^{(+)}_1,
\beta^{(-)}_1)
\Vert^{}_{{\cal D}^{0, \chi^{}_1}_{p,N}}
\Vert (\beta^{(+)}_2, \beta^{(-)}_2)
\Vert^{}_{{\cal D}^{0, \chi^{}_2}_{p,N}}
\Vert (\beta^{(+)}_3, \beta^{(-)}_3)
\Vert^{}_{{\cal D}^{d, \chi^{}_3}_{p,N}}\Big),\cr
}$$
$s \geq 0$,  $\chi = \chi^{}_1 + \chi^{}_2 + \chi^{}_3$,
where $N$ depends on
$p \geq 0$ and $l \geq 0$.
It follows from inequalities (4.106), (4.111) with $n=0$,
(4.117) with $j=0$
and (4.118), that
$$\eqalignno{
&\Vert {d^p \over ds^p}  P_{- \varepsilon} (-i \partial)
\sum_{0 \leq \mu \leq 3}  \gamma^0  \gamma^\mu
e^{-i \varepsilon \omega (-i \partial) s} H_\mu (s)  e^{i \varepsilon \omega
(-i \partial) s}  \beta^{(\varepsilon)}_3 (s) \Vert^{}_{D^{}_l}&(4.120)\cr
&\qquad{}\leq C_{l,p} \Big((1+s)^{\rho-3/2-p-\chi^{}_3}
\Vert (f, \dot{f}) \Vert^{}_{M^\rho_N}
 \Vert (\beta^{(+)}_3, \beta^{(-)}_3) \Vert^{}_{{\cal D}^{0,\chi^{}_3}_{p,N}}
\cr
&\qquad\quad{}+ (1+s)^{-1-p-\chi^{}_3+\delta}   \prod_{1 \leq i \leq 3}  \Vert
(\beta^{(+)}_i, \beta^{(-)}_i) \Vert^{}_{{\cal D}^{0,\chi^{}_i}_{p,N}}\Big),
\quad s \geq 0,  \chi = \chi^{}_1 + \chi^{}_2 + \chi^{}_3,\cr
}$$
where $N$ depends on $p \geq 0$ and $l \geq 0$.

Since $P_\varepsilon (-i \partial)  {\cal D} = i \varepsilon \omega (-i
\partial)
 P_\varepsilon (-i \partial)$ (cf. (1.2b) and (1.16)), we obtain according
to (4.102) that
$$\eqalignno{
&P_\varepsilon (-i \partial)  \lambda^\varepsilon (t)&(4.121\hbox{a})\cr
&\qquad{}= i \int^\infty_t  P_\varepsilon (-i \partial)  \sum_{0 \leq \mu \leq
3}
 \gamma^0  \gamma^\mu  e^{-i \varepsilon \omega (-i \partial) s}
 H_\mu (s)  e^{i \varepsilon \omega (-i \partial) s}
\beta^{(\varepsilon)}_3 (s)  ds\cr
\noalign{\hbox{and}}
&P_{- \varepsilon} (-i \partial)  \lambda^{(\varepsilon)} (t)&
(4.121\hbox{b})\cr
&\qquad{}= i \int^\infty_t  e^{- 2i \varepsilon \omega (-i \partial) (t-s)}
P_{- \varepsilon} (-i \partial)  \sum_{0 \leq \mu \leq 3}
\gamma^0  \gamma^\mu  e^{- i \varepsilon \omega (-i \partial) s}
 H_\mu (s)  e^{i \varepsilon \omega (-i \partial) s}
\beta^{(\varepsilon)}_3 (s)  ds,\cr
}$$
$t \geq 0$. It follows from (4.119) and (4.121a) that:
$$\eqalignno{
\hskip5mm\hbox{a)}\ & \hbox{if $(\beta^{(+)}_i, \beta^{(-)}_i) = 0$
for $i=1$ or $i=2$, then}\cr
&\Vert {d^p \over dt^p}  P_\varepsilon (-i \partial)  \lambda^{(\varepsilon)}
(t) \Vert^{}_{D^{}_l}&(4.122{\rm a}) \cr
&\quad{}\leq C_{l,p} (1+t)^{\rho-1/2-d-p-\chi^{}_3}     \Vert (f, \dot{f})
\Vert^{}_{M^\rho_N}  \Vert (\beta^{(+)}_3, \beta^{(-)}_3)
\Vert^{}_{{\cal D}^{d,\chi^{}_3}_{p,N}},\cr
\hbox{b)}\ & \hbox{if $(f, \dot{f}) = 0$, then}\cr
&\Vert {d^p \over dt^p}  P_\varepsilon (-i \partial)  \lambda^{(\varepsilon)}
(t) \Vert^{}_{D^{}_l}&(4.122{\rm b})\cr
&{}\leq C_{l,p} (1+t)^{-1-d-p-\chi^{}_3+\delta}
\Vert (\beta^{(+)}_1, \beta^{(-)}_1)
\Vert^{}_{{\cal D}^{0,\chi^{}_1}_{p,N}}  \Vert (\beta^{(+)}_2, \beta^{(-)}_2)
\Vert^{}_{{\cal D}^{0,\chi^{}_2}_{p,N}}  \Vert (\beta^{(+)}_3, \beta^{(-)}_3)
\Vert^{}_{{\cal D}^{d,\chi^{}_3}_{p,N}},\cr
&\hbox{where $t \geq 0,  \chi = \chi^{}_1 + \chi^{}_2 + \chi^{}_3$
and $N$ depends on $p \geq 0$ and $l \geq 0.$}\cr
}$$
Let
$$q^{}_\varepsilon (t) = P_{- \varepsilon} (-i \partial)
\sum_{0 \leq \mu \leq 3}
 \gamma^0  \gamma^\mu  e^{- i \varepsilon \omega
(-i \partial)t}  H_\mu (t)  e^{i \varepsilon \omega (-i \partial) t}
 \beta^{(\varepsilon)}_3 (t),\quad t \geq 0. \eqno{(4.123)}$$
Repeated  integration by parts  gives according to (4.121b)
$$\eqalignno{
&{d^p \over dt^p}  P_{- \varepsilon} (-i \partial)  \lambda^{(\varepsilon)}
(t)& (4.124)\cr
&\qquad{}= i  \sum_{0 \leq j \leq n}  (-2i \varepsilon
\omega (-i \partial))^{-j-1}
 {d^{j+p} \over dt^{j+p}}  q^{}_\varepsilon (t)\cr
&\qquad\qquad{}+ i  \int^\infty_t  (-2i \varepsilon
\omega (-i \partial))^{-n-1}
 e^{-2i \varepsilon \omega (-i \partial) (t-s)}  {d^{n+p+1}
\over ds^{n+p+1}}  q^{}_\varepsilon (s)  ds, \quad t \geq 0,
 p \geq 0,n \geq 0.\cr
}$$
It follows from the definition of the norm $\Vert\cdot  \Vert^{}_{D^{}_l}$,
that
$$\Vert e^{i \varepsilon \omega (-i \partial) t}  \alpha \Vert^{}_{D^{}_l}
\leq C_l (1+t)^l  \Vert \alpha \Vert^{}_{D^{}_l}, \quad l \geq 0.$$
Let $n \geq l$, then (4.124) gives
$$\eqalignno{
&\Vert {d^p \over dt^p}  P_{- \varepsilon} (-i \partial)
\lambda^{(\varepsilon)} (t) \Vert^{}_{D^{}_l}&(4.125)\cr
&\ {}\leq C_l  \sum_{0 \leq j \leq n}
 \Vert {d^{j+p} \over dt^{j+p}}  q^{}_\varepsilon (t) \Vert^{}_{D^{}_l}
+ C_l  \int^\infty_t (1+s)^l    \Vert {d^{n+p+1} \over ds^{n+p+1}}
 q^{}_\varepsilon (s) \Vert  ds,\quad
t \geq 0, p \geq 0,  n \geq l \geq 0.\cr
}$$
According to inequality (4.120) and definition (4.123) of
$q^{}_\varepsilon$ we obtain
from the last inequality with $n=l$, that:
$$\eqalignno{
\hskip8mm\hbox{c)}\ & \hbox{if $(\beta^{(+)}_i, \beta^{(-)}_i) = 0$
for $i=1$ or $i=2$, then}\cr
&\Vert {d^p \over dt^p}  P_{- \varepsilon} (-i \partial)
\lambda^{(\varepsilon)} (t) \Vert^{}_{D^{}_l}&(4.126{\rm a})\cr
&\qquad\leq C_{l,p} (1+t)^{\rho-3/2-p-\chi^{}_3}
\Vert (f,\dot{f}) \Vert^{}_{M^\rho_N}
 \Vert (\beta^{(+)}_3, \beta^{(-)}_3)
 \Vert^{}_{{\cal D}^{0, \chi^{}_3}_{p+l+1,N}},\cr
\hbox{d)}\ & \hbox{if $(f, \dot{f}) = 0$, then}\cr
&\Vert {d^p \over dt^p}  P_{- \varepsilon} (-i \partial)
\lambda^{(\varepsilon)} (t) \Vert^{}_{D^{}_l}&(4.126{\rm b})\cr
&\qquad\leq C_{l,p} (1+t)^{-1-p-\chi}  \prod_{1 \leq i \leq 3}
\Vert (\beta^{(+)}_i, \beta^{(-)}_i) \Vert^{}_{{\cal D}^{0,
\chi^{}_i}_{p+l+1,N}},\cr
&\hbox{where $t \geq 0$,  $\chi = \chi^{}_1 + \chi^{}_2 + \chi^{}_3$
and $N$ depends on
$p \geq 0$ and $l \geq 0$.}\hskip20mm
}$$
It follows from (4.122a) and (4.126a) that if $(\beta^{(+)}_i,
\beta^{(-)}_i) = 0$
for $i=1$ or $i=2$, then
$$\eqalignno{
&\sum_{\varepsilon = \pm}  \sum_{0 \leq p \leq n}  \Big( \sup_{t \geq 0}
 \big((1+t)^{1/2-\rho+d+p+\chi^{}_3}  \Vert {d^p \over dt^p}
P_\varepsilon (-i \partial)  \lambda^{(\varepsilon)} (t)
\Vert^{}_{D^{}_l}\big)&(4.127{\rm a})\cr
&\qquad{}+ \sup_{t \geq 0}  \big((1+t)^{3/2-\rho+p+\chi^{}_3}  \Vert {d^p
\over dt^p}  P_{- \varepsilon} (-i \partial)  \lambda^{(\varepsilon)}
(t) \Vert^{}_{D^{}_l}\big)\Big)\cr
&\qquad\qquad{}\leq C_{l,n}     \Vert (f, \dot{f}) \Vert^{}_{M^\rho_N}  \Vert
(\beta^{(+)}_3, \beta^{(-)}_3) \Vert^{}_{{\cal D}^{d, \chi^{}_3}_{n+l+1,N}},
\quad\chi^{}_3 + d > \rho - 1/2,\cr
}$$
where $N$ depends on $n \geq 0$ and $l \geq 0$. This proves statement i) of the
proposition.

It follows from (4.122b) and (4.126b) that, if $(f, \dot{f}) = 0$, then
$$\eqalignno{
&\sum_{\varepsilon = \pm}  \sum_{0 \leq p \leq n}  \Big(\sup_{t \geq 0}
 \big((1+t)^{d+p+\chi^{}_3-\delta}  \Vert {d^p \over dt^p}
P_\varepsilon (-i \partial)  \lambda^{(\varepsilon)} (t)
\Vert^{}_{D^{}_l}\big)&(4.127{\rm b})\cr
&\qquad{}+ \sup_{t \geq 0}
\big((1+t)^{1+p+\chi^{}_3-\delta}\Vert {d^p \over dt^p}
 P_{- \varepsilon} (-i \partial)\lambda^{(\varepsilon)} (t)
\Vert^{}_{D^{}_l}\big)\Big) \cr
&\qquad\qquad{}\leq C_{l,n}
\Vert (\beta^{(+)}_1, \beta^{(-)}_1)
\Vert^{}_{{\cal D}^{0,\chi^{}_1}_{l+n+1,N}}
 \Vert (\beta^{(+)}_2, \beta^{(-)}_2)
 \Vert^{}_{{\cal D}^{0,\chi^{}_2}_{l+n+1,N}}
 \Vert (\beta^{(+)}_3, \beta^{(-)}_3)
 \Vert^{}_{{\cal D}^{d,\chi^{}_3}_{l+n+1,N}},\cr
}$$
where $\chi+d > 0$ and $N$ depends on $n \geq 0$ and $l \geq 0$. This proves
statement ii) of the proposition.
\smallskip

As in \refFST\ (see formula (3.34) in \refFST) we can now construct
an approximate solution of the M-D equations by iterating the
functional $\lambda = (\lambda^{(+)},
\lambda^{(-)})$ a finite number of times. More precisely, let $(f, \dot{f}) \in
M^{\circ\rho}_\infty$,  $1/2 < \rho < 1$, $\alpha \in D^{}_\infty$
and let $\beta_n = (\beta^{(+)}_n, \beta^{(-)}_n)$,  $n \geq 0$, be the
sequence given by
$$\eqalignno{
\beta^{(\varepsilon)}_0 (t) &= P_\varepsilon (-i \partial) \alpha,
\quad \varepsilon = \pm,  t \geq 0,& (4.128{\rm a})\cr
\beta^{(\varepsilon)}_{n+1} (t) &= \beta^{(\varepsilon)}_0 (t) +
(\lambda^{(\varepsilon)}
 ((f, \dot{f}),  \beta_n, \beta_n, \beta_n)) (t),
\quad\varepsilon = \pm,  t \geq 0,& (4.128{\rm b})\cr
}$$
where $\lambda^{(\varepsilon)}$ is given by (4.91). We remind that
$M^{\circ\rho}_\infty$
is the subspace of $M^\rho_\infty$ satisfying the gauge condition (4.87b).
\saut
\noindent{\bf Proposition 4.8.}
{\it
Let $1/2 < \rho < 1$,  $Y \in \Pi' \cap U(\Rrm^4)$
and let the sequence $(\chi^{}_n, d_n)$,  $n \geq 0$, be defined by
$$\eqalignno{
&\chi^{}_{2p} = 2p(1-\rho),\quad  d_{2p} = 1-2p(1-\rho)\quad
\hbox{for $2(p-1) (1-\rho) < \rho - 1/2$, and $p \geq 0$,}\cr
&\chi^{}_{2p+1} = 3/2-\rho,\quad  d_{2p+1} = 2p(1-\rho)\quad
\hbox{for $2p(1-\rho) < \rho- 1/2$ and $p \geq 0$,}\cr
}$$
and $\chi^{}_n = 3/2-\rho$,  $d_n = \rho - 1/2$ otherwise. This sequence
satisfies $\chi^{}_n \geq 0$,  $0 \leq d_n \leq 1$ and
$\chi^{}_n + d_n > \rho-1/2$.
If $u = (f, \dot{f}, \alpha) \in E^{\circ\rho}_\infty$
then the element $\beta_n$,
 $n \geq 0$, in the sequence given by (4.128a) and (4.128b) defines a
continuous polynomial $u \mapsto \beta_n$ from
$E^{\circ\rho}_\infty$ to ${\cal D}^{3/2-\rho, 0}_{\infty, \infty}$
which satisfies, with $n \geq 0$,  $j \geq 0$,
 $l \geq 0$,  $\vert Y \vert \geq 0$,
\psaut
\noindent{\hbox{\rm  \phantom{i}i)}}
$\Vert \xi^{}_Y  \beta_n \Vert^{}_{{\cal D}^{3/2-\rho, 0}_{j,l}} \leq C
(1+ \Vert u \Vert^{}_{E^\rho_N})^N
\Vert u \Vert^{}_{E^\rho_{N+\vert Y \vert}},$
\psaut
\noindent{\hbox{\rm ii)}}
$\Vert \xi^{}_Y  (\beta_{n+1} - \beta_n)
\Vert^{}_{{\cal D}^{d_n,\chi^{}_n}_{j,l}}
\leq C (1+ \Vert u \Vert^{}_{E^\rho_N})^N  \Vert u \Vert^{n+1}_{E^\rho_N}
 \Vert u \Vert^{}_{E^\rho_{N+\vert Y \vert}}$,
\psaut
\noindent
where $C$ depends on $j$, $l$, $n$, $\rho$,$\vert Y \vert$,
and $N$ depends on $j$, $l$ and
$n$. Moreover if $\alpha = 0$, then $\beta_n = 0$ for $n\geq 0$.
}\saut
\noindent{Proof.}
We prove the two statements by induction. It follows from definition (4.90)
of
the norm in ${\cal D}^{d,\chi}_{j,l}$ and from definition
(4.128a) of $\beta_0$,
that
$$\Vert \xi^{}_Y  \beta_0 \Vert^{}_{{\cal D}^{3/2-\rho, 0}_{j,l}} \leq \Vert
\alpha \Vert^{}_{D^{}_{l+\vert Y \vert}} \leq
\Vert u \Vert^{}_{E^\rho_{l+\vert Y \vert}},
\quad Y \in \Pi' \cap U(\Rrm^4) \eqno{(4.129)}$$
since $T^{D1}_X$ commutes with $P_\varepsilon (-i \partial),  X \in \p$.
This proves that statement i) is true for $n = 0$.
Suppose it is true for $n \geq 0$.
According to definition (4.128b) of $\beta_{n+1}$
and according to (4.129) we have
$$\eqalignno{
\Vert \xi^{}_Y  \beta_{n+1} \Vert^{}_{{\cal D}^{3/2-\rho, 0}_{j,l}}
&\leq\Vert \alpha \Vert^{}_{D^{}_{l+\vert Y \vert}} + \Vert \xi^{}_Y
 \lambda((f, \dot{f}), \beta_n, \beta_n, \beta_n) \Vert^{}_{{\cal
 D}^{3/2-\rho, 0}_{j,l}}\cr &\leq \Vert u \Vert^{}_{E^\rho_{l+\vert Y
 \vert}} + \Vert \xi^{}_Y \lambda ((f, \dot{f}), \beta_n, \beta_n, \beta_n)
 \Vert^{}_{{\cal D}^{\rho-1/2, 2(1-\rho)}_{j,l}},\cr }$$ since it follows
 from definition (4.90) that ${\cal D}^{d, \chi}_{j,l}
\subset  {\cal D}^{d', \chi'}_{j,l}$ (topologically) if $\chi \geq \chi'$ and
$\chi + d \geq d' + \chi'$.
It follows from the last inequality and from statements i)
and ii) of Proposition 4.7, since $3/2 - \rho > \rho - 1/2 > 0$, that
$$\eqalignno{
&\Vert \xi^{}_Y  \beta_{n+1}
\Vert^{}_{{\cal D}^{3/2-\rho, 0}_{j,l}}&(4.130)\cr
&\qquad{}\leq
\Vert u \Vert^{}_{E^\rho_{l+\vert Y \vert}} + C'
\sum_{\vert Y_1 \vert + \vert Y_2 \vert \leq \vert Y \vert}
\Vert (f, \dot{f}) \Vert^{}_{M^\rho_{N'+\vert Y_1
\vert}} \Vert \xi^{}_{Y_2} \beta_n
\Vert^{}_{{\cal D}^{3/2-\rho, 0}_{L,N'}}\cr
&\qquad\qquad{}+ C' \sum_{\vert Y_1 \vert + \vert Y_2
\vert + \vert Y_3 \vert \leq \vert Y \vert}
 \Vert \xi^{}_{Y_1} \beta_n \Vert^{}_{{\cal D}^{0,0}_{L,N'}}  \Vert
\xi^{}_{Y_2} \beta_n \Vert^{}_{{\cal D}^{0,0}_{L,N'}}
\Vert \xi^{}_{Y_3} \beta_n
\Vert^{}_{{\cal D}^{3/2-\rho, 0}_{L,N'}}\cr
&\qquad{}\leq \Vert u \Vert^{}_{E^\rho_{N'+\vert Y \vert}} +
C' \sum_{\vert Y_1 \vert +  \vert Y_2 \vert \leq \vert Y \vert}
\Vert u \Vert^{}_{E^\rho_{N'+\vert Y_1 \vert}}
 \Vert \xi^{}_{Y_2} \beta_n \Vert^{}_{{\cal D}^{3/2-\rho, 0}_{L,N'}}\cr
&\qquad\qquad{}+ C' \sum_{\vert Y_1 \vert + \vert Y_2 \vert + \vert
Y_3 \vert\leq \vert Y\vert}
\Vert \xi^{}_{Y_1} \beta_n \Vert^{}_{{\cal D}^{3/2-\rho, 0}_{L,N'}}  \Vert
\xi^{}_{Y_2} \beta_n \Vert^{}_{{\cal D}^{3/2-\rho, 0}_{L,N'}}
\Vert \xi^{}_{Y_3} \beta_n \Vert^{}_{{\cal D}^{3/2-\rho, 0}_{L,N'}},\cr
}$$
where the summation is over $Y_j \in \Pi' \cap U(\Rrm^4)$, where $L$ and $N'$
depend on $j$, $l$, $n$ and where $C'$ depends on $j$, $l$, $n$, $\rho$.
It now follows from the induction hypothesis and from (4.130) with $L$ and $N'$
instead of $j$ and $L$ and, by choosing $N''$ sufficiently large, that
$$\eqalignno{
\Vert \beta_{n+1} \Vert^{}_{{\cal D}^{3/2-l, 0}_{j,l}}
 &\leq \Vert \alpha \Vert^{}_{D^{}_{N'}}
+ C' \big(\Vert u \Vert^{}_{E^\rho_{N'}} + C^2 (1+
\Vert u \Vert^{}_{E^\rho_N})^{2N}
 \Vert \alpha \Vert^{}_{D^{}_N}\big)  C (1+ \Vert u \Vert^{}_{E^\rho_N})^N
 \Vert \alpha \Vert^{}_{D^{}_N}\cr
&\leq C'' (1+ \Vert u \Vert^{}_{E^\rho_{N''}})^{N''}  \Vert \alpha
\Vert^{}_{D^{}_{N''}}.\cr
}$$
This proves statement i) of the proposition.

To prove statement ii) of the proposition we first observe that according to
definition (4.128a) and (4.128b), we have that
$$\Vert \beta_1 - \beta_0 \Vert^{}_{{\cal D}^{\rho-1/2,
\chi^{}_1}_{j,l}} = \Vert \lambda
((f, \dot{f}), \beta_0, \beta_0, \beta_0) \Vert^{}_{{\cal D}^{\rho-1/2,
\chi^{}_1}_{j,l}}.$$
Statements i) and ii) of Proposition 4.7 then give that
$$\Vert \beta_1 - \beta_0 \Vert^{}_{{\cal D}^{\rho-1/2, \chi^{}_1}_{j,l}}
\leq C \big(\Vert (f, \dot{f}) \Vert^{}_{M^\rho_N}
+ \Vert \beta_0 \Vert^2_{{\cal D}^{0,0}_{L,N}}\big)
 \Vert \beta_0 \Vert^{}_{{\cal D}^{3/2-\rho, 0}_{L,N}},$$
where $C$ depends on $j$, $l$, $\rho$ and, $L$ and $N$ depend on $j$ and $l$.
This proves together with (4.129) and the definition of the norm in
$E^\rho_N$ that
$$\Vert \beta_1 - \beta_0 \Vert^{}_{{\cal D}^{\rho-1/2,
\chi^{}_1}_{j,l}} \leq C (1+ \Vert u \Vert^{}_{E^{}_N})
 \Vert u \Vert^{}_{E^{}_N}  \Vert \alpha \Vert^{}_{D^{}_N},
\quad j,l \geq 0,$$
where $C$ depends on $\rho$, $j$ and $l$ and $N$ depend on $j$ and $l$.
This proves statement ii) for $n=0$. To prove statement ii) for $n \geq 1$,
we observe that for $g = (f, \dot{f})$, it follows from the definition of
$\beta_n$ and from definition (4.91) of $\lambda$ that
$$\eqalignno{
\beta_{n+1} - \beta_n &= \lambda (g, \beta_n, \beta_n, \beta_n) - \lambda (g,
\beta_{n-1}, \beta_{n-1}, \beta_{n-1})&(4.131)\cr
&= \lambda (g,0,0, \beta_n - \beta_{n-1}) + \lambda (0, \beta_n - \beta_{n-1},
\beta_n, \beta_n) \cr
&\qquad{}+ \lambda (0, \beta_{n-1}, \beta_n - \beta_{n-1}, \beta_n)
+ \lambda (0, \beta_{n-1},
\beta_{n-1}, \beta_n - \beta_{n-1}), \quad  n \geq 1.\cr
}$$
It follows from Proposition 4.7 and from (4.131) that
$$\eqalignno{
\Vert \beta_{n+1} - \beta_n \Vert^{}_{{\cal D}^{d_{n+1}, \chi^{}_{n+1}}_{j,l}}
&\leq C_1 \Big(\Vert (f, \dot{f}) \Vert^{}_{M^\rho_N}
\Vert \beta_n - \beta_{n-1} \Vert^{}_{{\cal D}^{d_n, \chi^{}_n}_{L,N}}
&(4.132)\cr
&\qquad{}+ \Vert \beta_n - \beta_{n-1} \Vert^{}_{{\cal D}^{0,
\chi^{}_n}_{L,N}}
\Vert \beta_n \Vert^{}_{{\cal D}^{0,0}_{L,N}}  \Vert \beta_n
\Vert^{}_{{\cal D}^{d_n,0}_{L,N}}\cr
&\qquad{}+ \Vert \beta_{n-1} \Vert^{}_{{\cal D}^{0,0}_{L,N}}
\Vert \beta_n -
\beta_{n-1} \Vert^{}_{{\cal D}^{0, \chi^{}_n}_{L,N}}  \Vert \beta_n
\Vert^{}_{{\cal D}^{d_n,0}_{L,N}}\cr
&\qquad{}+ \Vert \beta_{n-1} \Vert^{}_{{\cal D}^{0,0}_{L,N}}
\Vert \beta_{n-1}
\Vert^{}_{{\cal D}^{0,0}_{L,N}}  \Vert \beta_n - \beta_{n-1}
\Vert^{}_{{\cal D}^{d_n, \chi^{}_n}_{L,N}}\Big),\quad  n \geq 1.\cr
}$$
As a matter of fact, the hypothesis of statements i) and ii)
of Proposition 4.7 are satisfied since $d_n + \chi^{}_n > \rho - 1/2 > 0$
for $n \geq 0$. Moreover
$d_{n+1} = 1 - d_n$ and $\chi^{}_{n+1} = 1/2 - \rho + \chi^{}_n + d_n$
in agreement with statement i) of Proposition 4.7 and $\chi'$, given by
statement~ii) of Proposition 4.7, satisfies $\chi' \geq \chi^{}_{n+1}$.
Inequality (4.132) now follows from the topological inclusion
$${\cal D}^{d', \chi'}_{j,l}  \subset  {\cal D}^{d, \chi}_{j,l},
\quad \chi' \geq \chi,  d' + \chi' \geq d + \chi,  j \geq 0,
 l \geq 0.$$
Since $d_n \geq 3/2 - \rho$ for $n \geq 0$, inequality (4.132)
and this topological inclusion give
$$\eqalignno{
&\Vert \beta_{n+1} - \beta_n \Vert^{}_{{\cal D}^{d_{n+1},
\chi^{}_{n+1}}_{j,l}}& (4.133)\cr
&\qquad{}\leq C_2 \big(\Vert (f, \dot{f}) \Vert^{}_{M^\rho_N}
+ \Vert \beta_n \Vert^2_{{\cal D}^{3/2-\rho,0}_{L,N}}
+ \Vert \beta_{n-1} \Vert^2_{{\cal D}^{3/2-\rho,0}_{L,N}}\big)
\Vert \beta_n -\beta_{n-1}\Vert^{}_{{\cal D}^{d_n,\chi^{}_{n}}_{L,N}},\cr
}$$
$j \geq 0$,  $l \geq 0$,  $n \geq 1$,
where $C_2$ depends on $j$, $l$, $\rho$ and, $L$ and $N$ depends on $j$,
$l$.
This inequality
and statement i) of the proposition give that
$$\eqalignno{
&\Vert \beta_{n+1} - \beta_n \Vert^{}_{{\cal D}^{d_{n+1},
\chi^{}_{n+1}}_{j,l}}&(4.134)\cr
&\qquad{}\leq C' (1+ \Vert u \Vert^{}_{E^\rho_{N'}})^{N'}
\Vert u \Vert^{}_{E^\rho_{N'}}
 \Vert \beta_n - \beta_{n-1} \Vert^{}_{{\cal D}^{d_n, \chi^{}_n}_{L,N'}},
\quad n \geq 1,  j \geq 0,  l \geq 0,\cr
}$$
where $C'$ depending on $j$, $l$, $n$, $\rho$ and
$N'$ depending on $j$, $l$, $n$ are sufficiently
large. Since we already have proved that statement ii) of the proposition
is true for $n=0$, it follows from (4.134) by induction that it is true
for every $n \geq 0$.
This proves the proposition.

For $(f, \dot{f}, \alpha) \in E^{\circ\rho}_\infty$, $1/2 < \rho < 1$
and for the sequence $\beta_n$, $n \geq 0$, given by (4.128a) and (4.128b),
we introduce (cf. (4.87a))
$$\eqalignno{
A_{0, \mu} t) &= \cos ((- \Delta)^{1/2} t)  f_\mu + (- \Delta)^{- 1/2}
 \sin ((- \Delta)^{1/2})  \dot{f}_\mu,&(4.135{\rm a})\cr
A_{n+1, \mu} (t) &= A_{0, \mu} (t) + \sum_{\varepsilon = \pm}
(G_{\varepsilon, \mu}
(\beta^{(\varepsilon)}_n, \beta^{(\varepsilon)}_n)) (t),
\quad 0 \leq \mu \leq 3,
 t \geq 0,  n \geq 0,\cr
}$$
where $G_{\varepsilon, \mu}$ is given by (4.1a), and we introduce
$$\phi_n (t) = e^{- i \vartheta (A_n, t)}  \sum_{\varepsilon = \pm}
e^{i \varepsilon \omega (-i \partial) t}  \beta^{(\varepsilon)}_n (t),
\quad  n \geq 0,  t \geq 0, \eqno{(4.135{\rm b})}$$
where $(\vartheta (A_n,t)) (x) =\vartheta  (A_n, (t,x))$, $x \in \Rrm^3$,
with $\vartheta$ given by (1.19). For $n$ sufficiently large,
$(A_n, \phi_n)$ is an
approximate solution of the M-D equations (in the sense that inequality of
Theorem 4.11 is satisfied). To prove this, we need decay properties
for $A_{n, \mu} (t)$ and $\phi_n (t)$ which are direct consequences of Lemma
4.1
and Proposition 4.8. Before stating the result, we introduce
$${\cal R}^0_{N_0,L} = 0 \eqno{(4.136{\rm a})}$$
and
$${\cal R}^{l}_{N_0,L}  (v_1,\ldots,v_l) = \sum_{1 \leq j \leq l}
\prod_{i \neq j}  \Vert v_i \Vert^{}_{E^\rho_{N_0}}
\Vert v_j \Vert^{}_{E^\rho_{N_0+L}},
\eqno{(4.136{\rm b})}$$
where $l \geq 1$, $N_0 \geq 0$,  $L \geq 0$ and $ v_1,\ldots,
v_l \in E^\rho_\infty$.

We also introduce $A_{n,Y} (u)$,  $\dot{A}_{n, Y} (u)$,
$\phi'_{n,Y} (u)$, for $n \geq 0$,  $Y \in U(\p)$ and $u \in
E^{\circ\rho}_\infty$,
by
$$\eqalignno{
(A_{n,Y} (u)) (t) &= (\xi^M_Y  A_n) (t),\quad  (\dot{A}_{n,Y} (u)) (t) =
{d \over dt}  (\xi^M_Y  A_n) (t),& (4.137{\rm a})\cr
(\phi'_{n,Y} (u)) (t) &= (\xi^D_Y  \phi'_n) (t),\quad  \phi'_n (t) =
e^{i \vartheta (A_n, t)}  \phi_n (t),\cr
}$$
where $A_n (t)$,  $\phi_n (t)$ are given as functions of $u = (f, \dot{f},
\alpha)
\in E^{\circ\rho}_\infty$ by (4.128a), (4.128b), (4.135a) and (4.135b). We note
that $A_{n,Y} (u)$,  $\dot{A}_{n,Y} (u)$ and $\phi'_{n,Y} (u)$ are polynomials
in $u$ and we note for future reference that
$$A_{n+1, \mu} (t) = A_{0, \mu} (t) - \int^\infty_t  {\sin (\vert \nabla \vert
(t-s)) \over \vert \nabla \vert}  \sum_{\varepsilon = \pm}
((\phi'^\varepsilon_n)^+  \gamma^0  \gamma_\mu  \phi'^\varepsilon_n)
(s)  ds \eqno{(4.137{\rm b})}$$
and
$$\phi'^\varepsilon_{n+1} (t) = \phi'^\varepsilon_0 (t) + i \int^\infty_t
e^{(t-s) {\cal D}} ((A_{n, \mu} + B_{n, \mu})  \gamma^0  \gamma^\mu
 \phi'^\varepsilon_n) (s)  ds, \eqno{(4.137{\rm c})}$$
where $\phi'_n = \sum_\varepsilon  \phi'^\varepsilon_n$,
$\phi'_0 (t) = e^{t {\cal D}} \alpha$,  $\phi'^\varepsilon_0 (t) =
P_\varepsilon
 (-i \partial)  \phi'_0 (t)$,

$$A_{0, \mu} (t) = \cos ((- \Delta)^{1/2} t)  f_\mu + (- \Delta)^{1/2}
 \sin ((- \Delta)^{1/2} t)  \dot{f}_\mu, \quad u = (f, \dot{f}, \alpha),$$
 $\hbox{ and  } B_{n, \mu} (y) = -\partial_\mu \vartheta (A_{n, y}).$
\saut
\noindent{\bf Theorem 4.9.}
{\it
If $n \geq 0$ and $1/2 < \rho < 1$, then there exists $N_0 \geq 0$ such that
$$\eqalignno{
&\sup_{t \geq 0}  \Vert \big(D^l (A_{n,Y}, \dot{A}_{n,Y}, \phi'_{n,Y})
(u ; v_1,\ldots,v_l)\big) (t) \Vert^{}_{E^\rho_0}\cr
&\qquad{}+ \sup_{t \geq 0,  x \in \Rrm^3} \Big((1+t+\vert x \vert)^{3/2}
\vert \big((D^l  \phi'_{n,Y})  (u ; v_1,\ldots,v_l)\big) (t,x) \vert\cr
&\qquad{}+ (1+t+\vert x \vert)^{3/2-\rho}  \vert \big((D^l  A_{n,Y})
 (u ; v_1,\ldots,v_l)\big) (t,x) \vert\cr
&\qquad{}+ (1+t+\vert x \vert)
(1+\big\vert t - \vert x \vert \big\vert)^{3/2-\rho}
 \vert \big((D^l  A_{n, P_\mu Y})
 (u ; v_1,\ldots,v_l)\big) (t,x) \vert\Big)\cr
&\qquad\qquad{}\leq F_{L,l}(\Vert u \Vert^{}_{E^\rho_{N_0}})
{\cal R}^l_{N_0,L}
 (v_1,\ldots,v_l) + F'_{L,l} (\Vert u \Vert^{}_{E^\rho_{N_0}})
\Vert u \Vert^{}_{E^\rho_{N_0+L}}
 \Vert v_1 \Vert^{}_{E^\rho_{N_0}} \cdots \Vert v_l
\Vert^{}_{E^\rho_{N_0}},\cr
}$$
for every $L \geq 0$,  $l \geq 0$,  $Y \in \Pi'$,  $\vert Y \vert \leq L$,
$u, v_1,\ldots,v_l \in E^{\circ\rho}_\infty$, where $F_{L,l}$ and $F'_{L,l}$
are increasing polynomials on $[0,\infty[$.
Moreover $(A_{n,Y}, \dot{A}_{n,Y}, \phi'_{n,Y})$ is a
polynomial on $E^{\circ\rho}_\infty$ with vanishing constant term.
}\saut
\noindent{\it Proof.} This follows from Lemma 4.1, Proposition 4.8,
definitions (4.135a), (4.135b) of $A_{n, \mu} (t)$ and $\phi_n (t)$,
and from the covariance of the function $u \mapsto (A_{n, {\un}} (u),
\dot{A}_{n, {\un}} (u), \phi'_{n, {\un}} (u))$
from $E^\rho_\infty$ to $E^\rho_0$ under the Lorentz subgroup $SL (2, \Crm)$ of
${\cal P}_0$, i.e. $(A_{n, {\un}} (U^1_g(u))(y)=\Lambda_L A_{n, {\un}}
(\Lambda_L^{-1}(y-a))$, $g=(a, L)\in{\cal P}_0$,
and similarly for $\phi'_{n, {\un}} (u)$.
\saut
\noindent{\bf Theorem 4.10.}
{\it
If $1/2 < \rho < 1$,  $0 \leq \rho' \leq 1$, and $n \geq 0$ is such that
$\rho' - 1/2 + \chi^{}_n > 0$, where $\chi^{}_n$ is given in Proposition 4.8,
then there exists $N_0$ such that
$$\eqalignno{
&\sup_{t \geq 0} \Big((1+t)^{\rho' - 1/2 + \chi^{}_n}  \Vert ((D^l (A_{n+1,Y} -
A_{n,Y}, A_{n+1, P_0 Y} - A_{n, P_0 Y})) (u ; v_1,\ldots,v_l)) (t)
\Vert^{}_{M^{\rho'}}\cr
&\quad{}+ (1+t)^{\chi^{}_{n+1}}
\Vert \big((D^l (\phi'_{n+1, Y} - \phi'_{n,Y})) (u ;
v_1,\ldots,v_l)\big) (t) \Vert^{}_D\Big)\cr
&\quad\qquad{}\leq F_{L,l} (\Vert u \Vert^{}_{E^\rho_{N_0}})  {\cal R}^l_{N_0,
L}
 (v_1,\ldots,v_l) + F'_{L,l} (\Vert u \Vert^{}_{E^\rho_{N_0}})  \Vert
u \Vert^{}_{E^\rho_{N_0+L}}  \Vert v_1 \Vert^{}_{E^\rho_{N_0}}\cdots \Vert v_l
\Vert^{}_{E^\rho_{N_0}},\cr
}$$
for all $L \geq 0$,  $l \geq 0$,  $Y \in \Pi'$,  $\vert Y \vert
\leq L$, $u, v_1,\ldots,v_l \in E^{\circ\rho}_\infty$ and there exists $N$
depending on $\alpha \in \Nrm^4$ such that
$$\eqalignno{
&\vert \big((D^l (A_{n+1, P^\alpha Y} - A_{n, P^\alpha Y}))
(u ; v_1,\ldots,v_l)\big) (t,x)
\vert (1+t+\vert x \vert)^{1 + \chi^{}_n + \vert \alpha \vert - \varepsilon}
(1+t)^\varepsilon\cr
&\qquad{}+ \vert \big((D^l (\phi'_{n+1, Y} - \phi'_{n,Y}))
(u ; v_1,\ldots,v_l)\big) (t,x) \vert
 (1+t+\vert x\vert)^{3/2 + \chi^{}_{n+1}}\cr
&\qquad\qquad{}\leq G_{L,l,\varepsilon}
(\Vert u \Vert^{}_{E^\rho_N})  {\cal R}^l_{N,L}
(v_1,\ldots,v_l) + G'_{L,l,\varepsilon}  (\Vert u \Vert^{}_{E^\rho_N})
\Vert u \Vert^{}_{E^\rho_{N+L}}  \Vert v_1 \Vert^{}_{E^\rho_N}\cdots \Vert v_l
\Vert^{}_{E^\rho_N},\cr
}$$
for all $t \geq 0$,  $x \in \Rrm^3$,  $l \geq 0$,  $Y \in \Pi'$,
$\alpha \in \Nrm^4$,  $\vert Y \vert + \vert \alpha \vert \leq L$,
$ \varepsilon > 0$,  $u, v_1,\ldots,v_l \in
E^{\circ\rho}_\infty$.\penalty-10000
Here $F_{L,l}$, $F'_{L,l}$,  $G_{L,l,\varepsilon}$ and $G'_{L,l,\varepsilon}$
are increasing polynomials  on $[0,\infty[$ and\penalty-10000
 $P^\alpha = P^{\alpha_0}  P^{\alpha_1}
P^{\alpha_2}  P^{\alpha_3}$. Moreover $A_{n+1, Y} - A_{n,Y}$
and $\phi'_{n+1, Y} - \phi'_{n,Y}$ are polynomials on $E^{\circ\rho}_\infty$
with
constant and linear term vanishing.
}\saut
\noindent{\it Proof.}
The result follows as in the proof of Theorem 4.9.

We are now ready to prove
that $t \mapsto (A_n (t), \phi_n (t))$ where $A_n (t)$ and $\phi_n (t)$ are
given by (4.135a) and (4.135b), is an approximate solution of the M-D equations
(1.2a) and (1.2b) for $n$ sufficiently large. We introduce for
$u = (f, \dot{f}, \alpha) \in E^{\circ\rho}_\infty$:
$$\eqalignno{
\Delta^M_{n, \mu} (t) &= \cos (\vert \nabla \vert t)
f_\mu + \vert \nabla \vert^{-1}
 \sin (\vert \nabla \vert t)  \dot{f}_\mu& (4.138{\rm a})\cr
&\qquad{}- \int^\infty_t  \vert \nabla \vert^{-1}
\sin (\vert \nabla \vert (t-s))
 \overline{\phi'_n (s)}  \gamma_\mu  \phi'_n (s)
 ds - A_{n, \mu} (t)\cr
}$$
and
$$\Delta^D_n (t)
= e^{i {\cal D} t}\alpha + i \int^\infty_t
e^{i {\cal D} (t-s)}  \gamma^0  \gamma^\mu \big(A_{n, \mu} (s) +
B_{n, \mu} (s)\big)  \phi'_n (s)  ds - \phi'_n (t),\eqno{(4.138{\rm b})} $$
where $n \geq 0$,  $t \geq 0$,  $B_{n,i} (t) = - \partial_i
\vartheta  (A_n, t)$,  $1 \leq i \leq 3$, and $B_{n,0} (t) = - {d \over dt}
\vartheta (A_n,t)$. Similarly to (4.137),  we introduce $\Delta^M_{n,Y} (u)$,
$\Delta^D_{n,Y} (u)$ for $u \in E^{\circ\rho}_\infty$ and $Y \in U(\p)$ by
$$\eqalignno{
(\Delta^M_{n,Y} (u))_\mu (t) &= (\xi^M_Y  \Delta_n)_\mu (t),&(4.139)\cr
(\Delta^D_{n,Y} (u)) (t) &= (\xi^D_Y  \Delta^D_n) (t). \cr
}$$
\saut
\noindent{\bf Theorem 4.11.}
{\it
If $1/2 < \rho < 1$ and $n \geq n_0$, where $n_0$ is the integer part of
$3+1/(2(1-\rho))$, then there exists $N_0$ such that
$$\eqalignno{
&(1+t)^{1 - \rho}  \Vert \big((D^l\Delta^M_{n,Y}) (u ; v_1,\ldots,v_l)\big)
(t) \Vert^{}_{L^2(\Rrm^3, \Rrm^4)}\cr
&\quad{}+ (1+t)^{2 - \rho}  \Vert \big((D^l
\Delta^M_{n, P_\mu Y}) (u ; v_1,\ldots,v_l)\big)
(t) \Vert^{}_{L^2(\Rrm^3, \Rrm^4)}\cr
&\quad{}+ (1+t)^{3/2 - \rho}  \Vert \big((D^l
\Delta^D_{n,Y}) (u ; v_1,\ldots,v_l)\big)
(t) \Vert^{}_D\cr
&\quad{}+ (1+t+\vert x \vert)^{5/2 - \rho}  \vert \big((D^l\Delta^M_{n,Y})
(u ; v_1,\ldots,v_l)\big) (t,x) \vert\cr
&\quad{}+ (1+t+\vert x \vert)^{7/2 - \rho}
\vert \big((D^l\Delta^M_{n, P_\mu Y})
(u ; v_1,\ldots,v_l)\big) (t,x) \vert\cr
&\quad{}+ (1+t+\vert x \vert)^{3 - \rho}  \vert \big((D^l\Delta^D_{n,Y})
(u ; v_1,\ldots,v_l)\big) (t,x) \vert\cr
&\quad\quad{}\leq F_{L,l} (\Vert u \Vert^{}_{E^\rho_{N_0}})  {\cal R}^l_{N_0,L}
(v_1,\ldots,v_l) + F'_{L,l} (\Vert u \Vert^{}_{E^\rho_{N_0}})
\Vert u \Vert^{}_{E^\rho_{N_0+L}}
 \Vert v_1 \Vert^{}_{E^\rho_{N_0}} \cdots \Vert v_l \Vert^{}_{E^\rho_{N_0}},\cr
}$$
for all $L \geq 0$,  $l \geq 0$,  $Y \in \Pi'$,  $\vert Y \vert \leq L$,
$u, v_1,\ldots,v_l \in E^{\circ\rho}_\infty$, where $F_{L,l}$ and $F'_{L,l}$
are
increasing polynomials on $[0,\infty[$. Moreover
$\Delta^M_{n,Y}$ and $\Delta^D_{n,Y}$
are polynomials
on $E^{\circ\rho}_\infty$ with  constant and linear terms vanishing.
}\saut
\noindent{\it Proof.}
By the construction of $\Delta^M_{n, \mu} (t)$ and $\Delta^D_n (t)$ and by the
definition of $A_{n, \mu} (t)$, $B_{n, \mu} (t)$ and $\phi'_n (t)$ we
obtain
$$\eqalignno{
\Delta^M_{n, \mu} (t) &= A_{n+1, \mu} (t) - A_{n, \mu} (t) &(4.140{\rm a})\cr
&\quad{}- \int^\infty_t  \vert \nabla \vert^{-1}  \sin ((t-s)
\vert \nabla \vert)  \sum_{\varepsilon = \pm}  (e^{-i \varepsilon
\omega (-i \partial) s} \beta^{(- \varepsilon)}_n (s))^+  \gamma^0
\gamma_\mu  (e^{i \varepsilon \omega (-i \partial) s}
\beta^{(\varepsilon)}_n (s))ds\cr
}$$
and
$$\Delta^D_{n, \mu} (t) = \phi'_{n+1} (t) - \phi'_n (t). \eqno{(4.140{\rm
b})}$$
We set $\varepsilon_1 = - \varepsilon$, $\varepsilon_2 =
\varepsilon$, $m_1 = m_2 = m$, $f_1 (t) = \beta^{(- \varepsilon)}_n (t)$,
$f_2 (t) = \beta^{(\varepsilon)}_n (t)$
in definitions (4.48a) and (4.48c) of $(H_\mu (f_1, f_2)) (t)$,
which we denote by
$(H^{(\varepsilon)}_n (\phi'^{(- \varepsilon)}_n, \phi'^{(\varepsilon)}_n))_\mu
(t)$, where $\phi'^{(\varepsilon)}_n (t) = e^{i \varepsilon \omega (-i
\partial)}
\beta^{(\varepsilon)}_n (t)$. (4.140a) then gives
$$\Delta^M_{n, \mu} (t) = \sum_{\varepsilon = \pm}
(H_n (\phi'^{(- \varepsilon)}_n,
\phi'^{(\varepsilon)}_n))_\mu (t)+ A_{n+1, \mu} (t) - A_{n, \mu} (t).
\eqno{(4.141)}$$
The theorem now follows from applications of Lemma 4.3 to each term
$$H_n (\xi^D_{Y_1}  \phi'^{(- \varepsilon)}_n, \xi^D_{Y_2}
 \phi'^{(\varepsilon)}_n),\quad Y_1, Y_2 \in \Pi',  \vert Y_1 \vert +
\vert Y_2 \vert = \vert Y \vert, \eqno{(4.142)}$$
in the expansion of $\xi^D_Y  \Delta^M_n$ and from Theorem 4.9 and Theorem
4.10.

Let $B_{n,0} (t) = - {d \over dt} \vartheta (A_n,t), B_{n,i}
(t) = - \partial_i \vartheta(A_n, t)$, $1 \leq i \leq 3$,
$n \geq 0$. We introduce for $Y \in U(\p)$ and $u \in E^{\circ\rho}_\infty$,
$$\eqalignno{
\big(\vartheta_{n,Y} (u)\big)(t) &= \xi^{}_Y \vartheta \big(A_n, (t,x)\big),\cr
(B_{n,Y} (u)) (t) &= (\xi^M_Y  B_n) (t), \quad (\dot{B}_{n,Y} (u))
(t) = {d \over dt} (\xi^M_Y B_n) (t).\cr
}$$
\saut
\noindent{\bf Corollary 4.12.}
{\it
If $n \geq 0$ and $1/2 < \rho < 1$, then there exists $N_0 \geq 0$ such that
$$\eqalignno{
&\sup_{t \geq 0,  x \in \Rrm^3} (1+\vert x \vert+t)^{1/2-\rho}
\vert \big((D^l  \vartheta_{n,Y}) (u ; v_1,\ldots,v_l)\big) (t,x) \vert\cr
&\quad{}+ \sup_{t \geq 0}  \Vert (D^l (B_{n,Y}, \dot{B}_{n,Y})
(u ; v_1,\ldots,v_l))
(t) \Vert^{}_{M^\rho}\cr
&\quad{}+ \sup_{t \geq 0,  x \in \Rrm^3} ((1+t+\vert x \vert)^{3/2 - \rho}
 \vert \big((D^l  B_{n,Y}) (u ; v_1,\ldots,v_l)\big) (t,x) \vert\cr
&\quad{}+ (1+t+\vert x \vert) (1+\big\vert t -
\vert x \vert \big\vert)^{3/2 - \rho} \vert \big((D^l B_{n, P_\mu Y})
(u ; v_1,\ldots,v_l)\big) (t,x) \vert)\cr
&\quad\quad{}\leq F_{L,l} (\Vert u \Vert^{}_{E^\rho_{N_0}})  {\cal R}^l_{N_0,L}
(v_1,\ldots,v_l) + F'_{L,l} (\Vert u \Vert^{}_{E^\rho_{N_0}})
\Vert u \Vert^{}_{E^\rho_{N_0+L}}
\Vert v_1 \Vert^{}_{E^\rho_{N_0}}\cdots\Vert v_l \Vert^{}_{E^\rho_{N_0}},\cr
}$$
for every $L \geq 0$,  $l \geq 0$,  $Y \in \Pi'$,  $\vert Y \vert \leq L$, $u,
v_1,\ldots,v_l \in E^{\circ\rho}_\infty$, where $F_{L,l}$ and $F'_{L,l}$ are
increasing polynomials on $[0,\infty[$. Moreover $(B_{n,Y}, \dot{B}_{n,Y})$ is
a polynomial on $E^{\circ\rho}_\infty$ with vanishing constant term.
}\saut
\noindent{\it Proof.}
This is a direct consequence of Corollary 4.6 and Theorem 4.9.
\saut
\noindent{\bf Corollary 4.13.}
{\it
If $n \geq 1$ and $1/2 < \rho < 1$, then there exists $N_0$ such that
$$\sup_{t \geq 0}  (1+t)^{k (\rho)}  \Vert \big((A_{n,Y}, \dot{A}_{n,Y},
\phi'_{n,Y}) - (A_{0,Y}, \dot{A}_{0,Y}, \phi'_{0,Y})\big) (t)
\Vert^{}_{E^\rho_0} \leq F_L (\Vert u \Vert^{}_{E^\rho_{N_0}})
\Vert u \Vert^{}_{E^\rho_{N_0+L}}$$
for every $L \geq 0$,  $Y \in \Pi'$,  $\vert Y \vert \leq L$ and
$u \in E^{\circ\rho}_\infty$, where $F_L$ are increasing
polynomials on $[0,\infty[$ and $k (\rho) =\min (\rho - 1/2, 2 - 2 \rho)$.
}\saut
\noindent{\it Proof.}
This follows from Theorem 4.10 taking $\rho' = \rho$.
\vfill\eject
\noindent{\titre 5. Energy estimates and $L^2 - L^\infty$
estimates for the Dirac field.}
\saut
In this chapter we shall derive various new $L^2 - L^2$  estimates for
the inhomogeneous Dirac equation in $\Rrm^+ \times \Rrm^3$,
$$(i  \gamma^\mu  \partial_\mu + m) h - \gamma^\mu
G_\mu  h = g \eqno{(5.1{\rm a})}$$
for different choices of $g \in C^0 (\Rrm^+, D),  D = L^2 (\Rrm^3, \Crm^4)$,
and of electromagnetic potential $G$. To use these estimates we shall
also derive new $L^2 - L^\infty$ estimates. We shall suppose that
the initial data $h(t_0)$ at $t_0 \in \Rrm^+$ satisfies for some
$\tau \in {\Zrm}$
$$h(t_0) \in (1 - \Delta)^{- \tau/2} D, \eqno{(5.1{\rm b})}$$
where $(1 - \Delta)^{- \tau/2} D$ is the Hilbert space of
tempered distributions
$f$ such that\penalty-10000 $(1 - \Delta)^{\tau/2}  f \in D$.
We shall also consider the case $h(t_0) = 0$ where $t_0 = \infty$
or more precisely
$\displaystyle{\lim_{t \fl \infty}}  \Vert h(t) \Vert^{}_\tau \fl 0$, where
$\Vert\cdot\Vert^{}_\tau$ is the norm in $(1 - \Delta)^{- \tau/2} D$. We
shall suppose that
$$g \in C^0 (\Rrm^+, (1 - \Delta)^{- \tau/2} D) \eqno{(5.1{\rm c})}$$
and that the electromagnetic potential $G$ satisfies
$$(G,0) \in C^0 (\Rrm^+, (1 - \Delta)^{- (\vert \tau \vert + 1)/2} M^1).
\eqno{(5.1{\rm d})}$$
It follows from inequality (2.61) and from the definition of $M^1$ that
$G \in C^0 (\Rrm^+, W_{\vert \tau \vert + 1,6}$ $(\Rrm^3, \Rrm^4))$
and then that
$G \in C^0 (\Rrm^+, W_{\vert \tau \vert, \infty} (\Rrm^3, \Rrm^4))$ since
$W_{\vert \tau \vert + 1, 6}  (\Rrm^3) \subset W_{\vert \tau \vert, \infty}
 (\Rrm^3)$. If $B(t) = - i \gamma^0  \gamma^\mu
G_\mu (t)$, we therefore obtain that $B \in C^0 (\Rrm^+, L_b
((1 - \Delta)^{- n/2}  D))$,  $n \in \Nrm$,\penalty-10000
 $n \leq \vert \tau \vert$, where
$L_b (X)$ is the linear space, endowed with the norm topology,
of linear continuous operators on a Banach space $X$.
By duality we also have $B \in C^0 (\Rrm^+, L_b
((1 - \Delta)^{n/2}  D))$,  $n \in \Nrm$, $n \leq \vert
\tau \vert$. We define the operator ${\cal L} (t)$
in $(1 - \Delta)^{- \tau/2}
 D$ with domain $(1 - \Delta)^{- (\tau + 1)/2}  D$ by
$${\cal L} (t) = {\cal D} - i  \gamma^0  \gamma^\mu
G_\mu (t),\quad  t \geq 0, \eqno{(5.2)}$$
Since $B \in C^0 (\Rrm^+, L_b ((1 - \Delta)^{- \tau/2}  D))$ and ${\cal D}
\in L((1 - \Delta)^{- (\tau + 1)/2}  D,  (1 - \Delta)^{- \tau/2}
 D)$ it follows that ${\cal L} \in C^0 (\Rrm^+,
 L_b ((1 - \Delta)^{- (\tau + 1)/2}
 D,  (1 - \Delta)^{- \tau/2}  D))$. Moreover if
$$\widetilde{\Delta} (t) = (1 - \Delta)^{1/2}  {\cal L} (t)  (1 - \Delta)^{-
1/2}
- {\cal L} (t),$$
then $\widetilde{\Delta} \in C^0 (\Rrm^+, L_b ((1 - \Delta)^{- \tau/2}  D))$.
In
fact using lemma A.3 of \refKII\ it follows for $\tau \geq 0$
that the norm of the operators $\widetilde{\Delta} (t)$ (resp.
$\widetilde{\Delta} (t+\varepsilon) - \widetilde{\Delta} (t))$ are bounded by
$$C \Vert (1 - \Delta)^{(\tau + 1/2)/2}  \vert \nabla \vert
G(t) \Vert^{}_{L^2}\quad  (\hbox{resp.}\ C \Vert (1 - \Delta)^{(\tau + 1/2)/2}
 \vert \nabla \vert  (G (t + \varepsilon) - G(t)) \Vert^{}_{L^2}).$$
If $\tau \leq 0$, then considering the transposed of $\widetilde{\Delta}(t)$,
we obtain the above
bounds with $\tau + 1/2$ being replaced by $- \tau - 1/2$.
Thus the statement follows from condition (5.1d).
It now follows from theorem 1 of \refKI\ that there exists a
strongly continuous evolution operator $w(s,s')$,  $s,s' \geq 0$, in $
(1 - \Delta)^{- \tau/2}  D$. Moreover the function $(s,s') \mapsto w(s,s')
\alpha \in (1 - \Delta)^{- \tau/2}  D$ is $C^1$ for $\alpha \in
(1 - \Delta)^{- (\tau + 1)/2} D$ and
$$\eqalignno{
w(s,s) &= I,\qquad\hskip8mm  w(s,s')  w(s',s'') = w(s,s''),& (5.3\hbox{a})\cr
{d \over ds}    w(s,s') &= {\cal L} (s)  w(s,s'), \qquad
{d \over ds'}  w(s,s') = - w(s,s')  {\cal L}(s'), &(5.3\hbox{b})\cr
}$$
where equalities (5.3a) are defined on $(1 - \Delta)^{-\tau/2}D$
and equalities (5.3b) on \penalty-10000 $(1 - \Delta)^{- (\tau + 1)/2}D$.
Since ${\cal L} (t)$
is kew-adjoint on $D$ with domain $(1 - \Delta)^{-1/2}D$, it follows
that $w(s,s')$ is unitary on $D$.
The unique solution $h \in C^1 (\Rrm^+,(1 - \Delta)^{-\tau/2}D)$
of equation (5.1a) with electromagnetic potential $G$ satisfying (5.1d)
and satisfying, with $\tau$ replaced by $\tau+1$, conditions (5.1b) and (5.1c),
is given by
$$h(t) = w(t,t_0)  h(t_0) + \int^t_{t_0}  w(t,s)  (- i\gamma^0)g(s)ds,
\quad  t \geq 0. \eqno{(5.3\hbox{c})}$$

Let $h^{}_n (t_0) \in (1 - \Delta)^{- (\tau + 1)/2}  D$ (resp. $g^{}_n \in C^0
(\Rrm^+, (1 - \Delta)^{- (\tau + 1)/2}  D)),  n \geq 0$, be a
Cauchy sequence in $(1 - \Delta)^{- \tau/2}  D$ (resp. $C^0 (\Rrm^+,
(1 - \Delta)^{- \tau/2}  D))$ converging to $h (t_0)$ (resp. $g$) and
let $h^{}_n \in C^1 (\Rrm^+, (1 - \Delta)^{- \tau/2}  D)$ be the corresponding
sequence of solutions of (5.1a) given by (5.3c). It then follows from (5.1a)
that
$$\Vert {d \over dt} (h^{}_{n_1} (t) - h^{}_{n_2} (t))
\Vert^{}_{\tau - 1} \leq C(t) \Vert h^{}_{n_1} (t) - h^{}_{n_2} (t)
\Vert^{}_\tau + \Vert g^{}_{n_1} (t) - g^{}_{n_2} (t) \Vert^{}_{\tau - 1}$$
and from (5.3c), since $w(s,s')$ is a bounded operator on
$(1 - \Delta)^{- \tau/2} D$, that
$$\Vert h^{}_{n_1} (t) - h^{}_{n_2} (t) \Vert^{}_\tau
\leq C'(t) \big(\Vert h^{}_{n_1} (t_0) -
h^{}_{n_2} (t_0) \Vert^{}_\tau + \sup_{0 \leq s \leq t}  \Vert g^{}_{n_1} (s) -
g^{}_{n_2} (s) \Vert^{}_\tau\big),$$
for some constants $C(t)$ and $C'(t)$. This shows that if conditions
(5.1b), (5.1c) and (5.1d) are satisfied then equation (5.1a) has a
unique solution $h \in C^0 (\Rrm^+, (1 - \Delta)^{- \tau/2}  D)$.
This solution is given by (5.3c) and moreover $h \in C^1 (\Rrm^+,
(1 - \Delta)^{- (\tau - 1)/2}  D).$

We are mainly interested in the particular case of equation (5.1a) given by
$$(i  \gamma^\mu  \partial_\mu + m) h - G_\mu  \gamma^\mu
 h = F^{}_\mu  \gamma^\mu       r, \quad t \geq 0,
\eqno{(5.4\hbox{a})}$$
where $r$ satisfies
$$(i  \gamma^\mu  \partial_\mu + m)  r = q, \quad
t \geq 0, \eqno{(5.4\hbox{b})}$$
and where $F,r$ and $q$ satisfy conditions such that (5.1c)
is satisfied with $g = F_{\mu}\gamma^\mu r$. For example if $\tau = 0$ then
this is obviously the case if
$$\eqalignno{
F&\in W^{1, \infty}  (\Rrm^+ \times \Rrm^3,  \Rrm^4),
&(5.5)\cr
r &\in C^1 (\Rrm^+, D) \cap C^0 (\Rrm^+,(1 - \Delta)^{- 1/2}D),\cr
q &\in C^0 (\Rrm^+, D).\cr
}$$
For this case we can obtain a much better energy estimate than
$$\Vert h(t) \Vert^{}_D \leq \Vert h (t_0) \Vert^{}_D +
\int^{\max (t, t_0)}_{\min (t, t_0)}
 \Vert g(s) \Vert^{}_D\, ds, \eqno{(5.6)}$$
which follows from (5.3c) and the unitarity of $w(t,s)$ in $D$. We note that,
using $\gamma^\mu  \gamma^\nu + \gamma^\nu  \gamma^\mu = 2
g^{\mu\nu}$, one obtains
$$\eqalignno{
&(m - i \gamma^\mu  \partial_\mu + \gamma^\mu   G_\mu)
F^{}_\nu  \gamma^\nu  r &(5.7{\rm a})\cr
&\qquad {}= \gamma^\nu  F^{}_\nu (m + i
 \gamma^\mu  \partial_\mu - \gamma^\mu  G_\mu)r
- 2i F^\mu  \partial_\mu  r - i r
\partial_\mu  F^\mu\cr
&\qquad\qquad- {i \over 4} (\gamma^\mu  \gamma^\nu -
\gamma^\nu  \gamma^\mu)  r(\partial_\mu  F^{}_\nu -
\partial_\nu  F^{}_\mu)
+ 2  G_\mu  F^\mu  r.\cr
}$$
We also note for later reference, that if $(i \gamma^\mu  \partial_\mu +
m - \gamma^\mu  G_\mu)  h = \gamma^\mu  F^{}_\mu
r + g^{}_1$, then
$$\eqalignno{
&(i \gamma^\mu  \partial_\mu + m - \gamma^\mu   G_\mu)
(\partial_\nu  h + i G_\nu  h + i  F^{}_\nu
r) &(5.7{\hbox{b}})\cr
&\qquad{}= \gamma^\mu   h (\partial_\nu  G_\mu - \partial_\mu
G_\nu) + \gamma^\mu  r (\partial_\nu  F^{}_\mu - \partial_\mu
 F^{}_\nu) + \gamma^\mu F^{}_\mu  \partial_\nu r\cr
&\qquad\qquad{}+ i  \gamma^\mu (G_\nu  F^{}_\mu - G_\mu  F^{}_\nu) r +
i  F^{}_\nu (i  \gamma^\mu  \partial_\mu + m) r
+\partial_\nu   g^{}_1 + i  G_\nu  g^{}_1,\quad
0 \leq \nu \leq 3.\cr
}$$
This expression is useful for estimating $L^2$-norms of derivatives of $h$,
because of the gauge invariance of the electromagnetic fields in the
first two terms on the right-hand side. We also have that
$$\eqalignno{
&(i\gamma^\mu \partial_\mu + m - \gamma^\mu G_\mu)(h + i F^{}_\nu r)
&(5.7{\rm b'})\cr
&\qquad{}= \gamma^\mu r (\partial_\nu F^{}_\mu -\partial_\mu F^{}_\nu)
+F^{}_\nu (i\gamma^\mu \partial_\mu + m - \gamma^\mu G_\mu) r,\cr
}$$
if $(i\gamma^\mu \partial_\mu + m - \gamma^\mu G_\mu) h =
\gamma^\mu (\partial_\nu F^{}_\mu) r$.
Finally, we also note that if $y \in \Rrm^+ \times \Rrm^3$ and
$\nu \in \{0,1,2,3\}$ is given, then
$$(a + y^\nu) G_\mu  \partial^\mu = y^\mu  G_\mu
\partial^\nu + a  G_\mu  \partial^\mu + G_\mu (y^\nu
\partial^\mu - y^\mu  \partial^\nu), \eqno{(5.7{\rm c})}$$
where $a \in \Rrm$ and where the summation convention is used for $\mu$.
This gives that
$$\eqalignno{
&\Big(1 + \sum_\nu  \vert y^\nu \vert\Big) \big\vert \sum_\mu  G_\mu
 \partial^\mu  f \big\vert&(5.7{\rm d})\cr
&\qquad{}\leq C \Big(\big\vert \sum_\mu  G_\mu
 \partial^\mu  f \big\vert
+ \sum_\nu \big\vert \sum_\mu  y^\mu  G_\mu  \partial^\nu
 f + \sum_\mu  G_\mu (y^\nu  \partial^\mu
f - y^\mu  \partial^\nu  f) \big\vert\Big),\cr
}$$
where the sums are taken over $0 \leq \nu \leq 3,  0 \leq \mu \leq 3.$

\penalty-5000
\noindent{\bf Theorem 5.1.}
{\it
Let $Q = \gamma^\mu  \partial_\mu + i \gamma^\mu  G_\mu$,
(where the summation convention is used for $0 \leq \mu \leq 3$),
let $\dot{G}_\mu (t) = {d \over dt}  G_\mu (t)$ and let $t,t' \geq 0$.
\psaut
\noindent{\hbox{\rm ia)}}\ If $h \in C^1 (\Rrm^+, D) \cap C^0 (\Rrm^+,
(1 - \Delta)^{- 1/2} D)$ and
$(G,0) \in C^0 (\Rrm^+, (1- \Delta)^{- 1/2}  M^1)$ then
$$\Vert h(t) \Vert^2_D = \Vert h(t') \Vert^2_D + \int^t_{t'} 2  Re
\big(h(s), - i \gamma^0 ((m + i Q) h) (s)\big)^{}_D\, ds,$$
\noindent{\hbox{\rm ib)}}\ If $h \in C^2 (\Rrm^+, D) \cap C^1 (\Rrm^+,
(1 - \Delta)^{- 1/2} D) \cap C^0
(\Rrm^+, (1 - \Delta) D)$ and $(G, \dot{G}) \in C^0 (\Rrm^+, (1 - \Delta)^{1/2}
 M^1)$ then
$$\eqalignno{
\Vert h(t) \Vert^2_D &= \Vert h(t') \Vert^2_D + m^{-1} Re \Big(\big(h(t),
((m + iQ) h) (t)\big)^{}_D\cr
&\qquad{}- \big(h(t'), ((m + iQ) h) (t')\big)^{}_D + \int^t_{t'}  \big(h(s),
-i
\gamma^0 ((m^2 + Q^2) h) (s)\big)^{}_D\, ds\Big),\cr
}$$
\noindent{\hbox{\rm iia)}}\ If $(G,0) \in C^0 (\Rrm^+, (1 - \Delta)^{- 1/2}
M^1),
h(t_0) \in D,  g \in C^0 (\Rrm^+, D)$ and if $h \in C^0 (\Rrm^+, D)$
is the unique solution of equation (5.1a) then
$$\big\vert \Vert h(t) \Vert^{}_D - \Vert h(t') \Vert^{}_D \big\vert \leq
\int^{t'}_t
\Vert g(s) \Vert^{}_D\,  ds,\quad  \hbox{ for}\ 0 \leq t \leq t',$$
\noindent{\hbox{\rm iib)}}\ If $(G, \dot{G}) \in C^0 (\Rrm^+, (1 -
\Delta)^{1/2}  M^1),
h(t_0) \in D,  g = g^{(1)} + g^{(2)}$, $g^{(1)},  g^{(2)} \in
C^0 (\Rrm^+, D)$,  $\partial_\mu g^{(2)} \in L^1_{\rm loc} (\Rrm^+, D)$,
 $0 \leq \mu \leq 3$ and if $h \in C^0 (\Rrm^+, D)$ is the unique solution
of equation (5.1a) then
$$\eqalignno{
&\big\vert \Vert h(t) - (2m)^{-1}  g^{(2)} (t)
\Vert^{}_D - \Vert h(t') -(2m)^{-1}  g^{(2)} (t') \Vert^{}_D \big\vert\cr
&\qquad{}\leq \int^{t'}_t \Vert g^{(1)} (s) + (2m)^{-1} ((m - iQ)
g^{(2)}) (s) \Vert^{}_D\,
 ds, \quad\hbox{ for }\ 0 \leq t \leq t'.\cr
}$$
}\saut
\noindent{\it Proof.}
Since the operator ${\cal L} (t)$ defined in (5.2) is skew-adjoint on $D$ with
domain \penalty-10000 $(1 - \Delta)^{- 1/2}  D$, when the hypothesis on $G$ in
ia)
is satisfied, it follows that
$$\eqalign{
{d \over dt}  \Vert h(t) \Vert^2_D &= 2 Re \big(h(t), ({d \over dt} - {\cal L}
(t))
 h(t)\big)^{}_D \cr
&= 2 Re \big(h(t),-i \gamma^0 ((m + iQ) h) (t)\big)^{}_D, \cr
}$$
for $h \in C^1 (\Rrm^+, D) \cap C^0 (\Rrm^+, (1 - \Delta)^{- 1/2}  D).$
Integration of this equality from $t'$ to $t$ proves statement ia).

To prove the second statement introduce:
$$f = (m - iQ) h \quad\hbox{ and }\quad g = (m + iQ) h. \eqno{(5.8)}$$
Then $f,g \in C^1 (\Rrm^+, D) \cap C^0 (\Rrm^+, (1 - \Delta)^{- 1/2}  D),$
according to the hypothesis of statement ib), since $G \in C^0 (\Rrm^+,
W_{1, \infty} (\Rrm^3, \Rrm^4))$ which was proved after (5.1d) and since
$\dot{G} \in C^0 (\Rrm^+, L^\infty (\Rrm^3, \Rrm^4))$ by Sobolev embedding.
It follows from statement ia) that
$$\Vert f(t) \Vert^2_D = \Vert f(t') \Vert^2_D + \int^t_{t'}  2 Re
\big(f(s), - i \gamma^0 ((m + iQ) f) (s)\big)^{}_D\, ds. \eqno{(5.9)}$$
Substitution of the expressions
$$f = 2m  h - g\quad \hbox{ and }\quad (m + iQ)  f = (m - iQ) g,
\eqno{(5.10)}$$
which follows from (5.8), into the equality (5.9) gives
$$\eqalign{
&4 m^2  \Vert h(t) \Vert^2_D - 4 m  Re (h(t), g(t))^{}_D + \Vert
g(t) \Vert^2_D \cr
&\qquad{}= 4 m^2 \Vert h(t') \Vert^2_D - 4m  Re (h(t'),
g(t'))^{}_D + \Vert g(t')
\Vert^2_D \cr
&\qquad\qquad{}+ \int^t_{t'}  Re \big(4m  h(s) - 2 g(s),  -i
\gamma^0  ((m - iQ)  g) (s)\big)^{}_D\,  ds. \cr
}$$
This equality proves the equality of statement ib) since we obtain that
$$2  Re \big(g(s),i \gamma^0 ((m - iQ) g) (s)\big)^{}_D = {d \over ds}
 \Vert g(s) \Vert^2_D,$$
for $g \in C^1 (\Rrm^+, D) \cap C^0 (\Rrm^+, (1 - \Delta)^{- 1/2} D)$,
the operator
$$i \gamma^0  m + \sum^3_{j=1}  \gamma^0  \gamma^j
 \partial_j + i \gamma^0  \gamma^\mu  G_\mu (s)$$
being skew-adjoint on $D$ with domain $(1 - \Delta)^{- 1/2}  D$.

To prove the inequality of statement iia) we note that, according to the
introductory remarks of this chapter:
$$h(t) = w (t,t')  h(t') + \int^t_{t'}  w(t,s) (-i \gamma^0)
 g(s)  ds. \eqno{(5.11)}$$
Since $w(t_1, t_2), t_1, t_2 \geq 0$ is unitary in $D$ it follows that
$$\big\vert \Vert h(t) \Vert^{}_D - \Vert h(t')
\Vert^{}_D \big\vert \leq \int^{t'}_t
\Vert g(s) \Vert^{}_D  ds, \quad 0 \leq t \leq t',$$
which proves statement iia).

To prove iib) let $g^{(2)}_n \in S (\Rrm^4, \Crm^4),  n \geq 0$, be a
sequence such that
$$\sup_{0 \leq s \leq T}  \Big(\Vert g^{(2)}_n (s) - g^{(2)} (s) \Vert^{}_D +
\int^T_0  \sum_{0 \leq \mu \leq 3}  \Vert (\partial_\mu (g^{(2)}_n -
g^{(2)})) (s) \Vert^{}_D  ds\Big) \fl 0,$$
when $n \fl \infty$ for every $T \geq 0$. Since
$\Vert G (s) \Vert^{}_{L^\infty (\Rrm^3, \Rrm^4)} \leq C
\Vert (1 - \Delta)^{1/2}  (G(s), 0)) \Vert^{}_{M^1}$ it
then follows from the hypothesis of statement iib) that
$$\sup_{0 \leq s \leq T}  \Big(\Vert g^{(2)}_n (s) - g^{(2)} (s) \Vert^{}_D +
\int^T_0  \Vert ((m - iQ) (g^{(2)}_n - g^{(2)})) (s) \Vert^{}_D
ds\Big) \fl 0$$
when $n \fl \infty$ for every $T \geq 0$ and that $(m - iQ)  g^{(2)}_n
\in C^0 (\Rrm^+, D)$. Let $h^{(2)}_n \in C^0 (\Rrm^+, D)$
be the unique solution of
$(m + iQ)  h^{(2)}_n = g^{(2)}_n,  h^{(2)}_n (t_0) = 0$.
If $f^{}_n = (m - iQ)  h^{(2)}_n$, then $f^{}_n = 2m    h^{(2)}_n -
g^{(2)}_n$ and $(m + iQ)  f^{}_n = (m - iQ)  g^{(2)}_n$, according
to (5.10), so $(m + iQ)  (h^{(2)}_n - {(2m)}^{-1}  g^{(2)}_n) =
{(2m)}^{-1}  (m - iQ)g^{(2)}_n$. Let $h^{}_n \in C^0 (\Rrm^+, D)$
be the unique solution of equation $(m + iQ)  h^{}_n = g^{(1)} + g^{(2)}_n,
 h^{}_n (t_0) = h (t_0) \in D$. Then it follows from statement iia) and
the properties of $g^{}_n$ that
$$\Vert h^{}_n (t) - h(t) \Vert^{}_D \leq
\big\vert \int^t_{t_0} \Vert g^{(2)}_n (s) -
g^{(2)} (s) \Vert^{}_D  ds \big\vert,$$
so $\displaystyle{\sup_{0 \leq s \leq T}}  \Vert h^{}_n (t) - h(t) \Vert^{}_D
\fl 0$ when $n \fl \infty$ for every $T \geq 0.$
The definition of $h^{}_n$ gives that
$$\eqalignno{
(m + iQ) (h^{}_n -(2m)^{-1}    g^{(2)}_n) &= (m + iQ) ((h^{}_n - h^{(2)}_n)
+ (h^{(2)}_n - (2m)^{-1}  g^{(2)}_n))\cr
&= g^{(1)} + (2m)^{-1} (m - iQ)  g^{(2)}_n.\cr
}$$
Since $g^{(1)} + (2m)^{-1}  (m - iQ)  g^{(2)}_n \in C^0 (\Rrm^+, D)$
and
$$h^{}_n (t_0) - (2m)^{-1}  g^{(2)}_n (t_0) = h(t_0) - (2m)^{-1}
g^{(2)}_n (t_0) \in D,$$
it follows from statement iia) that
$$\eqalignno{
&\big\vert \Vert h^{}_n (t) - (2m)^{-1}  g^{(2)}_n (t)
\Vert^{}_D - \Vert h^{}_n (t') -
(2m)^{-1}  g^{(2)}_n (t') \Vert^{}_D \big\vert&(5.12) \cr
&\qquad{}\leq \int^{t'}_t  \Vert g^{(1)} (s) + (2m)^{-1} ((m - iQ)  g^{(2)}_n)
(s) \Vert^{}_D\, ds,\quad  0 \leq t \leq t'. \cr
}$$
The inequality in statement iib) follows from the convergence
properties, proved above, of $h^{}_n$ and $g^{(2)}_n$ and by taking
the limit $n\fl\infty$ in inequality
(5.12). This proves the theorem.
\saut
\noindent{\bf Corollary 5.2.}
{\it
Let $F^{(l)}, r^{(l)}, q^{(l)}$,  $l \geq 0$, be a finite number of
functions satisfying conditions (5.4b) and (5.5). Let $Q^{(l)} (t,x) = t
 F^{(l)}_0 (t,x) + \sum_{1 \leq i \leq 3}  x_i
F^{}_i (t,x)$ and let $\xi^D_X$,  $X \in \p$, be defined by (4.81d) and
(4.81e). If $(G, \dot{G}) \in C^0 (\Rrm^+, (1 - \Delta)^{-1}  M^1)$,
 $h (t_0) \in D$, \penalty-10000
 $g^{}_1 \in C^0 (\Rrm^+, D)$ and $0 \leq a^{}_l \leq 1$,
then the unique solution $h \in C^1 (\Rrm^+, (1 - \Delta)^{1/2}  D) \cap
C^0 (\Rrm^+, D)$ of equation (5.4a), with $g = g^{}_1 + \sum_l  \gamma^\mu
 F^{(l)}_\mu  r^{(l)}$ satisfies:

\penalty-5000
$$\eqalignno{
&\big\vert \Vert h(t) - {(2m)}^{-1}  \sum_l  \gamma^\mu
F^{(l)}_\mu (t)  r^{(l)} (t) \Vert^{}_D - \Vert h(t') - {(2m)}^{-1}
\sum_l  \gamma^\mu  F^{(l)}_\mu (t')
r^{(l)} (t') \Vert^{}_D \big\vert\cr
&\qquad{}\leq \int^{t'}_t  \Vert g^{}_1 (s) \Vert^{}_D  ds
+ {(2m)}^{-1}  \int^{t'}_t  \Big( 2\sum_l  \big\vert (1+s)^{a^{}_{l}-1}
 \big\vert Q^{(l)} (s) + s F^{(l)}_0 (s) \partial_0 r^{(l)} (s)\cr
&\qquad\qquad{}- \sum_{1 \leq i \leq 3}  F^{(l)}_i (s)  ((\xi^D_{M_{0i}}
 r^{(l)}) (s) - \sigma^{}_{0i}  r^{(l)} (s))  \big\vert^{1 - a^{}_l}
 \big\vert F^{(l)}_\mu (s)\partial^\mu  r^{(l)} (s)
\big\vert^{a^{}_l} \big\vert^{}_D\cr
&\qquad\qquad{}+ \big\vert \sum_l \Big( r^{(l)} (s)
\partial^\mu  F^{(l)}_\mu (s) + {1 \over 2}
 r^{(l)} (s)  \gamma^\mu  \gamma^\nu (\partial_\mu
 F^{(l)}_\nu (s) - \partial_\nu  F^{(l)}_\mu (s))\cr
&\qquad\qquad{}+ i  \gamma^\mu  \gamma^\nu  G_\mu (s)
F^{(l)}_\nu (s)  r^{(l)} (s) + i  \gamma^\mu  F^{(l)}_\mu
(s)  q^{(l)} (s)\Big) \big\vert^{}_D\Big)  ds,\cr
}$$
where $0 \leq t \leq t',  \partial_0$ is the time derivative and
the summation convention is used for repeated upper and lower
indices $\mu$ and
$\nu$.
}\saut
\noindent{\it Proof.}
Let $g^{(2)} = \sum_l  \gamma^\mu  F^{(l)}_\mu
r^{(l)}$. It follows from (5.5) that the hypotheses of statement iib)
of Theorem 5.1 are satisfied, which gives that
$$\eqalignno{
&\big\vert \Vert h(t) -{(2m)}^{-1}   g^{(2)} (t) \Vert^{}_D - \Vert h(t') -
{(2m)}^{-1}  g^{(2)} (t') \Vert^{}_D  \big\vert\cr
&\qquad{}\leq \int^{t'}_t
 \Vert g^{}_1 (s) \Vert^{}_D  ds
+ {(2m)}^{-1}  \int^{t'}_t  \Vert ((m - iQ) g^{(2)}) (s)
\Vert^{}_D  ds.\cr
}$$
The explicit expression of $(m - iQ)  g^{(2)}$ is given by (5.7a). Let
$\gamma^\mu  F^{}_\mu  r$ be one of the terms $\gamma^\mu
F^{(l)}_\mu  r^{(l)}$ in $g^{(2)}$. Consider the term $- 2i  F^\mu
 \partial_\mu  r$ on the right-hand side of (5.7a). Since
$$F^{}_0 (t,x) = (1 + t)^{-1}  \Big(Q (t,x) - \sum_{1 \leq i \leq 3}
x_i  F^{}_i (t,x) + F^{}_0 (t,x)\Big)$$
it follows from definition (4.81b) and (4.81e) of
$\xi^{}_{M_{0i}}$ and $\xi^D_{M_{0i}}$
that
$$\eqalignno{
&F^{}_0 (t,x)  {\partial \over \partial t} - \sum_{1 \leq i \leq 3}
F^{}_i (t,x)  \partial_i\cr
&\qquad{}= (1 + t)^{-1}  (Q (t,x) + F^{}_0 (t,x))  {\partial \over
\partial t} - (1 + t)^{-1}  \sum_{1 \leq i \leq 3}  F^{}_i (t,x)
 (x_i  {\partial \over \partial t} + t  \partial_i)\cr
&\qquad{}= (1 + t)^{-1}  \Big((Q (t,x) + F^{}_0 (t,x))
{\partial \over \partial t} - \sum_{1 \leq i \leq 3}  F^{}_i (t,x)
 (\xi^D_{M_{0i}} - \sigma^{}_{0i})\Big).\cr
}$$
Substitution of this expression into the factor $\vert F^\mu  \partial_\mu
r \vert^{1 - a}$ in inequality
$$\eqalignno{
&\vert (m - iQ)  \gamma^\mu  F^{}_\mu  r \vert \cr
&\quad{}\leq
2 \vert F^\mu  \partial_\mu  r \vert^{1 - a}  \vert
F^\mu  \partial_\mu  r \vert^{a}
+ \vert r  \partial_\mu  F^\mu + {1 \over 2}r
 \gamma^\mu  \gamma^\nu (\partial_\mu  F^{}_\nu - \partial_\nu
 F^{}_\mu)
+ i  \gamma^\mu  \gamma^\nu  G_\mu  F^\nu  r +
i  \gamma^\mu  F^{}_\mu  q \vert\cr
}$$
gives the announced result, which proves the corollary.

In order to use Corollary 5.2 with $L^\infty$-norms of the potentials we need
supplementary {\it decay properties of the Dirac field} $h$ in a conic
neighbourhood of
the light-cone as well as outside the light-cone. This will be proved in the
next
two theorems. To state the results we introduce the following notation:
$$(L_i (t)  f)( x) = x_i (({\cal D} f) (x) +
V (t,x)  f(x)) + t  \partial_i  f, \quad  t \geq 0,
 1 \leq i \leq 3, \eqno{(5.13{\rm a})}$$
$(l_i (t))(x) = x_i  F (t,x),$
where $f \in D_\infty$, $V \in C (\Rrm^+,  L^\infty (\Rrm^3,
 \hbox{Mat} (4,  \Crm)))$,  $F \in C (\Rrm^+,D)$,
$\hbox{Mat} (4,  \Crm)$ being the space of $4 \times 4$ complex matrices,
$$\eqalignno{
(R^{}_{ij}  f) (x) &= x_i  \partial_j  f(x) - x_j
\partial_i  f(x),\quad  1 \leq i \leq 3,  1 \leq j \leq 3;
&(5.13{\rm b})\cr
M_{n,t} (f) &= \Big(\Vert q^{n/2}_t  f \Vert^2_D + \sum_{1 \leq i \leq 3}
 \Vert q^{n/2}_t  \partial_i  f \Vert^2_D + \sum_{1 \leq i < j \leq 3}
 \Vert q^{n/2}_t  R^{}_{ij}  f \Vert^2_D &(5.13{\rm c})\cr
&\qquad{}+ \sum_{1 \leq i \leq 3}  \Vert q^{n/2}_t  (L_i (t)  f +
l_i (t)) \Vert^2_D\Big)^{1/2},\quad  t \geq 0,  n \geq 0,\cr
q_t (x) &= (1+t)  (1+ \big\vert t - \vert x \vert \big\vert)^{-1}.\cr
}$$
\saut
\noindent{\bf Theorem 5.3.}
{\it
If $V \in C^0 (\Rrm^+,  L^\infty (\Rrm^3,  \hbox{\rm Mat} (4,\Crm))),
F \in C^0 (\Rrm^+, D)$ and $n \in \Nrm$ then there exists $C_n > 0$
independent of $t,f,V,F$ such that
$$\eqalignno{
\Vert q^{(n+1)/2}_t f \Vert^{}_D &\leq C_n
\Big(\big(1+ \Vert q^{1/2}_t  V(t) \Vert^{}_{L^\infty}
+ (1+t)^{- 1/2}  \Vert V(t) \Vert^{}_{L^\infty}\big)  M_{n,t} (f)\cr
&\qquad+ \Vert q^{1+n/2}_t  F(t) \Vert^{}_D +
\Vert q^{n/2}_t  F(t) \Vert^{}_D\Big),\cr
}$$
for $f \in D_\infty,\quad  t \geq 0$.
}\saut
\noindent{\it Proof.}
Introduce $h^{}_t (x) = (1+ \delta^2
\big\vert t - \vert x \vert \big\vert^2)^{- 1/4},
 K_t = {\cal D} + V(t)$ and $N_{n,t} (f) = (\Vert h^n_t  {\cal D}
 f \Vert^2_D + (1+t)^{-n} M_{n,t} (f)^2)^{1/2}$ and let $\varepsilon > 0,
 \delta > 0$. Introduce also $\tau^{}_t (V) = (1+ \delta t)^{1/2} \Vert h^{}_t
 V(t) \Vert^{}_{L^\infty} +
 (1+ \delta t)^{- 1/2} \Vert V(t) \Vert^{}_{L^\infty},$
and $\lambda^{}_{n,t} (F) =
(1+ \delta t)  \Vert h^{n+2}_t F(t) \Vert^{}_D +
\Vert h^n_t  F(t) \Vert^{}_D.$

If $0 \leq t \leq 1$, then $\vert q_t (x) \vert \leq 2$
so the inequality of the theorem is trivially true with $C \geq 2^{1/2}.$
For a given $f \in D_\infty$ let now
$T_+ (f)$ (resp. $T_- (f)$) be the set of real
numbers $t \geq 1$ such that
$$\Vert h^n_t  {\cal D}  f \Vert^2_D + \sum_{1 \leq i \leq 3}
 \Vert h^n_t (L_i (t)  f + l_i (t)) \Vert^2_D - 2 \varepsilon t
 \Vert h^{n+1}_t  f \Vert^2_D \geq 0  \quad(\hbox{ resp } < 0).
\eqno{(5.14)}$$
Let $t \in T_+ (f)$. Since $t \geq 1$ it follows from (5.14) that
$$(1+t)  \Vert h^{n+1}_t  f \Vert^2_D \leq \varepsilon^{-1}
 N_{n,t} (f)^2,\quad  t \in T_+ (f). \eqno{(5.15)}$$
Let $g^{}_0 = K_t f + F(t),  g^{}_i = \partial_i  f,  1 \leq i \leq 3,
 g = (g^{}_1, g^{}_2, g^{}_3)$ and let $\vert g \vert = (\sum_{1 \leq i \leq 3}
 \vert g^{}_i \vert^2)^{1/2}$. Equality
$$(t- \vert x \vert)  \vert {\cal D} f \vert = t  (\vert {\cal D} f \vert -
\vert g \vert) - (\vert x \vert  \vert {\cal D} f \vert - t
\vert g \vert)$$
gives
$$\eqalignno{
\big\vert t - \vert x \vert \big\vert  \vert {\cal D}
f \vert &\geq t (\vert {\cal D} f \vert -
\vert g\vert) - \vert x {\cal D} f + t g \vert\cr
&\geq t (\vert {\cal D} f \vert - \vert g \vert) -
\vert x  g^{}_0 + t g \vert -
\big\vert t - \vert x \vert \big\vert  \vert V(t)
\vert  \vert f \vert\cr
&\qquad{}- t  \vert V(t) \vert  \vert f \vert -
\big\vert t - \vert x \vert \big\vert
\vert F(t) \vert - t \vert F(t) \vert.\cr
}$$
It follows that
$$\eqalignno{
&2 (1 + \delta^2 \big\vert t - \vert x \vert \big\vert^2)^{1/2}
\vert {\cal D} f \vert^2\cr
&\qquad{}\geq \vert {\cal D} f \vert^2 + \delta t  (\vert {\cal D} f \vert^2 -
\vert {\cal D} f \vert  \vert g \vert)
- \delta  \vert {\cal D} f \vert  \vert x  g^{}_0 + tg \vert\cr
&\qquad\qquad{}-\delta  \big\vert t - \vert x \vert \big\vert
(\vert V(t) \vert
\vert f \vert  \vert {\cal D} f \vert + \vert F(t) \vert
\vert {\cal D} f \vert)
- \delta t  (\vert V(t) \vert  \vert f \vert\vert
{\cal D} f \vert + \vert F(t) \vert  \vert {\cal D} f \vert).\cr
}$$
Since $0 \leq h^{}_t \leq 1$ and $\vert {\cal D} f \vert^2 -
\vert {\cal D} f \vert
 \vert g \vert \geq {1 \over 2}  (\vert {\cal D} f \vert^2 -
\vert g \vert^2)$ we get
$$\eqalignno{
&2 h^{2n}_t  \vert {\cal D} f \vert^2&(5.16)\cr
&\qquad{}\geq {1 \over 2} h^{2n+2}_t (1+ \delta t)
(\vert {\cal D} f \vert^2 - \vert g \vert^2)
- \delta  h^{2n}_t  \vert {\cal D} f \vert  \vert x  g^{}_0 + t g \vert\cr
&\qquad\qquad{}- h^{2n}_t (\vert V(t) \vert  \vert f \vert
 \vert {\cal D} f \vert + \vert F(t) \vert  \vert {\cal D} f \vert)
- \delta  t (h^{2n+2}_t \vert V(t) \vert  \vert {\cal D} f \vert
 \vert f \vert + h^{2n+2}_t \vert F(t) \vert  \vert {\cal D} f \vert).\cr
}$$
Integration of this inequality over $\Rrm^3$ and Schwarz inequality give that
$$\eqalignno{
&2 \Vert h^n_t  {\cal D} f \Vert^2_D &(5.17)\cr
&\qquad{}\geq {1 \over 2}(1+ \delta t)
 \Big(\Vert h^{n+1}_t  {\cal D} f \Vert^2_D - \sum_{1 \leq i \leq 3}
 \Vert h^{n+1}_t  \partial_i f \Vert^2_D\Big) \cr
&\qquad\qquad{}- \delta  \Vert h^n_t  {\cal D} f
\Vert^{}_D\Big(\sum_{1 \leq i \leq 3}
 \Vert h^n_t  (L_i (t)  f + l_i (t))\Vert^2_D\Big)^{1/2}\cr
&\qquad\qquad{}- (1+ \delta t)  \Big(\Vert h^{}_t  V(t) \Vert^{}_{L^\infty}
\Vert h^{n+1}_t  f \Vert^{}_D  \Vert h^n_t  {\cal D} f \Vert^{}_D
+ \Vert h^{n+2}_t  F (t) \Vert^{}_D  \Vert h^n_t  {\cal D} f \Vert^{}_D\Big)\cr
&\qquad\qquad{}- \Big(\Vert V(t) \Vert^{}_{L^\infty}  \Vert h^n_t
{\cal D} f \Vert^{}_D
 \Vert h^n_t  f \Vert^{}_D + \Vert h^n_t  F(t) \Vert^{}_D
 \Vert h^n_t  {\cal D} f \Vert^{}_D\Big).\cr
}$$
According to the definitions of $N_{n,t},  \tau^{}_t$ and $\lambda^{}_{n,t}$
we obtain, using $ab \leq {1 \over 2}  (a^2 + b^2)$,
$$\eqalignno{
&{1 \over 2}  (1+ \delta t)  \Big(\Vert h^{n+1}_t
{\cal D} f \Vert^2_D - \sum_{1 \leq i \leq 3}  \Vert h^{n+1}_t
\partial_i f \Vert^2_D\Big)&(5.18)\cr
&\qquad{}- \delta  \Vert h^n_t  {\cal D} f \Vert^{}_D   \Big(\sum_
{1 \leq i \leq 3}  \Vert h^n_t  (L_i (t)  f + l_i (t))
\Vert^2_D\Big)^{1/2}\cr
&\qquad\qquad{}\leq {5 \over 2}  N_{n,t} (f)^2 + {1 \over 2}
\lambda^{}_{n,t}(F)^2 + \tau^{}_t (V)    \Vert h^n_t  {\cal D} f \Vert^{}_D
(1+ \delta t)^{1/2}  \Vert h^{n+1}_t  f \Vert^{}_D.\cr
}$$
Let $t \in T_- (f)$. It then follows from inequality (5.14)
and inequality (5.18) that
$$\eqalignno{
&(1 + \delta t) \Big( \Vert h^{n+1}_t  {\cal D} f \Vert^2_D -
\sum_{1 \leq i \leq 3}  \Vert h^{n+1}_t  \partial_i f \Vert^2_D -
2 \varepsilon  \Vert h^{n+1}_t  f \Vert^2_D\Big)& (5.19)\cr
&\qquad{}\leq 5  N_{n,t} (f)^2 + \lambda^{}_{n,t} (F)^2 + 2  \tau^{}_t (V)
 \Vert h^n_t  {\cal D} f \Vert^{}_D  (1+ \delta t)^{1/2}
 \Vert h^{n+1}_t  f \Vert^{}_D.\cr
}$$
Let $j \in C^\infty (\Rrm^+),  0 \leq j (y) \leq 1,  j (y) = 0$
for $0 \leq y \leq 1/4$ and $j (y) = 1$ for $y \geq 1/2$. Let $\varphi^{}_t (x)
=
h^{2n+2}_t (x)  j (\vert x \vert / t)$ and $\psi_t (x) = h^{2n+2}_t (x)
 (1 - j (\vert x \vert / t)),  t > 0$. Since $0 \leq \psi_t (x)
\leq (2)^{1/2}  (1+ \delta t)^{-1}  h^{2n} (x)$, we obtain
that
$$\eqalignno{
&(1+ \delta t)  \Big(\Vert h^{n+1}_t  {\cal D} f \Vert^2_D -
\sum_{1 \leq i \leq 3}  \Vert h^{n+1}_t  \partial_i f \Vert^2_D\Big)\cr
&\qquad{}= (1+ \delta t)  \Big(({\cal D} f,
(\varphi^{}_t + \psi_t) {\cal D} f)^{}_D -
\sum_{1 \leq i \leq 3}  (\partial_i f,  (\varphi^{}_t + \psi_t)
\partial_i f)^{}_D\Big)\cr
&\qquad{}\geq (1+ \delta t)  \Big(({\cal D} f,  \varphi^{}_t
{\cal D} f)^{}_D - \sum_{1 \leq i \leq 3}  (\partial_i f,  \varphi^{}_t
 \partial_i f)^{}_D\Big)\cr
&\qquad\qquad{}- 2^{1/2} \Big(\Vert h^n_t  {\cal D} f
\Vert^2_D + \sum_{1 \leq i \leq 3}
 \Vert h^n_t  \partial_i f \Vert^2_D\Big).\cr
}$$
This inequality and inequality (5.19) give
$$\eqalignno{
&(1+ \delta t)  \Big(({\cal D} f,       \varphi^{}_t  {\cal D} f)^{}_D -
\sum_{1 \leq i \leq 3}  (\partial_i f,  \varphi^{}_t
\partial_i f)^{}_D - 2 \varepsilon  \Vert h^{n+1}_t f \Vert^2_D\Big)
&(5.20)\cr
&\qquad{}\leq C \Big(N_{n,t} (f)^2 +
\lambda^{}_{n,t} (F)^2 + 2\tau^{}_t (V)  \Vert h^n_t
 {\cal D} f \Vert^{}_D  (1+ \delta t)^{1/2}  \Vert h^{n+1}_t
 f \Vert^{}_D\Big),\quad  t \in T_- (f),\cr
}$$
where $C$ is a constant independent of $f,V,F$.
The function $\varphi^{}_t$ is $C^\infty$ and bounded together
with its derivatives for $t > 0$, so $\varphi^{}_t  {\cal D} f
\in D_\infty$. We add the operator
${\cal D}  \varphi^{}_t  {\cal D} = \varphi^{}_t  {\cal D}^2 +
[{\cal D},  \varphi^{}_t]       {\cal D}$ and its adjoint. Since ${\cal D}^2 =
\Delta - m^2$ and since ${\cal D}$ is skew-adjoint we obtain:
$${\cal D}  \varphi^{}_t  {\cal D} = {1 \over 2}  \big((\Delta -
m^2)  \varphi^{}_t + \varphi^{}_t  (\Delta - m^2) - [{\cal D}, [{\cal D},
 \varphi^{}_t]]\big)$$
and similarly we obtain that
$$\sum_{1 \leq i \leq 3}  \partial_i  \varphi^{}_t
\partial_i = {1 \over 2}  \Big(\Delta  \varphi^{}_t + \varphi^{}_t
\Delta - \sum_{1 \leq i \leq 3}  [\partial_i, [\partial_i,
\varphi^{}_t]]\Big).$$
These two equalities and inequality (5.20) give, when $t \in T_- (f)$:
$$\eqalignno{
&(1+ \delta t)  \big((m^2 - 2 \varepsilon)  \Vert h^{n+1}_t
f \Vert^2_D - m^2  (f,  \psi_t  f)^{}_D\big) &(5.21)\cr
&\qquad{} +(1+ \delta t)  {1 \over 2}   \big(f,  [{\cal D}, [
{\cal D},  \varphi^{}_t]] - \sum_{1 \leq i \leq 3}  [\partial_i,
[\partial_i,  \varphi^{}_t]] f\big)^{}_D\cr
&\qquad\qquad{}\leq C \Big(N_{n,t} (f)^2 + \lambda^{}_{n,t} (F)^2 + 2
\tau^{}_t (V) \Vert h^n_t  {\cal D} f \Vert^{}_D  (1+ \delta t)^{1/2}
\Vert h^{n+1}_t f \Vert^{}_D\Big).\cr
}$$
We shall calculate the commutators in expression (5.21). Since $[\partial_i,
\varphi^{}_t] = (\partial_i  \varphi^{}_t)$ we get
$$\sum_{1 \leq i \leq 3}  [\partial_i, [\partial_i,  \varphi^{}_t]] =
(\Delta  \varphi^{}_t)$$
and
$$
[{\cal D}, [{\cal D},  \varphi^{}_t]] = \sum_{\scr 1 \leq i \leq 3\atop\scr
1 \leq j \leq 3} [\gamma^0  \gamma^i  \partial_i,
\gamma^0  \gamma^j  (\partial_j  \varphi^{}_t)]
- [i    \gamma^0 m,  \sum_{1 \leq j \leq 3}  \gamma^0
 \gamma^j  (\partial_j  \varphi^{}_t)].$$
The last equalities and equalities $[\gamma^0,  \gamma^0  \gamma^j] =
2 \gamma^j,  1 \leq j \leq 3,$
$$\sum_{i,j}  [\gamma^0  \gamma^i  \partial_i,
\gamma^0  \gamma^j  (\partial_j  \varphi^{}_t)] = (\Delta
 \varphi^{}_t) + \sum_{i,j}  \gamma^i  \gamma^j
((\partial_i  \varphi^{}_t)  \partial_j - (\partial_j
\varphi^{}_t)  \partial_i)$$
and
$$
\partial_i  \varphi^{}_t = x_i  \vert x \vert^{-2}
\sum_{1 \leq j \leq 3}  x_j  \partial_j  \varphi^{}_t,$$
which follows from the spherical symmetry of $\varphi^{}_t$, give that
$$\eqalignno{
&\Big([{\cal D}, [{\cal D},  \varphi^{}_t]] f - \sum_{1 \leq i \leq 3}
[\partial_i, [\partial_i,  \varphi^{}_t]] f\Big)  (x)&(5.22)\cr
&\qquad{}= - 2im  \nu_t (x)  \sum_{1 \leq j \leq 3}  \gamma^j
\vert x \vert^{-1}  x_j  f(x)
+ 2  \vert x \vert^{-1}  \nu_t (x)  \sum_{1 \leq i<j \leq 3}
 \gamma^i  \gamma^j  (R^{}_{ij} f) (x), \quad t > 0,\cr
}$$
where $\nu_t (x) = \sum_{1 \leq l \leq 3}  \vert x \vert^{-1}
x_l  (\partial_l  \varphi^{}_t) (x)$. We observe that $\nu_t (x) = 0$
for $4 \vert x \vert \geq t > 0$ and $2 \vert x \vert \leq t$, according to the
definition of $\varphi^{}_t$. Since $\vert \sum  \gamma^j       x_j \vert =
\vert x \vert$ we obtain from (5.22) that
$$\eqalignno{
&\big\vert {1 \over 2} \big (f,  [{\cal D}, [{\cal D},
\varphi^{}_t]]f - \sum_{1 \leq i \leq 3}  [\partial_i,  [\partial_i,
 \varphi^{}_t]]f\Big)^{}_D \big\vert\cr
&\qquad{}\leq m  \Vert h^{-2-2n}_t  \nu_t \Vert^{}_{L^\infty}
\Vert h^{n+1}_t  f \Vert^2_D\cr
&\qquad\qquad{}+ 3^{1/2}  4     t^{-1}  \Vert h^{-2n}_t
\nu_t \Vert^{}_{L^\infty}  \Vert h^n_t  f \Vert^{}_D
\Big(\sum_{1 \leq i < j \leq 3}  \Vert h^n_t  R^{}_{ij}  f
\Vert^2_D\Big)^{1/2},\quad  t > 0.\cr
}$$
A direct calculation gives
$$\nu_t (x) = t^{-1}  h^{2+2n}_t  (x)  j' (\vert x \vert / t) +
(1+n)  \delta^2  (t - \vert x \vert)  h^{6+2n}_t
 (x)  j (\vert x \vert / t),$$
where $j'$ is the derivative of $j$. It follows that $\vert h^{-2n}_t
\nu_t  (x) \vert$ and $\vert h^{-2-2n}_t  \nu_t  (x) \vert$
are bounded by $(1+n) \delta + t^{-1}  \Vert j' \Vert^{}_{L^\infty}.$
This gives
$$\eqalignno{
&{1 \over 2} \big\vert  \Big(f,  [{\cal D}, [{\cal D},
\varphi^{}_t]]f - \sum_{1 \leq i \leq 3}  [\partial_i,  [\partial_i,
 \varphi^{}_t]]f\Big)^{}_D  \big\vert &(5.23)\cr
&\qquad{}\leq m  (1+n) \delta   \Vert h^{1+n}_t  f \Vert^2_D +
t^{-1}  C  N_{n,t}  (f)^2,\quad  t > 0,\cr
}$$
where the constant $C$ depends only on the function $j$ and on $n$. Since $(1+
\delta t)  (f,  \psi_t  f)^{}_D \leq 2^{3/2}
\Vert h^n_t  f \Vert^2_D$ it follows from inequalities (5.21) and (5.23)
that, with $t \in T_- (f)$,
$$\eqalignno{
&(1+ \delta t)  (m^2 - 2 \varepsilon - m (1+n) \delta)  \Vert
h^{n+1}_t  f \Vert^2_D&(5.24)\cr
&\qquad\leq C^2  (N_{n,t}  (f)^2 + \lambda^{}_{n,t}  (F)^2) +
2 C  N_{n,t}  (f)  \tau^{}_t  (V)
(1+ \delta t)^{1/2}  \Vert h^{n+1}_t  f \Vert^{}_D,\cr
}$$
where $C$ is a new constant. Let $\varepsilon = m^2 / 8$ and $(1+n) \delta =
m/4$.
It then follows from inequality (5.24) that (with a new constant $C$):
$$\eqalignno{
&(1+ \delta t)  \Vert h^{n+1}_t  f \Vert^2_D\cr
&\qquad{}\leq C^2 (N_{n,t}(f)^2 + \lambda^{}_{n,t} (F)^2)
+ 2C\tau^{}_t(V)  N_{n,t}  (f)
(1+ \delta t)^{1/2}  \Vert h^{n+1}_t  f \Vert^{}_D,
\quad t \in T_- (f),\cr
}$$
which gives
$$(1+ \delta t)^{1/2}  \Vert h^{n+1}_t  f \Vert^{}_D \leq 2C \big((1+
\tau^{}_t  (V))  N_{n,t}  (f) + \lambda^{}_{n,t}
(F)\big), \eqno{(5.25)}$$
with $(1+n) \delta = m/4$ and $\varepsilon = m^2 / 8$.

Since $T_+ (f)  \cup  T_- (f) = [1, \infty[$ it follows from
inequality (5.15) with $\varepsilon = m^2 / 8$, from inequality (5.25) and from
the definition of $\tau^{}_t$ and $\lambda^{}_{n,t}$ that
$$\eqalignno{
\Vert q^{(n+1)/2}_t  f \Vert^{}_D &\leq C'_n \big((1 + \tau^{}_t  (V))
 (1+t)^{n/2}  N_{n,t}  (f) + (1+t)^{n/2}
\lambda^{}_{n,t}  (F)\big)\cr
&\leq C''_n \big((1+ \Vert q^{1/2}_t  V(t) \Vert^{}_{L^\infty} + (1+t)^{-1/2}
 \Vert V(t) \Vert^{}_{L^\infty})  M_{n,t}  (f)\cr
&\qquad{}+ \Vert q^{1+n/2}_t  F(t) \Vert^{}_D + \Vert q^{n/2}_t  F(t)
\Vert^{}_D\big),\cr
}$$
for some constants $C'_n,  C''_n$. We have here used the fact that $(1+t)^{n/2}
 N_{n,t}  (f) \leq C_n  M_{n,t}  (f)$ for
some $C_n$. This proves the theorem.

In order to state the next theorem we introduce the following notation:
$$\eqalignno{
M^{(n)}_t  (f) &= \Big(\Vert r^{n/2}_t  f \Vert^2_D + \sum_{1 \leq i \leq 3}
 \Vert r^{n/2}_t  \partial_i  f \Vert^2_D + \sum_{1 \leq i < j \leq 3}
 \Vert r^{n/2}_t  R^{}_{ij}  f \Vert^2_D&(5.26) \cr
&\qquad{}+ \sum_{1 \leq i \leq 3}  \Vert r^{n/2}_t  (L_i (t)
f + l_i (t)) \Vert^2_D\Big)^{1/2},\quad n \geq 0,  t \geq 0, \cr
}$$
where $L_i (t),  l_i (t)$ and $R^{}_{ij}$ are defined in (5.13a) and (5.13b)
and where $r^{}_t (x) = 0$ if $\vert x \vert < t$ and $r^{}_t (x) = 1 +
\vert x \vert$
if $\vert x \vert \geq t$.
\saut
\noindent{\bf Theorem 5.4.}
{\it
If $V \in C^0 \big(\Rrm^+,  L^\infty (\Rrm^3,  \hbox{\rm Mat} (4, \Crm))\big)$,
 $F \in C^0 (\Rrm^+, D)$ and $\tau^{}_t (V) ={}$ $\sup_{x \in \Rrm^3}
((1 + \vert x \vert + t)^{1/2}  \vert V (t,x) \vert) < \infty$, then
there exist constants $C_n$ independent of $t,V,f, F$, such that
$$\eqalign{
&\Vert r^{(n+1)/2}_t  f \Vert^{}_D \cr
&\qquad{}\leq C_n (1 + \tau^{}_t (V))
\big(M_{n,t}  (f) + M^{(n)}_t  (f) + \Vert q^{1+n/2}_t
F(t) \Vert^{}_D + \Vert r^{1+n/2}_t  F(t) \Vert^{}_D\big)\cr
}$$
for $n \geq 0$ and $f \in D_\infty.$
}\saut
\noindent{\it Proof.}
We first consider the case where $F = 0$. Let $\psi_t \in C^\infty (\Rrm^3),
t \geq 0$, be a cut-off function defined by $\psi_t (x) = u
(\vert x \vert - t),$
where $u \in C^\infty (\Rrm^+),  0 \leq u(y) \leq 1$ for $y \in \Rrm^+,
 u(y) = 0$ for $0 \leq y \leq 1$ and $u(y) = 1$ for $2 \leq y$. Let
$K_t = {\cal D} + V(t),  g^{}_0 = K_t  f,  g^{}_i = \partial_i
 f,  1 \leq i \leq 3,  g = (g^{}_1, g^{}_2, g^{}_3)$, where
$f \in D_\infty.$

Since in the support of $r^{(n+1)/2}_t  (1- \psi_t)$ we have
$0 \leq t \leq \vert x \vert \leq 2+t$, it follows that
$$\eqalignno{
\vert r^{}_t (x)^{(n+1)/2}  (1 - \psi_t (x)) \vert &\leq C'_n ((1+t) (1+
\big\vert t - \vert x \vert \big\vert)^{-1})^{(n+1)/2} \cr
&= C'_n  q_t (x)^{(n+1)/2},\cr
}$$
where $q_t$ is defined in (5.13c). This gives for some constant $C_n$ that
$$\Vert r^{(n+1)/2}_t  (1 - \psi_t)  f \Vert^{}_D \leq C_n
\Vert q^{(n+1)/2}_t  f \Vert^{}_D,\quad  t \geq 0. \eqno{(5.27)}$$
Since $\vert x \vert - t \geq 1$ and $r^{}_t (x) = 1 + \vert x \vert$
in the support of $\psi_t$ it follows that
$$\eqalignno{
r^n_t  \psi^2_t  (\vert {\cal D} f \vert^2 + \vert x
g^{}_0 + t g \vert^2) &\geq 2 r^n_t  \psi^2_t  \vert {\cal D} f \vert
 \vert x  g^{}_0 + tg \vert\cr
&\geq 2  r^n_t  \psi^2_t  \vert {\cal D} f \vert
 \big(\vert x \vert (\vert {\cal D} f \vert - \vert g \vert) + (\vert x \vert -
t)
 \vert g \vert - \vert x \vert  \vert V(t) \vert
\vert f \vert\big)\cr
&\geq r^n_t  \psi^2_t   \vert x \vert  \big(\vert{\cal D} f \vert^2 -
\vert g \vert^2 - 2 \vert x \vert  \vert V(t) \vert  \vert {\cal D} f \vert
 \vert f \vert\big)\cr
&\geq {1 \over 2}  r^{n+1}_t  \psi^2_t  (\vert {\cal D} f \vert^2 -
\vert g \vert^2) - 2  r^{n+1}_t  \psi^2_t  \vert V(t) \vert
 \vert f \vert  \vert {\cal D} f \vert.\cr
}$$
Integration of this inequality over $\Rrm^3$ gives
$$\eqalignno{
&\Vert r^{(n+1)/2}_t  \psi_t  {\cal D} f \Vert^2_D - \sum_{1 \leq i \leq 3}
 \Vert r^{(n+1)/2}_t  \psi_t  \partial_i f \Vert^2_D
&(5.28)\cr
&\qquad{}\leq C^2  M^{(n)}_t  (f)^2 + 2C \Vert r^{1/2}_t  V(t)
\Vert^{}_{L^\infty}  \Vert r^{n/2}_t  {\cal D} f \Vert^{}_D
\Vert r^{(n+1)/2}_t  \psi_t  f \Vert^{}_D\cr
&\qquad{}\leq C^2_n  M^{(n)}_t  (f)^2 + 2 C_n   \Vert r^{1/2}_t
 V(t) \Vert^{}_{L^\infty}  M^{(n)}_t  (f)
\Vert r^{(n+1)/2}_t  \psi_t  f \Vert^{}_D,\cr
}$$
where $C$ and $C_n$ are constants depending only on the mass  $m$.

Let $\varphi^{}_t = \psi^2_t  r^{n+1}_t$. Following the proof of Theorem 5.3
from (5.20) to (5.22) we obtain, since $\varphi^{}_t$ is $C^\infty$,
$$\eqalignno{
&\Vert r^{(n+1)/2}_t  \psi_t  {\cal D} f \Vert^2_D - \sum_{1 \leq i \leq 3}
 \Vert r^{(n+1)/2}_t  \psi_t  \partial_i f \Vert^2_D\cr
&\qquad{}\geq m^2 \Vert r^{(n+1)/2}_t   \psi_t  f \Vert^2_D - m
(f, \nu^{}_t  f)^{}_D -
\sum_{1 \leq i < j \leq 3}  \vert (f,  \vert x \vert^{-1}
 \nu_t  R^{}_{ij}  f)^{}_D \vert,\cr
}$$
where $\nu^{}_t (x) = 0$ for $\vert x \vert \leq 1+t$ and $\nu_t
(x) = \sum_{1 \leq l \leq 3}
  \vert x \vert^{-1}
x_l  (\partial_l  \varphi^{}_t) (x)$ otherwise. It follows from the definition
of
$\varphi^{}_t$ that
$$\nu_t (x) = (n+1)     r^n_t  (x)  \psi^2_t (x) +
r^{n+1}_t  (x)  \psi_t  (x)  u' (\vert x \vert -t),$$
where $u'$ is the derivative of $u$. Because supp $u'\subset [1,2]$, we
have that
$$\eqalignno{
\vert r^{n+1}_t (x)  \psi_t (x)  u' (\vert x \vert - t) \vert
&\leq C_n (1+t)^{n+1}  (1 + \big\vert t - \vert x \vert \big\vert)^{-n-1}\cr
&= C_nq_t (x)^{n+1}\cr
}$$
for some constant $C_n$, so $\vert \nu_t \vert \leq (n+1)  r^n_t
\psi^2_t + C_n  q^{n+1}_t$. We also have $\vert x \vert^{-1}
\vert \nu_t (x) \vert \leq C_n  r^n_t$ for some constant $C_n$. This
gives
$$\eqalignno{
&\Vert r^{(n+1)/2}_t  \psi_t  {\cal D} f \Vert^2_D - \sum_{1 \leq i \leq 3}
 \Vert r^{(n+1)/2}_t  \psi_t  \partial_i f \Vert^2_D\cr
&\qquad{}\geq m^2  \Vert r^{(n+1)/2}_t  \psi_t  f \Vert^2_D -
C'_n  \Vert q^{(n+1)/2}_t  f \Vert^2_D
- C'_n  \Vert r^{n/2}_t  f \Vert^{}_D  \Big(\sum_{1 \leq i < j \leq 3}
 \Vert r^{n/2}_t  R^{}_{ij}  f \Vert^2_D\Big)^{1/2}.\cr
}$$
Applying the definition of $M^{(n)}_t$ to the last term on the right-hand side
of
this inequality, we obtain
$$\eqalignno{
&\Vert r^{(n+1)/2}_t  \psi_t  {\cal D} f \Vert^2_D - \sum_{1 \leq i \leq 3}
 \Vert r^{(n+1)/2}_t  \psi_t  \partial_i f \Vert^2_D&(5.29)\cr
&\qquad{}\geq m^2  \Vert r^{(n+1)/2}_t  \psi_t  f \Vert^2_D -
C^2_n  (\Vert q^{(n+1)/2}_t  f \Vert^2_D + M^{(n)}_t
(f)^2),\quad  t \geq 0,\cr
}$$
for some constant $C_n$. It follows from inequalities (5.28) and (5.29) that
(with
a new constant $C_n$)
$$\eqalignno{
&\Vert r^{(n+1)/2}_t  \psi_t  f \Vert^2_D \cr
&\qquad{}\leq C^2_n (M^{(n)}_t
 (f)^2 + \Vert q^{(n+1)/2}_t  f \Vert^2_D)
+ 2 C_n  \Vert r^{1/2}_t  V(t) \Vert^{}_{L^\infty}
M^{(n)}_t  (f)  \Vert r^{(n+1)/2}_t  \psi_t
f \Vert^{}_D.\cr
}$$
This inequality shows that (with a new constant $C_n$)
$$\eqalignno{
\Vert r^{(n+1)/2}_t  \psi_t  f \Vert^{}_D &\leq C_n \big(M^{(n)}_t
 (f) + \Vert q^{(n+1)/2}_t  f \Vert^{}_D + \Vert r^{1/2}_t
V(t) \Vert^{}_{L^\infty}  M^{(n)}_t  (f)\big)& (5.30)\cr
&= C_n \big((1+ \Vert r^{1/2}_t  V(t) \Vert^{}_{L^\infty})  M^{(n)}_t
 (f) + \Vert q^{(n+1)/2}_t  f \Vert^{}_D\big).\cr
}$$
It follows from inequalities (5.27) and (5.30) that, for $t\geq0$ and $F=0$,
$$\Vert r^{(n+1)/2}_t  f \Vert^{}_D \leq C_n \big((1 + \Vert r^{1/2}_t
V(t) \Vert^{}_{L^\infty})  M^{(n)}_t  (f) + \Vert q^{(n+1)/2}_t
 f \Vert^{}_D\big). \eqno{(5.31)}$$

To study the case where $F \neq 0$ we stress the dependence of $M^{(n)}_t
(f)$ on $F$ by the notation $M^{(n)}_t  (f,F)$. It follows from definition
(5.26) that
$$M^{(n)}_t  (f,0) \leq C \big(M^{(n)}_t  (f,F) + \Vert r^{1+n/2}_t
 F (t) \Vert^{}_D\big),$$
where $C$ is independent of $n,t,f,F$. This inequality and inequality (5.31)
give
(with new $C_n$):
$$\eqalignno{
&\Vert r^{(n+1)/2}_t  f \Vert^{}_D
\leq C_n \Big((1+ \Vert r^{1/2}_t
V(t) \Vert^{}_{L^\infty})  M^{(n)}_t  (f,F)&(5.32)\cr
&\quad{}+ \Vert q^{(n+1)/2}_t  f \Vert^{}_D + (1 + \Vert r^{1/2}_t  V(t)
\Vert^{}_{L^\infty})  \Vert r^{1+n/2}_t  F(t) \Vert^{}_D\Big).\cr
}$$
It follows from this inequality, Theorem 5.3 and from the definitions of $q_t$
and $r^{}_t$, that
$$\eqalignno{
\Vert r^{(n+1)/2}_t  f \Vert^{}_D &\leq C_n (1 + \sup_{x \in \Rrm^3}
 ((1+t+ \vert x \vert)^{1/2}) \vert V(t,x)\vert )\cr
&\qquad{}\big(M_{n,t}  (f,F) + M^{(n)}_t  (f,F) + \Vert q^{1+n/2}_t
F(t) \Vert^{}_D + \Vert r^{1+n/2}_t  F(t) \Vert^{}_D\big).\cr
}$$
This proves the theorem.

Before applying Theorem 5.3 and Theorem 5.4 to solutions of equation (5.1a), we
note that
$${1 + t + \vert x \vert \over 1 + \big\vert t - \vert x \vert \big\vert}
\leq 2(1 + q_t (x))
\leq 4 {1 + t + \vert x \vert \over 1 + \big\vert t - \vert x \vert \big\vert},
\quad t \geq 0,  x \in \Rrm^3, \eqno{(5.33)}$$
and we introduce the following notation:
$$h^{}_Y = \xi^D_Y  h,\quad  g^{}_Y = \xi^D_Y  g,\quad
G_{Y \mu} = (\xi^M_Y  G)_\mu, \eqno{(5.34{\rm a})}$$
where $Y \in U (\p),    0 \leq \mu \leq 3$ and where $h,g$ and $G$ are
the functions in equation (5.1a).

We now introduce the summation symbol $\suma_{Y_1, \ldots, Y_p}^{Y},
Y \in U(\p).$ Let $\cal V$ be a real vector space and $f: (U(\p)^p)
\rightarrow{\cal V}.$ We define inductively $\suma_{Y_1,\ldots,Y_p}^{Y}
f(Y_1, \ldots, Y_p),$ for $Y \in \Pi'$, by
$$\eqalignno{
&\suma_{Y_1,\ldots,Y_p}^{\un} f(Y_1, \ldots, Y_p) = f(\un, \ldots, \un)
&(5.34{\rm b})\cr
&\suma_{Y_1,\ldots,Y_p}^{XY} f(Y_1, \ldots, Y_p) &(5.34{\rm c})\cr
&\qquad{}= \suma_{1 \leq l \leq p}^{}
\sum_{Z_1, \ldots, Z_p}^{Y} f(Z_1, \ldots, Z_{l-1}, XZ_l,
Z_{l+1}, \ldots, Z_p),
X \in \Pi, XY \in \Pi'\cr
}$$
If $Y\in U(\p)$, we extend this definition by linearity in $Y$.
If ${\cal E} \subset (\Pi')^p$ and if in $\suma_{Y_1, \ldots, Y_p}^{Y}$
$f(Y_1, \ldots, Y_p)$ we add only the elements $f(Y_1,\ldots, Y_p)$
for which $(Y_1, \ldots, Y_p) \in {\cal E}$, the element of
${\cal V}$ so obtained is denoted by
$\suma_{(Y_1, \ldots, Y_p)\in {\cal E}}^{Y} f(Y_1, \ldots, Y_p).$

\saut
\noindent{\bf Theorem 5.5.}
{\it
Let $n \geq 0$, $k \geq 1$,$0 \leq L \leq n + k - 1$ be integers,
let $G_Y \!\!\in\!\! C^0 (\Rrm^+, L^\infty (\Rrm^3, \Rrm^4))$
for $0 \leq \vert Y \vert \leq L$,
 $Y \in \Pi'$ and let $G_Y \in C^0 (\Rrm^+, L^2_{\rm loc} (\Rrm^3, \Rrm^4))$
for $\vert Y \vert \leq n + k - 1$,  $Y \in \Pi'$. Let
$$\eqalign{
\tau^{(l)}_0 (t) &= \sum_{\scr \vert Y \vert \leq
l\atop\scr Y \in \Pi'} (1+t)^{1/2}
\Vert G_Y (t) \Vert^{}_{L^\infty},\quad l \geq 0, \cr
\tau^{(l)}_1 (t) &= \sum_{\scr \vert Y \vert \leq l\atop\scr
 Y \in \Pi'}  \sup_{x \in \Rrm^3} ((1+t+\vert x \vert)^{1/2}
\vert G_Y (t,x) \vert),\quad l \geq 0, \cr
}$$
and let
$$\tau^{}_{i,j} (t) = \sum_{1 \leq p \leq j}  \sum_{l_1+\cdots+l_p=j}
\prod_{1 \leq q \leq p}  \tau^{(l_q)}_i (t),\quad  i = 0, 1, j \geq 0.$$
For $h \in C^0 (\Rrm^+, D)$, let $g = (i  \gamma^\mu  \partial_\mu +
m - \gamma^\mu  G_\mu)  h$, and  let
$$(\lambda^{}_0 (t)) (x) = q_t
(x),  (\lambda^{}_1 (t)) (x) = q_t (x) + r^{}_t (x),\quad t \geq 0,
 x \in \Rrm^3.$$

If $h^{}_Y \in C^0 (\Rrm^+, D)$ for $0 \leq \vert Y \vert \leq n + k,  Y
\in \Pi'$, and if $G_{Y_1 \mu}  h^{}_{Y_2} \in C^0 (\Rrm^+, D)$ for $L+1 \leq
\vert Y_1 \vert \leq n + k - 1$, $0 \leq \vert Y_2 \vert \leq n + k - 2 - L,
Y_1, Y_2 \in \Pi'$, then
$$\eqalign{
&\wp^D_n \big((1 + \lambda^{}_i (t))^{k/2}  h(t)\big)\cr
&\quad{}\leq C'_k \Big(\wp^D_{n+k}  (h(t))^2 + \sum_{0 \leq j \leq k-1}
\Big(\wp^D_{n+j}\big((1+ \lambda^{}_i (t))^{(k+1-j)/2}  g(t)\big)^2\cr
&\quad\quad{}+
\sum_{{\sscr Y \in \Pi' \atop\sscr\vert Y \vert \geq L+1}\atop\sscr
\vert Y \vert = n+j}
 \Vert (1 + \lambda^{}_i (t))^{(k+1-j)/2}  \gamma^\mu
G_{Y \mu} (t)  h^{}_{\un} (t) \Vert^2_D\Big)\Big)^{1/2}\cr
&\quad\quad{}+ C'_{n+k} \Big(\wp^D_{n+k-1} (h(t)) +
\sum_{{\sscr 0 \leq j \leq k-1\atop\sscr Z_1, Z_2 \in \Pi' }\atop
{\sscr\vert Z_1 \vert + \vert Z_2 \vert =n+j
\atop\sscr 1 \leq \vert Z_2 \vert \leq n+j-L-1} }
\Vert (1+ \lambda^{}_i (t))^{(k+1-j)/2} \gamma^\mu  G_{Z_1 \mu} (t)
h^{}_{Z_2} (t) \Vert^{}_D\Big)\cr
&\quad\quad{}+ C'_{n+k}  \sum_{1 \leq l \leq L}  (1+\tau^{}_{i,l} (t))
\Big(\wp^D_{n+k-l}(h(t)) + \sum_{\sscr  0 \leq j \leq k-1\atop\sscr n+j-l\geq0
} \wp^D_{n+j-l}
 \big((1+ \lambda^{}_i (t))^{(k+1-j)/2}  g(t)\big)\cr
&\quad\quad{}+ \sum_{{\sscr 0 \leq j \leq k-1
\atop\sscr Z_1, Z_2 \in \Pi'}
\atop{\sscr \vert Z_2 \vert\leq n+j-L-1
\atop\sscr\vert Z_1 \vert + \vert Z_2 \vert \leq n+j-l }}
\Vert (1+ \lambda^{}_i (t))^{(k+1-j)/2}
 \gamma^\mu  G_{Z_1 \mu} (t)  h^{}_{Z_2} (t) \Vert^{}_D\Big),\cr
}$$
$i = 0,1,  t \geq 0$. The constants $C'_N,  N \geq 1$,
depend only on $\tau^{}_{i,0} (t).$
}\saut
\noindent{\it Proof.}
According to the definition of $g$, we obtain that
$$g^{}_Y = (i \gamma^\mu  \partial_\mu + m)  h^{}_Y - \suma_{Y_1, Y_2}^Y
 \gamma^\mu  G_{Y_1 \mu}  h^{}_{Y_2},\quad  Y \in \Pi',
\eqno{(5.35)}$$
$\vert Y \vert \leq n+k-1$. Since $\Vert (i \gamma^\mu  \partial_\mu + m)
 h^{}_Y (t) \Vert^{}_D \leq C  \wp^D_{n+k}  (h(t)),$
$\Vert G_{Y_1 \mu} (t)  h^{}_{Y_2} (t) \Vert^{}_D \leq \Vert G_{Y_1 \mu} (t)
\Vert^{}_{L^\infty}  \Vert h^{}_{Y_2} (t) \Vert^{}_D$ for $0 \leq
\vert Y_1 \vert \leq L,$
$0 \leq \vert Y_2 \vert \leq n+k-1$ and since $G_{Y_1 \mu}  h^{}_{Y_2} \in
C^0 (\Rrm^+, D)$ for $L+1 \leq \vert Y_1 \vert \leq n+k-1$,
$0 \leq \vert Y_2 \vert \leq n+k-2-L$ according to the hypothesis it
follows that $g^{}_Y \in C^0 (\Rrm^+, D)$
for $\vert Y \vert \leq n+k-1,  Y \in \Pi'.$

For given $Y \in \Pi',  l \geq 1,  \vert Y \vert + l \leq n+k$,
let
$$F = - i \suma_{\scr Y_1, Y_2\atop\scr \vert Y_2 \vert \leq \vert Y \vert -
1}^Y
 \gamma^0  \gamma^\mu  G_{Y_1 \mu}  h^{}_{Y_2}
 - i \gamma^0  g^{}_Y\quad \hbox{and}\quad  V = - i \gamma^0
\gamma^\mu  G_\mu. \eqno{(5.36)}$$
It follows as in the case of $g^{}_Y$ that $F\!\in\! C^0 (\Rrm^+\! , D)$.
Moreover $V\!\! \in\!
C^0 (\Rrm^+\!, L^\infty (\Rrm^3\!,\hbox{\rm Mat} (4, \Crm)))$,
where $\hbox{\rm Mat} (4, \Crm)$ is the
linear space of $4 \times 4$ complex matrices. According to the definition of
$\xi^D_{M_{0i}}$ it follows from (5.34a) and (5.36) that
$$\xi^D_{M_{0i}}  h^{}_Y = t \partial_i  h^{}_Y + x_i  V
h^{}_Y + x_i  F + \sigma^{}_{0i}  h^{}_Y, \quad 1 \leq i \leq 3.
\eqno{(5.37)}$$
It follows from (5.37), Theorem 5.3 and Theorem 5.4, that
$$\eqalignno{
&\Vert q^{l/2}_t  h^{}_Y (t) \Vert^{}_D& (5.38{\rm a})\cr
&\qquad{}\leq C_l \Big((1 + \tau^{}_{0,0} (t))
\wp^D_1 (q^{(l-1)/2}_t  h^{}_Y (t))
+\Vert q^{(l+1)/2}_t  F(t) \Vert^{}_D + \Vert q^{(l-1)/2}  F(t)
\Vert^{}_D\Big)\cr
\noalign{\hbox{\rm and}}
&\Vert r^{l/2}_t  h^{}_Y (t) \Vert^{}_D &(5.38{\rm b})\cr
&\qquad{}\leq C_l (1 + \tau^{}_{1,0} (t))
\Big(\wp^D_1 (q^{(l-1)/2}  h^{}_{\cdot Y} (t))
+ \wp^D_1 (r^{(l-1)/2}_t  h^{}_{\cdot Y} (t))\cr
&\qquad\qquad{}+ \Vert r^{(l+1)/2}_t  F(t)
\Vert^{}_D + \Vert r^{(l-1)/2}_t  F(t) \Vert^{}_D\Big),\cr
}$$
where $\wp^D_1$ is applied to the linear functions $Z \mapsto q^{(l-1)/2}_t
h^{}_{ZY} (t)$ and $Z \mapsto r^{(l-1)/2}_t  h^{}_{ZY} (t)$,
$Z \in \Pi'$.
Using that $(1 + q_t)^{l/2} \leq C_l (1 + q^{l/2}_t)$ and that
$(1+q_t+r^{}_t)^{l/2} \leq C_l (1+q^{l/2}_t + r^{l/2}_t)$ for some
constant $C_l$ it follows from (5.38a),
(5.38b) and the expression (5.36) of $F$ that
$$\eqalignno{
&\Vert (1 + \lambda^{}_i (t))^{l/2}  h^{}_Y (t) \Vert^{}_D&(5.39)\cr
&\qquad{} \leq C_l (1+\tau^{}_{i,0}
(t))  \Big(\wp^D_1 \big((1+\lambda^{}_i (t))^{(l-1)/2}  h^{}_{\cdot Y}
(t)\big)\cr
&\qquad\qquad{}+ \suma_{\scr Y_1,Y_2\atop\scr  \vert Y_2 \vert
\leq \vert Y \vert - 1}^Y \Vert (1 + \lambda^{}_i (t))^{(l+1)/2}
\gamma^\mu  G_{Y_1 \mu} (t)  h^{}_{Y_2} (t) \Vert^{}_D \cr
&\qquad\qquad{}+ \Vert (1 + \lambda^{}_i (t))^{(l+1)/2}  g^{}_Y (t)
\Vert^{}_D\Big), \quad i = 0,1,  t \geq 0,  l \geq 1,\cr
}$$
$Y \in \Pi',  \vert Y \vert + l \leq n + k$ for some constants $C_l$
depending only on $l.$

Let $l = 1$ and $Y = \un$, then it follows from (5.39) that
$$\eqalignno{
\Vert (1+\lambda^{}_i (t))^{1/2}  h^{}_{\un} (t) \Vert^{}_D &= \wp^D_0
\big((1+\lambda^{}_i
(t))^{1/2}  h(t)\big)&(5.40)\cr
&\leq C_1 (1+\tau^{}_{i,0} (t))
\big(\wp^D_1 (h(t)) + \wp^D_0 ((1+\lambda^{}_i (t))g(t))\big),
\quad i = 0,1, t \geq 0,\cr
}$$
which shows that the inequality of the theorem is true for $n = 0,  k = 1.$
Suppose that it is true for $0 \leq n \leq N$ and $k = 1$, for some $N \geq 0.$
Let $Y \in \Pi',  \vert Y \vert = N + 1$, and let $0 \leq L \leq N + 1.$
It follows from (5.39) that
$$\eqalignno{
&\Vert (1+\lambda^{}_i (t))^{1/2}  h^{}_Y (t) \Vert^{}_D& (5.41)\cr
&\qquad{}\leq C_1 (1 + \tau^{}_{i,0}
(t))  \Big(\wp^D_1 (h^{}_{\cdot Y} (t))
+ \suma_{{\sscr Y_1,Y_2\atop\sscr \vert Y_1 \vert \leq L}\atop\sscr
 \vert Y_2 \vert \leq \vert Y \vert - 1}^Y
 8 \tau^{}_{i, \vert Y_1 \vert} (t)  \wp^D_{\vert Y_2 \vert}
\big((1 + \lambda^{}_i (t))^{1/2}  h(t)\big)\cr
&\qquad\qquad{}+ \suma_{\scr Y_1,Y_2\atop\scr   \vert Y_2 \vert \leq N-L}^Y
\Vert (1 + \lambda^{}_i (t))
 \gamma^\mu  G_{Y_1 \mu} (t)  h^{}_{Y_2} (t) \Vert^{}_D
+ \Vert (1 + \lambda^{}_i (t))  g^{}_Y (t) \Vert^{}_D\Big),\cr
}$$
where we have used that $1 + \lambda^{}_0 (t) \leq 2 (1+t)$ and $1 +
(\lambda^{}_1 (t)) (x) \leq 3 (1 + t + \vert x \vert)$ and where we have
used that $\vert Y_1 \vert \geq L+1$ if and only if $\vert Y_2 \vert
\leq N-L$ and that $N-L \leq \vert Y \vert - 1.$
Let
$$I_Y (t) = C_1 (1 + \tau^{}_{i,0} (t))\hskip-.4cm
\suma_{{\sscr Y_1,Y_2\atop\sscr \vert Y_1 \vert \leq L}
\atop\sscr \vert Y_2 \vert \leq \vert Y \vert - 1}^Y \hskip-.4cm
8 \tau^{}_{i, \vert Y_1 \vert} (t)  \wp^D_{\vert Y_2 \vert}  \big((1+
\lambda^{}_i (t))^{1/2} h(t)\big), \eqno{(5.42)}$$
be the second term on the right-hand side of inequality (5.41). Since
$\vert Y \vert
= N+1$, we obtain that
$$I_Y (t) \leq C''_N  \sum_{1 \leq j \leq L}  \tau^{}_{i,j} (t)
 \wp^D_{N+1-j}  \big((1+ \lambda^{}_i (t))^{1/2}  h(t)\big),
\eqno{(5.43)}$$
where $C''_N$ is a constant depending only on $\tau^{}_{i,0} (t)$.
Inequality (5.41) and the definition of $I_Y (t)$ give that
$$\eqalignno{
&\Vert (1 + \lambda^{}_i (t))^{1/2}  h^{}_Y (t) \Vert^2_D\cr
&\quad{} \leq C^2
C^2_1 (1 + \tau^{}_{i,0} (t))^2  \Big(\wp^D_1 (h^{}_{\cdot Y} (t))^2\cr
&\quad\quad{}+ \Vert (1 + \lambda^{}_i (t))     \gamma^\mu  G_{Y \mu} (t)
 h^{}_{\un} (t) \Vert^2_D + \Vert (1 + \lambda^{}_i (t))  g^{}_Y (t)
\Vert^2_D\Big)\cr
&\quad\quad{}+ C^2 \Big(C_1 (1 + \tau^{}_{i,0} (t))
\hskip-.4cm\suma_{\scr Y_1,Y_2\atop\scr 1 \leq
\vert Y_2 \vert \leq N-L}^Y\hskip-.4cm  \Vert (1 + \lambda^{}_i (t))
\gamma^\mu
 G_{Y_1 \mu} (t)  h^{}_{Y_2} (t) \Vert^{}_D+ I_Y (t)\Big)^2,\cr
}$$
for some constant $C$ (independent of all the variables in the inequality).
Summation over $Y \in \Pi',  \vert Y \vert = N+1$ and the fact that
$$\sum_{\scr\vert Y \vert = N + 1\atop\scr      Y \in \Pi'}
\wp^D_1 (h^{}_{\cdot Y} (t))^2  \leq 2 \wp^D_{N+2} (h(t))^2 + C_N  \wp^D_{N+1}
(h(t))^2,$$
for some constant $C_N$, give that
$$\eqalignno{
&\sum_{\scr  Y \in \Pi'\atop\scr \vert Y \vert = N+1}
\Vert (1+ \lambda^{}_i (t))^{1/2}
 h^{}_Y (t) \Vert^2_D&(5.44)\cr
&\quad{}\leq C'^2_1 \Big(\wp^D_{N+2} (h(t))^2
+ \sum_{{\sscr Y \in \Pi'\atop\sscr \vert Y \vert = N+1}\atop\sscr
\vert Y \vert \geq L+1}
\Vert (1+ \lambda^{}_i (t))  \gamma^\mu  G_{Y \mu} (t)
h^{}_{\un} (t) \Vert^2_D + \wp^D_{N+1} \big((1+ \lambda^{}_i (t))
g(t)\big)^2\Big)\cr
&\quad{}+ C'^2_N  \Big(\wp^D_{N+1} (h(t)) +
\sum_{\scr Y \in \Pi'\atop\scr\vert Y \vert = N+1}
I_Y (t)
+ \sum_{{\sscr Y_1, Y_2 \in \Pi'\atop\sscr\vert Y_1
\vert + \vert Y_2 \vert = N+1}
\atop\sscr 1 \leq \vert Y_2 \vert \leq N-L}
\Vert (1 + \lambda^{}_i (t))  \gamma^\mu
 G_{Y_1 \mu} (t)  h^{}_{Y_2} (t) \Vert^{}_D\Big)^2,\cr
}$$
where $C'_1$ and $C'_N$ are constants  depending only on
$\tau^{}_{i,0} (t).$

According to the induction hypothesis, we obtain from the inequality of the
theorem that
$$\eqalignno{
&\wp^D_n \big((1 + \lambda^{}_i (t))^{1/2}  h(t)\big)&(5.45)\cr
&\qquad{}\leq C'_n  \sum_{0 \leq l \leq K}
 (1+ \tau^{}_{i,l} (t))
\Big(\wp^D_{n+1-l}  (h(t)) + \wp^D_{n-l}  \big((1+ \lambda^{}_i (t))
g(t)\big)\cr
&\qquad\qquad{}+ \sum_{{\sscr Z_1, Z_2 \in \Pi'\atop\sscr\vert Z_1
\vert + \vert Z_2 \vert
\leq n-l}\atop\sscr \vert Z_2 \vert \leq n-K-1}
\Vert (1 + \lambda^{}_i (t))  \gamma^\mu
 G_{Z_1 \mu} (t)  h^{}_{Z_2} (t) \Vert^{}_D\Big),\cr
}$$
for $0 \leq n \leq N$,  $0 \leq K \leq n$, where $C'_n$ is a constant
depending only on $\tau^{}_{i,0} (t)$. Since $\tau^{}_{i,l} (t)
\leq \tau^{}_{i,l+1} (t),$
it follows from (5.45) that
$$\eqalignno{
&\wp^D_N \big((1 + \lambda^{}_i (t))^{1/2}  h(t)\big)&(5.46{\rm a}) \cr
&\qquad{}\leq C'_N  \sum_{1 \leq l \leq L}
(1 + \tau^{}_{i,l} (t))
\Big(\wp^{D}_{N+2-l}  (h(t)) + \wp^D_{N+1-l} \big((1 +
\lambda^{}_i (t) g(t)\big)\cr
&\qquad\qquad{}+ \sum_{\scr \vert Z_1 \vert + \vert Z_2 \vert \leq N+1-l
\atop\scr \vert Z_2 \vert \leq N+1-L-1}
 \Vert (1 + \lambda^{}_i (t))  \gamma^\mu  G_{Z_1 \mu}
(t)  h^{}_{Z_2} (t) \Vert^{}_D\Big),\cr
}$$
if $1 \leq L \leq N+1$. According to the induction hypothesis the inequality of
the theorem (with $L=0$) gives that
$$\eqalignno{
&\wp^D_N \big((1+ \lambda^{}_i (t))^{1/2}  h(t)\big)&(5.46{\rm b})\cr
&\qquad{} \leq C'_1 \Big(\wp^D_{N+1} (h(t))^2 +
\wp^D_N \big((1 + \lambda^{}_i (t))  g(t)\big)^2\Big)^{1/2}\cr
&\qquad\qquad{}+ C'_N  \sum_{{\sscr Z_1, Z_2 \in \Pi'
\atop\sscr\vert Z_1 \vert + \vert Z_2 \vert \leq N}\atop\sscr \vert Z_2 \vert
\leq N-1}  \Vert (1 + \lambda^{}_i (t))
\gamma^\mu  G_{Z_1 \mu} (t)  h^{}_{Z_2} (t) \Vert^{}_D,\cr
}$$
where $C'_1$ and $C'_N$ are constants depending only on $\tau^{}_{i,0} (t).$

Adding $\big(\wp^D_N ((1 + \lambda^{}_i (t))^{1/2}  h(t)\big)^2$ to both sides
of inequality (5.44) and using (5.46b) for $L = 0$ and (5.46a) for
$1 \leq L \leq N+1$
we obtain that
$$\eqalignno{
&\wp^D_{N+1} \big((1 + \lambda^{}_i (t))^{1/2}  h(t)\big) & (5.47)\cr
&{}\leq C'_1 \Big(\wp^D_{N+2} (h(t))^2
+ \wp^D_{N+1} \big((1 + \lambda^{}_i (t))  g(t)\big)^2
+ \!\!\sum_{{\sscr Y \in \Pi'\atop\sscr\vert Y \vert = N+1}\atop\sscr
\vert Y \vert \geq L+1}\!\!  \Vert (1 + \lambda^{}_i (t))
\gamma^\mu  G_{Y \mu} (t)  h^{}_{\un} (t) \Vert^2_D\Big)^{1/2}\cr
&\quad\quad{}+ C'_N \Big(\wp^D_{N+1} (h(t))
+ \sum_{{\sscr Z_1, Z_2 \in \Pi'\atop\sscr\vert Z_1
\vert + \vert Z_2 \vert = N+1}\atop\sscr
1 \leq\vert Z_2 \vert \leq N-L}  \Vert (1 + \lambda^{}_i (t))
 \gamma^\mu  G_{Z_1 \mu} (t)  h^{}_{Z_2} (t) \Vert^{}_D\cr
&\quad\quad{}+\sum_{\scr Y\in \Pi'\atop\scr \vert Y\vert =N+1}I_Y(t)
+ \sum_{1 \leq l \leq L}  (1 + \tau^{}_{i,l} (t))
\big(\wp^D_{N+2-l} (h(t)) + \wp^D_{N+1-l} ((1 + \lambda^{}_i (t))  g(t))\cr
&\quad\quad{}+ \sum_{{\sscr Z_1, Z_2 \in \Pi'\atop\sscr\vert Z_2 \vert \leq
N-L}
\atop\sscr\vert Z_1 \vert + \vert Z_2 \vert \leq N+1-l}
\Vert (1 + \lambda^{}_i (t))  \gamma^\mu
 G_{Z_1 \mu} (t)  h^{}_{Z_2} (t) \Vert^{}_D\big)\Big),\cr
}$$
where $C'_1$ and $C'_N$ are constants depending only on $\tau^{}_{i,0} (t).$

Let $n = N+1-j$ and $K = L-j$ for $1 \leq j \leq L$, in inequality (5.45). Then
$0 \leq n \leq N$ and $0 \leq K \leq n$, so it follows from inequalities (5.43)
and (5.45) that
$$\eqalignno{
I_Y (t)&\leq C'_N  \sum_{1 \leq j \leq L}\  \sum_{0 \leq l \leq L-j}
 \tau^{}_{i, l+j} (t)
\Big(\wp^D_{N+2-l-j}  (h(t)) + \wp^D_{N+1-l-j}  \big((1 + \lambda^{}_i (t))
 g(t)\big)\cr
&\qquad{}+ \sum_{{\sscr Z_1, Z_2 \in \Pi'\atop\sscr \vert Z_2 \vert \leq N-L}
\atop\sscr \vert Z_1 \vert + \vert Z_2
\vert \leq N+1-l-j}  \Vert (1 + \lambda^{}_i (t))  \gamma^\mu
 G_{Z_1 \mu} (t)  h^{}_{Z_2} (t) \Vert^{}_D\Big),\cr
}$$
where $C'_N$ is a constant depending only on $\tau^{}_{i,0} (t)$. We have here
used
the fact that\penalty-10000
$(1 + \tau^{}_{i,l} (t))  \tau^{}_{i,j} (t) \leq C_{l,j}
\tau^{}_{i,l+j} (t)$ for some constant $C_{l,j}$. Substituting $j' = j,
l' = j+l$, we obtain that
$$\eqalignno{
I_Y (t) &\leq C'_N  \sum_{1 \leq l' \leq L}  \tau^{}_{i,l'} (t)
 \Big(\wp^D_{N+2-l'} (h(t))
+ \wp^D_{N+1-l'}  \big((1 + \lambda^{}_i (t))  g(t)\big)&(5.48)\cr
&\qquad{}+ \sum_{{\sscr Z_1,Z_2 \in \Pi'\atop\sscr
\vert Z_2 \vert\leq N-L} \atop\sscr
\vert Z_1 \vert + \vert Z_2 \vert \leq N + l - l'}  \Vert (1 + \lambda^{}_i
(t))
\gamma^\mu  G_{Z_1 \mu} (t)  h^{}_{Z_2} (t) \Vert^{}_D\Big),\cr
}$$
where $C'_N$ is a constant depending only on $\tau^{}_{i,0} (t)$. It follows
from
inequalities (5.47) and (5.48) that $\wp^D_{N+1}  \big((1 + \lambda^{}_i
(t))^{1/2}
 h(t)\big)$ is bounded by the left-hand side of inequality (5.47) without
the term $I_Y (t)$ and with a new constant $C'_N$ depending only on
$\tau^{}_{i,0}
(t)$. This proves the inequality of the theorem with $0 \leq n,  0 \leq
L \leq n$ and $k = 1$, by induction.

For $n \geq 0,  k \geq 1,  0 \leq L \leq n + k - 1$, let
$$\eqalignno{
&Q^{(k)}_{n,L} (t) &(5.49)\cr
&\quad{}= C'_k \Big(\wp^D_{n+k} (h(t))^2
+ \sum_{0 \leq j \leq k-1}  \Big(\wp^D_{n+j} \big((1 + \lambda^{}_i
(t))^{(k+1-j)/2}  g(t)\big)^2\cr
&\ \quad{}+\sum_{{\sscr Y \in \Pi'\atop\sscr \vert Y \vert = n+j}
\atop\sscr  \vert Y \vert \geq L+1}
 \Vert (1+\lambda^{}_i (t))^{(k+1-j)/2}  \gamma^\mu
G_{Y \mu} (t)  h^{}_{\un} (t) \Vert^2_D\Big)\Big)^{1/2}\cr
&\ \quad{}+ C'_{n+k}  \Big(\wp^D_{n+k-1} (h(t))
+\!\!\sum_{{\sscr0 \leq j \leq k-1\atop\sscr Z_1, Z_2 \in \Pi'}
\atop{\sscr \vert Z_1 \vert + \vert Z_2 \vert = n+j\atop\sscr  1 \leq
\vert Z_2 \vert
\leq n+j-L-1}}\!\!\Vert (1 + \lambda^{}_i (t))^{(k+1-j)/2}
 \gamma^\mu  G_{Z_1 \mu} (t)  h^{}_{Z_2} (t) \Vert^{}_D\Big)\cr
&\ \quad{}+ C'_{n+k}  \sum_{1 \leq l \leq L}  (1 + \tau^{}_{i,l} (t))
 \Big(\wp^D_{n+k-l} (h(t))
+ \!\sum_{\scr 0 \leq j \leq k-1\atop\scr  n+j-l \geq 0}\!  \wp^D_{n+j-l}
 \big((1 + \lambda^{}_i (t))^{(k+1-j)/2}  g(t)\big)\cr
&\ \quad{}+\sum_{{\sscr 0 \leq j \leq k-1\atop\sscr Z_1, Z_2 \in \Pi' }
\atop{\sscr \vert Z_1 \vert + \vert Z_2 \vert = n+j-l\atop\sscr \vert Z_2 \vert
\leq n+j-L-1}}  \Vert (1 + \lambda^{}_i (t))^{(k+1-j)/2}
 \gamma^\mu  G_{Z_1 \mu} (t)  h^{}_{Z_2} (t) \Vert^{}_D\Big),\quad t \geq 0,\cr
}$$
where $C'_N,  N \geq 1$, are constants depending only  on $\tau^{}_{i,0} (t).$
We note that with an appropriate choice of the constants $C'_N,  N \geq
1,$
$$\eqalignno{
Q^{(k)}_{n,L} (t) &\leq Q^{(k)}_{n+1, L} (t) \quad \hbox{ for } n \geq 0,
k \geq 1,  0 \leq L \leq n + k-1,& (5.50{\rm a})\cr
Q^{(k)}_{n+1,L} (t) &\leq Q^{(k+1)}_{n, L} (t)\quad \hbox{ for } n \geq 0,
k \geq 1,  0 \leq L \leq n + k, &(5.50{\rm b})\cr
Q^{(k)}_{n,L} (t) &\leq Q^{(k)}_{n+1, L+1} (t)\quad \hbox{ for } n \geq 0,
k \geq 1,  0 \leq L \leq n + k-1, &(5.50{\rm c})\cr
}$$
and moreover that the inequality of the theorem reads
$$\wp^D_n \big((1 + \lambda^{}_i (t))^{k/2}  h(t)\big) \leq Q^{(k)}_{n,L} (t),
\eqno{(5.50{\rm d})}$$
where $n \geq 0,  k \geq 1,  0 \leq L \leq n+k-1,
t \geq 0.$

We have proved that inequality (5.50d) is true for $n \geq 0,  k = 1,
 0 \leq L \leq n$ and we make the induction hypothesis $H_K$ that it
is true for $n \geq 0,  1 \leq k \leq K,  0 \leq L \leq n+k-1,$
where $K \geq 1.$

It follows from inequality (5.39) that
$$\eqalignno{
&\Vert (1+\lambda^{}_i (t))^{(k+1)/2}  h^{}_{\un} (t) \Vert^{}_D &(5.51)\cr
&\qquad{}\leq C'_{K+1}
 \Big(\wp^D_1 \big((1 + \lambda^{}_i (t))^{K/2}  h(t)\big)
+ \wp^D_0 \big((1 + \lambda^{}_i (t))^{(K+2)/2}  g(t)\big)\Big), \cr
}$$
where $C'_{K+1}$ is a constant depending only on $\tau^{}_{i,0} (t).$
Using the induction hypothesis $H_K$ for the first term on the right-hand side
of inequality (5.51) and the definition of $Q^{(K+1)}_{0,L} (t)$ for the second
term, we obtain, choosing $C'_{K+1}$ appropriately,
$$\wp^D_0 \big((1 + \lambda^{}_i (t))^{(K+1)/2}  h(t)\big) \leq
Q^{(K)}_{1,L} (t) + Q^{(K+1)}_{0,L} (t).$$
Inequality (5.50b) then gives that
$$\wp^D_0 \big((1 + \lambda^{}_i (t))^{(K+1)/2}  h(t)\big) \leq
Q^{(K+1)}_{0,L} (t), \quad 0 \leq L \leq K, \eqno{(5.52)}$$
after a redefinition of the constants $C'_N,  N \geq 1$. This shows that
inequality (5.50d) is true for $n = 0$, $k = K+1$, $0 \leq L \leq K$, if the
hypothesis $H_K$ is true. We now make the induction hypothesis $H_{K+1,N}$,
where $K \geq 1,  N \geq 0$, that (5.50d) is true for $0 \leq n \leq N,$
$1 \leq k \leq K+1,  0 \leq L \leq n+k-1$. Thus $H_{K+1,0}$ is true if
$H_K$ is true. Let $0 \leq L \leq N+K+1$ and let $Y \in \Pi',  \vert Y \vert =
N+1$. It then follows from inequality (5.39), in a similar way as (5.41)
was obtained, that
$$\eqalignno{
&\Vert (1 + \lambda^{}_i (t))^{(K+1)/2}  h^{}_Y (t) \Vert^{}_D&(5.53)\cr
&\qquad{}\leq C_{K+1}
 (1 + \tau^{}_{i,0} (t))
\Big(\wp^D_1 \big((1 + \lambda^{}_i (t))^{K/2}  h^{}_Y (t)\big) +
\Vert (1 + \lambda^{}_i
(t))^{(K+2)/2}  g^{}_Y (t) \Vert^{}_D\cr
&\qquad\qquad{}+
 \suma_{{\sscr Y_1,Y_2\atop\sscr \vert Y_1 \vert \leq L}
\atop\sscr \vert Y_2 \vert \leq \vert Y \vert - 1}^Y
 \tau^{}_{i, \vert Y_1 \vert} (t)  \wp^D_{\vert Y_2 \vert}
\big((1 + \lambda^{}_i (t))^{(K+1)/2}h(t)\big)\cr
&\qquad\qquad{}+ \suma_{\scr Y_1,Y_2 \atop\scr \vert Y_2 \vert \leq N-L}^Y
\Vert (1 +
\lambda^{}_i (t))^{(K+2)/2}  \gamma^\mu  G_{Y_1 \mu} (t)
h^{}_{Y_2} (t) \Vert^{}_D\Big),\cr
}$$
where we have redefined the constant $C_{K+1}$. Defining
$$\eqalignno{
&I^{(K+1)}_Y (t)&(5.54)\cr
&\qquad{}= C_{K+1} (1 + \tau^{}_{i,0} (t))
\sum_{{\sscr Y_1,Y_2\atop\sscr\vert Y_1 \vert \leq L}\atop\sscr
\vert Y_2 \vert \leq \vert Y \vert - 1}  \tau^{}_{i, \vert Y_1 \vert}
 \wp^D_{\vert Y_2 \vert}  \big((1+ \lambda^{}_i (t))^{(K+1)/2}
h(t)\big), \cr
}$$
where $\vert Y \vert = N+1,  Y \in \Pi',  0 \leq L \leq N +
K + 1$, we obtain from inequality (5.53) in the same way
as we obtained inequality  (5.44)
$$\eqalignno{
&\sum_{\scr Y   \in \Pi'\atop\scr \vert Y \vert = N+1}  \Vert (1 +
\lambda^{}_i (t))^{(K+1)/2}  h^{}_Y (t) \Vert^2_D &(5.55)\cr
&\qquad{}\leq C''^2_{K+1}  \Big(\wp^D_{N+2}  \big((1 +
\lambda^{}_i (t))^{K/2}  h (t)\big)^2\cr
&\qquad\qquad{}+\sum_{{\sscr Y \in \Pi'\atop\sscr \vert Y \vert = N+1}
\atop\sscr \vert Y \vert \geq L+1}  \Vert
(1 + \lambda^{}_i (t))^{(K+2)/2}  G_{Y \mu} (t)  h^{}_{\un} (t)
\Vert^2_D + \wp^D_{N+1} \big((1 + \lambda^{}_i
(t))^{(K+2)/2}g(t)\big)^2\Big)\cr
&\qquad\qquad{}+ (C''_{N+K+2})^2  \Big(\wp^D_{N+1}
\big((1 + \lambda^{}_i (t))^{K/2}
h(t)\big) + \sum_{\scr Y \in \Pi'\atop\scr \vert Y \vert = N+1}
I^{(K+1)}_Y (t)\cr
&\qquad\qquad{}
+ \sum_{{\sscr Y_1, Y_2 \in \Pi'\atop\sscr \vert Y_1 \vert +
\vert Y_2 \vert = N+1}
\atop\sscr 1 \leq \vert Y_2 \vert \leq N - L}  \Vert (1 +
\lambda^{}_i (t))^{(K+2)/2}
\gamma^\mu  G_{Y_1 \mu} (t)  h^{}_{Y_2} (t) \Vert^{}_D\Big)^2,\cr
}$$
where $C''_{K+1}$ and $C''_{N+K+2}$ are constants depending only on
$\tau^{}_{i,0} (t).$
Adding $\big(\wp^D_N ((1 + \lambda^{}_i (t))^{(K+1)/2}
h(t)\big)\big)^2$ to both sides
of inequality (5.55), redefining the constants
$C''_{K+1}$ and $C''_{N+K+2},$
using that $\wp^D_N \big((1 + \lambda^{}_i (t))^{(K+1)/2}
h(t)\big) \leq Q^{(K+1)}_{N,L'},$
for $0 \leq L' \leq N+K$, according to the hypothesis
$H_{K+1,N}$ and using that $\wp^D_{N+2} \big((1 + \lambda^{}_i (t))^{K/2}
h(t)\big) \leq Q^{(K)}_{N+2,L} (t)$
for $0 \leq L \leq N+K+1$ according to $H_K$, we obtain that
$$\eqalignno{
&\wp^D_{N+1} \big((1 + \lambda^{}_i (t))^{(K+1)/2}  h(t)\big)^2 &(5.56)\cr
&\quad{}\leq Q^{(K+1)}_{N,L'} (t)^2 + (C''_{K+1})^2
\Big(Q^{(K)}_{N+2,L} (t)^2\cr
&\qquad{}+\sum_{{\sscr Y \in \Pi'\atop\sscr \vert Y \vert = N+1}
\atop\sscr \vert Y \vert \geq L+1}  \Vert
(1 + \lambda^{}_i (t))^{(K+2)/2}  G_{Y \mu} (t)  h^{}_{\un} (t)
\Vert^2_D + \wp^D_{N+1} \big((1 + \lambda^{}_i
(t))^{(K+2)/2}g(t)\big)^2\Big)\cr
&\qquad{}+ (C''_{N+K+2})^2  \Big(Q^{(K)}_{N+1,L'} (t) + \sum_{\scr Y \in \Pi'
\atop\scr\vert Y \vert = N+1}  I^{(K+1)}_Y (t)\cr
&\qquad{}+\sum_{{\sscr Y_1, Y_2 \in \Pi'\atop\sscr \vert Y_1 \vert +
\vert Y_2 \vert = N+1} \atop\sscr  1 \leq \vert Y_2 \vert \leq N-L}
\Vert (1 + \lambda^{}_i (t))^{(K+2)/2} \gamma^\mu
 G_{Y_1 \mu} (t)  h^{}_{Y_2} (t) \Vert^{}_D\Big)^2,\cr
}$$
for $0 \leq L \leq N+K+1$ and $0 \leq L' \leq N+K.$

Using hypothesis $H_{K+1,N}$ it follows from definition (5.54) of $I^{(K+1)}_Y
(t)$, that
$$\sum_{\scr Y \in \Pi'\atop\scr \vert Y \vert = N+1}
I^{(K+1)}_Y (t) \leq C''_{K+1}
 \sum_{1 \leq l \leq L}  \tau^{}_{i,l} (t)  Q^{(K+1)}_
{N+1-l,L-l} (t),$$
$0 \leq L \leq N+K+1$. Since $\tau^{}_{i,l} (t)  (1 + \tau^{}_{i,j} (t)) \leq
\tau^{}_{i,l+j} (t)$ it follows from expression (5.49) of
$Q^{(k)}_{n,L} (t)$ that
$$\eqalignno{
&\sum_{\scr Y \in \Pi'\atop\scr\vert Y \vert = N+1}  I^{(K+1)}_Y (t) &(5.57)\cr
&\qquad{}\leq C''_{N+K+2}
\sum_{1 \leq l \leq L}  (1 + \tau^{}_{i,l} (t))
\Big(\wp^D_{N+K+2-l}  (h(t))\cr
&\qquad\qquad{}+ \sum_{\scr 0 \leq j \leq K\atop\scr N+1+j-l \geq 0}
\wp^D_{N+1+j-l}  \big((1+ \lambda^{}_i (t))^{(K+2-j)/2}  g(t)\big)\cr
&\qquad\qquad{}+ \sum_{{\sscr0 \leq j \leq K\atop\sscr Z_1, Z_2 \in \Pi'
 }\atop{\sscr\vert Z_1 \vert + \vert Z_2 \vert \leq N+1+j-l\atop\sscr
\vert Z_2 \vert \leq N+1+j-L-1}}  \Vert (1 + \lambda^{}_i (t))^{(K+
2-j)/2}  \gamma^\mu  G_{Z_1 \mu} (t)  h^{}_{Z_2} (t)
\Vert^{}_D\Big),\cr
}$$
where $C''_{N+K+2}$ is a constant depending only on $\tau^{}_{i,0} (t)$. If
$0 \leq L \leq N+K$, then it follows from inequality (5.56) with $L = L'$,
inequality (5.57) and expression (5.49) of $Q^{(K+1)}_{N+1,L}$, that
$$\wp^D_{N+1}  \big((1 + \lambda^{}_i (t))^{(K+1)/2}  h(t)\big)
\leq Q^{(K+1)}_{N+1,L}
(t),\quad  0 \leq L \leq N+K, \eqno{(5.58)}$$
after a suitable definition of $C'_{K+1}$ and $C'_{K+N+2}$.
If $L=N+k+1$, then it follows using (5.50c) for $Q^{(K+1)}_{N,L-1}$ and
inequalities (5.56) and (5.57), that
$$\wp^D_{N+1}  \big((1 + \lambda^{}_i (t))^{(K+1)/2}  h(t)\big)
\leq Q^{(K+1)}_{N+1,L}
(t),\quad  L=N+k+1, \eqno{(5.59)}$$
after a suitable definition of $C'_{K+1}$ and $C'_{K+N+2}$.
Inequalities (5.58)
and (5.59) prove that $H_{K+1,N+1}$ is true if $H_{K+1,N}$ and $H_K$ are
true. Since $H_{K+1,0}$ is true if $H_K$ is true, it follows by induction that
$H_{K+1,N}$ is true for all $N \geq 0$, i.e. $H_{K+1}$ is true if $H_K$ is
true.
Since $H_1$ is true it now follows by induction that $H_K$ is true for all
$K \geq 1.$ This proves the theorem.

In order to eliminate $L^\infty$-norms coming from the right-hand side of the
inequality of Theorem 5.5 and in later energy estimates we shall
{\it derive appropriate $L^2 - L^\infty$ estimates for the solution of the
 inhomogeneous Dirac equation and wave equation}.
 Let $u$, as in Proposition 2.15, be the solution of the wave
equation $\carre u = 0$ with initial data $(f, \dot{f})
\in M^\rho_\infty$. Since the evolution operator in $M^\rho_0$ defined by the
wave
equation is unitary in $M^\rho_0$ it follows that
$$\Vert (f, \dot{f}) \Vert^2_{M_{n}} = \sum_{\scr Y \in \Pi'\atop\scr
\vert Y \vert \leq n}
 \Vert T^{M1}_{Y(t)} (u(t), \dot{u} (t)) \Vert^2_{M_0},\quad n\geq 0,
 \eqno{(5.60)}$$
where $Y(t)$ is defined by (1.11). If $(g, \dot{g}) \in M^\rho_{\infty}$, is an
initial data for the wave equation at time $t$, if follows from Proposition
2.15 and (5.60) that
$$\eqalignno{
&(1 + \vert x \vert + t)^{3/2 - \rho}\vert g(x) \vert& (5.61)\cr
&\qquad{}+ (1 + \vert x \vert + t)  \sum_{0 \leq \vert \nu \vert \leq n-1}
(1 + \big\vert t - \vert x \vert \big\vert)^{3/2 - \rho + \vert \nu \vert}
(\vert \partial^\nu  \nabla g(x) \vert + \vert \partial^\nu
\dot{g} (x) \vert)\cr
&\qquad{}\qquad{}
\leq C_{n, \rho} \Big(\sum_{\scr Y \in \Pi'\atop\scr \vert Y \vert
\leq n + 2}  \Vert
T^{M1} _{Y(t)}  (g, \dot{g}) \Vert^2_{M_0}\Big)^{1/2},\quad  n \geq 1,
 t \geq 0,  1/2 < \rho < 1.\cr
}$$
Similarly, using the $L^1 - L^\infty$ estimate in \refVW, we
obtain
$$(1 + \vert x \vert + t)^{3/2}  \vert \alpha (x) \vert \leq C
\Big(\sum_{\scr Y \in \Pi'\atop\scr   \vert Y \vert \leq 3}  \Vert
T^{D1}_{Y(t)}
\alpha \Vert^2_D\Big)^{1/2},\quad  t \geq 0, \eqno{(5.62)}$$
$\alpha \in D_\infty$. We note that estimates (5.61) and (5.62) are relations
expressed directly on the space of initial data and that the bound is given
by the canonical seminorms (in the space of differentiable vectors)
composed by the time evolution defined in (1.11)
of the enveloping algebra of the Poincar\'e Lie
algebra. We shall generalize this to  a certain extent to the nonlinear case
by first expressing the derivatives $\partial_i$ and the second
derivatives $\partial_i  \partial_j,  1 \leq i \leq j$, in terms of the
nonlinear
generators for the representation of ${\frak{sl}}(2, \Crm)$ and then use the
Sobolev inequalities for weighted $L^p$ spaces developed in \refHSL.

In the case of the wave equation we introduce, for $F, \dot{F} \in C^\infty
(\Rrm^3), t \in \Rrm,  x \in \Rrm^3,  i,j \in \{ 1,2,3 \}$,
$$F^{}_{M_{0i}} (x) = x_i       \dot{F} (x),\quad  F^{}_{M_{ij}} (x) = (x_i
 \partial_j - x_j  \partial_i)  F(x),\quad
F^{}_{P_i} = \partial_i  F, \quad  F^{}_{P_0} = \dot{F}, \eqno{(5.63)}$$
where $M_{\mu \nu}$ are for $0 \leq \mu<\nu \leq 3$ the generators of
${\frak{sl}}(2,\Crm)$ and $M_{\mu \nu} = - M_{\nu \mu}$. Moreover for $f,
\dot{f}
\in C^\infty (\Rrm^3)$ let $K_Y (t, f, \dot{f}, F, \dot{F})$ and
$\dot{K}_Y (t, f, \dot{f}, F, \dot{F})$, where $Y \in U ({\frak{sl}}(2, \Crm))$
and the degree of $Y$ is at most two, be defined by
$$\eqalignno{
(K_{\un} (t),  \dot{K}_{\un} (t)) &= (f, \dot{f}),& (5.64{\rm a})\cr
(K_{M_{ij}} (t),  \dot{K}_{M_{ij}} (t)) (x) &= (x_i  \partial_j -
x_j  \partial_i)  (f(x), \dot{f} (x)),& (5.64{\rm b})\cr
(K_{M_{0i}} (t),  \dot{K}_{M_{0i}} (t)) (x)
&= (x_i  \dot{f} (x) + t \partial_i  f(x),  x_i
 \Delta  f(x) &(5.64{\rm c})\cr
&\quad{}+ \partial_i  f(x) + t \partial_i
\dot{f} (x) + x_i  F(x)),\cr
(K_{M_{ij}  M_{\mu \nu}} (t),  \dot{K}_{M_{ij} M_{\mu \nu}}
(t)) (x) &= (x_i  \partial_j - x_j  \partial_i) (K_{M_{\mu \nu}} (t),
\dot{K}_{M_{\mu \nu}} (t)) (x),& (5.64{\rm d})\cr
(K_{M_{0i} M_{0j}} (t)) (x) &= x_i (\dot{K}_{M_{0j}} (t)) (x) + t
\partial_i (K_{M_{0j}} (t)) (x),& (5.64{\rm e})\cr
(\dot{K}_{M_{0i} M_{0j}} (t)) (x) &= x_i  \Delta (K_{M_{0j}} (t)) (x)
+ (\partial_i  K_{M_{0j}} (t)) (x) &(5.64{\rm f})\cr
&\quad{}+ (t \partial_i  \dot{K}_{M_{0j}}
(t)) (x)
+ x_i  F^{}_{M_{0j} + t P_j} (x),\cr
}$$
where $\un$ is the unit element in $U ({\frak{sl}}(2, \Crm))$, $\mu, \nu \in
\{ 0,1,2,3 \}$
and $i,j \in \{ 1,2,3 \}.$

When there is no risk of confusion we omit the arguments
$f, \dot{f}, F, \dot{F}$
in $(K, \dot{K})$ and write $(K_Y (t), \dot{K}_Y (t))$. If $u$ is a
solution of the inhomogeneous wave equation $\carre u = G$, then it
follows from the definition of $\xi^{}_Y,  Y \in U{(\p)}$, that
$$\big((\xi^{}_Y u) (t),  {d \over dt} (\xi^{}_Y u) (t)\big)
= (K_Y,\dot{K}_Y)  \big(t, u(t), {d \over dt}  u(t),
 G(t),  {d \over dt}  G(t)\big),\eqno{(5.65)}$$
for $Y \in U({\frak{sl}}(2,\Crm))$ and $Y$ being of degree not greater  than
$2.$

$K$ and $\dot{K}$ defined by (5.64a)--(5.64f) satisfy
$$\eqalignno{
&\partial_i (t  \dot{f} (x) + \sum_{1 \leq j \leq 3}  x_j
 \partial_j  f(x)) &(5.66\hbox{a})\cr
&\qquad{}= (\dot{K}_{M_{0i}} (t)) (x)
- \sum_{1 \leq j \leq 3} (\partial_j    K_{M_{ij}} (t)) (x) - 2 \partial_i
 f(x) - x_i  F(x)\cr
\noalign{\hbox{and}}
&\partial_i \Big(t(\dot{K}_{M_{\mu \nu}} (t)) (x) + \sum_{1 \leq j \leq 3}
x_j (\partial_j  K_{M_{\mu \nu}} (t)) (x)\Big)&(5.66\hbox{b})\cr
&\qquad{}= \Big(\dot{K}_{M_{0i} M_{\mu \nu}} (t) -
\sum_{1 \leq j \leq 3}\partial_j  K_{M_{ij} M_{\mu \nu}} (t) - 2 \partial_i
K_{M_{\mu\nu}} (t)\Big) (x)
- x_i (F^{}_{M_{\mu \nu} (t)} (t)) (x),\cr
}$$
where $\mu, \nu \in \{ 0,1,2,3 \},  1 \leq i \leq 3$ and $M_{ij} (t) =
M_{ij}$ for $1 \leq i < j \leq 3$ and $M_{0i} (t) = M_{0i} + t P_i$.
These two formulas express the dilatation  generator $y^\mu \partial_\mu$
in terms of the action of ${\frak{sl}}(2, \Crm).$

In $4$-dimensional conventional notations with contravariant coordinates
$y^\mu,
 0 \leq \mu \leq 3$, we have, using the summation convention,
$$y^\nu  y_\nu  \partial_\mu = y_\mu  y^\nu
\partial_\nu + y^\nu (y_\nu  \partial_\mu - y_\mu  \partial_\nu),$$
which reads
$$\lambda  \partial_\mu = y_\mu  \overline{D} + y^\nu
\overline{M}_{\nu \mu},\quad  0 \leq \mu \leq 3, \eqno{(5.67)}$$
with $\overline{D} = y^\nu  \partial_\nu,  \lambda = y^\nu
y_\nu$ and $\overline{M}_{\mu \nu} = y_\mu  \partial_\nu - y_\nu
\partial_\mu.$
It follows from (5.67) that
$$\eqalignno{
\partial_\mu\partial_\nu&= \lambda^{-1} (y_\mu  y_\nu
 \carre + g^{}_{\mu \nu}  \overline{D})& (5.68)\cr
&\qquad{} + \lambda^{- 2} \Big(y^\alpha  y^\beta (\overline{M}_{\alpha \mu}
\overline{M}_{\beta \nu} + \overline{M}_{\alpha \nu}  \overline{M}_{\beta \mu})
+
y_\mu  y^\alpha  \overline{D}\, \overline{M}_{\alpha \nu} + y_\nu
y^\alpha  \overline{D}\, \overline{M}_{\alpha \mu}\cr
&\qquad{}- y_\mu  y^\alpha  \overline{M}_{\alpha \nu} - y_\nu
y^\alpha  \overline{M}_{\alpha \mu} - y_\mu  y_\nu ({1 \over 2}
 \overline{M}_{\alpha \beta}  \overline{M}^{\alpha \beta} + 4
 \overline{D})\Big),\cr
}$$
for $0 \leq \mu \leq 3,  0 \leq \nu \leq 3$. Formulas (5.66a), (5.66b),
(5.67) and (5.68) give $\partial_i, \partial_i  \partial_j$ expressed
in terms of $K_Y, \dot{K}_Y$, $F, \dot{F}$ since, on initial conditions,
$\carre$ can be
replaced by $F.$

For later reference we shall state particular $L^\infty$-estimates for the
electromagnetic potential.
\saut
\noindent{\bf Proposition 5.6.}
{\it
Let $t \geq 0,  (f, \dot{f}) \in M^0_\infty,  (F, \dot{F})
\in M^0_\infty$ and suppose that  the function $x \mapsto x_i (F(x),
\dot{F} (x))$ is an element of $M^0_\infty$, $1\leq i\leq 3$.

\noindent{\hbox{\rm i)}} If $a \in \Rrm$, then
$$(1 + \vert x \vert)^{1+a}  \vert f(x) \vert \leq C\Big(\sum_Y
\Vert (1 + \vert\cdot  \vert^2)^{a/2}  K_Y (t) \Vert^{}_{L^2} + \Vert
(1 - \Delta) f \Vert^{}_{L^2}\Big),\quad  x  \in \Rrm^3,$$
where the sum is taken over $Y \in \Pi' \cap U({\frak{su}}(2)),  \vert Y \vert
\leq 2$.

\noindent{\hbox{\rm ii)}} If $0 < \delta < 1$, then
$$\eqalignno{
(1+t)^{3/2}  \vert f(x) \vert &\leq C_\delta \Big(\sum_Y \Vert (K_Y (t),
 \dot{K}_Y (t)) \Vert^{}_{M^0} + \Vert (1 - \Delta) f \Vert^{}_{L^2}\cr
&\qquad{}+ \sum_Z  (1+t)  \Vert F^{}_{Z(t)} (t) \Vert^{}_{L^{6/5}}\Big),
 \quad 0 \leq \vert x \vert \leq \delta t,\cr
}$$
where $Z(t)$ is given by (1.11) and the sums are taken over $Y,Z \in \Pi'
\cap U ({\frak{sl}}(2, \Crm)),  \vert Y \vert \leq 2$ and $\vert Z \vert \leq
2$.

\noindent{\hbox{\rm iii)}} If $0 < \delta^{}_1 < 1 < \delta^{}_2$, then
$$\eqalignno{
(1+t) (1+\vert t - \big\vert x \vert \big\vert)^{1/2}  \vert f(x) \vert &\leq
C_{\delta^{}_1, \delta^{}_2} \Big(\sum_Y \Vert (K_Y (t),
\dot{K}_Y (t)) \Vert^{}_{M^0} + \Vert (1 - \Delta) f \Vert^{}_{L^2}\cr
&\qquad{} + \sum_Z (1+t)\Vert F^{}_{Z(t)} (t)
\Vert^{}_{L^{6/5}}\Big),\quad  \delta^{}_1 t \leq \vert x \vert \leq
\delta^{}_2 t,\cr
}$$
where the domains of summation are as in ii).

\noindent{\hbox{\rm iv)}} If $0 < \delta < 1$ and $a \in \Rrm$ then
$$\eqalignno{
(1+\vert x \vert)^{3/2 + a}  \vert f(x) \vert &\leq C_{\delta, a}
\Big(\sum_Y \Vert (1+ \vert \cdot  \vert^2)^{a/2} (K_Y (t), \dot{K}_Y (t))
\Vert^{}_{M^0}
+ \Vert (1 - \Delta) f \Vert^{}_{L^2}\cr
&\qquad{} + \sum_Z \Vert (1+ \vert\cdot  \vert)^{1+a}
F^{}_{Z(t)} (t) \Vert^{}_{L^{6/5}}\Big),\quad  0 \leq t \leq \delta
\vert x \vert,\cr
}$$
where domains of summation are as in ii).
}\saut\penalty-5000
\noindent{\it Proof.}
Statement i) of the proposition follows from [\refHSL, Theorem 3.1]. To prove
statement ii) let, for given $0 < \delta < 1,   \delta'$ be such that
$0 < \delta < \delta' < 1$ and let $\varphi \in C^\infty (\Rrm^3, \Rrm)$ be a
positive function with supp  $\varphi \subset \{x \big\vert
\vert x \vert \leq \delta' \},  \varphi (x) = 1$ for $\vert x \vert \leq
\delta$. For $t \geq 1$, let $\psi_t (x) = \varphi (x/t)$. It follows from
(5.66a)--(5.68) that
$$\eqalignno{
&\Vert \psi_t   f \Vert^{}_{L^2} + t \sum_{1 \leq i \leq 3}
\Vert \partial_i  \psi_t  f \Vert^{}_{L^2} + t^2
\sum_{\scr 1 \leq i \leq 3\atop\scr 1 \leq j \leq 3}  \Vert \partial_i
\partial_j  \psi_t  f \Vert^{}_{L^2}&(5.69)\cr
&\qquad{}\leq C_{\delta, \delta'}  \Big(\sum_Y \Vert (K_Y (t), \dot{K}_Y (t))
\Vert^{}_{M^0}
+ \sum_Z  (1+t)  \Vert \vert \nabla \vert^{-1}
F^{}_{Z(t)}  (t) \Vert^{}_{L^2}\Big),\quad      t \geq 1,\cr
}$$
where $Y,  Z \in \Pi' \cap U({\frak{sl}}(2, \Crm))$ and $\vert Y \vert \leq 2,
 \vert Z \vert \leq 2$. Let $g^{}_t (y) = \psi_t (ty)  f(ty),
 t \geq 1$. Since $\Vert \vert \nabla \vert^{-1}  F^{}_{Z(t)} (t)
\Vert^{}_{L^2} \leq C \Vert F^{}_{Z(t)} (t) \Vert^{}_{L^{6/5}}$ and
$$\eqalignno{
\sup_{\vert x\vert \leq \delta t}  t^{3/2}  \vert f(x) \vert
&\leq t^{3/2}  \Vert g^{}_t \Vert^{}_{L^\infty}&(5.70)\cr
&\leq C t^{3/2} \Vert (1 - \Delta)
 g^{}_t \Vert^{}_{L^2}\cr
&\leq C\Big(\Vert \psi_t  f \Vert^{}_{L^2} + t \sum_{1 \leq i \leq 3}
\Vert \partial_i  \psi_t f \Vert^{}_{L^2} + t^2 \sum_{\scr 1 \leq i
\leq 3\atop\scr 1
\leq j \leq 3}  \Vert \partial_i \partial_j  \psi_t f \Vert^{}_{L^2}\Big),
\quad t \geq 1,\cr
}$$
it follows from (5.69) that statement ii) is true for $t \geq 1$.
For $0 \leq t \leq 1$ it follows from $\Vert f \Vert^{}_{L^\infty}
\leq C \Vert (1 - \Delta) f \Vert^{}_{L^2}.$

To prove statement iii) we introduce the metric $ds^2 = (1+(R-t)^2)^{-1}
dR^2 + (1+R^2)^{-1}  ds'^2$ in $\Rrm^3-\{ x \big\vert \vert x \vert \leq
{1 \over 2}\}$ where $R = \vert x \vert$ and where $ds'^2$ is
the Euclidean metric on the unit sphere. It then follows from (5.66a)--(5.68)
and from [\refHSL, Proposition 2.1] that statement iii) is true.
Statement iv) follows similarly. This proves the proposition.

In the case of the Dirac equation, let $h$ be a solution in a time interval
containing $t$ of
$$(i  \gamma^\mu  \partial_\mu + m) h = g, \quad g \in
C^\infty (\Rrm^+ \times \Rrm^3). \eqno{(5.71)}$$
Let
$$\overline{{\cal M}}_{\mu \nu} = \overline{{M}}_{\mu \nu} +
{1 \over 4} (\gamma^{}_\mu
 \gamma^{}_\nu - \gamma^{}_\nu  \gamma^{}_\mu). \eqno{(5.72)}$$
It follows from (5.67) that
$$\lambda  \partial_\mu = y_\mu  \overline{D} - {1 \over 4}
y^\alpha (\gamma^{}_\alpha  \gamma^{}_\mu - \gamma^{}_\mu  \gamma^{}_\alpha) +
y^\alpha  \overline{{\cal M}}_{\alpha \mu}, \eqno{(5.73)}$$
and it follows from the Dirac equation (5.71) that
$$\overline{D} (e^{- i m  \gamma^\nu  y_\nu} h) = - e^{- i m
\gamma^\nu  y_\nu}  (- 3/2  h
+ \lambda^{-1}  \gamma^\mu  y_\mu  y^\alpha
\gamma^\beta  \overline{{\cal M}}_{\alpha \beta}  h + i \gamma^\alpha
 y_\alpha  g).\eqno{(5.74)}$$
We observe that
$$[\overline{{\cal M}}_{\mu \nu}, \gamma^\alpha  y_\alpha] = 0,\quad
 (\gamma^\nu  y_\nu)^2 = \lambda, \eqno{(5.75)}$$
\penalty9000

\noindent  which we get  by a direct calculation.

\penalty-9000
\saut
\noindent{\bf Theorem 5.7.}
{\it
Let $k \in \Nrm,  t \in \Rrm^+,  x \in \Rrm^3$ and let the
function $x \mapsto q_t (x)$ (resp. $x \mapsto r^{}_t (x))$
be defined as in (5.13c) (resp. (5.26)). If $h$ is a solution of equation
(5.71), then
$$\eqalignno{
&(1+t+\vert x \vert)^{3/2}  (1+q_t (x) + r^{}_t (x))^{k/2}  \vert
h(t,x) \vert\cr
&\ {}\leq C_k  \Big(\wp^D_{k+8} (h(t))
+ \sum_{0 \leq j \leq k+5}  \wp^D_{2+j} \big((1+q_t + r^{}_t)^{(k+7-j)/2}
g(t)\big) + t \wp^D_1 \big((1+q_t)^{(k+4)/2}  g(t)\big)\Big),\cr
}$$
for some constants $C_k$ independent of $t,x,h,g$.
}\saut
\noindent{\it Proof.}
Let $y_0 \geq 1 + \vert\vec{y}\vert,  \vec{y} = (y_1, y_2,
y_3) \in \Rrm^3$. Then $(\gamma^\mu  y_\nu)^2 = y_\nu  y^\nu
I$ and the operator $\exp (- i m  \gamma^\nu  y_\nu)$ is
unitary on $\Crm^4$. It follows from (5.73), (5.74) and (5.75) that
$$\eqalignno{
&y_0 \vert \partial_j e^{- im\gamma^\nu y_\nu}h(y) \vert &(5.76)\cr
&\qquad{}\leq C\Big(q_{y_0} (\vec{y}) \vert h (y) \vert + q_{y_0}
(\vec{y})^{3/2}
\sum_{\alpha, \beta}  \vert \overline{{\cal M}}_{\alpha \beta}
h(y) \vert + q_{y_0} (\vec{y})^{1/2}  y_0 \vert g (y) \vert\Big),
\quad 1 \leq j \leq 3.\cr
}$$
Let $\varphi \in C^\infty (\Rrm)$ be a positive function such that $\varphi (s)
= 0$
for $s \leq 1$ and $\varphi (s) = 0$ for $s \geq 2$ and let $\psi_{y_0}
(\vec{y}) =
\varphi (y_0 - \vert \vec{y} \vert)$. Since $\vert \partial_j  \psi_{y_0}
 (\vec{y}) \vert \leq C(1 + y_0)^{-1}  q_{y_0} (\vec{y})$, it
follows from (5.76), with $h^{}_1 (y) = e^{- i m \gamma^\nu y_\nu}
h(y)$, that
$$\eqalignno{
&y_0  \Vert \partial_j  q^{k/2}_{y_0}\psi_{y_0}
 h^{}_1 (y_0, \cdot) \Vert^{}_{L^p (\Rrm^3)}&(5.77)\cr
&\quad{}\leq C_k \Big(\Vert q^{k/2+1}_{y_0}  h(y_0, \cdot)
\Vert^{}_{L^p (Q_{y_0})} + \sum_{\alpha, \beta}  \Vert q^{k/2+3/2}_{y_0}
(\overline{{\cal M}}_{\alpha \beta}
 h)  (y_0, \cdot) \Vert^{}_{L^p (Q_{y_0})}\cr
&\qquad{}+ y_0  \Vert q^{1/2+k/2}_{y_0}  g(y_0, \cdot)
\Vert^{}_{L^p (Q_{y_0})}\Big), \quad  y_0 \geq 1,  1 \leq p \leq \infty,
1 \leq j \leq 3,  k \geq 0,\cr
}$$
where $Q_{y_0} = \{ \vec{y} \in \Rrm^3 \big\vert y_0 \geq 1 +
\vert \vec{y} \vert \}.$

It now follows by a substitution of variable as in the proof of statement ii)
of Proposition~5.6, using the Sobolev inequality
$\Vert\cdot  \Vert^{}_{L^\infty}
\leq C (\Vert\cdot  \Vert^{}_{L^p} + \sum_i \Vert \partial_i\cdot
\Vert^{}_{L^p}), p > 3$, that
$$\eqalignno{
&y^{3/p}_0  q^{k/2}_{y_0}  (\vec{y})  \vert h(y) \vert &(5.78)\cr
&\qquad{}\leq C_{k,p}  \Big(\Vert q^{k/2+1}_{y_0}  h(y_0, \cdot)
\Vert^{}_{L^p (Q_{y_0})}\cr
&\qquad\qquad{}
+ \sum_{\alpha, \beta}  \Vert q^{k/2+3/2}_{y_0}
(\overline{{\cal M}}_{\alpha, \beta} h) (y_0, \cdot)
\Vert^{}_{L^p (Q_{y_0})} + y_0 \Vert q^{k/2+1/2}_{y_0}
 g(y_0, \cdot) \Vert^{}_{L^p (Q_{y_0})}\Big),\cr
}$$
where $y_0 \geq 1,  p > 3,  1 \leq j \leq 3,  k \geq 0$
and $\vert \vec{y} \vert + 2 \leq y_0$. Similarly, as we obtained inequality
(5.77), we get that (with a new function $\varphi$)
$$\eqalignno{
&y_0 \Big(\Vert \partial_j  q^{k/2+1}_{y_0}  h^{}_1 (y_0, \cdot)
\Vert^{}_{L^p (Q_{y_0})} + \sum_{\alpha, \beta}  \Vert \partial_j
q^{k/2+3/2}_{y_0} (\overline{{\cal M}}_{\alpha \beta}  h^{}_1)
(y_0, \cdot) \Vert^{}_{L^p (Q_{y_0})}\Big)&(5.79)\cr
&\quad{}\leq C_k \Big(\Vert q^{k/2+2}_{y_0}  h(y_0, \cdot) \Vert^{}_{L^p
(Q'_{y_0})}
+ \sum_{\alpha, \beta}  \Vert q^{k/2+5/2}_{y_0}  (\overline{{\cal M}}_{\alpha
\beta}
 h) (y_0, \cdot) \Vert^{}_{L^p (Q'_{y_0})}\cr
&\qquad{}+ \sum_{\alpha, \beta,  \mu, \nu}  \Vert q^{k/2+3}_{y_0}
(\overline{{\cal M}}_{\alpha \beta}  \overline{{\cal M}}_{\mu \nu}  h)
(y_0, \cdot) \Vert^{}_{L^p (Q'_{y_0})}
+ y_0 \Vert q^{k/2+3/2}_{y_0}  g(y_0, \cdot) \Vert^{}_{L^p (Q'_{y_0})}\cr
&\qquad{}+ y_0 \sum_{\alpha, \beta} \Vert q^{k/2+2}_{y_0} (\overline{{\cal
M}}_{\alpha \beta}
 g) (y_0, \cdot) \Vert^{}_{L^p (Q'_{y_0})}\Big),\quad
y_0 \geq 1,  1 \leq p \leq \infty,  1 \leq j \leq 3,
k \geq 0,\cr
}$$
where $Q'_{y_0} = \{ \vec{y} \in \Rrm^3 \big\vert y_0 \geq 1/2 +
\vert \vec{y} \vert \}.$
Sobolev inequality $\Vert \cdot \Vert^{}_{L^p} \leq C (\Vert\cdot
\Vert^{}_{L^2}+$\penalty-10000
$ \sum_i \Vert \partial_i\cdot \Vert^{}_{L^2})$,  $2 \leq p \leq 6$,
and inequality (5.79) give, as in the case of (5.78), after a
change of variable
$$\eqalignno{
&y^{3/2-3/p}_0 \Big(\Vert q^{k/2+1}_{y_0}h(y_0, \cdot) \Vert^{}_{L^p (Q_{y_0})}
+
\sum_{\alpha, \beta} \Vert q^{k/2+3/2}_{y_0} (\overline{{\cal M}}_{\alpha
\beta}
h) (y_0, \cdot) \Vert^{}_{L^p (Q_{y_0})}\Big) &(5.80)\cr
&\quad{}\leq C_{k,p} \Big(\Vert q^{k/2+2}_{y_0}  h(y_0, \cdot)
\Vert^{}_{L^2 (Q'_{y_0})} +
\sum_{\alpha, \beta}  \Vert q^{k/2+5/2}_{y_0}
(\overline{{\cal M}}_{\alpha \beta}  h) (y_0, \cdot) \Vert^{}_{L^2
(Q'_{y_0})}\cr
&\qquad{}+ \sum_{\alpha, \beta,  \mu, \nu}  \Vert q^{k/2+3}_{y_0}
 (\overline{{\cal M}}_{\alpha \beta}  \overline{{\cal M}}_{\mu \nu}
 h) (y_0, \cdot) \Vert^{}_{L^2 (Q'_{y_0})}
+ y_0 \Vert q^{k/2+3/2}_{y_0}  g(y_0, \cdot) \Vert^{}_{L^2 (Q'_{y_0})}\cr
&\qquad{}+y_0  \sum_{\alpha, \beta}  \Vert q^{k/2+2}_{y_0}
(\overline{{\cal M}}_{\alpha \beta}  g) (y_0, \cdot) \Vert^{}_{L^2
(Q'_{y_0})}\Big),\cr
}$$
where $2 \leq p \leq 6,  y_0 \geq 1,  1 \leq j \leq 3$ and
$k \geq 0$. It follows now from inequalities (5.78) and (5.80) with $p=6$, that
$$\eqalignno{
&y^{3/2}_0  q^{k/2}_{y_0} (\vec{y})  \vert h(y) \vert&(5.81)\cr
&\ {} \leq
C_k  \Big(\Vert q^{k/2+2}_{y_0}  h(y_0, \cdot) \Vert^{}_{L^2 (Q'_{y_0})}
+ \sum_{\alpha^{}, \beta} \Vert q^{k/2+5/2}_{y_0}
(\overline{{\cal M}}_{\alpha \beta}
 h) (y_0, \cdot) \Vert^{}_{L^2 (Q'_{y_0})}\cr
&\quad{} + \sum_{\alpha,\beta, \mu, \nu}
 \Vert q^{k/2+3}_{y_0} (\overline{{\cal M}}_{\alpha \beta}
 \overline{{\cal M}}_{\mu \nu}
 h) (y_0, \cdot) \Vert^{}_{L^2 (Q'_{y_0})}
+ y_0 \Vert q^{k/2+3/2}_{y_0}  g(y_0, \cdot) \Vert^{}_{L^2 (Q'_{y_0})}\cr
&\quad{}+ y_0 \Vert q^{k/2+2}_{y_0}
(\overline{{\cal M}}_{\alpha \beta}  g) (y_0, \cdot)
\Vert^{}_{L^2 (Q'_{y_0})}
+ y_0 \Vert q^{k/2+1/2}_{y_0}  g(y_0, \cdot) \Vert^{}_{L^6 (Q_{y_0})}\Big),
\ y_0 \geq 1 + \vert \vec{y} \vert,  k \geq 0.\cr
}$$
By considering a time-translation in equation (5.71), by using Sobolev
embedding
for the $L^6$-term we obtain from (5.81), after changing the notation,
$$\eqalignno{
& (1+t)^{3/2}q^{k/2}_t (x) \vert h(t,x) \vert &(5.82)\cr
&\quad{}\leq C_k
\Big(\sum_{\scr Y \in \Pi'\atop\scr \vert Y \vert \leq 2}
\Vert q^{k/2+3}_t (\xi^D_Y  h) (t) \Vert^{}_{L^2 (Q''_t)}
+ t  \sum_{\scr Y \in \Pi'\atop\scr  \vert Y \vert \leq 1}
\Vert q^{k/2+2}_t (\xi^D_Y  g) (t) \Vert^{}_{L^2 (Q''_t)}\Big),\cr
}$$
$t \geq 0,\vert x \vert \leq t + 2,  k \geq 0$,
where $Q''_t = \{ x \in \Rrm^3 \big\vert  \vert x \vert \leq t + 3 \}.$

We next consider the decrease properties of $h$ outside the light-one.
According to the definition of $r^{}_t, t \geq 0$, we have, with
$\partial^\alpha = \partial^{\alpha^{}_1}_1  \partial^{\alpha^{}_2}_2
\partial^{\alpha^{}_3}_3$,
$$\sum_{\vert \alpha \vert \leq 2}  (1 + \vert x \vert)^{\vert \alpha \vert}
 \vert \partial^\alpha h(t,x) \vert = \sum_{\vert \alpha \vert \leq 2}
 r^{}_t (x)^{\vert \alpha \vert}  \vert \partial^\alpha h(t,x) \vert,$$
for $\vert x \vert \geq t$. Using a cut-off function and [\refHSL\ Proposition
2.1] we obtain
$$(1 + \vert x \vert)^{3/2+l/2}  \vert h(t,x) \vert \leq C
\sum_{\vert \alpha \vert \leq 2}  \Vert r^{\vert \alpha \vert + l/2}_t
 \partial^\alpha h(t) \Vert^{}_{L^2 (O_t)},\eqno{(5.83)}$$
$\vert x \vert \geq t+1,  l \geq 0$,
where $O_t = \{ x \in \Rrm^3 \vert  \vert x \vert \geq t \}.$

Inequality (5.82) and Theorem 5.5, with $G = 0,  i = 0,
n = 2,  L = 0$ and $k$ replaced by $k+6$, give that
$$\eqalignno{
&(1+t)^{3/2}  (1+q_t (x))^{k/2}  \vert h(t,x) \vert& (5.84)\cr
&\quad{}\leq
C_k \Big(\wp^D_{k+8} (h(t))
+ \sum_{0 \leq j \leq k+5}  \wp^D_{2+j} \big((1 + q_t)^{(k+7-j)/2} g(t)\big) +
t  \wp^D_1 \big((1 + q_t)^{(k+4)/2} g(t)\big)\Big),\cr
}$$
$\quad t \geq 0,  \vert x \vert \leq t + 2,  k \geq 0$.

Inequality (5.83) and Theorem 5.5 with $G=0$, $i=1,  n=2,
L=0$ and $k$ replaced by $l+4$, give that
$$\eqalignno{
(1 + \vert x \vert)^{(3+l)/2} \vert h(t,x) \vert
&\leq C  \wp^D_2 \big((1+q_t + r^{}_t)^{(l+4)/2} h(t)\big) &(5.85)\cr
&\leq C_l \Big(\wp^D_{l+6} (h(t)) + \sum_{0 \leq j \leq l+3}  \wp^D_{2+j}
 \big((1 + q_t + r^{}_t)^{(l+5-j)/2}  g(t)\big)\Big),\cr
}$$
$\vert x \vert \geq t + 1,  t \geq 0,  l \geq 0.$

The inequality of the theorem now follows from inequalities (5.84) and (5.85),
since $1 + q_t (x) + r^{}_t (x) \leq C(1 + q_t (x))$ for
$0 \leq \vert x \vert \leq t+1$ and $1 + q_t (x) + r^{}_t (x)
\leq C(1 + r^{}_t (x))$ for $\vert x \vert \geq t \geq 0.$
This proves the theorem.

To illustrate the use of the last theorem in our context, we shall apply it to
the equation
$$(i \gamma^\mu  \partial_\mu + m - \gamma^\mu  G_\mu) h = g.
\eqno{(5.86)}$$
In order to state the result we introduce certain notations.
Let $0 \leq a^{(0)} \leq\cdots \leq a^{(k)} \leq\cdots $ be a sequence
of real numbers and let
$$b^{}_n = \sum_{1 \leq p \leq n}
\sum_{\scr n_1+\cdots+n_p=n\atop\scr n_i\geq1}  \prod_{1 \leq j \leq p}
 a^{(n_j)},\quad n\geq1, b^{}_0=a^{(0)}. \eqno{(5.87)}$$
Let
$$\eqalignno{
T^{\infty (n)} (t)
&= \sum_{\scr Y \in \Pi'\atop\scr \vert Y \vert \leq n}
\Big(\Vert \delta (t)^{1/2}
 G_Y (t) \Vert^{}_{L^\infty} + (1+t)  \Vert \xi^{}_Y
\partial_\mu  G^\mu (t) \Vert^{}_{L^\infty} &(5.88{\rm a})\cr
&\qquad{}+ (1+t)  \Vert F^{}_Y (t) \Vert^{}_{L^\infty}
+ \Vert Q_Y (t) \Vert^{}_{L^\infty}\Big),
\quad t \geq 0,  n \geq 0\cr
}$$
where $(\delta(t))(x)=\delta (t,x) = 1 + t + \vert x \vert,
G_{Y \mu} = (\xi^M_Y  G)_\mu$ for $Y \in \Pi'$, $F^{}_{Y \mu \nu} =
\partial_\mu  G_{Y \nu} -
\partial_\nu  G_{Y \mu}$, $Q_Y (y) = y^\mu  G_{Y \mu} (y)$
and where $\xi^{}_Y$ is given by (4.81). Let
$$\eqalignno{
&T^{2(n)} (t) = \wp^{M^1}_n (G(t), \dot{G} (t)) &(5.88{\rm b})\cr
&\qquad{}+ \sum_{\scr Y \in \Pi'\atop\scr \vert Y \vert \leq n}
 \Big(\Vert \delta (t)^{- 3/2}  Q_Y (t) \Vert^{}_{L^2}
+ \Vert \delta (t)^{- 1/2}  F^{}_Y (t) \Vert^{}_{L^2} + \Vert \delta (t)^{-
1/2}
 \xi^{}_Y  \partial_\mu  G^\mu (t) \Vert^{}_{L^2}\Big),\cr
}$$
$t \geq 0,  n \geq 0$. We introduce the notation $T^\infty_n (t)$ and
$T^2_{N,n} (t),  N,n \geq 0$, by
$$T^\infty_n (t) = b^{}_n,\quad  (\hbox{resp. } T^2_{N,n} (t) = b^{}_n),
\eqno{(5.89{\rm a})}$$
where $b^{}_n$ is given by formula (5.87) for
$$a^{(n)} = T^{\infty (n)} (t),\quad  (\hbox{resp. } a^{(n)} = T^{2 (N+n)}
(t)).
\eqno{(5.89{\rm b})}$$
We define
$$T^2_n = T^2_{0,n} \eqno{(5.89{\rm c})}$$
and we note that it follows from (5.87) that $b^{}_{n_1}  b^{}_{n_2} \leq
b^{}_{n_1+n_2}$, $n_1,n_2\geq1$, which gives that
$$T^\infty_{n_1} (t)  T^\infty_{n_2} (t) \leq T^\infty_{n_1+n_2} (t),
 T^2_{N,n_1}  T^2_{N,n_2} (t) \leq T^2_{N,n_1+n_2} (t),\quad n_1,n_2 \geq 1.
\eqno{(5.89{\rm d})}$$
Moreover
$$T^2_{n+N} (t) \leq CT^2_{N,n} (t), \eqno{(5.89{\rm e})}$$
where $C$ is a function of $T^2_{N} (t)$.
\saut
\noindent{\bf Theorem 5.8.}
{\it
Let $N \geq 0,  h^{}_Y \in C^0 (\Rrm^+, D)$ for $\vert Y \vert \leq N+8,$
$(G_Y, \dot{G}_Y) \in C^0 (\Rrm^+, M^1)$ for $\vert Y \vert \leq N+10$, where
$Y \in \Pi'$, $h^{}_Y = \xi^D_Y  h,  G_Y = \xi^M_Y  G$
and $\dot{G}_Y (t) = {d \over dt}  G_Y (t)$. Let
$$H_n (t) = \sum_{\scr Y \in \Pi' \atop\scr\vert Y \vert + k \leq n}
\Vert \delta (t)^{3/2}  (1+q_t+r^{}_t)^{k/2}  h^{}_Y (t) \Vert^{}_{L^\infty},
\quad  n \geq 0,$$
where $q_t$ and $r^{}_t$ are as in (5.13c) and (5.26). Let
$$\eqalignno{
R^2_n (t) &= \sum_{l+k \leq n}  \wp^D_l
\big((1 + q_t + r^{}_t)^{k/2} g(t)\big),\cr
R^\infty_n (t) &= \sum_{\scr Y \in \Pi'\atop\scr\vert Y \vert + k \leq n}
\Vert\delta (t)^{3/2} (1+q_t+r^{}_t)^{k/2}  g^{}_Y (t) \Vert^{}_{L^\infty}\cr
\noalign{\hbox{and}}
R'_n (t) &= \sum_{l+k \leq n}  \wp^D_l
\big(\delta (t) (1 + q_t + r^{}_t)^{k/2}
g' (t)\big),\quad  n \geq 0,\cr
}$$
where $g' = {(2m)}^{-1} (m - i \gamma^\mu  \partial_\mu + \gamma^\mu
 G_\mu)  g, g^{}_Y = \xi^D_Y  g,  g'_Y =
\xi^D_Y  g'$. If $g$ is defined by formula (5.86) and if $R^2_{N+9} (t)
< \infty$, $R'_{N+7} (t) < \infty,  T^\infty_9 (t) < \infty,
T^2_{9,N} (t) < \infty$ then there is a constant
$a^{}_N$ depending only on $T^\infty_9 (t)$, such that
$$\eqalignno{
H_N (t) &\leq a^{}_N \Big(R'_{N+7} (t) + R^2_{N+9} (t) +
R^\infty_N (t) + \wp^D_{N+8}(h(t))\cr
&\qquad{}+ \sum_{\scr n_1 + n_2 \leq N\atop\scr n_2 \leq N-1}  T^2_{9,n_1} (t)
\big(R'_{n_2+7} (t) + R^2_{n_2+9} (t) + R^\infty_{n_2} (t) +
\wp^D_{n_2+8} (h(t))\big)\Big).\cr
}$$
}\saut
\noindent{\it Proof.}
We prove the theorem by induction in $N$. Suppose that the inequality of the
theorem, with $n$ instead of $N$, is true for
$0 \leq n \leq N-1$ and suppose that
the hypotheses of the theorem are true for $N.$

We first estimate $\wp^D_n \big((1 + \lambda^{}_i (t))^{k/2}  h(t)\big),
i=0$ or $i=1$, for $n+k \leq N+8$, where $\lambda^{}_i$ is as in
Theorem 5.5. According to the hypotheses, $h^{}_Y \in C^0 (\Rrm^+, D)$
for $\vert Y \vert \leq N+8, G_Y \in C^0 (\Rrm^+, L^2_{loc} (\Rrm^3,
\Rrm^4))$ (since $\Vert G_Y (t) \Vert^{}_{L^6}
\leq C \Vert (G_Y (t), 0) \Vert^{}_{M^1}$ and $L^6 (\Rrm^3) \subset
L^2_{\rm loc} (\Rrm^3))$
for $\vert Y \vert \leq N+10$ and $G_Y \in C^0 (\Rrm^+,
L^\infty (\Rrm^3, \Rrm^4))$
since $\Vert G_Y (t) \Vert^{}_{L^\infty} \leq C \Vert
(1 - \Delta)^{1/2} (G_Y (t), 0)
\Vert^{}_{M^1})$ for $\vert Y \vert \leq N+9,  Y \in \Pi'$. Moreover
$$\Vert
G_{Y_1 \mu} (t)  h^{}_{Y_2} (t) \Vert^{}_D \leq
\Vert G_{Y_1} (t) \Vert^{}_{L^6} \Vert h^{}_{Y_2} (t) \Vert^{}_{L^3}
\leq C  \wp^{M^1}_{\vert Y_1 \vert} (G(t), 0) \wp^D_{\vert
Y_2 \vert + 1}  (h(t)),$$
by Sobolev embedding.
This shows that $G_{Y_1 \mu} (t)  h^{}_{Y_2} (t) \in C^0 (\Rrm^+, D)$
for $Y_1, Y_2 \in \Pi'$ such that $\vert Y_1 \vert \leq N+7$ and
$\vert Y_2 \vert \leq N+6-L$, where $0 \leq L \leq N+7$.
Thus, the hypotheses of Theorem 5.5 are
satisfied for $n+k \leq N+8$. Using that
$$\eqalignno{
&\Vert (1 + \lambda^{}_i (t))^{(k+1-j)/2}  \gamma^\mu  G_{Z_1 \mu}
(t)  h^{}_{Z_2} (t) \Vert^{}_D\cr
&\qquad{}\leq \Vert G_{Z_1 \mu} (t) \Vert^{}_{L^6}
\Vert (1 + \lambda^{}_i (t))^{(k+1-j)/2}
 h^{}_{Z_2} (t) \Vert^{}_{L^3}\cr
&\qquad{}\leq C \Vert (G_{Z_1} (t), 0) \Vert^{}_{M^1}
\Vert \delta (t)^{- 3/2} \Vert^{}_{L^3}
 \Vert \delta (t)^{3/2}  (1 + \lambda^{}_i (t))^{(k+1-j)/2}
h^{}_{Z_2} (t) \Vert^{}_{L^\infty},\cr
}$$
for some constant $C$ and using that $\Vert \delta (t)^{- 3/2}
\Vert^{}_{L^3} \leq C' (1+t)^{- 1/2} \leq C'$ for some constant $C'$,
we obtain using Theorem 5.5 that
$$\eqalignno{
&\wp^D_n \big((1 + \lambda^{}_i (t))^{k/2}  h(t)\big)\cr
&\ {}\leq C'_{n+k}  \sum_{0 \leq l \leq L}  (1 + \tau^{}_{i,l} (t))
\sum_{0 \leq j \leq k-1}  \wp^{}_{n+j-l} \big((1 +
\lambda^{}_i (t))^{(k+1-j)/2}
 g(t)\big)\cr
&\quad{}+ C'_{n+k}  \sum_{0 \leq l \leq L} (1 + \tau^{}_{i,l} (t))
\wp^D_{n+k-l}  (h(t))\cr
&\quad{}+ C'_{n+k}  \sum_{0 \leq l \leq L} (1 + \tau^{}_{i,l} (t))\!\!
\sum_{{\sscr 0 \leq j \leq k-1\atop\sscr Z_2 \in \Pi'}
\atop {\sscr n_1 + \vert Z_2 \vert \leq n+j-l\atop\sscr \vert Z_2 \vert \leq
n+j-L-1}}\!\!  T^{2(n_1)} (t)
\Vert \delta (t)^{3/2} (1+\lambda^{}_i (t))^{(k+1-j)/2}    h^{}_{Z_2} (t)
\Vert^{}_{L^\infty},\cr
}$$
where $n+k \leq N+8,  0 \leq L \leq n+k-1$ and $C'_{n+k}$ is a constant
depending only on $\tau^{}_{i,0} (t)$. Using the definitions of $R^{}_n (t)$
and
$H_n (t)$, we obtain that
$$\eqalignno{
&\wp^D_n \big((1 + \lambda^{}_i (t))^{k/2}  h(t)\big) &(5.90)\cr
&\quad {}\leq C'_{n+k}  \sum_{\scr n_1 \leq L\atop\scr n_1 + n_2 = n+k+1}
 (1 + \tau^{}_{i, n_1} (t))  R^2_{n_2} (t)
+ C'_{n+k}  \sum_{\scr n_1 \leq L\atop\scr n_1 + n_2 = n+k}
(1 + \tau^{}_{i,n_1}(t))  \wp^D_{n_2}  (h(t))\cr
&\quad\quad{}+C'_{n+k}  \sum_{\scr n_1+n_2+n_3=n+k+1\atop\scr n_1
\leq L, n_3 \leq n+k-L}
(1 + \tau^{}_{i, n_1} (t))  T^{2 (n_2)} (t)  H_{n_3} (t),
\quad 0 \leq L \leq n+k-1.\cr
}$$
Let
$$h^{(1)} = h - (2m)^{-1}  (G_\mu  \gamma^\mu  h +
g). \eqno{(5.91)}$$
Then it follows that
$$\eqalignno{
\wp^D_n (h^{(1)} (t))
&\leq \wp^D_n (h(t)) + {(2m)}^{-1}  \wp^D_n (g(t))\cr
&\qquad{}+ C_n  \sum_{{\sscr Z \in\Pi'\atop\sscr\vert Z \vert + n_2 \leq n}
\atop\sscr \vert Z \vert \leq L}
\Vert G_Z (t) \Vert^{}_{L^\infty}  \wp^D_{n_2} (h(t))\cr
&\qquad{}+ C_n  \sum_{{\sscr Z_1, Z_2 \in \Pi'\atop\sscr
\vert Z_1 \vert + \vert Z_2 \vert \leq n}
\atop\sscr  \vert Z_2 \vert\leq n-L-1}  \Vert G_{Z_1} (t) \Vert^{}_{L^6}
\Vert h^{}_{Z_2} (t) \Vert^{}_{L^3}.\cr
}$$
Since $\Vert h^{}_{Z_2} (t) \Vert^{}_{L^3} \leq C (1+t)^{- 1/2}  \Vert \delta
(t)^{3/2}  h^{}_{Z_2} (t) \Vert^{}_{L^\infty} \leq C \Vert \delta (t)^{3/2}
 h^{}_{Z_2} (t) \Vert^{}_{L^\infty}$, we obtain that
$$\eqalignno{
\wp^D_n (h^{(1)} (t)) &\leq \wp^D_n (h(t)) + {(2m)}^{-1}  \wp^D_n (g(t))
&(5.92)\cr
&\qquad{}+ C_n \Big(\sum_{\scr n_1 + n_2 = n\atop\scr n_1
\leq L}  T^\infty_{n_1} (t)
\wp^D_{n_2} (h(t)) + \sum_{\scr n_1 + n_2 = n\atop\scr n_2
\leq n-L-1}  T^2_{n_1} (t)
 H_{n_2} (t)\Big),\cr
}$$
where $0 \leq L \leq n.$

Let
$$\eqalignno{
g^{(1)} &= {(2m)}^{-1}  \gamma^\mu  \gamma^\nu  G_\mu
 G_\nu  h - i{m}^{-1}  G^\mu    \partial_\mu
 h &(5.93{\rm a})\cr
&\qquad{}- i{(2m)}^{-1}  h  \partial_\mu  G^\mu - i{(8m)}^{-1}
 (\gamma^\mu  \gamma^\nu - \gamma^\nu  \gamma^\mu)
 h  F^{}_{\mu \nu},\cr
}$$
where
$$F^{}_{\mu \nu} = \partial_\mu  G_\nu - \partial_\nu  G_\mu, \quad
 0 \leq \mu \leq 3,  0 \leq \nu \leq 3. \eqno{(5.93{\rm b})}$$
It follows from (5.93a) and (5.93b) that
$$\eqalignno{
\xi^D_Y  g^{(1)} &= {(2m)}^{-1}  \gamma^\mu  \gamma^\nu
 \suma_{Y_1, Y_2, Y_3}^Y  G_{Y_1 \mu}  G_{Y_2 \nu}
 h^{}_{Y_3}& (5.94{\rm a})\cr
&\quad{}- i{m}^{-1}  \suma_{Y_1, Y_2}^Y  \Big(G^\mu_{Y_1}
\partial_\mu  h^{}_{Y_2} + {1 \over 2}  h^{}_{Y_2}  \partial_\mu
 G^\mu_{Y_1}
+ {1 \over 8} (\gamma^\mu  \gamma^\nu - \gamma^\nu  \gamma^\mu)
 h^{}_{Y_2}  F^{}_{Y_1 \mu \nu}\Big),\ Y \in U(\p),\cr
}$$
where
$$F^{}_{Y \mu \nu} = \partial_\mu       G_{Y \nu} - \partial_\nu
G_{Y \mu}, \quad G_{Y \mu} = (\xi^M_Y G)_\mu. \eqno{(5.94{\rm b})}$$
According to inequality (5.7d), we have
$$\eqalignno{
&\big\vert \sum_\mu  G^\mu_{Y_1}  \partial_\mu
h^{}_{Y_2} \big\vert\delta &(5.95)\cr
&\qquad{}\leq C\Big(\sum_\mu \vert G^\mu_{Y_1} \vert
\vert h^{}_{P_\mu Y_2} \vert
+ \vert Q_{Y_1} \vert  \sum_\nu  \vert h^{}_{P_\nu Y_2} \vert +
\sum_{\mu, \nu} \vert G^\mu_{Y_1} \vert
(\vert h^{}_{M_{\mu \nu} Y_2} \vert +
\vert h^{}_{Y_2} \vert)\Big),\cr
}$$
where the sums are taken over $0 \leq \mu \leq 3,  0 \leq \nu \leq 3.$
Equality (5.94a) and inequality (5.95) give
$$\eqalignno{
\vert \xi^D_Y  g^{(1)} \vert &\leq C\Big(\suma_{Y_1,Y_2,Y_3}^Y
\vert G_{Y_1} \vert  \vert G_{Y_2} \vert  \vert h^{}_{Y_3} \vert
&(5.96)\cr
&\qquad{}+ \suma_{Y_1,Y_2}^Y \Big(\delta^{-1} \big((\vert
G_{Y_1} \vert + \vert Q_{Y_1} \vert)
\sum_\mu  \vert h^{}_{P_\mu Y_2} \vert
+ \vert G_{Y_1} \vert  \big(\vert h^{}_{Y_2} \vert + \sum_{\mu, \nu}
\vert h^{}_{M_{\mu \nu} Y_2} \vert\big)\big)\cr
&\qquad{}+ \big\vert \sum_\mu  \partial_\mu  G^\mu_{Y_1} \big\vert
\vert h^{}_{Y_2} \vert + \vert F^{}_{Y_1} \vert  \vert h^{}_{Y_2}
\vert\Big)\Big),
\quad Y \in U(\p).\cr
}$$
Using that
$$\eqalignno{
\Vert G_{Y_1 \mu} (t)  G_{Y_2 \nu} (t)  h^{}_{Y_3} (t)
\Vert^{}_D &\leq \Vert G_{Y_1} (t) \Vert^{}_{L^6}
\Vert G_{Y_2} (t) \Vert^{}_{L^6}
\Vert h^{}_{Y_3} (t) \Vert^{}_{L^6}\cr
&\leq C \Vert (G_{Y_1}(t), 0) \Vert^{}_{M^1}\Vert
(G_{Y_2} (t), 0) \Vert^{}_{M^1}\Vert h^{}_{Y_3} (t) \Vert^{}_{L^6},\cr
}$$
by H\"older inequality
and Sobolev embedding, using the argument in the derivation of (5.92) and using
definitions (5.88a) and (5.88b) of $T^{\infty (n)}$ and $T^{2 (n)}$, we obtain
from (5.96)
$$\eqalignno{
&\wp^D_n \big(\delta (t)  (1 + \lambda^{}_1 (t))^{k/2}  g^{(1)} (t)\big)\cr
&\qquad{}\leq C_{n+k}   \sum_{\scr n_1+n_2+n_3 \leq n\atop\scr n_1+n_2 \leq L}
T^\infty_{n_1} (t)  T^\infty_{n_2} (t)  \wp^D_{n_3}
\big((1+\lambda^{}_1 (t))^{k/2}
 h(t)\big)\cr
&\qquad\qquad{}+ C_{n+k}  \sum_{{Z\sscr
\in \Pi'\atop\sscr n_1+n_2+\vert Z \vert \leq n}
\atop\sscr n_1+n_2 \geq L+1}
 T^2_{n_1} (t)  T^2_{n_2} (t)  \Vert \delta (t)
(1 + \lambda^{}_1 (t))^{k/2}  h^{}_Z (t) \Vert^{}_{L^6}\cr
&\qquad\qquad{}+ C_{n+k}  \sum_{\scr n_1+n_2 \leq n\atop\scr
n_1 \leq L}T^\infty_{n_1} (t)
 \wp^D_{n_2+1} \big((1+\lambda^{}_1 (t))^{k/2}  h(t)\big)\cr
&\qquad\qquad{}+ C_{n+k}  \sum_{{\sscr Z \in \Pi',\vert Z \vert \geq 1}\atop
{\sscr n_1+\vert Z \vert \leq n+1\atop\sscr n_1 \geq L+1}}
T^2_{n_1} (t)  \Vert \delta(t)^{3/2}
(1 + \lambda^{}_1 (t))^{k/2}  h^{}_Z (t) \Vert^{}_{L^\infty},\cr
}$$
where $0 \leq L \leq n$. Using inequality (5.89d) we obtain that
$$\eqalignno{
&\wp^D_n \big(\delta (t) (1 + \lambda^{}_1 (t))^{k/2}  g^{(1)} (t)\big)\cr
&\qquad{}\leq C_{n+k} \Big(\sum_{\scr n_1+n_2
\leq n\atop\scr  n_1 \leq L}  T^\infty_{n_1} (t)
 \wp^D_{n_2+1} \big((1 + \lambda^{}_1 (t))^{k/2}  h(t)\big)\cr
&\qquad\qquad{}+ \sum_{{\sscr n_1 \leq n, Z \in \Pi'}\atop
{\sscr n_1 + \vert Z \vert \leq n+1\atop\sscr n_1 \geq L+1}}
 T^2_{n_1} (t)  \Vert \delta (t)^{3/2}  (1 + \lambda^{}_1
(t))^{k/2}  h^{}_Z (t) \Vert^{}_{L^\infty}\cr
&\qquad\qquad{}+ \sum_{\scr n_1 + \vert Z \vert \leq n\atop\scr
n_1 \geq L+1}  T^2_{n_1} (t)
 \Vert \delta (t)  (1 + \lambda^{}_1 (t))^{k/2}  h^{}_Z (t)
\Vert^{}_{L^6}\Big),\cr
}$$
where $0 \leq L \leq n$. Since
$$\Vert \delta (t)  f \Vert^{}_{L^6} \leq
\Vert f \Vert^{1/3}_{L^2}\big(\Vert (\delta (t))^{3/2}f
\Vert^{}_{L^\infty}\big)^{2/3}
\leq C\big(\Vert f \Vert^{}_{L^2} +
\Vert (\delta (t))^{3/2}  f \Vert^{}_{L^\infty}\big),$$
it follows that
$$\eqalignno{
&\wp^D_n (\delta (t)  (1 + \lambda^{}_1 (t))^{k/2}  g^{(1)} (t))\cr
&\qquad{}\leq C_{n+k} \Big(\sum_{\scr n_1+n_2
\leq n\atop\scr n_1 \leq L }  T^\infty_{n_1} (t)
 \wp^D_{n_2+1} \big((1 + \lambda^{}_1 (t))^{k/2}  h(t)\big)\cr
&\qquad\qquad{}+ \sum_{\scr n_1+n_2 \leq n\atop\scr  n_1 \geq L+1}  T^2_{n_1}
(t)
\wp^D_{n_2} \big((1 + \lambda^{}_1 (t))^{k/2}h(t)\big)\cr
&\qquad\qquad{}+ \sum_{{\sscr n_1 + \vert Z \vert \leq n+1}
\atop{\sscr n_1 \leq n, Z \in \Pi'\atop\sscr  n_1 \geq L+1}}
 T^2_{n_1} (t)  \Vert \delta (t)^{3/2} (1 + \lambda^{}_1 (t))^{k/2}
 h^{}_Z (t) \Vert^{}_{L^\infty}\Big),\cr
}$$
where $0 \leq L$. This inequality and the definition of $H_n$, give that
$$\eqalignno{
&\sum_{n+k \leq M}  \wp^D_n \big(\delta (t) (1 + \lambda^{}_1 (t))^{k/2}
g^{(1)} (t)\big)& (5.97)\cr
&\qquad{}\leq C_M \Big(\sum_{\scr n_1+n_2+k \leq M\atop\scr n_1 \leq L }
T^\infty_{n_1} (t)
 \wp^D_{n_2+1} \big((1 + \lambda^{}_1 (t))^{k/2}  h(t)\big)\cr
&\qquad\qquad{}+ \sum_{\scr n_1+n_2+k \leq M\atop\scr n_1 \geq L+1}
T^2_{n_1} (t)
\wp^D_{n_2} \big((1 + \lambda^{}_1 (t))^{k/2}h(t)\big)\cr
&\qquad\qquad{}+ \sum_{\scr n_1+n_2 \leq M\atop\scr  n_1 \geq L+1}
T^2_{n_1} (t)H_{n_2+1}
(t)\Big),\cr
}$$
where $L \geq 0$ is an integer. With $L = L_0$ and $i = 1$ in inequality
(5.90),
we obtain, using that $\tau^{}_{1,n} (t) \leq T^\infty_n (t)$,
$$\eqalignno{
&\wp^D_n \big((1 + \lambda^{}_1 (t))^{k/2}  h(t)\big)&(5.98)\cr
&\qquad{}\leq C''_{n+k} \Big(R^2_{n+k+1} (t) + \wp^D_{n+k}  (h(t))
+ \sum_{\scr n_1+n_2 \leq n+k+1\atop\scr n_2 \leq n+k-L_0}  T^2_{n_1} (t)
H_{n_2} (t)\Big),\cr
}$$
where $C''_{n+k}$ is a constant depending only on $T^\infty_{L_0} (t)$. It
follows
from inequalities (5.97) and (5.98) that

\penalty-9900

$$\eqalignno{
&\sum_{\scr n_1+n_2+k \leq M\atop\scr n_1 \leq L }  T^\infty_{n_1} (t)
\wp^D_{n_2+1} \big((1 + \lambda^{}_1 (t))^{k/2}  h(t)\big)&(5.99)\cr
&\qquad{}\leq a^{}_{M, L_0}  \Big(\sum_{\scr n_1+n_2
\leq M+2 \atop\scr n_1 \leq L }
T^\infty_{n_1} (t)  R^2_{n_2} (t) + \sum_{\scr n_1+n_2
\leq M+1\atop\scr n_1 \leq L}
 T^\infty_{n_1} (t)  \wp^D_{n_2} (h(t))\cr
&\qquad\qquad{}+ \sum_{\scr n_1+n_2+n_3 \leq M+2\atop\scr n_1
\leq L, n_3 \leq M+1-L_0}
T^\infty_{n_1}
(t)  T^2_{n_2} (t)  H_{n_3} (t)\Big), \quad  L \geq 0,
L_0 \geq 0,  M \geq 0,\cr
}$$
where $a^{}_{M,L_0}$ is a constant depending only on
$T^\infty_{L_0} (t)$. Similarly
we obtain that
$$\eqalignno{
&\sum_{\scr n_1+n_2+k \leq M \atop\scr  n_1 \geq L+1}
T^2_{n_1} (t)    \wp^D_{n_2}
\big((1 + \lambda^{}_1 (t))^{k/2}  h(t)\big) &(5.100)\cr
&\qquad{}\leq a^{}_{M,L_0}
\Big(\sum_{\scr n_1 \geq L+1\atop\scr n_1+n_2 \leq M+2}
T^2_{n_1}(t)  R^2_{n_2} (t) +
\sum_{\scr n_1+n_2 \leq M+1\atop\scr n_1 \geq L+1}
T^2_{n_1} (t)  \wp^D_{n_2} (h(t))\cr
&\qquad{}\qquad{}+ \sum_{\sscr n_1+n_2 \leq M+2\atop{\sscr n_1 \geq L+1
\atop\sscr  n_2 \leq M-L-L_0}}  T^2_{n_1}(t)
H_{n_2} (t)\Big), \quad L \geq 0, L_0 \geq 0,  M \geq 0,\cr
}$$
where $a^{}_{M,L_0}$ is a constant depending only on $T^\infty_{L_0} (t).$
It follows from inequalities (5.97) and (5.99), and from inequality (5.100)
with
$N_0$ instead of $L_0$, that
$$\eqalignno{
&\sum_{n+k \leq M}  \wp^D_n \big(\delta (t)  (1 + \lambda^{}_1 (t))^{k/2}
 g^{(1)} (t)\big)&(5.101)\cr
&\qquad{}\leq a^{}_{M,L_0,N_0}
\Big(\sum_{\scr n_1+n_2 \leq M+2\atop\scr n_1 \leq L}
T^\infty_{n_1} (t)  R^2_{n_2} (t) +
\sum_{\scr n_1+n_2 \leq M+2 \atop\scr n_1 \geq L+1}
 T^2_{n_1} (t)  R^2_{n_2} (t)\cr
&\qquad\qquad{}+ \sum_{\scr n_1+n_2
\leq M+1\atop\scr n_1 \leq L }  T^\infty_{n_1} (t)
\wp^D_{n_2} (h(t)) + \sum_{\scr n_1+n_2
\leq M+1\atop\scr n_1 \geq L+1 }  T^2_{n_1} (t)
 \wp^D_{n_2}  (h(t))\cr
&\qquad\qquad{}+ \sum_{\sscr n_1+n_2+n_3 \leq M+2 \atop{\sscr n_1 \leq L
\atop\sscr n_3 \leq M+1-L_0}}  T^\infty_{n_1}(t)  T^2_{n_2} (t)  H_{n_3} (t)
+ \sum_{\sscr n_1+n_2 \leq M+2
\atop {\sscr n_1 \geq L+1\atop\sscr n_2 \leq M-L-N_0}}  T^2_{n_1}
(t)  H_{n_2} (t)\Big),\cr
}$$
where $M,L,L_0,N_0$ are nonnegative integers and
$a^{}_{M,L_0,N_0}$ is a constant depending only on
$T^\infty_{L_0} (t)$ and $T^\infty_{N_0} (t).$

Since $h$ and $g$ satisfy equation (5.86) it follows, as in the proof of
Theorem 5.1 (cf. (5.8) and (5.9)), that
$$(i  \gamma^\mu  \partial_\mu + m)  h^{(1)} = g' +
g^{(1)}, \eqno{(5.102{\rm a})}$$
where $h^{(1)}$ is given by (5.91), $g^{(1)}$ by (5.93a) and $g' = {(2m)}^{-1}
 (m - i  \gamma^\mu  \partial_\mu + \gamma^\mu
 G_\mu)  g$. It follows from (5.102) that
$$(i  \gamma^\mu  \partial_\mu + m)  h^{(1)}_Y =
g'_Y + g^{(1)}_Y,\quad  Y \in U(\p). \eqno{(5.102\hbox{b})}$$
It follows from Theorem 5.7 and equation (5.102b) that
$$\eqalignno{
&\sum_{\scr Y \in \Pi'\atop\scr \vert Y \vert + k \leq N}  (\delta (t,x))^{3/2}
 (1 + \lambda^{}_1 (t,x))^{k/2}  \vert h^{(1)} (t,x) \vert&(5.103)\cr
&\qquad{}\leq C_N \Big(\wp^D_{N+8}
(h^{(1)} (t)) + R'_{N+7} (t) + \sum_{n+k \leq N+7}
 \wp^D_n \big(\delta (t)  (1 + \lambda^{}_1 (t))^{k/2}
g^{(1)}(t)\big)\Big),\cr
}$$
for some constant $C_N$. Let $n = N+8,  L=8$ in (5.92) and let $M =
N+7,  L=9,  L_0 = 9,  N_0 = 0$ in (5.101). It then
follows from inequalities (5.92), (5.101) and (5.103) that
$$\sum_{\scr Y \in \Pi'\atop\scr \vert Y \vert + k \leq N}  (\delta
(t,x))^{3/2}
 (1 + \lambda^{}_1 (t,x))^{k/2}  \vert h^{(1)}_Y (t,x) \vert \leq
b^{}_N (t), \eqno{(5.104)}$$
where
$$\eqalignno{
\hskip-5mm b^{}_N (t) &= a^{}_N \Big(R'_{N+7} (t) + R^2_{N+9} (t) +
\wp^D_{N+8} (h(t))
+ \sum_{\scr n_1+n_2\leq N+9\atop\scr n_2 \leq N-1}  T^2_{n_1}
(t)R^2_{n_2} (t)&(5.105{\rm a})\cr
&\qquad{}+ \sum_{\scr n_1+n_2\leq N+8\atop\scr n_2 \leq N-2}
T^2_{n_1} (t)
\wp^D_{n_2} (h(t)) + \sum_{\scr n_1+n_2\leq N+9\atop\scr n_2 \leq N-1}
T^2_{n_1} (t)
 H_{n_2} (t)\Big),\cr
}$$
for some constant $a^{}_N$ depending only on $T^\infty_9 (t)$.
According to definition (5.89a) of $T^2_{N,n}$ and according to inequality
(5.89e), it follows from (5.104a)
that
$$\eqalignno{
b^{}_N (t) &\leq a^{}_N \Big(R'_{N+7} (t) + R^2_{N+9} (t) +
\wp^D_{N+8} (h(t))&(5.105{\rm b})\cr
&\qquad{}+ \sum_{\scr n_1+n_2 \leq N\atop\scr  n_2 \leq N-1}  T^2_{9,n_1} (t)
\big(R^2_{n_2} (t) + \wp^D_{n_2} (h(t)) + H_{n_2} (t)\big)\Big).\cr
}$$

Definition (5.91) of $h^{(1)}$ gives that
$$h^{(1)}_Y = h^{}_Y - {(2m)}^{-1}  G_{\un \mu}  \gamma^\mu
 h^{}_Y - {(2m)}^{-1}  g^{}_Y - {(2m)}^{-1}  \suma_{\scr Y_1,Y_2\atop\scr
\vert Y_2 \vert \leq \vert Y \vert - 1}^Y  G_{Y_1 \mu}  \gamma^\mu
 h^{}_{Y_2},$$
$Y \in \Pi'$, which together with inequality (5.104) shows that
$$\eqalignno{
&\sum_{\scr Y \in \Pi' \atop\scr \vert Y \vert + k \leq N}  \delta (t,x)^{3/2}
(1 + \lambda^{}_1 (t,x))^{k/2}  \big(\vert h^{}_Y (t,x) \vert - {(2m)}^{-1}
 \vert G_{\un \mu} (t,x) \vert  \vert h^{}_Y (t,x) \vert\big)\cr
&\ \leq {2m}^{-1} R^\infty_N (t) + {(2m)}^{-1}
\!\!\!\sum_{{\sscr Y_1, Y_2 \in \Pi'\atop\sscr\vert Y \vert + k \leq N}\atop
{\sscr \vert Y_1 \vert + \vert Y_2 \vert = \vert Y \vert
\atop\sscr \vert Y_2 \vert \leq \vert Y \vert -1}}\!\!\!
 \Vert G_{Y_1} (t) \Vert^{}_{L^\infty}
\Vert \delta(t)^{3/2} (1 + \lambda^{}_1 (t))^{k/2} h^{}_{Y_2} (t)
\Vert^{}_{L^\infty} + b^{}_N (t).\cr
}$$
Using that $\Vert G_{Y_1} (t) \Vert^{}_{L^\infty} \leq \Vert (1 - \Delta)^{1/2}
\nabla G_{Y_1} (t) \Vert^{}_{L^2} \leq \wp^{M^1}_{\vert Y_1 \vert +
1}\big((G(t),0)\big),$ that
$$\eqalignno{
&\delta (t,x)^{3/2}  (1 + \lambda^{}_1 (t,x))^{k/2}
\vert G_\mu (t,x) \vert  \vert h^{}_Y (t,x) \vert\cr
&\qquad{}\leq \vert \delta (t,x)^{1/2}  G (t,x) \vert   \big(\delta (t,x)^{3/2}
 (1 + \lambda^{}_1 (t,x))^{k/2}  \vert h^{}_Y (t,x) \vert\big)^{\varepsilon_k}
 \vert h^{}_Y (t,x) \vert^{1 - \varepsilon_k}\cr
}$$
where $\varepsilon_k = (2+k)/(3+k)$ and that $\vert h^{}_Y (t,x) \vert \leq C
\wp^D_{\vert Y \vert + 2}  (h(t))$, we obtain that
$$\eqalignno{
&\delta (t,x)^{3/2}  (1+ \lambda^{}_1 (t,x))^{k/2}  \vert h^{}_Y
(t,x) \vert& (5.106)\cr
&\qquad{}- (2m)^{-1}  T^\infty_0 (t)  \big(\delta (t,x)^{3/2}
(1 + \lambda^{}_1 (t,x))^{k/2}  \vert h^{}_Y (t,x) \vert\big)^{\varepsilon_k}
 (\wp^D_{\vert Y \vert + 2}  (h(t))^{1 - \varepsilon_k}\cr
&\qquad\qquad{}\leq (2m)^{-1}  R^\infty_N (t) + (2m)^{-1}  \sum_{\scr n_1+n_2
\leq N+1\atop\scr n_2 \leq N-1}  T^2_{n_1} (t)  H_{n_2} (t) + b^{}_N (t).\cr
}$$
For fixed $t,x,k$ and $Y$, let $a = \delta (t,x)^{3/2}  (1 + \lambda^{}_1
(t,x))^{k/2}  \vert h^{}_Y (t,x) \vert$ and let $b'_N$ be the right-hand
side of inequality (5.106). Then
$$a - (2m)^{-1}  T^\infty_0 (t)  \big(\wp^D_{\vert Y \vert + 2}
 (h(t))\big)^{1 - \varepsilon_k}  a^{\varepsilon_k} \leq b'_N.$$
Since $0 \leq \varepsilon_k < 1$, it follows that
$$a \leq C_1  \wp^D_{\vert Y \vert + 2}  (h(t)) + C_2
b'_N,$$
where $C_1$ and $C_2$ are constants depending on $\varepsilon_k$ and $(2m)^{-1}
 T^\infty_0 (t)$. Thus, it follows from (5.106) that
$$\eqalignno{
&\delta (t,x)(1 + \lambda^{}_1 (t,x))^{k/2}  \vert h^{}_Y (t,x) \vert\cr
&\qquad{}\leq C_{k, \vert Y \vert}  \Big(R^\infty_N (t) +
\sum_{\scr n_1+n_2 \leq N\atop\scr
n_2 \leq N-1}  T^2_{1, n_1} (t)  H_{n_2} (t) + b^{}_N (t)\Big),\cr
}$$
$Y \in \Pi',  \vert Y \vert + k \leq N,$
which together with inequalities (5.104) and (5.105b) give that
$$\eqalignno{
H_N (t) &\leq a^{}_N \Big(R'_{N+7} (t) + R^2_{N+9} (t) +
R^\infty_N (t) + \wp^D_{N+8}  (h(t))&(5.107)\cr
&\qquad{}+ \sum_{n_1+n_2 \leq N, n_2 \leq N-1}  T^2_{9,n_1} (t)
\big(R^2_{n_2} (t) + \wp^D_{n_2} (h(t)) + H_{n_2} (t)\big)\Big),\cr
}$$
where $a^{}_N$ is a constant depending only on $T^\infty_9 (t).$

The inequality of the theorem for $N=0$ follows by taking $N=0$ in (5.107). For
$N \geq 1$ it follows from inequality (5.107), the induction hypothesis and the
inequality of the theorem, that
$$\eqalignno{
H_N (t) &\leq a^{}_N \Big(R'_{N+7} (t) + R^2_{N+9} (t) + R^\infty_N (t) +
\wp^D_{N+8}  (h(t))\cr
&\qquad{}+ \sum_{\scr n_1+n_2 \leq N\atop\scr n_2 \leq N-1}  T^2_{9,n_1} (t)
\big(R^2_{n_2}
(t) + \wp^D_{n_2}  (h(t))\big)\cr
&\qquad{}+ \sum_{\scr n_1+n_2 \leq N\atop\scr n_2 \leq N-1}  T^2_{9,n_1} (t)
a^{}_{n_2}
\Big(R'_{n_2+7} (t) + R^2_{n_2+9} (t) + R^\infty_{n_2} (t) + \wp^D_{n_2+8}
(h(t))\cr
&\qquad{}+ \sum_{\scr n_3+n_4 \leq n_2\atop\scr  n_4 \leq n_2-1}  T^2_{9,n_3}
(t)
\big(R'_{n_4+7} (t) + R^2_{n_4+9} (t) + R^\infty_{n_4} (t) + \wp^D_{n_4+8}
(h(t))\big)\Big)\Big),\cr
}$$
$N \geq 1$, where $a^{}_N$ depends only on $T^\infty_9 (t)$ and where
we have chosen
$a^{}_0 \leq a^{}_1 \leq \cdots\leq a^{}_N.$

This inequality and the fact that
$$\eqalignno{
&a^{}_N  \sum_{\scr n_1+n_2 \leq N\atop\scr n_2 \leq N-1}
\sum_{\scr n_3+n_4 \leq n_2\atop\scr
n_4 \leq n_2-1}  T^2_{9,n_1} (t)  a^{}_{n_2} \big(R'_{n_4+7} (t) +
R^2_{n_4+9} (t) + R^\infty_{n_4} (t)+ \wp^D_{n_4+8}  (h(t))\big)\cr
&\qquad{}\leq a^{}_N  a^{}_{N-1}
\sum_{\scr n_1+n_2 \leq N\atop\scr n_2 \leq N-1}
 T^2_{9,n_1} (t)  \big(R'_{n_2+7} (t) + R^2_{n_2+9} (t) + R^\infty_{n_2}
(t) + \wp^D_{n_2+8}  (h(t))\big),\cr
}$$
obtained by using inequality (5.89d), prove the theorem, after redefinition of
the constant~$a^{}_N.$

Theorem 5.5 and Theorem 5.8 give an estimate of $\wp^D_n
\big((1 + q_t + r^{}_t)^{k/2} (h(t)\big)$ not containing $L^\infty$-norms
of $h^{}_Y (t), Y \in \Pi'$. To state the result we adapt the notation
of these theorems.
\saut
\noindent{\bf Corollary 5.9.}
{\it
Let $n \geq 0$, $k \geq 1$,  $L \geq 3$, let $h^{}_Y \in C^0 (\Rrm^+,D)$
for $Y \in \Pi'$,  $\vert Y \vert \leq \max (n+k, n+k+8-L)$, let $G_Y \in
C^0 (\Rrm^+, L^\infty (\Rrm^3, \Rrm^4))$ for $0 \leq \vert Y \vert \leq L$,
 $Y \in \Pi'$ and let $\delta^{-1}  G_Y \in C^0 (\Rrm^+,
L^2 (\Rrm^3, \Rrm^4))$, for $\vert Y \vert \leq n+k-1$,  $Y \in \Pi'$, if
$L \leq n+k-2$, where $(\delta (t)) (x) = (1+t+\vert x \vert)$. Let
$$R^i_{p,q} (t) = \Big(\sum_{\scr 0 \leq j \leq q-1\atop\scr p+j\geq0}
\big(\wp^D_{p+j} \big((1+\lambda^{}_i
(t))^{(q+1-j)/2}  g(t)\big)\big)^2\Big)^{1/2},\quad i=0,1,$$
for integers $p \geq -q+1,  q \geq 1$ and $R^i_{p,q} = 0$ otherwise, let
$$\Gamma^{}_p (t) = \Big(\sum_{\scr Y \in \Pi'\atop\scr  \vert Y \vert \leq p}
\Vert \delta (t)^{-1}  G_Y (t) \Vert^2_{L^2 (\Rrm^3, \Rrm^4)}\Big)^{1/2},
\quad  p \geq 0$$
and let $\chi^{}_+ (s) = 0$ for $s < 0$ and $\chi^{}_+ (s) = 1$ for $s \geq 0,
s \in \Rrm$. Let $(G_Y, \dot{G}_Y) \in C^0 (\Rrm^+, M^1)$ for $Y \in \Pi',
\vert Y \vert \leq n+k-L+10$. If $R^2_{N+9} (t) + R'_{N+7} (t) + T^2_{9,N} (t)
+
T^\infty_9 (t) < \infty$ for $N = \max(1, n+k-L)$, then
$$\eqalignno{
&\wp^D_n \big((1+ \lambda^{}_i (t))^{k/2}h(t)\big)\cr
&\ {}\leq C'_k \big(\wp^D_{n+k} (h(t)) +
R^i_{n,k} (t)\big)\cr
&\quad{}+ C'_{n+k}  \sum_{\scr n_1+n_2=n+k\atop\scr  1 \leq n_1 \leq L}
(1 + \tau^{}_{1,n_1} (t)) \big(\wp^D_{n_2} (h(t)) + R^i_{n_2-k,k} (t)\big)\cr
&\quad{}+ C''_k  \chi^{}_+ (n+k-L-2)  \Gamma^{}_{n+k-1} (t)  \big(1+
T^2_{9,0} (t)) (R'_8 (t) + R^2_{10} (t) + R^\infty_1 (t) + \wp^D_9
(h(t))\big)\cr
&\quad{}+ C''_{n+k}  \sum_{{\sscr n_1+n_2+n_3+n_4=n+k}
\atop{\sscr  n_1 \leq L, n_2 \leq n+k-2
\atop\sscr n_3+n_4 \leq n+k-L-2}} (1+\tau^{}_{1,n_1} (t))  \Gamma^{}_{n_2} (t)
(1+T^2_{9,n_3} (t))\cr
&\quad\qquad\qquad{}\big(R'_{n_4+7} (t) + R^2_{n_4+9} (t) + R^\infty_{n_4} (t)
+
\wp^D_{n_4+8} (h(t))\big),\cr
}$$
where the constants $C'_q$ depend only on $\tau^{}_{1,0} (t)$ and the constants
$C''_q$ on $T^\infty_9 (t).$
}\saut
\penalty-9000
\noindent{\it Proof.}
The hypotheses of Theorem 5.8 are satisfied for $N = \max (1, n+k-L)$,
so $H_N (t)$
is finite. Therefore and according to the hypotheses of the corollary,
the hypotheses of Theorem 5.5 are satisfied. It follows from theorem 5.5 that
$$\eqalignno{
&\wp^D_n \big((1+\lambda^{}_i (t))^{k/2}  h(t)\big)\cr
&\quad{}\leq C'_k \Big(\wp^D_{n+k} (h(t)) +R^i_{n,k} (t)
+ \sum_{\sscr n_1+n_2=n+k\atop {\sscr n_1 \leq n+k-1
\atop\sscr  n_2 \leq n+k-L-2}} \Gamma^{}_{n_1} (t)
H_{n_2} (t)\Big)\cr
&\qquad\quad{}+ C'_{n+k}  \sum_{\sscr n_1+n_2=n+k\atop{\sscr  n_1 \leq n+k-2
\atop\sscr n_2 \leq n+k-L-1}}
 \Gamma^{}_{n_1} (t)  H_{n_2} (t)\cr
&\qquad\quad{}+C'_{n+k}  \sum_{1 \leq l \leq L}  (1+\tau^{}_{1,l} (t))
\Big(\wp^D_{n+k-l} (h(t)) + R^i_{n-l,k} (t)
+ \sum_{\sscr n_1+n_2=n+k-l\atop{\sscr  n_1 \leq n+k-l-1
\atop\sscr n_2 \leq n+k-L-1}}  \Gamma^{}_{n_1}
(t)  H_{n_2} (t)\Big),\cr
}$$
where $C'_k$ and $C'_{n+k}$ are constants depending only on $\tau^{}_{1,0}
(t).$
This inequality gives
$$\eqalignno{
&\wp^D_n \big((1+\lambda^{}_i (t))^{k/2}  h(t)\big)&(5.108)\cr
&\qquad{} \leq C'_k \Big(\wp^D_{n+k} (h(t)) +
R^i_{n,k} (t) + \chi^{}_+ (n+k-L-2)  \Gamma^{}_{n+k-1} (t)  H_1 (t)\Big)\cr
&\qquad\qquad{}+ C'_{n+k}  \sum_{1 \leq l \leq L}  (1+\tau^{}_{1,l} (t))
\big(\wp^D_{n+k-l} (h(t)) + R^i_{n-l,k} (t)\big)\cr
&\qquad\qquad{}+C'_{n+k}  \sum_{0 \leq l \leq L} (1 + \tau^{}_{1,l} (t))
\sum_{\sscr n_1+n_2=n+k-l\atop{\sscr  n_1 \leq n+k-2\atop\sscr
n_2 \leq n+k-L-2}}
\Gamma^{}_{n_1} (t)H_{n_2} (t),\cr
}$$
where $C'_k$ and $C'_{n+k}$ are constants depending only on
$\tau^{}_{1,0} (t).$

It follows from Theorem 5.8 that
$$\eqalignno{
&\sum_{0 \leq l \leq L}  (1 + \tau^{}_{1,l} (t))  \sum_{\sscr n_1+n_2=
n+k-l\atop{\sscr  n_1 \leq n+k-2\atop\sscr  n_2 \leq n+k-L-2}}
\Gamma^{}_{n_1} (t)
H_{n_2} (t)&(5.109)\cr
&\qquad{}= \sum_{\sscr n_1+n_2+n_3=n+k\atop{\sscr  n_1 \leq L, n_2 \leq n+k-2
\atop\sscr  n_3 \leq n+k-L-2}}
(1+\tau^{}_{1,n_1} (t))  \Gamma^{}_{n_2} (t)  H_{n_3} (t)\cr
&\qquad{}\leq a^{}_{n+k}\Big(\sum_{\sscr n_1+n_2+n_3+n_4=n+k+1\atop{
\sscr  n_1 \leq L, n_2 \leq n+k-2\atop
\sscr n_3+n_4 \leq n+k-L-2}}  (1+\tau^{}_{1,n_1} (t))   \Gamma^{}_{n_2} (t)\cr
&\hskip40mm(1+T^2_{9,n_3} (t))
\big(R'_{n_4+7} (t) + R^2_{n_4+9} (t) + R^\infty_{n_4} (t) +
\wp^D_{n_4+8} (h(t))\big)\Big),\cr
}$$
where $a^{}_{n+k}$ is a constant depending only on $T^\infty_9 (t)$.
Theorem 5.8 also gives that
$$H_1 (t) \leq a^{}_1 (1+T^2_{9,0} (t))  \big(R'_8 (t) + R^2_{10} (t) +
R^\infty_1 (t) + \wp^D_9 (h(t))\big), \eqno{(5.110)}$$
where $a^{}_1$ depends only on $T^\infty_9 (t)$. The inequality of the
corollary
follows from inequalities (5.108), (5.109) and (5.110), and by suitably
defining
the constants $C''_k$ and $C''_{n+k}$. This proves the corollary.

The preceding results of this chapter permit to establish $L^2$-estimates of
$\xi^D_Y  h$, where $h$ is a solution of equation (5.1) and $Y \in \Pi'.$
Suppose that $\xi^D_Y  g \in C^0 (\Rrm^+, D),  Y \in \Pi'$, and
that $(\xi^M_Y  G,  \xi^M_{P_0 Y}  G) \in C^0 (\Rrm^+,
M^\rho),  Y \in \Pi'$, for some $1/2 < \rho \leq 1$. We recall that
$$h^{}_Y = \xi^D_Y  h,\quad g^{}_Y = \xi^D_Y  g,\quad
G_{Y \mu} = (\xi^M_Y  G)_\mu, \quad  Y \in U (\p), \eqno{(5.111{\rm a})}$$
and introduce
$$f^{}_Y (t) = \int^t_{t_0}  w(t,s) (- i \gamma^0)      g^{}_Y (s) ds,
\quad  Y \in U (\p). \eqno{(5.111{\rm b})}$$
We also introduce the subset ${\sg}^n, n \geq 0$, of $\Pi'$ defined by
$${\sg}^n = I_n \cap \Pi', \eqno{(5.112)}$$
where $I_n$ is the ideal in $U (\p)$ generated by the elements of order $n$
in $U (\Rrm^4)$.

According to the definition of $\Pi',{\sg}^n$ is a
basis of $I_n$. We note that $\Pi'={\sg}^0 \supset{\sg}^1 \supset
\cdots\supset{\sg}^n \supset{\sg}^{n+1} \supset\cdots$, and that
${\sg}^1 \cap U({\frak{sl}}(2, \Crm)) =
\{ 0 \}$. When
$\Pi$ is given the standard ordering, $P_0 < P_1 < P_2 < P_3 < M_{23} <
M_{13} < M_{12} < M_{01} < M_{02} < M_{03},$
we note that, if $Y \in{\sg}^n$, then
$Y = XZ$, where $X \in \Pi' \cap U(\Rrm^4)$,  $\vert X \vert = n,
Z \in \Pi'$, and that $\Pi' = {\sg}^1 \cup(U({\frak{sl}}(2, \Crm)) \cap \Pi')$.
When not specified, $\Pi$ will always be given the standard ordering.
\saut
\noindent{\bf Proposition 5.10.}
{\it
Let $Y, Y_2, Y_3 \in \Pi'$, let $0 \leq a(Y_1,Y_2) \leq 1$, let
$h^{}_Y, G_{Y \mu}, g^{}_Y, f^{}_Y$ be given by (5.111a) and (5.111b),
let $\theta^{}_Z = \xi^D_Z (\gamma^\mu
 G_\mu  h) + g^{}_Z,  Z \in \Pi'$, let $h$ be given by
(5.3c), let $x^\mu  G_\mu = 0$ and let $t, t_0 \geq 0.$

\noindent{\hbox{\rm i)}}
If $h^{}_Z (t_0) \in D$ for $Z \in \Pi', \vert Z \vert \leq \vert Y \vert$
and if $h'_Y = h^{}_Y - {(2m)}^{-1}\! \ds{\suma_{\sscr Y_1,Y_2
\atop{\sscr \vert Y_2 \vert \leq
\vert Y \vert - 1\atop\sscr Y_1 \in \Pi' \cap U({\frak{sl}}(2, \Crm))}}^Y}\!
\gamma^\mu
G_{Y_1 \mu}  H_{Y_2}$, then
$$\eqalignno{
&\big\vert \Vert h'_Y (t) \Vert^{}_D - \Vert h'_Y (t_0) \Vert^{}_D \big\vert
- \Vert f^{}_Y (t)\Vert^{}_D\cr
&{}\leq \suma_{\sscr Y_1,Y_2\atop{\sscr  \vert Y_2 \vert
\leq \vert Y \vert - 1\atop \sscr  Y_1 \in {\sssg}^1}}^Y
\int^{\max(t,t_0)}_{\min(t,t_0)}  \Vert G_{Y_1 \mu} (s)
\gamma^\mu  h^{}_{Y_2} (s) \Vert^{}_D  ds\cr
&\ {}+ 2m^{-1}\ds{\suma_{\sscr Y_1,Y_2\atop{\sscr \vert Y_2 \vert \leq
\vert Y \vert - 1\atop\sscr Y_1 \in \Pi' \cap U({\frak{sl}}(2, \Crm))}}^Y}
\int^{\max(t,t_0)}_{\min(t,t_0)}
\Big(\Big((1+s)^{-1} \big(\Vert G_{Y_1 0} (s)  h^{}_{P_0 Y_2} (s) \Vert^{}_D\cr
&\ {}+ \!\!\sum_{1 \leq i \leq 3}\!\! \big(\Vert G_{Y_1 i} (s)
h^{}_{M_{0i} Y_2} (s)
\Vert^{}_D
+ \Vert G_{Y_1 i} (s)  h^{}_{Y_2} (s) \Vert^{}_D\big)\big)\Big)^{1-a(Y_1,Y_2)}
\Vert G^\mu_{Y_1} (s)  \partial_\mu  h^{}_{Y_2} (s) \Vert^{a(Y_1,Y_2)}_D\cr
&\ {}+ \Vert (\partial_\mu  G^\mu_{Y_1} (s))  h^{}_{Y_2} (s) \Vert^{}_D +
{1 \over 2}  \Vert \gamma^\mu  \gamma^\nu \big(\partial_\mu
G_{Y_1 \nu} (s) - \partial_\nu  G_{Y_1 \mu} (s)\big)
h^{}_{Y_2} (s) \Vert^{}_D\cr
&\ {}+ \Vert \gamma^\mu  \gamma^\nu  G_\mu (s)  G_{Y_1 \nu}
(s)  h^{}_{Y_2} (s) \Vert^{}_D + \Vert \gamma^\mu  G_{Y_1 \mu} (s)
 \theta^{}_{Y_2} (s) \Vert^{}_D\Big)  ds,\cr
}$$

\noindent{\hbox{\rm ii)}} If $h^{}_Z (t_0) \in D,  \partial_\nu
h^{}_Z (t_0) \in D$ for
$Z \in \Pi',  0 \leq \nu \leq 3,  \vert Z \vert \leq \vert Y \vert$
and if $h^{(\nu)}_Z = \partial_\nu  h^{}_Z + i \ds{\suma_{Z_1,Z_2}^Z}
G_{Z_1 \nu}  h^{}_{Z_2}$,
$$\eqalignno{
g^{(\nu)}_{1 Z } &= - \gamma^\mu  \suma_{Z_1,Z_2}^Z
(\partial_\mu  G_{Z_1 \nu} - \partial_\nu  G_{Z_1 \mu}) h^{}_{Z_2}
+ \gamma^\mu\suma_{{\sscr Z_1,Z_2\atop\sscr \vert Z_2 \vert \leq \vert Z \vert-
1}\atop\sscr Z_1 \in \sssg^1}^Z G_{Z_1 \mu}(\partial_\nu  h^{}_{Z_2} + i
G_\nu h^{}_{Z_2})\cr
&\qquad{}+ i
\suma_{\scr Z_1,Z_2\atop\scr \vert Z_2 \vert \leq \vert Z \vert - 1}^Z
G_{Z_1 \nu}
(i \gamma^\mu  \partial_\mu + m - \gamma^\mu  G_\mu) h^{}_{Z_2}\cr
&\qquad{}+ \suma_{\sscr Z_1,Z_2\atop{\sscr \vert Z_2
\vert \leq \vert Z \vert - 1
\atop\sscr  Z_1 \in \Pi' \cap
U({\frak{sl}}(2, \Crm))}}^Z i \gamma^\mu  G_{Z_1 \mu} G_\nu  h^{}_{Z_2} +
i G_\nu g^{}_Z,\cr
g^{(\nu)}_{2Z} &= \gamma^\mu  \suma_{\sscr Z_1,Z_2\atop{\sscr
\vert Z_2 \vert \leq
\vert Z \vert - 1\atop\sscr Z_1 \in \Pi' \cap U({\frak{sl}}(2,\Crm))}}^Z
G_{Z_1 \mu}
\partial_\nu  h^{}_{Z_2},\cr
}$$
then
$$\eqalignno{
&\big\vert \Vert h^{(\nu)}_Y (t) - {(2m)}^{-1}  g^{(\nu)}_{2Y} (t)
\Vert^{}_D
- \Vert h^\nu_Y (t_0) - {(2m)}^{-1}  g^{(\nu)}_{2Y}
(t_0) \Vert^{}_D \big\vert\cr
&\quad {}\leq \Vert f^{}_{P_\nu Y} (t) \Vert^{}_D +
\int^{\max (t,t_0)}_{\min (t,t_0)}
\Vert g^{(\nu)}_{1Y} (s) \Vert^{}_D ds\cr
&\qquad{}+ {(2m)}^{-1}  \suma_{\sscr Y_1,Y_2\atop{\sscr \vert Y_2
\vert \leq \vert Y \vert -1
\atop\sscr
Y_1 \in \Pi' \cap U({\frak{sl}}(2, \Crm))}}^Y  \int^{\max (t,t_0)}_{\min
(t,t_0)}
 \Big(\Big((1+s)^{-1}  \big(\Vert G_{Y_1 0} (s)  h^{}_{P_0 P_\nu Y_2}
(s) \Vert^{}_D\cr
&\qquad{}+ \sum_{1 \leq i \leq 3}  \big(\Vert G_{Y_1 i} (s)  h^{}_{M_{0i}
P_\nu Y_2} (s) \Vert^{}_D\cr
&\qquad{}+ \Vert G_{Y_1 i} (s)  h^{}_{P_\nu Y_2} (s)
\Vert^{}_D\big)\big)\Big)^{1-a
(Y_1,Y_2)}  \Vert G^\mu_{Y_1} (s)  \partial_\mu
h^{}_{P_\nu Y_2} (s) \Vert^{a (Y_1,Y_2)}_D\cr
&\qquad{}+ \Vert (\partial_\mu  G^\mu_{Y_1} (s))  h^{}_{P_\nu Y_2} (s)
\Vert^{}_D + {1 \over 2} \Vert \gamma^\alpha  \gamma^\beta \big(\partial_\alpha
 G_{Y_1 \beta} (s) - \partial_\beta  G_{Y_1 \alpha} (s)\big)
 h^{}_{P_\nu Y_2} (s) \Vert^{}_D\cr
&\qquad{}+ \Vert \gamma^\alpha  \gamma^\beta  G_\alpha (s)
G_{Y_1 \beta} (s)  h^{}_{P_\nu Y_2} (s) \Vert^{}_D + \Vert \gamma^\mu
G_{Y_1 \mu} (s)  \theta^{}_{P_\nu Y_2} (s) \Vert^{}_D\Big)ds.\cr
}$$
}\saut
\noindent{\it Proof.}
Since
$$\xi^D_Y  (\gamma^\mu  G_\mu  h) = \suma_{Y_1,Y_2}^Y
 \gamma^\mu  G_{Y_1}  h^{}_{Y_2},$$
it follows from equation (5.1) and definition (5.111b) of $f^{}_Y$ that
$$(i \gamma^\mu  \partial_\mu + m - \gamma^\mu  G_\mu) (h^{}_Y -
f^{}_Y) = g^{}_{1 Y} + g^{}_{2Y}, \eqno{(5.113{\rm a})}$$
where
$$g^{}_{1 Y} = \suma_{\sscr Y_1,Y_2\atop {\sscr \vert Y_2 \vert \leq
\vert Y \vert - 1\atop\sscr  Y_1 \in \sssg^1}}^Y
 \gamma^\mu  G_{Y_1 \mu}  h^{}_{Y_2},  g^{}_{2Y},\quad
g^{}_{2 Y}= \suma_{\sscr Y_1,Y_2\atop{\sscr  \vert Y_2 \vert \leq \vert
Y \vert - 1
\atop\sscr  Y_1 \in \Pi' \cap U(sl(
2, \Crm))}}^Y  \gamma^\mu G_{Y_1 \mu}  h^{}_{Y_2}. \eqno{(5.113{\rm b})}$$
We have here used that $\Pi' = \sg^1 \cup \big(\Pi'
\cap U({\frak{sl}}(2, \Crm))\big)$ and that
$\sg^1 \cap \big(\Pi' \cap U({\frak{sl}}(2, \Crm))\big) = \emptyset$.
We note that $f^{}_Y (t_0) = 0$
and that $h'_Y = h^{}_Y - {(2m)}^{-1}  g^{}_{2Y}$, which gives that
$$\eqalignno{
&\big\vert \Vert h'_Y (t) \Vert^{}_D - \Vert h'_Y (t_0) \Vert^{}_D\vert
\leq \Vert f^{}_Y (t)\Vert^{}_D\big\vert \cr
&\quad{}
+ \big\vert \Vert h^{}_Y (t) - f^{}_Y (t) - {(2m)}^{-1}  g^{}_{2Y}
(t) \Vert^{}_D - \Vert h^{}_Y (t_0) - f^{}_Y (t_0) - {(2m)}^{-1}
g^{}_{2Y} (t_0) \Vert^{}_D\big\vert.\cr
}$$
The inequality in statement i) of the proposition now follows from Corollary
5.2 whith $h,g,g^{}_1$ replaced by $h^{}_Y - f^{}_Y, g^{}_{1 Y} +
g^{}_{2Y}, g^{}_{1 Y}$,
respectively, and from the fact that $x_\mu  G^\mu_{Y_1} (x) = 0$ for
$Y_1 \in \Pi' \cap U({\frak{sl}}(2, \Crm))$, which in its turn follows
from the ${\frak{sl}}(2, \Crm)$ covariance of the condition $x_\mu
G^\mu_\un (x) = 0$.

The definitions in statement ii) of the proposition give
$$(i \gamma^\mu  \partial_\mu + m - \gamma^\mu  G_\mu) (h^{(\nu)}_Y -
f^{}_{P_\nu Y}) = g^{(\nu)}_{1 Y} + g^{(\nu)}_{2Y}.$$
The inequality in statement ii) follows, similarly as in the case
of the inequality in statement~i), from this equation and from
Corollary 5.2 with $h,g,g^{}_1$ replaced by $h^{(\nu)}_Y - f^{}_{P_\nu Y}$,
$g^{(\nu)}_{1 Y} + g^{(\nu)}_{2Y}, g^{(\nu)}_{1 Y}$,
respectively. This proves the proposition.

In the sequel of this chapter we shall suppose that
$$G_\mu (y) = A_\mu (y) - \partial_\mu  {\vartheta} (A,y),\quad
0 \leq \mu \leq 3, \eqno{(5.114)}$$
where $\xi^M_Y  A \in C^0 (\Rrm^+, M^\rho_0)$ for some $0 \leq \rho \leq 1,$
for $Y \in \Pi'$, and $\vert Y \vert \leq N$ for some $N \geq 0$. We introduce
the notation
$$\eqalignno{
S^Y (t)
&= \sup_{0 \leq s \leq t}  \Big((1+s)^{\rho - 1}  \Vert
(A_Y (s), A_{P_0 Y} (s)) \Vert^{}_{M^\rho}
+ \Vert (A_Y (s), A_{P_0 Y} (s)) \Vert^{}_{M^1}& (5.115{\rm a})\cr
&\qquad{}+ (1+s)^{1/2-\rho}  \Vert
\delta (s)\carre A_Y (s) \Vert^{}_{L^2 (\Rrm^3, \Rrm^4)}
+ (1+s)^{1/2 - \rho}\Vert (\partial^\mu  A_{Y \mu} (s)
\Vert^{}_{L^2 (\Rrm^3, \Rrm)}\Big),\cr
}$$
for $0 \leq \rho \leq 1$, $Y \in \Pi'$, and
$$S^{\rho, n} (t) = \Big(\sum_{\scr Y \in \Pi'\atop\scr
\vert Y \vert \leq n} (S^Y (t))^2 +
\sum_{\scr Y \in \Pi'\atop\scr \vert Y \vert \leq n-1 }
\Vert (B_Y (s), B_{P_0 Y} (s))
\Vert^2_{M^1}\Big)^{1/2}, \eqno{(5.115{\rm b})}$$
where $0 \leq \rho \leq 1,  n \geq 0$ and $B_\mu (y) = y_\mu
\partial^\nu  A_\nu (y)$, $A_{Y \mu} = (\xi^M_Y A)_\mu,  B_{Y \mu} =
(\xi^M_Y B)_\mu.$
We also introduce
$$\eqalignno{
[A]^n  (t)&
= \sum_{\scr Y\in \Pi'\atop\scr  \vert Y \vert \leq n }
\sup_{\scr x \in \Rrm^3\atop\scr  0 \leq s \leq t}
\big((1+s+\vert x \vert)^{3/2 - \rho}  (\vert A_Y (s,x) \vert +
\vert B_Y (s,x) \vert)\big)&(5.116{\rm a})\cr
&\qquad{}+ \sum_{\scr Y \in \ssg^1\atop\scr \vert Y \vert \leq n }
\sup_{\scr x \in \Rrm^3\atop\scr  0 \leq s \leq t}
\big((1+s+\vert x \vert) (1+\big\vert s - \vert x \vert \big\vert)^{1/2}
 (\vert A_Y (s,x) \vert + \vert B_Y (s,x) \vert)\big),\cr
}$$
$n \geq 0,  0 \leq \rho \leq 1$, where $B_Y$ is as in (5.115b), and let
$$\eqalignno{
[A]'^n (t) &= \sum_{\scr Y \in \Pi'\atop\scr \vert Y \vert \leq n}
\sup_{x \in \Rrm^3}
 \big((1+t+\vert x \vert)^{3/2 - \rho}  \vert A_Y (t,x) \vert\big)
&(5.116{\rm b})\cr
&\qquad{}+ \sum_{\scr Y \in \ssg^1\atop\scr \vert Y \vert \leq n}
\sup_{x \in \Rrm^3}
 \big((1+t+\vert x \vert)  (1+\big\vert t-\vert x \vert \big\vert)^{1/2}
 \vert A_Y (t,x) \vert\big),\cr
}$$
$n \geq 0,  0 \leq \rho \leq 1$. We note that the second sum in (5.116a)
and (5.116b) is absent for $n=0$, since $\vert Y \vert
\geq 1$ if $Y \in \sg^1.$
It follows from Lemma 4.4 that
$$[G]'^n (t) \leq C_n  [A]^{n+1} (t),\quad  n \geq 0,  t \geq 0,
 1/2 < \rho \leq 1, \eqno{(5.116{\rm c})}$$
where the constant $C_n$ depends only on $\rho.$

Let $S^\rho_{N,n} (t)$ (resp. $[A]_{N,n} (t)$) be defined by $b^{}_n$ in
formula
(5.87) with $a^{(n)} = S^{\rho, N+n} (t)$ (resp. $[A]^{N+n}
(t)$) and let $S^\rho_n (t) =
S^\rho_{0,n} (t)$ (resp. $[A]_n (t) = [A]_{0,n}(t)$). We introduce the
notation:
$$\eqalignno{
\wp^D_{n,i} (a) &= \Big(\sum_{\scr Y \in \ssg^i\atop\scr \vert Y \vert
\leq n}\Vert a^{}_Y
\Vert^2_D\Big)^{1/2}, \quad  0 \leq i \leq n, &(5.117{\rm a})\cr
\noalign{\hbox{and}}
\wp^{M^\rho}_{n,i} (b) &=\Big (\sum_{\scr Y \in \ssg^i\atop\scr
\vert Y \vert \leq n}
\Vert b^{}_Y \Vert^2_{M^\rho}\Big)^{1/2},\quad  0 \leq i \leq n,& (5.117{\rm
b})\cr
}$$
where $Y \mapsto a^{}_Y$ (resp. $b^{}_Y$) is a function from
$\sg^i$ to $D$ (resp. $M^\rho$).

In the next lemma we shall estimate the sum over
$Y_1 \in \Pi' \cap U({\frak{sl}}(2, \Crm))$
on the right-hand side in the inequalities of statements i) and ii)
of Proposition 5.10. To state the result, let us introduce the following
notations:
$$\eqalignno{
J_Y (t) &= 2m^{-1}  \suma_{\sscr Y_1,Y_2\atop{\sscr
\vert Y_2 \vert \leq \vert Y \vert - 1\atop\sscr
Y_1 \in \Pi' \cap U({\frak{sl}}(2, \Crm))}}^Y \Big(\Big((1+t)^{-1}
(\Vert G_{Y_1 0} (t)
 h^{}_{P_0 Y_2} (t) \Vert^{}_D& (5.118)\cr
&\qquad{}+ \sum_{1 \leq i \leq 3}  \big(\Vert G_{Y_1 i} (t)
h^{}_{M_{0i} Y_2} (t) \Vert^{}_D + \Vert G_{Y_1 i} (t)  h^{}_{Y_2} (t)
\Vert^{}_D\big)\big)\Big)^{1-a(Y_1,Y_2)}\cr
&\qquad\qquad{}\Vert G^\mu_{Y_1} (t)  \partial_\mu
h^{}_{Y_2} (t) \Vert^{a(Y_1,Y_2)}_D
+ \Vert (\partial_\mu  G^\mu_{Y_1} (t))h^{}_{Y_2} (t) \Vert^{}_D\cr
&\qquad{}+ {1 \over 2} \Vert
\gamma^\mu  \gamma^\nu \big(\partial_\mu  G_{Y_1 \nu} (t) - \partial_\nu
 G_{Y_1 \mu} (t)\big) h^{}_{Y_2} (t) \Vert^{}_D\cr
&\qquad{}+ \Vert \gamma^\mu  \gamma^\nu  G_\mu (t)  G_{Y_1 \nu}
(t)  h^{}_{Y_2} (t) \Vert^{}_D + \Vert \gamma^\mu  G_{Y_1 \mu} (t)
 \theta^{}_{Y_2} (t) \Vert^{}_D\Big),\cr
}$$
$Y \in \Pi',  t \geq 0$, where $\theta^{}_Z$ is defined as in Proposition
5.10, and where $a(Y_1,Y_2) = 1/2$, if $\vert Y_1 \vert = 1$ and
$\vert Y_2 \vert = \vert Y \vert - 1$, and $a(Y_1,Y_2) = 0$ otherwise;
$$\eqalignno{
k^{(1)}_n (L,t) &= \sum_{n_1+n_2=n\atop\scr 1 \leq n_1 \leq L}
\big([A]^{n_1+1}
(t)  (1+t)^{1/2}  \wp^D_{n_2} (g(t))
+ [A]_{n_1+3} (t)  \wp^D_{n_2}  (h(t))\big),\hskip3mm&(5.119{\rm a})\cr
l^{(1)}_n (L,t) &= \sum_{\sscr n_1+n_2+n_3=n\atop{\sscr  n_2 \leq n_1 \leq n-1
\atop\sscr n_3 \leq n-L-1}}
 S^{\rho,n_1} (t)  (1+[A]^{n_2+1} (t))  H_{n_3+1}
(t)&(5.119{\rm b})\cr
&\qquad{}+ \sum_{\sscr Y_1,Y_2\in \Pi'
\atop{\sscr \vert Y_1\vert+\vert Y_2\vert =n
\atop\sscr L+1 \leq\vert Y_1\vert \leq n-1}}
 (1+t)^{2 - \rho} \Vert \gamma^\mu G_{Y_1\mu}g^{}_{Y_2}(t)
\Vert^{}_{D},\cr
k^{(2)}_n (L,t) &= \sum_{n_1+n_2=n\atop\scr 1 \leq n_1 \leq L} [A]_{3,n_1}
(t)  (\wp^D_{n_2} (h(t)) + R^0_{n_2-1,1} (t))& (5.119{\rm c})\cr
\noalign{\hbox{and}}
l^{(2)}_n (L,t) &= \sum_{\sscr n_1+n_2+n_3+n_4=n\atop{\sscr
n_1 \leq L, n_2 \leq n-1
\atop\sscr  n_3+n_4 \leq
n-L-2}} (1+[A]_{2,n_1} (t))  S^{\rho,n_2} (t) (1+S^\rho_{10,n_3}
(t)) &(5.119{\rm d})\cr
&\quad\big(R'_{n_4+7} (t) + R^2_{n_4+9} (t) + R^\infty_{n_4} (t) +
\wp^D_{n_4+8}
(h(t))\big),\cr
}$$
where $n \geq 0,  L \geq 0,  t \geq 0$, where $H_n, R^0_{n,k},
R'_n, R^2_n, R^\infty_n$ are defined in Theorem 5.8 and Corollary 5.9.
\saut
\noindent{\bf Lemma 5.11.}
{\it
Let $1/2 < \rho <1,  h^{}_Z \in C^0 (\Rrm^+, D)$, let $G$ be given by
(5.114) and $(A_Z, A_{P_0 Z}) \in C^0 (\Rrm^+, M^\rho) \cap C^0
(\Rrm^+, M^1)$ for
$Z \in \Pi'$ and $\vert Z \vert$ sufficiently large. Let $L \geq 1,  g =
(i  \gamma^\mu  \partial_\mu + m - \gamma^\mu  G_\mu)
h,  \chi^{}_+ (s) = 0$ for $s < 0,  \chi^{}_+ (s) = 1$ for $s \geq 0$
and let $Y \in \sg^i,   Y \notin \sg^{i+1},  i \geq 0.$
If $i = \vert Y \vert$, then $J_Y (t) = 0$, if $1 \leq i \leq
\vert Y \vert - 1,$
then
$$J_Y (t) \leq C'_{\vert Y \vert}  (1+t)^{-2+\rho}
J^{(i)}_{\vert Y \vert} (L,t),$$
where
$$\eqalignno{
J^{(l)}_n (L,t) & = [A]_3 (t)  \Big(\wp^D_n  (h(t))^\varepsilon
 \wp^D_{n,l+1}  (h(t))^{1 - \varepsilon}
+ R^0_{n-1,1} (t)^{2(1-\rho)}  \wp^D_{n-1}  (h(t))^{2 \rho-1}\Big)\cr
&\qquad{}+ k^{(1)}_n (L,t) + l^{(1)}_n (L,t)
+ k^{(2)}_n (L,t) + C''_n  l^{(2)}_n (L,t),\cr
}$$
$n \geq 1$,  $0\leq l \leq n$,  $\varepsilon = \max (1/2, 2 (1 - \rho))$,
$\wp^{}_{n,n+1}=0$,
$J^{(l)}_n=0$ for $l\geq n+1$, and if $i=0,  \vert Y \vert \geq 1$, then
$$\eqalignno{
J_Y (t) &\leq (1+t)^{-2 + \rho}  \Big(C'_{\vert Y \vert}
J^{(0)}_{\vert Y \vert} (L,t)\cr
&\quad{}
+ C  \chi^{}_+  (\vert Y \vert - L - 1)
\big( S^Y (t) (1+[A]^1 (t))  H_1 (t) + (1+t)^{2 - \rho}
\Vert\gamma^\mu G_{Y\mu}(t)g(t) \Vert^{}_{D}\big)\Big),\cr
}$$
for some constant $C$ depending only on $\rho$, a constant $C'_{\vert Y \vert}$
depending only on $\rho$ and $[A]^3 (t)$ and a constant $C''_{\vert Y \vert}$
depending only on $\rho$ and $[A]^9 (t).$
}\saut
\noindent{\it Proof.}
Let $Y \in \sg^i,  i \geq 0$. Then the sum in  definition (5.118)
of $J_Y (t)$ runs over a subset of $\{ (Y_1, Y_2) \in \Pi' \times \Pi'
\big\vert Y_1
\in \Pi' \cap U({\frak{sl}}(2, \Crm))$, $Y_2 \in \sg^i,  i \leq \vert Y_2 \vert
\leq \vert Y \vert - 1 \}$, which shows that $1 \leq \vert Y_1
\vert\leq \vert Y \vert -i.$
If $i = \vert Y \vert \geq 0$, then it follows that $J_Y = 0$.
We therefore only need to consider the case
$0 \leq i \leq \vert Y \vert - 1.$

Let $Y \in \sg^i,  0 \leq i \leq \vert Y \vert - 1$ and let $L \geq 1.$
Since $F^{}_{Z \mu \nu} \equiv \partial_\mu  G_{Z \nu} - \partial_\nu
 G_{Z \mu} = \partial_\mu  A_{Z \nu} - \partial_{\nu}
A_{Z \mu}$ for $Z \in \Pi'$, it follows from definition (5.118) of $J_Y$ and
from the definitions of norms, that
$$\eqalignno{
 J_Y (t)&\leq C  \suma_{\sscr Y_1,Y_2\atop{ 1 \leq \vert Y_1 \vert
\leq L\atop\sscr Y \in \Pi' \cap U({\frak{sl}}(2, \Crm))}}^Y
\Big((1+t)^{- {5/2} + \rho + a(Y_1,Y_2)}  [G]'^{\vert Y_1 \vert}
(t)&(5.120)\cr
&\hskip25mm\wp^D_{\vert Y_2 \vert + 1,i}  (h(t))^{1 - a(Y_1,Y_2)}
 \wp^D_{\vert Y_2 \vert + 1, i+1}  (h(t))^{a(Y_1,Y_2)}\cr
&\qquad{}+ (1+t)^{- 2 + \rho}[G]'^{\vert Y_1 \vert +1} (t)
\wp^D_{\vert Y_2 \vert,i}
 ((1+\lambda^{}_0 (t))^{1/2}  h(t))^{2(1-\rho)}  \wp^D_{
\vert Y_2 \vert,i}  (h(t))^{2 \rho -1}\cr
&\qquad{}+ (1+t)^{- 3+2 \rho}   [G]'^0 (t)  [G]'^{\vert Y_1 \vert} (t)
 \wp^D_{\vert Y_2 \vert, i}  (h(t))\cr
&\qquad{}+ (1+t)^{- 3/2 + \rho}  [G]'^{\vert Y_1 \vert} (t)  \wp^D_{
\vert Y_2 \vert,i}  (g(t))\Big)\cr
&\quad{}+ C  \suma_{\sscr Y_1,Y_2\atop{\sscr  1+L \leq \vert Y_1 \vert
\atop\sscr Y_1 \in \Pi' \cap U
({\frak{sl}}(2, \Crm))}}^Y\Big((1+t)^{- 3/2}     \Vert \delta (t)^{-1}
 G_{Y_1} (t) \Vert^{}_{L^2}  H_{\vert Y_2 \vert +1} (t)\cr
&\qquad{}+ (1+t)^{- 1/2}  \Vert \delta (t)^{-1}  \partial^\mu
 G_{Y_1 \mu} (t) \Vert^{}_{L^2}  H_{\vert Y_2 \vert} (t)\cr
&\qquad{}+ (1+t)^{- 3/2}  \Vert (A_{Y_1} (t),   A_{P_0 Y_1} (t)) \Vert^{}_{M^1}
 H_{\vert Y_2 \vert} (t)\cr
&\qquad{}+ (1+t)^{- 2 + \rho}   [G]'^0 (t)  \Vert \delta (t)^{-1}
 G_{Y_1} (t) \Vert^{}_{L^2}  H_{\vert Y_2 \vert} (t)\Big)\cr
&\quad{}+ 2m^{-1}  \suma_{Y_1,Y_2\atop\scr i \leq \vert Y_2
\vert \leq \vert Y \vert-1}^Y
\Vert \gamma^\mu  G_{Y_1 \mu} (t)\xi^D_{Y_2}
(\gamma^\nu  G_\nu  h+g) (t) \Vert^{}_D,\cr
}$$
$1/2 < \rho <1, t \geq 0, 0 \leq i \leq \vert Y \vert - 1, Y \in \sg^i,$
for some constant $C$ depending only on $\rho$.

We shall estimate the different terms on the right-hand side of inequality
(5.120). According to definition (5.114) of $G$, it follows that
$$G_{Y \mu} (y) = A_{Y \mu} (t) - \partial_\mu  {\vartheta} (A_Y, y),
\quad 0 \leq \mu \leq 3, \eqno{(5.121{\rm a})}$$
$y \in \Rrm^+ \times \Rrm^3$, for $Y \in U({\frak{sl}}(2, \Crm))$. This gives
that
$$G_{Y \mu} (y) = A_{Y \mu} (y) - I_{Y \mu} (y) - y^\nu  \partial_\mu
 I_{Y \nu} (y), \eqno{(5.121{\rm b})}$$
$Y \in U({\frak{sl}}(2, \Crm))$, where
$$I_\mu (y) = \int^1_0  A_\mu (sy)  ds,\quad  0 \leq \mu
\leq 3,$$
and $I_{Y \mu} = (\xi^M_Y  I)_\mu$ for $Y \in U(\p)$. It follows
from equality (5.121b) that for $Z \in \Pi' \cap U(\Rrm^4)$ and $Y \in \Pi'
\cap
U({\frak{sl}}(2, \Crm))$:
$$
G_{Z Y\mu} (y) = A_{ZY\mu} (y) - y^\nu  \partial_\mu
I_{ZY\nu} (y) - I_{ZY\mu} (y)
- \vert Z \vert  \sum_{0 \leq \nu \leq 3}  C_\nu (Z)
 I_{Z_\nu P_\mu Y\nu} (y), \eqno{(5.121{\rm c})}$$
for some positive integers $C_\nu (Z)$ and some elements $Z_\nu \in \Pi' \cap U
(\Rrm^4),  \vert Z_\nu \vert = \vert Z \vert - 1$. Since
$$\eqalignno{
\Vert \delta(t)^{-1}  A_{ZY} (t) \Vert^{}_{L^2}
&\leq \Vert \delta (t)^{-1} \Vert^{}_{L^{3/\rho}}  \Vert A_{ZY} (t)
\Vert^{}_{L^{6/(3-2\rho)}}\cr
&\leq C_\rho  (1+t)^{-1+\rho}   \Vert \vert \nabla \vert^\rho
 A_{ZY} (t) \Vert^{}_{L^2},\quad  0 \leq \rho < 1,\cr
}$$
it follows from statement ii) of Lemma 4.5 that
$$\eqalignno{
&\Vert  \delta (t)^{-1}  G_{ZY} (t) \Vert^{}_{L^2 (\Rrm^3, \Rrm^4)}
&(5.122)\cr
&\quad{}\leq C (1+t)^{-a}  \sup_{0 \leq s \leq t}  \Big((1+s)^{b+\rho-1}
 \Vert \vert \nabla \vert^\rho  A_{ZY} (s) \Vert^{}_{L^2}
+ (1+s)^b  \Vert (A_{ZY} (s),  A_{P_0 ZY} (s)) \Vert^{}_{M^1}\cr
&\qquad{}+ (1+s)^{b-1}  \vert Z \vert   C_{\vert Z \vert}
\sum_{\scr X \in \Pi' \cap U(\Rrm^4)\atop\scr \vert X \vert = \vert Z \vert -
1}
\Vert (A_{XY} (s),  A_{P_0 XY} (s)) \Vert^{}_{M^1}\Big),\cr
}$$
$0 \leq \rho < 1,  Z \in \Pi' \cap U (\Rrm^4),  Y \in \Pi'
\cap U({\frak{sl}}(2, \Crm)),$
where $a=b$ for $b < \vert Z \vert + 1/2,  a = \vert Z \vert + 1/2$
for $b > \vert Z \vert + 1/2$, and where the constants $C, C_{\vert Z \vert}$
depend only on $\rho, a$ and $b.$

We introduce the notation
$$\Gamma^{}_n (t) = \Big(\sum_{\scr Y \in \Pi'\atop\scr  \vert Y \vert \leq n}
\Vert \delta (t)^{-1}  G_Y (t) \Vert^2_{L^2 (\Rrm^3, \Rrm^4)}\Big)^{1/2},
\quad n \geq 0.
\eqno{(5.123)}$$
It follows from (5.122) that
$$\eqalignno{
\Gamma^{}_n (t) &\leq (1+t)^{-a} C \sup_{0 \leq s \leq t}
\Big((1+s)^{b+\rho-1}
 \wp^{M^\rho}_n  ((A(s), 0))&(5.124)\cr
&\quad{}+ (1+s)^b  \wp^{M^1}_n  \big((A (s), \dot{A} (s))\big) + (1+s)^{b-1}
n  C_n  \wp^{M^1}_{n-1}  \big((A(s), \dot{A} (s))\big)\Big),\cr
}$$
$n \geq 0,  0 \leq \rho < 1,$
where $a=b$ for $b < 1/2,  a = 1/2$ for $b > 1/2$ and where the constants
$C,C_n$ depend only on $\rho, a,b$. Inequalities (5.122) and (5.124) with $b=0$
give, according to (5.115a) and (5.115b),
$$\eqalignno{
&\Vert \delta (t)^{-1}  G_{ZY} (t) \Vert^{}_{L^2} &(5.125{\rm a})\cr
&\quad{}\leq C  S^{ZY}(t) + \vert Z \vert  C_{\vert Z \vert}
S^{1,\vert ZY \vert} (t),
\quad Z \in \Pi' \cap U (\Rrm^4),  Y \in \Pi' \cap U ({\frak{sl}}(2, \Crm)),\cr
}$$
and
$$\Gamma^{}_n (t) \leq C  S^{\rho,n} (t) + n  C_n  S^{1,n-1}
(t),\quad  n \geq 0, \eqno{(5.125{\rm b})}$$
where $0 \leq \rho < 1$ and $C, C_n$ depends only on $\rho.$

It follows from the definition of $G$ that for $Z \in \Pi' \cap U(\Rrm^4),
Y \in \Pi' \cap U({\frak{sl}}(2, \Crm))$:
$$\eqalignno{
\partial^\mu    G_{ZY\mu} (y) &= \xi^{}_Z  \partial^\mu
G_{Y\mu} (y)&(5.126)\cr
&= \xi^{}_Z \Big(\partial^\mu  A_{Y\mu} (y) - \carre  \int^1_0
 y^\nu  A_{Y\nu} (sy)   ds\Big)\cr
&= \partial^\mu  A_{ZY\mu} (y) - I_1 (y) - I_2 (y) - I_3 (y),\cr
}$$
where
$$\eqalignno{
I_1 (y) &= 2  \int^1_0  s^{1+\vert Z \vert}  (\partial^\mu
 A_{ZY\mu}) (sy)  ds,& (5.127{\rm a})\cr
I_2 (y) &= \int^1_0  y^\mu  s^{2+\vert Z \vert}
(\carre  A_{ZY\mu}) (sy)  ds,& (5.127{\rm b})\cr
\noalign{\hbox{and}}
I_3 (y) &= \vert Z \vert  \sum_{0 \leq \nu \leq 3}  C_\nu (Z)
 \int^1_0  s^{1+\vert Z \vert}  (\carre
A_{Z_\nu Y}) (sy)  ds, &(5.127{\rm c})\cr
}$$
for some constants $C_\nu (Z)$ and elements $Z_\nu \in \Pi' \cap U(\Rrm^4),
\vert Z_\nu \vert = \vert Z \vert - 1$. The result, given by (4.86a),
(4.86b) and (4.86c) in the proof of Lemma 4.5, implies that
$$\Vert I_1 (t) \Vert^{}_{L^2} \leq C_{a,b}  (1+t)^{- a^{}_1}
\sup_{0 \leq s \leq t}  \big((1+s)^{b^{}_1}  \Vert \partial^\mu
A_{ZY\mu} (s) \Vert^{}_{L^2}\big), \eqno{(5.128{\rm a})}$$
where $a^{}_1 = b^{}_1$ for $b^{}_1 < 1/2 + \vert Z \vert$ and
$a^{}_1 = 1/2 + \vert Z \vert$
for $b > 1/2 + \vert Z \vert,$
$$\Vert \delta (t)^{- \varepsilon}I_2 (t) \Vert^{}_{L^2}
\leq C_{a,b, \varepsilon}  (1+t)^{- a^{}_2}  \sup_{0 \leq s \leq t}
 \big((1+s)^{b^{}_2}  \Vert \delta (s)^{1 - \varepsilon}
(\carre  A_{ZY}) (s) \Vert^{}_{L^2}\big),\eqno{(5.128{\rm b})}$$
where $0 \leq \varepsilon \leq 1,  a^{}_2 = b^{}_2$ for
$b^{}_2 < 1/2 + \varepsilon + \vert Z \vert$ and $a^{}_2 = 1/2 +
\varepsilon + \vert Z \vert$ for $b^{}_2 > 1/2 +
\varepsilon + \vert Z \vert,$
$$\Vert I_3 (t) \Vert^{}_{L^2} \leq \vert Z \vert  C_{a,b,\vert Z \vert}
 (1+t)^{- a^{}_3}
\sum_{\scr X \in \Pi' \cap U(\Rrm^4)\atop\scr  \vert X \vert = \vert Z \vert -
1}
 \sup_{0 \leq s \leq t}  \big((1+s)^{b^{}_3}  \Vert (\carre
 A_{XY}) (s) \Vert^{}_{L^2}\big),\eqno{(5.128{\rm c})}$$
where $a^{}_3 = b^{}_3$ for $b^{}_3 < 1/2 + \vert Z \vert$ and
$a^{}_3 = 1/2 + \vert Z \vert$
for $b^{}_3 > 1/2 + \vert Z \vert$. Let $a^{}_1 = a - \varepsilon,  a^{}_2 = a,
 a^{}_3 = a - \varepsilon$, $b^{}_1 = b - \varepsilon,  b^{}_2 = b$
and $b^{}_3 = b - \varepsilon$. It then follows from (5.126), (5.128a),
(5.128b)
and (5.128c) that, for $Z \in \Pi' \cap U(\Rrm^4),  Y \in \Pi' \cap U
({\frak{sl}}(2, \Crm))$,
$$\eqalignno{
&\Vert \delta (t)^{- \varepsilon}  \partial^\mu G_{ZY\mu}
(t) \Vert^{}_{L^2 (\Rrm^3, \Rrm)}& (5.129)\cr
&\quad\leq C_{a,b,\varepsilon}  (1+t)^{-a}\sup_{0 \leq s \leq t}
 \Big((1+s)^{b - \varepsilon}  \Vert \partial^\mu  A_{ZY\mu}
(s) \Vert^{}_{L^2}
+ (1+s)^b  \Vert \delta (s)^{1 - \varepsilon}  \carre
A_{ZY} (s) \Vert^{}_{L^2}\cr
&\qquad{}+ \vert Z \vert  C_{\vert Z \vert}  \sum_{\scr X \in \Pi' \cap U
(\Rrm^4)\atop\scr  \vert X \vert = \vert Z \vert - 1}  (1+s)^{b - \varepsilon}
 \Vert \carre  A_{XY} (s) \Vert^{}_{L^2}\Big),\cr
}$$
where $a=b$ for $b < 1/2 + \varepsilon + \vert Z \vert$ and $a = 1/2 +
\varepsilon + \vert Z \vert$ for $b > 1/2 + \varepsilon + \vert Z \vert$
and where $0 \leq \varepsilon
\leq 1$. Inequality (5.129) gives that
$$\eqalignno{
&\Big(\sum_{\scr Y \in \Pi'\atop\scr  \vert Y \vert \leq n}
\Vert \delta (t)^{- \varepsilon}
 \partial^\mu  G_{Y\mu} (t) \Vert^2_{L^2 (\Rrm^3, \Rrm)}\Big)^{1/2}
&(5.130{\rm a})\cr
&\qquad{}\leq C_{a,b,\varepsilon}  (1+t)^{-a}   \sup_{0 \leq s \leq t}
 \Big((1+s)^{b - \varepsilon}  \Big(\sum_{\scr Y \in \Pi'\atop\scr  \vert
Y \vert \leq n}  \Vert \partial^\mu  A_{Y\mu} (s) \Vert^2_{L^2}\Big)^{1/2}\cr
&\qquad\qquad{}+ (1+s)^b  \Big(\sum_{\scr Y \in \Pi'\atop\scr
\vert Y \vert \leq n} \Vert \delta (s)^{1 - \varepsilon}  \carre  A_Y (s)
\Vert^2_{L^2}\Big)^{1/2}\cr
&\qquad\qquad{}+ n  C_n  (1+s)^{b - \varepsilon}
\Big(\sum_{\scr Y \in \Pi'\atop\scr
 \vert Y \vert \leq n-1}  \Vert \carre  A_Y (s)
\Vert^2_{L^2}\Big)^{1/2}\Big),\cr
}$$
where $a=b$ for $b < 1/2 + \varepsilon,  a = 1/2 + \varepsilon$ for
$b > 1/2 + \varepsilon$ and $0 \leq \varepsilon \leq 1$. Moreover,
with $\varepsilon = 1$ and $b = 3/2 - \rho$, $0 < \rho \leq 1$
in inequality (5.129), we obtain that
$$\Vert \delta (t)^{-1}  \partial^\mu  G_{Y\mu} (t) \Vert^{}_{L^2}
\leq C (1+t)^{- 3/2 + \rho}  S^Y (t), \eqno{(5.130{\rm b})}$$
$Y \in \Pi' \cap U ({\frak{sl}}(2, \Crm)),  0 < \rho \leq 1,$
where $C$ is a constant depending only on $\rho.$

In order to estimate the last sum on the right-hand side of inequality (5.120),
we note that according to (5.116c)
$$\eqalignno{
&\suma_{\scr Y_1,Y_2\atop\scr i \leq \vert Y_2 \vert \leq \vert Y \vert - 1}^Y
\suma_{\scr Z_1,Z_2\atop\scr  \vert Y_1 \vert + \vert Z_1 \vert \leq
L}^{Y_2}\Vert
G_{Y_1 \mu} (t)  G_{Z_1 \nu} (t)  h^{}_{Z_2} (t) \Vert^{}_{L^2}
&(5.131{\rm a})\cr
&\qquad{}\leq C_{\vert Y \vert}  (1+t)^{- 3+2 \rho}  \sum_{\sscr n_1+n_2+n_3=
\vert Y \vert\atop{\sscr 1 \leq n_1+n_2 \leq L\atop i \leq n_2+n_3}}
[A]^{n_1+1} (t)  [A]^{n_2+1} (t)  \wp^D_{n_3} (h(t)),\cr
}$$
$Y \in \Pi',  1/2 < \rho \leq 1$. Moreover, according to (5.120), (5.125a)
and (5.125b), we obtain that
$$\eqalignno{
&\suma_{\scr Y_1,Y_2\atop \scr i \leq \vert Y_2 \vert \leq \vert Y \vert - 1}^Y
\suma_{\scr Z_1,Z_2 \atop \scr\vert Y_1 \vert + \vert Z_1 \vert \geq L+1}^{Y_2}
\Vert G_{Y_1 \mu} (t)  G_{Z_1 \nu} (t)  h^{}_{Z_2} (t) \Vert^{}_{L^2}
&(5.131{\rm b})\cr
&\qquad{}\leq (1+t)^{- 2+\rho}  \suma_{\scr Y_1,Y_2\atop \scr  i \leq
\vert Y_2 \vert \leq
\vert Y \vert - 1}^Y  \Big(\suma_{\sscr Z_1,Z_2\atop {\sscr
\vert Y_1 \vert + \vert Z_1
\vert \geq L+1\atop\sscr  \vert Y_1 \vert \geq \vert Z_1 \vert}}^{Y_2}
\Vert \delta (t)^{-1}  G_{Y_1} (t) \Vert^{}_{L^2}
[G]'^{\vert Z_1 \vert} (t)  H_{\vert Z_2 \vert} (t)\cr
&\qquad\qquad{}+\suma_{\sscr Z_1,Z_2\atop{ \sscr\vert Y_1
\vert + \vert Z_1 \vert \geq L+1
 \atop\sscr \vert Y_1 \vert < \vert Z_1 \vert}}^{Y_2}
[G]'^{\vert Y_1 \vert} (t)  \Vert \delta(t)^{-1}  G_{Z_1} (t) \Vert^{}_{L^2}
H_{\vert Z_2 \vert} (t)\Big)\cr
&\qquad{}\leq (1+t)^{- 2+\rho}  \Big(C  \chi^{}_+ (-i)  \chi^{}_+ (\vert Y
\vert
-L-1)  S^Y (t)  [A]^1 (t)  H_0 (t)\cr
&\qquad\qquad{}+ C_{\vert Y \vert}
\sum_{{\sscr n_1+n_2+n_3=\vert Y \vert\atop\sscr  L+1 \leq n_1+n_2}\atop
{\sscr n_2 \leq n_1 \leq \vert Y \vert -1\atop\sscr
i \leq n_2+n_3 \leq \vert Y \vert -1}} S^{\rho,n_1}
(t)  [A]^{n_2+1} (t)  H_{n_3} (t)\Big),\cr
}$$
$Y \in \sg^i,   Y \notin \sg^{i+1},  0 \leq i \leq \vert Y \vert
-1,  1/2 < \rho < 1$, where the constants $C, C_{\vert Y \vert}$
depend only on $\rho$ and where $\chi^{}_+ (s) = 0$ for
$s < 0$ and $\chi^{}_+ (s) = 1$ for
$s \geq 0$. It follows from the definition of $[A]_n$ and from
(5.131a) and (5.131b), that
$$\eqalignno{
&\suma_{\sscr Y_1,Y_2\atop{\sscr  \vert Y_2 \vert \leq \vert Y \vert - 1
\atop\sscr  Y_1 \in \Pi' \cap U
({\frak{sl}}(2, \Crm))}}^Y\suma_{Z_1,Z_2}^{Y_2}  \Vert G_{Y_1 \mu} (t)
 G_{Z_1 \nu} (t)  h^{}_{Z_2} (t) \Vert^{}_D&(5.132{\rm a})\cr
&\qquad{}\leq C_{\vert Y \vert} (1+t)^{- 3 + 2\rho}
\sum_{n_1+n_2=\vert Y \vert\atop\scr
1 \leq n_1 \leq L}  [A]_{n_1+2} (t)  \wp^D_{n_2} (h(t))\cr
&\qquad\qquad{}+ (1+t)^{- 2+\rho}  \Big(C \chi^{}_+ (-i)
\chi^{}_+ (\vert Y \vert - L-1)
 S^Y (t)  [A]^1 (t)  H_0 (t)\cr
&\qquad\qquad{}+ C_{\vert Y \vert}  \sum_{\sscr n_1+n_2+n_3=\vert Y \vert
\atop{\sscr
 n_1+n_2 \geq L+1\atop\sscr
n_2 \leq n_1 \leq \vert Y \vert - 1}}  S^{\rho,n_1} (t)
[A]^{n_2+1} (t)  H_{n_3} (t)\Big),\cr
}$$
$Y \in \sg^i,   Y \notin \sg^{i+1},  0 \leq i \leq \vert Y \vert - 1,
 1/2 < \rho < 1,  L \geq 0$. The first sum on the right-hand
side of inequality (5.132a) is bounded by $k^{(1)}_{\vert Y \vert}  (L,t)$
and the second by $l^{(1)}_{\vert Y \vert} (L,t)$. This gives
$$\eqalignno{
&\suma_{\sscr Y_1,Y_2\atop{\sscr  \vert Y_2 \vert \leq \vert Y \vert - 1
\atop\sscr  Y_1 \in \Pi' \cap U
({\frak{sl}}(2, \Crm))}}^Y       \suma_{Z_1,Z_2}^{Y_2}  \Vert G_{Y_1 \mu} (t)
 G_{Z_1 \nu} (t)  h^{}_{Z_2} (t) \Vert^{}_D &(5.132{\rm b})\cr
&\qquad{}\leq (1+t)^{- 2+\rho}  C \chi^{}_+ (-i)  \chi^{}_+ (\vert Y \vert
-L-1)  S^Y (t)  [A]^1 (t)  H_0 (t)\cr
&\qquad\qquad{}+ C_{\vert Y \vert}  \big((1+t)^{- 3+2\rho}
k^{(1)}_{\vert Y \vert}
 (L,t) + (1+t)^{- 2+\rho}  l^{(1)}_{\vert Y \vert} (L,t)\big),\cr
}$$
$Y \in \sg^i,   Y \notin \sg^{i+1},  0 \leq i \leq \vert Y \vert - 1,
 1/2 < \rho < 1,  L \geq 0$, where $C$ and $C_{\vert Y \vert}$
are constants depending only on $\rho.$

To use Corollary 5.9, in order to estimate $\wp^D_{\vert Y_2 \vert}
((1+\lambda^{}_1 (t))^{1/2}
 h(t))$ in inequality (5.120), we shall first express the quantities
$T^2_n (t)$, defined by (5.89c), in terms of $S^Y$ defined by (5.115a). If
$Y \in \Pi' \cap U ({\frak{sl}}(2, \Crm))$, then $y^\mu  G_{Y\mu} = 0$ since
$y^\mu  G_\mu = 0$. Therefore, if $Z \in \Pi' \cap U(\Rrm^4)$, then
$$0 = \xi^{}_Z  y^\mu  G_{Y\mu} (y) = y^\mu  G_{ZY\mu}
(y) + \vert Z \vert  \sum_{0 \leq \mu \leq 3}  C_\mu (Z)
G_{Z_\mu Y\mu} (y),$$
for some constants $C_\mu (Z)$ and elements $Z_\mu \in \Pi' \cap U (\Rrm^4)$,
$\vert Z_\mu \vert = \vert Z \vert - 1$.
Let $Q_X (y) = y^\mu  G_{X\mu}
(y)$,  $X \in \Pi'$. Then we obtain that
$$Q_{ZY} = - \vert Z\vert\sum_{0 \leq \mu \leq 3}  C_\mu (Z)  G_{Z_\mu Y\mu},
\eqno{(5.133)}$$
for $Z \in \Pi' \cap U(\Rrm^4),  Y \in \Pi' \cap U({\frak{sl}}(2, \Crm))$. It
follows from inequality (5.122) and from equality (5.133) that
$$\eqalignno{
&\Vert \delta (t)^{-1}  Q_{ZY} (t) \Vert^{}_{L^2}\leq C_{a,b,\vert Z \vert}
\vert Z \vert (1+t)^{-a}&(5.134)\cr
&\quad\sup_{0 \leq s \leq t} \Big(\sum_{0 \leq \mu \leq 3}
\Big((1+s)^{b+\rho-1}
 \Vert \vert \nabla \vert^\rho  A_{Z_\mu Y} (s) \Vert^{}_{L^2}
+(1+s)^b \Vert (A_{Z_\mu Y} (s), A_{P_0 Z_\mu Y} (s)) \Vert^{}_{M^1}\Big)\cr
&\qquad{}+ (1+s)^{b-1}  (\vert Z \vert - 1) \sum_{\scr X \in \Pi' \cap
U(\Rrm^4)
\atop\scr
\vert X \vert = \vert Z \vert - 2}  \Vert (A_{XY} (s),  A_{P_0 XY}
(s)) \Vert^{}_{M^1}\Big),\cr
}$$
where $Z \in \Pi' \cap U(\Rrm^4),  Y \in \Pi' \cap U({\frak{sl}}(2, \Crm)),
 0 \leq \rho < 1,  a=b$ for $b < \vert Z \vert - 1/2$ and $a=
\vert Z \vert - 1/2$ for $b > \vert Z \vert - 1/2$.
Since
$$\sum_{\scr\vert Y \vert \leq n\atop\scr  Y \in \Pi'}  \Vert \delta
(t)^{-1}  Q_Y (t) \Vert^{}_{L^2}
\leq \sum_{n_1+n_2 = n}\  \sum_{\sscr Y \in \Pi' \cap U({\frak{sl}}(2,
\Crm))\atop{\sscr
Z \in \Pi' \cap U(\Rrm^4)\atop\sscr  \vert Y \vert = n_1, \vert Z \vert = n_2}}
 \Vert \delta (t)^{-1}  Q_{ZY} (t) \Vert^{}_{L^2},$$
it follows from (5.134), choosing $\rho=0$ for $\vert Z \vert \geq 2$, that
$$\eqalignno{
&\sum_{\scr Y \in \Pi'\atop\scr \vert Y \vert \leq n }  \Vert \delta
(t)^{-1}  Q_Y (t) \Vert^{}_{L^2}&(5.135)\cr
&\qquad{} \leq n  C_n (1+t)^{-a}
\sup_{0 \leq s \leq t}  \Big((1+s)^{b+\rho-1}  \wp^{M^\rho}_{n-1}
 ((A(s), 0)) + (1+s)^b  \wp^{M^1}_{n-1}  ((A(s),
\dot{A} (s)))\Big),\cr
}$$
where $a=b$ for $b < 1/2,  a = 1/2$ for $b > 1/2,  0 \leq \rho < 1,$
$n \geq 0$ and where the constant $C_n$ depends only on $a,b,\rho$. Statement
i)
of Lemma 4.5 gives that
$$\eqalignno{
&\wp^{M^1}_n  \big((G(t),  \dot{G} (t))\big)&(5.136)\cr
&\qquad{} \leq C (1+t)^{-a}
\sup_{0 \leq s \leq t}  \Big((1+s)^b
\big(\wp^{M^1}_{n+1}  ((A(s),  \dot{A} (s))) \cr
&\qquad\qquad{}+ \wp^{M^1}_n ((B(s),
\dot{B} (s))) + C_n  \wp^{M^1}_n ((A(s),  \dot{A} (s)))\big)\Big),\cr
}$$
where $a=b$ for $b < 1/2,  a = 1/2$ for $b > 1/2,  B_\mu =
y_\mu  \partial^\nu  A_\nu$ and where the constants $C, C_n$
depend only on $a$ and $b$. It follows from inequality (5.130a) with
$\varepsilon = 1/2,  b = 1-\rho,  0 < \rho \leq 1$, that
$$\eqalignno{
&\Big(\sum_{\scr Y \in \Pi'\atop\scr  \vert Y \vert \leq n}
\Vert \delta (t)^{- 1/2}
 \partial^\mu  G_{Y\mu} (t) \Vert^2_{L^2}\Big)^{1/2} &(5.137)\cr
&\qquad{} \leq (1+t)^{- 1+\rho}  C \big(S^{\rho,n} (t) + n      C_n
S^{\rho, n-1} (t)\big),\quad  0 < \rho \leq 1,\cr
}$$
where the constants depend only on $\rho$. It follows from (5.88b),
(5.115a), (5.115b),
(5.135), (5.136) and (5.137), that
$$T^{2(n)} (t) \leq C  S^{\rho, n+1} (t) + C_n  S^{\rho, n} (t),
\quad 0 < \rho < 1, \eqno{(5.138)}$$
where the constants $C, C_n$ depend only on $\rho.$

Corollary 5.9, (5.116c), (5.125b), (5.138), give that
$$\eqalignno{
&\wp^D_n \big((1+\lambda^{}_i (t))^{1/2}  h(t)\big)&(5.139{\rm a})\cr
&\ {} \leq C'_1 \big(\wp^D_{n+1} (h(t)) +
R^i_{n,1} (t)\big)
+ C'_{n+1}\!\!  \sum_{\scr n_1+n_2=n+1\atop\scr  1 \leq n_1 \leq L_0} \!\!
(1 + [A]^{n_1+1} (t))  \big(\wp^D_{n_2} (h(t)) + R^i_{n_2-1,1} (t)\big)\cr
&\quad{}+ C''_1  \chi^{}_+ (n - L_0 - 1) \big(S^{\rho,n} (t) + n  C_{n-1}
 S^{\rho, n-1} (t)\big)\cr
&\hskip30mm \big(1+S^{\rho, 10} (t) + C_9  S^{\rho, 9} (t)\big)
\big(R'_8 (t) + R^2_{10} (t) + R^\infty_1 (t) + \wp^D_9 (h(t)\big)\cr
&\quad{}+ C''_{n+1}  \sum_{\sscr n_1+n_2+n_3+n_4=n+1\atop{\sscr n_1
\leq L_0, n_2 \leq n-1
\atop\sscr n_3+n_4 \leq n-L_0-1}}  (1+[A]^{n_1+1} (t))  S^{\rho, n_2} (t)
(1 + S^\rho_{10, n_3} (t))\cr
&\hskip50mm\big(R'_{n_4+7} (t) + R^2_{n_4+9} (t) + R^\infty_{n_4} (t)
+ \wp^D_{n_4+8} (t)\big),\cr
}$$
where $n \geq 0,  L_0 \geq 3,  1/2 < \rho < 1$, $i=0,1$, where the
constants $C_l$  depend only on $\rho, C'_l$ depends only on $\rho$ and
$[A]^1 (t),$
and $C''_l$ depend only on $\rho$ and $[A]^9 (t)$. The last inequality shows
that
$$\eqalignno{
&\wp^D_n \big((1 + \lambda^{}_0 (t))^{1/2}  h(t)\big)&(5.139{\rm b})\cr
&\quad{} \leq C'_1 \big(\wp^D_{n+1} (h(t)) +
R^0_{n,1} (t)\big)
+C'_{n+1} \big(\wp^D_{n} (h(t)) +
R^0_{n-1,1} (t)\big) \cr
&\qquad{}+ C'_{n+1}  k^{(2)}_{n+1}  (L_0, t) + C''_{n+1}
l^{(2)}_{n+1}  (L_0, t),\quad  n \geq 0,  L_0 \geq 3,1/2 < \rho < 1,\cr
}$$
where the constants $C'_1,  C'_{n+1}$ depend only
on $\rho$ and $[A]^1 (t)$, the constant $C''_{n+1}$ depends only on $\rho$ and
$[A]^9 (t)$, and where $k^{(2)}_{n+1}$, respectively $l^{(2)}_{n+1}$,
are given by (5.119c) and (5.119d).

Let $Y \in \sg^i,  Y \notin \sg^{i+1},  0 \leq i \leq \vert Y \vert - 1.$
It then follows from inequalities (5.116c), (5.120), (5.125a), (5.125b),
(5.130b),
from the fact that the domain of summation for the sums on the left-hand
side of inequality (5.120) is a subset of $\{ (Y_1, Y_2) \in \Pi' \times \Pi'
\big\vert Y_1 \in U({\frak{sl}}(2, \Crm)), Y_2 \in \sg^i, Y_2 \notin \sg^{i+1}
\}$
and from the definition of $a (Y_1, Y_2)$ in (5.118),
$$\eqalignno{
& J_Y (t) \cr
&\quad{}\leq C_{\vert Y \vert}  \suma_{\sscr Y_1,Y_2\atop{\sscr
 1 \leq \vert Y_1 \vert \leq L\atop\sscr  Y_1 \in \Pi' \cap U({\frak{sl}}(2,
\Crm))}}^Y
\Big((1+t)^{- 2+\rho}
\big([A]^{\vert Y_1 \vert + 1} (t) + [A]^{\vert Y_1 \vert +2}
(t) + [A]^1 (t)  [A]^{\vert Y_1 \vert + 1} (t)\big)\cr
&\qquad\qquad{}\big(\wp^D_{\vert Y_2 \vert + 1, i} (h(t))^{1-a(Y_1,Y_2)}
\wp^D_{\vert Y_2 \vert+1, i+1} (h(t))^{a(Y_1,Y_2)} \cr
&\qquad\qquad{}+\wp^D_{\vert Y_2 \vert,i}  ((1+ \lambda^{}_0 (t))^{1/2}
h(t))^{2 (1-\rho)} \wp^D_{\vert Y_2 \vert, i}  (h(t))^{2 \rho - 1} +
\wp^D_{\vert Y_2 \vert, i} (h(t))\big)\cr
&\qquad\qquad{}+ (1+t)^{- 3/2 + \rho}  [A]^{\vert Y_1 \vert + 1} (t)
\wp^D_{\vert Y_2 \vert, i}  (g(t))\Big)\cr
&\qquad{}+C  \suma_{\sscr Y_1,Y_2\atop{\sscr 1 + L \leq \vert Y_1
\vert\atop\sscr
 Y_1 \in \Pi' \cap U ({\frak{sl}}(2, \Crm))}}^Y (1+t)^{- 2 + \rho}  S^{Y_1} (t)
\big(H_{\vert Y_2 \vert + 1} (t) + H_{\vert Y_2 \vert} (t)
+[A]^1 (t)  H_{\vert Y_2 \vert} (t)\big) \cr
&\qquad{}+2m^{-1} \suma_{\sscr Y_1,Y_2\atop{\sscr \vert Y_2 \vert \leq
\vert Y \vert - 1
\atop\sscr
Y_1 \in \Pi' \cap U({\frak{sl}}(2, \Crm))}}^Y  \Vert \gamma^\mu  G_{Y_1 \mu}
(t)  \xi^D_{Y_2} (\gamma^\nu  G_\nu  h+g) (t) \Vert^{}_D,\quad
 1/2 < \rho < 1,\cr
}$$
where the constants $C, C_{\vert Y \vert}$ depend only on $\rho.$
This inequality gives, since
$$\wp^D_{\vert Y_2 \vert, i} (h(t)) \leq \wp^D_{\vert Y_2 \vert,i}
 ((1 + \lambda^{}_0 (t))^{1/2}  h(t))^{2 (1-\rho)}
\wp^D_{\vert Y_2 \vert,i}
(h(t))^{2 \rho-1},$$
that
$$\eqalignno{
 J_Y (t)
&\leq C_{\vert Y \vert}  \suma_{\sscr Y_1,Y_2 \atop{\sscr 1  \leq
\vert Y_1 \vert \leq L\atop\sscr  Y_1 \in \Pi' \cap U ({\frak{sl}}(2,
\Crm))}}^Y
\Big((1+t)^{-2+\rho}  [A]_{\vert Y_1 \vert + 2} (t) &(5.140)\cr
&\qquad\qquad\big(\wp^D_{\vert Y_2 \vert +1,i}
(h(t))^{1 - a(Y_1,Y_2)}  \wp^D_{\vert Y_2 \vert +1, i+1} (h(t))^{a(Y_1,Y_2)}\cr
&\qquad\qquad{}+ \wp^D_{\vert Y_2 \vert, i}
((1+\lambda^{}_0 (t))^{1/2} h(t))^{2(1-\rho)}
 \wp^D_{\vert Y_2 \vert, i}  (h(t))^{2 \rho-1}\big)\cr
&\qquad\qquad{}+ (1+t)^{- 3/2 + \rho}  [A]^{\vert Y_1 \vert + 1} (t)
\wp^D_{\vert Y_2 \vert, i} (g(t))\Big)\cr
&\qquad{}+C \suma_{\sscr Y_1,Y_2\atop{\sscr  1+L \leq \vert Y_1 \vert\atop\sscr
 Y_1 \in \Pi' \cap U ({\frak{sl}}(2, \Crm))}}^Y
 (1+t)^{- 2+\rho}  S^{Y_1} (t)  (1+[A]^1 (t))
H_{\vert Y_2 \vert + 1} (t)\cr
&\qquad{}+ 2m^{-1}  \suma_{\sscr Y_1,Y_2\atop{\sscr \vert Y_2 \vert \leq
\vert Y \vert - 1
\atop\sscr
Y_1 \in \Pi' \cap U ({\frak{sl}}(2, \Crm))}}^Y   \Vert \gamma^\mu
G_{Y_1 \mu} (t)  \xi^D_{Y_2} (\gamma^\nu  G_\nu h+g) (t) \Vert^{}_D,\quad
 1/2 < \rho < 1,\cr
}$$
where $C$ and $C_{\vert Y \vert}$ are constants depending only on $\rho.$

According to the definition of $a(Y_1,Y_2)$ in (5.118) and according
to definitions (5.119a) and (5.119b) of $k^{(1)}_n$ and $l^{(1)}_n$,
we obtain from (5.132b) and
(5.140) that
$$\eqalignno{
 J_Y (t)
&\leq C_{\vert Y \vert} (1+t)^{- 2+\rho} \Big([A]_3 (t)
\big(\wp^D_{\vert Y \vert, i}  (h(t))^{1/2}  \wp^D_{\vert Y \vert, i+1}
(h(t))^{1/2}&(5.141)\cr
&\qquad{}+ \wp^D_{\vert Y \vert - 1,i}  ((1+\lambda^{}_0 (t))^{1/2}
h(t))^{2(1-\rho)}
 \wp^D_{\vert Y \vert - 1, i}  (h(t))^{2 \rho - 1}\big)\cr
&\qquad{}+ \sum_{\scr n_1+n_2=\vert Y \vert\atop\scr 2 \leq n_1 \leq L,
i \leq n_2}  [A]_{n_1+2}
(t)  \wp^D_{n_2,i}  ((1+\lambda^{}_0 (t))^{1/2}  h(t))
+ k^{(1)}_{\vert Y \vert} (L,t) + l^{(1)}_{\vert Y \vert} (L,t)\Big)\cr
&\qquad{}+ C (1+t)^{- 2 + \rho}  \chi^{}_+ (\vert Y \vert - L - 1)
\chi^{}_+ (-i)  \cr
&\qquad\qquad{}\big(S^Y (t)
(1 + [A]^1 (t))  H_1 (t) + (1+t)^{2-\rho}
\Vert \gamma^\mu G_{Y\mu}(t)  g^{}_{\un} (t) \Vert^{}_{D}\big),\cr
}$$
where $Y \in \sg^i,  Y \notin \sg^{i+1},  0 \leq i \leq
\vert Y \vert - 1,  1/2 < \rho < 1,  L \geq 1$, and where $C$
and $C_{\vert Y \vert}$ are constants depending only on $\rho.$

It follows from inequality (5.139a) that
$$\eqalignno{
&\wp^D_n  \big((1+\lambda^{}_i (t))^{1/2}h(t)\big)&(5.142)\cr
&\quad{}\leq C'_{n+1}  \sum_{\scr n_1+n_2 = n+1\atop\scr  0 \leq n_1 \leq L_0}
 (1 + [A]^{n_1+1} (t))  \big(\wp^D_{n_2} (h(t)) + R^i_{n_2-1,1} (t)\big)\cr
&\qquad{}+ C''_{n+1}  \sum_{\sscr n_1+n_2+n_3+n_4=n+1\atop{\sscr  n_1
\leq L_0, n_2 \leq n
\atop\sscr  n_3+n_4
\leq n-L_0-1}}  (1+[A]^{n_1+1} (t))  S^{\rho,n_2} (t)\cr
&\qquad\qquad{}(1 + S^\rho_{10, n_3} (t))
\big(R'_{n_4+7} (t) + R^2_{n_4+9} (t) + R^\infty_{n_4} (t) + \wp^D_{n_4+8}
(t)\big),\cr
}$$
where $n \geq 0,  L_0 \geq 3,  1/2 < \rho <1$, $i=0,1$, and where the
constant $C'_{n+1}$ depends only on $\rho$ and $[A]^1 (t)$ and $C''_{n+1}$
depends only on $\rho$ and $[A]^9 (t)$. It follows from (5.142) that
$$\eqalignno{
&\sum_{\scr n_1+n_2= \vert Y \vert\atop\scr  2 \leq n_1 \leq L}  [A]_{n_1+2}
(t)  \wp^D_{n_2,i}  \big((1+\lambda^{}_0 (t))^{1/2}  h(t)\big)\cr
&\quad{}\leq C'_{\vert Y \vert}  \sum_{\scr n_1+n_2= \vert Y \vert\atop\scr 2
\leq n_1 \leq L}  \sum_{\scr n_3+n_4=n_2+1\atop\scr     0 \leq n_3 \leq
L_0(n_1)}
[A]_{n_1+2} (t)  (1+[A]^{n_3+1} (t))    \big(\wp^D_{n_4}
(h(t)) + R^0_{n_4,1} (t)\big)\cr
&\qquad{}+ C''_{\vert Y \vert}  \sum_{\scr n_1+n_2= \vert Y \vert\atop\scr  2
\leq
n_1 \leq L}  \sum_{\sscr n_3+n_4+n_5+n_6=n_2+1\atop{\sscr  n_3 \leq L_0 (n_1)
\atop\sscr  n_4 \leq
n_2+1, n_5+n_6 \leq n_2+1-L_0 (n_1) - 1}}
[A]_{n_1+2} (t)  (1+[A]^{n_3+1} (t))    S^{\rho,n_4} (t)
 \cr
&\hskip40mm{}(1+S^\rho_{10,n_5} (t))\big(R'_{n_6+7} (t) + R^2_{n_6+9} (t)
+ R^\infty_{n_6} (t) + \wp^D_{n_6+8} (t)\big),\cr
}$$
where we have chosen $L_0 (n_1) = L-n_1$ for $n_1 \leq L-3$ and $L_0 = 3$ for
$n_1 \geq L-2$ and where $1/2 < \rho < 1$, the constant $C'_{\vert Y \vert}$
depends on $[A]^1 (t)$ and the constant $C''_{\vert Y \vert}$ on $[A]^9 (t)$.
It
follows from definition (5.119c) of $k^{(2)}_n$ and the convexity properties
(analoguous to those of (5.89d) and (5.89c)) for $[A]_{N,n}$, that
$$
\sum_{\scr n_1+n_2 = \vert Y \vert\atop\scr 2 \leq n_1 \leq L}  [A]_{n_1+2}
(t)  \wp^D_{n_2,i}  \big((1+\lambda^{}_0 (t))^{1/2}  h(t)\big)
\leq C'_{\vert Y \vert}  k^{(2)}_{\vert Y \vert} (L,t) + C''_{\vert Y \vert}
 l^{(2)}_{\vert Y \vert}  (L,t),
\eqno{(5.143)}$$
where $1/2 < \rho < 1,  Y \in \Pi',  L \geq 0$ and where $C'_{\vert Y \vert}$
is a constant depending only on $\rho,L,[A]^3 (t)$, and $C''_{\vert Y \vert}$
is a
constant depending only on $\rho,L,[A]^9 (t).$

Using that $(a+b)^e  c^{1-e} \leq (a^e + b^e)  c^{1-e} \leq
a^e  c^{1-e} + eb + (1-e) c$ for $a,b,c \geq 0$,  $0 \leq e \leq 1$,
we obtain from (5.139b) that
$$\eqalignno{
&\wp^D_{\vert Y \vert - 1}
\big((1+\lambda^{}_0 (t))^{1/2}  h(t)\big)^{2(1 - \rho)}
 \wp^D_{\vert Y \vert - 1,i}  (h(t))^{2 \rho - 1} &(5.144)\cr
&\quad{}\leq \big(C'_1 (\wp^D_{\vert Y \vert} (h(t))
+ R^0_{\vert Y \vert - 1,1} (t))\big)^{2(1-\rho)}
 \wp^D_{\vert Y \vert - 1,i}  (h(t))^{2 \rho - 1}\cr
&\qquad{}+ C'_{\vert Y \vert}  k^{(2)}_{\vert Y \vert} (L,t) +
C''_{\vert Y \vert}
 l^{(2)}_{\vert Y \vert} (L,t)
+C'_{\vert Y \vert}\wp^D_{\vert Y \vert-1} (h(t))
+C'_{\vert Y \vert}R^0_{\vert Y \vert - 2,1} (t) ,\cr
}$$
where $1/2 < \rho < 1$ and where the constants $C'_1,  C'_{\vert Y \vert}$
depend only on $\rho$ and $[A]^1 (t)$, and the constant
$C''_{\vert Y \vert}$ depends only
on $\rho$ and $[A]^9 (t).$

It follows from inequalities (5.141), (5.143) and (5.144), that
$$\eqalignno{
 J_Y (t)  &\leq C(1+t)^{- 2 + \rho}  \chi^{}_+ (-i)
\chi^{}_+ (\vert Y \vert - L - 1)&(5.145)\cr
&\quad{}\big(S^Y (t)(1+[A]^1 (t))  H_1 (t) + (1+t)^{2 - \rho}
\Vert \gamma^\mu G_{Y\mu}
(t)  g^{}_{\un} (t) \Vert^{}_{D}\big)\cr
&\qquad{}+ C^{(1)}_{\vert Y \vert}  (1+t)^{- 2+\rho} \Big([A]_3 (t)
\wp^D_{\vert Y \vert} (h(t))^\varepsilon  \wp^D_{\vert Y \vert,i}
(h(t))^{1-\varepsilon}\cr
&\qquad{}+ [A]_3 (t)  R^0_{\vert Y \vert-1,1} (t)^{2(1-\rho)}   \wp^D_{\vert
Y \vert - 1,i} (h(t))^{2 \rho-1}\cr
&\qquad{}+ k^{(1)}_{\vert Y \vert} (L,t) + l^{(1)}_{\vert Y \vert} (L,t)
+ C^{(3)}_{\vert Y \vert}
 k^{(2)}_{\vert Y \vert} (L,t) + C^{(9)}_{\vert Y \vert}
l^{(2)}_{\vert Y \vert} (L,t)\Big),\cr
}$$
where $Y \in \sg^i,  Y \notin \sg^{i+1},  0 \leq i \leq
\vert Y \vert - 1,  1/2 < \rho < 1,  L \geq 1$ and where
$C^{(n)}_m$ are constants depending only on $[A]^n (t)$ and $\rho$. This proves
the lemma.

To obtain $L^2$-estimates of solutions $h$ of equation (5.1), we next estimate
the terms on the right-hand side of the inequalities in Proposition 5.10, which
have not already been considered in Lemma 5.11.
\saut
\noindent{\bf Lemma 5.12.}
{\it
Let $1/2 < \rho < 1$,  $h^{}_Z \in C^0 (\Rrm^+, D)$ and $(A_Z, A_{P_0 Z}) \in
C^0 (\Rrm^+, M^\rho) \cap C^0 (\Rrm^+, M^1)$ for $Z \in \Pi'$ and
$\vert Z \vert$ sufficiently large. Let $G$ be given by (5.114),
$L \geq 1$,  $g = (i \gamma^\mu  \partial_\mu + m - \gamma^\mu  G_\mu) h$,
$\chi^{}_+ (s) = 0$
for $s < 0$,  $\chi^{}_+ (s) = 1$ for $s \geq 0$, let $Y \in \sg^i$,
$Y \notin \sg^{i+1}$,  $i \geq 0$, and let $J^{(l)}_n$ be defined as in Lemma
5.11.

\noindent\hbox{\rm a)} If $i=0$, then
$$\eqalignno{
\Vert h^{}_Y (t) \Vert^{}_D
&\leq \Vert h^{}_Y (t_0) \Vert^{}_D + (1+t_0)^{- 3/2+\rho}
C_{\vert Y \vert}  \sum_{\scr n_1+n_2 = \vert Y \vert\atop\scr
1 \leq n_1 \leq L}
 [A]^{n_1+1} (t_0)  \wp^D_{n_2} (h(t_0))\cr
&\qquad{}+ (1+t_0)^{- 1/2}  C_{\vert Y \vert}
\sum_{\scr n_1+n_2=\vert Y \vert\atop\scr
n_1 \geq L+1}  S^{\rho,n_1} (t_0)  H_{n_2} (t_0) + \Vert f^{}_Y (t)
\Vert^{}_D\cr
&\qquad{}+ (1+t)^{- 3/2 + \rho}  C_{\vert Y \vert}  \sum_{\scr n_1+n_2=
\vert Y \vert\atop\scr 1 \leq n_1 \leq L}  [A]^{n_1+1} (t)
\wp^D_{n_2} (h(t))\cr
&\qquad{}+ (1+t)^{- 1/2}  C_0   \chi^{}_+ (\vert Y \vert - L - 1)
S^Y (t)  H_0 (t)\cr
&\qquad{}+ (1+t)^{- 1/2}  C_{\vert Y \vert}
\sum_{\scr n_1+n_2=\vert Y \vert\atop\scr
L+1 \leq n_1 \leq \vert Y \vert - 1}  S^{\rho,n_1} (t)  H_{n_2}
(t)\cr
&\qquad{}+ \int^{\max (t,t_0)}_{\min (t,t_0)}   (1+s)^{- 2+\rho} \Big(C'_{\vert
Y
\vert}  J^{(0)}_{\vert Y \vert} (L,s)+  \chi^{}_+  (\vert Y \vert -L-1)\cr
&\qquad\qquad{} C_0\big( S^Y (s)(1+[A]^1 (s))
H_1 (s) + (1+s)^{2-\rho}  \Vert\gamma^\mu G_{Y\mu}(s)
 g^{} (s) \Vert^{}_{D}\big)\Big)  ds.\cr
}$$
\noindent\hbox{\rm b)} If $1 \leq i \leq \vert Y \vert$, then $Y = P_\nu  Z$
for some
$Z \in \Pi'$, and
$$\eqalignno{
\Vert h^{}_Y (t) \Vert^{}_D &\leq \Vert h^{}_Y (t_0) \Vert^{}_D\cr
&\qquad{}+ (1+t_0)^{- 3/2 + \rho}  C_{\vert Y \vert}  \sum_{n_1+n_2=
\vert Y \vert - 1\atop\scr  0 \leq n_1 \leq L}  [A]^{2+n_1} (t_0)
\wp^D_{n_2} (h(t_0))\cr
&\qquad{}+ (1+t_0)^{- 1/2}  C_{\vert Y \vert}
\sum_{\scr n_1+n_2=\vert Y \vert -1\atop\scr
n_1 \geq L+1}  S^{\rho,n_1} (t_0)  H_{n_2+1} (t_0)\cr
&\qquad{}+ (1+t)^{- 3/2 + \rho}  C_{\vert Y \vert}  \sum_{\scr n_1+n_2=
\vert Y \vert - 1\atop\scr  0 \leq n_1 \leq L}  [A]^{2+n_1} (t) \wp^D_{n_2}
(h(t))\cr
&\qquad{}+ (1+t)^{- 1/2}  C_{\vert Y \vert}  \sum_{\scr n_1+n_2=
\vert Y \vert - 1\atop\scr
n_1 \geq L+1}  S^{\rho,n_1} (t)  H_{n_2+1} (t)\cr
&\qquad{}+ \Vert f^{}_Y (t) \Vert^{}_D +
\int^{\max (t,t_0)}_{\min (t,t_0)} (1+s)^{- 2+\rho}
 C'_{\vert Y \vert}  J^{(i)}_{\vert Y \vert} (L,s) ds.\cr
}$$
\penalty-9000
The constants $C_n, n \geq 0$, depend only on $\rho$ and the constants $C'_n$
depend only on $\rho$ and $[A]^3 (\max(t_0,t)).$
}\saut
\noindent{\it Proof.}
To prove statement a), let $h'_Y (t)$ be as in statement i) of Proposition
5.10.
It follows from inequalities (5.116c), (5.125a) and (5.125b), that
$$\eqalignno{
\Vert h^{}_Y (t) - h'_Y (t) \Vert^{}_D &\leq (1+t)^{- 3/2+\rho}
C_{\vert Y \vert}
 \sum_{\scr n_1+n_2= \vert Y \vert\atop\scr 1 \leq n_1 \leq L}  [A]^{n_1+1}
(t)  \wp^D_{n_2} (h(t))&(5.146)\cr
&\quad{}+ (1+t)^{- 1/2}  C  \chi^{}_+ (\vert Y \vert -L-1)
S^Y (t)  H_0 (t)\cr
&\quad{}+ (1+t)^{- 1/2}  C_{\vert Y \vert}  \sum_{\scr n_1+n_2=
\vert Y \vert\atop\scr
L+1 \leq n_1 \leq \vert Y \vert - 1}  S^{\rho,n_1} (t)  H_{n_2}
(t),\cr
}$$
where the constants $C$ and $C_{\vert Y \vert}$ depend only on $\rho$. Since
$Y \in \sg^0$ and $Y \notin \sg^1$, the sum over $Y_1 \in \sg^1$ on the
right-hand side in the inequality of statement i) of Proposition 5.10, vanish.
This inequality then gives, according to definition (5.118) of $J_Y$, that
$$\Vert h'_Y (t) \Vert^{}_D \leq \Vert h'_Y (t_0) \Vert^{}_D +
\Vert f^{}_Y (t) \Vert^{}_D
+ \int^{\max (t,t_0)}_{\min (t,t_0)}    J_Y (s)  ds.$$
This inequality, the use of inequality (5.146) to estimate
$\Vert h'_Y (t) \Vert^{}_D$ and $\Vert h'_Y (t_0) \Vert^{}_D$ and
the estimate of $J_Y (s)$ in Lemma 5.11, give the estimate of statement a).

To prove statement b), we observe that it follows from the definition of
$\sg^i$ that $Y = P_\nu  Z$ for some $Z \in \sg^{i-1}$ and $Z \notin \sg^i$.
Let $h^{(\nu)}_Z, g^{(\nu)}_{1 Z}$ and $g^{(\nu)}_{2 Z}$ be as in statement
ii) of Proposition 5.10. It follows that
$$\eqalignno{
&\Vert h^{}_Y (t) - h^{(\nu)}_Z (t) + {(2m)}^{-1}  g^{(\nu)}_Z (t)
\Vert^{}_D &(5.147{\rm a})\cr
&\qquad{}\leq \suma_{Z_1,Z_2}^Z  \Vert G_{Z_1 \nu} (t)  h^{}_{Z_2} (t)
\Vert^{}_D + {(2m)}^{-1}  \suma_{\sscr Z_1,Z_2\atop{\sscr
 \vert Z_2 \vert \leq \vert Z \vert - 1\atop\sscr
Z_1 \in \Pi' \cap U ({\frak{sl}}(2, \Crm))}}^Z  \Vert \gamma^\mu  G_{Z_1 \mu}
(t)  h^{}_{P_\nu Z_2} (t) \Vert^{}_D.\cr
}$$
This inequality, inequalities (5.116c), (5.125b) and the fact that $S^{1,n} (t)
\leq S^{\rho,n} (t)$ for $\rho \leq 1$, give that
$$\eqalignno{
&\Vert h^{}_Y (t) - h^{(\nu)}_Z (t) + {(2m)}^{-1}  g^{(\nu)}_Z (t) \Vert^{}_D
&(5.147{\rm b})\cr
&\quad{}\leq C_{\vert Y \vert}  (1+t)^{- 3/2 + \rho}    \sum_{\scr n_1+n_2=
\vert Y \vert - 1\atop\scr 0 \leq n_1 \leq L}  [A]^{2+n_1} (t)  \wp^D_{n_2}
(h(t))\cr
&\qquad{}+ C_{\vert Y \vert}  (1+t)^{- 1/2}  \sum_{\scr n_1+n_2=
\vert Y \vert -1\atop\scr
n_1 \geq L+1}  S^{\rho,n_1} (t)  H_{n_2+1} (t),\cr
}$$
where $C_{\vert Y \vert}$ depends only on $\rho.$

Using the gauge invariance of the electromagnetic field, and
$$(i \gamma^\mu  \partial_\mu + m - \gamma^\mu  G_\mu)
 h^{}_{Z_2} = g^{}_{Z_2}
+ \suma_{\scr Z_3,Z_4\atop\scr \vert Z_4 \vert \leq \vert Z_2 \vert - 1}^{Z_2}
 \gamma^\mu  G_{Z_3 \mu}  h^{}_{Z_4},$$
we obtain from the definition of $g^{(\nu)}_{1Z}$ that
$$\eqalignno{
\Vert g^{(\nu)}_{1Z} (t) \Vert^{}_D &\leq \suma_{Z_1,Z_2}^Z  \Vert
\gamma^\mu (\partial_\mu  A_{Z_1 \nu} (t) - \partial_\nu  A_{Z_1 \mu}
(t))  h^{}_{Z_2} (t) \Vert^{}_D&(5.148)\cr
&\quad{}+ \suma_{\sscr Z_1,Z_2\atop{\sscr  \vert Z_2
\vert \leq \vert Z \vert - 1
\atop\sscr  Z_1 \in \sssg^1}}^Z
\Vert \gamma^\mu  G_{Z_1 \mu} (t)  h^{}_{P_\nu Z_2} (t) \Vert^{}_D\cr
&\quad{}+ C  \suma_{\scr Z_1,Z_2,Z_3\atop\scr  \vert Z_3 \vert
\leq \vert Z \vert
-1}^Z   \sum_{\alpha, \beta} \Vert G_{Z_1 \alpha} (t)
G_{Z_2 \beta} (t)  h^{}_{Z_3} (t) \Vert^{}_D
+ \suma_{Z_1,Z_2}^Z  \Vert G_{Z_1 \nu} (t)  g^{}_{Z_2} (t) \Vert.\cr
}$$

Let $X \in \Pi'$. Then according to statement i) of Lemma 4.5, with $\rho=1$,
$$\eqalignno{
\Vert G_{P_\mu X} (t) \Vert^{}_{L^2}
&\leq \Vert (G_X (t), G_{P_0 X} (t)) \Vert^{}_{M^1}\cr
&\leq C \sup_{0 \leq s \leq t}  \Big(\Vert (A_X (s),  A_{P_0 X}
(s)) \Vert^{}_{M^1} + \Vert (B_X (s),  B_{P_0 X} (s)) \Vert^{}_{M^1}\cr
&\quad{}+ \wp^{M^1}_{\vert X \vert + 1}  ((A(s),  \dot{A} (s))) +
C_{\vert X \vert}  \wp^{M^1}_{\vert X \vert}  ((A(s),
\dot{A} (s)))\Big),\cr
}$$
where $B_\mu (y) = y_\mu  \partial^\nu  A_\nu (y)$ and $\dot{A}_X =
A_{P_0 X}$ for $X \in \Pi'$. This gives that
$$\Vert G_X (t) \Vert^{}_{L^2} \leq C_0  S^{1, \vert X \vert} (t) +
C_{\vert X \vert}  S^{1, \vert X \vert -1} (t), \quad\hbox{ for } X \in \sg^1,
\eqno{(5.149)}$$
for two numerical constants $C_0$ and $C_{\vert X \vert}.$

For the first term on the right-hand side of inequality (5.148),
we obtain that
$$\eqalignno{
&\suma_{Z_1,Z_2}^Z  \Vert \gamma^\mu (\partial_\mu  A_{Z_1 \nu}
(t) - \partial_\nu  A_{Z_1 \mu} (t))  h^{}_{Z_2} (t) \Vert^{}_D
&(5.150)\cr
&\quad{}\leq (1+t)^{- 2+\rho}  C_{\vert Z \vert}  \sum_{\scr n_1+n_2=
\vert Z \vert\atop\scr
0 \leq n_1 \leq L}  [A]^{n_1+1} (t)  \wp^D_{n_2} ((1+ \lambda^{}_0
(t))^{1/2}  h(t))^{2 (1-\rho)}  \wp^D_{n_2} (h(t))^{2 \rho - 1}\cr
&\qquad{}+ (1+t)^{- 3/2}  C_{\vert Z \vert}  \sum_{\scr n_1+n_2=
\vert Z \vert\atop\scr
n_1 \geq L+1}  \wp^{M^1}_{n_1} ((A(t), \dot{A} (t)))
H_{n_2} (t)\cr
&\quad{}\leq (1+t)^{- 2+\rho}  C_{\vert Z \vert}  \sum_{\scr n_1+n_2=
\vert Z \vert\atop\scr 0 \leq n_1 \leq L}  [A]^{n_1+1} (t)  \wp^D_{n_2}
((1+\lambda^{}_0 (t))^{1/2}  h(t))^{2(1-\rho)}  \wp^D_{n_2}
(h(t))^{2\rho-1}\cr
&\qquad{}+ (1+t)^{- 3/2}  C_{\vert Z \vert}  l^{(1)}_{\vert Z \vert+1}
 (L+1,t),\quad L \geq 0,  Z \in \Pi'.\cr
}$$
For the second term on the right-hand side of inequality (5.148), we obtain,
using (5.149),
$$\eqalignno{
&\suma_{\sscr Z_1,Z_2\atop{\sscr \vert Z_2 \vert \leq
\vert Z \vert - 1\atop\sscr
 Z_1 \in \sssg^1}}^Z
\Vert \gamma^\mu  G_{Z_1 \mu} (t)  h^{}_{P_\nu Z_2} (t) \Vert^{}_D
&(5.151)\cr
&\qquad{}\leq (1+t)^{-2+\rho}   C_{\vert Z \vert}  \chi^{}_+ (i-2)
 \sum_{\scr n_1+n_2=\vert Z \vert\atop\scr 1 \leq n_1 \leq L}  [A]^{n_1+2}
(t)\cr
&\qquad\qquad{}\wp^D_{n_2+1} ((1+\lambda^{}_0 (t))^{1/2}  h(t))^{2(1-\rho)}
\wp^D_{n_2+1} (h(t))^{2 \rho-1}\cr
&\qquad\qquad{}+ (1+t)^{- 3/2}  C_{\vert Z \vert}  \chi^{}_+ (i-2)
\sum_{\scr n_1+n_2=\vert Z \vert\atop\scr n_1 \geq L+1}  S^{1,n_1} (t)
H_{n_2+1} (t)\cr
&\qquad{}\leq (1+t)^{- 2+\rho}  C_{\vert Z \vert}  \chi^{}_+ (i-2)
 \sum_{\scr n_1+n_2=\vert Z \vert\atop\scr 1 \leq n_1 \leq L}  [A]^{n_1+2}
(t)\cr
&\qquad\qquad{}\wp^D_{n_2+1} ((1+\lambda^{}_0 (t))^{1/2} h(t))^{2(1-\rho)}
\wp^D_{n_2+1}
(h(t))^{2 \rho - 1}\cr
&\qquad\qquad{}+ (1+t)^{- 3/2}  C_{\vert Z \vert}  \chi^{}_+ (i-2)
l^{(1)}_{\vert Z \vert + 1}  (L+1,t),\cr
}$$
where $Z \in \sg^{i-1}$ and $Z \notin \sg^i,1 \leq i \leq \vert Z \vert + 1.$

Proceeding like in (5.131a) and (5.131b), we obtain
$$\eqalignno{
&\suma_{\scr Z_1,Z_2,Z_3\atop\scr \vert Z_3 \vert \leq \vert Z \vert - 1}^Z
\Vert
G_{Z_1 \alpha} (t)  G_{Z_2 \beta} (t)  h^{}_{Z_3} (t) \Vert^{}_D
&(5.152)\cr
&\qquad{}\leq (1+t)^{- 3+2\rho}  C_{\vert Z \vert}
\sum_{\scr n_1+n_2=\vert Z \vert\atop\scr
1 \leq n_1 \leq L}  [A]_{n_1+2} (t)  \wp^D_{n_2} (h(t))\cr
&\qquad\qquad{}
+ (1+t)^{- 2+\rho}  C_{\vert Z \vert}  \sum_{\sscr n_1+n_2+n_3=
\vert Z \vert\atop{
\sscr n_2 \leq n_1\atop\sscr  n_3 \leq \vert Z \vert -L-1}} S^{\rho,n_1} (t)
[A]^{n_2+1} (t)  H_{n_3} (t)\cr
&\qquad{}\leq (1+t)^{- 3+2\rho}  C_{\vert Z \vert}  k^{(1)}_{\vert
Z \vert + 1} (L,t) + (1+t)^{- 2+\rho}  C_{\vert Z \vert}
l^{(1)}_{\vert Z \vert + 1} (L,t),\cr
}$$
for $Z \in \Pi'$ and $L \geq 0$. It follows from definitions
(5.119a) and (5.119b), that
$$\eqalignno{
&\suma_{Z_1,Z_2}^Z  \Vert G_{Z_1 \nu} (t)  g^{}_{Z_2} (t) \Vert^{}_D
&(5.153)\cr
&\qquad{}\leq (1+t)^{- 3/2+\rho}  C_{\vert Z \vert}  \sum_{\scr n_1+n_2=
\vert Z \vert\atop\scr
0 \leq n_1 \leq L}  [A]^{n_1+1} (t)  \wp^D_{n_2} (g(t))\cr
&\qquad\qquad{}+ C_{\vert Z \vert}  \sum_{\sscr \vert Z_1\vert +
\vert Z_2\vert =\vert Z \vert
\atop{\sscr  \vert Z_1\vert  \geq L+1\atop\sscr Z_1, Z_2
\in \Pi'}}  \Vert \gamma^\mu G_{Z_1\mu}(t)
g^{}_{Z_2} (t) \Vert^{}_{D}\cr
&\qquad{}\leq (1+t)^{- 2+\rho}C_{\vert Z \vert}
\big(k^{(1)}_{\vert Z \vert + 1}
 (L+1,t) + l^{(1)}_{\vert Z \vert + 1}  (L+1,t)\big),
\quad L \geq 0,  Z \in \Pi'.\cr
}$$

Inequality (5.148), inequalities (5.150), (5.151) and (5.153),
with $L$ instead of $L+1$, and inequality (5.152), give that
$$\eqalignno{
&\Vert g^{(\nu)}_{1Z} (t) \Vert^{}_D&(5.154)\cr
&\qquad{}\leq (1+t)^{- 2+\rho}  C_{\vert Y \vert}
 [A]^3 (t)  \wp^D_{\vert Y \vert - 1} ((1+\lambda^{}_0 (t))^{1/2}
 h(t))^{2(1-\rho)}
\wp^D_{\vert Y \vert - 1} (h(t))^{2 \rho-1}\cr
&\qquad\qquad{}+ (1+t)^{- 2+\rho}  C_{\vert Y \vert}
\sum_{\scr n_1+n_2=\vert Y \vert -1\atop\scr
1 \leq n_1 \leq L-1}  [A]^{n_1+1} (t)  \wp^D_{n_2} ((1+\lambda^{}_0
(t))^{1/2}  h(t))\cr
&\qquad\qquad{}+ (1+t)^{- 2+\rho}  C_{\vert Y \vert}
\sum_{\scr n_1+n_2=\vert Y \vert - 1\atop\scr
2 \leq n_1 \leq L-1}  [A]^{n_1+2} (t)  \wp^D_{n_2+1} ((1+\lambda^{}_0
(t))^{1/2}  h(t))\cr
&\qquad\qquad{}+ (1+t)^{- 2+\rho}  C_{\vert Y \vert}
(k^{(1)}_{\vert Y \vert}
 (L,t) + l^{(1)}_{\vert Y \vert} (L,t)),\cr
}$$
where $Z \in \Pi',  1/2 < \rho < 1$ and where $C_{\vert Y \vert}$ is
a constant depending only on $\rho$. Majorizing the second and the
third term on the right-hand side of this inequality with the help of
(5.143), we obtain that
$$\eqalignno{
&\Vert g^{(\nu)}_{1Z} (t) \Vert^{}_D&(5.155)\cr
&\qquad{}\leq (1+t)^{- 2+\rho}  C_{\vert Y \vert}
[A]^3 (t)  \wp^D_{\vert Y \vert -1} ((1+\lambda^{}_0 (t))^{1/2}
h(t))^{2 (1-\rho)}
\wp^D_{\vert Y \vert - 1} (h(t))^{2 \rho-1}\cr
&\qquad\qquad{}+ (1+t)^{- 2+\rho}  C_{\vert Y \vert}
\big(k^{(1)}_{\vert Y \vert} (L,t)
+ l^{(1)}_{\vert Y \vert} (L,t)\big)\cr
&\qquad\qquad{}+ (1+t)^{- 2+\rho}
\big(C'_{\vert Y \vert}  k^{(2)}_{\vert Y \vert}
(L,t) + C''_{\vert Y \vert}  l^{(2)}_{\vert Y \vert} (L,t)\big),\cr
}$$
where $Z \in \Pi',  L \geq 1,  1/2 < \rho < 1$ and where the
constant $C_{\vert Y \vert}$ depends only on $\rho$, the constant
$C'_{\vert Y \vert}$
only on $\rho$ and $[A]^3 (t)$ and the constant $C''_{\vert Y \vert}$
only on $\rho$ and $[A]^9 (t)$. Inequalities (5.144) and (5.155) and
the definition of $J^{(l)}_n (L,t)$ in Lemma 5.11, give that
$$\eqalignno{
&\Vert g^{(\nu)}_{1 Z} (t) \Vert^{}_D&(5.156)\cr
&\qquad{}\leq (1+t)^{- 2+\rho} C'_{\vert Y \vert}  [A]^3 (t)
 \big(\wp^D_{\vert Y \vert} (h(t)) +
 R^0_{\vert Y \vert -1,1} (t)\big)^{2(1-\rho)}
 \wp^{}_{\vert Y \vert -1}  (h(t))^{2 \rho-1}\cr
&\qquad\qquad{}+ (1+t)^{- 2+\rho}
C_{\vert Y \vert}  \big(k^{(1)}_{\vert Y \vert}
(L,t) + l^{(1)}_{\vert Y \vert} (L,t)\big)\cr
&\qquad\qquad{}+ (1+t)^{- 2+\rho}
\big(C'_{\vert Y \vert}  k^{(2)}_{\vert Y \vert}
(L,t) + C''_{\vert Y \vert}  l^{(2)}_{\vert Y \vert} (L,t)\big)\cr
&\qquad{}\leq (1+t)^{- 2+\rho}  C'_{\vert Y \vert}  J^{(i)}_{\vert Y \vert}
(t),\quad  i \geq 1,\cr
}$$
where $Z \in \Pi',  Y = P_\nu  Z,  L \geq 1,  1/2
< \rho < 1$ and where the constant $C_{\vert Y \vert}$ depends only on $\rho,$
$C'_{\vert Y \vert}$ only of $\rho$ and $[A]^3 (t)$
and the constant $C''_{\vert Y \vert}$ only on $\rho$ and $[A]^9 (t).$

Since, according to statement ii) of Proposition 5.10,
$$\eqalignno{
\Vert h^{}_Y (t) \Vert^{}_D &\leq \Vert h^{}_Y (t) -
h^{(\nu)}_Z (t) + {(2m)}^{-1}
g^{(\nu)}_{2Z} (t) \Vert^{}_D + \Vert h^{(\nu)}_Z (t) - {(2m)}^{-1}
g^{(\nu)}_{2Z} (t) \Vert^{}_D\cr
&\leq \Vert h^{}_Y (t) - h^{(\nu)}_Z (t) + {(2m)}^{-1}  g^{(\nu)}_{2Z}
(t) \Vert^{}_D + \Vert h^{(\nu)}_Z (t_0) - {(2m)}^{-1}  g^{(\nu)}_{2Z}
(t_0) \Vert^{}_D\cr
&\qquad{}+ \Vert f^{}_Y (t) \Vert^{}_D + \int^{\max (t,t_0)}_{\min (t,t_0)}
\big(\Vert
g^{(\nu)}_{1 Z} (s) \Vert^{}_D + J_Y (s)\big)  ds,\cr
}$$
where $J_Y (s)$ is given by (5.118), the inequality
of statement b) follows from (5.147b), (5.156) and Lemma 5.11.
This proves the lemma.

Lemma 5.11 and Lemma 5.12 lead to $L^2$-estimates of
$h^{}_Y (t),  Y \in \Pi',$
where $h$ is a solution of (5.1), in terms of $L^2$ and $L^\infty$ norms of
$h^{}_Z (t_0),  f^{}_Z (s),  g^{}_Z (s)$ and $A_Z (s),  Z \in
\Pi'$. We recall that $h^{}_Y (t_0)$ is a function of $h^{}_{\un} (t_0)$
for a fixed potential $G$ and an inhomogeneous term $g$ in equation (5.1),
which is obtained by eliminating time derivatives in $h^{}_Y (t_0)$
by equation (5.1).
\saut
\noindent{\bf Theorem 5.13.}
{\it
Let $n \geq 0$, $1/2 < \rho < 1$,  $(1-\Delta)^{1/2}
(A_X, A_{P_0 X}) \in C^0(\Rrm^+, M^1)$ for $X = \un$,  $X \in {\frak{sl}}(2,
\Crm)$,
let $(1-\Delta)^{1/2}  (B, \dot{B}) \in C^0(\Rrm^+, M^1)$, where $B_\mu
(y) = y_\mu  \partial^\nu  A_\nu (y)$,  $\dot{B} =
{d \over dt}  B$, let $A_Y \in C^0(\Rrm^+, L^\infty (\Rrm^3, \Rrm^4))$
for $Y \in \Pi'$,  $\vert Y \vert \leq n+3$, let $B_Y \in C^0 (\Rrm^+, L^\infty
(\Rrm^3, \Rrm^4))$ for $Y \in \Pi'$,  $\vert Y \vert \leq n+2$, let $G_\mu$
be given by (5.144), $f$ be given by (5.111b), let $g^{}_Y = \xi^D_Y  g \in
C^0(\Rrm^+, D)$ for $Y \in \Pi'$,  $\vert Y \vert \leq n$, let
$$\eqalignno{
Q_n (t) &= \sup_{t_0 \leq s \leq t} \wp^D_n (f(s)) + \sum_{0 \leq l \leq n-1}
 [A]_{3,n-l} (t)\cr
&\quad{}\Big(\sup_{t_0 \leq s \leq t}  \wp^D_l(f(s))
+ \int^t_{t_0} (1+s)^{- 3/2+\rho}  \wp^D_l ((1+\lambda^{}_0 (s))^{1/2}
 g(s))  ds\Big),\cr
}$$
for $0 \leq t_0 \leq t$, where $\lambda^{}_0$ is as in Theorem 5.5
and for $0 \leq t < t_0$ let $Q_n$
be given by the same expression, but with $t$ and $t_0$ interchanged on the
right-hand side. If $h^{}_Y (t_0) \in D$ for $Y \in \Pi'$,
$\vert Y \vert \leq n$, then $h$ given by (5.3c) is the unique solution of
equation (5.1a) in $C^0(\Rrm^+,D)$, with initial data $h(t_0)$.
This solution satisfies
$h^{}_Y \in C^0(\Rrm^+,D)$ for $Y \in \Pi'$,  $\vert Y \vert \leq n$,
$h^{}_{P_\mu Y} \in C^0(\Rrm^+, (1 - \Delta)^{1/2} D)$ for $Y \in \Pi'$,
$\vert Y \vert \leq n$, and
$$\wp^D_n (h(t)) \leq C_n \Big(\wp^D_n (h(t_0)) + \sum_{0 \leq l \leq n-1}
[A]_{3,n-l} (t')  \wp^D_l (h(t_0)) + Q_n (t)\Big),$$
for $t,t_0 \geq 0$, where $t' = \max (t,t_0)$ and where the constant $C_n$
depends only on
$[A]^3 (t')$ and $\rho$.

If moreover the function $t\! \mapsto\! (1+t)^{- 3/2+\rho}
 (1+\lambda^{}_0 (t))^{1/2}  g^{}_Y (t)$ is an element of $L^1 (\Rrm^+\!,D)$
for $Y \in \Pi'$,  $\vert Y \vert \leq n$, and if for each
$Y \in \Pi'$,  $\vert Y \vert \leq n$, there exists $g^{}_{1Y}$ and
$g^{}_{2Y}$ such that $g^{}_{Y}=g^{}_{1Y}+g^{}_{2Y}$, and such that:\psaut

\hbox{{\rm a)} $g^{}_{1Y} \in L^1 (\Rrm^+,D)$,}\psaut

\hbox{{\rm b)} $(m - i \gamma^\mu  \partial_\mu + \gamma^\mu  G_\mu)
 g^{}_{2Y} \in L^1 (\Rrm^+,D)$ and $\displaystyle{\lim_{t \fl \infty}}
\Vert g^{}_{2Y}(t) \Vert^{}_D = 0$,}\psaut

\noindent
then there exists a unique solution $h \in C^0(\Rrm^+,D)$ of equation (5.1a)
such that $\Vert h(t) \Vert^{}_{D} \fl 0$, when $t \fl \infty$.
This solution satisfies
$\wp^D_n (h(t)) \leq C_n  Q^\infty_n (t)$, $t \geq 0$, where $Q^\infty_n$
is given by the above expression of $Q_n (t)$, with $t_0 = \infty$. Further,
$f^{}_Y \in C^0 (\Rrm^+,D)$, $f^{}_Y (t) \fl 0$ as $t \fl \infty$ and
$f^{}_Y (t)$ is the limit of $\int^T_t  w(t,s)  i \gamma^0
g^{}_Y (s)  ds$ in $D$ as $T \fl \infty$.
}\saut
\noindent{\it Proof.}
First let $t_0 < \infty$. It follows from statement i) of Lemma 4.5,
with $\rho=1$, that $(G, \dot{G}) \in C^0 (\Rrm^+, (1-\Delta)^{- 1/2} M^1)$,
where $\dot{G} (t) = {d \over dt}  G(t)$, since $(1-\Delta)^{1/2}
(A_X, A_{P_0 X})
\in C^0 (\Rrm^+,M^1)$ for $X = \un$,  $X \in {\frak{sl}}(2,\Crm))$ and
$(1-\Delta)^{1/2}
 (B,\dot{B}) \in C^0 (\Rrm^+,M^1)$. Since $h(t_0) \in D$ and $g \in C^0
(\Rrm^+,D)$, it follows that conditions (5.1b), (5.1c) and
(5.1d) hold  for $\tau = 0$, which proves that equation (5.1a) has a
unique solution $h \in C^0 (\Rrm^+,D)$,
that this solution is given by (5.3c) and that $h^{}_{P_\mu}
\in C^0 (\Rrm^+, (1-\Delta)^{1/2}
D)$ for $0\leq \mu\leq 3$.

Application of $\xi^M_Y$ to equation (5.1a) gives that $(i \gamma^\mu
\partial_\mu + m - \gamma^\mu  G_\mu)  h^{}_Y = g'_Y$,
$Y \in \Pi'$, where
$$g'_Y = \suma_{\scr Y_1,Y_2\atop\scr  \vert Y_2 \vert
\leq \vert Y \vert - 1}^Y
 \gamma^\mu  G_{Y_1 \mu}  h^{}_{Y_2} + g^{}_Y.$$
Since, according to inequality (4.82)
$$\Vert G_Z (t) \Vert^{}_{L^\infty} \leq C  \sup_{0 \leq s \leq t}
\Big(\sum_{\vert Y \vert \leq \vert Z \vert + 1}  \Vert A_Y (s)
\Vert^{}_{L^\infty} + \sum_{\vert Y \vert \leq \vert Z \vert}
\Vert B_Y (s) \Vert^{}_{L^\infty}\Big),$$
for $Z \in \Pi'$, it follows from the hypothesis that $G_Z \in
C^0 (\Rrm^+, L^\infty (\Rrm^3, \Rrm^4))$ for $\vert Z \vert \leq n$.
This shows that, if $\vert Y \vert \leq n$,
then $g'_Y \in C^0 (\Rrm^+,D)$ if $h^{}_Z \in C^0 (\Rrm^+,D)$ for
$\vert Z \vert \leq \vert Y \vert - 1$. Repeating the argument which
proved that $h^{}_{\un} \in C^0 (\Rrm^+,D)$ and that $h^{}_{P_\mu} \in C^0
(\Rrm^+, (1-\Delta)^{1/2} D)$, it follows that $h^{}_Y \in C^0 (\Rrm^+,D)$
and that $h^{}_{P_\mu Y} \in C^0 (\Rrm^+, (1-\Delta)^{1/2}
D)$ for $\vert Y \vert \leq n$, $Y \in \Pi'$.

It follows from the definition of $J^{(l)}_n (L,t)$ in Lemma 5.11
and definitions (5.119a), (5.119b), (5.119c) and (5.119d) of
$k^{(1)}_n, l^{(1)}_n, k^{(2)}_n$
and $l^{(2)}_n$, that
$$J^{(i)}_n (n,t) \leq 2[A]_3 (t)  (\wp^D_n (h(t)) + R^0_{n-1,1} (t)) +
k^{(1)}_n (n,t) + k^{(2)}_n (n,t),
\quad n \geq 1,  i \geq 0.$$
This inequality and definitions (5.119a) and (5.119c) of $k^{(1)}_n$ and
$k^{(2)}_n$ give that
$$\eqalignno{
J^{(i)}_n (n,t) &\leq C  \sum_{n_1+n_2=n}  [A]_{3,n_1}(t)
 \big(\wp^D_{n_2} (h(t)) + R^0_{n_2-1,1} (t)\big)&(5.157)\cr
&\qquad{}+\sum_{\scr n_1+n_2=n\atop\scr  n_2 \leq n-1}  [A]_{3,n_1} (t)
(1+t)^{1/2}  \wp^D_{n_2} (g(t)),\quad  n \geq 1, i \geq 0,\cr
}$$
where $C$ is a numerical constant. It follows from this inequality,
from statement
a) of Lemma 4.12 in the case $Y \in \sg^0,  Y \notin \sg^1$ and from
statement b) in the case $Y \in \sg^1$, with $L = \vert Y \vert$, that
$$\eqalignno{
\Vert h^{}_Y (t) \Vert^{}_D &\leq \Vert h^{}_Y (t_0) \Vert^{}_D +
\Vert f^{}_Y (t) \Vert^{}_D
&(5.158)\cr
&\quad{}+ C_{\vert Y \vert}  (1+t_0)^{- 3/2+\rho}  \sum_{\scr n_1+n_2=
\vert Y \vert\atop\scr
 n_2 \leq \vert Y \vert - 1}  [A]^{n_1+1} (t_0)
\wp^D_{n_2} (h(t_0))\cr
&\quad{}+ C_{\vert Y \vert}  (1+t)^{- 3/2+\rho}  \sum_{\scr n_1+n_2=
\vert Y \vert\atop\scr
 n_2 \leq \vert Y \vert - 1}  [A]^{n_1+1} (t)
\wp^D_{n_2} (h(t))\cr
&\quad{}+ \vert Y \vert  C'_{\vert Y \vert}  \int^{\max(t,t_0)}_{\min(t,t_0)}
 (1+s)^{- 2+\rho}  \Big(\sum_{n_1+n_2=\vert Y \vert}
[A]_{3,n_1} (s)  \big(\wp^D_{n_2} (h(s)) + R^0_{n_2-1,1} (s)\big)\cr
&\quad{}+ \sum_{n_1+n_2=\vert Y \vert,  n_2 \leq \vert Y \vert - 1}
[A]_{3,n_1} (s)  (1+s)^{1/2}  \wp^D_{n_2} (h(s))\Big)ds,\quad
Y \in \Pi',  1/2 < \rho < 1,\cr
}$$
where $C_{\vert Y \vert}$ is a constant depending only on
$\rho$ and $C'_{\vert Y \vert}$
a constant depending only on $\rho$ and $[A]^3 (t')$, $t' = \max (t,t_0)$.
Summation over $\vert Y \vert \leq k \leq n$, give that
$$\eqalignno{
\wp^D_k (h(t)) &\leq C  \wp^D_k (h(t_0)) + C  \wp^D_k (f(t))
&(5.159)\cr
&\quad{}+ C_k  \sum_{\scr n_1+n_2=k\atop\scr  n_2 \leq k-1}
\Big((1+t_0)^{- 3/2+\rho}
 [A]^{n_1+1} (t_0)  \wp^D_{n_2} (h(t_0))\cr
&\qquad{}+ (1+t)^{- 3/2+\rho}   [A]^{n_1+1} (t)  \wp^D_{n_2} (h(t))\Big)\cr
&\quad{}+ k  C'_k  \int^{\max(t,t_0)}_{\min(t,t_0)}  \Big((1+s)^{- 2+\rho}
 \sum_{n_1+n_2=k}  [A]_{3,n_1} (s)  \big(\wp^D_{n_2}
(h(s)) + R^0_{n_2-1,1} (s)\big)\cr
&\qquad{}+ (1+s)^{- 3/2+\rho}   \sum_{\scr n_1+n_2=k\atop\scr  n_2 \leq k-1}
[A]_{3,n_1} (s)  \wp^D_{n_2} (g(s))\Big)  ds,\cr
}$$
$0 \leq k \leq n,  1/2 < \rho < 1$, where $C$ is a numerical constant,
$C_k$ depends only on $\rho$ and $C'_k$ only on $\rho$ and $[A]^3 (t').$

Let, for $0 \leq k \leq n,  t' = \min (t,t_0),  t'' = \max (t,t_0)$,
$$\eqalignno{
Q^{(k)} (t) &= \sup_{t' \leq s \leq t''}  \Big(C  \wp^D_k (h(t_0)) +
C  \wp^D_k (f(s))&(5.160)\cr
&\quad{}+ C_k  \sum_{\scr n_1+n_2=k\atop\scr
n_2 \leq k-1} \Big((1+t_0)^{- 3/2+\rho}
 [A]^{n_1+1} (t_0)  \wp^D_{n_2} (h(t_0))\cr
&\qquad{}+ (1+s)^{- 3/2+\rho}   [A]^{n_1+1} (s)
\wp^D_{n_2} (h(s))\Big)\Big)\cr
&\quad{}+ k  C'_k  \int^{t''}_{t'}
\Big(\sum_{\scr n_1+n_2=k\atop\scr  n_2 \leq k-1}
 [A]_{3,n_1} (s)  \Big((1+s)^{- 2+\rho}  \big(\wp^D_{n_2}
(h(s)) + R^0_{n_2-1,1} (s)\big)\cr
&\qquad{}+ (1+s)^{- 3/2+\rho}   \wp^D_{n_2} (g(s))\Big) + (1+s)^{- 2+\rho}
R^0_{k-1,1} (s)\Big)  ds,\cr
}$$
where the constants $C,C_k,C'_k$ are as in (5.159).
Inequality (5.159) now reads
$$\wp^D_k (h(t)) \leq Q^{(k)} (t) + k  C'_k  \int^{\max(t,t_0)}_{
\min(t,t_0)} (1+s)^{- 2+\rho}  [A]_{3,0} (s)  \wp^D_k (h(s))
 ds, \eqno{(5.161)}$$
$0 \leq k \leq n,  t \geq 0$, where $C'_k$ is a constant depending only
on $\rho$ and $[A]^3 (\max (t,t_0))$. Since $Q^{(k)}$
is increasing on the interval
$[t_0,t]$, when $t_0 \leq t$ and decreasing on the interval $[t,t_0]$,
when $t < t_0,$ it follows from Gr\"onwall lemma that
$$\wp^D_k (h(t)) \leq C'_k  Q^{(k)} (t),\quad  t \geq 0,
k \geq 0, \eqno{(5.162)}$$
where $C'_k$ is a new constant depending only on $\rho$ and
$[A]^3 (\max(t,t_0)).$

Inequality (5.162) gives for $k=0$ that
$$\wp^D_0 (h(t)) \leq C'_0 \Big(\wp^D_0 (h(t_0)) + \sup_{t' \leq s \leq t''}
 \wp^D_0 (f(s))\Big),$$
where $t' = \min (t,t_0),  t'' = \max (t,t_0)$ and $C'_0$ a constant
depending only on $\rho$ and $[A]^3 (t'')$, which shows that the
statement of the theorem is true for $n=0$. Suppose that the inequality
of the theorem is true with $n-1$ instead of $n$. Using the convexity
property for the functions $[A]_{N,l} (t)$ (analog to (5.89d), (5.89e)), we
then obtain that $\wp^D_n (h(t))$ satisfies the inequality of the theorem
for $n$, since
$$\eqalignno{
&\sum_{n_1+n_2=n}  (1+s)^{- 2+\rho}  [A]_{3,n_1} (s)
R^0_{n_2-1,1} (s) \cr
&\qquad{}= (1+s)^{- 2+\rho}  \sum_{\scr n_1+n_2=n\atop\scr      n_2 \geq 1}
 [A]_{3,n_1} (s)
\wp^D_{n_2-1}  ((1+\lambda^{}_0 (s))g(s)) \cr
&\qquad{}\leq C(1+s)^{- 3/2+\rho}
 \sum_{0 \leq l \leq n-1}  [A]_{3,n-l-1} (s)  \wp^D_l
((1+\lambda^{}_0 (s))^{1/2}  g(s)),\cr
}$$
according to the definition of $R^0_{p,q}$ in Corollary 5.9.
This proves the statement of the theorem for $0 \leq t_0 < \infty.$

Let $t_0 = \infty$. If $h^{}_1, h^{}_2 \in C^0(\Rrm^+,D)$ are
two solutions of equation (5.1a) such that $\Vert h^{}_i (t) \Vert^{}_D
\fl 0,  i = 1,2$, when $t \fl \infty,$
then due to the unitarity of $w(t,s)$ in $D$:
$$0 = \lim_{s \fl \infty}  \Vert h^{}_1 (s) - h^{}_2 (s)
\Vert^{}_D = \Vert h^{}_1
(t) - h^{}_2 (t) \Vert^{}_D,\quad  t \geq 0.$$
This proves the uniqueness of the solution $h$.
To prove the existence of $h$ we first prove the existence of
$f^{}_Y$ for $Y \in \Pi'$,  $\vert Y \vert
\leq n$,  $T \geq 0$. Introduce $f^T_Y,f^T_{(i)Y} \in C^0(\Rrm^+,D)$,
$i\in\{0,1,2\}$, by
$f^T_Y=f^T_{(0)Y}$.
$$\eqalignno{
f^T_{(i)Y} (t) &= 0 \qquad\hbox{for}\ 0 \leq T \leq t,&(5.163)\cr
  f^T_{(i)Y} (t) &= \int^T_t
 w(t,s)  i \gamma^0  g^{}_{(i)Y} (s)  ds \qquad\hbox{for}\
0 \leq t < T,\cr
}$$
where $ g^{}_{(i)Y}= g^{}_{iY}$ are given by conditions a) and b) of the
theorem for $i\in\{1,2\}$ and where $ g^{}_{(0)Y}= g^{}_{Y}$.
$f^T_Y$ has the properties:
$$f^{T_1}_Y (t) - f^{T_2}_Y (t) = w(t,T_2)  f^{T_1}_Y,\quad
0 \leq t \leq T_2 \leq T_1,$$
which together with the definition of $f^T_{(i)Y}$ give that
$$\sup_{s \geq 0}  \Vert f^{T_1}_{(i)Y} (s) - f^{T_2}_{(i)Y} (s)
\Vert^{}_D \leq
\sup_{T_2 \leq s \leq T_1}  \Vert f^{T_1}_{(i)Y} (s) \Vert^{}_D,
\quad i\in\{0,1,2\}, \eqno{(5.164)}$$
for $0 \leq T_2 \leq T_1$. In the case of condition  a) we obtain that
$$\sup_{s \geq 0}  \Vert f^{T_1}_{(1)Y} (s) - f^{T_2}_{(1)Y}
(s) \Vert^{}_D \leq
\int^{T_2}_{T_1}  \Vert g^{}_{(1)Y} (s) \Vert^{}_D  ds,\quad  0
\leq T_2 \leq T_1. \eqno{(5.165{\rm a})}$$
In the case of condition  b), statement iib) of Theorem 5.1 gives, since
$f^{T_1}_{(2)Y}$
is a solution of equation (5.1a), (with $g^{}_{(2)Y}$ instead of
$g$) in the interval
$[0,T_1]$ and which can be extended to a solution
$\overline{h} \in C^0(\Rrm^+,D)$,
$\overline{h} (t) = \int^{T_1}_t  w(t,s)  i \gamma^0
g^{}_{(2)Y} (s)  ds$, with $\overline{h} (T_1) = 0$:
$$\eqalignno{
&\sup_{s \geq 0}  \Vert f^{T_1}_{(2)Y} (s) - f^{T_2}_{(2)Y} (s)
\Vert^{}_D&(5.165{\rm b}) \cr
&\quad{}\leq
\sup_{T_2 \leq s \leq T_1}  (m^{-1}  \Vert g^{}_{(2)Y} (s) \Vert^{}_D)
+(2m)^{-1}  \int^{T_1}_{T_2}  \Vert ((m - i \gamma^\mu
 \partial_\mu + \gamma^\mu  G_\mu) g^{}_{(2)Y}) (s) \Vert^{}_D
ds,\cr
}$$
$0 \leq T_2 \leq T_1$. Inequalities (5.165a) and (5.165b), conditions
a) and  b), prove
that $f^T_Y$,  $T \in \Nrm$, is a Cauchy sequence in $L^\infty (\Rrm^+,D)$
converging to an element $f^{}_Y \in C^0(\Rrm^+,D)$ satisfying
$\Vert f^{}_Y (t) \Vert^{}_D
\fl 0$ when $t \fl \infty$, for $Y \in \Pi'$,  $\vert Y \vert \leq n$.
Let us define $h(t) = f^{}_{\un} (t)$.

We next construct a sequence of solutions $h^{}_k$, vanishing for sufficiently
large $t$, of equation (5.1a) converging to $h$ and to which we can apply
the estimate of the theorem for finite $t_0$. Let $\varphi \in C^\infty_0
(\Rrm),
\varphi (s) = 1$ for $\vert s \vert \leq 1$,  $\varphi (s) = 0$ for
$\vert s \vert \geq 2$ and let $\varphi^{}_k (t,x) = \varphi (k^{-1}
(1+t^2+\vert x \vert^2)^{1/2})$,
 $k \geq 1$. Then $\vert (\xi^{}_Y  \varphi^{}_k) (t,x) \vert \leq
C_{\vert Y \vert}
(1+t^2+\vert x \vert^2)^{\vert Y \vert / 2}  k^{- \vert Y \vert}$,
 $Y \in \Pi'$, where $C_{\vert Y \vert}$ is independent of $t$ and $x$.
If $(t,x)$ belongs to the support of $\xi^{}_Y  \varphi^{}_k$, then $k^{-1}
(1+t^2+\vert x \vert)^{1/2} \leq 2$, which shows that $\vert (\xi^{}_Y
\varphi) (t,x) \vert \leq C_{\vert Y \vert}  2^{\vert Y \vert}$.
Let $g^{(i)}_{kY}=
\suma_{Y_1,Y_2}^{Y}(\xi^{}_{Y_1}\varphi^{}_k)g^{}_{(i)Y_2}$
for $Y\in\Pi'$, $\vert Y\vert\leq n$, $i\in\{0,1,2\}$.
Then $\xi^D_Y(\varphi^{}_k g)
=g^{(0)}_{kY}=g^{(1)}_{kY}+g^{(2)}_{kY}$ and
$$\Vert g^{(i)}_{kY} (t) \Vert^{}_D \leq C_{\vert Y \vert}
 \sum_{\scr Z \in \Pi'\atop\scr\vert Z \vert \leq \vert Y \vert  }
\Vert g^{}_{(i)Z} (t) \Vert^{}_D,\quad  Y \in \Pi',  \vert Y \vert
\leq n,  t \geq 0,i\in\{0,1,2\}, \eqno{(5.166{\rm a})}$$
for some constant $C_{\vert Y \vert}$ independent of $t$. Similarly we obtain
that
$$\eqalignno{
&\Vert (1+\lambda^{}_j (t))^{1/2}  (\xi^D_Y (\varphi^{}_k  g)) (t)
\Vert^{}_D&(5.166{\rm b})\cr
&\quad{}\leq C_{\vert Y \vert}  \sum_{\scr Z \in \Pi'\atop\scr
\vert Z \vert \leq \vert Y \vert}
 \Vert (1+\lambda^{}_j (t))^{1/2}  (\xi^D_Z  g) (t) \Vert^{}_D,\quad
 Y \in \Pi',  \vert Y \vert \leq n,  t \geq 0,j=0,1,\cr
}$$
where $C_{\vert Y \vert}$ is independent of $t$. Moreover
$$\vert (\xi^{}_{P_\mu Y}  \varphi^{}_k) (t,x) \vert \leq C_{\vert Y \vert}
 (1+t^2+\vert x \vert^2)^{- 1/2},$$
where $C_{\vert Y \vert}$ is independent of $t$, gives that
$$\eqalignno{
&\Vert \big((m - i \gamma^\mu  \partial_\mu + \gamma^\mu  G_\mu)g^{(2)}_{kY}
 \big)(t) \Vert^{}_D&(5.166{\rm c})\cr
&\quad{}\leq C_{\vert Y \vert}  \sum_{\scr Z \in \Pi'
\atop\scr \vert Z \vert \leq \vert Y \vert }
 \Big((1+t)^{-1}  \Vert g^{}_{(2)Z}(t) \Vert^{}_D +
\Vert \big((m-i \gamma^\mu      \partial_\mu + \gamma^\mu  G_\mu)g^{}_{(2)Z}
\big) (t) \Vert^{}_D\Big),\cr
}$$
$t \geq 0,  Y \in \Pi',  \vert Y \vert \leq n$, where $C_{\vert Y \vert}$
is independent of $t$.

Let $g^{}_k = \varphi^{}_k  g$,  $n \geq 1$. Since $\varphi^{}_k (t,x) = 1$
for $(1+t^2+\vert x \vert^2)^{1/2} \leq k$ and since $\vert (\xi^{}_Y
\varphi^{}_k)  (t,x) \vert$ is uniformly bounded in $k,t,x$, it follows
from the dominated convergence theorem that $\Vert
(g^{(i)}_{kY}-g^{}_{(i)Y}) (t) \Vert^{}_D
\fl 0$ for $i\in\{0,1,2\}$,
$t \geq 0,  \Vert (1+\lambda^{}_0 (t))^{1/2} (g^{(0)}_{kY}-g^{}_{(0)Y})
(t) \Vert^{}_D \fl 0$, a.e. $t \geq 0$ and $\Vert \big((m - i \gamma^\mu
 \partial_\mu + \gamma^\mu
 G_\mu) (g^{(2)}_{kY}-g^{}_{(2)Y})\big)(t)\Vert^{}_D
\fl 0$, a.e. $t \geq 0$, for $Y \in \Pi'$,  $\vert Y \vert \leq n$. This
shows, together with the bound (5.166a) that
$$\lim_{k \fl \infty}  (\sup_{s \geq 0}  \Vert (g^{(i)}_{kY}-g^{}_{(i)Y})
(s) \Vert^{}_D) = 0,\quad i\in\{0,1,2\}, \eqno{(5.167{\rm a})}$$
together with the bound (5.166b) that
$$\lim_{k \fl \infty}  \int^\infty_0  (1+s)^{- 3/2+\rho}
 \Vert (1+\lambda^{}_0 (s))^{1/2}  (g^{(0)}_{kY}-g^{}_{(0)Y}) (s) \Vert^{}_D
 ds = 0, \eqno{(5.167{\rm b})}$$
together with the bound (5.166a) and the condition  a) of the theorem that
$$\lim_{k \fl \infty}  \int^\infty_0  \Vert (g^{(1)}_{kY}-g^{}_{(1)Y})
(s) \Vert^{}_D  ds = 0, \eqno{(5.167{\rm c})}$$
and together with the bounds (5.166a), (5.166c) and the condition  b) of the
theorem that
$$\lim_{k \fl \infty}  \int^\infty_0  \Vert \big((m - i \gamma^\mu
 \partial_\mu + \gamma^\mu  G_\mu) (g^{(2)}_{kY}-g^{}_{(2)Y})\big)
(s) \Vert^{}_D ds = 0, \eqno{(5.167{\rm d})}$$
and that
$$\lim_{k \fl \infty}  \lim_{s \fl \infty}
\Vert (g^{(2)}_{kY}-g^{}_{(2)Y}) (s) \Vert^{}_D = 0, \eqno{(5.167{\rm e})}$$
for $Y \in \Pi'$ and $\vert Y \vert \leq n$. Let $f^{}_{k,Y}
\in C^0 (\Rrm^+,D)$ be
the unique solution of equation (5.1a), with $  g^{(0)}_{kY}$ instead
of $g$ and with initial data $f^{}_{k,Y} (t_0) = 0,  k \geq 1$, where
$(1+t^2_0)^{1/2} \geq 2k$. Then $ g^{(0)}_{kY} (t) = 0$ for $t \geq t_0$,
so $f^{}_{k,Y} (t)
=0$ for $t \geq t_0$. It follows from statement iib) of Theorem 5.1
and from limits (5.167a), (5.167c), (5.167d) and (5.167e)
that $f^{}_{k,Y}$ is a Cauchy sequence for the uniform convergence
topology in $C^0(\Rrm^+,D).$
$f^{}_{k,Y}$ therefore converges to $f^{}_Y$ in this topology, for $Y \in \Pi',
\vert Y \vert \leq n$. In particular $f^{}_{k, \un}$ converges to
$f^{}_{\un} = h.$

Let us define $h^{}_k = f^{}_{k, \un},  k \geq 1$. The support of the
function $f^{}_{k, \un} \colon \Rrm^+ \fl D$ is a subset of $[0,2k[,
k \geq 1.$
Taking $t_0$, sufficiently large, it now follows from the inequality
of the theorem for finite $t_0$, that
$$\eqalignno{
&\wp^D_n (h^{}_{k_1} (t) - h^{}_{k_2} (t))&(5.168)\cr
&\qquad{}\leq C_n \Big(\sup_{s \geq t}  \wp^D_n
(f^{}_{k_1} (s) - f^{}_{k_2} (s))
+ \sum_{0 \leq l \leq n-1}  [A]_{3,n-l} (\infty)  \Big(\sup_{s \geq t}
 \wp^D_l (f^{}_{k_1} (s) - f^{}_{k_2} (s))\cr
&\qquad\qquad{}+ \int^\infty_t  (1+s)^{- 3/2 + \rho}
\wp^D_l \big((1+\lambda^{}_0
(s))^{1/2}  (g^{}_{k_1} (s) - g^{}_{k_2} (s))\big)      ds\Big)\Big),\cr
}$$
where $C_n$ depends only on $[A]^3 (\infty)$. Since $f^{}_{k,Y}$ is a
Cauchy sequence in $C^0 (\Rrm^+,D)$ for the uniform convergence topology,
it follows from the limit (5.167b) and inequality (5.168) that
$\xi^D_Y  h^{}_k \in C^0(\Rrm^+,D)$,
is a Cauchy sequence, in this topology, of functions with compact support,
$\vert Y \vert \leq n,  Y \in \Pi'$. This proves that $\xi^D_Y
h \in C^0(\Rrm^+,D)$
and $\Vert \xi^D_Y  h(t) \Vert^{}_D \fl 0$, when $t \fl \infty$, for $Y \in
\Pi',  \vert Y \vert \leq n$. Finally, it follows from inequality (5.168)
that $\wp^D_n (h(t)) \leq C_n  Q^\infty_n (t)$, where $C_n$ depends only
on $[A]^3 (\infty)$. This proves the theorem.

In Theorem 5.13 we loose several orders in the seminorms
of the electromagnetic potential. We shall derive a result,
based on Theorem 5.8, Lemma 5.11, Lemma 5.12
and Theorem 5.13, where no seminorms are lost. We introduce the notation
$$\eqalignno{
\overline{H}_n (t_0,t) &= \sum_{n_1+n_2=n}   (1+S^\rho_{10,n_1} (t))
\Big(R'_{n_2+7} (t) + R^2_{n_2+9} (t)&(5.169)\cr
&\qquad{}+ R^\infty_{n_2} (t) + \wp^D_{n_2+8} (h(t_0)) +
\sum_{0 \leq l \leq n_2+7}
[A]_{3,n_2+8-l} (t'')  \wp^D_l (h(t_0)) + Q_{n_2+8} (t)\Big),\cr
}$$
where $n \geq 0,  t_0,  t \geq 0,  t'' = \max (t,t_0)$
and where $R'_n$, $R^2_n$, $R^\infty_n$ are given in Theorem 5.8 and
$Q_n$ is given in Theorem 5.13. It follows from Theorem 5.8, definition
(5.88a) of $T^{\infty (n)},$
inequality (5.116c), $x_\mu  G^\mu (x) = 0$ for $G_\mu$ given by (5.114)
and inequality (5.138) that
$$H_n (t) \leq a^{}_n  \overline{H}_n (t_0,t),\quad  n \geq 0,
t_0,t \geq 0, \eqno{(5.170)}$$
where $a^{}_n$ depends only on $[A]^{11} (\max (t,t_0))$. Let
$$\overline{\wp}^D_n (t_0,t) = \wp^D_n (h(t_0)) + Q_n (t)
+ \sum_{0 \leq l \leq n-1}  [A]_{3,n-l} (t'')  \wp^D_l (h(t_0)),
\eqno{(5.171)}$$
where $t_0, t \geq 0,  n \geq 0,  t'' = \max (t_0,t)$ and where
$Q_n$ is given in Theorem 5.13.
In the situation of Theorem 5.13 it follows that
$$\wp^D_n (h(t)) \leq C_n  \overline{\wp}^D_n (t_0,t),\quad  n \geq 0,
t_0,t \geq 0. \eqno{(5.172)}$$
We introduce, for $n \geq 0,  t_0,t \geq 0,  t' = \min (t_0,t),
 t'' = \max (t_0,t),  L \geq 0$, the notation
$$\eqalignno{
&k'_n (L,t_0,t) &(5.173{\rm a})\cr
&= (1 + t_0)^{- 3/2 + \rho}  \sum_{\scr n_1+n_2=n\atop\scr
1 \leq n_1 \leq L}  [A]^{n_1+1} (t_0)  \wp^D_{n_2} (h(t_0))\cr
&\quad{}+ (1+t_0)^{-1/2}\!\!  \sum_{\scr n_1+n_2=n\atop\scr n_1 \geq L+1}\!\!
S^{\rho, n_1} (t_0)  \overline{H}_{n_2}  (t_0,t_0)
+ (1+t)^{- 3/2 + \rho}\!\!  \sum_{\sscr n_1+n_2=n\atop{\sscr
1 \leq n_1 \leq L\atop\sscr
 n_2 \leq L}}  [A]^{n_1+1} (t)  \overline{\wp}^D_{n_2}
(t_0,t)\cr
&\quad{}+ (1+t)^{-1/2}  \sum_{\scr n_1+n_2=n\atop\scr L+1 \leq n_1 \leq n-1}
 S^{\rho, n_1} (t)  \overline{H}_{n_2} (t_0,t)
+ \int^{t''}_{t'}  (1+s)^{- 2 + \rho}  \overline{k}_n
(L,t_0,s)  ds,\cr
}$$
where
$$\eqalignno{
&\overline{k}_n  (L,t_0,t) &(5.173{\rm b})\cr
&= \sum_{\sscr n_1+n_2=n\atop{\sscr   1 \leq n_1 \leq L\atop\sscr
 n_2 \leq L}}  \Big([A]_{3,n_1} (t)\big(\overline{\wp}^D_{n_2}
(t_0,t) + R^0_{n_2-1,1} (t)\big)
+ (1+t)^{1/2}  [A]^{n_1+1} (t)  \wp^D_{n_2}  (g(t))\Big)\cr
&\quad{}+ \sum_{\sscr n_1+n_2+n_3+n_4=n\atop{\sscr   n_1 \leq L-1, n_2 \leq n-1
\atop\sscr  n_3+
n_4 \leq n-L}}  (1+[A]_{2,n_1} (t))  S^{\rho,n_2} (t)
(1+S^\rho_{10,n_3} (t))\cr
&\hskip50mm\big(R'_{n_4+7} (t) + R^2_{n_4+9} (t)
+ R^\infty_{n_4} (t) + \overline{\wp}_{n_4+8} (t_0,t)\big)\cr
&\quad{}+ \sum_{\sscr Y_1,Y_2 \in \Pi'
\atop{\sscr \vert Y_1\vert +\vert Y_2\vert=n\atop
\sscr L\leq\vert Y_1\vert=n-1 }}
(1+t)^{2 - \rho} \Vert \gamma^\mu G_{Y_1\mu}(t) g^{}_{Y_2}(t)\Vert^{}_{D}.\cr
}$$
Here $R^0_{n_2-1,1}$ is given in Corollary 5.9.
\saut
\noindent{\bf Theorem 5.14.}
{\it
Let $1/2 < \rho < 1,  18 \leq L+9 \leq n+8 \leq 2L$, $(1-\Delta)^{1/2}
 (A_X, A_{P_0 X}) \in C^0(\Rrm^+, M^1)$ for $X = \un$ or  $X \in {\frak{sl}}(2,
\Crm)$,
let $(1-\Delta)^{1/2}  (B, \dot{B}) \in C^0(\Rrm^+,M^1)$, where $B_\mu (y) =
y_\mu  \partial^\nu  A_\nu$, $\dot{B} = {d \over dt}  B$,
let $(A_Y, A_{P_0 Y}) \in C^0(\Rrm^+, M^\rho) \cap C^0(\Rrm^+, M^1)$ for
$Y \in \Pi'$,
 $\vert Y \vert \leq n$, let $\carre  A_Y \in C^0(\Rrm^+, L^2 (\Rrm^3,
\Rrm^4))$ for $Y \in \Pi'$,  $\vert Y \vert \leq n$, let $(B_Y, B_{P_0 Y})
\in C^0(\Rrm^+,M^1)$ for $Y \in \Pi'$,  $\vert Y \vert \leq n-1$, let
$\delta^{3/2-\rho}
 A_Y \in C^0(\Rrm^+, L^\infty (\Rrm^3, \Rrm^4))$ for $Y \in \Pi'$,
$\vert Y \vert \leq L+3$, let the funtion $(t,x) \mapsto (1+ \vert x \vert + t)
 (1+ \big\vert t - \vert x \vert \big\vert)^{1/2}  A_Y (t,x)$ be an
element of $C^0(\Rrm^+, L^\infty (\Rrm^3, \Rrm^4))$ for $Y \in \sg^1$,  $\vert
Y \vert \leq L+3$, let $\delta^{3/2 - \rho}  B_Y \in C^0(\Rrm^+, L^\infty
(\Rrm^3, \Rrm^4))$ for $Y \in \Pi'$,  $\vert Y \vert \leq L+2$, let the
function $(t,x) \mapsto (1+t+\vert x \vert)  (1+\big\vert t -
\vert x \vert \big\vert)^{1/2}
 B_Y (t,x)$ be an element of $C^0(\Rrm^+, L^\infty (\Rrm^3, \Rrm^4))$ for
 $Y \in \sg^1$,
 $\vert Y \vert \leq L+2$, let $g^{}_Y \in C^0(\Rrm^+,D)$ for $Y \in \Pi'$,
 $\vert Y \vert \leq n$, let $R'_{L-1} (s),  R^2_{L+1} (s)$,
$R^\infty_{L-8} (s)$ be finite for $t' \leq s \leq t''$, where
$t' = \min (t,t_0)$
and $t'' = \max (t,t_0)$, let $G_\mu$ be given by (5.114), let $f$ be given by
(5.111b) and let
$$\eqalignno{
&k_n (L,t_0,t) \cr
&\quad{}= k'_n (L, t_0,t) + (1+t)^{- 1/2}  S^{\rho, n} (t)
\overline{H}_0 (t_0,t)
+ \int^{t''}_{t'}  (1+s)^{- 2+\rho}  \Big([A]^3 (s)
R^0_{n-1,1} (s)\cr
&\qquad{}+ S^{\rho,n} (s)  (1+[A]^1 (s))
\overline{H}_1 (t_0,s) + (1+s)^{2-\rho}
\Big( \sum_{\vert Y\vert=n}
\Vert\gamma^\mu G_{Y\mu}(s) g(s) \Vert^{2}_{D}\Big)^{1/2}\Big)  ds.\cr
}$$
If $h^{}_Y (t_0) \in D$ for $Y \in \Pi',  \vert Y \vert \leq n$ then $h$
given by (5.3c) is the unique solution of equation (5.1a) in $C^0(\Rrm^+,D)$,
with initial data $h(t_0)$. This solution satisfies
$$h^{}_Y \in C^0(\Rrm^+,D),\quad  h^{}_{P_\mu Y} \in C^0(\Rrm^+,
(1-\Delta)^{1/2} D)
\quad \hbox{ for}\ Y \in \Pi',  \vert Y \vert \leq n,$$
and
$$\eqalignno{
&\wp^D_n (h(t)) \leq C_n \Big(\wp^D_n (h(t_0)) + \sup_{t' \leq s \leq t''}
\big(\wp^D_n (f(s)) + a^{}_n  k_n (L,t_0,s)\big)\cr
&\quad{}+ \sum_{\sscr n_1+n_2=n\atop{\sscr   1 \leq n_1 \leq L\atop\sscr
n_2 \geq L+1}}
[A]_{3,n_1} (t'')  \Big(\wp^D_{n_2} (h(t_0)) + \sup_{t' \leq s \leq t''}
\big(\wp^D_{n_2} (f(s)) + a^{}_{n_2}  k_{n_2} (L,t_0,s)\big)\Big)\Big),\cr
}$$
where $C_n$ depends only on $\rho$ and $[A]^3 (t'')$ and $a^{}_l$,
$0 \leq l \leq n$,
depends only on $\rho$ and $[A]^{11} (t'')$.

Moreover if $k^\infty_n (L,t)$ is given
by $k_n (L,t_0,t)$ with $\wp^D_n (h(t_0)) = 0$
and $t_0 = \infty$, if $k^\infty_n
(L) = \sup_{t \geq 0} k^\infty_n (L,t) < \infty$ and
if for each
$Y \in \Pi'$,  $\vert Y \vert \leq n$, there exists $g^{}_{1Y}$ and
$g^{}_{2Y}$ such that
$g^{}_{Y}=g^{}_{1Y}+g^{}_{2Y}$, and such that:\psaut

\hbox{{\rm a)} $g^{}_{1Y} \in L^1 (\Rrm^+,D)$,}\psaut

\hbox{{\rm b)} $(m - i \gamma^\mu  \partial_\mu + \gamma^\mu  G_\mu)
 g^{}_{2Y} \in L^1 (\Rrm^+,D)$ and $\displaystyle{\lim_{t \fl \infty}}
\Vert g^{}_{2Y}(t) \Vert^{}_D = 0$,}\psaut

\noindent
then there exists a unique solution
$h \in C^0(\Rrm^+,D)$ of equation (5.1a) such
that $\Vert h(t) \Vert^{}_D \fl 0$, when $t \fl \infty$.
This solution satisfies
$$\eqalignno{
\wp^D_n (h(t)) &\leq C_n \Big(\sup_{t \leq s}  \big(\wp^D_n (f(s)) + a^{}_n
k^\infty_n (L,s)\big)\cr
&\qquad{}+ \sum_{\sscr n_1+n_2=n\atop{\sscr
1 \leq n_1 \leq L\atop\sscr   n_2 \geq L+1} }
[A]_{3, n_1} (\infty)  \sup_{t \leq s}  \big(\wp^D_{n_2} (f(s)) +
a^{}_{n_2}  k^\infty_{n_2} (L,s)\big)\Big),\cr
}$$
where the constants $C_n$ and $a^{}_l,  0 \leq l \leq n$ are as above.
Further, $f^{}_Y \in C^0(\Rrm^+,D)$, $f^{}_Y (t) \fl 0$ as $t \fl \infty$ and
$f^{}_Y (t)$ is the limit of $\int^T_t  w(t,s)  i \gamma^0
g^{}_Y (s)  ds$ in $D$ when $T \fl \infty$.
}\saut
\noindent{\it Proof.}
First let $0 \leq t_0 < \infty$. According to the hypothesis, it follows from
Theorem 5.13 that equation (5.1a) has a unique solution
$h \in C^0 (\Rrm^+,D)$, that
$$h^{}_Y \in C^0(\Rrm^+,D),\quad  (1-\Delta)^{- 1/2}  h^{}_{P_\mu Y} \in
C^0(\Rrm^+,D) \quad\hbox{ for}\ Y \in \Pi',  \vert Y \vert \leq L,
0 \leq \mu \leq 3,$$
and that
$$\wp^D_j (h(t)) \leq C_j (\wp^D_j (h(t_0)) + \sum_{0 \leq l \leq j-1}
[A]_{3,j-l} (t'')  \wp^D_l (h(t_0)) + Q_n (t)),\quad  j \leq L,$$
where $t'' = \max (t,t_0)$ and where $C_j$ depends only on $[A]^3 (t'').$

Like in the beginning of the proof of Theorem 5.13, let
$$g'_Y = \suma_{\scr Y_1,Y_2\atop\scr
\vert Y_2 \vert \leq \vert Y \vert - 1}^Y
 \gamma^\mu  G_{Y_1 \mu}  h^{}_{Y_2} + g^{}_Y.$$
As in the proof of Theorem 5.13 it follows that
$G_Y \in C^0(\Rrm^+, L^\infty
(\Rrm^3, \Rrm^4))$ for $Y \in \Pi',  \vert Y \vert \leq L$. It follows
from the hypothesis of the theorem, from definition (5.115d) of $S^{\rho,n}$
and from inequality (5.125b) that $\delta^{-1}  G_Y \in C^0(\Rrm^+, L^2
(\Rrm^3, \Rrm^4))$ for $Y \in \Pi',  \vert Y \vert \leq n$. Moreover
since $\Vert \delta (t)  h^{}_Z (t) \Vert^{}_{L^\infty} \leq H_{\vert Z \vert}
(t) \leq a^{}_{\vert Z \vert}  \overline{H}_{\vert Z \vert} (t_0,t)$, where
$a^{}_{\vert Z \vert}$ depends only on $[A]^{11} (\max (t,t_0))$, according to
(5.170), it follows from the hypothesis that
$\delta  h^{}_Z \in C^0(\Rrm^+, L^\infty
(\Rrm^3, \Crm^4))$ for $Z \in \Pi',  \vert Z \vert \leq n - L \leq L-8$.
Since $g^{}_Y \in C^0(\Rrm^+,D)$ for $\vert Y \vert \leq n$,  $Y \in \Pi'$, it
follows that $g'_Y \in C^0(\Rrm^+,D)$ for $Y \in \Pi'$,
$\vert Y \vert \leq n$,
which together with $h^{}_Y (t_0) \in D$ for $Y \in \Pi',  \vert Y \vert
\leq n$, give that $h^{}_Y \in C^0(\Rrm^+,D)$ and $h^{}_{P_\mu Y}
\in C^0(\Rrm^+, (1-\Delta)^{1/2}
D)$ for $Y \in \Pi'$,  $\vert Y \vert \leq n$.

Statement a) of Lemma 5.12 and statement b), but with $L-1$ instead of $L$,
give, for $0 \leq i \leq n$,
$$\eqalignno{
&\wp^D_{n,i} (h(t))&(5.174)\cr
& \leq \wp^D_{n,i}  (h(t_0)) + \wp^D_{n,i} (f(t))
+ (1+t_0)^{- 3/2 + \rho}  C_n  \sum_{\scr n_1+n_2=n\atop\scr
1 \leq n_1 \leq L}  [A]^{n_1+1} (t_0)  \wp^D_{n_2} (h(t_0))\cr
&\quad{}+ (1+t_0)^{- 1/2}  C_n \!\!\! \sum_{\scr n_1+n_2=n\atop\scr
n_1 \geq L+1}\!\!\!  S^{\rho, n_1} (t_0)  H_{n_2} (t_0)
+ (1+t)^{- 3/2 + \rho}  C_n \!\!\! \sum_{\scr n_1+n_2=n\atop\scr
1 \leq n_1 \leq L} \!\!\! [A]^{n_1+1} (t)  \wp^D_{n_2} (h(t))\cr
&\quad{}+ (1+t)^{- 1/2}  C_n  \sum_{\scr n_1+n_2=n\atop\scr
L+1 \leq n_1 \leq n-1}  S^{\rho, n_1} (t)  H_{n_2} (t)
+ (1+t)^{- 1/2}  \chi^{}_+ (-i)  C_0    S^{\rho, n}
(t)  H_0 (t)\cr
&\quad{}+ C'_n  \int^{t''}_{t'}  (1+s)^{- 2+\rho}
\Big(\sum_{l \geq \max (i,1)}
 J^{(l)}_n (L-1,s) + \chi^{}_+ (-i)  J^{(0)}_n (L,s)\Big) ds\cr
&\quad{}+ \chi^{}_+ (-i)  C_0  \int^{t''}_{t'}  (1+s)^{- 2+\rho}
 \Big(S^{\rho,n} (s)  (1+[A]^1 (s))  H_1 (s)\cr
&\quad{}+ (1+s)^{2-\rho} \Big(\sum_{\vert Y \vert= n}
\Vert \gamma^\mu G_{Y\mu} (s)  g(s) \Vert^{2}_{D}\Big)^{1/2}\Big)
 ds,\cr
}$$
where $C_n$ depends only on $\rho$ and $C'_n$ depends on
$\rho$ and $[A]^3 (t'').$
It follows from definitions (5.119a)--(5.119d) of
$k^{(1)}_n, l^{(1)}_n, k^{(2)}_n$
and $l^{(2)}_n$ and from definition (5.173b) of $\overline{k}_n$, that
$$\eqalignno{
&k^{(1)}_n (L,t) + l^{(1)}_n (L,t) +
k^{(2)}_n (L,t) + l^{(2)}_n (L,t)&(5.175)\cr
&\quad{}+ k^{(1)}_n (L-1,t) + l^{(1)}_n (L-1,t) + k^{(2)}_n
(L-1,t) + l^{(2)}_n (L-1,t)\cr
&\qquad{}\leq 2 \Big(k^{(1)}_n (L,t) + l^{(1)}_n (L-1,t)
+ k^{(2)}_n (L,t) + l^{(2)}_n (L-1,t)\Big)\cr
&\qquad{}\leq C  \sum_{\sscr n_1+n_2=n\atop{\sscr   1 \leq n_1
\leq L\atop\sscr n_2
\geq L+1}}  [A]_{3,n_1} (t)  \wp^D_{n_2} (h(t)) + a^{}_n
\overline{k}_n (L,t_0,t),\cr
}$$
where $C$ is a numerical constant and $a^{}_n$ depends only on
$[A]^{11} (t'')$. The definition of $J^{(l)}_n$ in Lemma 5.11,
inequalities (5.170) and (5.172), definition
(5.173a) of $k'_n (L,t_0,t)$ and inequalities (5.174) and (5.175) give
$$\eqalignno{
&\wp^D_n (h(t))&(5.176)\cr
&\quad{}\leq \wp^D_n (h(t_0)) + \wp^D_n (f(t)) + a^{}_n  k'_n (L,t_0,t)
+ (1+t)^{- 1/2}  a^{}_0  S^{\rho, n_1} (t)  \overline{H}_0
(t_0,t)\cr
&\qquad{}+ (1+t)^{- 3/2 + \rho}  C_n  \sum_{\sscr n_1+n_2=n\atop{\sscr
1 \leq n_1 \leq L\atop\sscr  n_2 \geq L+1}}  [A]^{n_1+1} (t)
\wp^D_{n_2} (h(t))\cr
&\qquad{}+ C_0  \int^{t''}_{t'}  (1+s)^{- 2+\rho}  \Big(S^{\rho,n}
(s)  (1+[A]^1 (s))  a^{}_1  \overline{H}_1 (t_0,s) \cr
&\qquad{}+ (1+s)^{2-\rho}\Big(\sum_{\vert Y\vert=n}
 \Vert \gamma^\mu G_{Y\mu} (s)   g(s) \Vert^{2}_{D}\Big)^{1/2}\Big)  ds\cr
&\qquad{}+ C_n  \int^{t''}_{t'}  (1+s)^{- 2+\rho}
\Big(\sum_{\sscr n_1+n_2=n\atop{\sscr
 1 \leq n_1 \leq L\atop\sscr  n_2 \geq L+1}}  [A]_{3,n_1} (s)
 \wp^D_{n_2} (h(s))\cr
&\qquad{}+ C'_n  [A]^3 (s)  \big(\wp^D_n (h(s))^\varepsilon
\wp^D_{n,1}  (h(s))^{1 - \varepsilon} + R^0_{n-1,1} (s)^{2 (1-\rho)}
 \wp^D_{n_1} (h(s))^{2 \rho - 1}\big)\Big)  ds,\cr
}$$
and for $1 \leq i \leq n$, that
$$\eqalignno{
&\wp^D_{n,i} (h(t))&(5.177)\cr
&\quad{}\leq \wp^D_{n,i} (h(t_0)) + \wp^D_{n,i} (f(t)) + a^{}_n
k'_n (L,t_0,t)\cr
&\qquad{}+ (1+t)^{- 3/2 + \rho}  C_n  \sum_{\sscr n_1+n_2=n\atop{\sscr
1 \leq n_1 \leq L\atop\sscr  n_2 \geq L+1}}  [A]^{n_1+1} (t)
\wp^D_{n_2} (h(t))\cr
&\qquad{}+ C'_n  \int^{t''}_{t'}  (1+s)^{- 2+\rho}  \Big(\sum_{\sscr n_1+
n_2=n \atop{\sscr 1 \leq n_1 \leq L\atop\sscr  n_2 \geq L+1}}  [A]_{3,n_1}
(s)  \wp^D_{n_2} (h(s))\cr
&\qquad{}+ [A]^3 (s)  \big(\wp^D_n (h(s))^\varepsilon
\wp^D_{n,i+1} (h(s))^{1-\varepsilon}
+ R^0_{n-1,1} (s)^{2(1-\rho)}  \wp^D_{n_1} (h(s))^{2 \rho-1}\big)\Big)
ds,\cr
}$$
where $18 \leq L+9 \leq n+8 \leq 2L,  \varepsilon = \max (1/2,2(1-\rho)),
 1/2 < \rho < 1$, $t' = \min (t,t_0),  t'' = \max (t,t_0)$ and
where $C_n$ depends only on $\rho$, $C'$ only on $\rho$ and $[A]^3 (t'')$
and $a^{}_n$
only on $\rho$ and $[A]^{11} (t'')$. Let
$$\eqalignno{
k_n (L,t_0,t) &= k'_n (L,t_0,t) + (1+t)^{- 1/2}  S^{\rho,n} (t)
 \overline{H}_0 (t_0,t)&(5.178)\cr
&\qquad{}+ \int^{t''}_{t'}  (1+s)^{- 2+\rho}  \Big(S^{\rho,n} (s)
(1+[A]^1 (s))  \overline{H}_1 (t_0,s)\cr
&\qquad{}+ (1+s)^{2-\rho}  \Big(\sum_{\vert Y\vert=n}
 \Vert \gamma^\mu G_{Y\mu} (s) g(s) \Vert^{2}_{D}\Big)^{1/2}
+ [A]^3 (s)  R^0_{n_1,1} (s)\Big)  ds,\cr
}$$
for $n \geq 0,  L \geq 0,  t,t_0 \geq 0,  t' = \min
(t,t_0),  t'' = \max (t,t_0)$. It follows from inequalities (5.176)
and (5.177) that
$$\eqalignno{
\wp^D_{n,i} (h(t)) &\leq \wp^D_{n,i} (h(t_0)) + \wp^D_{n,i} (f(t)) + a^{}_n
k_n (L,t_0,t)&(5.179)\cr
&\qquad{}+ C_n (1+t)^{- 3/2 + \rho}  \sum_{\sscr n_1+n_2=n\atop{\sscr 1 \leq
n_1
\leq L\atop\sscr  n_2 \geq L+1}}[A]^{n_1+1} (t)  \wp^D_{n_2}
(h(t))\cr
&\qquad{}+ C'_n  \int^{t''}_{t'}  (1+s)^{- 2+\rho}  \Big([A]^3
(s)  \wp^D_n (h(s))^\varepsilon  \wp^D_{n,i+1} (h(s))^{1-\varepsilon}\cr
&\qquad{}+ \sum_{\sscr n_1+n_2=n\atop{\sscr
1 \leq n_1 \leq L\atop\sscr  n_2 \geq L+1}}
[A]_{3,n_1} (s)  \wp^D_{n_2} (h(s))\Big)  ds,\cr
}$$
where $0 \leq i \leq n,  18 \leq L+9 \leq n+8 \leq 2L,
\varepsilon = \max (1/2, 2(1-\rho))$, $1/2 < \rho < 1$
and where $C_n$ depends only on $\rho$, $C'$ only on $\rho$ and
$[A]^3 (t'')$ and $a^{}_n$ only on $\rho$ and $[A]^{11} (t'').$

To solve the system of inequalities (5.179) in the variables
$\wp^D_{n,i} (h(t)),$
we introduce the variables
$$\eqalignno{
\xi^{}_{l,j} &= \sup_{t' \leq s \leq t''}  \wp^D_{l,j} (h(s)),
\quad \hbox{for}\ 0 \leq j \leq l \leq n, & (5.180{\rm a})\cr
\xi^{}_{l,j} &= 0\quad  \hbox{for}\ j \geq l+1,\cr
}$$
where $t' = \min (t,t_0),  t'' = \max (t,t_0)$,
and we introduce the positive real numbers
$$\eta^{}_l = \wp^D_l (h(t_0)) + \sup_{t' \leq s \leq t''} \big(\wp^D_l
(f(s)) + a^{}_l k_l (L,t_0,s)\big),\quad 0 \leq l \leq n. \eqno{(5.180{\rm
b})}$$
It then follows from (5.179) that
$$\xi^{}_{n,i} \leq \eta^{}_n + b^{}_n  \sum_{\sscr n_1+n_2=n\atop{\sscr
1 \leq n_1
\leq L\atop\sscr  n_2 \geq L+1}}[A]_{3,n_1} (t'')  \xi^{}_{n_2,0}
+ b^{}_n  [A]^3 (t'')  \xi^\varepsilon_{n,0}
\xi^{1-\varepsilon}_{n,i+1},\eqno{(5.181)}$$
where $0 \leq i \leq n,  L+9 \leq n+8 \leq 2L$ and where $b^{}_n$ is a
constant depending only on $\rho$ and on $[A]^3 (t'')$. Let
$$\eqalignno{
\alpha^{}_n &= \eta^{}_n + b^{}_n  \sum_{\sscr n_1+n_2=n\atop{\sscr
1 \leq n_1 \leq
L\atop\sscr  n_2 \geq L+1}}  [A]_{3,n_1} (t'')  \xi^{}_{n_2,0},
&(5.182{\rm a})\cr
\noalign{\hbox{\rm and}}
\beta^{}_n &= b^{}_n  [A]^3 (t'').&(5.182{\rm b})\cr
}$$
Then inequalities (5.181) give that
$$\xi^{}_{n,i} \leq \alpha^{}_n + \beta^{}_n  \xi^\varepsilon_{n,0}
\xi^{1-\varepsilon}_{n,i+1}, \eqno{(5.182{\rm c})}$$
where $0 \leq i \leq n,  L+9 \leq n+8 \leq 2L$.
We note that $\alpha^{}_{L+1} = \eta^{}_{L+1}$ because of (5.182a).
The solutions $x \geq 0$ of the inequality
$$x \leq a + b  x^\varepsilon  y^{1-\varepsilon}, \eqno{(5.183{\rm a})}$$
where $y \geq 0,  a \geq 0,  b \geq 0,  0 \leq \varepsilon
< 1$ satisfy
$$x \leq 2a + 2b (1+2b)^{k-1}  y,\quad  k \in \Nrm,
k \geq 1 / (1 - \varepsilon) \geq 1. \eqno{(5.183{\rm b})}$$
As a matter of fact for $0 < \varepsilon < 1$:
\psaut
\item{\hbox{ i)}} if $\varepsilon b \leq 1/2$ then
$x \leq a + b  x^\varepsilon  y^{1-\varepsilon} \leq a + \varepsilon
 bx + (1-\varepsilon)  by$ gives that $x \leq 2a + 2(1-\varepsilon)
 by$,
\psaut
\item{\hbox{ ii)}} if $\varepsilon b > 1/2$ and $(2b)^{1/(1-\varepsilon)}
y \leq x$ then $b  x^\varepsilon  y^{1-\varepsilon} \leq x/2$
and inequality (5.183a) give that $x \leq 2a$,
\psaut
\item{\hbox{ iii)}} if $\varepsilon b > 1/2$ and
$(2b)^{1/(1-\varepsilon)}  y > x$ then $x < (2b)^k  y$ since
$2b > 1/\varepsilon > 1$.
\psaut
Applying the result (5.183b) to the solutions $\xi^{}_{n,0}$
of inequality (5.182c), with $i=0$, it follows that
$\xi^{}_{n,0} \leq 2 \alpha^{}_n + 2 \beta^{}_n  (1+2 \beta^{}_n)^{k-1}
\xi^{}_{n,1}$. Majorizing $\xi^{}_{n,1}$
by the right-hand side of inequality (5.182c), with $i=1$, we obtain that
$$\xi^{}_{n,0} \leq 2 \alpha^{}_n + 2 \beta^{}_n  (1+2 \beta^{}_n)^{k-1}
\alpha^{}_n + 2 \beta^{}_n  (1+2 \beta^{}_n)^{k-1} \beta^{}_n
\xi^\varepsilon_{n,0}  \xi^{1-\varepsilon}_{n,2},$$
to which we apply inequality (5.183b). Continuing this iteration, we obtain
after a finite number of steps, since
$\xi^{}_{n,n+1} = 0$, that $\xi^{}_{n,0} \leq
\gamma^{}_n  \alpha^{}_n$, where $\gamma^{}_n$ is a polynomial in $\beta^{}_n.$
Since $\alpha^{}_{L+1} = \eta^{}_{L+1}$, we obtain, using
$\xi^{}_{l,0} \leq \gamma^{}_l
 \alpha^{}_l$ for $L+1 \leq l \leq n$ and using expression (5.182a) of
$\alpha^{}_n$ and (5.182b) of $\beta^{}_n$, that
$$\alpha^{}_l \leq \eta^{}_l + b^{}_l  \sum_{\sscr n_1+n_2=l\atop{\sscr
1 \leq n_1 \leq L\atop\sscr
 n_2 \geq L+1}} [A]_{3,n_1} (t'')  \alpha^{}_{n_2},
 \alpha^{}_{L+1} = \eta^{}_{L+1},$$
where $L+10 \leq l+8 \leq 2L$ and where $b^{}_n$ depends only on $\rho$ and
$[A]^3 (t'')$. Iteration of this inequality for $L+2 \leq l \leq n$,
the convexity property
$[A]_{3,n_1} (t'')  [A]_{3,n_2} (t'') \leq C_{n_1+n_2}  [A]_{3,
n_1+n_2} (t'')$, where $C_{n_1+n_2}$  depends only on
$[A]_{3,0} (t'')$ and the fact that $n-L-1 \leq L-9 \leq L$ give that
$$\alpha^{}_n \leq \eta^{}_n + b^{}_n  \sum_{\sscr n_1+n_2=n\atop{\sscr
  1 \leq n_1 \leq L\atop\sscr
 n_2 \geq L+1}} [A]_{3,n_1} (t'')  \eta^{}_{n_2}, \quad
L+9 \leq n+8 \leq 2L,$$
for some constants $b^{}_n$ depending only on $\rho$ and $[A]^3 (t'')$.
The last inequality and the fact that $\xi^{}_{n,0} \leq
\gamma^{}_n  \alpha^{}_n$ prove the inequality of the theorem, when
$t_0 < \infty$. The proof of the case $t_0 = \infty$, is done by the same
limit procedure as in the proof of the case $t_0 = \infty$ of Theorem 5.13.
We omit the details since they are so similar to
those of that proof. This proves the theorem.
\saut
\noindent{\bf Remark 5.15.}
For the case of $t_0 = \infty$ in Theorem 5.14 expressions (5.173a) of $k'_n$
and (5.173b) of $\overline{k}_n$ are simplified. Let
$$\overline{H}^\infty_n (t) = \sum_{n_1+n_2=n}  (1+S^\rho_{10,n_1} (t))
 \big(R'_{n_2+7} (t) + R^2_{n_2+9} (t)
+ R^\infty_{n_2} (t) + Q^\infty_{n_2+8} (t)\big),\eqno{(5.184)}$$
where $Q^\infty_l (t)$ is given in Theorem 5.13 and let
$$\eqalignno{
k'^\infty_n (L,t) &= (1+t)^{- 3/2+\rho}  \sum_{\sscr n_1+n_2=n\atop{\sscr
1 \leq n_1 \leq L\atop\sscr   n_2 \leq L}}  [A]^{n_1+1} (t)
Q^\infty_{n_2} (t)&(5.185)\cr
&\qquad{}+ (1+t)^{- 1/2}  \sum_{\scr n_1+n_2=n\atop\scr  L+1 \leq n_1 \leq n-1}
 S^{\rho,n_1} (t)  \overline{H}^\infty_{n_2} (t) + \int^\infty_t
 (1+s)^{- 2+\rho}  \overline{k}^\infty_n (L,s)  ds,\cr
}$$
where $\overline{k}^\infty_n (L,t)$ is given by expression (5.173b) of
$\overline{k}_n (L,t_0,t)$, with $\overline{\wp}^D_l (t_0,t)$ replaced by
$Q^\infty_l (t)$ and with $t_0 = \infty$. Inequalities (5.176) and (5.177)
are then true with $\overline{H}_l (t_0,t)$ replaced by
$\overline{H}^\infty_l (t),  k'_n (L,t_0,t)$ replaced
by $k'^\infty_n (L,t)$, with $\wp^D_n (h(t_0)) = 0$
and with $t_0 = \infty$. These inequalities will be useful for the nonlinear
case, where $A$ is a function of the Dirac field.

In order to use Theorem 5.13 and 5.14 we need an estimate to
$f^{}_Y$, $Y \in \Pi'$,
given by (5.111b). To state the result we introduce first
certain notations. Let $1/2 < \rho < 1,  \eta \in [0, \rho[,
\eta \neq 1/2$, let $\varepsilon = \eta$ if $\eta < 1/2$ and $\varepsilon =
1/2$ if $\eta > 1/2$, let $0 \leq \rho' < 1,$
let
$$\eqalignno{
\tau^M_n (t_0,t) &= (1+t_0)^{- 1/2 - \varepsilon}  \sup_{0 \leq s \leq t_0}
\big((1+s)^\eta  \wp^{M^1}_n (a(s), \dot{a} (s))\big)& (5.186)\cr
&\qquad{}+ \int^t_{t_0}  (1+s)^{-2+\rho-\varepsilon}  \sup_{0 \leq s' \leq s}
 \Big((1+s')^{\eta + \rho'-1}  \wp^{M^{\rho'}}_n (a(s'), \dot{a} (s'))\cr
&\qquad{}+ (1+s')^\eta  \wp^{M^1}_n (a(s'),
\dot{a} (s')) + (1+s')^{\varepsilon +  1/2 - \rho}  \wp^{M^1}_n
(0, \partial_\mu  a^\mu (s'))\cr
&\qquad{}+ (1+s')^{\eta + 3/2 - \rho}  \wp^{M^1}_n (0, \carre a(s'))\Big)
ds,\cr
}$$
for $n \geq 0,  0 \leq t_0 \leq t < \infty$, let $\tau^M_n (t_0,t) =
\tau^M_n (t,t_0)$ for $0 \leq t < t_0 < \infty$ and let $\tau^M_n (t,\infty) =
\lim_{t_0 \fl \infty}  \tau^M_n (t,t_0)$ when this limit exists. Here
$a^{}_Y = \xi^M_Y a,  Y \in \Pi'.$
Let
$$\eqalignno{
\tau^D_n (t_0,t) &= (1+t_0)^{\rho - 3/2}  \wp^D_n (r(t_0)) &(5.187)\cr
&\quad{}+ \int^{\max (t,t_0)}_{\min (t,t_0)}    \Big((1+s)^{- 5/2 + \rho}
\wp^D_{n+1} (r(s)) + (1+s)^{- 3/2}  \wp^D_n
((1+\lambda^{}_0 (s))^{1/2} r(s))\cr
&\quad{}+ (1+s)^{- 3+2 \rho}\wp^D_n (r(s)) + (1+s)^{- 3/2 + \rho}
\wp^D_n (((i  \gamma^\mu  \partial_\mu + m) r) (s))\Big) ds,\cr
}$$
for $n \geq 0,  t \geq 0,  t_0 \geq 0,  1/2 < \rho < 1$
and let $\tau^D_n (\infty,t)$ be given by (5.187),
but without the first term on the right-hand side. When
$t_0 \in \Rrm^+ \cup\{ \infty \}$ is fixed we shall
write $\tau^M_n (t)$ and $\tau^D_n (t)$ instead of
$\tau^M_n (t_0,t)$ and $\tau^D_n (t_0,t).$
\saut
\noindent{\bf Proposition 5.16.}
{\it
Let $t_0 \in \Rrm^+  \cup \{ \infty \},  \rho \in ]1/2,1[,
\eta \in [0,1/2 [\cup] 1/2,\rho[,\rho' \in [0,1[$ and let $\varepsilon = \eta$
if $\eta < 1/2$ and $\varepsilon = 1/2$ if $\eta > 1/2$.
Let $g = \gamma^\mu (a^{}_\mu -
\partial_\mu  {\vartheta} (a)) r,g^{}_Y = \xi^D_Y  g$ for
$Y \in \Pi'$, let $f^{}_Y,  Y \in \Pi'$ be given by (5.111b) and let $\tau^M_n$
and $\tau^D_n$ be defined by (5.186) and (5.187). Let $(1-\Delta)^{1/2}
(A_X, A_{P_0 X}) \in C^0(\Rrm^+,M^1)$ for $X = \un$ or  $X \in sl (2,\Crm),$
let $(1-\Delta)^{1/2}  (B,\dot{B}) \in C^0(\Rrm^+, M^1)$ and let $G_\mu$
be given by (5.114). Let $B_\mu(y)=y_\mu\partial_\nu A^\nu(y)$, and let
$$\eqalignno{
\theta^D_n (t) &= \sum_{\scr Y \in \Pi'\atop\scr  \vert Y \vert \leq n+1}
\sup_{t' \leq s \leq t''} \big((1+[A]^1 (s))  \Vert \delta (s)^{3/2}
 r^{}_Y (s) \Vert^{}_{L^\infty}\big)\cr
&\quad{}+ \sum_{\scr Y \in \Pi'\atop\scr \vert Y \vert \leq n}
\sup_{t' \leq s \leq t''}
\big((1+s)^{2 - \rho}
\Vert \delta (s) ((i \gamma^\mu \partial_\mu + m)
r^{}_Y) (s) \Vert^{}_{L^\infty}\big),\quad n \geq 0,  t \geq 0,\cr
}$$
where $t' = \min (t,t_0),  t'' = \max (t,t_0)$ and let $\theta^M_n (t) =
[a]^{n+2} (t'')  (1+[A]^1 (t'')).$
\psaut
\noindent\hbox{\rm  i)} If $(a^{}_Y, a^{}_{P_0 Y}) \in C^0 (\Rrm^+, M^1)
\cap C^0(\Rrm^+, M^{\rho'})$ for $Y \in \Pi'$,
$\vert Y \vert \leq n$, $\delta^{3/2}  r^{}_Y \in C^0(\Rrm^+$,
$L^\infty (\Rrm^3, \Crm^4))$ for $\vert Y \vert \leq n$, if
$\delta (i \gamma^\mu  \partial_\mu + m) r^{}_Y \in C^0(\Rrm^+$,
$L^\infty (\Rrm^3, \Rrm^4))$ for
$\vert Y \vert \leq n$, where $\delta(t,x)=(\delta(t))(x)$,
and if $0\leq j\leq n$, then
$$\wp^D_{n,j} (f(t)) \leq C^{(j)}  \tau^M_n (t)  \theta^D_0 (t) + C_n
 \sum_{0 \leq l \leq n-1}  \tau^M_l (t)  \theta^D_{n-l}
(t),$$
where $C^{(j)}=0$ if $1\leq j\leq n$.
\psaut
\noindent\hbox{\rm  ii)} If $r^{}_Y \in C^0(\Rrm^+, D)$ for $Y \in \Pi'$,
$\vert Y \vert \leq n+1$,
$(1+\lambda^{}_1)^{1/2}  r^{}_Y \in C^0(\Rrm^+,D)$ for $\vert Y \vert \leq n$
and, if $a^{}_Y\in C^0(\Rrm^+,L^\infty (\Rrm^3, \Crm^4))$ for $Y\in\Pi'$,
$\vert Y\vert\leq n+2$ and if $[a]^{n+2}(t'')<\infty$, then
$$\wp^D_n (f(t)) \leq C_n  \sum_{0 \leq l \leq n}  \theta^M_{n-l}
(t) \tau^D_l (t),\quad  n \geq 0.$$
\psaut
\noindent\hbox{\rm  iii)} If $0 \leq L \leq n,
 (a^{}_Y, a^{}_{P_0 Y}) \in C^0(\Rrm^+,M^1)
\cap C^0(\Rrm^+,M^{\rho'})$ for $Y \in \Pi'$,
$\vert Y \vert \leq n$,  $r^{}_Y \in C^0(\Rrm^+,D)$
for $\vert Y \vert \leq n+1$,
$(1+\lambda^{}_1)^{1/2}  r^{}_Y \in C^0(\Rrm^+,D)$
for $\vert Y \vert \leq n$, if
$a^{}_Y\in C^0(\Rrm^+,L^\infty (\Rrm^3, \Crm^4))$
for $Y\in\Pi'$, $\vert Y\vert\leq n+2$ and if $[a]^{L+2} (t'')<\infty$,
if $\delta^{3/2}  r^{}_Y \in C^0(\Rrm^+, L^\infty (\Rrm^3, \Crm^4))$ for
$\vert Y \vert \leq n-L$, $\delta (i \gamma^\mu
 \partial_\mu + m) r^{}_Y \in C^0(\Rrm^+, L^\infty (\Rrm^3, \Rrm^4))$ for
$\vert Y \vert \leq n$,  then
$$\eqalignno{
&\wp^D_{n, j} (f(t)) \cr
&\quad{}\leq C^{(j)}  \tau^M_n (t)  \theta^D_0 (t)
+ C_n \Big(\sum_{\scr n_1+n_2=n\atop\scr 0 \leq n_1 \leq L}  \theta^M_{n_1}
(t)  \tau^D_{n_2} (t) + \sum_{\scr n_1+n_2=n\atop\scr  L+1 \leq n_1 \leq
n-1}  \tau^M_{n_1} (t)  \theta^D_{n_2} (t)\Big),\quad   n \geq 0,\cr
}$$
where $C^{(j)},C_n$ are constants depending only on $\rho, \rho', \eta$, and
$C^{(j)}=0$ for $1\leq j\leq n$.
}\saut
\noindent{\it Proof.}
Let $b^{}_\mu = a^{}_\mu - \partial_\mu  {\vartheta} (a),0 \leq \mu \leq 3$
and $b^{}_{Y \mu} = (\xi^M_Y b)_\mu,  Y \in \Pi'$. If
$$f^{}_{Y_1,Y_2} (t) = \int^t_{t_0} w(t,s) (- i \gamma^0)
\gamma^\mu  b^{}_{Y_1 \mu}
(s)  r^{}_{Y_2} (s) ds, \eqno{(5.188{\rm a})}$$
then by definition (5.111b) of $f^{}_Y$
$$f^{}_Y (t) = \suma_{Y_1,Y_2}^Y  f^{}_{Y_1,Y_2} (t). \eqno{(5.188{\rm b})}$$
According to the hypothesis, it follows that $\gamma^\mu  b^{}_{Y_1 \mu}
 r^{}_{Y_2} \in C^0 (\Rrm^+,D)$, when $Y_1,Y_2$ are in the domain of summation
in (5.188b) and $\vert Y \vert \leq n,  Y \in \Pi'.$

Let $t_0 < \infty$. Since the two cases $0 \leq t_0 \leq t$
and $0 \leq t < t_0$ are so similar, we only consider the situation where
$0 \leq t_0 \leq t$. Like in the beginning of the
proof of Theorem 5.13, it follows that $f^{}_{Y_1,Y_2}
\in C^0(\Rrm^+,D)$ is the unique
solution of $(i \gamma^\mu  \partial_\mu + m - \gamma^\mu  G_\mu)
 f^{}_{Y_1,Y_2} = \gamma^\mu  b^{}_{Y_1 \mu}  r^{}_{Y_2},$
with $f^{}_{Y_1,Y_2} (t_0) = 0.$

For $Y_1 \in \Pi' \cap U({\frak{sl}}(2, \Crm))$ denote by
$$\eqalignno{
&I_{Y_1,Y_2} (t) &(5.189{\rm a})\cr
&\quad{}= \big\vert  \Vert f^{}_{Y_1,Y_2} (t) - (2m)^{-1}  \gamma^\mu
 b^{}_{Y_1\mu} (t)  r^{}_{Y_2} (t) \Vert^{}_D - \Vert (2m)^{-1}
\gamma^\mu  b^{}_{Y_1\mu} (t_0)  r^{}_{Y_2} (t_0) \Vert^{}_D \big\vert,\cr
}$$
where $(Y_1,Y_2)$ is in the domain of summation in (5.188b) and $Y \in \Pi',
\vert Y \vert \leq n$. It follows from Corollary 5.2, with
$a^{}_l = 0$ and using the gauge
invariance of the electromagnetic field that
$$\eqalignno{
&I_{Y_1,Y_2} (t)&(5.189{\rm b})\cr
&\leq C  \int^{t''}_{t'}\!\! \Big((1+s)^{-1} \Big(\Vert b^{}_{Y_1 0} (s)
 r^{}_{P_0 Y_2} (s) \Vert^{}_D
+\!\! \sum_{1 \leq i \leq 3}\!\!  \big(\Vert b^{}_{Y_1 i} (s)
r^{}_{M_{0i} Y_2} (s) \Vert^{}_D + \Vert b^{}_{Y_1 i} (s)  r^{}_{Y_2} (s)
\Vert^{}_D\big)\Big)\cr
&\quad{}+\Vert (\partial_\mu  b^\mu_{Y_1} (s))  r^{}_{Y_2} (s) \Vert^{}_D +
\Vert \gamma^\mu  \gamma^\nu (\partial_\mu  a^{}_{Y_1 \nu} (s) -
\partial_\nu  a^{}_{Y_1 \mu} (s))  r^{}_{Y_2} (s) \Vert^{}_D\cr
&\quad{}+ \Vert \gamma^\mu  \gamma^\nu  G_\mu (s)  b^{}_{Y_1 \nu} (s)
 r^{}_{Y_2} (s) \Vert^{}_D + \Vert \gamma^\mu  b^{}_{Y_1 \mu} (s)
R^{}_{Y_2} (s) \Vert^{}_D\Big)  ds,\cr
}$$
where $R^{}_{Y_2} = (i \gamma^\mu       \partial_\mu + m)  r^{}_{Y_2}$
and $C$ is a numerical constant. Since $[G]'^0 (t) \leq C_0  [A]^1 (t),$
according to (5.116c), it follows from (5.189b) that
$$\eqalignno{
&I_{Y_1,Y_2} (t)\cr
&\quad{}\leq C  \int^{t''}_{t'} (1+s)^{- 2+\rho-\varepsilon}
\Big((1+s)^{1/2 - \rho + \varepsilon}
\Vert (a^{}_{Y_1} (s), a^{}_{P_0 Y_1} (s))
\Vert^{}_{M^1}  \Vert \delta (s)^{3/2}  r^{}_{Y_2} (s)
\Vert^{}_{L^\infty}\cr
&\qquad{}+ (1+s)^\varepsilon  \Vert \delta (s)^{-1}  b^{}_{Y_1} (s)
\Vert^{}_{L^2}  \Big((1+s)^{1/2 - \rho}  \Big(\Vert \delta (s)^{3/2}
 r^{}_{P_0 Y_2} (s) \Vert^{}_{L^\infty}\cr
&\qquad{}+ \sum_{1 \leq i \leq 3}
\big(\Vert \delta (s)^{3/2}  r^{}_{M_{0i} Y_2}
(s) \Vert^{}_{L^\infty} + \Vert \delta (s)^{3/2}  r^{}_{Y_2} (s)
\Vert^{}_{L^\infty}\big)\Big)\cr
&\qquad{}+ [A]^1 (s)  \Vert \delta (s)^{3/2}
r^{}_{Y_2} (s) \Vert^{}_{L^\infty}
+ (1+s)^{2-\rho}  \Vert \delta (s)R^{}_{Y_2} (s)
\Vert^{}_{L^\infty}\Big)\cr
&\qquad{}+ (1+s)^{3/2 - \rho + \varepsilon}  \Vert \delta (s)^{-1}
\partial_\mu  b^\mu_{Y_1} (s) \Vert^{}_{L^2}  \Vert \delta (s)^{3/2}
 r^{}_{Y_2} (s) \Vert^{}_{L^\infty}\Big)  ds,\cr
}$$
where $C$ is a constant depending only on
$\rho$. Since $1/2 < \rho < 1$, this
inequality gives, according to the definition of $\theta^D_n$,
$$\eqalignno{
&I_{Y_1,Y_2} (t)&(5.190)\cr
&{} \leq C  \int^{t''}_{t'}\!\!  (1+s)^{-2+\rho-\varepsilon}
\Big((1+s)^{1/2 - \rho + \varepsilon}
\Vert (a^{}_{Y_1} (s), a^{}_{P_0 Y_1} (s))
\Vert^{}_{M^1} + (1+s)^\varepsilon  \Vert \delta (s)^{-1}
b^{}_{Y_1} (s) \Vert^{}_{L^2}\cr
&\quad{}+ (1+s)^{3/2 - \rho + \varepsilon}  \Vert \delta (s)^{-1}
\partial_\mu  b^\mu_{Y_1} (s) \Vert^{}_{L^2}\Big)  ds
\theta^D_{\vert Y_2 \vert} (t),\cr
}$$
where $C$ is a constant depending only on $\rho$.
Inequality (5.122) gives, since
$Y_1 \in \Pi' \cap U({\frak{sl}}(2, \Crm))$, that
$$\eqalignno{
&(1+t)^\varepsilon  \Vert \delta (t)^{-1}b^{}_{Y_1} (t)
\Vert^{}_{L^2} &(5.191{\rm a})\cr
&\quad{}\leq C  \sup_{0 \leq s \leq t}  \Big((1+s)^{\eta+\rho'-1}
 \Vert \vert \nabla \vert^{\rho'}  a^{}_{Y_1} (s) \Vert^{}_{L^2}
+ (1+s)^\eta    \Vert (a^{}_{Y_1} (s), a^{}_{P_0 Y_1} (s))
\Vert^{}_{M^1}\Big),\cr
}$$
since $\varepsilon = \eta$ for $\eta < 1/2,  \varepsilon = 1/2$ for
$\eta > 1/2$ and $0 \leq \rho' < 1$.
Moreover inequality (5.129) gives that
$$\eqalignno{
&(1+t)^{\varepsilon + 3/2 - \rho}  \Vert \delta (t)^{-1}
\partial_\mu  b^\mu_{Y_1} (t) \Vert^{}_{L^2}&(5.191{\rm b})\cr
&\quad{}\leq C  \sup_{0 \leq s \leq t}
\Big((1+s)^{\varepsilon + 1/2 - \rho}
\Vert \partial_\mu  a^\mu_{Y_1} (s) \Vert^{}_{L^2} + (1+s)^{\varepsilon +
3/2 - \rho}  \Vert \carre  a^{}_{Y_1} (s) \Vert^{}_{L^2}\Big),\cr
}$$
since $\varepsilon + 3/2 - \rho < 3/2$.
In inequality (5.191a), $C$ depends only on
$\varepsilon, \eta, \rho'$ and in (5.191b) only on
$\varepsilon, \rho$. It follows
from inequalities (5.190), (5.191a) and (5.191b),
since $1/2 - \rho + \varepsilon <
\varepsilon \leq \eta$, that
$$I_{Y_1,Y_2} (t) \leq C  \tau^{1M}_{Y_1} (t)  \theta^D_{\vert Y_2 \vert}
(t), \eqno{(5.192{\rm a})}$$
where $C$ depends only on $\varepsilon, \eta, \rho', \rho$,
where $(Y_1,Y_2)$ is in
the domain of summation in (5.188b), $Y \in \Pi',  \vert Y \vert \leq n,$
$Y_1 \in \Pi' \cap U({\frak{sl}}(2, \Crm))$, and where
$$\eqalignno{
\tau^{1M}_{Y_1} (t) &= \int^{t''}_{t'}  (1+s)^{- 2 + \rho - \varepsilon}
 \sup_{0 \leq s' \leq s}  \Big((1+s')^{\eta + \rho' - 1}
\Vert (a^{}_{Y_1} (s'), a^{}_{P_0 Y_1} (s')) \Vert^{}_{M^{\rho'}}\cr
&\quad{}+ (1+s')^\eta  \Vert (a^{}_{Y_1} (s'),
a^{}_{P_0 Y_1} (s')) \Vert^{}_{M^1} +
(1+s')^{\varepsilon + 1/2 - \rho}  \Vert \partial_\mu  a^\mu_{Y_1}
(s') \Vert^{}_{L^2}\cr
&\quad{}+ (1+s')^{\eta + 3/2 - \rho}
\Vert \carre  a^{}_{Y_1} (s') \Vert^{}_{L^2}\Big) ds. &(5.192{\rm b})\cr
}$$
It follows from (5.116c) and (5.189b) that
$$\eqalignno{
I_{Y_1, Y_2} (t) &\leq C_{\vert Y_1 \vert}  \int^{t''}_{t'}
\Big((1+s)^{- 5/2 + \rho}  [a]^{\vert Y_1 \vert + 1} (s)
\wp^D_{\vert Y_2 \vert + 1}
(r(s))\cr
&\quad{}+ (1+s)^{- 3/2}  [a]^{\vert Y_1 \vert + 2} (s)
\wp^D_{\vert Y_2 \vert}
((1+\lambda^{}_1 (s))^{1/2}  r(s))\cr
&\quad{}+ (1+s)^{- 3/2}  [a]^{\vert Y_1 \vert + 1} (s)
\wp^D_{\vert Y_2 \vert}
 ((1+\lambda^{}_1 (s))^{1/2}  r(s))\cr
&\quad{}+ (1+s)^{- 3+2 \rho}    [A]^1 (s)  [a]^{\vert Y_1 \vert + 1}
(s)  \wp^D_{\vert Y_2 \vert} (r(s))\cr
&\quad{}+ (1+s)^{- 3/2 + \rho}  [a]^{\vert Y_1 \vert +1}
\wp^D_{\vert Y_2 \vert}
 (R (s))\Big)  ds,\cr
}$$
which shows that
$$\eqalignno{
&I_{Y_1,Y_2} (t)&(5.193)\cr
&\quad{} \leq C_{\vert Y_1 \vert}  \theta^M_{\vert Y_1 \vert}
(t)  \int^{t''}_{t'} \Big((1+s)^{- 5/2 + \rho}  \wp^D_{\vert Y_2 \vert + 1}
(r(s))+ (1+s)^{- 3/2}  \wp^D_{\vert Y_2 \vert} ((1+\lambda^{}_1 (s))^{1/2}
r(s)) \cr
&\qquad{}+ (1+s)^{- 3 + 2 \rho}  \wp^D_{\vert Y_2 \vert} (r(s))
+ (1+s)^{- 3/2 + \rho}  \wp^D_{\vert Y_2 \vert} (R(s))\Big)  ds,\cr
}$$
where $C_{\vert Y_1 \vert}$ depends only on $\rho$ and where
$(Y_1,Y_2)$ is in the domain of summation in (5.88b), $Y \in \Pi'$,
$\vert Y \vert \leq n.$

Let $q^{}_\mu (y) = \int^1_0  a^{}_\mu (sy)  ds,  0 \leq
\mu \leq 3$, and let $q^{}_{Y \mu} = (\xi^M_Y  q)_\mu$ for $Y \in U(\p).$
Similarly as (5.121{\rm c}) was obtained, it follows that
$$b^{}_{ZY \mu} (y) = a^{}_{ZY \mu} (y) - y^\nu  \partial_\mu  q^{}_{ZY \nu}
(y) - q^{}_{ZY \mu} (y)
- \vert Z \vert  \sum_{0 \leq \nu \leq 3}  C_\nu (Z)
q^{}_{Z_\nu P_\mu Y^\nu} (y), \eqno{(5.194)}$$
for some positive integers $C_\nu (Z)$ and some elements $Z_\nu \in \Pi' \cap
U({\frak{sl}}(2, \Crm)),  \vert Z_\nu \vert = \vert Z \vert - 1$, when $Z \in
\Pi'
\cap U(\Rrm^4)$ and $Y \in \Pi' \cap U({\frak{sl}}(2, \Crm))$. Using the
estimate
$$\eqalignno{
\Vert \delta (t)^{- 3/2} \varphi \Vert^{}_{L^2} &
\leq \Vert \delta (t)^{- 3/2}
\Vert^{}_{L^3}  \Vert \varphi \Vert^{}_{L^6}\cr
&\leq C' (1+t)^{- 1/2}  \Vert \vert \nabla \vert \varphi \Vert^{}_{L^2}\cr
&\leq C (1+t)^{- 1/2}  \sum_{0 \leq i \leq 3}  \Vert \partial_i
 \varphi \Vert^{}_{L^2}\cr
}$$
for the first and third term on the right-hand side of inequality (5.194), we
obtain that
$$\eqalignno{
&\Vert \delta (t)^{- 3/2}b^{}_{ZY} (t) \Vert^{}_{L^2}\cr
&\qquad{}\leq C (1+t)^{- 1/2}   \sum_{0 \leq \mu \leq 3}
\big(\Vert a^{}_{P_\mu ZY}
(t) \Vert^{}_{L^2} + \Vert q_{P_\mu ZY} (t) \Vert^{}_{L^2}\big)\cr
&\qquad\qquad{}+ \vert Z \vert  (1+t)^{- 3/2}
\sum_{\scr 0 \leq \nu \leq 3\atop\scr
 0 \leq \mu \leq 3}  C_\nu (Z)  \Vert q_{Z_\nu P_\mu Y}
(t) \Vert^{}_{L^2}.\cr
}$$
This inequality and the result given by (4.86a)--(4.86b), show that
$$\eqalignno{
&\Vert \delta (t)^{- 3/2} b^{}_{ZY} (t) \Vert^{}_{L^2} &(5.195)\cr
&\quad{}\leq C (1+t)^{- 1/2 - \chi}
\sup_{0 \leq s \leq t}  \big((1+s)^{\chi'}
 \Vert (a^{}_{ZY} (s), a^{}_{P_0 ZY} (s)) \Vert^{}_{M^1}\big)\cr
&\qquad{}+ \vert Z \vert  C  (1+t)^{- 1/2 - \chi}  \sum_{0 \leq \nu \leq 3}
 C_\nu (Z)  \sup_{0 \leq s \leq t} \big((1+s)^{\chi'-1}
\Vert (a^{}_{Z_\nu Y} (s), a^{}_{P_0 Z_\nu Y} (s)) \Vert^{}_{M^1}\big),\cr
}$$
where $\chi = \chi'$ for $\chi' < \vert Z \vert + 1/2,
\chi = \vert Z \vert + 1/2$
for $\chi' > \vert Z \vert + 1/2$, where $C$
depends only on $\chi$ and $\chi'$
and where $Z \in \Pi' \cap U(\Rrm^4),  Y \in \Pi' \cap U ({\frak{sl}}(2,
\Crm)).$

Due to the unitarity of the linear time evolution in $M^1$, we have
$$\Vert (a^{}_Y (t), a^{}_{P_0 Y} (t)) \Vert^{}_{M^1}
\leq \Vert (a^{}_Y (t_0), a^{}_{P_0 Y} (t_0))\Vert^{}_{M^1}
+ \int^{t''}_{t'}  \Vert (0, J_Y (s)) \Vert^{}_{M^1}  ds,\eqno{(5.196)}$$
where $J_Y = \carre  a^{}_Y$ and $Y \in \Pi'$. It follows from inequalities
(5.195) and (5.196) that
$$\eqalignno{
&\Vert \delta (t)^{- 3/2}b^{}_Y (t) \Vert^{}_{L^2}
\leq C (1+t)^{- 1/2 - \chi}&(5.197)\cr
&\quad{}  \sup_{0 \leq s \leq t_0} \Big((1+s)^{\chi'}
 \Vert (a^{}_Y (s), a^{}_{P_0 Y} (s)) \Vert^{}_{M^1}
+ \vert Y \vert  (1+s)^{\chi' - 1}  \wp^{M^1}_{\vert Y \vert - 1}
 ((a(s), \dot{a} (s)))\Big)\cr
&\qquad{}+ C  \int^{t''}_{t'} \Big((1+s)^{- 1/2 - \chi + \chi'}  \Vert
\carre a^{}_Y (s) \Vert^{}_{L^2} + \vert Y \vert (1+s)^{- 3/2 - \chi + \chi'}
\wp^{M^1}_{\vert Y \vert - 1} ((0, \carre a(s)))\Big)ds,\cr
}$$
where $Y \in \Pi'$, $ \chi = \chi'$ for $\chi' < 1/2$, $\chi = 1/2$
for $\chi' > 1/2$
and $- 1/2 - \chi + \chi' \leq 0$. Since these conditions are
satisfied with $\chi = \varepsilon$
and $\chi' = \eta$ it follows from (5.186), (5.192b) and (5.197) that
$$\Vert \delta (t)^{- 3/2}b^{}_Y (t) \Vert^{}_{L^2}
\leq C (\tau^{0 M}_Y (t) + \tau^{1M}_Y (t) + \vert Y \vert
\tau^M_{\vert Y \vert - 1}
(t)),\quad  Y \in \Pi',\eqno{(5.198)}$$
where
$$\tau^{0 M}_Y (t) = (1+t_0)^{- 1/2 - \varepsilon}  \sup_{0 \leq s \leq t_0}
 \big((1+s)^\eta  \Vert (a^{}_Y (s), a^{}_{P_0 Y} (s)) \Vert^{}_{M^1}\big).
\eqno{(5.199)}$$
(We note that inequality (5.197) is also true for $t_0 > t$).

Since ${d \over dt}  e^{- {\cal D} t}  r^{}_Y (t) = e^{- i {\cal D}t}
 (- i \gamma^0)  (i \gamma^\mu  \partial_\mu + m)
 r^{}_Y (t),$
we obtain using (5.187) that
$$\eqalignno{
&(1+t)^{- 3/2 + \rho}   \wp^D_n (r(t))& (5.200)\cr
&\quad{}\leq (1+t_0)^{- 3/2 + \rho}
\wp^D_n (r(t_0))
+ \int^{t''}_{t'} \Big((1+s)^{- 5/2 + \rho}  (\rho - 3/2)  \wp^D_n
(r(s))\cr
&\qquad{}+ (1+s)^{- 3/2 + \rho}  \wp^D_n ((i \gamma^\mu  \partial_\mu
+ m) r(s))\Big)  ds\cr
&\quad{}\leq \tau^D_n (t).\cr
}$$
It follows from inequalities (5.189a), (5.192a) and (5.198), that
$$\Vert f^{}_{Y_1,Y_2} (t) \Vert^{}_D \leq C\big(\tau^{0 M}_{Y_1} (t)
+ \tau^{1 M}_{Y_1} (t) +
\vert Y_1 \vert  \tau^M_{\vert Y_1 \vert - 1} (t)\big)
\theta^D_{\vert Y_2 \vert}
(t) \eqno{(5.201{\rm a})}$$
and from (5.187), (5.189a), (5.193) and (5.200), that
$$\Vert f^{}_{Y_1,Y_2} (t) \Vert^{}_D \leq
C_{\vert Y_1 \vert, \vert Y_2 \vert}
\theta^M_{\vert Y_1 \vert} (t)
\tau^D_{\vert Y_2 \vert} (t), \eqno{(5.201{\rm b})}$$
where $Y_1 \in \Pi' \cap U({\frak{sl}}(2, \Crm)),  (Y_1,Y_2)$ are in the domain
of summation in (5.188b) and $Y \in \Pi',  \vert Y \vert \leq n.$

For $Y_1 \in \sg^1$, there is $\nu$ such that $Y_1 = P_\nu Z$, $Z \in \Pi',
\vert Z \vert = \vert Y_1 \vert -1$. It follows from
(5.7b$'$) and from the gauge invariance of the electromagnetic field, that
$$\eqalignno{
&(i \gamma^\mu  \partial_\mu + m - \gamma^\mu   G_\mu)
(f^{}_{Y_1,Y_2} + i b^{}_{Z \nu}  r^{}_{Y_2})& (5.202)\cr
&\qquad{}= \gamma^\mu (\partial_\nu  a^{}_{Z \mu} -
\partial_\mu  a^{}_{Z \nu})
 r^{}_{Y_2} + b^{}_{Z \nu} (i \gamma^\mu  \partial_\mu + m - \gamma^\mu
 G_\mu)  r^{}_{Y_2}.\cr
}$$
With the notation
$$I_{Y_1,Y_2} (t) = \Vert f^{}_{Y_1,Y_2} (t) + i b^{}_{Z \nu} (t)
r^{}_{Y_2} (t) \Vert^{}_D -
\Vert b^{}_{Z \nu} (t_0)  r^{}_{Y_2} (t_0) \Vert^{}_D, \eqno{(5.203)}$$
it follows from (5.202) that
$$\eqalignno{
I_{Y_1,Y_2} (t) &\leq \int^{t''}_{t'}  \Big(\Vert \gamma^\mu (\partial_\nu
 a^{}_{Z \mu} (s) - \partial_\mu  a^{}_{Z \nu} (s))  r^{}_{Y_2}
(s) \Vert^{}_D &(5.204)\cr
&\qquad{}+ \Vert b^{}_{Z \nu}   (s)  R^{}_{Y_2} (s) \Vert^{}_D +
\Vert b^{}_{Z \nu} (s)
 \gamma^\mu  G_\mu (s)  r^{}_{Y_2} (s) \Vert^{}_D\Big) ds,\cr
}$$
for $Y_1 = P_\nu Z,  Z \in \Pi'$. This inequality gives that
$$\eqalignno{
I_{Y_1,Y_2} (t) &\leq C \int^{t''}_{t'} ((1+s)^{- 3/2}  \Vert (a^{}_Z (s),
a^{}_{P_0 Z} (s)) \Vert^{}_{M^1}\cr
&\qquad{}+ (1+s)^{- 2 + \rho}   \Vert \delta (s)^{-1}  b^{}_{Z^\nu}
(s) \Vert^{}_{L^2})  ds  \theta^D_{Y_2} (t).\cr
}$$
It follows then from inequality (5.122), that
$$\eqalignno{
I_{Y_1,Y_2} (t) &\leq C\theta^D_{\vert Y_2 \vert} (t)
\int^{t''}_{t'}  \Big((1+s)^{- 3/2}  \Vert
(a^{}_Z (s), a^{}_{P_0 Z} (s)) \Vert^{}_{M^1}\cr
&\quad{}+ (1+s)^{- 2+\rho-\varepsilon}  \sup_{0 \leq s' \leq s}
\Big((1+s')^{\eta + \rho'-1}
 \Vert \vert \nabla \vert^{\rho'}  a^{}_Z (s') \Vert^{}_{L^2}\cr
&\quad{}+ (1+s')^\eta  \Vert (a^{}_Z (s'),
a^{}_{P_0 Z} (s')) \Vert^{}_{M^1}\cr
&\quad{}+ (1+s')^{\eta - 1}  C_{\vert Z \vert}  \vert Z \vert
\wp^{M^1}_{\vert Z \vert - 1} ((a(s'), \dot{a} (s')))\Big)\Big)  ds,\cr
}$$
and then from definition (5.186) of $\tau^M$, that
$$I_{Y_1,Y_2} (t) \leq C_{\vert Y_1 \vert}  \tau^M_{\vert Y_1 \vert - 1}
(t)  \theta^D_{\vert Y_2 \vert} (t), \eqno{(5.205{\rm a})}$$
where $Y_1 \in \sg^1$. It also follows from inequalities (5.116c) and (5.204)
that
$$\eqalignno{
I_{Y_1,Y_2} (t) &\leq C_{\vert Y_1 \vert}  \int^{t''}_{t'} \Big((1+s)^{- 3/2}
 [a]^{\vert Y_1 \vert} (s)  \wp^D_{\vert Y_2 \vert} ((1+\lambda^{}_1
(s))^{1/2}  r(s))\cr
&\qquad{}+ (1+s)^{- 3/2 + \rho}  [a]^{\vert Y_1 \vert} (s)
\wp^D_{\vert Y_2 \vert}
 (R(s))\cr
&\qquad{}+ (1+s)^{- 3 + 2 \rho}  [A]^1 (s)  [a]^{\vert Y_1 \vert} (s)
 \wp^D_{\vert Y_2 \vert}  (r(s))\Big)  ds,\cr
}$$
which gives that
$$I_{Y_1,Y_2} (t) \leq C_{\vert Y_1 \vert}  \theta^M_{\vert Y_1 \vert - 1}
(t)  \tau^D_{\vert Y_2 \vert} (t), \eqno{(5.205{\rm b})}$$
where $Y_1 \in \sg^1$. It follows from definition (5.203) of $I_{Y_1,Y_2}$ and
from inequalities (5.198) and (5.205a), that
$$f^{}_{Y_1,Y_2} (t) \leq C_{\vert Y_1 \vert,
\vert Y_2 \vert}  \tau^M_{\vert Y_1 \vert - 1}
(t)  \theta^D_{\vert Y_2 \vert} \eqno{(5.206{\rm a})}$$
and that
$$f^{}_{Y_1,Y_2} (t) \leq C_{\vert Y_1 \vert, \vert Y_2 \vert}
\theta^M_{\vert Y_1 \vert - 1}
(t)  \tau^D_{\vert Y_2 \vert} (t), \eqno{(5.206{\rm b})}$$
where $Y_1 \in \sg^1.$

Decomposition (5.188b), inequality (5.201a) and inequality (5.206a) prove
statement~i) of the proposition, since
$$\Vert f^{}_Y (t) \Vert^{}_D \leq \Vert
f^{}_{Y, \un} (t) \Vert + C_{\vert Y \vert}
\suma_{\scr Y_1,Y_2\atop\scr  \vert Y_1 \vert \leq \vert Y \vert - 1}^Y
\Vert f^{}_{Y_1,Y_2} (t) \Vert^{}_D.$$
Statement ii) follows from inequalities (5.201b) and (5.206b). Statement iii)
follows by using (5.201a) and (5.206a) for $\vert Y_1
\vert \geq L+1$ and inequalities
(5.201b) and (5.206b) for $0 \leq \vert Y_1 \vert \leq L$.
This proves the proposition.

Finally in this chapter we shall prove two corollaries,
which are particular cases of Theorem 5.13 and Proposition 5.16 and which
are obtained by using the convexity property of the seminorms $\Vert\cdot
\Vert^{}_{E_n}$ given by Corollary 2.6.
\saut
\noindent{\bf Corollary 5.17.}
{\it
Let $t_0 \in [0,\infty],  n \geq 0, 1/2 < \rho < 1,
(1-\Delta)^{1/2}  (A_X, A_{P_0 X}) \in C^0(\Rrm^+, M^1)$ for $X = \un$ or
 $X \in {\frak{sl}}(2, \Crm)$, let $(1-\Delta)^{1/2}  (B, \dot{B}) \in
C^0(\Rrm^+, M^1)$, where $B_\mu (y) = y_\mu  \partial^\nu  A_\nu
(y)$,  $\dot{B} = {d \over dt}  B$, let $A_Y \in C^0(\Rrm^+, L^\infty
(\Rrm^3, \Rrm^4))$,  $Y \in \Pi'$, let
$$[A]^l (t) \leq C_l  \Vert u \Vert^{}_{E_{N_0+l}},\quad  l \geq 0,
 t \geq 0,$$
for some element $u \in E^\rho_\infty$, integers $N_0$ and
$L_0$ independent of $u$ and $l,$
and some constants $C_l$ depending only on $\Vert u \Vert^{}_{E_{L_0}}$,
$L_0\geq N_0$, let $G_\mu$
be given by (5.114), $f$ be given by (5.111b), let
$$\eqalignno{
Q_n (t) &= \sup_{t' \leq t \leq t''}
\wp^D_n (f(s)) + \sum_{0 \leq l \leq n-1}
 \Vert u \Vert^{}_{E_{N_0+3+n-l}}\cr
&\quad{}\Big(\sup_{t' \leq s \leq t''}  \wp^D_l (f(s))
+ \int^{t''}_{t'} (1+s)^{- 3/2 + \rho}
 \wp^D_l ((1+\lambda^{}_0 (s))^{1/2}  g(s))  ds\Big),\cr
}$$
where $t' = \min (t,t_0)$,  $t'' = \max (t,t_0)$. If $t_0 \in \Rrm^+$ and
$h^{}_Y (t_0) \in D$ for $Y \in \Pi'$,  $\vert Y \vert \leq n$, then $h$
given by (5.3c) is the unique solution of equation (5.1a) in
$C^0(\Rrm^+,D)$ with initial data $h(t_0)$ at $t_0$. This solution satisfies
$h^{}_Y, (1-\Delta)^{- 1/2}
 h^{}_{P_\mu Y} \in C^0(\Rrm^+,D), 0\leq \mu \leq 3$, and
$$\wp^D_n (h(t)) \leq C_n \Big(\wp^D_n (h(t_0)) +
\sum_{0 \leq l \leq n-1}  \Vert u \Vert^{}_{E_{N_0+3+n-l}}
\wp^D_l (h(t_0)) + Q_n (t)\Big),$$
for $t \in \Rrm^+$, where the constant $C_n$ depends only
on $\Vert u \Vert^{}_{E_{L}}$, $L=\max(L_0, N_0 +3)$ and $\rho$.
If $t_0 = \infty$, if the function
$t \mapsto (1+t)^{- 3/2 + \rho}  (1+\lambda^{}_0 (t))^{1/2}  g^{}_Y (t)$
is an element of $L^1 (\Rrm^+,D)$ for $Y \in \Pi',  \vert Y \vert \leq n$,
and if for each
$Y \in \Pi'$,  $\vert Y \vert \leq n$, there exists
$g^{}_{1Y}$ and $g^{}_{2Y}$ such that
$g^{}_{Y}=g^{}_{1Y}+g^{}_{2Y}$, and such that:\psaut

\hbox{{\rm a)} $g^{}_{1Y} \in L^1 (\Rrm^+,D)$,}\psaut

\hbox{{\rm b)} $(m - i \gamma^\mu  \partial_\mu + \gamma^\mu  G_\mu)
 g^{}_{2Y} \in L^1 (\Rrm^+,D)$ and $\displaystyle{\lim_{t \fl \infty}}
\Vert g^{}_{2Y}(t) \Vert^{}_D = 0$,}\psaut

\noindent then there exists a unique solution $h \in C^0(\Rrm^+,D)$ of
equation (5.1a) such
that $\Vert h(t) \Vert^{}_D \fl 0$, when $t \fl \infty$.
This solution satisfies
$\wp^D_n (h(t)) \leq C_n  Q^\infty_n (t)$, where $Q^\infty_n (t)$ is given
by the above expression of $Q_n (t)$ with $t_0 = \infty$. Further $f^{}_Y \in
C^0(\Rrm^+,D),$
$f^{}_Y (t) \fl 0$ in $D$ when $t \fl \infty$ and $f^{}_Y (t)$
is the limit of $\int^T_t w(t,s)  i \gamma^0  g^{}_Y (s)ds$ in $D$,
when $T \fl \infty$,
for $Y \in \Pi'$,  $\vert Y \vert \leq n$.
}\saut
\noindent{\it Proof.}
Since $[A]^{q+l} (t) \leq C_{l+q}  \Vert u \Vert^{}_{E_{N_0+q+l}},
l \geq 0,  q \geq 0$, according to the hypothesis, it follows from the
definition of $[A]_{q,l}$ before formula (5.117a) and from (5.87), that
$$\eqalignno{
[A]_{q,0} (t) &\leq C_q  \Vert u \Vert^{}_{E_{N_0+q}}\cr
\noalign{\hbox{and that}}
[A]_{q,l} (t) &\leq C^{(l)}_q   \sum_{1 \leq p \leq l}
\sum_{\scr n_1+\cdots+n_p = l \atop\scr  n_i \geq 1}
\Vert u \Vert^{}_{E_{N_0+q+n_1}} \cdots \Vert u \Vert^{}_{E_{N_0+q+n_p}},
\quad l \geq 1,\cr
}$$
where $C^{(l)}_q$ is a constant depending only on
$\rho$ and $\Vert u \Vert^{}_{E_{L_0}}.$
Repeated use of Corollary 2.6 shows that
$$\Vert u \Vert^{}_{E_{N_0+q+n_1}}\cdots \Vert u \Vert^{}_{E_{N_0+q+n_p}}
\leq K_{N_0+q+l}  (\Vert u \Vert^{}_{E_{N_0+q}})^{p-1}  \Vert u
\Vert^{}_{E_{N_0+q+l}},$$
since $n_1+\cdots+n_p = l$, where $K_{N_0+q+l}$ is a constant
independent of $u.$ This shows that
$$[A]_{q,l} (t) \leq C_{q,l}  \Vert u \Vert^{}_{E_{N_0+q+l}},
\eqno{(5.207)}$$
where $C_{q,l}$ is a constant depending only on
 $\rho$ and $\Vert u \Vert^{}_{E_{M}}$, $M=\max(L_0,N+q)$.
Using inequality (5.207) with $q=3$, the corollary
follows from Theorem 5.13 by
redefining $Q_N (t)$. This proves the corollary.
\saut
\noindent{\bf Corollary 5.18.}
{\it
Let $n \geq 0,  t_0 \in [0,\infty]$, let $A$,  $h(t_0)$, $f$
and $G$ be as in Corollary 5.17, let $\rho \in ]1/2,1[,  \eta \in [0,
1/2[ \cup ]1/2,\rho[,  \rho' \in [0,1[$ and let $\varepsilon = \eta$ if
$\eta < 1/2$ and $\varepsilon = 1/2$ if $\eta > 1/2$. Let
$g = \gamma^\mu (a^{}_\mu - \partial_\mu  {\vartheta} (a)) r$ and $g^{}_Y =
\xi^D_Y g$ for $Y \in \Pi'.$

\noindent\hbox{\rm i)} If $(a^{}_Y, a^{}_{P_0 Y}) \in C^0(\Rrm^+, M^1)
\cap C^0(\Rrm^+, M^{\rho'})$ for $Y \in \Pi',
 \vert Y \vert \leq n,  \partial_\mu  a^\mu = 0,$
$$\Vert \delta (t)^{3/2}  (1+\lambda^{}_1 (t))^{\rho - 1/2}
r^{}_Y (t) \Vert^{}_{L^\infty} +
\Vert (1+\lambda^{}_1 (t))^{\rho - 1/2}  r^{}_Y (t)
\Vert^{}_D \leq C_{\vert Y \vert}  \Vert u
\Vert^{}_{E_{N_0+\vert Y \vert}}$$
and
$$\Vert \delta (t)^{3 - \rho}  (i  \gamma^\mu
\partial_\mu + m) r^{}_Y (t) \Vert^{}_{L^\infty} \leq C_{\vert Y \vert}
\Vert u \Vert^{}_{E_{N_0}}  \Vert u \Vert^{}_{E_{N_0+\vert Y \vert}},\quad
t \geq 0,  Y \in \Pi'$$
where $C_{\vert Y \vert}, N_0$ and $u$ are as
in Corollary 5.17, then
$$\eqalignno{
&\wp^D_n (h(t))\cr
&\quad \leq C'_n \Big(\wp^D_n (h(t_0)) + \sum_{0 \leq l\leq n-1}
\Vert u \Vert^{}_{E_{N_0+3+n-l}}\wp^D_l (h(t_0))
+ \sum_{0 \leq l \leq n}  \Vert u \Vert^{}_{E_{N_0+3+n-l}}
\tau^M_l (t)\Big),\cr
}$$
for $t_0 < \infty$, and
$$\wp^D_n (h(t)) \leq C'_n  \sum_{0 \leq l \leq n}
\Vert u \Vert^{}_{E_{N_0+5+n-l}}
 \tau^M_l (t)$$
for $t_0 = \infty$, where $C'_n$ is a constant depending only on
 $\Vert u \Vert^{}_{E_{N_0+3}}.$

\noindent\hbox{\rm ii)}
 If $r^{}_Y \in C^0(\Rrm^+,D)$ for $Y \in \Pi',  \vert Y \vert \leq n+1,$
$(1+\lambda^{}_1)^{1/2}  r^{}_Y \in C^0(\Rrm^+,D)$ for $\vert Y \vert \leq n$
and if $[a]^l (t) \leq C_l  \Vert u \Vert^{}_{E_{N_0+l}},  l \geq 0,$
where $C_l, N_0$ and $u$ are as in Corollary 5.17, then
$$\eqalignno{
&\wp^D_n (h(t))\cr
&\quad \leq C'_n  \Big(\wp^D_n (h(t_0)) + \sum_{0 \leq l \leq n-1}
 \Vert u \Vert^{}_{E_{N_0+3+n-l}}  \wp^D_l (h(t_0))
+ \sum_{0 \leq l \leq n}  \Vert u \Vert^{}_{E_{N_0+3+n-l}}
\tau^D_l (t)\Big),\cr}
$$
for $t_0 < \infty$, and
$$\wp^D_n (h(t)) \leq C'_n  \sum_{0 \leq l \leq n}  \Vert u \Vert^{}_{E_{N_0
+3+n-l}}  \tau^D_l (t),$$
for $t_0 = \infty$, where $C'_n$ is a constant depending only on
 $\Vert u \Vert^{}_{E_{N_0+3}}.$
}\saut\penalty-9000
\noindent{\it Proof.}
It follows from the definitions of $\theta^D_j$ in Proposition 5.16 that
$$\theta^D_j (t) \leq K_j \Big((1+C_1  \Vert u \Vert^{}_{E_{N_0+1}})
C_{j+1}  \Vert u \Vert^{}_{E_{N_0+j+1}}
+ C_j  \Vert u \Vert^{}_{E_{N_0}}  \Vert u \Vert^{}_{E_{N_0+j}}\Big),$$
when the hypothesis of statement i) are satisfied.
Here $K_j$ is a numerical constant
independent of $u$ and $C_l,  l \geq 0$, are constants depending only
on $\Vert u \Vert^{}_{E_{N_0}}$. Thus for new constants depending only on
 $\Vert u \Vert^{}_{E_{N_0+1}},$
we have that
$\theta^D_j(t) \leq C_j  \Vert u \Vert^{}_{E_{N_0+j+1}}$, which together with
statement i) of Proposition 5.16 gives that
$$\wp^D_j  (f(t)) \leq C_n  \sum_{0 \leq l \leq j}
\Vert u \Vert^{}_{E_{N_0+j+1-l}}  \tau^M_l (t). \eqno{(5.208)}$$
With $b^{}_\mu = a^{}_\mu - \partial_\mu  {\vartheta} (a)$,
H\"older inequality give that
$$\Vert (1+\lambda^{}_1 (t))^{1/2}  g^{}_Y (t)
\Vert^{}_D \leq \suma_{Y_1,Y_2}^Y
 \Vert b^{}_{Y_1} (t) \Vert^{}_{L^p}  \Vert (1+\lambda^{}_1 (t))^{1/2}
 r^{}_{Y_2} (t) \Vert^{}_{L^q},$$
$p = 6/(3 - 2 \rho),  q = 3/\rho$. We have that
$$\eqalignno{
&\Vert (1+\lambda^{}_1 (t))^{1/2}  r^{}_{Y_2} (t) \Vert^{}_{L^{3/\rho}}\cr
&\quad\leq \Vert (1+\lambda^{}_1 (t))^{\rho - 1/2}  \delta (t)^{1 - \rho}
r^{}_{Y_2} (t) \Vert^{}_{L^{3/\rho}}\cr
&\quad\leq \big(\Vert \delta (t)^{(1-\rho) 3/\rho}
\vert (1+\lambda^{}_1 (t))^{\rho - 1/2}
 r^{}_{Y_2} (t) \vert^{3/\rho - 2} \Vert^{}_{L^\infty}
\Vert (1+\lambda^{}_1 (t))^{\rho - 1/2}  r^{}_{Y_2} (t)
\Vert^2_{L^2}\big)^{\rho/3}\cr
&\quad\leq \big(\Vert \delta (t)^{- 3/2\rho}  \vert \delta (t)^{3/2}
 (1+\lambda^{}_1 (t))^{\rho - 1/2}  r^{}_{Y_2} (t) \vert^{3/\rho - 2}
\Vert^{}_{L^\infty}
\Vert (1+\lambda^{}_1 (t))^{\rho - 1/2}  r^{}_{Y_2} (t)
\Vert^2_D\big)^{\rho / 3}\cr
&\quad\leq (1+t)^{- 1/2}  \Vert \delta (t)^{3/2}  (1+\lambda^{}_1
(t))^{\rho - 1/2}  r^{}_{Y_2} (t) \Vert^{1 - 2 \rho / 3}_{L^\infty}
\Vert (1+\lambda^{}_1 (t))^{\rho - 1/2}
r^{}_{Y_2} (t) \Vert^{2 \rho / 3}_D\cr
&\quad\leq C_{\vert Y_2 \vert}  \Vert u \Vert^{}_{E_{N_0+\vert Y_2 \vert}}
 (1+t)^{- 1/2},\cr
}$$
where $C_{\vert Y_2 \vert}$ depends only on
$\Vert u \Vert^{}_{E_{N_0}}$ and where
the last step follows from the hypothesis of statement i).
Since $\Vert b^{}_{Y_1}
(t) \Vert^{}_{L^p} \leq C_\rho  \Vert \vert \nabla \vert^\rho
b^{}_{Y_1} (t) \Vert^{}_{L^2}$, it now follows from statement i) of
Lemma 4.5 (with $\rho=1$) that
$$\eqalignno{
&\Vert (1+\lambda^{}_1 (t))^{1/2}  g^{}_Y (t) \Vert^{}_D \cr
&\quad\leq C_n
\sum_{n_1+n_2=\vert Y \vert}  (1+t)^{- 1/2 - \varepsilon}
\sup_{0 \leq s \leq t}  \Big((1+s)^\eta  \wp^{M^1}_{n_1+1}
 \big((a(s), \dot{a} (s))\big)\Big)  \Vert u \Vert^{}_{E_{N_0+n_2}},\cr}
$$
where we have used that $\partial_\mu  a^\mu = 0$. This shows that
$$\int^{t''}_{t'}  (1+s)^{- 3/2 + \rho}  \wp^D_j ((1+\lambda^{}_1 (s))^{1/2}
 g(s))  ds \leq C_j  \sum_{n_1+n_2=j}
\tau^M_{n_1+1} (t)  \Vert u \Vert^{}_{E_{N_0+n_2}}, \eqno{(5.209)}$$
where $C_j$ depends only on $\Vert u \Vert^{}_{E_{N_0}}$.
It follows from inequalities
(5.208) and (5.209) that $Q_n$ defined in Corollary 5.17 satisfies
$$\eqalignno{
Q_n (t) &\leq C_n \Big(\sum_{0 \leq l \leq n}
\Vert u \Vert^{}_{E_{N_0+n+1-l}}
 \tau^M_l (t)\cr
&\quad{}+ \sum_{0 \leq l \leq n-1}  \Vert u \Vert^{}_{E_{N_0+3+n-l}}
\Big(\sum_{0 \leq j \leq l}  \Vert u \Vert^{}_{E_{N_0+l+1-j}}  \tau^M_j
(t)
+ \sum_{0 \leq j \leq l}  \Vert u \Vert^{}_{N_0+l-j}    \tau^M_{j+1}
(t)\Big)\Big)\cr
&\leq C'_n  \sum_{0 \leq l \leq n}  \Vert u \Vert^{}_{E_{N_0+3+n-l}}
 \tau^M_l (t),\cr
}$$
where we have used Corollary 2.6. Here $C_n$ depends only on
 $\Vert u \Vert^{}_{E_{N_0}}$
and $C'_n$ only on $\Vert u \Vert^{}_{E_{N_0+3}}$. Statement i) of the
corollary now follows from Corollary 5.17. Since the proof of statement
ii) is so similar, we omit it.
\vfill\eject
\noindent{\titre 6. Construction of the modified wave operator and its
inverse.}
\saut
The properties, given by Theorem 4.9, Theorem 4.10 and Theorem 4.11 of $A_J,
\phi'_J$, where we choose $J \geq (3/2 - \rho) / (1 - \rho),  1/2 <
\rho < 1$, of the approximate solution $A_J, \phi^{}_J$ defined by
(4.135a) and (4.135b), permit to prove the existence of modified wave
operators for the M-D equations. To do this we shall first introduce {\it new
functions} $A^*_n$, $\phi^*_n$,
 $n \geq 0$, slightly different from the above approximate solutions,
but having the advantage of satisfying the {\it Lorentz gauge
condition} $\partial_\mu  A^{* \mu}_n = 0$. Then, we shall prove that $(A^*_n,
\exp (i {\vartheta} (A^*_n))  \phi^*_n),  n \geq 0$, {\it converges to
a solution of the M-D equations}.

Let us fix once for all $J \geq (3/2 - \rho) / (1 - \rho) + 2$. Given $u = (f,
\dot{f}, \alpha) \in E^{\circ\rho}_\infty$, we shall prove that there is a
unique solution $\phi^*_0 \in C^0(\Rrm^+,D)$ such that
$\Vert \phi^*_0 (t) - e^{t {\cal D}}
\alpha \Vert^{}_D \fl 0$,  when  $t\fl \infty$, of the equation
$$\phi^*_0 (t) = e^{t {\cal D}} \alpha + i  \int^\infty_t
e^{(t-s) {\cal D}} ((A_{J,\mu} + B_{J,\mu})  \gamma^0  \gamma^\mu
 \phi^*_0) (s)	ds,\quad  t \geq 0, \eqno{(6.1{\rm a})}$$
where $B_{J,\mu} = - \partial_\mu  {\vartheta} (A_J),  0 \leq \mu \leq 3.$
$A^*_{0,\mu}$ is defined by
$$\eqalignno{
A^*_{0,\mu} (t) &= \cos (\vert \nabla \vert t)	f_\mu + \vert \nabla \vert^{-1}
 \sin (t \vert \nabla \vert)  \dot{f}_\mu &(6.1{\rm b})\cr
&\qquad{}- \int^\infty_t  \vert \nabla \vert^{-1}  \sin (\vert \nabla \vert
(t-s))	((\phi^*_0)^+  \gamma^0  \gamma_\mu
\phi^*_0) (s)  ds,\quad  t \geq 0.\cr
}$$
For $n \geq 0$, we then prove that the equations
$$\phi^*_{n+1} (t) = e^{t {\cal D}} \alpha + i	\int^\infty_t
e^{(t-s) {\cal D}}  ((A^*_{n,\mu} + B^*_{n,\mu}) \gamma^0
\gamma^\mu  \phi^*_{n+1}) (s)  ds,\quad  t \geq 0, \eqno{(6.2{\rm a})}$$
where $B^*_{n,\mu} = - \partial_\mu  {\vartheta} (A^*_n)$ and
$$\eqalignno{
A^*_{n+1,\mu} (t) &= \cos (\vert \nabla \vert t)
f_\mu + \vert \nabla \vert^{-1}
 \sin (t \vert \nabla \vert)  \dot{f}_\mu &(6.2{\rm b})\cr
&\qquad{}- \int^\infty_t  \vert \nabla \vert^{-1}
\sin (\vert \nabla \vert
 (t-s))  ((\phi^*_{n+1})^+  \gamma^0  \gamma_\mu
 \phi_{n+1}) (s)  ds,\quad  t \geq 0, \cr
}$$
have a unique solution $(\phi^*_{n+1},  A^*_{n+1},  {d \over dt}
 A^*_{n+1}) \in C^0(\Rrm^+, E^\rho_0)$ and that $(\Delta^{*M}_n,
\Delta^{*D}_n)$, where
$$
\Delta^{*M}_{n, \mu} = A^*_{n+1,\mu} - A^*_{n,\mu}, \quad
  \Delta^{*D}_n =\phi^*_{n+1} - \phi^*_n, \quad n \geq 0,\eqno{(6.3)}
$$
converges to zero in an appropriate space when $n \fl \infty.$

To begin with we complete Theorem 4.9 and Theorem 5.10 by {\it decrease
properties}, established in chapter 5, of solutions of the
inhomogeneous Dirac equation. We adapt the notation used in Theorem 4.9,
Theorem 4.10, Corollary 5.9 and use
$(\delta (t)) (x) = 1+t + \vert x \vert$. We recall that $\lambda_0$ and
$\lambda_1$ are defined  in Theorem 5.5.
\saut
\noindent{\bf Lemma 6.1.}
{\it
If $n \geq 0$, and $1/2 < \rho < 1$ then there exists $N_0 \geq 0$ such that
$$\eqalignno{
&\sup_{t \geq 0}  \Big(\Vert (1+\lambda_1 (t))^{k/2}  (D^l
\phi'_{n,Y} (u ; v_1,\ldots,v_l)) (t) \Vert^{}_D\cr
&\hskip2mm{}+ (1+t)^{\chi^{}_{n+1}}    \Vert (1+\lambda_1 (t))^{k/2}  (D^l
(\phi'_{n+1,Y} - \phi'_{n,Y}))  (u ; v_1,\ldots,v_l)) (t) \Vert^{}_D\cr
&\hskip2mm{}+ \Vert (\delta (t))^{3/2}	    (1+\lambda_1 (t))^{k/2}  (D^l
 \phi'_{n,Y} (u ; v_1,\ldots,v_l)) (t) \Vert^{}_{L^\infty}\cr
&\hskip2mm{}+ \Vert (\delta (t))^{3/2 + \chi^{}_{n+1}}	(1+\lambda_1 (t))^{k/2}
(D^l (\phi'_{n+1,Y} - \phi'_{n,Y})) (u ; v_1,\ldots,v_l)) (t)
\Vert^{}_{L^\infty}\Big)\cr
&\quad{} \leq F_{L,l,k} (\Vert u \Vert^{}_{E^\rho_{N_0}})  {\cal R}^l_{N_0,L+k}
 (v_1,\ldots,v_l) + F'_{L,l,k} (\Vert u \Vert^{}_{E^\rho_{N_0}})
\Vert u \Vert^{}_{E^\rho_{N_0+L+k}}  \Vert v_1 \Vert^{}_{E^\rho_{N_0}}\cdots
\Vert v_l \Vert^{}_{E^\rho_{N_0}},\cr
}$$
for all $L \geq 0,  l \geq 0,  k \geq 0,  Y \in \Pi',
 \vert Y \vert \leq L,	u, v_1,\ldots,v_l \in E^{\circ\rho}_\infty,$
where $F_{L,l,k}$ and $F'_{L,l,k}$ are increasing polynomials on $[0,\infty[$.
}\saut
\noindent{\it Proof.}
If $k = 0$ then the statement follows from Theorem 4.9 and Theorem 4.10.
Let $k \geq 1.$ It follows from (4.137c) that $(i \gamma^\mu \partial_\mu + m)
\phi'_0 = 0$ and that
$$(i \gamma^\mu  \partial_\mu + m) \phi'_n = \gamma^\mu (A_{n-1,\mu} +
B_{n-1,\mu}) \phi'_{n-1},\quad  n \geq 1. \eqno{(6.4)}$$
Corollary 5.9 (with $G=0,  g = g^{}_n = \gamma^\mu (A_{n-1,\mu} + B_{n-1,\mu})
\phi'_{n-1}$ for $n \geq 1$ and $g = g^{}_0 = 0$ for $n=0$ and $L \geq
\max (3,n+k-1))$
gives
$$\eqalignno{
&\wp^D_L \big((1+\lambda_1 (t))^{k/2}  \phi'_n (t)\big) &(6.5)\cr
&\quad\leq C_{L+k}
\Big(\wp^D_{L+k} (\phi'_n (t))
+ \sum_{0 \leq j \leq k-1}  \wp^D_{L+j} \big((1+\lambda_1 (t))^{(k+1-j)/2}
 g^{}_n (t)\big)\Big),\cr
}$$
$n \geq 0$, $L \geq 0$, $k \geq 1$, where $C_{L+k}$ is a numerical constant.
This gives that
$$\wp^D_L \big((1+\lambda_1 (t))^{k/2}  \phi'_0 (t)\big) \leq C_{L+k}
\wp^D_{L+k} (\phi'_0 (t)) = C_{L+k}  \Vert \alpha \Vert^{}_{D_{L+k}},
\eqno{(6.6)}$$
$L \geq 0,  k \geq 1$, where we have used that $\phi'_0 (t) = e^{t {\cal D}}
\alpha$. Moreover using, for $n \geq 1$, that
$$\eqalignno{
&\wp^D_{L+j} \big((1+\lambda_1 (t))^{(k+1-j)/2}  g^{}_n\big)&(6.7)\cr
&\qquad{}\leq C'_{L+j}
\sum_{i_1+i_2=L+j}
\Vert u \Vert^{}_{E^\rho_{N_0+i_1}}  \wp^D_{i_2} \big((1+\lambda_1
(t))^{(k-j)/2}
\phi'_{n-1} (t)\big),\cr
}$$
$C'_{L+j}$ being a polynomial in $\Vert u \Vert^{}_{E^\rho_{N_0}}$
which follows from Theorem 4.9 and Corollary 4.12, inequality (6.5) gives that
$$\eqalignno{
&\wp^D_L \big((1+\lambda_1 (t))^{k/2}  \phi'_n (t)\big)&(6.8)\cr
&\qquad{} \leq C^{(n)}_{L+k}
\Big(\wp^D_{L+k}  (\phi'_n (t))
+ \sum_{\scr i_1 + i_2 = L+j\atop\scr 0 \leq j \leq k-1}
\Vert u \Vert^{}_{E^\rho_{N_0+i_1}}
 \wp^D_{i_2}  \big((1+\lambda_1 (t))^{(k-j)/2}  \phi'_{n-1} (t)\big)\Big),\cr
}$$
$n\geq 1$, $L \geq 0$, $k \geq 1$,
where $C^{(n)}_{L+k}$ is a polynomial in $\Vert u \Vert^{}_{E^\rho_{N_0}}$. We
make the induction hypothesis,
$$\wp^D_L \big((1+\lambda_1 (t))^{k/2}  \phi'_N (t)\big) \leq C^{(N)}_{L+k}
\Vert u \Vert^{}_{E^\rho_{N_0+L+k}},\quad  L \geq 0,  k \geq 1, \eqno{(6.9)}$$
for $0 \leq N \leq n-1$, where $C^N_{L+k}$ is some polynomial in
$\Vert u \Vert^{}_{E^\rho_{N_0}}.$
According  to inequality~(6.6) the hypothesis is true for $N=0$ since $\Vert
\alpha \Vert^{}_{D_{L+k}} \leq \Vert u \Vert^{}_{E^\rho_{N_0+L+k}}$.
It follows from inequality~(6.8) and Theorem~4.9 that (for a new polynomial
$C^{(n)}_{L+k})$
$$\eqalignno{
&\wp^D_L \big((1 + \lambda_1 (t))^{k/2}  \phi'_n (t)\big)\cr
&\qquad{} \leq C^{(n)}_{L+k}
 \Big(\Vert u \Vert^{}_{E^\rho_{N_0+L+k}}
+ \sum_{\scr i_1+i_2 = L+j\atop\scr 0 \leq j \leq k-1}
\Vert u \Vert^{}_{E^\rho_{N_0+i_1}}
 \Vert u \Vert^{}_{E^\rho_{N_0+i_2+k-j}}\Big),\quad  n \geq 1,\cr
}$$
which together with Corollary 2.6 proves that inequality (6.9)
is true for every $N \geq 0.$

Similarly differentiation of both members of equation (6.4) $l$
times and induction give
that
$$\eqalignno{
&\wp^D_L \big((1+\lambda_1 (t))^{k/2}  (D^l  \phi'_n (u ; v_1,\ldots,v_l))
(t)\big)&(6.10)\cr
&\quad{}\leq F_{L,l,k}  (\Vert u \Vert^{}_{E^\rho_{N_0}})  {\cal R}^l_{N_0,L+k}
(v_1,\ldots,v_l)
+ F'_{L,l,k}  (\Vert u \Vert^{}_{E^\rho_{N_0}})
\Vert u \Vert^{}_{E^\rho_{N_0+L+k}}
 \Vert v_1 \Vert^{}_{E^\rho_{N_0}}\cdots\Vert v_l \Vert^{}_{E^\rho_{N_0}},\cr
}$$
$L \geq 0,  l \geq 0,  k \geq 1$, where $F_{L,l,k}$ and
$F'_{L,l,k}$ are polynomials depending on $n \geq 0.$

To prove the estimate of $\Vert (\delta (t))^{3/2}  (1+\lambda_1 (t))^{k/2}
 \phi'_{n,Y} (t) \Vert^{}_{L^\infty},  Y \in \Pi'$, we use
Theorem~5.8 with $G = 0$, equation (6.4) and note that the hypothesis of
Theorem~5.8 are satisfied due to Theorem 4.9. With the notation
$$
Q_{n,k,L} (t) = \sum_{\scr  Y \in \Pi'\atop\scr\vert Y \vert \leq L}
\Vert (\delta (t))^{3/2}
 (1+\lambda_1 (t))^{k/2}  \phi'_{n,Y} (t) \Vert^{}_{L^\infty},
\eqno{(6.11)}$$
$n \geq 0,  k \geq 0,  L \geq 0$, we then obtain that
$$\eqalignno{
Q_{n,k,L} (t) &\leq C_{k+L}  \Big(\wp^D_{k+L} (\phi'_n (t))
+ \sum_{i+j \leq k+L+7}  \wp^D_i \big(\delta (t)  (1+\lambda_1
(t))^{j/2}  g'_n (t)\big)&(6.12)\cr
&\qquad{}+ \sum_{i+j \leq k+L+9}  \wp^D_i ((1+\lambda_1 (t))^{j/2}
g^{}_n (t))\cr
&\qquad{}+ \sum_{\scr Y\in\Pi'\atop\scr\vert Y \vert + j \leq k+L}
\Vert (\delta (t))^{3/2}
(1+\lambda_1 (t))^{j/2}  g^{}_{n,Y} (t) \Vert^{}_{L^\infty}\Big),\quad	n
\geq 0,  k \geq 0,  L \geq 0,\cr
}$$
where $g'_n = (2m)^{-1}  (m - i \gamma^\mu  \partial_\mu)
 g^{}_n$,  $g^{}_{n,Y} = \xi^D_Yg^{}_n$,  $g'_{n,Y}
= \xi^D_Y  g'_n$ and where $C_{k+L}$ is a numerical constant. Similarly
as we obtained (6.7) it follows that
$$\eqalignno{
&\sum_{i+j \leq k+L+9}	\wp^D_i \big((1+\lambda_1 (t))^{j/2}  g^{}_n
(t)\big)\cr
&\qquad{}\leq C'_{L+k}  \sum_{i+j \leq k+L+9}\  \sum_{i_1+i_2=i}
 \Vert u \Vert^{}_{E^\rho_{N_0+i_1}}  \wp^D_{i_2}
 \big((1+\lambda_1 (t))^{(j-1)/2}  \phi'_{n-1} (t)\big),\cr
}$$
where $C'_{L+k}$ is a polynomial in $\Vert u \Vert^{}_{E^\rho_{N_0}}$.
This inequality, Corollary 2.6 and inequality (6.9), which we have proved
is true for all $N \geq 0,$ give that
$$\sum_{i+j \leq L+k+9}  \wp^D_i \big((1+\lambda_1 (t))^{j/2}
g^{}_n (t)\big)
\leq C^{(n)}_{L+k} \Vert u \Vert^{}_{E^\rho_{N_0}}
\Vert u \Vert^{}_{E^\rho_{N_0+k+L+9}}, \quad
n \geq 0,  k \geq 0,  L \geq 0,\eqno{(6.13)}$$
where $C^{(n)}_{L+k}$ is a polynomial in $\Vert u \Vert^{}_{E^\rho_{N_0}}$.
Similarly it follows that
$$\eqalignno{
&\sum_{\scr Y\in\Pi'\atop\scr\vert Y \vert + j \leq k+L}
\Vert (\delta (t))^{3/2}
(1+\lambda_1 (t))^{j/2}  g^{}_{n,Y} (t) \Vert^{}_{L^\infty}&(6.14)\cr
&\qquad{}\leq C'_{k+L}  \sum_{i+j \leq k+L}\ \sum_{i_1 + \vert Y_2 \vert = i}
 \Vert u \Vert^{}_{E^\rho_{N_0+i_1}}  \Vert \delta (t)
(1+\lambda_1 (t))^{j/2}  \phi'_{n-1, Y_2} (t) \Vert^{}_{L^\infty},\cr
}$$
$n \geq 0,  k \geq 0,  L \geq 0$, where $C'_{k+L}$ is a
polynomial in $E^\rho_{N_0}$. To estimate the term with
$g'_n$ in (6.12), we note that according to equality (5.7a) (with $G = 0$)
$$\eqalignno{
(m - i \gamma^\mu  \partial_\mu)  g^{}_n &= \gamma^\nu (A_{n-1, \nu}
+ B_{n-1, \nu})  (m + i \gamma^\mu  \partial_\mu)
\phi'_{n-1}& (6.15)\cr
&\qquad{}- 2i (A^\mu_{n-1} + B^\mu_{n-1})  \partial_\mu  \phi'_{n-1} - i
 \phi'_{n-1}  \partial_\mu (A^\mu_{n-1} + B^\mu_{n-1})\cr
&\qquad{}- {i \over 4}	(\gamma^\mu  \gamma^\nu - \gamma^\nu
\gamma^\mu)  \phi'_{n-1}  (\partial_\mu  A_{n-1, \nu} -
\partial_\nu  A_{n-1, \mu}), \quad n \geq 1,\cr
}$$
where we have used the gauge invariance of the last term. The first term on the
right-hand side vanish when $n=1$, since $(m + i \gamma^\mu  \partial_\mu)
 \phi'_0 = 0$. According to Theorem 4.9 and Corollary~4.12 it follows
from equality (6.15) that
$$\eqalignno{
&\Vert \delta (t)(1+\lambda_1 (t))^{j/2}  g'_{n,Y} (t)
\Vert^{}_D&(6.16)\cr
&\quad{}\leq C^{(n)}_{j, \vert Y \vert}  \suma_{Y_1,Y_2}^Y
\Big(\Vert u \Vert^{}_{E^\rho_{N_0 + \vert Y_1 \vert}}
 \Vert (\delta (t))^{1/2} (1+\lambda_1 (t))^{j/2}  (i \gamma^\mu
 \partial_\mu + m)  \phi'_{n-1, Y_2} (t) \Vert^{}_D\cr
&\qquad{}+ \Vert u \Vert^{}_{E^\rho_{N_0 + \vert Y_1 \vert}}
\Vert (1+\lambda_1 (t))^{j/2}
 \phi'_{n-1, Y_2} (t) \Vert^{}_D\Big)\cr
&\qquad{}+ 2  \suma_{Y_1,Y_2}^Y  \Vert \delta (t) (1+\lambda_1 (t))^{j/2}
 (A^\mu_{n-1, Y_1} (t) + B^\mu_{n-1, Y_1} (t))	\partial_\mu
 \phi'_{n-1, Y_2} (t) \Vert^{}_D,\cr
}$$
where $Y \in \Pi',  n \geq 1$, and where $C^{(n)}_{j, \vert Y \vert}$
is a polynomial in $\Vert u \vert_{E^\rho_{N_0}}$. If $Y_1 \in \sg^1$ in the
last term in the right-hand side of inequality (6.16),
then it follows from Theorem 4.9 and Corollary 4.12 that
$$\eqalignno{
&\Vert \delta (t)(1+\lambda_1 (t))^{j/2}  (A^\mu_{n-1,Y_1} (t)
+ B^\mu_{n-1,Y_1} (t))	\partial_\mu  \phi'_{n-1,Y_2} (t) \Vert^{}_D&(6.17)\cr
&\qquad{}\leq C^{(n)}_{\vert Y_1 \vert}
\Vert u \Vert^{}_{E^\rho_{N_0 + \vert Y_1 \vert}}
 \sum_{0 \leq \mu \leq 3}  \Vert (1+\lambda_1 (t))^{j/2}
\phi'_{n-1, P_\mu Y_2} (t) \Vert^{}_D, \cr
}$$
where $Y_1 \in \sg^1$ and where $C^{(n)}_{\vert Y_1 \vert}$
is a polynomial in $\Vert u \Vert^{}_{E^\rho_{N_0}}$. If $Y_1 \in \Pi'
\cap U({\frak{sl}}(2, \Crm))$, then $y_\mu
 (A_{n-1, Y_1} + B_{n-1, Y_1})^\mu  (y) = 0$. It therefore
follows from inequality (5.7d), the definition of $\xi^D_{M_{\mu \nu}}$,
Theorem 4.9 and Corollary 4.12 that
$$\eqalignno{
&\Vert \delta (t)(1+\lambda_1 (t))^{j/2}  (A^\mu_{n-1, Y_1}
(t) + B^\mu_{n-1,Y_1} (t))  \partial_\mu  \phi'_{n-1, Y_2}
(t) \Vert^{}_D&(6.18)\cr
&\qquad{}\leq C^{(n)}_{\vert Y_1 \vert}
\Vert u \Vert^{}_{E^\rho_{N_0 + \vert Y_1 \vert}}
 \Big(\sum_{0 \leq \mu \leq 3}	\Vert (1+\lambda_1 (t))^{j/2}
 \phi'_{n-1,P_\mu Y_2} (t) \Vert^{}_D\cr
&\qquad{}\qquad{}+ \sum_{0 \leq \mu < \nu \leq 3}
\Vert (1+\lambda_1 (t))^{j/2}
\phi'_{n-1, M_{\mu \nu} Y_2} (t) \Vert^{}_D
+ \Vert (1+\lambda_1 (t))^{j/2}  \phi'_{n-1, Y_2} (t) \Vert^{}_D\Big),\cr
}$$
where $Y_1 \in \Pi' \cap U({\frak{sl}}(2, \Crm))$ and where
$C^{(n)}_{\vert  Y_1 \vert}$ is a
polynomial in $\Vert u \Vert^{}_{E^\rho_{N_0}}$. Since $(i \gamma^\mu
\partial_\mu + m)  \phi'_0 = 0$, it follows from equation (6.4), Theorem
4.9 and Corollary 4.12 that
$$\eqalignno{
&\Vert (\delta (t))^{1/2}  (1+\lambda_1 (t))^{j/2}  (i
 \gamma^\mu  \partial_\mu + m)	\phi'_{n-1,Z} (t)
\Vert^{}_D &(6.19)\cr
&\qquad{}\leq C^{(n)}_{\vert Z \vert}	\sum_{i_1+i_2 = \vert Z \vert}
\vert u \vert_{E^\rho_{N_0+i_1}}  \wp^D_{i_2} \big((1+\lambda_1 (t))^{j/2}
 \phi'_{n-2} (t)\big),\quad  n \geq 1,\cr
}$$
where $Z \in \Pi',  C^{(n)}_{\vert Z \vert}$ is a polynomial in $\Vert
u \Vert^{}_{E^\rho_{N_0}}$ and where the right-hand side is given
the value zero when $n=1$. Inequalities (6.16), (6.17), (6.18) and
(6.19) give that
$$\eqalignno{
&\Vert \delta (t)(1+\lambda_1 (t))^{j/2}  g'_{n,Y}
(t) \Vert^{}_D& (6.20)\cr
&\quad{}\leq C^{(n)}_{j, \vert Y \vert}
\Big(\sum_{i_1+i_2+i_3 = \vert Y \vert}
 \Vert u \Vert^{}_{E^\rho_{N_0+i_1}}  \Vert u \Vert^{}_{E^\rho_{N_0+i_2}}
 \wp^D_{i_3} \big((1+\lambda_1 (t))^{j/2}  \phi'_{n-2} (t)\big)\cr
&\qquad{}+ \sum_{i_1+i_2 = \vert Y \vert}  \Vert u \Vert^{}_{E^\rho_{N_0+i_1}}
 \wp^D_{i_2} \big((1+\lambda_1 (t))^{j/2}  \phi'_{n-1} (t)\big)\Big), \quad
n \geq 1,  Y \in \Pi',\cr
}$$
where the sum over $i_1+i_2+i_3 = \vert Y \vert$
is absent when $n=1$ and where
$C^{(n)}_{j, \vert Y \vert}$ is a polynomial in
$\Vert u \Vert^{}_{E^\rho_{N_0}}.$
The already proved inequality (6.9),
inequality (6.20) and Corollary 2.6 give that
$$\Vert \delta (t)(1+\lambda_1 (t))^{j/2}  g'_{n,Y} (t)
\Vert^{}_D
\leq C^{(n)}_{j + \vert Y \vert}  \Vert u \Vert^{}_{E^\rho_{N_0}}
\Vert u \Vert^{}_{E^\rho_{N_0+j+\vert Y \vert + 1}},
\eqno{(6.21)}$$
$Y \in \Pi'$, $n \geq 1$,  $j \geq 0$,
where $C^{(n)}_{j + \vert Y \vert}$ is a polynomial
in $\Vert u \Vert^{}_{E^\rho_{N_0}}.$
It follows from inequalities (6.9), (6.12), (6.13), (6.14) and (6.21) that
$$\eqalignno{
&Q_{n,k,L} (t)& (6.22{\rm a})\cr
&\quad{}\leq C^{(n)}_{k+L}  \Big(\Vert u \Vert^{}_{E^\rho_{N_0+L+k+9}}
+ \sum_{\scr Z \in \Pi'\atop\scr i+j+\vert Z \vert \leq k+L}  \Vert u
\Vert^{}_{E^\rho_{N_0+i}}  \Vert \delta (t) (1+\lambda_1 (t))^{j/2}
\phi'_{n-1,Z} (t) \Vert^{}_{L^\infty}\Big),\cr
}$$
$n \geq 1$, and
$$Q_{0,k,L} (t) \leq C^{(0)}_{k+L}  \Vert u \Vert^{}_{E^\rho_{N_0+L+k+9}}.
\eqno{(6.22{\rm b})}$$
It follows by induction from (6.22a) and (6.22b) that
$$Q_{n,k,L} (t) \leq C^{(n)}_{k+L}  \Vert u \Vert^{}_{E^\rho_{N_0 (n)+L+k}},
\eqno{(6.23)}$$
where $n \geq 0,  k \geq 0,  L \geq 0$, where $C^{(n)}_{k+L}$
is a polynomial in $\Vert u \Vert^{}_{E^\rho_{N_0}}$ and where
$N_0 (n) \in \Nrm$
depends on $n$. Inequality (6.23) shows that the estimate of
$$\Vert (\delta (t))^{3/2}  (1+\lambda_1 (t))^{k/2}
\phi'_{n,Y} (t) \Vert^{}_{L^\infty}$$
given by the lemma is true.

The proof of the announced $L^\infty$-estimates of the derivatives
$D^l  \phi'_{n,Y}$ is so similar to the proof of inequality (6.23)
that we omit it. Also the $L^2$- and $L^\infty$-estimates of
$D^l (\phi'_{n+1, Y} - \phi'_{n,Y})$ are obtained
so similarly by using Theorem 4.10, that we omit the proofs.
This ends the proof of Lemma 6.1.

Next we shall prove the existence of $\phi^*_j$ and
$A^*_j$ for $j=0$ and $j=1,$ which permits to prove the existence for
$j \geq 2$  by a contraction theorem.
Denote $A^*_{j,Y} = \xi^M_Y  A^*_j$ and $\phi^*_{j,Y} = \xi^D_Y
\phi^*_j$, (c.f. (4.81d)).
\saut
\noindent{\bf Proposition 6.2.}
{\it
Let $u = (f, \dot{f}, \alpha) \in E^{\circ\rho}_\infty,  1/2 < \rho < 1$,
and $0 \leq \rho' \leq 1$. Then equation (6.1a) has a unique solution
$\phi^*_0 \in C^0(\Rrm^+,D)$ and equation (6.2a) has a unique
solution $\phi^*_1 \in C^0(\Rrm^+,D).$
Moreover there exists $N_0$ such that
$$\eqalignno{
&\sup_{t \geq 0}  \Big(\Vert \big((D^l (A^*_{j,Y},  A^*_{j, P_0 Y}))
 (u ; v_1,\ldots,v_l)\big) (t) \Vert^{}_{M^\rho_0}\cr
&\quad{}+ \sum_{n+k \leq L}  \wp^D_n \big((1+\lambda_1 (t))^{k/2}  \big((D^l
 \phi^*_j)  (u ; v_1,\ldots,v_l)\big) (t)\big)\cr
&\quad{}+ (1+t)^{1+\rho'-\rho}  \Vert \big((D^l
(\Delta^{*M}_{0,Y}, \Delta^{*M}_{0,P_0 Y}))
 (u ; v_1,\ldots,v_l)\big) (t) \Vert^{}_{M^{\rho'}_0}\cr
&\quad{}+ (1+t)^{3/2 - \rho}	\Vert \big((D^l  \Delta^{*D}_{0,Y})
 (u ; v_1,\ldots,v_l)\big) (t) \Vert^{}_D\cr
&\quad{}+ (1+t)^{2 - \rho}  \Vert (1+t+\vert\cdot\vert)
\big(\carre \big((D^l  \Delta^{*M}_{0,Y})  (u ; v_1,\ldots,v_l)\big)\big) (t)
\Vert^{}_{L^2}\cr
&\quad{}+ (1+t)^{1 - \rho}  \Vert (1+t+\vert\cdot\vert)
\big(\carre \big((D^l  \Delta^{*M}_{0,Y})  (u ; v_1,\ldots,v_l)\big)\big) (t)
\Vert^{}_{L^{6/5}}\Big)\cr
&\hskip-6.75mm +\sup_{t \geq 0,  x \in \Rrm^3}  \Big((1+t+\vert x
\vert)^{3/2-\rho}
 \vert \big((D^l  A^*_{j,Y})  (u ; v_1,\ldots,v_l)\big)(t,x) \vert\cr
&\quad{}+ (1+t+\vert x \vert)  (1+\big\vert t - \vert x \vert \big\vert)^{1/2}
(1+\big\vert t - \vert x \vert \big\vert_\delta)^{1 - \rho}  \vert \big((D^l
A^*_{j,P_0 Y})	(u ; v_1,\ldots,v_l)\big) (t,x) \vert\cr
&\quad{}+ \sum_{\scr Z \in \Pi'\atop\scr\vert Z \vert + k \leq L}
\big\vert (1+t+\vert x \vert)^{3/2}  (1+(\lambda_1 (t)) (x))^{k/2}  \big((D^l
 \phi^*_{j,Z})	(u ; v_1,\ldots,v_l)\big) (t,x) \big\vert\Big)\cr
&\qquad{}\leq C_{L,l}	\Big({\cal R}^l_{N_0,L}  (v_1,\ldots,v_l) + \Vert u
\Vert^{}_{E^\rho_{N_0+L}}  \Vert v_1 \Vert_{E^\rho_{N_0}}\cdots\Vert v_l
\Vert^{}_{E^\rho_{N_0}}\Big),\cr
}$$
for every $j=0,1,  \delta > 0,  Y \in \Pi',  \vert
Y \vert \leq L,  l \geq 0$, $u, v_1,\ldots,v_l \in E^{\circ\rho}_\infty,$
where $C_{L,l}$ is a constant depending only on
$\Vert u \Vert^{}_{E^\rho_{N_0}}$
and $\delta$. Here $\big\vert t - \vert x \vert \big\vert_\delta
= \big\vert t - \vert x \vert \big\vert$
for $\vert x \vert \leq \delta t$ and $\big\vert t -
\vert x \vert \big\vert_\delta = 0$
for $\vert x \vert > \delta t$. Moreover $\partial_\mu	A^{* \mu}_j = 0,$
$j = 0,1$, and, as functions of $u \in E^{0\rho}_{\infty}$,
$(A^*,\phi^*)$ has a zero of order at least $1$ and
$(\Delta^{*M},\Delta^{*D})$ has a zero of order at least $2$ at $u=0$.
}\saut
\noindent{\it Proof.}
If $\phi^*_0 \in C^0 (\Rrm^+,D)$, satisfying
$\Vert \phi^*_0 (t) - e^{t {\cal D}}
\alpha \Vert^{}_D \fl 0$ when $t \fl \infty$, exists, then it is unique.
 In fact, if $w$ is the difference
between two solutions of equation (6.1a) then
$(i \gamma^\mu	\partial_\mu + m - \gamma^\mu (A_{J,\mu} + B_{J,\mu}))
 w= 0$ and $\Vert w (t) \Vert^{}_D \fl 0$ when $t \fl \infty.$
It then follows from Theorem 4.9 and Corollary 5.17 that $w = 0.$

It follows from equation (4.137c) that
$$(\phi^*_0 - \phi'_{J+1}) (t) = i \int^\infty_t  e^{(t-s) {\cal D}}
 \big((A_{J,\mu} + B_{J,\mu}) \gamma^0	\gamma^\mu
(\phi^*_0 - \phi'_J)\big) (s)  ds,$$
and then from the relation (4.140b) that
$$\eqalignno{
(\phi^*_0 - \phi'_J - \Delta^D_J) (t) &= i \int^\infty_t  e^{(t-s) {\cal D}}
 \big((A_{J,\mu} + B_{J,\mu})  \gamma^0  \gamma^\mu
 \Delta^D_J\big) (s)  ds\cr
&\quad{}+ i \int^\infty_t  e^{(t-s) {\cal D}}  \big((A_{J,\mu} + B_{J,\mu})
 \gamma^0  \gamma^\mu  (\phi^*_0 - \phi'_J - \Delta^D_J)\big)
(s)  ds.\cr
}$$
This shows that
$$\big(i  \gamma^\mu  \partial_\mu + m - \gamma^\mu (A_{J,\mu} +
B_{J,\mu})\big)  (\phi^*_0 - \phi'_J - \Delta^D_J)
= \gamma^\mu (A_{J,\mu} + B_{J,\mu})\Delta^D_J. \eqno{(6.24)}$$
It follows from Theorem 4.9 that $[A_J]^l (t) \leq C_l
\Vert u \Vert^{}_{E_{N_0+l}},$
where $N_0$ is an integer independent of $u$ and $l$ and where
$C_l$ is a constant
depending only on $\Vert u \Vert^{}_{E_{N_0}}$. Statement ii) of
Corollary~5.18, equation
(6.24) and the relation $\phi^*_0 - \phi'_J - \Delta^D_J =
\phi^*_0 - \phi'_{J+1}$,
then give
$$\eqalignno{
\hskip-10mm\wp^D_n  \big(\phi^*_0 (t) - \phi'_{J+1} (t)\big)
&\leq C_n  \sum_{0 \leq l \leq n}
 \Vert u \Vert^{}_{E_{N_0+5+n-l}}
\int^\infty_t \Big((1+s)^{- 5/2 + \rho}  \wp^D_{l+1}  (\Delta^D_J
(s))&(6.25)\cr
&\qquad{}+ (1+s)^{- 3/2}  \wp^D_l \big((1+\lambda_1 (s))^{1/2}	\Delta^D_J
(s)\big)\cr
&\qquad{}+ (1+s)^{- 3+2 \rho}	\wp^D_l (\Delta^D_J (s))\cr
&\qquad{}+ (1+s)^{- 3/2+\rho}
 \wp^D_l \big(((i \gamma^\mu  \partial_\mu + m) \Delta^D_J) (s)\big)\Big)
 ds.\cr
}$$
Since $\Delta^D_J = \phi'_{J+1} - \phi'_J$, it follows from (4.137c) that
$$(i \gamma^\mu  \partial_\mu + m)  \Delta^D_J = \gamma^\mu
(A_{J,\mu} + B_{J,\mu})  \phi'_J - \gamma^\mu (A_{J-1,\mu} + B_{J-1, \mu})
 \phi'_{J-1}.$$
This equality, Theorem 4.9, Theorem 4.10, statement ii) of
Lemma 4.5 and Corollary 2.6 give that
$$\wp^D_n \big(((i \gamma^\mu  \partial_\mu + m) \Delta^D_J) (t)\big) \leq C_n
 (1+t)^{-1}  \Vert u \Vert^{}_{E_{N_0}}  \Vert u \Vert^{}_{E_{N_0+n}},$$
where $C_n$ is a polynomial in $\Vert u \Vert^{}_{E_{N_0}}$. This inequality,
inequality (6.25), Theorem 4.9, Lemma 6.1 and Corollary 2.6 then give
(after redefining
$N_0$)
$$\wp^D_n \big(\phi^*_0 (t) - \phi'_{J+1} (t)\big) \leq (1+t)^{- 3/2 + \rho}
C_n  \Vert u \Vert^{}_{E_{N_0}}  \Vert u \Vert^{}_{E_{N_0+n}},
\quad n \geq 0,  t \geq 0.\eqno{(6.26{\rm a})}$$
Differentiation of equation (6.24) gives similarly by induction
$$\eqalignno{
&\wp^D_n \big(\big((D^l (\phi^*_0 - \phi'_{J+1}))
(u ; v_1,\ldots,v_l)\big) (t)\big)&(6.26{\rm b})\cr
&\quad{} \leq C_{n,l}	(1+t)^{- 3/2 + \rho}  \Big({\cal R}^l_{N_0,n}
 (v_1,\ldots,v_l)
+ \Vert u \Vert^{}_{E_{N_0+n}}  \Vert v_1 \Vert^{}_{E_{N_0}}\cdots
 \Vert v_l \Vert^{}_{E_{N_0}}\Big),\cr
}$$
$n \geq 0,  l \geq 0,  t \geq 0$, where $C_{n,l}$ depends only on
$\Vert u \Vert^{}_{E_{N_0}}$. Inequality (6.26b),
the estimates of $\wp^D_n \big(\big((D^l  \phi'_J)
(u ; v_1,\ldots,v_l)\big) (t)\big)$
given by Theorem 4.9 prove the estimate of $\Vert \big((D^l  \phi^*_{0,Y})
 (u ; v_1,\ldots,v_l)\big) (t) \Vert^{}_D$ given by the inequality of
 the proposition.

The announced $L^2$- and $L^\infty$-estimates of $(1+\lambda_1 (t))^{k/2}
\big((D^l  \phi^*_{0,Y})  (u ; v_1,\ldots,v_l)\big) (t)$ follow in a
such a similar way as in the proof of Lemma 6.1 that we omit the details. The
$M^\rho$-estimate of $(A^*_{0,Y}, A^*_{0, P_0 Y})$ follows from the definition
of $A^*_0$ and from the already proved estimates of $\phi^*$.
It follows from definition (4.138a) that
$$\eqalignno{
&A^*_{0,\mu} (t) - A_{J+1,\mu} (t) - \Delta^M_{J+1,\mu} (t)& (6.27)\cr
&\qquad{}= - \int^\infty_t  \vert \nabla \vert^{-1}  \sin (\vert \nabla \vert
 (t-s))  (\overline{\phi}^{*}_0  \gamma_\mu  \phi^*_0 -
\overline{\phi}'_{J+1}  \gamma_\mu  \phi'_{J+1}) (s)
ds.\cr
}$$
This gives
$$\eqalignno{
&\wp^{M^{\rho'}}_n \big((A^*_0 - A_{J+1} - \Delta^M_{J+1}) (t),
{d \over dt}  (A^*_0 - A_{J+1} - \Delta^M_{J+1}) (t)\big)&(6.28)\cr
&\qquad{}\leq C_n (1+t)^{-1-\rho'+\rho}  \Vert u \Vert^{}_{E_{N_0+n}},\quad
t \geq 0,  0 \leq \rho \leq \rho',\cr
}$$
where $C_n$ depends only on $\Vert u \Vert^{}_{E_{N_0}}$ and where
we have used (6.26a),
(6.27), the already proved $L^2$- and $L^\infty$-estimates of $\phi^{*}_{0}$,
those given by
Theorem 4.9 of $\phi'_{J+1}$ and the convexity properties of the seminorms
given
by Corollary 2.6. Estimate (6.28), Theorem 4.9 and Theorem 4.11 give the
$L^2$-estimates of $\Delta^{*M}_{0,Y}$ in the proposition.
Using statement ii), statement
iii) and statement iv) of Proposition 5.6, we obtain that
$$\eqalignno{
&(1+t+\vert x \vert)  (1+\big\vert t - \vert x \vert \big\vert)^{1/2}
\vert (A^*_{0,Y} - A_{J+1,Y} - \Delta^M_{J+1,Y}) (t,x) \vert\cr
&\qquad{}\leq C \Big(\wp^M_{\vert Y \vert + 2}
\big((A^*_0 - A_{J+1} - \Delta^M_{J+1}) (t),
{d\over dt}(A^*_0 - A_{J+1} - \Delta^M_{J+1}) (t)\big)\cr
&\qquad\qquad{}+ \sum_{\vert Z \vert \leq \vert Y \vert + 2}  (1+t)
\Vert (\xi^M_Z	G) (t) \Vert^{}_{L^{6/5} (\Rrm^3, \Rrm^4)}\Big),\cr
}$$
where $G_\mu = \overline{\phi}^*_0  \gamma_\mu	\phi^*_0 -
\overline{\phi}'_{J+1}
 \gamma_\mu  \overline{\phi}'_{J+1}$ is obtained using equation
(6.27). The estimates for $\phi^*_0$ and $\phi'_{J+1}$ give
$$(1+t)  \Vert (\xi^M_Z  G) (t) \Vert^{}_{L^{6/5}} \leq C_{\vert Z \vert}
 \Vert u \Vert^{}_{E_{N_0+\vert Z \vert + 2}}  (1+t)^{\rho - 1}.$$
It then follows from inequality (6.28) that
$$\eqalignno{
&(1+t)^{1 - \rho}  (1+t+\vert x \vert)  (1+\big\vert t - \vert x \vert
\big\vert)^{1/2}  \vert (A^*_{0,Y} - A_{J+1,Y} - \Delta^M_{J+1,Y}) (t,x)
\vert&(6.29)\cr
&\quad{}\leq C_{\vert Y \vert}  \Vert u \Vert^{}_{E_{N_0+\vert Y \vert}},
\quad  t \geq 0,  x \in \Rrm^3,  Y \in \Pi',\cr
}$$
where $C_{\vert Y \vert}$ depends only on $\Vert u \Vert^{}_{E_{N_0}}$,
and where $N_0+2$ has been redefined by $N_0$. Since $(1+t)^{1-\rho}
(1+t+\vert x \vert)	(1+\big\vert t - \vert x \vert \big\vert)^{1/2}
\geq (1+t+\vert x \vert)^{3/2-\rho}$
for $(t,x) \in \Rrm^+ \times \Rrm^3$ the estimate for $L^\infty$-norms of
$A^*_{0,Y}$ and $A^*_{0,P_\mu Y}$ in the proposition follows from inequality
(6.29) and from Theorem 4.9 and Theorem 4.11. Using Theorem 4.10,
Theorem 4.11 with interpolation and inequality (6.8) we obtain that
$$\wp^{M^{\rho'}}_n \big((A^*_0 - A_J) (t),  {d \over dt} (A^*_0 - A_J)
(t)\big)
\leq C_n  (1+t)^{-1-\rho'+\rho}  \Vert u \Vert^{}_{E_{N_0+n}},$$
$n \geq 0,  t \geq 0,  0 \leq \rho' \leq 1$. The proof of
the existence and the properties of $\phi^*_1, A^*_1$ and $\Delta^*_0$ are so
similar to that of $\phi^*_0$ and $A^*_0$ that we omit it.
That the gauge conditions
$\partial_\mu  A^{*\mu}_j  = 0$ is satisfied follows directly from the fact
that
$$\carre  i \partial_\mu  A^{*\mu}_j= (\overline{i \gamma^\mu
\partial_\mu  \phi^*_j})  \phi^*_j + \overline{\phi}^*_j (i \gamma^\mu
 \partial_\mu  \phi^*_j) = 0,$$
by the equation satisfied by $\phi^*_j$ and by the fact that $(f, \dot{f}) \in
M^{\circ\rho}$. This proves the proposition.

We now make the following variable transformation in the M-D equations:
$$\psi' = e^{i {\vartheta} (A)}  \psi,\quad  \Phi = \psi' - \phi^*,\quad
K = A - A^*, \quad  \phi^* = \phi^*_1, \quad  A^* = A^*_1, \eqno{(6.30)}$$
and denote $\Delta^{*M}_\mu = \Delta^{*M}_{0,\mu}$. The M-D equations are then
transformed into
$$\eqalignno{
&\big(i \gamma^\mu  \partial_\mu + m - \gamma^\mu (K_\mu + A^*_\mu -
\partial_\mu {\vartheta} (K+A^*))\big)\Phi&(6.31{\rm a})\cr
&\qquad{}= \gamma^\mu	\big(K_\mu + \Delta^{*M}_\mu - \partial_\mu
{\vartheta} (K + \Delta^{*M})\big)  \phi^*,\cr
&\carre  K_\mu = \overline{\Phi}  \gamma_\mu  \Phi +
\overline{\Phi}  \gamma_\mu  \phi^* + \overline{\phi}^*
\gamma_\mu  \Phi& (6.31{\rm b})\cr
}$$
and
$$\partial_\mu K^\mu = 0.\eqno(6.31{\rm c})$$
We introduce the Banach space ${\cal F}_N,  N \geq 0$, which is the completion
of the space of all functions $(K,\Phi)$ such that $K_Y \in C^0(\Rrm^+,L^2),$
$K_{P_0 Y} \in C^0(\Rrm^+, \vert \nabla \vert  L^2)$, $K_{P_\mu Y} \in
C^0(\Rrm^+, L^2)$, $\delta \carre  K_Y \in C^0(\Rrm^+, L^2)$,$\delta \carre
 K_Y \in C^0(\Rrm^+,L^{6/5})$, $\Phi_Y
\in C^0(\Rrm^+,D)$ for $Y \in \Pi',  \vert Y \vert \leq N$, and such that
$\Vert (K, \Phi) \Vert^{}_{{\cal F}_N} < \infty$, with respect to the norm
$\Vert \cdot \Vert^{}_{{\cal F}_N}$ defined by
$$\eqalignno{
&(e^{}_N (K, \Phi)) (t)&(6.32{\rm a})\cr
&\quad{} = (1+t)^{1 - \rho}  \wp^{M^0}_N (K(t), \dot{K} (t))
+ (1+t)^{2 - \rho}  \wp^{M^1}_N (K(t), \dot{K} (t)) + (1+t)^{3/2 - \rho}
 \wp^D_N (\Phi (t))\cr
&\qquad{}+ \Big(\sum_{\scr Y \in \Pi'\atop\scr  \vert Y \vert \leq N}
\big((1+t)^{2 - \rho}
 \Vert \delta (t)  \carre K_Y (t) \Vert^{}_{L^2} + (1+t)^{1 - \rho}
 \Vert \delta (t)  \carre K_Y (t) \Vert^{}_{L^{6/5}}\big)^2\Big)^{1/2}\cr
}$$
and
$$\Vert (K,\Phi) \Vert^{}_{{\cal F}_N} = \sup_{t \geq 0}  (e^{}_N (K, \Phi))
(t), \eqno{(6.32{\rm b})}$$
where $(\delta (t)) (x) = 1+t + \vert x \vert,  1/2 < \rho < 1,
K_Y = \xi^M_Y K,  \dot{K}_Y = K_{P_0 Y},  \Phi_Y = \xi^D_Y
 \Phi$ for $Y \in \Pi'$ (c.f. (4.81d)). We denote by ${\cal F}^M_N$
 the subspace of elements
$(K,0)$ and by ${\cal F}^D_N$ the subspace of elements $(0, \Phi).$

If $N$ is sufficiently large and $K \in {\cal F}^M_N$, then equation (6.31a)
has a unique solution $\Phi \in C^0(\Rrm^+,D)$ satisfying
$\displaystyle{\lim_{t \fl \infty}}
 \Vert \Phi (t) \Vert^{}_D = 0$. It will be proved that the equation
$$\carre  K'_\mu = \overline{\Phi}  \gamma_\mu  \Phi +
\overline{\Phi}  \gamma_\mu  \phi^* + \overline{\phi}^*
\gamma_\mu  \Phi,\quad	K' \in {\cal F}^M_N, \eqno{(6.33)}$$
has a unique solution for $N$ sufficiently large and that $(u,K) \mapsto K'=
{\cal N}(u, K)$ defines a map ${\cal N}\colon
 E^{\circ\rho}_\infty \times {\cal F}^M_N
\fl {\cal F}^M_N.$ $K'$ depends on $K$ via $\Phi$ in equation (6.33) and
depends on $u$ via $A^*, \phi^*$ and $\Phi$. For $u$ and $K$
sufficiently small this map turns out to be a contraction map
(in the variable $K$).

The definition of the space ${\cal F}^M_N$, gives space-time decrease
properties
of the absolute value of its elements.
\saut
\noindent{\bf Lemma 6.3.}
{\it
There exists $C_\rho > 0$ such that
$$(1+t+\vert x \vert)  (1+t)^{1 - \rho}  (1+\big\vert t -
\vert x \vert \big\vert)^{1/2}
 \vert K_Y (t,x) \vert \leq C_\rho  \Vert K \Vert^{}_{{\cal F}^M_{\vert
Y \vert +2}},$$
for $Y \in \Pi',  \vert Y \vert \leq N,  t \in \Rrm^+,
x \in \Rrm^3,  K \in {\cal F}^M_{N+2}.$
}\saut
\noindent{\it Proof.}
Statement ii) of Proposition 5.6, with $F_{Y (t)} = \carre  K_Y (t),$
gives that
$$\eqalignno{
(1+t)^{3/2}  \vert K_Y (t,x) \vert
&\leq C_\delta  \Big(
\sum_{\scr \vert Y\vert \leq 2\atop\scr
Z \in\Pi' \cap U({\frak{sl}}(2, \Crm))}  \big(\Vert (K_{ZY}
(t),  K_{P_0 ZY} (t)) \Vert^{}_{M^0}&(6.34)\cr
&\qquad{}+ (1+t)  \Vert \carre  K_{ZY} (t) \Vert^{}_{L^{6/5}}\big) + \Vert
(1 - \Delta)  K_Y (t) \Vert^{}_{L^2}\Big),\cr
}$$
for $0 \leq \vert x \vert \leq \delta t,  t \geq 0$, where $0 < \delta < 1$
and where $C_\delta$ is a numerical constant. Definitions (6.32a) and (6.32b)
of
the norm in ${\cal F}^M_N$, then give that
$$(1+t)^{3/2}  \vert K_Y (t,x) \vert \leq C_\delta  (1+t)^{- 1 + \rho}
 \Vert K \Vert^{}_{{\cal F}^M_{\vert Y \vert + 2}}, \eqno{(6.35{\rm a})}$$
for $0 \leq \vert x \vert \leq \delta t,  t \geq 0$, where $C_\delta$
is a new constant.

Similarly it follows from statement iii) and iv) of Proposition 5.6 that
$$(1+t)  (1+\big\vert t - \vert x \vert \big\vert)^{1/2}  \vert K_Y (t,x)
\vert \leq C_{\delta_1, \delta_2}  (1+t)^{-1 + \rho}  \Vert
K \Vert^{}_{{\cal F}^M_{\vert Y \vert + 2}}, \eqno{(6.35{\rm b})}$$
for $\delta_1 t \leq \vert x \vert \leq \delta_2 t,  t \geq 0$, where
$0 < \delta_1 < 1 < \delta_2$, and that
$$(1+\vert x \vert)^{3/2}  \vert K_Y (t,x) \vert \leq C_\delta
(1+t)^{- 1 + \rho}  \Vert K \Vert^{}_{{\cal F}^M_{\vert Y \vert + 2}},
\eqno{(6.35{\rm c})}$$
for $0 \leq t \leq \delta \vert x \vert,  \vert x \vert \geq 0$, where
$0 < \delta < 1$. The inequality of the lemma now follows by choosing suitably
$\delta, \delta_1$ and $\delta_2$. This proves the lemma.

In order to study the map ${\cal N}$ we consider the equation
$$\big(i \gamma^\mu  \partial_\mu + m - \gamma^\mu (K_\mu + A^*_\mu -
\partial_\mu
 {\vartheta} (K+A^*))\big)\Psi = g, \eqno{(6.36)}$$
for $\Psi \in {\cal F}^D_N$, where $K \in {\cal F}^M_N$. Most often $g =
\gamma^\mu$
$(L_\mu - \partial_\mu	{\vartheta} (L)  \phi^*,  L
\in {\cal F}^M_N$. We shall use
in next proposition the notation
$$\chi^{(n)} = \Vert u \Vert^{}_{E^\rho_{N_0+n}} +
\Vert K \Vert^{}_{{\cal F}^M_n},
\quad n \geq 0, \eqno{(6.37)}$$
where $N_0$ is given by Proposition 6.2 and where $u \in E^\rho_\infty$ and $K
\in {\cal F}^M_n$ and $\chi^{}_{N,n}$,  $N \geq 0$,  $n \geq 0$ is
defined by $\chi^{}_{N,n} = b_n$ given by formula (5.87) with $a^{(n)} =
\chi^{(N+n)}.$
It follows from this definition that $\chi^{}_{N,n} < \infty$ if
$u \in E^\rho_\infty$
and $K \in {\cal F}^M_{N+n}$. We note that according to Proposition
6.2 and Lemma 6.3
$$[A^*]^n (t) \leq C_n	\Vert u \Vert^{}_{E^\rho_{N_0+n}}, \quad
[K]^n (t) \leq C  \Vert K \Vert^{}_{{\cal F}^M_{n+2}},\quad t \geq 0,
\eqno{(6.38{\rm a})}$$
where $C_n$ depends only on $\rho$ and (polynomially) on
$\Vert u \Vert^{}_{E^\rho_{N_0}}$
and $C$ depends only on $\rho$,	$1/2 < \rho < 1$. Using Corollary 2.6
we then obtain that
$$[A^*]_{N,n} (t) \leq C_{n+N}	\Vert u \Vert^{}_{E^\rho_{N_0+N+n}}, \quad
N,n \geq 0,  t \geq 0, \eqno{(6.38{\rm b})}$$
where $C_{n+N}$ depends only on $\rho$ and (polynomially) on
$\Vert u \Vert^{}_{E^\rho_{N_0+N}}.$
Moreover, it follows from inequalities (6.38a) that
$$[A^* + K]^n (t) \leq C_n  \chi^{(n+2)}, \quad [A^* + K]_{N,n} (t)
\leq C_{N+n}  \chi^{}_{N+2,n}, \eqno{(6.38{\rm c})}$$
for $t \geq 0,	n \geq 0,  N \geq 0$, where $C_n$ and $C_{N+n}$
depends only on $\rho$ and (polynomially) on $\Vert u \Vert^{}_{E^\rho_{N_0}}.$

In order to state next proposition we introduce the following notations:
$$\eqalignno{
\overline{Q}_n (t) &= \sup_{s \geq t}  \big((1+s)^{3/2 - \rho}\wp^D_n
(f(s))\big) + Q'_n (t),& (6.39{\rm a})\cr
Q'_n (t) &= \sum_{\scr n_1+n_2=n\atop\scr  n_2 \leq n - 1}  \chi^{}_{5,n_1}
\sup_{s \geq t}  \big((1+s)^{3/2-\rho}  \wp^D_{n_2} (f(s))& (6.39{\rm b})\cr
&\qquad{}+ (1+s) \wp^D_{n_2} ((1+\lambda_0 (s))^{1/2}  g(s))\big),\quad n \geq
0,
 t \geq 0,\cr
}$$
where $f$ is defined by (5.111b) for equation (6.36),
$$R^{(1)}_n (t) = \sup_{s \geq t}  \big((1+s)^{3/2-\rho}
(R'_{n+7} (s) + R^2_{n+9}
(s) + R^\infty_n (s))\big)+ \overline{Q}_{n+8} (t),  \eqno{(6.39{\rm c})}$$
$n \geq 0$,  $t \geq 0$, where $R'_n,  R^2_n,  R^\infty_n$
are defined in Theorem 5.8,
$$\eqalignno{
h'_n (L,t) &= \sum_{\sscr n_1+n_2+n_3+n_4=
n\atop{ \sscr n_1 \leq n-1, n_2 \leq L-1
\atop\sscr  n_3+n_4 \leq n-L}}
 \overline{\chi}^{(n_1)}  (1+\chi^{}_{4,n_2}) (1+\chi^{}_{10,n_3})
R^{(1)}_{n_4} (t)& (6.39{\rm d})\cr
&\qquad{}+ \sum_{\sscr n_1+n_2=n\atop{\sscr1 \leq n_1 \leq L
\atop\sscr n_2 \leq L}}  \chi^{}_{5,n_1}
\big(\overline{Q}_{n_2} (t) + \sup_{s \geq t}  \big((1+s)  \wp^D_{n_2}
((1+\lambda_0 (s))^{1/2}  g(s))\big)\big),\cr
}$$
$n \geq 0,  L \geq 0,  t \geq 0$, where $\overline{\chi}^{(n)} =
\Vert u \Vert_{E^\rho_{N_0+n+1}} + \Vert K \Vert^{}_{{\cal F}^M_n},  n
\geq 0,$
$$\eqalignno{
h''_n (t) &= \Vert K \Vert^{}_{{\cal F}^M_n}  (1+\chi^{(3)})
\sum_{n_1+n_2=1}  (1+\chi^{}_{10,n_1})	R^{(1)}_{n_2} (t), \quad
 n \geq 0,&(6.39{\rm e})\cr
k''_n (t) &= (1+t)^{- 1/2}  S^{\rho,n} (t)  \overline{H}_0 (t)
+ \int^\infty_t  \Big((1+s)^{-2 + \rho}  S^{\rho,n} (s)
(1+[A]^1  (s))	\overline{H}_1 (s)\hskip10mm &(6.39{\rm f})\cr
&\qquad{}+ \big(\sum_{\scr Y \in \Pi'\atop\scr \vert Y \vert = n }
\Vert \gamma^\mu
 G_{Y \mu} (s)	g(s) \Vert^2_D \big)^{1/2}\Big)  ds, \quad
n \geq 0,\cr
}$$
where $\overline{H}_n (t) = \overline{H}_n (\infty, t)$,
which makes sense by formula (5.169) since $\Psi \in {\cal F}^D_{N}.$

\saut
\noindent{\bf Proposition 6.4.}
{\it
Let $1/2 < \rho < 1,  n \geq 19,  u \in E^{\circ\rho}_\infty,
 K \in {\cal F}^M_n$ and let $n_0$ be the integer part
of $n/2 + 5$. Let $g^{}_Y \in C^0(\Rrm^+,D)$ for $Y \in \Pi',  \vert Y
\vert \leq n$, and let
$$\sup_{t \geq 0}  \big((1+t)^{3/2 - \rho}	(R'_{n_0-1} (t) +
R^2_{n_0+1} (t) + R^\infty_{n_0} (t))\big) < \infty.$$
If for each $Y \in \Pi',  \vert Y \vert \leq n$, there are two functions
$e^{}_1,e^{}_2\in C^0(\Rrm^+,D)$ such that $g^{}_Y=e^{}_1+e^{}_2$, where:
\psaut
\noindent\hbox{\rm a)} $e^{}_1 \in L^1 (\Rrm^+,D)$
\psaut
\noindent\hbox{\rm b)} $(m - i \gamma^\mu\partial_\mu + \gamma^\mu   G_\mu)
 e^{}_2 \in L^1 (\Rrm^+,D)$ and $\lim_{t \fl \infty}  \Vert e^{}_2
(t) \Vert^{}_D = 0$,
\psaut
\noindent and if $\sup_{t \geq 0}  \big((1+t)
\wp^D_{n-1}  \big((1+\lambda_0 (t))^{1/2}  g(t)\big)\big) < \infty$ and
$\sup_{t \geq 0}  \big((1+t)^{3/2-\rho}  \wp^D_n
(f(t))\big) <\penalty10000 \infty$, then there
exists a unique solution $\Psi \in C^0(\Rrm^+,D)$
of equation (6.36) such that\penalty-10000
$\Vert \Psi (t) \Vert^{}_D \fl\penalty10000 0$, when $t \fl
\infty$. Moreover $\Psi \in {\cal F}^D_n,  k'_l (n_0,t) \leq (1+t)^{-
3/2 + \rho}  C_l  h'_l (n_0,t)$, for $0 \leq l \leq n,$
$k'_l (n_0,t) + k''_l (t) \leq (1+t)^{- 3/2 + \rho}  (C_l
h'_l (n_0,t) + C  h''_l (t))$ for $n_0+1 \leq l \leq n$, where $C_l$
and $C$ depend only on $\rho$ and $\Vert u \Vert^{}_{E^\rho_{N_0}}$, the
functions $t \mapsto h'_l (n_0,t),  0 \leq l \leq n$, and $h''_l$, $0
\leq l \leq n$, are uniformly bounded on $\Rrm^+$ and the following estimates
are satisfied:
$$\eqalignno{
&\hbox{\rm i)}\ \wp^{}_l  (\Psi (t)) \leq (1+t)^{- 3/2 + \rho}  C_l
\overline{Q}_l (t),\cr
\noalign{\noindent \hbox{for $0 \leq l \leq n-5$, where $C_l$
depends only on $\rho$~and~$\chi^{(5)}$,}}
&\hbox{\rm ii)}\ \wp^D_l (\Psi (t)) \leq (1+t)^{- 3/2 + \rho}  C_l
\Big(\ds{\sup_{s \geq t}}  \big((1+s)^{3/2 - \rho}  \wp^D_l (f(s))\big) +
a_l  h^\infty_l (n_0,t)\hskip25.5mm\cr
&\hskip30mm{}+ \sum_{\sscr n_1+n_2=l\atop{\sscr  1 \leq n_1
\leq n_0\atop\sscr  n_2 \geq n_0+1}}
 (1+\chi^{}_{5,n_1})  \big(\ds{\sup_{s \geq t}} \big((1+s)^{3/2 - \rho}
\wp^D_{n_2} (f(s))\big) + a_{n_2}  h^\infty_{n_2} (n_0,t)\big)\Big)\cr
}$$
and $k^\infty (n_0,t) \leq (1+t)^{-3/2 + \rho}  C  h^\infty
(n_0,t)$, for $n_0 + 1 \leq l \leq n$, where $h^\infty_j (n_0,t) = h'_j (n_0,t)
+ h''_j (t) + \chi^{(5)}  \sup_{s \geq t} \big((1+s)
\wp^D_{j-1} ((1+\lambda_0 (s))^{1/2}  g(s))\big)$ and where $C_j$
depends only on $\rho$ and $\chi^{(5)}$, and $a_j$ depends only on $\rho$ and
$\chi^{(13)},$
$$\eqalignno{
&\hbox{\rm iii)} \sum_{\scr Y \in \Pi' \atop\scr\vert Y \vert + k \leq l}
\Vert
(\delta (t))^{3/2}  (1+\lambda_1 (t))^{k/2}  \Psi_Y (t)
\Vert^{}_{L^\infty}\leq (1+t)^{- 3/2 + \rho}  a_l  \ds{\sum_{n_1+n_2=l}}
(1+ \chi^{}_{10,n_1})R^{(1)}_{n_2} (t),\cr
}$$
for $0 \leq l \leq n-10$, where $a_l$ depends only on $\rho$ and $\chi^{(13)}$
and where it is supposed that
$$\sup_{t \geq 0}  \big((1+t)^{3/2 - \rho}(R'_{l+7} (t) +
R^2_{l+9} (t) + R^\infty_l (t))\big) < \infty,$$
$$\eqalignno{
\hbox{\rm iv)}\
\wp^D_{l,i} (\Psi (t)) &\leq \wp^D_{l,i} (f(t)) + (1+t)^{- 3/2+\rho}
(a_l  h'_l (n_0,t) + C  h''_{l,i} (t))\cr
&\qquad{}+ (1+t)^{- 3/2+\rho}C_l  \sum_{\sscr n_1+n_2=l\atop{\sscr  1
\leq n_1 \leq n_0\atop\sscr  n_2 \geq n_0+1}}  \chi^{}_{5,n_1}
 \sup_{s \geq t}  \big((1+s)^{1/2}\wp^D_{n_2} (\Psi(s))\big)\cr
&\qquad{}+ (1+t)^{- 3/2+\rho}	C_l  \chi^{(5)}  \sup_{s\geq t}
\big((1+s)^{1/2}\cr
&\qquad{}\qquad{}\big(\wp^D_l (\Psi (s))^{\varepsilon}  \wp^D_{l,i+1}  (\Psi
(s))^{1 - \varepsilon} + R^0_{l-1,1} (s)^{2 (1-\rho)}  \wp^D_{l-1}
 (\Psi (s))^{2 \rho - 1}\big)\big),\hskip7mm\cr
}$$
for $n_0 + 1 \leq l \leq n,  0 \leq i \leq l$, where $h''_{l,0} =
h''_l$, $h''_{l,i} = 0$ for $1 \leq i \leq l$, where $C$ depends only on
$\rho$ and $\chi^{(13)}$, $C_l$ depends only on $\rho$ and $\chi^{(5)}$ and
$a_l$ depends only on $\rho$ and $\chi^{(13)}$, and where $\varepsilon=
\max(1/2,2(1-l))$.
}\saut
\noindent{\it Proof.}
To prove the inequality of statement i) of the proposition, we first note that
the hypothesis on $A = A^* + K$ of Theorem 5.13, with $l$ instead of $n$, are
satisfied for $0 \leq l \leq n-5$, according to the definition of the norm
$\Vert\cdot\Vert^{}_{{\cal F}^M_n}$, according to Proposition 6.2 and according
to inequalities (6.38c). Since also the hypothesis on $g$ are satisfied for $0
\leq l \leq n-1$ it follows from Theorem 5.13 that there exists a unique
solution
$\Psi \in C^0(\Rrm^+,D)$ of equation (6.36), such that
$\Vert \Psi (t) \Vert^{}_D \fl 0$,
when $t \fl \infty$, and that this solution satisfies $\wp^D_l (\Psi (t)) \leq
C_l
 Q^\infty_l (t)$, $0 \leq l \leq n-5$, where $Q^\infty_l (t)$ is defined
in Theorem 5.13, and where $C_l$ depends only on $\rho$ and $[A]^3 (\infty).$
The inequality of statement i) now follows from definition (6.39a) of
$\overline{Q}_n (t)$
and from inequalities (6.38c).

Next we shall use Theorem 5.14 with $n_0$ instead of $L,l$ instead of $n$ and
with $A = A^* + K$. Since $n \geq 18$, $n_0 + 1 \leq l \leq n$, it follows from
the definition of $n_0$ that $18 \leq n_0 + 9 \leq l + 8 \leq 2 n_0$. We note
that $\partial^\mu  A_\mu = 0$, according to Proposition 6.2 and the
definition of the space ${\cal F}^M_n$. Since $[A]^{n_0+2} (t) \leq C_{n_0}$
$\chi^{(n_0+4)}$, according to (6.38c), and $n_0+4 \leq n$, the hypothesis in
Theorem 5.14, that $A_Y$ is a continuous map from $\Rrm^+$ to weighted
$L^\infty$ spaces, are satisfied. To prove also that the other hypothesis on
$A$ are satisfied,
we estimate $S^{\rho, N},  0 \leq N \leq n$. Since $\carre
A^*_\mu = \overline{\phi}^*  \gamma_\mu  \phi^*$ according to
equation (6.2b) and substitution (6.30) it follows from Proposition 6.2 and
Corollary 2.6 that
$$\Vert (\delta (t))^{3/2}  \carre  A^*_Y (t) \Vert^{}_{L^2(\Rrm^3,
\Rrm^4)} \leq C_{\vert Y \vert}  \Vert u \Vert^{}_{E^\rho_{N_0}}
\Vert u \Vert^{}_{E^\rho_{N_0+\vert Y \vert}}, \eqno{(6.40)}$$
for $Y \in \Pi'$, where $C_{\vert Y \vert}$ depends only on $\rho$ and
$\Vert u \Vert^{}_{E^\rho_{N_0}}.$
The gauge conditions $\partial^\mu  A^*_\mu = 0$ and $\partial^\mu
K_\mu = 0$, the inequality $\Vert \vert \nabla \vert^\varepsilon
h \Vert^{}_{L^2 (\Rrm^3, \Crm)} \leq \Vert h \Vert^{1 - \varepsilon}_{L^2}
\Vert \vert \nabla \vert h \Vert^\varepsilon_{L^2}$ for $0 \leq \varepsilon
\leq 1$, equalities
(5.115a) and (5.115b) and inequality (6.40) give that
$$S^{\rho,N} (t) \leq C_N  \Vert u \Vert^{}_{E^\rho_{N_0+N}} + C \Vert K
\Vert^{}_{{\cal F}^M_N}, \eqno{(6.41{\rm a})}$$
for $1/2 < \rho < 1,  N \geq 0,  t \geq 0$, where $C_N$
depends only on $\rho$ and $\Vert u \Vert^{}_{E^\rho_{N_0}}$ and where $C$
depends only on $\rho$. It now follows from definition (6.37) of $\chi^{(N)}$
that
$$S^{\rho,N} (t) \leq C_N  \chi^{(N)},\quad S^\rho_{k,N} (t) \leq
C_{k+N}  \chi^{}_{k,N}, \eqno{(6.41{\rm b})}$$
for $1/2 < \rho < 1,  N \geq 0,  k \geq 0,  t \geq 0,$
where $C_N$ and $C_{k+N}$ are constants depending only on $\rho$ and
$\Vert u \Vert^{}_{E^\rho_{N_0}}.$
It follows from (6.41a), with $N = l$, that the rest of the hypothesis on $A,$
in Theorem 5.14, are satisfied. The hypothesis concerning $g$ in Theorem 5.14
are
satisfied and so are the hypothesis on $R'_{n_0-1}, R^2_{n_0+1}$ and
$R^\infty_{n_0-8}.$

In order to estimate $k^\infty_j (n_0,t)$, we begin by estimating
$\overline{H}_j (t) = \overline{H}_j (\infty, t)$ defined in (5.169):
$$\overline{H}_j (t) = \sum_{n_1+n_2=j}  (1+S^\rho_{10,n_1} (t))
\big(R^2_{n_2+9} (t) + R^\infty_{n_2} (t)
+ R'_{n_2+7} (t) + Q^\infty_{n_2+8} (t)\big),\quad  0 \leq j,\eqno{(6.42)}$$
where $Q^\infty_j$, defined in Theorem 5.13, is given by
$$\eqalignno{
Q^\infty_j (t) &= \sup_{s \geq t}  \big(\wp^D_j (f(s))\big) +
\sum_{\scr n_1+n_2=n\atop\scr
 n_2 \leq j-1}	[A]_{3,n_1} (\infty) &(6.43)\cr
&\quad{}\Big(\sup_{s \geq t}  \big(\wp^D_{n_2} (f(s))\big) + \int^\infty_t
(1+s)^{- 3/2+\rho}
 \wp^D_{n_2}  \big((1+\lambda_0 (s))^{1/2}g(s)\big) ds\Big),\cr
}$$
$0 \leq j$ and where $R'_j,  R^2_j$ and $R^\infty_j$
are given in Theorem 5.8. It follows from (6.38c), (6.39a), (6.39b) and (6.43)
that
$$\eqalignno{
Q^\infty_j (t) &\leq (1+t)^{- 3/2 + \rho}  \Big(\sup_{s \geq t}
\big((1+s)^{3/2 - \rho}
 \wp^D_j (f(s))\big) + C_j Q'_j (t)\Big)& (6.44)\cr
&\leq (1+t)^{- 3/2 + \rho}  C_j  \overline{Q}_j (t),\cr
}$$
for $0 \leq j \leq n$, where $C_j \geq 1$ depends only on $\rho$
and $\Vert u \Vert^{}_{E^\rho_{N_0}}.$
Inequalities (6.41b) and (6.44) together with equality (6.42) give that
$$\eqalignno{
\overline{H}_j (t) &\leq (1+t)^{- 3/2 + \rho} C_j\sum_{n_1+n_2=j}
(1+\chi^{}_{10,n_1})\cr
&\qquad{}\Big(\overline{Q}_{n_2+8} (t)
+ \sup_{s \geq t}  \big((1+s)^{3/2 - \rho}  \big(R^2_{n_2+9} (s) +
R^\infty_{n_2} (s) + R'_{n_2+7} (s)\big)\big)\Big),  \cr
}$$
$0 \leq j \leq n-10$, where $C_j$ depends only on $\rho$
and $\Vert u \Vert^{}_{E^\rho_{N_0}}$. Definition
(6.39c) of $R^{(1)}_j$, then gives that
$$\overline{H}_j (t) \leq (1+t)^{- 3/2 + \rho}	C_j  \sum_{n_1+n_2=j}
 (1+\chi^{}_{10,n_1})  R^{(1)}_{n_2} (t),\quad  0 \leq j \leq
n-10, \eqno{(6.45)}$$
where $C_j$ depends only on $\rho$ and $\Vert u \Vert^{}_{E^\rho_{N_0}}.$

We can now estimate $\overline{k}_n (n_0, \infty,t)$,
which we denote $\overline{k}_j (n_0,t)$
and which according to definition (5.173b) of $\overline{k}_j$,
definition (5.171) of $\overline{\wp}^{}_j$ and expression (6.42) of
$\overline{H}_j$ reads
$$\eqalignno{
&\overline{k}_j (n_0,t)&(6.46)\cr
&\quad{} = \sum_{\sscr n_1+n_2=j\atop{\sscr  1 \leq n_1 \leq n_0
\atop\sscr n_2 \leq n_0}}
\Big([A]_{3,n_1} (t)  (Q^\infty_{n_2} (t) + R^0_{n_2-1,1} (t))
+ (1+t)^{1/2}  [A]^{n_1+1} (t)  \wp^D_{n_2} (g(t))\Big) \cr
&\qquad{} + \sum_{{\sscr n_1+n_2+n_3=j\atop\sscr  n_1 \leq j-1}
\atop{\sscr  n_2 \leq n_0-1\atop\sscr  n_3 \leq j-n_0}}
S^{\rho,n_1} (t)  (1+[A]_{2,n_2} (t))  \overline{H}_{n_3} (t)\cr
&\qquad{} + \sum_{\sscr Y_1,Y_2 \in \Pi'\atop{\sscr \vert Y_1 \vert +
\vert Y_2 \vert = j
\atop\sscr  n_0 \leq \vert Y_1 \vert \leq j-1}}  (1+t)^{2-\rho}
\Vert \gamma^\mu
G_{Y_1 \mu} (t)  g^{}_{Y_2} (t) \Vert^{}_D,\quad  0 \leq j \leq n.\cr
}$$
Since $G_\mu = A^*_\mu - \partial_\mu  {\vartheta} (A^*) + M_\mu$, with
$M_\mu = K_\mu - \partial_\mu  {\vartheta} (K)$, it follows from inequality
(5.116c) that
$$\eqalignno{
&\Vert \gamma^\mu  G_{Y_1 \mu} (t)  g^{}_{Y_2} (t) \Vert^{}_D&(6.47)\cr
&\quad{} \leq
C_{\vert Y_1 \vert}  [A^*]^{\vert Y_1 \vert + 1}  (1+t)^{- 3/2+\rho}
 \Vert g^{}_{Y_2} (t) \Vert^{}_D
+ C \Vert (\delta (t))^{-1}  M_{Y_1} (t) \Vert^{}_{L^2}
\Vert \delta (t)  g^{}_{Y_2} (t) \Vert^{}_{L^\infty},\cr
}$$
$Y_1,Y_2\in \Pi'$, where $C$ and $C_{\vert Y \vert}$ depend only on $\rho$. Let
$$\Gamma_j (K,t) = \Big(\sum_{\scr Y \in \Pi'\atop\scr \vert Y \vert \leq j}
\Vert (\delta (t))^{-1}  M_Y (t) \Vert^2_{L^2 (\Rrm^3, \Rrm^4)}\Big)^{1/2}
\eqno{(6.48)}$$
for $j \geq 0$. It follows from inequality (5.124), with $M_\mu$ instead of
$G_\mu$, $\dot{K}_Y = K_{P_0 Y},  b > 1/2$, and $a = 1/2$, that
$$\eqalignno{
\Gamma_j (K,t) &\leq (1+t)^{- 1/2}  C\sup_{0 \leq s \leq t}
 \Big((1+s)^{b-1}  \wp^{M^0}_j \big((K(s), 0)\big)& (6.49{\rm a})\cr
&\quad{}+ (1+s)^b  \wp^{M^1}_j \big((L(s), \dot{L} (s))\big) + (1+s)^{b-1} j
C_j  \wp^{M^1}_{j-1} \big((L(s), \dot{L} (s))\big)\Big), \quad j \geq 0,\cr
}$$
where $C$ and $C_j$ depend only on $b$. With $b = 3/4$, we obtain that
$$\Gamma_j (K,t) \leq (1+t)^{- 1/2}  \big(C \Vert K \Vert^{}_{{\cal F}^M_j} +
j  C_j \Vert K \Vert^{}_{{\cal F}^M_{j-1}}\big),\quad  j \geq 0,
\eqno{(6.49{\rm b})}$$
where $C$ and $C_j$ are as in (6.49a). The first of inequalities (6.38a) and
inequality (6.49b) give that
$$\eqalignno{
&\sum_{\sscr Y_1,Y_2 \in \Pi'\atop{\sscr \vert Y_1 \vert + \vert Y_2 \vert = j
\atop\sscr  n_0 \leq \vert Y_1 \vert \leq j-1}}  \Vert \gamma^\mu
G_{Y_1 \mu} (t) g^{}_{Y_2} (t) \Vert^{}_D&(6.50{\rm a})\cr
&\qquad{}\leq (1+t)^{- 3/2 + \rho}  C'_j  \sum_{\scr n_1+n_2=j\atop\scr
n_0 \leq
n_1 \leq j-1}  \Vert u \Vert^{}_{E^\rho_{N_0+n_1+1}}  \wp^D_{n_2}
(g(t))\cr
&\qquad\qquad{}+ (1+t)^{- 1/2}  C_j
\sum_{\sscr Z \in \Pi'\atop{\sscr n_1 + \vert Z \vert = j\atop
\sscr  n_0 \leq n_1 \leq j-1}}  \Vert K \Vert^{}_{{\cal F}^M_{n_1}}  \Vert
\delta (t) g^{}_Z (t) \Vert^{}_{L^\infty}, \quad j \geq 0,\cr
}$$
where $C'_j$ depends only on $\rho$ and $\Vert u \Vert^{}_{E^\rho_{N_0}}$
and $C_j$ is constant. We note that the right-hand side of this inequality
vanishes if $0 \leq j
\leq n_0$. If $\overline{\chi}^{(j)} = \Vert u \Vert^{}_{E^\rho_{N_0+j+1}} +
\Vert K \Vert^{}_{{\cal F}^M_j},$
it then follows from the definition of $R^\infty_j (t)$ and from inequality
(6.50a) that
$$\eqalignno{
&\sum_{\sscr Y_1,Y_2 \in \Pi'\atop{\sscr \vert Y_1 \vert + \vert Y_2 \vert = j
\atop\sscr  n_0 \leq \vert Y_1 \vert \leq j-1}}
\Vert \gamma^\mu  G_{Y_1 \mu} (t)
g^{}_{Y_2} (t) \Vert^{}_D& (6.50{\rm b})\cr
&\qquad{}\leq C_j  \sum_{\sscr n_1+n_2=j\atop{\sscr  n_1 \leq j-1\atop\sscr
n_2 \leq j-n_0}}
\overline{\chi}^{(n_1)}  \big((1+t)^{- 3/2+\rho}  \wp^D_{n_2} (g(t)) +
(1+t)^{-1}  R^\infty_{n_2} (t)\big),\quad  j \geq 0,\cr
}$$
where $C_j$ depends only on $\rho$ and $\Vert u \Vert^{}_{E^\rho_{N_0}}$.
Inequalities
(6.38c), (6.41b), (6.45) and (6.50b), and equality (6.46) give
$$\eqalignno{
&\overline{k}_j (n_0,t)&(6.51)\cr
&\quad{}\leq C_j  \sum_{\sscr n_1+n_2=j\atop{\sscr  1 \leq n_1
\leq n_0\atop\sscr n_2 \leq n_0}}
\chi^{}_{5,n_1} \big(Q^\infty_{n_2} (t) + R^0_{n_2-1,1} (t) + (1+t)^{1/2}
\wp^D_{n_2} (g(t))\big)\cr
&\qquad{}+ (1+t)^{- 3/2 + \rho}  C_j
\sum_{{\sscr n_1+n_2+n_3+n_4=j\atop\sscr n_1
\leq j-1}\atop{\sscr  n_2 \leq n_0-1\atop\sscr  n_3+n_4 \leq j-n_0}}
\chi^{(n_1)}
(1+\chi^{}_{4,n_2}) (1+\chi^{}_{10,n_3})  R^{(1)}_{n_4} (t)\cr
&\qquad{}+ C_j  \sum_{\sscr n_1+n_2=j\atop{\sscr  n_1
\leq j-1\atop\sscr n_2 \leq j-n_0} }
\overline{\chi}^{(n_1)}  \big((1+t)^{- 3/2 + \rho}  \wp^D_{n_2} (g(t)) +
(1+t)^{-1}  R^\infty_{n_2} (t)\big), \quad j \geq 0,\cr
}$$
where $C_j$ depends only on $\rho$ and $\Vert u \Vert^{}_{E^\rho_{N_0}}$.
According to definition (6.39) of $R^{(1)}_j$ it follows from inequality
(6.51) that\hfil
\penalty-10000
$$\eqalignno{
&\overline{k}_j (n_0,t)&(6.52)\cr
&\quad{}\leq (1+t)^{- 1/2}	C_j
\Big(\sum_{\sscr n_1+n_2=j\atop{\sscr  1 \leq n_1 \leq n_0\atop\sscr
 n_2 \leq n_0}}	\chi^{}_{5,n_1}
 (1+t)^{1/2}  \big(Q^\infty_{n_2} (t) + R^0_{n_2-1,1} (t) + (1+t)^{1/2}
 \wp^D_{n_2} (g(t))\big)\cr
&\qquad{}+ (1+t)^{-1 + \rho}  \sum_{\sscr n_1+n_2+n_3+n_4=j
\atop{\sscr n_1 \leq j-1, n_2 \leq
n_0-1\atop\sscr  n_3+n_4 \leq j-n_0} } \overline{\chi}^{(n_1)}
(1+\chi^{}_{4,n_2})
(1+\chi^{}_{10,n_3})  R^{(1)}_{n_4} (t)\Big),\quad  j \geq 0,\cr
}$$
where $C_j$ depends only on $\rho$ and $\Vert u \Vert^{}_{E^\rho_{N_0}}.$

We can now estimate $k'_j (n_0, \infty, t)$ defined by (5.173a).
With the notation
$k'_j (n_0,t) = k'_j (n_0, \infty,t)$, it follows from definitions (5.171) and
(5.173a) that
$$\eqalignno{
k'_j (n_0,t) &= (1+t)^{- 3/2 + \rho}  \sum_{\sscr n_1+n_2=j\atop{\sscr
 1 \leq n_1 \leq n_0\atop\sscr  n_2 \leq n_0}}  [A]^{n_1+1} (t)
 Q^\infty_{n_2} (t)
&(6.53)\cr
&\quad{}+ (1+t)^{- 1/2}  \sum_{\scr n_1+n_2=j\atop\scr n_0+1
\leq n_1 \leq j-1}
S^{\rho,n_1}  \overline{H}_{n_2} (t) + \int^\infty_t (1+s)^{-2+\rho}
\overline{k}_j (n_0,s)	ds,  \quad j \geq 0.\cr
}$$
Inequalities (6.38c), (6.41b), (6.44), (6.45), (6.52) and (6.53) give that
$$\eqalignno{
&k'_j (n_0,t)\cr
&\quad{} \leq (1+t)^{- 3+2 \rho}  C_j  \sum_{\sscr n_1+n_2=j\atop{\sscr
1 \leq n_1 \leq n_0\atop\sscr  n_2 \leq n_0}}  \chi^{(n_1+3)}
\overline{Q}_{n_2}
(t)\cr
&\qquad{}+ (1+t)^{- 2+\rho}  C_j  \sum_{\sscr n_1+n_2+n_3=j
\atop{\sscr n_1 \leq j-1\atop\sscr
n_2+n_3 \leq j-n_0-1}}  \chi^{(n_1)}  (1+\chi^{}_{10,n_2})
R^{(1)}_{n_3} (t)\cr
&\qquad{}+ (1+t)^{- 3/2 + \rho}  C_j  \sup_{s \geq t}
\Big(\sum_{\sscr n_1+n_2=j\atop{ \sscr 1
\leq n_1 \leq n_0\atop \sscr n_2 \leq n_0}}
\chi^{}_{5,n_1}  (1+s)^{1/2}\cr
&\qquad\qquad\big(Q^\infty_{n_2} (s) + R^0_{n_2-1,1}
(s) + (1+s)^{1/2}  \wp^D_{n_2} (g(s))\big)\cr
&\qquad{}+ \sum_{{\sscr n_1+n_2+n_3+n_4=j\atop\sscr  n_1 \leq j-1}
\atop{\sscr  n_2 \leq n_0-1\atop\sscr n_3+n_4 \leq j-n_0}}
 \overline{\chi}^{(n_1)}  (1+\chi^{}_{4,n_2})
 (1+\chi^{}_{10,n_3}) (1+s)^{-1+\rho}
 R^{(1)}_{n_4} (s)\Big),\quad  j \geq 0,\cr
}$$
where $C_j$ depends only on $\rho$ and
$\Vert u \Vert^{}_{E^\rho_{N_0}}$. It follows
from this inequality, from inequality (6.44) for $Q^\infty_j$ and by regrouping
the second and the fourth sum on the right-hand side that
$$\eqalignno{
k'_j (n_0,t) &\leq (1+t)^{- 3/2 + \rho}  C_j&(6.54)\cr
&\quad\Big(\sum_{\sscr n_1+n_2=j\atop{\sscr  1 \leq n_1 \leq n_0\atop\sscr
n_2 \leq n_0}}
\chi^{}_{5,n_1}
 \big(\overline{Q}_{n_2} (t) + \sup_{s \geq t} \big((1+s)^{1/2}  R^0_{n_2-1,1}
(s) + (1+s)  \wp^D_{n_2} (g(s))\big)\big)\cr
&\quad{}+ \sum_{\sscr n_1+n_2+n_3+n_4=j\atop{\sscr  n_1 \leq j-1, n_2 \leq
n_0-1
\atop\sscr n_3+n_4 \leq j-n_0}}
 \overline{\chi}^{(n_1)}  (1+\chi^{}_{4,n_2}) (1+\chi^{}_{10,n_3})
 R^{(1)}_{n_4} (t)\Big),\quad  j \geq 0,\cr
}$$
where $C_j$ depends only on $\rho$ and $\Vert u \Vert^{}_{E^\rho_{N_0}}.$

To estimate $k_j (n_0,t)$, where $k_j (n_0,t) = k_j (n_0, \infty,t)$ is defined
in Theorem 5.14, let us introduce the notation
$$\eqalignno{
&k''_j (n_0,t)&(6.55)\cr
&\quad{}= (1+t)^{- 1/2}  S^{\rho,j} (t)  \overline{H}_0 (t)\cr
&\qquad{}+ \int^\infty_t  \Big((1+s)^{-2+\rho}  S^{\rho,j} (s)
(1+[A]^1 (s))  \overline{H}_1 (s)
+ \big(\sum_{\scr Y \in \Pi'\atop\scr \vert Y \vert = j}  \Vert \gamma^\mu
G_{Y \mu} (s)  g(s) \Vert^2_D\big)^{1/2}\Big)  ds,\cr
}$$
$j \geq 0$. It follows from inequalities (6.38c), (6.41a) and definition
(6.55) of $k''_j$ that
$$\eqalignno{
k''_j (n_0,t)
&\leq (1+t)^{-1+\rho}\big(C \Vert K \Vert^{}_{{\cal F}^M_j} + C_j \Vert u
\Vert^{}_{E^\rho_{N_0+j}}\big)  \sup_{s \geq t}  \big((1+C_1
\chi^{(3)})  \overline{H}_1 (s)\big)&(6.56)\cr
&\quad{}+ \int^\infty_t  \big(\sum_{\scr Y \in \Pi'\atop\scr \vert Y \vert = j}
\Vert \gamma^\mu  G_{Y \mu} (s)  g(s) \Vert^2_D\big)^{1/2}
ds,\quad  j \geq 0,\cr
}$$
where $C$ depends only on $\rho$ and $C_j$, $j \geq 0$ depends only on $\rho$
and $\Vert u \Vert^{}_{E^\rho_{N_0}}$. Here we have used that $- 1/2 < -1 +
\rho$
for $1/2 < \rho < 1$. Inequalities (6.38a), (6.45), (6.47), (6.49b) and (6.56)
give that
$$\eqalignno{
&k''_j  (n_0,t)&(6.57)\cr
&\quad{}\leq (1+t)^{- 3/2+\rho}  C_1 \big(\Vert K \Vert^{}_{{\cal F}^M_j} + C_j
 \Vert u \Vert^{}_{E^\rho_{N_0+j}}\big)  (1+\chi^{(3)})
\sum_{n_1+n_2=1}  (1+\chi^{}_{10,n_1})	R^{(1)}_{n_2} (t)\cr
&\qquad{}+ (1+t)^{- 3/2+ \rho}  \sup_{s \geq t}  \Big(C_j
\Vert u \Vert^{}_{E^\rho_{N_0+j}}  (1+s)  \Vert g(s) \Vert^{}_D\cr
&\qquad{}+\big(C \Vert K \Vert^{}_{{\cal F}^M_j} + j
C_j \Vert K \Vert^{}_{{\cal F}^M_{j-1}}\big)
 (1+s)^{1/2}  \Vert \delta (s)  g(s) \Vert^{}_{L^\infty}\Big),
\quad j \geq 0,\cr
}$$
where $C$ depends only on $\rho$ and $C_j,  j \geq 0$, on $\rho$ and
$\Vert u \Vert^{}_{E^\rho_{N_0}}$. By adding the terms, proportional to $\Vert
u
\Vert^{}_{E^\rho_{N_0+j}}$ and $\Vert K \Vert^{}_{{\cal F}^M_{j-1}}$, on the
right-hand side of inequality (6.57) to the second sum on the right-hand side
of
inequality (6.54), we obtain that
$$\eqalignno{
&k'_j (n_0,t) + k''_j (n_0,t)&(6.58)\cr
&\quad{} \leq (1+t)^{- 3/2+\rho}  C \Big(\Big(\sup_{s \geq t}
 \big((1+s)^{1/2}  \Vert \delta (s) g(s) \Vert^{}_{L^\infty}\big)\cr
&\qquad{}+ (1+\chi^{(3)})  \sum_{n_1+n_2=1}  (1+\chi^{}_{10,n_1})
 R^{(1)}_{n_2} (t)\Big)  \Vert K \Vert^{}_{{\cal F}^M_j}\cr
&\qquad{}+ C_j  \sum_{\sscr n_1+n_2=j\atop{\sscr  1 \leq n_1 \leq n_0
\atop\sscr n_2 \leq n_0}}
\chi^{}_{5, n_1}  \big(\overline{Q}_{n_2} (t) + \sup_{s \geq t}\big((1+s)^{1/2}
 R^0_{n_2-1,1} (s) + (1+s)  \wp^D_{n_2} (g(s))\big)\big)\cr
&\qquad{}+ C_j  \sum_{\sscr n_1+n_2+n_3+n_4=j
\atop{\sscr  n_1 \leq j-1, n_2 \leq n_0-1
\atop\sscr  n_3+n_4\leq j-n_0} } \overline{\chi}^{(n_1)}  (1+\chi^{}_{4,n_2})
(1+\chi^{}_{10,n_3})  R^{(1)}_{n_4} (t)\Big),\quad  j \geq n_0+1,\cr
}$$
where $C$ and $C_j$ depend only on $\rho$ and $\Vert u
\Vert^{}_{E^\rho_{N_0}}.$
Since
$$R^0_{j-1,1} (t) \leq (1+t)^{1/2}  \wp^D_{j-1} \big((1+\lambda_0 (t))^{1/2}
 g(t)\big),\quad  j \geq 1, \eqno{(6.59)}$$
it follows from inequality (6.54) that
$$k'_j (n_0,t) \leq (1+t)^{- 3/2 + \rho}  C_j  h'_j (n_0,t),
\quad j \geq 0, \eqno{(6.60{\rm a})}$$
and from inequality (6.58) that
$$k'_j (n_0,t) + k''_j (n_0,t) \leq (1+t)^{- 3/2}  \big(C_j  h'_j
 (n_0,t) + C  h''_j (t)\big),\quad j \geq n_0+1, \eqno{(6.60{\rm b})}$$
where $C$ and $C_j$ depend only on $\rho$ and
$\Vert u \Vert^{}_{E^\rho_{N_0}}.$
Using inequalities (6.38c), (6.59) and (6.60b), we obtain that
$$\eqalignno{
k^\infty_j (n_0,t) &\leq (1+t)^{- 3/2 + \rho}	C_j \Big(h'_j (n_0,t) +
h''_j (t)&(6.61)\cr
&\qquad{}+ \chi^{(5)}	\sup_{s \geq t}  \big((1+s)  \wp^D_{j-1}
\big((1+\lambda_0 (s))^{1/2}  g(s)\big)\big)\Big)\cr
&= (1+t)^{- 3/2 + \rho}  C_j  h^\infty_j (n_0,t),\quad
j \geq n_0+1,\cr
}$$
where $h^\infty_j (n_0,t)$ is defined in statement i) of the proposition and
where $C_j$ depends only on $\rho$ and $\Vert u \Vert^{}_{E^\rho_{N_0}}$. Now,
 $\chi^{(j)}< \infty$ and $\overline{\chi}^{(j)} < \infty$ for
$0 \leq j \leq n$, since $u \in E^{\circ\rho}_\infty$
and $K \in {\cal F}^M_n$ according to the hypothesis. It follows that
$\sup_{t \geq 0} (Q'_j (t)) < \infty$ for $0 \leq j \leq n-5$
by using expression (6.39b) and that $\sup_{t \geq 0}
\big((1+t)^{3/2 - \rho}  \wp^D_j (f(t))\big) < \infty$ for $0 \leq j \leq n$
and
$\sup_{t \geq 0} \big((1+t) \wp^D_j \big((1+\lambda_0 (t))^{1/2}
g(t)\big)\big) < \infty$ for $0 \leq j \leq n-1$. Definition (6.39a) then gives
that
$\sup_{t \geq 0}  (\overline{Q}_j (t)) < \infty$ for $0 \leq
j \leq n-5$. Definition (6.39c) and the hypothesis on $R'_{n_0-1},
R^2_{n_0+1}$ and $R^\infty_{n_0}$, now give that $\sup_{t \geq 0}
 (R^{(1)}_j (t)) < \infty$ for $0 \leq j \leq n_0-8$, since $n_0 \leq
n-5$ for $n \geq 19$. The function $t \mapsto h'_j (n_0,t)$ is uniformly
bounded
on $\Rrm^+$ for $0 \leq j \leq n$, since in its definition (6.39d) $n_0 \leq
n-5,
 n-n_0 \leq n_0-9$. The function $h''_j$, defined in (6.39e) is also
uniformly bounded on $\Rrm^+$ for $0 \leq j \leq n$. This proves, together with
inequalities (6.60a), (6.60b) and (6.61) the claimed properties of the
functions
$k'_j (n_0,\cdot )$, $k''_j$, $k^\infty_j (n_0,\cdot )$, $h'_j (n_0,\cdot )$,
$h''_j$, $h^\infty_j (n_0,\cdot )$. The estimates in statement ii) follows from
inequality (6.61), Theorem 5.14 and inequality (6.38c). Statement iii) follows
from inequalities (5.170) and (6.45). Statement iv) follows from inequalities
(5.176), (5.177), (6.38c) and from the already proved inequalities for
$k'_l (n_0,t)$ and $k''_l (t)$ in terms of $h'_l (n_0,t)$ and $h''_l (t)$.
This proves the proposition.

In order to use Proposition 6.4 for the particular case $g = \gamma^\mu (L_\mu
-
\partial_\mu  {\vartheta} (L))	\phi^*$,  $L \in {\cal F}^M_n$,
we group together {\it preliminary estimates} of $R'_j, R^\infty_j, R^2_j$
defined in Theorem 5.8.
\saut
\noindent{\bf Lemma 6.5.}
{\it
Let $1/2 < \rho < 1,  n \geq 5,  K \in {\cal F}^M_n,
L \in {\cal F}^M_n$, $r^{}_Y \in C^0 (\Rrm^+,D),  r^{}_Y = \xi^D_Y  r$
for $Y \in \Pi'$ and let
$$\sup_{t \geq 0}  \big(\wp^D_j ((1+\lambda_1 (t))^{k/2}  r(t)) +
H_l (r,t) + u^{}_l (t)\big) < \infty$$
for $j + k = l,  l \geq 0$, where
$$H_l (r,t) = \sum_{\scr Y \in \Pi'\atop\scr k + \vert Y \vert \leq l}
\Vert (\delta (t))^{3/2}  (1+\lambda_1 (t))^{k/2}  r^{}_Y (t)
\Vert^{}_{L^\infty},
\quad  l \geq 0$$
and
$$u^{}_l (t) = H_{l+1} (r,t) + \sum_{\scr Y \in \Pi'\atop\scr k +
\vert Y \vert \leq l }
\Vert (\delta (t))^{3 - \rho}  (1+\lambda_1 (t))^{k/2}  ((m+
i \gamma^\mu  \partial_\mu) r^{}_Y)  (t) \Vert^{}_{L^\infty},
\quad l \geq 0.$$
Let $A = A^* + K$ and let $G_\mu = A_\mu - \partial_\mu  {\vartheta} (A).$
If
$$g = \gamma^\mu (L_\mu - \partial_\mu	{\vartheta} (L))  r,$$
then $g^{}_Y \in C^0 (\Rrm^+,D)$, $\Vert g^{}_Y (t) \Vert^{}_D \fl 0$ when
$t \fl \infty$, $(m - i \gamma^\mu  \partial_\mu + \gamma^\mu	G_\mu)
 g^{}_Y \in L^1 (\Rrm^+,D)$ for $Y \in \Pi'$,  $\vert Y \vert \leq n$
and:
$$\eqalignno{
&\hbox{\rm i)}\ R'_l (t) \leq (1+t)^{- 3/2+\rho}  C'_l
\Big(\sum_{n_1+n_2+n_3=l}
(1+\Vert u \Vert^{}_{E^\rho_{N_0+n_1+1}})  \Vert L \Vert^{}_{{\cal F}^M_{n_2}}
u^{}_{n_3} (t)\cr
&\hskip15mm{}+ \sum_{n_1+n_2+n_3=l}\min\big(\Vert K \Vert^{}_{{\cal
F}^M_{n_1+3}}
 \Vert L \Vert^{}_{{\cal F}^M_{n_2}} , \Vert K \Vert^{}_{{\cal F}^M_{n_1}}
 \Vert L \Vert^{}_{{\cal F}^M_{n_2+3}}\big) u^{}_{n_3} (t)\Big),
\quad 0 \leq l \leq n,\hskip5mm\cr
&\hbox{\rm ii)}\
R^\infty_l (t) \leq (1+t)^{-1}  C_l  \sum_{n_1+n_2=l}
 \Vert L \Vert^{}_{{\cal F}^M_{n_1+3}}	H_{n_2} (r,t), \quad
0 \leq l \leq n-3,\cr
&\hbox{\rm iii)}\
\wp^D_l \big((\delta(t))^{1/2}(1+\lambda_1 (t))^{k/2}  g(t)\big)\cr
&\qquad\leq (1+t)^{-1/2}\sum_{n_1+n_2=l} C_{n_1,n_2} \Vert L
\Vert^{}_{{\cal F}^M_{n_1}}  H_{n_2+k} (r,t),\quad  0 \leq l \leq n,
 k \geq 0,\cr
&\hbox{\rm iv)}\
\wp^D_{l,i} (f(t)) \leq (1+t)^{- 3/2 + \rho}(1+\chi^{(3)})\cr
&\hskip15mm\Big(C^{(i)}\Vert L \Vert^{}_{{\cal F}^M_l}
\sup_{s \geq t} (u^{}_1 (s)) + C_l
\sum_{\scr n_1+n_2=l\atop\scr n_1 \leq l-1}\Vert L \Vert^{}_{{\cal F}^M_{n_1}}
\sup_{s \geq t}(u^{}_{n_2+1} (s))\Big),\quad 0 \leq l \leq n,\cr
}$$
$0\leq i\leq n$, where $C^{(i)}=0$ for $1\leq i\leq n$ and
$C^{(0)}=C_{l,0}=C_{0,l}=C$.
Here $C$, $C_l$ and $C_{n_1,n_2}$ depend only on $\rho$, and $C'_l$
depends only on $\rho$ and
$\Vert u \Vert^{}_{E^\rho_{N_0}}.$
}\saut
\noindent{\it Proof.}
Let $B_\mu = - \partial_\mu  {\vartheta} (L),  B_{Y,\mu} =
(B_Y)_\mu = (\xi^M_Y  B)_\mu$ for $Y \in \Pi',  0 \leq \mu
\leq 3$. Since, by covariance, $B_{Y,	\mu} = - \partial_\mu  {\vartheta}
(L_Y)$ for $Y \in \Pi' \cap U({\frak{sl}}(2, \Crm))$, it follows from equality
(4.83),
using that $\partial_\mu  L^\mu = 0$, that
$$\vert B_{ZY} (y) \vert \leq C  \int^1_0  s^{\vert Z \vert}
 \Big(\sum_{0 \leq \mu < \nu \leq 3}  \vert L_{ZM_{\mu \nu}Y}
(sy) \vert
+ \vert L_{ZY} (sy) \vert\Big)  ds, \eqno{(6.62)}$$
for $Z \in \Pi' \cap U(\Rrm^4),  Y \in \Pi' \cap U({\frak{sl}}(2, \Crm))$ and
$y \in \Rrm^+ \times \Rrm^3$. Let
$$I_l (t,x) = \int^1_0	s^l (1+s(t+ \vert x \vert))^{-1}
(1+st)^{-1 + \rho}
(1+s \big\vert t - \vert x \vert \big\vert)^{- 1/2}  ds,\quad  l \in
\Nrm.\eqno{(6.63)}$$
It follows from Lemma 6.3 and inequality (6.62) that
$$\vert B_{ZY} (t,x) \vert \leq C  \Vert L \Vert^{}_{{\cal F}^M_{\vert Z \vert
+
\vert Y \vert + 3}}  I_{\vert Z \vert} (t,x), \quad (t,x) \in \Rrm^+
\times \Rrm^3, \eqno{(6.64)}$$
for $Z \in \Pi' \cap U(\Rrm^4),  Y \in \Pi' \cap U({\frak{sl}}(2, \Crm))$,
where
$C$ depends only on $\rho$. To estimate $I_0 (t,x)$ we observe that
$$\eqalignno{
&(1+s(t + \vert x \vert))^{-1}  (1+st)^{- 1+\rho}  (1+s
\big\vert t - \vert x \vert \big\vert)^{- 1/2}\cr
&\qquad{}\leq C \big(1+s(\delta(t))(x)\big)^{- 2 + \rho},\quad  t \geq 0,
x \in \Rrm^3,\cr
}$$
where $C$ depends only on $\rho$. Integration of this inequality gives, with a
new constant, that
$$
I_0 (t,x) \leq C ((\delta(t))(x))^{-1}, \quad t \geq 0,  x \in
\Rrm^3,\eqno{(6.65{\rm a})}$$
where $C$ depends only on $\rho$. For $l \geq 1$, proceeding as in the end of
the
proof of Lemma 4.4, it follows that
$$\eqalignno{
I_l (t,x) &\leq C ((\delta(t))(x))^{-1}  (1+t)^{- 1+\rho}
(1+\big\vert t - \vert x \vert \big\vert)^{- 1/2}\cr
&\quad{}\int^1_0  {s((\delta(t))(x)) \over 1+s((\delta(t))(x))}
\biggl( {s(1+t) \over 1+s(1+t)} \biggr)^{1-\rho}  \biggl( {s(1+\big\vert t-
\vert x \vert \big\vert) \over 1+s(1+\big\vert t - \vert x \vert
\big\vert)} \biggr)^{1/2}
 s^{l-5/2+\rho}  ds.\cr
}$$
Since $l - 5/2 + \rho > - 1$ for $l \geq 1$ and $1/2 < \rho < 1$, we obtain
that
$$I_l (t,x) \leq C((\delta(t))(x))^{-1}  (1+t)^{-1+\rho}
(1+\big\vert t - \vert x \vert \big\vert)^{-1/2}, \quad l \geq 1,
\eqno{(6.65{\rm b})}$$
$t \geq 0,  x \in \Rrm^3$, where $C$ depends only on $\rho$. It follows
from inequalities (6.64), (6.65a) and (6.65b) that
$$\vert B_Y (t,x) \vert \leq C((\delta(t))(x))^{-1}	\Vert L \Vert^{}_{
{\cal F}^M_{\vert Y \vert + 3}},\quad  Y \in \Pi' \eqno{(6.66{\rm a})}$$
and that
$$\vert B_Y (t,x) \vert \leq C ((\delta(t))(x))^{-1}  (1+t)^{-1+\rho}
 (1+\big\vert t - \vert x \vert \big\vert)^{-1/2}  \Vert L \Vert^{}_{{\cal
F}^M_{
\vert Y \vert + 3}}, \eqno{(6.66b)}$$
$Y \in \sg^1,  1/2 < \rho < 1$, where $C$ depends only on $\rho.$

Let $A = A^* + K$,  $G_\mu = A_\mu - \partial_\mu
{\vartheta} (A)$ and $F_\mu = L_\mu - \partial_\mu  {\vartheta} (L)$. To
estimate $\wp^D_l \big(\delta (t)  (1+\lambda_1 (t))^{k/2}
g' (t)\big)$,  $l \geq 0$,  $k \geq 0$, where $g' = (2m)^{-1}
(m - i \gamma^\mu  \partial_\mu + \gamma^\mu  G_\mu)  g$,
we use equality (5.7a), which gives that
$$\eqalignno{
g'_Y &= \xi^D_Y  g'&(6.67)\cr
&= (2m)^{-1}  \suma_{Y_1,Y_2}^Y  \Big(\gamma^\nu
F_{Y_1 \nu}  \xi^D_{Y_2} \big((m + i \gamma^\mu  \partial_\mu -
\gamma^\mu  G_\mu)  r\big)\cr
&\qquad{}- 2i	F^\mu_{Y_1}  \partial_\mu  r^{}_{Y_2} + i
 (\partial_\mu	B^\mu_{Y_1})  r^{}_{Y_2}\cr
&\qquad{}- {i \over 4}  (\gamma^\mu  \gamma^\nu - \gamma^\nu
\gamma^\mu)  \big((\partial_\mu  L_{Y_1 \nu}) - (\partial_\nu
 L_{Y_1 \mu})\big)r^{}_{Y_2}\Big)\cr
&\qquad{}+ m^{-1}  \suma_{Y_1,Y_2,Y_3}^Y  G_{Y_1 \mu}
F^\mu_{Y_2}  r^{}_{Y_3}, \quad  Y \in \Pi',\cr
}$$
where $G_Y = \xi^M_Y  G,  F_Y = \xi^M_Y  F$ for
$Y \in \Pi'$ and where we have used that $\partial_\mu  L^\mu = 0$ and
$\partial_\mu  A^\mu = 0$, according to the definition of ${\cal F}^M_n$
and Proposition 6.2. It follows that
$$\eqalignno{
g'_Y &= (2m)^{-1} \suma_{Y_1,Y_2}^Y  \Big(\gamma^\nu
F_{Y_1 \nu}  (m + i \gamma^\mu	\partial_\mu)  r^{}_{Y_2}
- 2i F^\mu_{Y_1}  \partial_\mu  r^{}_{Y_2} + i
(\partial_\mu  B^\mu_{Y_1})  r^{}_{Y_2}&(6.68)\cr
&\qquad{}- {i \over 4}  (\gamma^\mu  \gamma^\nu - \gamma^\nu
\gamma^\mu)  ((\partial_\mu  L_{Y_1 \nu}) - (\partial_\nu
 L_{Y_1 \mu}))	r^{}_{Y_2}\Big)\cr
&\qquad{}+ (2m)^{-1} \suma_{Y_1,Y_2,Y_3}^Y\big(G_{Y_1 \mu}
F^\mu_{Y_2}  r^{}_{Y_3} + {1 \over 2}  G_{Y_1 \mu}
F_{Y_2 \nu}  (\gamma^\mu  \gamma^\nu - \gamma^\nu
\gamma^\mu)  r^{}_{Y_3}\big),\quad Y \in \Pi'.\cr
}$$
Inequality (5.130a), with $\varepsilon = 1/2,  b = 5/2 -
\rho$ and $a=1$, gives that
$$\Big(\sum_{\scr Y \in \Pi'\atop\scr \vert Y \vert \leq j}
\Vert (\delta (t))^{-1/2}
 \partial_\mu  B^\mu_Y (t) \Vert^2_{L^2}\Big)^{1/2}
\leq (1+t)^{-1}  \big(C \Vert L \Vert^{}_{{\cal F}^M_j} + j  C_j
 \Vert L \Vert^{}_{{\cal F}^M_{j-1}}\big),\quad j \geq 0,\eqno{(6.69)}$$
where $C$ and $C_j$ depend only on $\rho$. Let $Q_Y (y) = y^\mu
F_{Y \mu} (y)$, $Y \in \Pi',  y \in \Rrm^+ \times \Rrm^3$. It follows
from inequality (5.135) and the definition of $\Vert\cdot
\Vert^{}_{{\cal F}^M_j}$
that
$$\sum_{\scr Y \in \Pi'\atop\scr \vert Y \vert \leq j}  \Vert (\delta (t))^{-1}
 Q_Y (t) \Vert^{}_{L^2 (\Rrm^3, \Rrm)} \leq j  C_j
(1+t)^{-1/2}  \Vert L \Vert^{}_{{\cal F}^M_j},\quad j \geq 0, \eqno{(6.70)}$$
where $C_j$ depends only on $\rho$. Inequalities (5.7d) and (6.70)
give that
$$\eqalignno{
&\suma_{Y_1,Y_2}^Y  \Vert \delta (t) (1+\lambda_1 (t))^{k/2}
 F^\mu_{Y_1} (t)  \partial_\mu	r^{}_Y (t) \Vert^{}_D
&(6.71)\cr
&\qquad{}\leq C_{\vert Y \vert}  (1+t)^{-1}  \sum_{\sscr Z \in \Pi'\atop{
\sscr n_1+n_2=\vert Y \vert\atop\sscr\vert Z \vert \leq n_2+1 }}
\Vert L \Vert^{}_{{\cal F}^M_{n_1}}
 \Vert (\delta (t))^{3/2} (1+\lambda_1 (t))^{k/2} r^{}_Z (t)
 \Vert^{}_{L^\infty},\cr
}$$
$Y \in \Pi'$, where $C_{\vert Y \vert}$ depends only on $\rho$.
Since $A = A^*+K,$
it follows from inequalities (5.116c), (6.38a), (6.49b) and (6.66a) that
$$\eqalignno{
&\suma_{Y_1,Y_2,Y_3}^Y	\Vert \delta (t) (1+\lambda_1 (t))^{k/2}
 G_{Y_1 \mu} (t)  F_{Y_2 \nu} (t)  r^{}_{Y_3} (t)
\Vert^{}_D &(6.72)\cr
&\quad{}\leq C'_{\vert Y \vert}  (1+t)^{-3/2+\rho}  \suma_{Y_1,Y_2,Y_3}^Y
 \Vert u \Vert^{}_{E^\rho_{N_0+\vert Y_1 \vert + 1}}  \Vert L
\Vert^{}_{{\cal F}^M_{\vert Y_{2} \vert}}
\Vert (\delta (t))^{3/2}  (1+\lambda_1 (t))^{k/2}  r^{}_{Y_3}
(t) \Vert^{}_{L^\infty}\cr
&\qquad{}+ C_{\vert Y \vert}  (1+t)^{-1}  \suma_{Y_1,Y_2,Y_3}^Y
\min \big(\Vert K \Vert^{}_{{\cal F}^M_{\vert Y_1 \vert+3}}
\Vert L \Vert^{}_{{\cal F}^M_{\vert Y_2 \vert}},
\Vert K \Vert^{}_{{\cal F}^M_{\vert Y_1 \vert}}
\Vert L \Vert^{}_{{\cal F}^M_{\vert Y_2 \vert +3}}\big)\cr
&\qquad{}\qquad{}\Vert (\delta (t))^{3/2}  (1+\lambda_1 (t))^{k/2}  r^{}_{Y_3}
(t) \Vert^{}_{L^\infty},\cr
}$$
$\quad Y \in \Pi'$, where $C_{\vert Y \vert}$ depends only on
$\rho$ and $C'_{\vert Y \vert}$
depends only on $\Vert u \Vert^{}_{E^\rho_{N_0}}$ and $\rho$. Inequalities
(6.49b),
(6.69), (6.71) and (6.72) give together with equality (6.68) that
$$\eqalignno{
&\Vert \delta (t)(1+\lambda_1 (t))^{k/2}  g'_Y (t)
\Vert^{}_D&(6.73)\cr
&\quad{}\leq (1+t)^{- 3/2 + \rho}  C_{\vert Y \vert}	\suma_{Y_1,Y_2}^Y
 \Vert L \Vert^{}_{{\cal F}^M_{\vert Y_1 \vert}}
\Big(\Vert (\delta (t))^{3-\rho}  (1+\lambda_1 (t))^{k/2}
(m + i \gamma^\mu  \partial_\mu)  r^{}_Y (t) \Vert^{}_{L^\infty}\cr
&\qquad{}+ \sum_{\scr Z \in \Pi'\atop\scr \vert Z \vert \leq \vert
Y_2 \vert + 1}  \Vert
(\delta (t))^{3/2}  (1+\lambda_1 (t))^{k/2}  r^{}_Z (t)
\Vert^{}_{L^\infty}\Big)\cr
&\qquad{}+ (1+t)^{-1}	C_{\vert Y \vert}  \suma_{Y_1,Y_2,Y_3}^Y
\min\big(\Vert K \Vert^{}_{{\cal F}^M_{\vert Y_1 \vert+3}}
\Vert L \Vert^{}_{{\cal F}^M_{\vert Y_2 \vert}},
\Vert K\Vert^{}_{{\cal F}^M_{\vert Y_1 \vert}}
\Vert L\Vert^{}_{{\cal F}^M_{\vert Y_2 \vert+3}}\big)\cr
&\quad\qquad{}\Vert (\delta (t))^{3/2}  (1+\lambda_1 (t))^{k/2}  r^{}_{Y_3}
(t) \Vert^{}_{L^\infty}\cr
&\qquad{}+ (1+t)^{-3/2 + \rho}  C'_{\vert Y \vert}  \suma_{Y_1,Y_2,Y_3}^Y
 \Vert u \Vert^{}_{E^\rho_{N_0+\vert Y_1 \vert + 1}}  \Vert L
\Vert^{}_{{\cal F}^M_{\vert Y_2 \vert}}
\Vert (\delta (t))^{3/2}  (1+\lambda_1 (t))^{k/2}  r^{}_{Y_3}
(t) \Vert^{}_{L^\infty},\cr
}$$
$Y \in \Pi'$, where $C_{\vert Y \vert}$ only depends $\rho$ and
$C'_{\vert Y \vert}$ only on
$\rho$ and $\Vert u \Vert^{}_{E^\rho_{N_0}}$. The estimate of $R'_l (t)$
in statement
i) of the lemma follows from the definition of $u^{}_j (t)$, the definition of
$R'_j$ in Theorem 5.8 and from inequality (6.73). If on the right-hand side of
equality (6.68) we restrict the domain of summation in the last sum over
$Y_1,Y_2,Y_3$
to $Y_1 = \un$, then the right-hand side of this equality is equal to
$(2m)^{-1}
 (m-i \gamma^\mu  \partial_\mu + \gamma^\mu  G_\mu)
 g^{}_Y$. The method of the proof of inequality (6.73) then gives also that
$\Vert \delta (t) ((m-i \gamma^\mu\partial_\mu + \gamma^\mu
 G_\mu)  g^{}_Y) (t) \Vert^{}_D$ is majorized by the right-hand
side of inequality (6.73). This shows that
$$(m - i \gamma^\mu  \partial_\mu + \gamma^\mu	G_\mu)
g^{}_Y \in L^1 (\Rrm^+,D), \quad  Y \in \Pi',  \vert Y \vert \leq n.$$
Since
$$g^{}_Y = \suma_{Y_1,Y_2}^Y  \gamma^\mu  F_{Y_1 \mu}
r^{}_{Y_2}, \quad Y \in \Pi',$$
it follows from Lemma 6.3 and inequality (6.66a) that
$$\eqalignno{
&\Vert (\delta (t))^{3/2}  (1+\lambda_1 (t))^{k/2}  g^{}_Y (t)
\Vert^{}_{L^\infty}&(6.74)\cr
&\qquad{}\leq C_{\vert Y \vert}  (1+t)^{-1}  \sum_{\scr Z \in \Pi'\atop\scr
n_1+\vert Z \vert= \vert Y \vert}\Vert L \Vert^{}_{{\cal F}^M_{n_1+3}}
\Vert (\delta (t))^{3/2} (1+\lambda_1 (t))^{k/2}  r^{}_Z (t)
\Vert^{}_{L^\infty},\cr
}$$
$Y \in \Pi',  \vert Y \vert \leq n-3$, where $C_{\vert Y \vert}$
depends only on $\rho$. This inequality and the definition of $R^\infty_j$
in Theorem 5.8 prove statement ii) of the lemma. According to inequality
(6.49b) and the definition of ${\cal F}^M_j$ we obtain that
$$\eqalignno{
&\wp^D_j  \big((\delta(t))^{1/2} (1+\lambda_1 (t))^{k/2}  g(t)\big)&(6.75)\cr
&\quad{} \leq (1+t)^{-1/2}
\sum_{\scr Z \in \Pi'\atop\scr n_1+\vert Z \vert \leq j}
C_{n_1, \vert Z \vert} \Vert L \Vert^{}_{{\cal F}^M_{n_1}}
 \Vert (\delta (t))^{3/2}  (1+\lambda_1 (t))^{k/2}
r^{}_Z (t) \Vert^{}_{L^\infty},\quad  0 \leq j \leq n,\cr
}$$
where $C_{n_1,n_2}$ depends only on $\rho$. This inequality proves
statement iii) of the lemma and it also proves that $\Vert g^{}_Y (t)
\Vert^{}_D \fl 0$, when $t \fl \infty$ for $Y \in \Pi',
\vert Y \vert \leq n.$

To prove statement iv) of the lemma we shall use statement i) of Proposition
5.16, with $t_0 = \infty, a_\mu = L_\mu, \rho' = 0, 1/2 < \rho < 1, \eta \in ]
1/2,\rho[$ and $\varepsilon = 1/2$. It follows from inequality (6.41a) and from
definitions (5.115a) and (5.115b) of $S^{\rho,j}$ that $(A_Y, A_{P_0 Y}) \in
C^0
(\Rrm^+, M^1)$ for $Y \in \Pi',  \vert Y \vert \leq n$. Since, $\partial_\mu
 A^\mu = 0$ and $n \geq 2$, the hypothesis on $A$ in Proposition 5.16
are satisfied. Due to the hypothesis on $r$ and since $L \in {\cal F}^M_n$, the
hypothesis of statement~i) of Proposition 5.16 are satisfied. To use the
estimate
of $\wp^D_l (f(t))$ in that statement for $0 \leq l \leq n$, we first estimate
$\theta^D_j (t)$ and $\tau^M_j (\infty,t)$, which we denote by $\tau^M_j (t).$
Here $\tau^M_j (\infty,t)$ is defined in (5.186) and the sequel. Definitions
(6.32b) and (6.32c) of $\Vert\cdot  \Vert^{}_{{\cal F}^M_j}$ show that, if
$L \in {\cal F}^M_j$, then
$$\tau^M_j (t) \leq (1+t)^{- 3/2 + \rho}  C \Vert L \Vert^{}_{{\cal F}^M_j},
\quad  j \geq 0, \eqno{(6.76)}$$
where $C$ depends only on $\rho$. It follows from inequality (6.38c) that
$$\eqalignno{
\theta^D_j (t) &\leq \sum_{\scr Y \in \Pi'\atop \scr \vert Y \vert \leq j+1}
 \sup_{s \geq t}\big((1+C_1  \chi^{(3)})  \Vert (\delta (s))^{3/2}
 r^{}_Y (s) \Vert^{}_{L^\infty}\big)&(6.77)\cr
&\qquad{}+ \sum_{\scr Y \in \Pi'  \atop\scr \vert Y \vert \leq j}
\sup_{s \geq t}
\big(\Vert (\delta (s))^{3-\rho}((i \gamma^\mu	\partial_\mu +
m) r^{}_Y)  (s) \Vert^{}_{L^\infty}\big)\cr
&\leq C'  (1+\chi^{(3)})  \sup_{s \geq 0}  u^{}_{j+1}
(s),\quad  j \geq 0,\cr
}$$
where $C_1$ and $C'$ depend only on $\rho$ and
$\Vert u \Vert^{}_{E^\rho_{N_0}}.$
Inequalities (6.76) and (6.77) and the estimate in statement i)
of Proposition 5.16 give that
$$\eqalignno{
\wp^D_{l,j}  (f(t)) &\leq  (1+\chi^{(3)})  (1+t)^{- 3/2+\rho}
\Big(C^{(j)}\Vert L \Vert^{}_{{\cal F}^M_l}  \sup_{s \geq t}
(u^{}_1 (s))& (6.78)\cr
&\qquad{}+ C_l \sum_{\scr n_1+n_2=l\atop\scr n_1 \leq l-1}
\Vert L \Vert^{}_{{\cal F}^M_{n_1}}
\sup_{s \geq t} (u^{}_{n_2+1} (s))\Big), \quad 0 \leq l \leq n,\cr
}$$
where $C^{(0)}$ and $C_l$ depend only on $\rho$ and
$\Vert u \Vert^{}_{E^\rho_{N_0}}$,
and $C^{(j)}=0$ for $1\leq j\leq n$.
This proves the proposition.

We shall prove a result, analog to Lemma 6.5, for the case where
$g = \gamma^\mu (L_\mu - \partial_\mu  \vartheta (L))  \Phi$,  $L \in
{\cal F}^M_n$ and $\Phi \in {\cal F}^D_n.$
\saut
\noindent{\bf Lemma 6.6.}
{\it
Let $1/2 < \rho < 1$, $n\geq5$, $K \in {\cal F}^M_n$,
$L \in {\cal F}^M_n$, $\Phi\in {\cal F}^D_n$ and let
$$\eqalignno{
H_l (\Phi,t) & = \sum_{\scr Y \in \Pi'\atop \scr
k+\vert Y \vert \leq l }
\Vert \delta (t)^{3/2} (1+\lambda_1 (t))^{k/2} \Phi_Y (t)
\Vert^{}_{L^\infty},\cr
u^{}_l(\Phi, t) & = H_{l+1} (\Phi,t) + \!
\sum_{\scr Y \in \Pi' \atop \scr k+\vert Y \vert \leq l}\!
\Vert \delta (t)^{3-\rho} (1+\lambda_1 (t))^{k/2}
\big(\xi^{M}_{Y}(i\gamma^\mu\partial_\mu +m-\gamma^\mu
G_\mu)\Phi\big)(t)\Vert^{}_{L^\infty},\cr
U_{l,k}(\Phi,t) & =\wp^{D}_{l+1}\big((1+\lambda_{1}(t))^{k/2}\Phi(t)\big)+
\wp^{D}_{l}\big((1+\lambda_{0}(t))^{1/2}(1+\lambda_{1}(t))^{k/2}\Phi(t)\big)+
\overline{U}_{l,k}(\Phi,t),\cr
& \cr
\overline{U}_{l,k}(\Phi,t) & =
\wp^{D}_{l}\big(\delta(t)^{3/2-\rho}(1+\lambda_{1}(t))^{k/2}
\big((i\gamma^\mu\partial_\mu +m-\gamma^\mu G_\mu)\Phi\big)(t)\big)\cr
}$$
and
$$U_l(\Phi,t) = \sum_{i+j=l}U_{i,j}(\Phi,t).$$
Then:
$$\eqalignno{
&\hbox{\rm i)}\
R'_l (t) \leq (1+t)^{-3/2 + \rho} \Big(C'_l \sum_{n_1+n_2+n_3=l}
(1+\Vert u \Vert^{}_{E^\rho_{N_0+n_1+1}})\cr
&\qquad\qquad\qquad{}\min \big(\Vert L \Vert^{}_{{\cal F}^M_{n_2}}
u^{}_{n_3}(\Phi,t), \Vert L\Vert^{}_{{\cal F}^M_{n_2 +4}}U_{n_3}
(\Phi,t)\big)\cr
&\qquad\qquad{}+ C_l \sum_{n_1+n_2+n_3=l} \min
\big(\Vert K \Vert^{}_{{\cal F}^M_{n_1}}
\Vert L \Vert^{}_{{\cal F}^M_{n_2 +3}} u^{}_{n_3}(\Phi,t),\cr
&\qquad\qquad\qquad{}\Vert K \Vert^{}_{{\cal F}^M_{n_1 +3}}
\Vert L \Vert^{}_{{\cal F}_{n_2}}u^{}_{n_3} (\Phi,t),
\Vert K \Vert^{}_{{\cal F}^M_{n_1+3}}\Vert L \Vert^{}_{{\cal F}^M_{n_2+3}}
U_{n_3} (\Phi,t)\big)\Big),\quad 0 \leq l,\hskip14.6mm\cr
&\hbox{\rm ii)}\
R^\infty_l (t) \leq (1+t)^{-1} C_l\sum_{n_1+n_2=l}
\Vert L \Vert^{}_{{\cal F}^M_{n_1+3}} H_{n_2} (\Phi,t),\quad 0 \leq l,\cr
}$$
$$\eqalignno{
&\hbox{\rm iii)}\ \wp^D_l\big(\delta(t)(1+\lambda_1 (t))^{k/2} g(t)\big)
\leq \sum_{n_1+n_2=l} C_{n_1,n_2}
\min \big(\Vert L \Vert^{}_{{\cal F}^M_{n_1}} H_{n_2+k+1} (\Phi,t),\cr
&\qquad{}\Vert L \Vert^{}_{{\cal F}^M_{n_1+3}}\wp^D_{n_2}
\big((1+\lambda_1 (t))^{k/2}
\Phi(t)\big)\big),\quad 0 \leq l,\cr
&\hbox{\rm iv)}\
\wp^D_l(f(t)) + (1+t) \wp^D_l(g(t))
\leq (1+t)^{- 3/2+\rho}\cr
&\qquad{}\sum_{n_1+n_2=l}
C_{n_1,n_2}
\min \big(\Vert L \Vert^{}_{{\cal F}^M_{n_1+3}}
\Vert \Phi \Vert^{}_{{\cal F}^D_{n_2}},
\Vert L \Vert^{}_{{\cal F}^M_{n_1}}
\sup_{s\geq t}\big((1+s)^{3/2-\rho} H_{n_2} (\Phi,s)\big)\big)\hskip7mm\cr
\noalign{\noindent and}
& \wp^{D}_{l,i}(f(t))\leq (1+t)^{-3/2 +\rho}\sum_{n_1+n_2=l-1}C'_{n_1,n_2}
\sup_{s\geq t}\cr
&\min\big(\Vert L\Vert^{}_{{\cal F}^{M}_{n_1}}\big(u^{}_{n_2}
(\Phi,s) +(1+s)^{3/2-\rho}
H_{n_2+1}(\Phi,s)\big), \Vert L\Vert^{}_{{\cal F}^{M}_{n_1+3}}
\big(\Vert \Phi\Vert^{}_{{\cal F}^{M}_{n_2+1}}+
\overline{U}_{n_2,0}(\Phi,s)\big)\big),\cr
}$$
$l\geq 0, 1\leq i\leq l.$
Here $ C_{l,0}=C_{0,l}=C'_{l-1,0}=C'_{0,l-1}=C$, $C_{n_1,n_2}$,
$C'_{n_1,n_2}$ depend only on $\rho$ and $C'_{l}$ depends only on
$\rho$ and $\Vert u \Vert^{}_{E^{\rho}_{N_0}}$.
}\saut
\noindent{\it Proof.}
Since
$$
g^{}_Y = \suma_{Y_1,Y_2}^Y \gamma^\mu F_{Y_1 \mu} \Phi_{Y_2},
\quad Y \in \Pi'$$
where $F_\mu = L_\mu - \partial_\mu \vartheta (L),
(\xi^M_Y F)_\mu = F_{Y\mu}$, and since
$\delta(t)\leq	C(1+\lambda_{1}(t)+t)$ where $C$ is independent of $t$, it
follows that
$$\eqalignno{
&\Vert \delta(t) (1+\lambda_{1}(t))^{k/2}g^{}_{Y}(t)\Vert^{}_{D}\cr
&\qquad{}\leq C\suma_{Y_1,Y_2}^{Y}
\min\big(\Vert\delta(t)F_{Y_1} (t) \Vert^{}_{L^\infty}
\Vert(1+\lambda_{1}(t))^{k/2}
\Phi_{Y_2} (t) \Vert^{}_D,\cr
&\qquad\qquad{}(1+t)^{1/2} \Vert \delta(t)^{-1}  F_{Y_1} (t) \Vert^{}_{L^2}
H_{\vert Y_2 \vert+k+1}  (\Phi,t)\big),\cr
}$$
where $C$ is independent of $t \geq 0$, $Y \in \Pi'$, $L$ and $\Phi.$
Inequality (6.49b), Lemma 6.3 and inequality (6.66a) then give that
$$\eqalignno{
&\wp^D_l\big(\delta(t)(1+\lambda_{1}(t))^{k/2} (g(t)\big) \cr
&\qquad{}\leq  \sum_{n_1+n_2=l}C_{n_1,n_2}
\min \big(\Vert L\Vert^{}_{{\cal F}^{M}_{n_1+3}}\wp^{D}_{n_2}
((1+\lambda_{1}(t))^{k/2}\Phi(t)), \Vert L\Vert^{}_{{\cal F}^{M}_{n_1}}
H_{n_2+k+1}  (\Phi,t)\big),\cr
}$$
where $ C_{l,0}=C_{0,l}=C$ and $C_{n_1,n_2}$ depends only on $\rho$.
This proves statement iii) of the lemma.
Statement iv) follows from statement iii) and the definition of the spaces
${\cal F}^D_j$ and by using (5.7b$'$) when $i\geq 1$.
Statement ii)  follows as in the proof of Lemma 6.5. To prove statement
i) let
$\overline{g}= (i\gamma^\mu\partial_\mu +m-\gamma^\mu G_\mu)\Phi$ and
$\overline{g}^{}_{Y}=\xi^{D}_{Y}\overline{g}$, ${g}'_{Y}=(2m)^{-1}\xi^{D}_{Y}
(m - i\gamma^\mu\partial_\mu +\gamma^\mu G_\mu)g$, for $Y\in\Pi'$.
It follows from equality (6.67) that
$$\eqalignno{
g'_{Y}&=(2m)^{-1}\suma_{Y_1,Y_2}^{Y}\Big(\gamma^{\nu}F_{Y_1\nu}\overline{g}_{Y_2}
- 2iF^{\mu}_{Y_1}\partial_{\mu}\Phi_{Y_2} + i (\partial_\mu
B^{\mu}_{Y_1})\Phi_{Y_2}\cr
&\qquad{}-{i\over 4}(\gamma^{\mu}\gamma^{\nu}-\gamma^{\nu}\gamma^{\mu})
\big((\partial_{\mu}L_{Y_1\nu})-(\partial_{\nu}L_{Y_1\mu})\big)
\Phi_{Y_2}\Big)\cr
&\qquad{}+m^{-1}\suma_{Y_1,Y_2,Y_3}^{Y}G_{Y_1\mu}F^{\mu}_{Y_2}\Phi_{Y_3},
\quad Y\in \Pi',\cr
}$$
where $B_{\mu}= -\partial_{\mu}\vartheta(L)$. We denote $S_{Y_1,Y_2}$
the terms
of the first sum and $T_{Y_1,Y_2,Y_3}$ the terms of the second sum of this
expression. It follows by the same argument as in the proof of Lemma 6.5 led
from (6.68) to (6.73) that
$$ \Vert \delta(t)(1+\lambda_{1}(t))^{k/2}S_{Y_1,Y_2}(t)\Vert^{}_{D}
\leq (1+t)^{-3/2 + \rho}a_{\vert Y \vert}
\Vert L \Vert^{}_{{\cal F}^M_{\vert Y_1\vert}}
u^{}_{\vert Y_2\vert +k}(\Phi,t)$$
and that
$$\eqalignno{
& \Vert \delta(t)(1+\lambda_{1}(t))^{k/2}T_{Y_1,Y_2,Y_3}(t)\Vert^{}_{D}\cr
&\quad{} \leq (1+t)^{-3/2 + \rho}b'_{\vert Y \vert}
\Vert u \Vert^{}_{E^\rho_{N_0+\vert Y_1\vert+1}}
\Vert L \Vert^{}_{{\cal F}^M_{\vert Y_2\vert}}
H_{\vert Y_3\vert +k}(\Phi,t)\cr
&\qquad{}+(1+t)^{-1}b_{\vert Y \vert}\min\big(
\Vert K \Vert^{}_{{\cal F}^M_{\vert Y_1\vert}}
\Vert L \Vert^{}_{{\cal F}^M_{\vert Y_2\vert+3}},
\Vert K \Vert^{}_{{\cal F}^M_{\vert Y_1\vert+3}}
\Vert L \Vert^{}_{{\cal F}^M_{\vert Y_2\vert}}\big)
H_{\vert Y_3\vert +k}(\Phi,t),\cr
}$$
where $a^{}_{\vert Y\vert}$ and $b^{}_{\vert Y\vert}$ depend only on $\rho$
and $b'_{\vert Y\vert}$ depends only on $\rho$ and
$\Vert u\Vert^{}_{E^{\rho}_{N_0}}$. Thus, to prove statement i) of the lemma we
only have to show that the last argument of the minima functions in the
estimate of $R'_{l}$ in statement i) majorizes $S_{Y_1,Y_2}$
and $T_{Y_1,Y_2,Y_3}$. It follows from inequality (5.7d) and equality (5.133)
that
$$\eqalignno{
&\Vert \delta(t)(1+\lambda_{1}(t))^{k/2}F_{Y_1}^{\mu}\partial_{\mu}
\Phi_{Y_2}(t)\Vert^{}_{D}\cr
&\quad{}\leq C_{Y_1}
 \Big(\sum_{\scr Z \in \Pi'\atop \scr \vert Z \vert \leq \vert Y_1\vert }
\Vert F_{Z}(t) \Vert^{}_{L^{\infty}}\Big)
\wp^{D}_{\vert Y_2\vert +1}\big((1+\lambda_{1}(t))^{k/2}\Phi(t)\big)\cr
}$$
and it then follows that
$$\eqalignno{
&\Vert \delta(t)(1+\lambda_{1}(t))^{k/2}S_{Y_1,Y_2}\Vert^{}_{D}\cr
&\quad{}\leq C_{Y}\Big(\Vert \delta(t)F_{Y_1}(t)\Vert^{}_{L^{\infty}}
\Vert (1+\lambda_{1}(t))^{k/2}\overline{g}_{Y_2}(t)\Vert^{}_{D}\cr
&\qquad{}+ \sum_{\scr Z \in \Pi'\atop \scr\vert Z \vert \leq \vert Y_1\vert }
\Vert F_{Z}(t) \Vert^{}_{L^{\infty}}
\wp^{D}_{\vert Y_2\vert +1}\big((1+\lambda_{1}(t))^{k/2}\Phi(t)\big)\cr
&\qquad{}+\Big(\Vert \delta(t)(1+\lambda_{0}(t))^{-1/2}
\partial_{\mu} B^{\mu}_{Y_1}(t)\Vert^{}_{L^{\infty}}
+\sum_{\mu,\nu}\Vert \delta(t)(1+\lambda_{0}(t))^{-1/2}
\partial_{\mu}L_{Y_1\nu}(t)\Vert^{}_{L^{\infty}}\Big)\cr
&\qquad\qquad{}\wp^{D}_{\vert Y_2\vert}\big((1+\lambda_{0}(t))^{1/2}
(1+\lambda_{1}(t))^{k/2}\Phi(t)\big)\Big),\cr
}$$
where $\lambda_{0}$ is given in Theorem 5.5. It now follows from inequalities
(5.33), (6.66a), (6.66b) and from Lemma 6.3 that
$$\eqalignno{
&\Vert \delta(t)(1+\lambda_{1}(t))^{k/2}S_{Y_1,Y_2}(t)\Vert^{}_{D}\cr
&\quad{}\leq C_{\vert Y\vert}\Big(\Vert L
\Vert^{}_{{\cal F}^M_{\vert Y_1\vert+3}}
\Vert (1+\lambda_{1}(t))^{k/2}\overline{g}_{Y_2}(t)\Vert^{}_{D}\cr
&\qquad{}+(1+t)^{-1}\Vert L \Vert^{}_{{\cal F}^M_{\vert Y_1\vert+3}}
\wp^{D}_{\vert Y_2\vert +1}\big((1+\lambda_{1}(t))^{k/2}\Phi(t)\big)\cr
&\qquad{}+(1+t)^{-3/2 +\rho}\Vert L \Vert^{}_{{\cal F}^M_{\vert Y_1\vert+4}}
\wp^{D}_{\vert Y_2\vert}\big((1+\lambda_{0}(t))^{1/2}
(1+\lambda_{1}(t))^{k/2}\Phi(t)\big)\Big).\cr
}$$
This gives that
$$\Vert \delta(t)(1+\lambda_{1}(t))^{k/2}S_{Y_1,Y_2}(t)\Vert^{}_{D}
\leq C_{\vert Y\vert}
(1+t)^{-3/2 +\rho}
\Vert L \Vert^{}_{{\cal F}^M_{\vert Y_1\vert+4}}
U_{\vert Y_2\vert, k}(\Phi,t),$$
where $C_{\vert Y\vert}$ depends only on $\rho$. Proceeding as in the proof
of Lemma 6.5, we obtain that
$$\eqalignno{
&\Vert \delta(t)(1+\lambda_{1}(t))^{k/2}T_{Y_1,Y_2,Y_3}(t)\Vert^{}_{D}\cr
&\quad{}\leq C'_{\vert Y_1\vert}
\Vert u \Vert^{}_{E^\rho_{N_0+\vert Y_1\vert +1}}
\Vert
\delta(t)^{\rho-1/2}(1+\lambda_{1}(t))^{k/2}F_{Y_2}(t)
\Phi_{Y_3}(t)\Vert^{}_{D}\cr
&\qquad{}+C \Vert K \Vert^{}_{{\cal F}^M_{\vert Y_1\vert+3}}
\Vert F_{Y_2}(t)\Vert^{}_{L^{\infty}}
\wp^{D}_{\vert Y_3\vert}\big((1+\lambda_{1}(t))^{k/2}\Phi(t)\big).\cr
}$$
Lemma 6.3 and inequality (6.66a) then give, changing constants, that
$$\eqalignno{
&\Vert \delta(t)(1+\lambda_{1}(t))^{k/2}T_{Y_1,Y_2,Y_3}(t)\Vert^{}_{D}\cr
&\quad{}\leq (1+t)^{-3/2 +\rho}
\big(C'_{\vert Y_1\vert}\Vert u \Vert^{}_{E^\rho_{N_0+\vert Y_1\vert +1}}
+ C (1+t)^{-\rho+1/2} \Vert K \Vert^{}_{{\cal F}^M_{\vert Y_1\vert+3}}\big)\cr
&\qquad{} \Vert L \Vert^{}_{{\cal F}^M_{\vert Y_2\vert+3}}
\wp^{D}_{\vert Y_3\vert}\big((1+\lambda_{1}(t))^{k/2}\Phi(t)\big),\cr
}$$
where $C'_{\vert Y_1\vert}$  depends only on $\rho$ and
$\Vert u\Vert^{}_{E^{\rho}_{N_0}}$ and $C$  depends only on $\rho$.
This proves statement~i) of the lemma and therefore the lemma.

We now return to the {\it study of the map} ${\cal N}$ introduced after
equation (6.33).
\saut
\noindent{\bf Proposition 6.7.}
{\it
If $1/2 < \rho < 1$ and $n \geq 50$, then ${\cal N}$ is a map from
$E^{\circ\rho}_\infty
\times {\cal F}^M_n$ to ${\cal F}^M_n,$
$$\Vert {\cal N} (u,K) \Vert^{}_{{\cal F}^M_n}
\leq C_n  \Vert u \Vert^{}_{E^\rho_{N_0+n+3}}
 (\Vert u \Vert^{}_{E^\rho_{N_0+n}} + \Vert K \Vert^{}_{{\cal F}^M_n})$$
and
$$\Vert {\cal N} (u,K^{(1)}) - {\cal N} (u,K^{(2)}) \Vert^{}_{{\cal F}^M_n}
\leq
C'_n  \Vert u \Vert^{}_{E^\rho_{N_0+n+3}}  \Vert K^{(1)} -
K^{(2)} \Vert^{}_{{\cal F}^M_n},$$
for $u \in E^{\circ\rho}_\infty, K, K^{(1)}, K^{(2)} \in {\cal F}^M_n$, where
$C_n$ depends only on $\rho, \Vert u \Vert^{}_{E^\rho_{N_0+n+2}}$ and $\Vert K
\Vert^{}_{{\cal F}^M_n}$, and $C'_n$ depends only on $\rho,
\Vert u \Vert^{}_{E^\rho_{N_0+n+2}}$
and $\Vert K^{(i)} \Vert^{}_{{\cal F}^M_n},  i=1,2.$
}\saut
\noindent{\it Proof.}
Let $u \in E^{\circ\rho}_\infty,  K \in {\cal F}^M_n,  L \in
{\cal F}^M_n$, $A = A^* + K$ and let $G_\mu = A_\mu - \partial_\mu
{\vartheta} (A)$ for $0 \leq \mu \leq 3.$

We first consider the equation
$$(i \gamma^\mu  \partial_\mu + m - \gamma^\mu	G_\mu)
\Psi = \gamma^\mu \big(L_\mu - (\partial_\mu  {\vartheta} (L))\big)  r,
\eqno{(6.79)}$$
with unknown $\Psi \in {\cal F}^M_n$, where $r = (D^j  \phi^*)
(u ; v_1,\ldots,v_j)$
for some $j \geq 0$ and $v_1,\ldots,v_j \in E^{\circ\rho}_\infty$.
(For this proof
we only need the case $j=0$ and we consider $j \geq 0$ only to prepare later
proofs). It follows from Proposition 6.2 that
$$H_l (r,t) \leq C_{l,j} \big(R^j_{N_0,l}  (v_1,\ldots,v_j)
+ \Vert u \Vert^{}_{E^\rho_{N_0+l}}
 \Vert v_1 \Vert^{}_{E^\rho_{N_0}}\cdots\Vert v_j
 \Vert^{}_{E^{\rho}_{N_0}}\big),
\eqno{(6.80)}$$
$l \geq 0$, where $H_l (r,t)$ is defined in Lemma 6.5 and $C_{l,j}$
depends only on $\rho$ and $\Vert u \Vert^{}_{E^\rho_{N_0}}$. Equations
(6.2a) and substitution (6.30) give that
$$(i \gamma^\mu  \partial_\mu + m)  \phi^* = \gamma^\mu
 (A^*_{0, \mu} + B^*_{0, \mu})	\phi^*.$$
Derivation of this equation $j$ times in $u$, inequality (5.116c), Proposition
6.2 and the convexity property of the seminorms $\Vert\cdot\Vert^{}_{E^\rho_N}$
in Corollary 2.6 show together with inequality (6.80) that
$$u^{}_l (t) \leq C_{l,j}  \big(R^j_{N_0, l+1}  (v_1,\ldots,v_j) +
\Vert u \Vert^{}_{E^\rho_{N_0+l+1}}  \Vert v_1 \Vert^{}_{E^\rho_{N_0}}\cdots
\Vert v_j \Vert^{}_{E^\rho_{N_0}}\big), \eqno{(6.81)}$$
$l \geq 0$, where $u^{}_l (t)$ is defined in Lemma 6.5 and where $C_{l,j}$
depends only on $\rho$ and $\Vert u \Vert^{}_{E^\rho_{N_0}}$.
It follows from Proposition 6.2 that
$$\eqalignno{
&\sum_{i+k \leq l}  \wp^D_i  \big((1+\lambda_1 (t))^{k/2}
r(t)\big) &(6.82)\cr
&\qquad{}\leq C_{l,j}\big(R^j_{N_0,l}(v_1,\ldots,v_j)
+ \Vert u \Vert^{}_{E^\rho_{N_0+l}}
 \Vert v_1 \Vert^{}_{E^\rho_{N_0}}\cdots\Vert v_l
 \Vert^{}_{E^\rho_{N_0}}\big),\cr
}$$
$l \geq 0$, where $C_{l,j}$ depends only on $\rho$ and
$\Vert u \Vert^{}_{E^\rho_{N_0}}.$
According to inequalities (6.80), (6.81) and (6.82), the hypotheses of
Lemma 6.5 are satisfied, and it follows from this lemma that, if
$g = \gamma^\mu (L_\mu - \partial_\mu  {\vartheta} (L))$, $g^{}_Y =
\xi^D_Y  g$ for $Y \in \Pi',$
then $g^{}_Y \in C^0(\Rrm^+,D)$, $\Vert g^{}_Y (t)
\Vert^{}_D \fl 0$ when $t \fl \infty$
and $(m - i \gamma^\mu	\partial_\mu + \gamma^\mu  G_\mu)
 g^{}_Y \in L^1 (\Rrm^+,D)$ for $Y \in \Pi',  \vert Y \vert \leq n.$
Statement i)--iv) of Lemma 6.5 and the convexity property of the seminorms
$\Vert\cdot\Vert^{}_{E^\rho_N}$ give, with the notation
$$\tau^j_{N_0,l} = R^j_{N_0,l} (v_1,\ldots,v_j) +
\Vert u \Vert^{}_{E^\rho_{N_0+l}}
 \Vert v_1 \Vert^{}_{E^\rho_{N_0}}\cdots\Vert v_j \Vert^{}_{E^\rho_{N_0}},$$
that
$$\eqalignno{
R'_l (t) &\leq (1+t)^{- 3/2+\rho}  C_{l,j}  \Big(\sum_{n_1+n_2=l}
 \tau^j_{N_0,n_1+2}  \Vert L \Vert^{}_{{\cal F}_{n_2}}
&(6.83{\rm a})\cr
&\quad{}+ \sum_{\scr n_1+n_2+n_3=l\atop\scr  n_2 \leq n_3}	\tau^j_{N_0,n_1+1}
\big(\Vert K \Vert^{}_{{\cal F}^M_{n_2+3}}\Vert L \Vert^{}_{{\cal F}^M_{n_3}} +
\Vert L     \Vert^{}_{{\cal F}^M_{n_2+3}}\Vert K
\Vert^{}_{{\cal F}^M_{n_3}}\big)\Big),
\quad 0 \leq l \leq n,\cr
R^\infty_l (t) &\leq (1+t)^{-1}  C_{l,j}  \sum_{n_1+n_2=l}
 \tau^j_{N_0,n_1}  \Vert L \Vert^{}_{{\cal F}^M_{n_2+3}},\quad
 0 \leq l \leq n-3,&(6.83{\rm b})\cr
}$$
$$\eqalignno{
&\wp^D_l  \big((1+\lambda_1 (t))^{k/2}  g(t)\big)&(6.83{\rm c})\cr
&\qquad{}\leq (1+t)^{-1}  C_{l,k,j}  \sum_{n_1+n_2=l}
\tau^j_{N_0,n_1+k}  \Vert L \Vert^{}_{{\cal F}^M_{n_2}},\quad 0 \leq
l \leq n,  k \geq 0,
}$$
and
$$\eqalignno{
\wp^D_l  (f(t)) &\leq (1+t)^{- 3/2 + \rho}  C_j
(1+\chi^{(3)})& (6.83{\rm d})\cr
&\qquad{}\Big(\tau^j_{N_0,2}  \Vert L \Vert^{}_{{\cal F}^M_l} + C_{l,j}
\sum_{\scr n_1+n_2=l\atop\scr n_2 \leq l-1}\tau^j_{N_0,n_1+2}  \Vert L
\Vert^{}_{{\cal F}^M_{n_2}}\Big),\quad  0 \leq l \leq n,\cr
}$$
where $C_j,  C_{l,j}$ and $C_{l,k,j}$ depend only on $\rho$ and
$\Vert u \Vert^{}_{E^\rho_{N_0}}$. This proves that the hypothesis of
Proposition 6.4 are satisfied and therefore we can conclude that equation
(6.79) has a unique solution $\Psi \in {\cal F}^D_n$, satisfying the
inequalities of statement i)--iv) of Proposition 6.4.

Proposition 6.2 shows that $\Delta^{*M} \in {\cal F}^M_l$ for $l \geq 0$. Let
$K^{(1)}, K^{(2)} \in {\cal F}^M_n$ and let $\Phi^{(i)} \in {\cal F}^M_n$ be
the solution of equation (6.79) when $L = K^{(i)} + \Delta^{*M},  K =
K^{(i)}$ and $r = \phi^*$, i.e. the solution of equation (6.31b)
with $K = K^{(i)}.$
According to expression (6.33) let
$$K'^{(i)}_\mu (t) = - \int^\infty_t  \vert \nabla \vert^{-1}
\sin (\vert \nabla \vert (t-s)) (\overline{\Phi}^{(i)}
\gamma_\mu \Phi^{(i)} + \overline{\Phi}^{(i)}
\gamma_\mu  \phi^* + \overline{\phi}^* \gamma_\mu \Phi^{(i)}) (s) ds,
\eqno{(6.84)}$$
for $0 \leq \mu \leq 3,  t \geq 0,  i = 1,2$. It
follows from equation (6.2a), with $\phi_1 = \phi^*$, from equation (6.79) and
from expression (6.84) that $\partial^\mu  K'^{(i)}_\mu = 0.$
It follows from expression (6.84) that
$$\eqalignno{
&(K'^{(1)}_Y - K'^{(2)}_Y)_\mu (t) &(6.85)\cr
&\quad{}= - \int^\infty_t  \vert \nabla
\vert^{-1}  \sin (\vert \nabla \vert (t-s))\cr
&\qquad{}\suma_{Y_1,Y_2}^Y  \big((\overline{\Phi}^{(1)}_{Y_1} +
\overline{\phi}^*_{Y_1})
\gamma_\mu (\Phi^{(1)}_{Y_2} - \Phi^{(2)}_{Y_2}) +
(\overline{\Phi}^{(1)}_{Y_1} -
\overline{\Phi}^{(2)}_{Y_1}) \gamma_\mu (\Phi^{(2)}_{Y_2} +
\phi^*_{Y_2})\big) (s) ds,\cr
}$$
for $Y \in \Pi'$. Using the notation
$$H_l (\Phi,t) = \sum_{\scr Y \in \Pi'\atop\scr k+\vert Y \vert \leq l}
\Vert (\delta (t))^{3/2}  (1+\lambda_1 (t))^{k/2}  \Phi_Y (t)
\Vert^{}_{L^\infty}, \eqno{(6.86)}$$
for $l \geq 0,	N_Y = K'^{(1)}_Y - K'^{(2)}_Y,  \dot{N}_Y =
N_{P_0 Y}$, equality (6.85) gives that
$$\eqalignno{
&(1+t)^{1-\rho}  \wp^{M^0}_l (N(t), \dot{N} (t)) + (1+t)^{2-\rho}
\wp^{M^1}_l (N(t), \dot{N} (t))&(6.87)\cr
&\quad{}\leq \sup_{s \geq t}	\Big((1+s)^{3/2-\rho}  \Big(C\big(H_0
(\Phi^{(1)},s)
+ H_0 (\Phi^{(2)},s) + H_0 (\phi^*,s)\big)\wp^D_l (\Phi^{(1)} (s) -
\Phi^{(2)} (s))\cr
&\qquad{}+ C \big(\wp^D_l (\Phi^{(1)} (s)) + \wp^D_l (\Phi^{(2)} (s))\big)
  H_0 (\Phi^{(1)} -\Phi^{(2)},s)\cr
&\qquad{}+ C_l  \sum_{\scr n_1+n_2=l\atop\scr  n_1 \leq n_2 \leq l-1}
\Big(\big(H_{n_1} (\Phi^{(1)},s) + H_{n_1} (\Phi^{(2)},s)\big)
\wp^D_{n_2} (\Phi^{(1)} (s) - \Phi^{(2)} (s))\cr
&\qquad\qquad{}+ H_{n_1} (\Phi^{(1)} - \Phi^{(2)},s) \big(\wp^D_{n_2}
(\Phi^{(1)} (s)) +
\wp^D_{n_2} (\Phi^{(2)} (s))\big)\Big)\cr
&\qquad{}+ C_l  \sum_{\scr n_1+n_2=l\atop\scr  n_2 \leq l-1}  H_{n_1}
(\phi^*,s)
 \wp^D_{n_2} (\Phi^{(1)} (s) - \Phi^{(2)} (s))\Big)\Big), \quad 0 \leq l \leq
n,
 t \geq 0,\cr
}$$
where $C$ and $C_l$ are independent of $t$, $\phi^*$ and $\Phi^{(i)}$.
Using that there are two positive real numbers $C$ and $C'$ such that
$(\delta (t)) (x) \leq C(1+t+(\lambda_1 (t)) (x)) \leq C''(\delta (t))(x)$
and that $(\delta (t)) (x) = 1+t + \vert x \vert$ we obtain by H\"older
inequality and with a new constant:
$$\eqalignno{
&(1+t)^{1-\rho}  \Vert \delta (t)f_1 (t)
f_2 (t) \Vert^{}_{L^{6/5}} &(6.88)\cr
&\qquad{}\leq (1+t)^{1-\rho}  \Vert f_1 (t) \Vert^{}_{L^2}  \Vert
\delta (t)  f_2 (t) \Vert^{}_{L^3}\cr
&\qquad{}\leq (1+t)^{3/2 - \rho}  C \Vert f_1 (t) \Vert^{}_{L^2}  \Vert
f_2 (t) \Vert^{2/3}_{L^2}  \Vert(\delta (t))^{3/2}  (1+
\lambda_1 (t))^{3/2} f_2 (t) \Vert^{1/3}_{L^\infty}\cr
&\qquad{}\leq (1+t)^{3/2 - \rho}  C \Vert f_1 (t) \Vert^{}_{L^2}  \big(\Vert
f_2 (t) \Vert^{}_{L^2} + \Vert (\delta (t))^{3/2}  (1+ \lambda_1
(t))^{3/2}  f_2 (t) \Vert^{}_{L^\infty}\big).\cr
}$$
This inequality and the fact that $\carre  K'^{(i)}_\mu = \overline{\Phi}^{(i)}
 \gamma_\mu  \Phi^{(i)} + \overline{\Phi}^{(i)}  \gamma_\mu
 \phi^* + \overline{\phi}^*  \gamma^\mu  \Phi^{(i)}$
give, similarly as inequality (6.87) was obtained, that
$$\eqalignno{
&\sum_{\scr Y \in \Pi'\atop\scr \vert Y \vert \leq l}  ((1+t)^{2-\rho}
\Vert \delta (t)  \carre N_Y (t) \Vert^{}_{L^2}
+ (1+t)^{1-\rho}  \Vert \delta (t)  \carre N_Y (t)
\Vert^{}_{L^{6/5}}&(6.89)\cr
&\quad{}\leq \sup_{s \geq t}	\Big((1+s)^{3/2-\rho}  \Big(C\big(H_3
(\Phi^{(1)},s)
+ H_3 (\Phi^{(2)},s) + H_3 (\phi^*,s)\cr
&\qquad{}+ \wp^D_0 (\Phi^{(1)} (s))
+ \wp^D_0 (\Phi^{(2)} (s)) + \wp^D_0 (\phi^* (s))\big)
\wp^D_l (\Phi^{(1)} (s) - \Phi^{(2)} (s))\cr
&\qquad{}+ C \big(\wp^D_l (\Phi^{(1)} (s)) + \wp^D_l (\Phi^{(2)} (s))\big)
\big(\wp^D_0 (\Phi^{(1)}
(s) - \Phi^{(2)} (s)) + H_3 (\Phi^{(1)} - \Phi^{(2)},s)\big)\cr
&\qquad{}+ C_l  \sum_{\scr n_1+n_2=l\atop\scr n_1 \leq n_2 \leq l-1}
\Big(\big(H_{n_1+3}(\Phi^{(1)},s) + H_{n_1+3}  (\Phi^{(2)},s) +
H_{n_1+3} (\phi^*,s)\cr
&\qquad{}+ \wp^D_{n_1} (\Phi^{(1)} (s))
+ \wp^D_{n_1} (\Phi^{(2)} (s)) + \wp^D_{n_1} (\phi^*
(s))\big)  \wp^D_{n_2} (\Phi^{(1)} (s) - \Phi^{(2)} (s))\cr
&\qquad{}+ \big(H_{n_1+3} (\Phi^{(1)} - \Phi^{(2)},s)
+ \wp^D_{n_1} (\Phi^{(1)} (s) - \Phi^{(2)}
(s))\big)\big(\wp^D_{n_2} (\Phi^{(1)} (s))
+ \wp^D_{n_2} (\Phi^{(1)} (s))\big)\Big)\cr
&\qquad{}+ C_l  \sum_{\scr n_1+n_2=l\atop\scr n_2 \leq l-1}
\big(H_{n_1+3} (\phi^*,s) +
\wp^D_{n_1} (\phi^* (s))\big) \wp^D_{n_2}
(\Phi^{(1)} (s) - \Phi^{(2)} (s))\Big)\Big),\cr
}$$
$0 \leq l \leq n,  t \geq 0$, where $C$ and $C_l$ are independent of
$t,\phi^*$ and $\Phi^{(i)}$. It follows from inequalities (6.87) and (6.89)
that
$$\eqalignno{
&\Vert K'^{(1)} - K'^{(2)} \Vert^{}_{{\cal F}^M_l}&(6.90)\cr
&\quad{}\leq \sup_{t \geq 0} \Big(
C\big(H_3 (\phi^*,t) + \wp^D_0 (\phi^* (t)) + \sum^2_{i=1}  \big(H_3
(\Phi^{(i)},t)
+ \wp^D_0 (\Phi^{(i)} (t))\big)\big)
\Vert \Phi^{(1)} - \Phi^{(2)} \Vert^{}_{{\cal F}^D_l}\cr
&\qquad{}+ C\big(\wp^D_0 (\Phi^{(1)} (t) - \Phi^{(2)} (t))
+ H_3 (\Phi^{(1)} - \Phi^{(2)},t)\big)
\big(\Vert \Phi^{(1)} \Vert^{}_{{\cal F}^D_l}
+ \Vert \Phi^{(2)} \Vert^{}_{{\cal F}^D_l}\big)\cr
&\qquad{}+ C_l  \sum_{\scr n_1+n_2=l\atop\scr n_1 \leq n_2 \leq l-1}
\Big(\big(H_{n_1+3}
 (\phi^*,t) + \wp^D_{n_1} (\phi^* (t))\cr
&\qquad\quad{}+ \sum^2_{i=1}  \big(H_{n_1+3} (\Phi^{(i)}, t) +
\wp^D_{n_1} (\Phi^{(i)}
(t))\big)\big)\Vert \Phi^{(1)} - \Phi^{(2)} \Vert^{}_{{\cal F}^D_{n_2}}\cr
&\qquad{}+ \big(H_{n_1+3} (\Phi^{(1)} - \Phi^{(2)}, t)
+ \wp^D_{n_1} (\Phi^{(1)} (t) - \Phi^{(2)}
(t))\big)\big(\Vert \Phi^{(1)} \Vert^{}_{{\cal F}^D_{n_2}} + \Vert \Phi^{(2)}
\Vert^{}_{{\cal F}^D_{n_2}}\big)\Big)\cr
&\qquad{}+ C_l  \sum_{\scr n_1+n_2=l\atop\scr n_2 \leq l-1}
\big(H_{n_1+3} (\phi^*,t)
+ \wp^D_{n_1} (\phi^* (t))\big)  \Vert \Phi^{(1)}
- \Phi^{(2)} \Vert^{}_{{\cal F}^D_{n_2}}\Big),
\quad  0 \leq l \leq n,  t \geq 0,
}$$
where $C$ and $C_l$ are independent of $t, \phi^*$ and $\Phi^{(i)}.$

Let us denote by $R'_l (t)$ (resp. $R^\infty_l (t), R^2_l (t),
\wp^D_l (f(t)))$ on the
left-hand side of inequality (6.83a) (resp. (6.83b), (6.83c), (6.83d)) by $R'_l
(K,L,t)$ (resp. $R^\infty_l (L,t)$, $(\wp^D_l ((1+\lambda_1 (t))^{k/2}
g(L,t)),$
$\wp^D_l (f(K,L,t)))$ to indicate the dependence of these quantities on
$K,L \in {\cal F}^M_n.$
It follows from inequalities (6.83a)--(6.83d), in the case $j=0$, that
$$\eqalignno{
&\sup_{t \geq 0}  \big((1+t)^{3/2 - \rho}\big(R'_l (K,L,t) +
\wp^D_l (f(K,L,t))\big)\big)&(6.91{\rm a})\cr
&\qquad{}\leq C_l  \big(1 + \Vert u \Vert^{}_{E^\rho_{N_0+l_0}}
+ \Vert K \Vert^{}_{{\cal F}^M_{l_0}}\big)
 \Vert u \Vert^{}_{E^\rho_{N_0+l+2}}  \Vert M \Vert^{}_{{\cal F}^M_l},\cr
}$$
$0 \leq l \leq n$, where $l_0$ is the integer part of $l/2 + 3$ and where $C_l$
depends only on $\rho$ and $\Vert u \Vert^{}_{E^\rho_{N_0}}$, it follows that
$$\sup_{t \geq 0}  ((1+t)  R^\infty_l (L,t))
\leq C_l  \Vert u \Vert^{}_{E^\rho_{N_0+l}}  \Vert L \Vert^{}_{{\cal
F}^M_{l+3}},
\quad 0 \leq l \leq n-3,\eqno{(6.91{\rm b})}$$
where $C_l$ depends only on $\rho$ and $\Vert u \Vert^{}_{E^\rho_{N_0}}$
and it follows
that
$$\sup_{t \geq 0}  \big((1+t)  \wp^D_l \big((1+\lambda_1 (t))^{k/2}
 g(L,t)\big)\big)
\leq C_{l,k}	\Vert u \Vert^{}_{E^\rho_{N_0+l+k}}  \Vert L
\Vert^{}_{{\cal F}^M_l}, \eqno{(6.91{\rm c})}$$
$0 \leq l \leq n,  k \geq 0$, where $C_{l,k}$ depends only on $\rho$ and
$\Vert u \Vert^{}_{E^\rho_{N_0}}$.
Introducing also, for the left-hand side of inequalities (6.39a)--(6.39e) the
notation $\overline{Q}_n (K,L,t)$, $Q'_n (K,L,t)$, $R^{(1)}_n(K,L,t)$,
$h'_n(K,L,n_0,t)$
and $h''_n(K,L,t)$ to stress the dependence on $K$, $L$ and $n_0$,
it follows from inequalities (6.91a), (6.91b)
and (6.91c) that, if $n_0$ is the integer part of $n/2 + 5$, then
$$\eqalignno{
&\overline{Q}_l (K,L,t) &(6.92{\rm a})\cr
&\quad{}\leq C_l \big(1 + \Vert u \Vert^{}_{E^\rho_{N_0+l+5}} + \Vert K
\Vert^{}_{{\cal F}^M_{l+5}}\big)^2
\Vert u \Vert^{}_{E^\rho_{N_0+l+2}}  \Vert L \Vert^{}_{{\cal F}^M_l},
\quad  0 \leq l \leq n-5,\cr
&R^{(1)}_l (K,L,t) &(6.92{\rm b})\cr
&\quad{}\leq C_l \big(1+\Vert u \Vert^{}_{E^\rho_{N_0+l+13}} + \Vert K
\Vert^{}_{{\cal F}^M_{l+13}}\big)^2
\Vert u \Vert^{}_{E^\rho_{N_0+l+10}}	\Vert L \Vert^{}_{{\cal F}^M_{l+9}},
\quad  0 \leq l \leq n-13,\cr
&h'_l (K,L,n_0,t) + h''_l (K,L,t)& (6.92{\rm c})\cr
&\quad{}\leq C'_l  \big(\Vert u \Vert^{}_{E^\rho_{N_0+n_0+5}}
+ \Vert K \Vert^{}_{{\cal F}^M_{n_0+5}}
+ \Vert u \Vert^{}_{E^\rho_{N_0+l}} + \Vert K \Vert^{}_{{\cal F}^M_l}\big)
\Vert u \Vert^{}_{E^\rho_{N_0+n_0+2}}  \Vert L \Vert^{}_{{\cal F}^M_{n_0}},
\ 0 \leq l \leq n,\cr
}$$
and
$$\eqalignno{
h^\infty_l (K,L,n_0,t) &\leq C'_l \big(\Vert u \Vert^{}_{E^\rho_{N_0+n_0+2}}
\Vert L \Vert^{}_{{\cal F}^M_{n_0}} + \Vert u \Vert^{}_{E^\rho_{N_0+l}}
\Vert L \Vert^{}_{{\cal F}^M_{l-1}}\big)&(6.92{\rm d})\cr
&\qquad{}\big(\Vert u \Vert^{}_{E^\rho_{N_0+n_0+5}}
+ \Vert K \Vert^{}_{{\cal F}^M_{n_0+5}} +
\Vert u \Vert^{}_{E^\rho_{N_0+l}} + \Vert K \Vert^{}_{{\cal F}^M_l}\big),\quad
0 \leq l \leq n,
}$$
where $C_l$ depends only on $\rho$ and $\Vert u \Vert^{}_{E^\rho_{N_0}}$,
where $C'_l$
depends only on $\rho$, $\Vert u \Vert^{}_{E^\rho_{N_0+n_0+5}}$
and $\Vert K \Vert^{}_{{\cal F}^M_{n_0+5}}$,
and where $h^\infty_l$ is defined in statement ii) of Proposition 6.4.
In deducing (6.92c) we have used that $n - n_0 \leq n_0 - 9$. It follows
from statements ii) and iii) of Proposition 6.4 and from inequalities
(6.91a), (6.92b) and (6.92d) that, if $r = \phi^*$ then the solution of
equation (6.79) satisfies
$$\Vert \Psi \Vert^{}_{{\cal F}^D_n} \leq C_n
\Vert u \Vert^{}_{E^\rho_{N_0+n+2}}
 \Vert L \Vert^{}_{{\cal F}^M_n} \eqno{(6.93{\rm a})}$$
and
$$H_l (\Psi,t) \leq (1+t)^{- 3/2+\rho}	C_l
\Vert u \Vert^{}_{E^\rho_{N_0+l+10}}
 \Vert L \Vert^{}_{{\cal F}^M_{l+9}}, \eqno{(6.93{\rm b})}$$
$0 \leq l \leq n-13$, where $C_n$ and $C_l$ depend only on
$\Vert u \Vert^{}_{E^\rho_{N_0+n}}$
and $\Vert K \Vert^{}_{{\cal F}^M_n}$, and where we have used
that $n-n_0 \leq n_0-9.$

It follows from inequalities (6.93a) and (6.93b), with $\Psi = \Phi^{(i)}$,
 $K = K^{(i)}$,  $L = K^{(i)} + \Delta^{*M}$ and from Proposition
6.2 that
$$\eqalignno{
\Vert \Phi^{(i)} \Vert^{}_{{\cal F}^D_n}
&\leq C^{(i)}_n  \Vert u \Vert^{}_{E^\rho_{N_0+n+2}}
\Vert K^{(i)} + \Delta^{*M} \Vert^{}_{{\cal F}^M_n}&(6.94{\rm a})\cr
&\leq C'^{(i)}_n  \Vert u \Vert^{}_{E^\rho_{N_0+n+2}}  (\Vert
u \Vert^{}_{E^\rho_{N_0+n}} + \Vert K^{(i)} \Vert^{}_{{\cal F}^M_n})\cr
}$$
and that
$$\eqalignno{
H_l (\Phi^{(i)},t) &\leq (1+t)^{- 3/2 + \rho}C^{(i)}_l
\Vert u \Vert^{}_{E^\rho_{N_0+l+10}}  \Vert K^{(i)}
+ \Delta^{*M} \Vert^{}_{{\cal F}^M_{l+9}}&(6.94{\rm b})\cr
&\leq (1+t)^{- 3/2+\rho}  C'^{(i)}_l  \Vert u \Vert^{}_{E^\rho_{N_0+l+10}}
 (\Vert u \Vert^{}_{E^\rho_{N_0+l+9}} + \Vert K^{(i)}
 \Vert^{}_{{\cal F}^M_{l+9}}),\cr
}$$
$0 \leq l \leq n-13$, where $C^{(i)}_l, C'^{(i)}_l, C^{(i)}_n$ and $C'^{(i)}_n$
depend only on $\rho$, $\Vert u \Vert^{}_{E^\rho_{N_0+n}}$ and
$\Vert K \Vert^{}_{{\cal F}^M_n}.$
Using that $n \geq 32$, it follows from inequalities (6.90), (6.94a) and
(6.94b)
that
$$\eqalignno{
\Vert K'^{(1)} - K'^{(2)} \Vert^{}_{{\cal F}^M_n} &\leq C_n  \Vert u
\Vert^{}_{E^\rho_{N_0+n+3}}  \Vert \Phi^{(1)} - \Phi^{(2)}
\Vert^{}_{{\cal F}^D_n}
&(6.95)\cr
&\ {}+ C_n  \Vert u \Vert^{}_{E^\rho_{N_0+n+2}}  \sum_{i \leq n/2}
 \big(H_{i+3} (\Phi^{(1)} - \Phi^{(2)},t) + \wp^D_i (\Phi^{(1)} (t) -
 \Phi^{(2)}
(t))\big),\cr
}$$
where $C_n$ depends only on $\rho, \Vert u \Vert^{}_{E^\rho_{N_0+n+2}}$ and
$\Vert K^{(i)} \Vert^{}_{{\cal F}^M_n},  i = 1,2.$

In particular if $K^{(2)} = - \Delta^{*M}$ then $\Phi^{(2)} = 0$ and
$K'^{(2)} = 0$.
If moreover $K^{(1)} = K$,  $\Phi = \Phi^{(1)}$ and $K' = K'^{(1)}$, then
$K' = {\cal N} (u,K)$. It therefore follows from inequalities (6.94a), (6.94b)
and (6.95) and from Proposition 6.2 that
$$\Vert {\cal N} (u,K) \Vert^{}_{{\cal F}^M_N}
\leq C_n  \Vert u \Vert^{}_{N_0+n+3}(\Vert u \Vert^{}_{E^\rho_{N_0+n}}
+ \Vert K \Vert^{}_{{\cal F}^M_n}), \eqno{(6.96)}$$
where $C_n$ depends only on $\rho$, $\Vert u \Vert^{}_{E^\rho_{N_0+n+2}}$ and
$\Vert K \Vert^{}_{{\cal F}^M_n}$. This proves that ${\cal N}$ is a map from
$E^{\circ\rho}_\infty
\times {\cal F}^M_n$ to ${\cal F}^M_n.$

To estimate $\Vert \Phi^{(1)} - \Phi^{(2)} \Vert^{}_{{\cal F}^D_n}$, we note
that
according to the definition of $\Phi^{(i)}$, it follows from equation (6.79)
that
$$\eqalignno{
&\big(i  \gamma^\mu  \partial_\mu + m - \gamma^\mu (A^*_\mu +
K^{(1)}_\mu - \partial_\mu  {\vartheta} (A^* + K^{(1)}))\big)
(\Phi^{(1)} - \Phi^{(2)})&(6.97)\cr
&\qquad{}= \gamma^\mu \big(K^{(1)}_\mu - K^{(2)}_\mu - \partial_\mu
{\vartheta}
(K^{(1)} - K^{(2)})\big)  (\phi^* + \Phi^{(2)}),\cr
}$$
where $\Phi^{(i)} \in {\cal F}^D_n,  K^{(i)} \in {\cal F}^M_n$ for
$i=1$ and $i=2$. Let $\Delta$ be the unique solution in ${\cal F}^D_n$ of the
equation
$$\eqalignno{
&\big(i \gamma^\mu  \partial_\mu + m - \gamma^\mu (A^*_\mu + K^{(1)}_\mu -
\partial_\mu  {\vartheta} (A^* + K^{(1)}))\big)  \Delta&(6.98)\cr
&\qquad{}= \gamma^\mu	\big(K^{(1)}_\mu - K^{(2)}_\mu - \partial_\mu
{\vartheta} (K^{(1)} - K^{(2)})\big)  \phi^*.\cr
}$$
According to inequalities (6.93a) and (6.93b) we obtain that
$$\Vert \Delta \Vert^{}_{{\cal F}^D_n} \leq C_n
\Vert u \Vert^{}_{E^\rho_{N_0+n+2}}
 \Vert K^{(1)} - K^{(2)} \Vert^{}_{{\cal F}^M_n} \eqno{(6.99{\rm a})}$$
and
$$H_l (\Delta,t) \leq (1+t)^{- 3/2 + \rho}  C_l  \Vert u
\Vert^{}_{E^\rho_{N_0+l+10}}  \Vert K^{(1)} - K^{(2)} \Vert^{}_{{\cal F}^M_n},
\eqno{(6.99{\rm b})}$$
$0 \leq l \leq n-13$, where $C_l$ depends only
on $\rho$, $\Vert u \Vert^{}_{E^\rho_{N_0+n}}$
and $\Vert K^{(1)} \Vert^{}_{{\cal F}^M_n}$. We next prove that the equation
$$\eqalignno{
&\big(i \gamma^\mu  \partial_\mu + m - \gamma^\mu (A^*_\mu + K^{(1)}_\mu -
\partial_\mu  {\vartheta} (A^* + K^{(1)}))\big)  \Delta'&(6.100)\cr
&\qquad{}= \gamma^\mu \big(K^{(1)}_\mu - K^{(2)}_\mu - \partial_\mu
{\vartheta}
(K^{(1)} - K^{(2)})\big)  \Phi^{(2)}\cr
}$$
has a unique solution $\Delta' \in {\cal F}^D_n$. To do this we first use Lemma
6.6. Let
$$u^{}_l (\Phi^{(2)},t) = H_l (\Phi^{(2)},t)
+ \sum_{\scr Y \in \Pi'\atop\scr k+\vert Y \vert \leq l+1}
 \Vert (\delta (t))^{3-\rho}	(1+\lambda_1 (t))^{k/2}
 ((m + i \gamma^\mu  \partial_\mu) \Phi^{(2)}_Y) (t) \Vert^{}_{L^\infty}.$$
Since
$$\eqalignno{
(i \gamma^\mu  \partial_\mu  + m)  \Phi^{(2)} &= \gamma^\mu
\big(A^*_\mu + K^{(2)}_\mu - \partial_\mu  {\vartheta} (A^* + K^{(2)})\big)
 \Phi^{(2)}\cr
&\qquad{}+ \gamma^\mu	\big(K^{(2)}_\mu + \Delta^{*M}_\mu - \partial_\mu
{\vartheta} (K^{(2)} + \Delta^{*M})\big)  \phi^*,\cr
}$$
it follows from inequalities (5.116c), (6.38a), from Lemma 6.3 and
from inequality (6.66a) that
$$\eqalignno{
u^{}_l (\Phi^{(2)},t) &\leq H_l (\Phi^{(2)},t) + C_l
(\Vert u \Vert^{}_{E^\rho_{l+4}} +
\Vert K^{(2)} \Vert^{}_{{\cal F}^M_{l+4}})  H_{l+1}  (\Phi^{(2)},t)\cr
&\qquad{}+ (1+t)^{1/2 - \rho}	C_l  \Vert K^{(2)} + \Delta^{*M}
\Vert^{}_{{\cal F}^M_{l+4}}  H_{l+1}  (\phi^*,t),\cr
}$$
where $C_l$ depends only on $\rho$ and $\Vert u \Vert^{}_{E^\rho_{N_0}}$.
It then follows from Proposition 6.2 and inequality (6.94b) that
$$\eqalignno{
u^{}_l (\Phi^{(2)},t) &\leq (1+t)^{- 3/2 + \rho}  C'_l  \Vert
u \Vert^{}_{E^\rho_{N_0+11+l}}	\Vert K^{(2)} + \Delta^{*M}
\Vert^{}_{{\cal F}^M_{10+l}}
&(6.101{\rm a})\cr
&\qquad{}+ (1+t)^{(1/2 - \rho)}	C_l  \Vert u \Vert^{}_{E^\rho_{N_0+l+1}}
 \Vert K^{(2)} + \Delta^{*M} \Vert^{}_{{\cal F}^M_{l+4}},
\quad 0 \leq l \leq n-14,\cr
}$$
where $C_l$ depends only on $\rho$ and $\Vert u \Vert^{}_{E^\rho_{N_0}}$ and
$C'_l$ depends only on $\rho$, $\Vert u \Vert^{}_{E^\rho_{N_0+n}}$ and $\Vert
K^{(2)} \Vert^{}_{{\cal F}^M_n}$. Proposition 6.2, inequality (6.101a) and the
inequality $- 3/2 + \rho < 1/2 - \rho$ give that
$$u^{}_l (\Phi^{(2)},t) \leq (1+t)^{1/2-\rho}  C_l  \Vert u
\Vert^{}_{E^\rho_{N_0+l+11}}  (\Vert u \Vert^{}_{E^\rho_{N_0+l+10}} + \Vert
K^{(2)} \Vert^{}_{{\cal F}^M_{l+10}}), \eqno{(6.101{\rm b})}$$
for $0 \leq l \leq n-14$, where $C_l$ depends only on $\rho$,
$\Vert u \Vert^{}_{E^\rho_{N_0+n}}$
and $\Vert K^{(2)} \Vert^{}_{{\cal F}^M_n}$. This inequality and inequality
(6.94b) show that the hypothesis of Lemma 6.6 are satisfied for
$p_1 = n-14$, $p_2 = n-13$, since $n/2 \leq p_2$. We can therefore conclude
that $g^{}_Y \in C^0(\Rrm^+,D) \cap
L^1 (\Rrm^+,D)$,  $Y \in \Pi'$,  $\vert Y \vert \leq n$, where
$g$ is given by the right-hand side of equation (6.100), and by using (6.94a)
and (6.94b) that
$$R'_l (t) \leq (1+t)^{- 3/2+\rho}  C_l  \Vert u \Vert^{}_{E^\rho_{N_0+l+11}}
 \Vert K^{(1)} - K^{(2)} \Vert^{}_{{\cal F}^M_{n-14}}, \eqno{(6.102{\rm a})}$$
for $0 \leq l \leq n-14,$
$$R^\infty_l (t) \leq (1+t)^{- 5/2 + \rho}  C_l  \Vert u
\Vert^{}_{E^\rho_{N_0+l+10}}  \Vert K^{(1)} - K^{(2)} \Vert^{}_{{\cal
F}^M_{l+3}},
\eqno{(6.102{\rm b})}$$
for $0 \leq l \leq n-13,$
$$\wp^D_l  \big((1+\lambda_1 (t))^{k/2}  g(t)\big)
\leq (1+t)^{- 5/2 + \rho}  C_l  \Vert u \Vert^{}_{E^\rho_{N_0+l+k+10}}
 \Vert K^{(1)} - K^{(2)} \Vert^{}_{{\cal F}^M_l}, \eqno{(6.102{\rm c})}$$
for $0 \leq l+k \leq n-13$ and if $l_0$ is the integer part of $l/2+3$ then
$$\eqalignno{
&\wp^D_l (f(t)) + (1+t)  \wp^D_l (g(t))&(6.102{\rm d})\cr
&\quad{}\leq (1+t)^{- 3/2 + \rho}
C_l\big(\Vert u \Vert^{}_{E^\rho_{N_0+l_0+7}}
\Vert K^{(1)} - K^{(2)} \Vert^{}_{{\cal F}^M_l}
+ \Vert u \Vert^{}_{E^\rho_{N_0+l+2}}
\Vert K^{(1)} - K^{(2)} \Vert^{}_{{\cal F}^M_{l_0}}\big),\cr
}$$
for $0 \leq l \leq n$, where $C_l$ depends only on $\rho$,
$\Vert u \Vert^{}_{E^\rho_{N_0+N}}$
and $\Vert K^{(2)} \Vert^{}_{{\cal F}^M_n}$. Let $n_0$ be the integer part
of $n/2+5.$ Since $n_0 \leq n-14$ for $n \geq 38$ and since
$(1+\lambda_0 (t)) (x) \leq C (1+t),$
with $C$ independent of $t \in \Rrm^+,	x \in \Rrm^3$, if follows from
inequalities (6.102a)--(6.102d) that the hypotheses of Proposition 6.4 applied
to equation (6.100) are satisfied, which proves that $\Delta' \in {\cal
F}^M_n.$
Moreover statements ii) and iii) of Proposition 6.4 and inequalities
(6.102a)--(6.102d) show that
$$\Vert \Delta' \Vert^{}_{{\cal F}^D_n} \leq C_n
\Vert u \Vert^{}_{E^\rho_{N_0+n+1}}
 \Vert K^{(1)} - K^{(2)} \Vert^{}_{{\cal F}^M_n} \eqno{(6.103{\rm a})}$$
and that
$$H_l (\Delta',t) \leq (1+t)^{- 3/2+\rho}  C_l  \Vert u
\Vert^{}_{E^\rho_{N_0+n}}  \Vert K^{(1)} - K^{(2)} \Vert^{}_{{\cal F}^M_n},
\eqno{(6.103{\rm b})}$$
$0 \leq l \leq n-22$, where $C_l$ and $C_n$ depend only on $\rho$, $\Vert u
\Vert^{}_{E^\rho_{N_0+n}}$ and $\Vert K^{(i)} \Vert^{}_{{\cal F}^M_n}$,
$i=1,2$.
Since
$\Delta \in {\cal F}^D_n$ and $\Delta' \in {\cal F}^D_n$ satisfy respectively
equations (6.98) and (6.100), it follows that the unique solution in ${\cal
F}_n$
of equation (6.97) is given by $\Phi^{(1)} - \Phi^{(2)} = \Delta + \Delta'.$
Inequalities (6.99a), (6.99b), (6.103a) and (6.103b) give that
$$\Vert \Phi^{(1)} - \Phi^{(2)} \Vert^{}_{{\cal F}^D_n} \leq C_n  \Vert u
\Vert^{}_{E^\rho_{N_0+n+2}}  \Vert K^{(1)} - K^{(2)} \Vert^{}_{{\cal F}^M_n}
\eqno{(6.104{\rm a})}$$
and that
$$H_l  (\Phi^{(1)} - \Phi^{(2)},t) \leq (1+t)^{- 3/2+\rho}
C_n  \Vert u \Vert^{}_{E^\rho_{N_0+n}}	\Vert K^{(1)} - K^{(2)}
\Vert^{}_{{\cal F}^M_n}, \eqno{(6.104{\rm b})}$$
$0 \leq l \leq n-22$, where $C_n$ depends only on $\rho$,
$\Vert u \Vert^{}_{E^\rho_{N_0+n}}$
and $\Vert K^{(i)} \Vert^{}_{{\cal F}^M_n}$, $i=1,2$.
Since $n/2+3 \leq n-22$ for $n \geq 50,$
the inequality of the proposition follows from inequalities (6.95), (6.104a)
and (6.104b). This proves the proposition.

We are now ready to prove the {\it existence of solutions} $(K, \Phi)$ of
equations (6.31a), (6.31b)
and (6.31c). We recall  that $\phi^*, A^*$ and $\Delta^{*M}$ are
functions of $ u \in E^{\circ\rho}_\infty.$
\saut
\noindent{\bf Corollary 6.8.}
{\it
Let $n \in \Nrm$, $\varepsilon \in ]0, \infty[, \rho \in ]1/2,1[$ and
let ${\cal O}_{n+3}$ (resp. $Q_n$) be the open ball with center at the
origin and radius $\varepsilon$ (resp.$2 \varepsilon$) in
$E^{\circ\rho}_{n+3}$ (resp. ${\cal F}_n$). Let ${\cal O}_\infty =
{\cal O}_{n+3} \cap E^{\circ\rho}_\infty$. If $n \geq 50$, then there exists
$\varepsilon$ such that equations (6.31a), (6.31b) and (6.31c) have a
unique solution $(K,\Phi) \in Q_n$ for each $u \in {\cal O}_\infty$.
This solution satisfies
$$\Vert (K, \Phi) \Vert^{}_{{\cal F}_n} \leq C_n
\Vert u \Vert^2_{E^\rho_{N_0+n+3}}$$
and
$$(1+t)^{3/2 - \rho}  H_{n-13} (\Phi,t) + (1+t)^{\rho - 1/2}  u^{}_{n-14}
(\Phi,t) \leq C_n  \Vert u \Vert^2_{E^\rho_{N_0+n-1}}$$
for $t \geq 0$, where $C_n$ depends only on $\rho$ and $\varepsilon$,
and where $H_j
(\Phi,t)$ and $u_j (\Phi,t)$ are as in Lemma 6.6.
}\saut
\noindent{\it Proof.}
Let $n \geq 50$ and let $Q^M_n$ be the closed ball with center at the
origin and of radius
$\varepsilon$ in ${\cal F}^M_n$. Let $C$ be an upper bound of the two
constants $C_n$ and
$C'_n$ in Proposition 6.7 and the two constants $C'^{(1)}_n$ and $C'^{(2)}_n$
in inequality
(6.94a), for all $u \in {\cal O}_\infty$ and $K, K^{(1)}, K^{(2)} \in Q^M_n$.
If $u \in
{\cal O}_\infty$, then it follows from Proposition 6.7 that
 $K \mapsto {\cal N} (u,K)$ is a
contraction mapping from $Q^M_n$ into $Q^M_n$ if $C \varepsilon < 1$ and
$2 \varepsilon^2 C \leq \varepsilon$. Since $C$ is bounded for $\varepsilon$
belonging to a fixed bounded interval we can choose $\varepsilon$ such that
$C \varepsilon \leq 1/2$. It then follows that $\varepsilon$ satisfies the
above two inequalities. This
proves that, for such $\varepsilon$ the equation $K = {\cal N} (u,K)$ has a
unique solution
$K \in Q^M_n$ for each $u \in {\cal O}_\infty$ and, according to the first
of the inequalities in Proposition 6.7, that
$\Vert K \Vert^{}_{{\cal F}^M_n} \leq {1 \over 2} (\Vert
u \Vert^{}_{E^\rho_{N_0+n}} + \Vert K \Vert^{}_{{\cal F}^M_n})$.
This shows that $\Vert K\Vert^{}_{{\cal F}^M_n} \leq
\Vert u \Vert^{}_{E^\rho_{N_0+n}}$,
which introduced into the first
inequality of Proposition 6.7 gives that
$$
\Vert K \Vert^{}_{{\cal F}^M_n} \leq 2C  \Vert u \Vert^2_{E^\rho_{N_0+n+3}}
\leq
\varepsilon \eqno{(6.105{\rm a})}$$
It follows from inequality (6.94a) and from the fact that
$\Vert K \Vert^{}_{{\cal F}^M_n} \leq \Vert u \Vert^{}_{E^\rho_{N_0+n}}$
that the unique solution $\Phi \in {\cal F}^D_n$ of
equation (6.31a) satisfies
$$\eqalignno{
\Vert \Phi \Vert^{}_{{\cal F}^D_n} &\leq C \Vert u \Vert^{}_{E^\rho_{N_0+n+2}}
(\Vert u \Vert^{}_{E^\rho_{N_0+n}} + \Vert K \Vert^{}_{{\cal
F}^M_n})&(6.105{\rm b})\cr
&\leq 2 C  \Vert u \Vert^2_{E^\rho_{N_0+n+2}}\leq \varepsilon.\cr}$$
$(K, \Phi) \in Q_n$, since inequalities (6.105a) and (6.105b) shows that
$$\Vert (K, \Phi) \Vert^{}_{{\cal F}_n} \leq 2 \sqrt 2	C  \Vert u
\Vert^2_{E^\rho_{N_0+n+3}} \leq \sqrt 2  \varepsilon < 2 \varepsilon,$$
which also gives the first inequality of the corollary.
The second inequality of the corollary follows from inequalities (6.94b),
with $\Phi$ and $K$ instead of $\Phi^{(i)}$
and $K^{(i)}$, and (6.101b), with $\Phi$ and $K$ instead of
$\Phi^{(2)}$ and $K^{(2)}.$
This proves the corollary, since equation (6.31c) is satisfied,
by the definition of the space ${\cal F}_n.$

According to Corollary 6.8, $v(u) = (K, \Phi)$ defines a mapping
$v\colon {\cal O}_\infty \fl
Q_{50}	\subset  {\cal F}_{50}$ into solutions of equations (6.31a), (6.31b)
and (6.31c). In the next theorem we give supplementary
{\it differentiability properties} of this map.
\saut
\noindent{\bf Theorem 6.9.}
{\it
Let $1/2 < \rho < 1$, $L_0 = N_0 + 53$, let ${\cal O}_{L_0}$
(resp. $Q_{50}$) be the open ball with center at the origin and radius
$\varepsilon$ (resp. $2 \varepsilon$) in
$E^{\circ\rho}_{L_0} (\hbox{resp.} {\cal F}_{50}$), let ${\cal O}_\infty =
E^{\circ\rho}_\infty \cap {\cal O}_{L_0}$ and let
$Q_\infty = (\displaystyle{\cap_{n \geq 0}}
 {\cal F}_n) \cap Q_{50}$. If $\varepsilon > 0$ is sufficiently small, then
equations (6.31a), (6.31b) and (6.31c) have a unique solution
$(K, \Phi) = v(u) \in Q_{50}$
for each $u \in {\cal O}_\infty$, the image of the map
$v\colon {\cal O}_\infty \fl Q_{50}$ is a
subset of $Q_\infty$, the map $v\colon {\cal O}_\infty \fl Q_\infty$
is $C^\infty$ and if moreover
$(K^{(l)}, \Phi^{(l)}) = (D^l v) (u ; u^{}_1,\ldots,u^{}_l), l \geq 0,
n \geq 0, u \in {\cal
O}_\infty, u^{}_1,\ldots,u^{}_l \in E^{\circ\rho}_\infty$ then:
\psaut
\noindent\hbox{\rm i)} $\Vert K^{(l)} \Vert^{}_{{\cal F}^M_n} \leq C_{n,l}
\overline{{\cal
R}}^l_{L_0,N_0+3+n+l}  (u^{}_1,\ldots,u^{}_l)
+ C'_{n,l}  \Vert u \Vert^{}_{E^\rho_{N_0+3+n+l}}  \Vert u^{}_1
\Vert^{}_{E^\rho_{L_0}}\cdots \Vert u^{}_l \Vert^{}_{E^\rho_{L_0}}$,
\psaut
\noindent where $C_{n,l}$ and $C'_{n,l}$ depend only
on $\rho$ and $\Vert u \Vert^{}_{E^\rho_{L_0}},$
where $C_{n,l} \leq C_n  \Vert u \Vert^{3-l}_{E^\rho_{L_0}}$ for
$0 \leq l \leq 3$ and $C'_{n,l} \leq C_n  \Vert u \Vert^{2-l}_{E^\rho_{L_0}}$
for $0 \leq l \leq 2,$
with $C_n$ depending only on $\rho$ and $\Vert u \Vert^{}_{E^\rho_{L_0}}$
and where
${\overline{{\cal R}}}^0_{i,j} = 0$ and
$${\overline{{\cal R}}}^l_{i,j} (u^{}_1,\ldots,u^{}_l) =
\sum_{1 \leq q \leq l}  \Vert u^{}_1
\Vert^{}_{E^\rho_i}\cdots \Vert u^{}_{q-1} \Vert^{}_{E^\rho_i}
\Vert u^{}_q \Vert^{}_{E^\rho_j}
\Vert u^{}_{q+1} \Vert^{}_{E^\rho_i}\cdots \Vert u^{}_l \Vert^{}_{E^\rho_i},$$
for $i,j \geq 0,$
\psaut
\noindent\hbox{\rm ii)}
$\Vert \Phi^{(l)} \Vert^{}_{{\cal F}^D_n} \leq C_{n,l}
\overline{{\cal R}}^l_{L_0,
N_0+3+n+l} (u^{}_1,\ldots,u^{}_l)
+ C'_{n,l}  \Vert u \Vert^{}_{E^\rho_{N_0+3+n+l}}  \Vert u^{}_1
\Vert^{}_{E^\rho_{L_0}}\cdots \Vert u^{}_l \Vert^{}_{E^\rho_{L_0}}$,
\psaut
\noindent where $C_{n,l}$ and $C'_{n,l}$ depend only on $\rho$ and
$\Vert u \Vert^{}_{E^\rho_{L_0}}$, and
where $C_{n,l} \leq C_n  \Vert u \Vert^{2-l}_{E^\rho_{L_0}}$ for
$0 \leq l \leq 2$, and $C'_{n,0} \leq C_n  \Vert u \Vert^{}_{E^\rho_{L_0}}$
with $C_n$
depending only on $\rho$ and $\Vert u \Vert^{}_{E^\rho_{L_0}},$
$$\eqalignno{
&\hbox{\rm iii)}\
\sup_{t \geq 0}  \Big((1+t)^{3/2-\rho}
\sum_{\scr Y\in \Pi'\atop\scr k + \vert Y \vert \leq n}
\Vert (\delta (t))^{3/2}  (1+ \lambda_1 (t)^{k/2}
 \Phi^{(l)}_Y (t) \Vert^{}_{L^\infty}\cr
&\quad{}+(1+t)^{\rho-1/2}  \sum_{\scr Y \in \Pi'\atop\scr k +
\vert Y \vert \leq n}
\Vert(\delta (t))^{3 - \rho}  (1+ \lambda_1 (t))^{k/2}  (m + i \gamma^\mu
 \partial_\mu - \gamma^\mu  G_\mu) \Phi^{(l)}_Y (t)
 \Vert^{}_{L^\infty}\Big)\cr
&\qquad{}\leq C_{n,l}	\overline{{\cal R}}^l_{L_0, N_0+3+n+l+13}
(u^{}_1,\ldots,u^{}_l)
+ C'_{n,l}  \Vert u \Vert^{}_{E^\rho_{N_0+16+n+l}}  \Vert u^{}_1
\Vert^{}_{E^\rho_{L_0}}\cdots \Vert u^{}_l \Vert^{}_{E^\rho_{L_0}},
\hskip13mm\cr
}$$
where $C_{n,l}$ and $C'_{n,l}$ are as in statement ii),
$$\eqalignno{
&\hbox{\rm iv)}\ \sup_{t \geq 0}  \Big((1+t)^{3/2 - \rho}\sum_{j+k \leq n}
\wp^D_j \big((1+\lambda_1 (t))^{k/2} \Phi^{(l)} (t)\big)\Big)\cr
&\qquad{}\leq C_{n,l}	\overline{{\cal R}}^l_{L_0, N_0+3+n+l}
(u^{}_1,\ldots,u^{}_l)
+ C'_{n,l}  \Vert u \Vert^{}_{E^\rho_{N_0+3+n+l}}  \Vert u^{}_1
\Vert^{}_{E^\rho_{L_0}} \cdots \Vert u^{}_l \Vert^{}_{E^\rho_{L_0}},
\hskip18mm\cr
}$$
where $C_{n,l}$ and $C'_{n,l}$ are as in statement ii).
}\saut
\noindent{\it Proof.}
According to Corollary 6.8, with $n=50$, there exists $\varepsilon > 0$
such that equations (6.31a), (6.31b) and (6.31c) have a unique solution
$(K, \Phi) \in Q_{50}$ for each $u \in
{\cal O}_\infty$. This establishes the existence of the map
 $v\colon {\cal O}_\infty \fl Q_{50}$
where $v(u) = (K, \Phi).$

We shall estimate the solution $\Phi$ of equation (6.31a) by using
Proposition 6.4. To do this we use the results (6.83a)--(6.83d), with
$L = K + \Delta^{*M}$ and $j = 0$, wich
were obtained from Lemma 6.5 for the equation (6.79). This gives with
$g = \gamma^\mu (K+
\Delta^{*M} - \partial_\mu  {\vartheta} (K + \Delta^{*M}))$:
$$\eqalignno{
R'_l (t) &\leq (1+t)^{- 3/2 + \rho}  C_l \Vert u \Vert^2_{E^\rho_{L_0}}
 \Vert u \Vert^{}_{E^\rho_{N_0+l+2}},\quad  0 \leq l \leq 50,&(6.106{\rm a})\cr
R^\infty_l (t) &\leq (1+t)^{-1}	C_l  \Vert u \Vert^2_{E^\rho_{L_0}}
 \Vert u \Vert^{}_{E^\rho_{N_0+l}},\quad   0 \leq l \leq 47,&(6.106{\rm b})\cr
\wp^D_l \big((1 +& \lambda_1 (t))^{k/2}  g(t)\big)&(6.106{\rm c})\cr
&\leq (1+t)^{-1}  C_{l,k}
 \Vert u \Vert^2_{E^\rho_{L_0}}  \Vert u \Vert^{}_{E^\rho_{N_0+l+k}},
\quad  0 \leq l \leq 50,  k \geq 0, \cr
\noalign{\noindent and}
\wp^D_l (f(t)) &\leq (1+t)^{- 3/2 + \rho}  C_l  \Vert u
\Vert^2_{E^\rho_{L_0}}	\Vert u \Vert^{}_{E^\rho_{N_0+l+2}},\quad   0 \leq l
\leq 50, &(6.106{\rm d})\cr
}$$
where $C_l$ and $C_{l,k}$ depend only on $\rho$ and
$\Vert u \Vert^{}_{E^\rho_{L_0}}$ and where $f_Y, Y \in \Pi'$, is given
by (5.111b) in the context of equation (6.31a). It follows
from inequalities (6.39a)--(6.39e) and (6.106a)--(6.106b) that in
the context of equation (6.31a):
$$\eqalignno{
\overline{Q}_l (t) &\leq C_l  \Vert u \Vert^2_{E^\rho_{L_0}}  \Vert u
\Vert^{}_{E^\rho_{N_0+l+1}}, \quad\hbox{for}\ 0 \leq l \leq 45,&(6.107{\rm
a})\cr
R^{(1)}_l (t) &\leq C_l  \Vert u \Vert^2_{E^\rho_{L_0}}  \Vert u
\Vert^{}_{E^\rho_{N_0+l+10}}, \quad\hbox{for}\ 0 \leq l \leq 45,&(6.107{\rm
b})\cr
h'_l (j_0,t) &\leq C_l  \Vert u \Vert^3_{E^\rho_{L_0}}  \Vert u
\Vert^{}_{E^\rho_{N_0+j_0+1}}, \quad\hbox{for}\ 19 \leq j \leq 51,&(6.107{\rm
c})\cr
}$$
$j_0+1 \leq l \leq j$, where $j_0$ is the integer part of $j/2 + 5,$
$$h''_l (t) \leq C_l  \Vert u \Vert^4_{E^\rho_{L_0}}  \Vert u
\Vert^{}_{E^\rho_{N_0+11}},\qquad \hbox{for}\ 0 \leq l \leq 50,
\eqno{(6.107{\rm d})}$$
and
$$h^\infty_l (j_0,t) \leq C_l  \Vert u \Vert^3_{E^\rho_{L_0}}  \Vert u
\Vert^{}_{E^\rho_{N_0+l}},\qquad \hbox{for}\ j_0+1 \leq l \leq \min (50,j),
\eqno{(6.107{\rm e})}$$
$19 \leq j \leq 51$. Here $C_l$ is a constant depending only on $\rho$ and
$\Vert u \Vert_{E^\rho_{L_0}}$. It follows from statement i) of Proposition
6.4 and inequality (6.107a), for $0 \leq l \leq 45$, from statement ii) of
Proposition 6.4 and inequality
(6.107e) with $j=50$, for $46 \leq l \leq 50$ that
$$\Vert \Phi \Vert^{}_{{\cal F}^D_l} \leq C_l  \Vert u \Vert^2_{E^\rho_{L_0}}
 \Vert u \Vert^{}_{E^\rho_{N_0+l+2}},\quad  0 \leq l \leq 50, \eqno{(6.108{\rm
a})}$$
where $C_l$ depends only on $\rho$ and $\Vert u \Vert^{}_{E^\rho_{L_0}}$.
Statement iii) of
Proposition 6.4 and inequality (6.107b) show that
$$\eqalignno{
H_l (\Phi,t) &= \sum_{\scr Y \in \Pi'\atop\scr \vert Y \vert + k \leq l }
\Vert (\delta(t))^{3/2}  (1 + \lambda_1 (t))^{k/2}  \Phi_Y (t)
\Vert^{}_{L^\infty}
&(6.108{\rm b})\cr
&\leq (1+t)^{- 3/2 + \rho}  C_l  \Vert u \Vert^2_{E^\rho_{L_0}}
 \Vert u \Vert^{}_{E^\rho_{N_0+l+10}},	\quad 0 \leq l \leq 37,\cr
}$$
where $C_l$ depends only on $\rho$ and $\Vert u \Vert^{}_{E^\rho_{L_0}}$.
Equation (6.31b) and
inequality (6.90), (with $K^{(2)} = - \Delta^{*M},  \Phi^{(2)} = 0,
K^{(1)} = K$, $\Phi^{(1)} = \Phi,  K'^{(2)} = 0,  K'^{(1)} = K)$, give
that
$$\eqalignno{
\Vert K \Vert^{}_{{\cal F}^M_l} &\leq C_l  \sup_{t \geq 0}
\Big(\sum_{\scr n_1+n_2=l\atop\scr  n_1 \leq n_2}  \big(H_{n_1+3} (\Phi,t) +
\wp^D_{n_1} (\Phi
(t))\big)  \Vert \Phi \Vert^{}_{{\cal F}^D_{n_2}}\cr
&\qquad{}+ \sum_{n_1+n_2=l}  \big(H_{n_1+3} (\phi^*, t) + \wp^D_{n_1} (\phi^*
(t))\big)
 \Vert \Phi \Vert^{}_{{\cal F}^D_{n_2}}\Big),\quad  0 \leq l \leq 50.\cr
}$$
This inequality and inequalities (6.108a) and (6.108b) show that
$$\Vert K \Vert^{}_{{\cal F}^M_l} \leq C_l  \Vert u \Vert^3_{E^\rho_{L_0}}
 \Vert u \Vert^{}_{E^\rho_{N_0+l+3}},\quad  0 \leq l \leq 50, \eqno{(6.109)}$$
where $C_l$ depends only on $\rho$ and $\Vert u \Vert^{}_{E^\rho_{L_0}}.$

We shall prove, by induction, that the theorem is true for the case of $l=0$.
Let $n \geq
50$ and suppose that
$$\Vert K \Vert^{}_{{\cal F}^M_j} \leq C_j  \Vert u \Vert^3_{E^\rho_{L_0}}
 \Vert u \Vert^{}_{E^\rho_{N_0+j+3}}, \quad \hbox{for}\ 0 \leq j \leq n
\eqno{(6.110{\rm a})}$$
and that
$$\Vert \Phi \Vert^{}_{{\cal F}^D_j} \leq C_j  \Vert u \Vert^2_{E^\rho_{L_0}}
 \Vert u \Vert^{}_{E^\rho_{N_0+j+3}},  \quad\hbox{for}\ 0 \leq j \leq n,
\eqno{(6.110{\rm b})}$$
where $C_j$ depends only on $\rho$ and $\Vert u \Vert^{}_{E^\rho_{L_0}}$.
The hypothesis is true for $n=50$, according to inequalities (6.108a) and
(6.109). The use of Corollary 2.6
and Proposition 6.2 will not be systematically mentionned in the sequel of this
proof. It follows from inequalities (6.83a)--(6.83d), inequalities
(6.106a)--(6.106d) and hypotheses (6.110a) and (6.110b) that
$$\eqalignno{
R'_j (t) &\leq (1+t)^{- 3/2 + \rho}  C_j  \Vert u
\Vert^2_{E^\rho_{L_0}}	\Vert u \Vert^{}_{E^\rho_{N_0+j+3}},\quad  0 \leq j
\leq n, &(6.111{\rm a})\cr
R^\infty_j (t) &\leq (1+t)^{- 1}  C_j	\Vert u \Vert^2_{E^\rho_{L_0}}
 \Vert u \Vert^{}_{E^\rho_{N_0+j+3}},\quad   0 \leq j \leq n-3,&(6.111{\rm
b})\cr
\wp^D_j  \big((1+\lambda_1 (t))^{k/2}  g(t)\big)&\leq (1+t)^{-1}
C_{j,k}  \Vert u \Vert^{}_{E^\rho_{L_0}}  & (6.111{\rm c})\cr
&\big(\Vert u \Vert^{}_{E^\rho_{N_0+3}}
\Vert u \Vert^{}_{E^\rho_{N_0+j+k}} + \Vert u
\Vert^{}_{E^\rho_{N_0+j+3}}  \Vert u \Vert^{}_{E^\rho_{N_0+k}}\big),
\quad  0 \leq j
\leq n,  k \geq 0,\cr
\wp^D_j (f(t)) &\leq (1+t)^{- 3/2 + \rho}  C_j  \Vert u
\Vert^2_{E^\rho_{L_0}}	\Vert u \Vert^{}_{E^\rho_{N_0+j+3}},\quad   0 \leq j
\leq n,&(6.111{\rm d})\cr
}$$
where $C_j$ and $C_{j,k}$ depend only on $\rho$ and
$\Vert u \Vert^{}_{E^\rho_{L_0}}$. It follows from definition (6.37) of
$\chi^{(j)}$ and from hypothesis (6.110a) that
$$\chi^{}_{k,j} \leq C_{k,j}  \Vert u \Vert^{}_{E^\rho_{N_0+j+k+3}},
\quad  0 \leq
j+k \leq n, \eqno{(6.112)}$$
where $C_{k,j}$ depends only on $\rho$ and $\Vert u \Vert^{}_{E^\rho_{L_0}}$.
Using definitions (6.39a)--(6.39d) of $\overline{Q}_j$, $R^{(1)}_j$, $h'_j$
in the context of equation (6.31a), using hypothesis (6.110a) and using
inequalities (6.107a)--(6.107c), (6.111a)--(6.111d) and (6.112) we obtain that
$$\eqalignno{
\overline{Q}_j(t) &\leq C_j  \Vert u \Vert^2_{E^\rho_{L_0}}  \Vert u
\Vert^{}_{E^\rho_{N_0+j+8}}, \quad 0 \leq j \leq n-5,&(6.113{\rm a}) \cr
R^{(1)}_j (t) &\leq C_j  \Vert u \Vert^2_{E^\rho_{L_0}}  \Vert u
\Vert^{}_{E^\rho_{N_0+j+16}}, \quad  0 \leq j \leq n-13,&(6.113{\rm b})\cr
h'_j (j_0,t) &\leq C_j  \Vert u \Vert^2_{E^\rho_{L_0}}  \Vert u
\Vert^{}_{E^\rho_{N_0+j+2}}, \quad  50 \leq j \leq n+1,&(6.113{\rm c})\cr
}$$
where $j_0$ is the integer part of $j/2 + 5$ and where $C_j$
depends only on $\rho$ and $\Vert u \Vert^{}_{E^\rho_{L_0}}$. Statement iii)
of Proposition
6.4 and inequalities (6.112) and (6.113b) give that
$$H_j (\Phi, t) \leq C_j  \Vert u \Vert^2_{E^\rho_{L_0}}  \Vert u
\Vert^{}_{E^\rho_{N_0+16+j}},\quad  0 \leq j \leq n-13, \eqno{(6.114)}$$
where $C_j$ depends only on $\rho$ and $\Vert u \Vert^{}_{E^\rho_{L_0}}$.
For $\vert Y \vert = n+1,
 Y \in \Pi'$, the existence of $\Phi_Y \in C^0(\Rrm^+,D)$ such that
 $\sup_{t \geq 0}  ((1+t)^{3/2-\rho}  \Vert \Phi_Y (t) \Vert^{}_D) < \infty$
 follows from the linear inhomogeneous equation for $\Phi_Y$ obtained by
 substitution of the solution $K$
of equation (6.31b) into equation (6.31a), from hypotheses (6.110a) and
(6.110b) and from
the energy estimates in Theorem 5.1. The existence of
$\Phi \in {\cal F}^D_{n+1}$ then
follows from hypothesis (6.110b). To prove that inequality (6.110b)
is true also for $n+1,$
we note that it follows from inequality (6.90), with $(K^{(2)} = - \Delta^{*M},
\Phi^{(2)} = 0,  K^{(1)} = K,  \Phi^{(1)} = \Phi,  K'^{(2)} =
0,  K'^{(1)} = K)$, hypothesis (6.10b) and inequality (6.114) that
$$\Vert K \Vert^{}_{{\cal F}^M_{n+1}} \leq C  \Vert u
\Vert^{}_{E^\rho_{N_0+19}}
 \Vert \Phi \Vert^{}_{{\cal F}^D_{n+1}}
+ C_n  \Vert u \Vert^3_{E^\rho_{L_0}}  \Vert u
\Vert^{}_{E^\rho_{N_0+n+4}}, \eqno{(6.115)}$$
where $C$ and $C_n$ depend only on $\rho$ and $\Vert u
\Vert^{}_{E^\rho_{L_0}}$.
It then follows from inequalities (6.81) (with $j=0$) and (6.115) and from
statement iv) of Proposition 6.5 that
$$\wp^D_{n+1} (f(t)) \leq (1+t)^{- 3/2 + \rho}	\big(C
\Vert u \Vert^2_{E^\rho_{L_0}}
 \Vert \Phi \Vert^{}_{{\cal F}_{n+1}}
+ C_n  \Vert u \Vert^4_{E^\rho_{L_0}}  \Vert u
\Vert^{}_{E^\rho_{N_0+n+4}}\big)\eqno{(6.116{\rm a})}$$
and that
$$\wp^D_{n+1,i} (f(t)) \leq (1+t)^{- 3/2 + \rho}  C_n  \Vert u
\Vert^4_{E^\rho_{L_0}}	\Vert u \Vert^{}_{E^\rho_{N_0+n+4}},\quad  1 \leq i
\leq n+1, \eqno{(6.116{\rm b})}$$
where $C$ and $C_n$ depend only on $\rho$ and
$\Vert u \Vert^{}_{E^\rho_{L_0}}$. We also
obtain, using (6.39e), (6.113b) and (6.115) that
$$h''_{n+1} (t) \leq C \Vert u \Vert^4_{E^\rho_{L_0}}
\Vert \Phi \Vert^{}_{{\cal
F}^D_{n+1}} + C_n  \Vert u \Vert^6_{E^\rho_{L_0}}  \Vert u
\Vert^{}_{E^\rho_{N_0+n+4}}, \eqno{(6.117)}$$
where $C$ and $C_n$ depend only on $\rho$ and
$\Vert u \Vert^{}_{E^\rho_{L_0}}$. According to
the definition of $R^0_{n,1}$ in Corollary 5.9 if follows that
$$\eqalignno{
&(1+t)^{1/2}  (R^0_{n,1} (t))^{2(1-\rho)}  \big(\wp^D_n (\Phi (t))\big)^{2
\rho-1}\cr
&\qquad{}\leq (1+t)^{3/2 - \rho}  \big(\wp^D_n \big((1 + \lambda_0 (t))^{1/2}
g(t)\big)\big)^{2(1-\rho)}  \big(\wp^D_n (\Phi (t))\big)^{2 \rho - 1}\cr
&\qquad{}\leq C (1+t)^{3/2 - \rho}  \big(\wp^D_n \big((1 + \lambda_0(t))^{1/2}
g(t)\big) + \wp^D_n (\Phi (t))\big),\cr
}$$
where $C$ depends only on $\rho$. Hypothesis (6.110b) and inequality (6.111c)
then give that
$$(1+t)^{1/2}  R^0_{n,1} (t)^{2(1-\rho)}  \wp^D_n (\Phi (t))^{2 \rho-1}
\leq C_n  \Vert u \Vert^2_{E^\rho_{L_0}}  \Vert u
\Vert^{}_{E^\rho_{N_0+n+4}}, \eqno{(6.118)}$$
where $C_n$ depends only on $\rho$ and $\Vert u \Vert^{}_{E^\rho_{L_0}}$.
We also have
$$\sum_{\sscr n_1+n_2=n+1\atop{\sscr 1 \leq n_1 \leq j_0\atop\sscr n_2
\geq j_0+1}}
\chi^{}_{5,n_1}
 \Vert \Phi \Vert^{}_{{\cal F}^D_{n_2}} \leq C_n
 \Vert u \Vert^3_{E^\rho_{L_0}}
 \Vert u \Vert^{}_{E^\rho_{N_0+n+4}}, \eqno{(6.119)}$$
where $j_0$ is the integer part of $(n+1)/2 + 5$ and where $C_n$
depends only on
$\rho$ and
$\Vert u \Vert^{}_{E^\rho_{L_0}}$. Statement iv) of Proposition 6.4,
inequality (6.113c) with
$j=n+1$, inequalities (6.116a), (6.116b), (6.117), (6.118), (6.119)
and the notation
$$\xi^{}_{n+1,i} = \sup_{t \geq 0}  \big((1+t)^{3/2 - \rho}  \wp^D_{n+1,i}
 (\Phi (t))\big),\quad 0 \leq i \leq n+1, \eqno{(6.120)}$$
and $\xi^{}_{n+1,i} = 0$ for $i \geq n+2$ give
$$\eqalignno{
&\xi^{}_{n+1,0}& (6.121{\rm a})\cr
&\quad{}\leq C  \Vert u \Vert^2_{E^\rho_{L_0}}  \xi^{}_{n+1,0}
+ C_{n+1}  (\xi^{}_{n+1,0})^{\varepsilon'}  (\xi^{}_{n+1,1})^{1 -
\varepsilon'}
+ C_{n+1}  \Vert u \Vert^2_{E^\rho_{L_0}}  \Vert u
\Vert^{}_{E^\rho_{N_0+n+4}}\cr}$$
and
$$\xi^{}_{n+1,i} \leq C_{n+1}  (\xi^{}_{n+1,0})^{\varepsilon'}  (\xi^{}_{n+1,
i+1})^{1 - \varepsilon'}
+ C_{n+1}  \Vert u \Vert^2_{E^\rho_{L_0}}  \Vert u
\Vert^{}_{E^\rho_{N_0+n+4}},\quad  1 \leq i \leq n+1, \eqno{(6.121{\rm b})}$$
where $\varepsilon' = \max (1/2, 2 (1-\rho))$ and $C$ and $C_{n+1}$
depend only on $\rho$ and $\Vert u \Vert^{}_{E^\rho_{L_0}}$. We choose
$\varepsilon > 0$ sufficiently small, such
that $C \Vert u \Vert^2_{E^\rho_{L_0}} \leq 1/2$ for $u \in {\cal O}_\infty$.
With the notation $\alpha_{n+1} = 2  C_{n+1}  \Vert u \Vert^2_{E^\rho_{L_0}}
 \Vert u \Vert^{}_{E^\rho_{N_0+n+4}}$ and $\beta_{n+1} = 2 C_{n+1}$,
 it follows then from inequalities (6.121a) and (6.121b) that
$$\xi^{}_{n+1,i} \leq \alpha_{n+1} + \beta_{n+1}
(\xi^{}_{n+1,0})^{\varepsilon'}
 (\xi^{}_{n+1,i+1})^{1 - \varepsilon'}, \quad 0 \leq i \leq n+1.
\eqno{(6.122)}$$
Proceeding as in solving the system (5.182c) we obtain that
$\xi^{}_{n+1,0} \leq C'_{n+1}
 \alpha_{n+1}$, where $C'_{n+1}$ is a polynomial in $\beta_{n+1}$. This proves
that
$$\Vert \Phi \Vert^{}_{{\cal F}^D_{n+1}} \leq C_{n+1}  \Vert u
\Vert^2_{E^\rho_{L_0}}	\Vert u \Vert^{}_{E^\rho_{N_0+n+4}}, \eqno{(6.123{\rm
a})}$$
for some $C_{n+1}$ depending only on $\rho$ and
$\Vert u \Vert^{}_{E^\rho_{L_0}}$.
Inequality
(6.115) then gives that
$$\Vert K \Vert^{}_{{\cal F}^M_{n+1}} \leq C_{n+1}
\Vert u \Vert^3_{E^\rho_{L_0}}
 \Vert u \Vert^{}_{E^\rho_{N_0+n+4}}, \eqno{(6.123{\rm b})}$$
where $C_{n+1}$ depends only on $\rho$ and $\Vert u \Vert^{}_{E^\rho_{L_0}}$.
It now follows,
by induction, from the hypothesies (6.110a) and (6.110b) and from
inequalities (6.123a) and
(6.123b) that inequalities (6.110a) and (6.110b) are true for every
$n \in \Nrm$. This
proves statements i) and ii) of the theorem for the case of $l=0$.
It also proves that inequalities (6.111a)--(6.111d), (6.113a), (6.113b)
and (6.114) are true for every $j \geq
0$ and that inequality (6.112) is true for every $j \geq 0, k \geq 0$.
Statement iii), with $l=0$, follows by considering equation (6.31c)
and by using inequality (6.114). To prove statement iv) of the theorem,
for the case of $l=0$, we note that $\tau^{}_{i,j} (t),
 i = 0,1$, defined in Theorem 5.5 satisfy $\tau^{}_{i,j} (t) \leq C_j
\chi^{}_{3,j}$, according to inequalities (5.116c) and (6.38c),
that $T^2_{N,j}$ defined in (5.89a) satisfies $T^2_{N,j} \leq C_{N,j}
\chi^{}_{N+1,j}$ according to inequalities (5.138) and (6.41b) and the
fact that $\Gamma_j$ in Corollary 5.9 satisfies $\Gamma_j
(t) \leq C_j  \chi^{(j+1)}$, according to inequalities (5.125b) and (6.41b),
where $C_j$ and $C_{N,j}$ depends only on $\Vert u \Vert^{}_{E^\rho_{N_0}}$.
This gives, together with Corollary 5.9, if $J_0 \geq 3$, that
$$\eqalignno{
&\wp^D_j \big((1 + \lambda_1 (t))^{k/2}  \Phi (t)\big)&(6.124)\cr
&\quad{}\leq C_{j+k}\sum_{\scr n_1+n_2=j+k\atop\scr n_1 \leq J_0}
(1 + \chi^{}_{3,n_1}) (\wp^D_{n_2} (\Phi (t)) + R^1_{n_2-k,k} (t))\cr
&\qquad{}+ C'_{j+k}  \sum_{\sscr n_1+n_2+n_3+n_4=j+k\atop{\sscr n_1 \leq J_0,
n_2 \leq j+k-1
\atop\sscr n_3+n_4\leq j+k-J_0-1}}  (1 + \chi^{}_{3,n_1})\chi^{(n_2)}  (1 +
\chi^{}_{10, n_3})\cr
&\qquad{}\big(R'_{n_4+7} (t) + R^2_{n_4+9} (t) + R^\infty_{n_4} (t) +
\wp^D_{n_4+8}  (\Phi (t))\big),\quad j \geq 0,  k \geq 1,\cr
}$$
where $C_{j+k}$ depends only on $\rho$ and $\chi^{(3)}$ and $C'_{j+k}$
depends only on $\rho$ and $\chi^{(11)}$. It follows from inequalities
(6.110b), (6.111a)--(6.111c), (6.112) and (6.124), and from Corollary 2.6 that
$$\eqalignno{
\wp^D_j  \big((1 + \lambda_1 (t))^{k/2}  \Phi (t)\big) &\leq (1+t)^{- 3/2 +
\rho}  C_{j+k}	\Vert u \Vert^{}_{E^\rho_{L_0}}&(6.125)\cr
&\quad\big(\Vert u \Vert^{}_{E^\rho_{N_0+j+3}}
\Vert u \Vert^{}_{E^\rho_{N_0+k}} + \Vert u
\Vert^{}_{E^\rho_{N_0+3}}  \Vert u \Vert^{}_{E^\rho_{N_0+j+k}}\big),
\quad  j \geq 0, k \geq 1,\cr
}$$
where $C_{j+k}$ depends only on $\rho$ and $\Vert u \Vert^{}_{E^\rho_{L_0}}$.
This proves statement iv) for $l=0.$

Let $\overline{l} \in \Nrm$. We suppose the induction hypothesis that
statements i)--iv) of the theorem are true for $n \geq 0$ and $0 \leq l \leq
\overline{l}$, with the exeption of the decrease properties of $C_{n,l}$ and
$C'_{n,l}$ which we only suppose for $\overline{l} = 0$. We
have proved that the hypothesis is true for $\overline{l} = 0$ and we shall
prove that it is true for $0 \leq l \leq \overline{l} + 1$ if it is true for
$l \leq \overline{l}$. Derivation of equations (6.31a), (6.31b) and (6.31c)
give the following equations for
$(K^{(l)}, \Phi^{(l)})$ in ${\cal F}_n$:
$$\eqalignno{
(i  \gamma^\mu  \partial_\mu + m - \gamma^\mu
G_\mu)	\Phi^{(l)} &= \sum^4_{i=0}  g^{(l)}_i,
&(6.126{\rm a})\cr
\carre  K^{(l)}_\mu &= J^{(l)}_\mu &(6.126{\rm b})\cr
\noalign{\noindent and}
\partial^\mu  J^{(l)}_\mu &= 0,&(6.126{\rm c})\cr
}$$
for $0 \leq l \leq \overline{l} + 1$, where $G_\mu = A^*_\mu + K_\mu -
\partial_\mu  {\vartheta} (A^* + K)$, where
$$\eqalignno{
g^{(0)}_0 &= \gamma^\mu (K_\mu - \partial_\mu	{\vartheta} (K))
 \phi^*, &(6.127{\rm a})\cr
g^{(l)}_0 &= \gamma^\mu (K^{(l)}_\mu -
\partial_\mu  {\vartheta} (K^{(l)}))  (\phi^* + \Phi),
\quad\hbox{for} l \geq 1,\cr
g^{(l)}_1 &= \sum_{\scr i_1+i_2=l\atop\scr i_2 \leq l-1}  C_{i_1,i_2}
\gamma^\mu (A^{* i_1}_\mu - \partial_\mu  {\vartheta} (A^{* i_1}))
 \Phi^{i_2}  \sigma^{}_l,&(6.127{\rm b})\cr
g^{(l)}_2 &= \sum_{\scr i_1+i_2=l\atop\scr  1 \leq i_2 \leq l-1}  C_{i_1,i_2}
 \gamma^\mu (K^{i_1}_\mu - \partial_\mu  {\vartheta}
(K^{i_1}))  \Phi^{i_2}	\sigma^{}_l,&(6.127{\rm c})\cr
g^{(l)}_3 &= \sum_{\scr i_1+i_2=l\atop\scr  i_1 \leq l-1}  C_{i_1,i_2}
\gamma^\mu (K^{i_1}_\mu - \partial_\mu	{\vartheta} (K^{i_1}))
 \phi^{*i_2}  \sigma^{}_l,& (6.127{\rm d})\cr
g^{(l)}_4 &= \sum_{i_1+i_2=l}	C_{i_1,i_2}
\gamma^\mu (\Delta^{*M i_1}_\mu - \partial_\mu	{\vartheta} (\Delta^{*M
i_1}))	\phi^{* i_2}  \sigma^{}_l,& (6.127{_rm e})\cr
}$$
with $K^i$ (resp. $\Phi^i, A^{*i}, \phi^{*i}, \Delta^{* Mi})$ being equal to
the
$l$-linear symmetric map $D^i  K(u)$ (resp. $D^i  \Phi (u)$
$D^i  A^* (u)$, $D^i  \phi^* (u)$, $D^i
\Delta^{*M} (u)$), $\sigma^{}_l$ being the normalized symmetrization of
$(u^{}_1,\ldots,u^{}_l)$ and $C_{i_1,i_2}$ being the binomial coefficients
and where
$$J^{(l)}_\mu = \sum_{i_1+i_2=l}  C_{i_1,i_2}
\big(\overline{\Phi}^{i_1}\gamma_\mu  \phi^{* i_2} +
\overline{\phi}^{*i_1}	\gamma_\mu  \Phi^{i_2} + \overline{\Phi}^{i_1}
\gamma_\mu  \Phi^{i_2}\big)  \sigma^{}_l. \eqno{(6.127{\rm f})}$$
To prove the existence of a unique solution $(K^{(l)}, \Phi^{(l)}) \in {\cal
F}_n,  n \geq 0$, for $l = \overline{l} + 1$, of the linear inhomogeneous
system (6.126a)--(6.126c), we shall use Proposition 6.4. In the context of
this proposition let $R'_j (g^{(l)}_i, t)$, (resp. $R^\infty_j (g^{(l)}_i, t)$,
$f^{(l)}_j, \cdots$) denotes $R'_j (t)$, (resp. $R^\infty_j (t)$, $f(t)$)
in the case of the equation
$$(i  \gamma^\mu  \partial_\mu + m - \gamma^\mu
G_\mu)	\Phi^{(l)}_i = g^{(l)}_i,\quad  0 \leq i \leq 4,
 l = \overline{l} + 1. \eqno{(6.128)}$$
We first study this equation for $1 \leq i \leq 4$ and for $0 \leq  l  \leq
\overline{l} + 1$. To simplify the notation we shall write $A^{* i_1}$ (resp.
$K^{i_1}, \Delta^{*M i_1}$) instead of $A^{* i_1} (u^{}_1,\ldots,u^{}_{i_1})$
(resp.
$K^{i_1} (u^{}_1,\ldots,u^{}_{i_1})$,  $\Delta^{*M i_1} (u^{}_1,
\ldots,u^{}_{i_1}))$ and
$\phi^{* i_2}$ (resp. $\Phi^{i_2}$) instead of $\phi^{* i_2}
(u^{}_{i_1+1},\ldots,u^{}_l)$ \penalty-10000(resp. $\Phi^{i_2} (u^{}_{i_1+1},
\ldots,u^{}_l))$  where
$i_1+i_2=l$. Let
$$\overline{{\cal R}}^{(0)}_{i,j} (u) = \Vert u \Vert^{}_{E^\rho_j}
\quad\hbox{for} i,j\geq 0 \eqno{(6.129{\rm a})}$$
and
$$\overline{{\cal R}}^{(l)}_{i,j} (u ; u^{}_1,\ldots,u^{}_l)
= \overline{{\cal R}}^l_{i,j}
(u^{}_1,\ldots,u^{}_l)
+ \Vert u \Vert^{}_{E^\rho_j}  \Vert u^{}_1
\Vert^{}_{E^\rho_i}\cdots\Vert u^{}_l
\Vert^{}_{E^\rho_i},\eqno{(6.129{\rm b})}$$
$ i,j \geq 0,  l \geq 1. $If $g = \gamma^\mu (A^{* i_1}_\mu - \partial_\mu
{\vartheta} (A^{*
i_1}))	\Phi^{i_2},  i_1 + i_2 = l \leq \overline{l} + 1,
 i_2 \leq l-1$ then it follows from inequality (5.116c), Proposition
6.2 and the induction hypothesis (statement iv) with
$l \leq \overline{l})$ that
$$\eqalignno{
&\wp^D_j  \big((\delta (t))^{3/2 - \rho}  (1+\lambda_1 (t))^{k/2}
 g(t)\big)\cr
&\quad{}\leq (1+t)^{- 3/2 + \rho}  C_{j+k,l}\cr
&\qquad{}\sum_{n_1+n_2=j}
\overline{{\cal R}}^{(i_1)}_{N_0, N_0+n_1+1} (u ; u^{}_1,
\ldots,u^{}_{i_1})
\overline{{\cal R}}^{(i_2)}_{L_0, N_0+3+n_2+k+i_2} (u ;u^{}_{i_1+1},
\ldots,u^{}_l),\cr
}$$
where $C_{j+k,l}$ depending only on $\rho, l$ and
$\Vert u \Vert^{}_{E^\rho_{L_0}}$. Using
Corollary 2.6 it follows that
$$\eqalignno{
&\wp^D_j  \big((\delta (t))^{3/2 - \rho}  (1+\lambda_1 (t))^{k/2}
 g(t)\big) &(6.130)\cr
&\quad{}\leq C_{j+k,l}  (1+t)^{- 3/2 + \rho}
\overline{{\cal R}}^{(l)}_{L_0, N_0+3+j+k+l}(u ; u^{}_1,
\ldots,u^{}_l), \quad i_1 +
i_2 = l \leq \overline{l} + 1,	i_2 \leq l-1,\cr
}$$
where $C_{j+k,l}$ is as in the last inequality.
Applying inequality (6.130) to each term
in the sum defining $g^{(l)}_1$ we obtain that
$$\eqalignno{
&\wp^D_j \big((\delta (t))^{3/2 - \rho}  (1 + \lambda_1 (t))^{k/2}
g^{(l)}_1 (t)\big)&(6.131)\cr
&\quad{} \leq (1+t)^{- 3/2 + \rho}  C_{j+k,l}
\overline{{\cal R}}^{(l)}_{L_0, N_0+3+j+k+l} (u ; u^{}_1,\ldots,u^{}_l),
\quad j \geq 0,  k \geq 0,  0 \leq l \leq \overline{l} + 1,\cr
}$$
where $C_{j+k,l}$ depends only on $\rho$ and $\Vert u \Vert^{}_{E^\rho_{L_0}}$.
If
$g = \gamma^\mu (K^{i_1}_\mu - \partial_\mu  {\vartheta} (K^{i_1}))
\Phi^{i_2}$, $i_1 + i_2 = l \leq \overline{l} + 1$,
$1 \leq i_1 \leq l-1$ then it follows from
statement iii) of Lemma 6.6 and from the induction hypothesis that
$$\eqalignno{
&\wp^D_j \big(\delta (t) (1+\lambda_1 (t))^{k/2}
g(t)\big)\cr
&\ \leq (1+t)^{- 3/2 + \rho}  C_{j+k} \cr
&\quad\sum_{n_1+n_2=j}\min\Big(\overline{{\cal R}}^{(i_1)}_{L_0, N_0+3+n_1+i_1}
(u ; u^{}_1,\ldots,u^{}_{i_1})
\overline{{\cal R}}^{(i_2)}_{L_0, N_0+3+n_2+k+1+i_2+13}
(u ;u^{}_{i_1+1},\ldots,u^{}_l),\cr
&\hskip30mm\overline{{\cal R}}^{(i_1)}_{L_0, N_0+3+n_1+3+i_1}
(u ; u^{}_1,\ldots,u^{}_{i_1})
\overline{{\cal R}}^{(i_2)}_{L_0, N_0+3+n_2+k+i_2}
(u ;u^{}_1,\ldots,u^{}_l)\Big).\cr
}$$
For terms with $n_1+3+i_1 \leq j+l+k$ we choose the second argument in
the minimum and use Corollary 2.6 together with $N_0+6 \leq
\min(N_0+6+n_1+i_1, N_0+3+n_2+i_2+k) \leq
\max(N_0+6+n_1+i_1, N_0+3+n_2+i_2+k) \leq N_0+3+j+k+l$. For terms with
$n_1+3+i_1 > j+l+k$
we choose the first argument in the minimum and use that
$N_0+n_2+k+i_2+17 \leq N_0+19 \leq L_0$. This gives that
$$\eqalignno{
&\wp^D_j  \big(\delta (t)  (1+\lambda_1 (t))^{k/2}
g(t)\big) &(6.132)\cr
&\quad{}\leq (1+t)^{- 3/2 + \rho}  C_{j+k}
\overline{{\cal R}}^{(l)}_{L_0, N_0+3+j+k+l} (u ; u^{}_1,\ldots,u^{}_l),
\quad  j,k \geq 0,
 0 \leq l \leq \overline{l} + 1.\cr
}$$
Applying inequality (6.132) to each term in the sum defining $g^{(l)}_2$
we obtain that
$$\eqalignno{
&\wp^D_j  \big(\delta (t)  (1+\lambda_1 (t))^{k/2}
g^{(l)}_2 (t)\big)&(6.133)\cr
&\quad{}\leq (1+t)^{- 3/2 + \rho}  C_{j+k,l}
\overline{{\cal R}}^{(l)}_{L_0, N_0+3+j+k+l} (u ; u^{}_1,\ldots,u^{}_l),
\quad j,k \geq 0,  0 \leq l \leq \overline{l} + 1,\cr
}$$
where $C_{j+k,l}$ depends only on $\rho$ and
$\Vert u \Vert^{}_{E^\rho_{L_0}}$. If $g =
\gamma^\mu (K^{i_1}_\mu - \partial_\mu	{\vartheta} (K^{i_1}))
\phi^{*i_2}$, $i_1+i_2 = l$,  $i_1 \leq l-1$, then using Lemma 6.5 and $\delta
(t) \leq C (1 + \lambda_1 (t) + t)$ we obtain that
$$\wp^D_j  \big(\delta (t)  (1 + \lambda_1 (t))^{k/2}
g(t)\big)
\leq C_j  \sum_{n_1+n_2=j}  \Vert K^{i_1} \Vert^{}_{{\cal F}^M_{n_1}}
 H_{n_2+k+1}  (\phi^{* i_2}, t), \eqno{(6.134)}$$
where $C_j$ depends only on $\rho$. The induction hypothesis and
Proposition 6.2 give that
$$\eqalignno{
&\wp^D_j  \big(\delta (t)  (1+\lambda_1 (t))^{k/2}
g(t)\big)&(6.135)\cr
&\quad{}\leq C_{j+k,l}  \sum_{n_1+n_2=j}
\overline{{\cal R}}^{(i_1)}_{L_0, N_0+3+n_1+i_1} (u ; u^{}_1,
\ldots,u^{}_{i_1})	 \overline{{\cal R}}^{(i_2)}_{N_0, N_0+n_2+k+1}
(u ;u^{}_{i_{1+1}},\ldots,u^{}_l)\cr
&\quad{}\leq (1+t)^{- 3/2 + \rho}  C'_{j+k,l}\overline{{\cal R}}^{(l)}_{L_0,
N_0+3+j+k+l} (u ; u^{}_1,\ldots,u^{}_l), \quad j,k \geq 0,  0
\leq l \leq \overline{l} +
1.\cr
}$$
Applying inequality (6.135) to each term in the sum defining $g^{(l)}_3$
we obtain that
$$\eqalignno{
&\wp^D_j  \big(\delta (t)  (1+\lambda_1 (t))^{k/2}
g^{(l)}_3 (t)\big) &(6.136)\cr
&\quad{}\leq C_{j+k,l}
\overline{{\cal R}}^{(l)}_{L_0, N_0+3+j+k+l} (u ; u^{}_1,\ldots,u^{}_l),
\quad  j,k \geq 0, 0 \leq l \leq \overline{l} + 1,\cr
}$$
where $C_{j+k,l}$ depends only on $\rho$ and $\Vert u \Vert^{}_{E^\rho_{L_0}}$.
Using also inequality (6.134) in the case of $g^{(l)}_4$ we obtain from
inequalities (6.131), (6.133) and (6.136) that
$$\eqalignno{
&\sum^4_{i=1}\wp^D_j  \big((\delta (t))^{3/2 - \rho}
(1+\lambda_1 (t))^{k/2}  g^{(l)}_i (t)\big)&(6.137)\cr
&\quad{}\leq (1+t)^{- \rho + 1/2} C_{j+k,l}
\overline{{\cal R}}^{(l)}_{L_0, N_0+3+j+k+l} (u ; u^{}_1,\ldots,u^{}_l),
\quad  j,k \geq 0, 0 \leq l \leq \overline{l} + 1,\cr
}$$
where $C_{j+k,l}$ depends only on $\rho$ and $\Vert u \Vert^{}_{E^\rho_{L_0}}$.
Similarly,
it follows from statement iii) of Lemma 6.5 and statement iii) of Lemma 6.6
that
$$\eqalignno{
&\wp^D_j  \big(\delta (t)  (1+\lambda_1 (t))^{k/2}
g^{(l)}_0 (t)\big)&(6.138)\cr
&\quad{} \leq C_{j+k,l}
\overline{{\cal R}}^{(l)}_{L_0, N_0+3+j+k+l} (u ; u^{}_1,\ldots,u^{}_l),
\quad  j,k \geq 0, 0 \leq l \leq \overline{l},\cr
}$$
where $C_{j+k,l}$ depends only on $\rho$ and $\Vert u \Vert^{}_{E^\rho_{L_0}}$.
To estimate $R'_j (g^{(l)}_1, t)$, where $R'_j$ is defined in Theorem 5.8,
let $g = \gamma^\mu (a_\mu
- \partial_\mu	{\vartheta} (a)) r$ and $\partial_\mu  a^\mu = 0.$
Changing the notation in inequality (6.68) we obtain with $g' = (2m)^{-1} (m-i
\gamma^\mu  \partial_\mu + \gamma^\mu  G_\mu) g,  F_\mu =
a_\mu - \partial_\mu  {\vartheta} (a)$:
$$\eqalignno{
g'_Y &= (2m)^{-1}  \suma_{Y_1,Y_2}^Y  \Big(\gamma^\nu
F_{Y_1 \nu}  \xi^M_{Y_2}  ((m+i  \gamma^\mu
\partial_\mu - \gamma^\mu  G_\mu) r)& (6.139)\cr
&\quad{}- 2i	F^\mu_{Y_1}  \partial_\mu  r^{}_{Y_2} + i
(\partial_\mu  F^\mu_{Y_1})  r^{}_{Y_2}\cr
&\quad{}- {i \over 4} (\gamma^\mu  \gamma^\nu - \gamma^\nu
\gamma^\mu)  ((\partial_\mu  a_{Y_1 \nu}) - (\partial_\nu
 a_{Y_1 \mu}) ) r^{}_{Y_2}\Big)\cr
&\quad{}+ m^{-1}  \suma_{Y_1,Y_2,Y_3}^Y  G_{Y_1 \mu}
 F^\mu_{Y_2}  r^{}_{Y_3},\quad  Y \in \Pi',  k\geq 0.\cr
}$$
Taking weighted supremum norms of $F_Z$ and using inequalities (5.7d) and
(5.116c), the facts that $A = A^* + K$ and $y^\mu  F_\mu (y) = 0,$
Lemma 6.3 and inequalities (6.49b) and (6.66a) we obtain that
$$\eqalignno{
&\Vert \delta (t)(1 + \lambda_1 (t))^{k/2}  g'_Y (t)
\Vert^{}_D& (6.140)\cr
&\ {}\leq C_{\vert Y \vert}  \suma_{Y_1,Y_2}^Y  [a]^{\vert Y_1
\vert + 2} (t)	\Big(\Vert (\delta (t))^{\rho-1/2}  (1 +
\lambda_1 (t))^{k/2}  (\xi^M_{Y_2} (i \gamma^\mu
\partial_\mu + m - \gamma^\mu  G_\mu) r) (t) \Vert^{}_D\cr
&\quad{}+ (1+t)^{- 3/2 + \rho}  \wp^D_{\vert Y_2 \vert + 1}  ((1 +
\lambda_1 (t))^{k/2}  r(t))
+ (1+t)^{- 1/2}  \wp^D_{\vert Y_2
\vert}	((1 + \lambda_1 (t))^{(k+1)/2}	r(t))\Big)\cr
&\quad{}+C_{\vert Y \vert}  \suma_{Y_1,Y_2,Y_3}^Y  \Big((1+t)^{- 2 +
2 \rho}  [A^*]^{\vert Y_1 \vert + 1} (t)  [a]^{\vert Y_2
\vert +1} (t)  \wp^D_{\vert Y_3 \vert}	((1 + \lambda_1
(t))^{k/2}  r(t))\cr
&\qquad\qquad{}+ (1+t)^{- 3/2 + \rho}  [a]^{\vert Y_2 \vert + 1}  (t)\cr
&\qquad\qquad{}\min \big(\Vert K \Vert^{}_{{\cal F}^M_{\vert Y_1 \vert + 3}}
\wp^D_{\vert Y_3
\vert}	((1 + \lambda_1 (t))^{k/2}  r(t)),  \Vert
K \Vert^{}_{{\cal F}^M_{\vert Y_1 \vert}}  H_{\vert Y_3 \vert + k}
(r,t)\big)\Big),\cr
}$$
$Y \in \Pi'$,  $k \geq 0$, where $C_{\vert Y \vert}$ depends only on
$\rho$. The use of inequality (6.140), with $a = A^{* i_1}$, $r =
\Phi^{i_2}$,  $i_1 + i_2 = l$, $i_2 \leq l-1$, $0 \leq l \leq \overline{l} +
1$, the use of equation (6.126a), with $l \leq \overline{l}$, and inequalities
(6.137) and (6.138) to estimate the weighted $D$-norms of $\xi^M_{Y_2}$
$(i \gamma^\mu	\partial_\mu + m -\gamma^\mu  G_\mu)
\Phi^{i_2}$ and the use of Proposition 6.2 and the induction hypothesis give
$$\eqalignno{
&\wp^D_j  \big(\delta (t)  (1+\lambda_1 (t))^{k/2}
g'^{(l)}_1 (t)\big) &(6.141)\cr
&\quad{}\leq (1+t)^{- 3/2 + \rho}  C_{j+k,l}
\overline{{\cal R}}^{(l)}_{L_0, N_0+3+j+k+l+1} (u ; u^{}_1,\ldots,u^{}_l),
\quad j,k\geq 0,   0 \leq l \leq \overline{l}+1,\cr
}$$
where $g'^{(l)}_1 = (2m)^{-1}  (m - i \gamma^\mu
\partial_\mu + \gamma^\mu  G_\mu)  g^{(l)}_1$, where
$C_{j+k,l}$ depends only on $\rho$ and $\Vert u \Vert^{}_{E^\rho_{L_0}}.$
Statement i) of Lemma 6.5, statement i) of Lemma 6.6 and inequality (6.141)
give that
$$\eqalignno{
&R'_j  (g^{(l)}_i, t) &(6.142{\rm a})\cr
&\quad{}\leq (1+t)^{- 3/2 + \rho}  C_{j+k,l}
\overline{{\cal R}}^{(l)}_{L_0, N_0+3+j+l+1} (u ; u^{}_1,\ldots,u^{}_l),
\quad  j \geq 0,  0 \leq l \leq \overline{l}+1,  1 \leq i \leq 4,\cr
}$$
where $C_{j+k,l}$ depends only on $\rho$ and $\Vert u \Vert^{}_{E^\rho_{L_0}}$.
It follows from inequality (6.137) that
$$\eqalignno{
&R^2_j  (g^{(l)}_i, t)& (6.142{\rm b})\cr
&\quad{}\leq (1+t)^{-1}  C_{j,l}
\overline{{\cal R}}^{(l)}_{L_0, N_0+3+j+l} (u ; u^{}_1,\ldots,u^{}_l),
\quad j \geq 0,  0 \leq l \leq \overline{l} + 1,  1 \leq i \leq 4,\cr
}$$
where $C_{j,l}$ depends only on $\rho$ and $\Vert u \Vert^{}_{E^\rho_{L_0}}.$
Moreover using statement ii) of Lemma 6.5, statement ii) of Lemma 6.6 and
Proposition 6.2 we obtain that
$$\eqalignno{
&R^\infty_j  (g^{(l)}_i, t)&(6.142{\rm c})\cr
&\quad{}\leq (1+t)^{-1}  C_{j,l}
 \overline{{\cal R}}^{(l)}_{L_0, N_0+3+j+l+13} (u ; u^{}_1,\ldots,u^{}_l),
\quad j \geq 0,  0 \leq l \leq \overline{l} +1,  1 \leq i \leq 4,\cr
}$$
where $C_{j,l}$ is as in (6.142b). Proposition 5.16, in the case where $i=1$
and
Lemma 6.5 and Lemma 6.6 in the case where $2 \leq i \leq 4$ give that
$$\eqalignno{
&\wp^D_j (f^{(l)}_i (t))& (6.142{\rm d})\cr
&\quad{}\leq C_{j,l}	(1+t)^{- 3/2 + \rho}
\overline{{\cal R}}^{(l)}_{L_0, N_0+3+j+l} (u ; u^{}_1,\ldots,u^{}_l),
\quad j \geq 0,  0 \leq l \leq \overline{l} +1,  1 \leq i \leq 4,\cr}
$$
where $C_{j,l}$ is as in (6.142b). Definition (6.39a) of $\overline{Q}_n$ and
inequalities (6.142b) and (6.142d) give, for $1 \leq i \leq 4,	0
\leq l \leq \overline{l} + 1$, that
$$\overline{Q}_j (g^{(l)}_i, t) \leq C_{j,l}  \overline{{\cal R}}^{(l)}_{L_0,
N_0+3+j+l} (u ; u^{}_1,\ldots,u^{}_l), \eqno{(6.143{\rm a})}$$
if $0 \leq j \leq 45$ and
$$\overline{Q}_j (g^{(l)}_i, t) \leq C_{j,l}  \overline{{\cal R}}^{(l)}_{L_0,
N_0+3+j+l+5} (u ; u^{}_1,\ldots,u^{}_l), \eqno{(6.143{\rm b})}$$
if $0 \leq j$, where $C_{j,l}$ is as in (6.142b). It follows from inequalities
(6.142a), (6.142b), (6.142c) and (6.143b) and from definition (6.39c) of
$R^{(1)}_j$ that
$$\eqalignno{
&R^{(1)}_j  (g^{(l)}_i, t)&(6.143{\rm c})\cr
&\quad{} \leq C_{j,l}  \overline{{\cal
R}}^{(l)}_{L_0, N_0+3+j+l+13} (u ; u^{}_1,\ldots,u^{}_l),
\quad j \geq 0,  0 \leq l \leq \overline{l} +1,  1 \leq i \leq 4,\cr
}$$
where $C_{j,l}$ is as in (6.142b). Definitions (6.39d) of $h'_j$ and (6.39e) of
$h''_j$ and the definition of $h^\infty_j$ in statement ii) of Proposition 6.4
give, $j_0$ being the integer part of $j/2 + 5$, that
$$\eqalignno{
&h^\infty_j  (g^{(l)}_i,j_0, t)& (6.143{\rm d})\cr
&\quad{}\leq C_{j,l}  \overline{{\cal
R}}^{(l)}_{L_0, N_0+3+j+l} (u ; u^{}_1,\ldots,u^{}_l),
\quad j \geq 46,  0 \leq l \leq \overline{l} +1,  1 \leq i \leq 4,\cr
}$$
where $C_{j,l}$ depends only on $\rho$ and $\Vert u \Vert^{}_{E^\rho_{L_0}}.$
Using first statement i) of Proposition 6.4 with $n=50$ and inequality
(6.143a) and using secondly statement ii) of Proposition 6.4 with $n \geq 82$
and inequality (6.143d) we obtain that the solution $\Phi^{(l)}_i$ of equation
(6.128), with $1 \leq i \leq 4$ and $0 \leq l \leq \overline{l} + 1$ exists and
satisfies
$$\eqalignno{
&\Vert \Phi^{(l)}_i \Vert^{}_{{\cal F}^D_j}&(6.144{\rm a})\cr
&\quad{} \leq C_{j,l}  \overline{{\cal
R}}^{(l)}_{L_0, N_0+3+j+l} (u ; u^{}_1,\ldots,u^{}_l),
\quad j \geq 0,  0 \leq l \leq \overline{l} +1,  1 \leq i \leq 4,\cr
}$$
where $C_{j,l}$ depends only on $\rho$ and $\Vert u \Vert^{}_{E^\rho_{L_0}}.$
Inequality (6.143c) and statement iii) of Proposition 6.4 give that
$$\eqalignno{
&H_j  (\Phi^{(l)}_i, t)& (6.144{\rm b})\cr
&\quad{}\leq (1+t)^{- 3/2}  C_{j,l}
\overline{{\cal R}}^{(l)}_{L_0, N_0+3+j+l+13} (u ; u^{}_1,\ldots,u^{}_l),
\quad j \geq 0,  0 \leq l \leq \overline{l} +1,  1 \leq i \leq 4,\cr
}$$
where $C_{j,l}$ depends only on $\rho$ and $\Vert u \Vert^{}_{E^\rho_{L_0}}$.
In (6.144a) and (6.144b) $C_{j,0} \leq C'_j  \Vert u
\Vert^2_{E^\rho_{L_0}}$, $C_{j,l} \leq C'_j  \Vert u
\Vert^{}_{E^\rho_{L_0}}$.

To study equation (6.128), with $i=0$ and $l = \overline{l} + 1$, we introduce
the equation
$$(i \gamma^\mu  \partial_\mu + m - \gamma^\mu	G_\mu)
 \Psi (L) = \overline{g} \eqno{(6.145)}$$
for $\Psi (L) \in {\cal F}^D_n$, where $\overline{g} = \gamma^\mu (L_\mu -
\partial_\mu  {\vartheta} (L))	(\Phi + \phi^*)$ and $L \in
{\cal F}^M_n$. If $n = 37$, then it follows from ii) and iii), with $l=0$, of
the present theorem and from Lemma 6.5 and Lemma 6.6 that
$$\eqalignno{
R'_j (\overline{g}, t) &\leq (1+t)^{- 3/2 + \rho}  C_j  \Vert u
\Vert^{}_{E^\rho_{L_0}}  \Vert L \Vert^{}_{{\cal F}^M_{j+1}},\quad  0
\leq j \leq 36, &(6.146{\rm a})\cr
R^\infty_j (\overline{g}, t) &\leq (1+t)^{- 1}  C_j  \Vert u
\Vert^{}_{E^\rho_{L_0}}  \Vert L \Vert^{}_{{\cal F}^M_{j+3}},\quad  0
\leq j \leq 34,&(6.146{\rm b})\cr
R^2_j (\overline{g}, t) &\leq (1+t)^{- 1}  C_j  \Vert u
\Vert^{}_{E^\rho_{L_0}}  \Vert L \Vert^{}_{{\cal F}^M_j},\quad  0
\leq j \leq 37,&(6.146{\rm c})\cr
\wp^D_j (\overline{f} (t)) &\leq (1+t)^{- 3/2 + \rho}	C_j  \Vert
u \Vert^{}_{E^\rho_{L_0}}  \Vert u \Vert^{}_{{\cal F}^M_j},\quad  0
\leq j \leq 37,&(6.146{\rm d})\cr
}$$
where $C_j$ depends only on $\rho$ and $\Vert u \Vert^{}_{E^\rho_{L_0}}$.
These inequalities give that
$$\eqalignno{
\overline{Q}_j (\overline{g}, t) &\leq C_j  \Vert u \Vert^{}_{E^\rho_{L_0}}
 \Vert L \Vert^{}_{{\cal F}^M_j},\quad  0 \leq j \leq 37,
&(6.147{\rm a})\cr
R^{(1)}_j (\overline{g}, t) &\leq C_j	\Vert u \Vert^{}_{E^\rho_{L_0}}
 \Vert L \Vert^{}_{{\cal F}^M_{j+9}}, \quad  0 \leq j \leq 28,
&(6.147{\rm b})\cr
h^\infty_j (\overline{g}, n_0, t) &\leq C_j  \Vert u
\Vert^2_{E^\rho_{L_0}}	\Vert L \Vert^{}_{{\cal F}^M_j},\quad   n_0+1
\leq j \leq n = 37,&(6.147{\rm c})\cr
}$$
where $n_0$ is the integer part of $n/2 + 5$ and where $C_j$ depend only on
$\rho$ and $\Vert u \Vert^{}_{E^\rho_{L_0}}$. Statements i), ii) and iii) of
Proposition 6.4 and inequalities (6.147a)--(6.147c) then give that
$$\Vert \Psi (L) \Vert^{}_{{\cal F}^D_j} \leq C_j  \Vert u
\Vert^{}_{E^\rho_{L_0}}  \Vert L \Vert^{}_{{\cal F}^M_j},\quad   0
\leq j \leq 37, \eqno{(6.148{\rm a})}$$
and that
$$H_j (\Psi (L), t) \leq (1+t)^{- 3/2 + \rho}  C_j  \Vert u
\Vert^{}_{E^\rho_{L_0}}  \Vert L \Vert^{}_{{\cal F}^M_{j+9}},\quad   0
\leq j \leq 27, \eqno{(6.148{\rm b})}$$
where $C_j$ depends only on $\rho$ and $\Vert u \Vert^{}_{E^\rho_{L_0}}.$

For given solutions $\Phi^{(l)}_i,  1 \leq i \leq 4,  0
\leq l \leq \overline{l} + 1$, of equation (6.128) and $\Psi (L)$
of equation (6.145) we introduce the current
$\overline{J}^{(\overline{l} + 1)}$ by
$$\overline{J}^{(\overline{l} + 1)}_\mu (L) =
\overline{J}^{(\overline{l} + 1)}_{0 \mu} (L) +
\overline{J}^{(\overline{l} + 1)}_{1 \mu}, \eqno{(6.149{\rm a})}$$
where
$$\overline{J}^{(\overline{l} + 1)}_{0 \mu} (L) = \overline{\Psi} (L)
\gamma_\mu  (\phi^* + \Phi) + (\overline{\phi^* + \Phi})
\gamma_\mu  \Psi (L) \eqno{(6.149{\rm b})}$$
and
$$\eqalignno{
\overline{J}^{(\overline{l} + 1)}_{1 \mu} &=
\sum_{\scr i_1+i_2= \overline{l} +1\atop\scr
1 \leq i_1 \leq\overline{l}}  C_{i_1,i_2}  \big(\overline{\Phi}^{i_1}
\gamma_\mu  \phi^{* i_2} + \overline{\phi}^{* i_1}  \gamma_\mu
 \Phi^{i_2}
+ \overline{\Phi}^{i_1}  \gamma_\mu  \Phi^{i_2}\big)
\sigma^{}_{\overline{l} + 1}&(6.149{\rm c})\cr
&\qquad{}+ \sum^4_{i=1}  (\overline{\Phi}^{(\overline{l}+1)}_i
\gamma_\mu  (\phi^* + \Phi) + (\overline{\phi^* + \Phi})
\gamma_\mu  \Phi^{(\overline{l}+1)}_i),\cr
}$$
where $C_{i_1,i_2}$ are the binomial coefficients and where we have used the
notation of (6.127b)--(6.127e). Since equation (6.126a) is satisfied for $0
\leq l \leq \overline{l}$, according to the induction hypothesis, since we have
proved that there is a (unique) solution $\Phi_i \in {\cal F}^D_j$,  $j
\geq 0$, of equation (6.128) for $1 \leq i \leq 4,  0 \leq l \leq
\overline{l} + 1$ and since we have proved that equation (6.145) has a (unique)
solution $\Psi (L) \in {\cal F}^D_0$, it follows that
$$\partial^\mu	\overline{J}^{(\overline{l} + 1)}_\mu (L) = 0. \eqno{(6.149{\rm
d})}$$
We shall estimate the ${\cal F}^M_j$ norm of the unique solution $N(L) \in
{\cal F}^M_j$ of the equation
$$\carre  N_\mu (L) = \overline{J}^{(\overline{l} + 1)}_\mu (L).
\eqno{(6.150)}$$
It follows from the definition of $\Vert \cdot \Vert^{}_{{\cal F}^M_j}$ that
$$\eqalignno{
\Vert N (L) \Vert^{}_{{\cal F}^M_j}
&\leq C \Big(\sum_{\vert Y \vert \leq j}
 \sup_{t \geq 0} \big((1+t)^{2 - \rho}  \Vert \delta (t)
 (\xi^M_Y  \overline{J}^{(\overline{l} + 1)} (L)) (t) \Vert^{}_{L^2}&(6.151)\cr
&\qquad{}+ (1+t)^{1 - \rho}  \Vert \delta (t)  (\xi^M_Y
 \overline{J}^{(\overline{l} + 1)} (L)) (t)
 \Vert^{}_{L^{6/5}}\big)^2\Big)^{1/2}.\cr
}$$
Using inequalities (6.88), (6.144a), (6.144b) and the induction hypothesis we
obtain that
$$\eqalignno{
&\Big(\sum_{\vert Y \vert \leq j}\sup_{t \geq 0} \big((1+t)^{2 - \rho}
 \Vert \delta (t)  (\xi^M_Y  \overline{J}^{(\overline{l}
+ 1)}_1) (t) \Vert^{}_{L^2}
+ (1+t)^{1 - \rho}  \Vert \delta (t)  (\xi^M_Y
 \overline{J}^{(\overline{l} + 1)}_1) (t)
\Vert^{}_{L^{6/5}}\big)^2\Big)^{1/2}\cr
&\qquad{}\leq C_{j, \overline{l}}  \overline{{\cal R}}^{(\overline{l} +
1)}_{L_0,
N_0+3+j+\overline{l}+1} (u ; u^{}_1,\ldots,u^{}_{\overline{l}+1}), \quad
 j \geq 0,&(6.152)\cr
}$$
where $C_{j, \overline{l}}$ depends only on $\rho$ and $\Vert u
\Vert^{}_{E^\rho_{L_0}}$. Using once more inequality (6.88) we obtain that
$$\eqalignno{
&\Big(\sum_{\vert Y \vert \leq j}\sup_{t \geq 0} \big((1+t)^{2 - \rho}
 \Vert \delta (t)  (\xi^M_Y  \overline{J}^{(\overline{l}
+ 1)}_0 (L)) (t) \Vert^{}_{L^2}& (6.153)\cr
&\qquad\qquad{}+ (1+t)^{1 - \rho}  \Vert \delta (t)  (\xi^M_Y
\overline{J}^{(\overline{l} + 1)}_0 (L)) (t)
\Vert^{}_{L^{6/5}}\big)^2\Big)^{1/2}\cr
&\quad{}\leq \sum_{n_1+n_2=j}  C_{n_1,n_2} \Big(\big(H_{3+n_1} (\phi^*,t) +
\wp^D_{n_1} (\phi^* (t))\big)  \Vert \Psi (L) \Vert^{}_{{\cal F}^D_{n_2}}\cr
&\qquad\qquad{}+ \min \Big(\big(H_{3+n_1} (\Phi,t)
+ \wp^D_{n_1} (\Phi (t))\big)  \Vert \Psi (L)
\Vert^{}_{{\cal F}^D_{n_2}},\cr
&\qquad\qquad\qquad{}\Vert \Phi \Vert^{}_{{\cal F}^M_{n_1}}
\big(H_{3+n_2} (\Psi (L),t) +
\wp^D_{n_2}  ((\Psi (L)) (t))\big)\Big)\Big),\cr
}$$
where $C_{0,j} = C_{j,0} = C$ and $C_{n_1,n_2}$ depend only on $\rho$. The
induction hypothesis (for $\overline{l} = 0$), Proposition 6.2 and inequalities
(6.148a), (6.151), (6.152) and (6.153) give that
$$\eqalignno{
&\Vert N (L) \Vert^{}_{{\cal F}^M_j}& (6.154)\cr
&\quad{}\leq C \Vert u \Vert^{}_{E^\rho_{L_0}}
 \Vert L \Vert^{}_{{\cal F}^M_j}
+ C_{\overline{l}}  \overline{{\cal R}}^{(\overline{l} + 1)}_{L_0,
N_0+3+j+\overline{l}+1} (u ; u^{}_1,\ldots,u^{}_{\overline{l}+1}),
\quad  j \leq 37,\cr
}$$
where $C$ and $C_{\overline{l}}$ depend only on $\rho$ and $\Vert u
\Vert^{}_{E^\rho_{L_0}}$. If $\varepsilon > 0$ is sufficiently small, then $C
 \Vert u \Vert^{}_{E^\rho_{L_0}} \leq 1/2$ for $u \in {\cal O}_\infty$, which
together with inequality (6.154) shows that the linear inhomogeneous equation
in ${\cal F}^M_{37}$
$$K^{(\overline{l} + 1)} = N (K^{(\overline{l} + 1)}) \eqno{(6.155)}$$
has a unique solution $K^{(\overline{l} + 1)} \in {\cal F}^M_{37}$ and
that this solution satisfies
$$\Vert K^{(\overline{l} + 1)} \Vert^{}_{{\cal F}^M_j} \leq C_{\overline{l}}
\overline{{\cal R}}^{(\overline{l} + 1)}_{L_0, N_0+3+j+\overline{l}+1} (u ;
u^{}_1,\ldots,u^{}_{\overline{l}+1}),\quad j \leq 37 = n, \eqno{(6.156{\rm
a})}$$
where $C_{\overline{l}}$ depends only on $\rho$ and $\Vert u
\Vert^{}_{E^\rho_{L_0}}$. Inequalities (6.148a) and (6.148b) then show that
$$\Vert \Phi^{(\overline{l} + 1)}_0 \Vert^{}_{{\cal F}^M_j} \leq
C_{\overline{l}}
\overline{{\cal R}}^{(\overline{l} + 1)}_{L_0, N_0+3+j+\overline{l}+1} (u ;
u^{}_1,\ldots,u^{}_{\overline{l}+1}),\quad j \leq 37 = n, \eqno{(6.156{\rm
b})}$$
and that
$$\eqalignno{
&H_j (\Phi^{(\overline{l} + 1)}_0, t)&(6.156{\rm c})\cr
&\quad{} \leq (1+t)^{- 3/2 + \rho}
C_{\overline{l}} \overline{
{\cal R}}^{(\overline{l} + 1)}_{L_0, N_0+3+j+\overline{l}+10}
(u ; u^{}_1,\ldots,u^{}_{\overline{l}+1}),\quad j \leq 27 = n-10, \cr
}$$
where $C_{\overline{l}}$ depends only on $\rho$ and
$\Vert u\Vert^{}_{E^\rho_{L_0}}.$

We now suppose that (6.156a) and (6.156b) are true for $j \leq n$ when  $n \in
\Nrm$,  $n \geq 37$. We study equation (6.128) for $i=0$ and $l=
\overline{l} + 1$. It follows from statements i) and ii), with $l=0$, of the
present theorem and from Lemma 6.5 and Lemma 6.6 that
$$\eqalignno{
R'_j (g^{(\overline{l} + 1)}_0, t) &\leq (1+t)^{- 3/2 + \rho}
C_{j, \overline{l}}   \overline{{\cal R}}^{(\overline{l} + 1)}_{L_0,
N_0+3+j+1+\overline{l}+1} (u ; u^{}_1,\ldots,u^{}_{\overline{l}+1}),
\  j \leq n-14,\hskip12mm&(6.157{\rm a})\cr
R^\infty_j (g^{(\overline{l} + 1)}_0, t) &\leq (1+t)^{- 1}  C_{j, \overline{l}}
 \overline{{\cal R}}^{(\overline{l} + 1)}_{L_0, N_0+3+j+4+\overline{l}+1} (u ;
u^{}_1,\ldots,u^{}_{\overline{l}+1}),\quad   j \leq n-13, &(6.157{\rm b})\cr
R^2_j (g^{(\overline{l} + 1)}_0, t) &\leq (1+t)^{- 1}	C_{j, \overline{l}}
 \overline{{\cal R}}^{(\overline{l} + 1)}_{L_0, N_0+3+j+\overline{l}+1} (u ;
u^{}_1,\ldots,u^{}_{\overline{l}+1}), \quad  j \leq n,& (6.157{\rm c})\cr
\wp^D_j (f^{(\overline{l} + 1)}_0 (t)) &\leq (1+t)^{- 3/2 + \rho}
C_{j, \overline{l}}   \overline{{\cal R}}^{(\overline{l} + 1)}_{L_0,
N_0+3+j+\overline{l}+1} (u ; u^{}_1,\ldots,u^{}_{\overline{l}+1}), \quad j \leq
n,
&(6.157{\rm d})\cr
}$$
where $C_{j, \overline{l}}$ depends only on $\rho$
and $\Vert u \Vert^{}_{E^\rho_{L_0}}.$
These inequalities give that
$$\eqalignno{
R^{(1)}_j (g^{(\overline{l} + 1)}_0, t)
&\leq C_{j, \overline{l}}  \overline{{\cal R}}^{(\overline{l} + 1)}_{L_0,
N_0+3+j+9+\overline{l}+1} (u ; u^{}_1,\ldots,u^{}_{\overline{l}+1}),
\quad  j \leq n-21,&(6.158{\rm a})\cr
h'_j (g^{(\overline{l} + 1)}_0, j_0, t)
&\leq C_{j, \overline{l}}  \overline{{\cal
R}}^{(\overline{l} + 1)}_{L_0, N_0+3+j+\overline{l}+1}
(u ; u^{}_1,\ldots,u^{}_{\overline{l}+1}),\quad
 j = n+1,&(6.158{\rm b})\cr
}$$
where $j_0$ is the integer part of $j/2 + 5,$
$$h''_j (g^{(\overline{l} + 1)}_0, t) \leq C_{j, \overline{l}}  \overline{{\cal
R}}^{(\overline{l} + 1)}_{L_0, N_0+3+j+\overline{l}+1}
(u ; u^{}_1,\ldots,u^{}_{\overline{l}+1}),\quad
 j = n+1, \eqno{(6.158{\rm c})}$$
where $C_{j, \overline{l}}$ depends only on $\rho$ and $\Vert u
\Vert^{}_{E^\rho_{L_0}}$. It follows from Lemma 6.5 and Lemma 6.6 that
$$\eqalignno{
&\wp^D_{n+1} (f^{(\overline{l} + 1)}_0 (t)) + (1+t)  \wp^D_n
((1+\lambda_1 (t))^{1/2}  g^{(\overline{l} + 1)}_0 (t)) &(6.158{\rm d})\cr
&\quad{}\leq (1+t)^{- 3/2 + \rho}  \big(C \Vert u \Vert^{}_{E^\rho_{L_0}}
 \Vert K^{(\overline{l} + 1)} \Vert^{}_{{\cal F}^M_{n+1}}
+ C_{n, \overline{l}}  \overline{{\cal R}}^{(\overline{l} + 1)}_{L_0,
N_0+3+n+1+\overline{l}+1} (u ; u^{}_1,\ldots,u^{}_{\overline{l}+1})\big)\cr
}$$
and
$$\eqalignno{
&\wp^D_{n+1,i}  (f^{(\overline{l} + 1)}_0 (t)) &(6.158{\rm e})\cr
&\quad{}\leq C_{n, \overline{l}}  (1+t)^{- 3/2 + \rho}  \overline{{\cal
R}}^{(\overline{l} + 1)}_{L_0, N_0+3+n+1+\overline{l}+1}
(u ; u^{}_1,\ldots,u^{}_{\overline{l}+1}),
\quad i \geq 1,\cr
}$$
where $C$ and $C_{n, \overline{l}}$ depend only on $\rho$ and $\Vert u
\Vert^{}_{E^\rho_{L_0}}$. It follows from the induction hypothesis in
$n$ and from inequalities (6.151), (6.152) and (6.153) that
$$\Vert K^{(\overline{l} + 1)} \Vert^{}_{{\cal F}^M_{n+1}} \leq C  \Vert u
\Vert^{}_{E^\rho_{L_0}}  \Vert \Phi^{(\overline{l} + 1)}_0 \Vert^{}_{{\cal
F}^D_{n+1}}
+ C_{j, \overline{l}}  \overline{{\cal R}}^{(\overline{l} + 1)}_{L_0,
N_0+3+n+1+\overline{l}+1} (u ; u^{}_1,\ldots,u^{}_{\overline{l}+1}),
\eqno{(6.159)}$$
where $C$ and $C_{n, \overline{l}}$ depend only on $\rho$ and $\Vert u
\Vert^{}_{E^\rho_{L_0}}$.
Applying statement iv) of Proposition 6.4 to equation
(6.128) with $i=0$ and $l = \overline{l} + 1$, using inequalities
(6.158b)--(6.159) and using the convention $\xi^{}_{n+1,i} = 0$ for
$i \geq n+2$ and
$$\xi^{}_{n+1,i} = \sup_{t \geq 0} \big((1+t)^{3/2 - \rho}	\wp^D_{n+1,i}
(\Phi^{(\overline{l}+1)}_0 (t))\big),\quad  0 \leq i \leq n+1, \eqno{(6.160)}$$
we obtain that
$$\eqalignno{
&\xi^{}_{n+1,0}& (6.161{\rm a})\cr
&\ {}\leq C \Vert u \Vert^{}_{E^\rho_{L_0}}  \xi^{}_{n+1,0} +
C_{n, \overline{l}}  (\xi^{}_{n+1,0})^{\varepsilon'}
(\xi^{}_{n+1,1})^{1 - \varepsilon'}
+ C_{n, \overline{l}}  \overline{{\cal R}}^{(\overline{l} + 1)}_{L_0,
N_0+3+n+1+\overline{l}+1} (u ; u^{}_1,\ldots,u^{}_{\overline{l}+1})\cr
}$$
and
$$\eqalignno{
&\xi^{}_{n+1,i}&(6.161{\rm b})\cr
&\ {}\leq C_{n, \overline{l}} (\xi^{}_{n+1,0})^{\varepsilon'}
(\xi^{}_{n+1,i+1})^{1 - \varepsilon'}
+ C_{n, \overline{l}}  \overline{{\cal R}}^{(\overline{l} + 1)}_{L_0,
N_0+3+n+1+\overline{l}+1} (u ; u^{}_1,\ldots,u^{}_{\overline{l}+1}),
\quad 1 \leq i \leq n+1,\cr
}$$
where $\varepsilon' = \min (1/2, 2(1 - \rho))$ and where $C$ and
$C_{n, \overline{l}}$ depend only on $\rho$ and $\Vert u
\Vert^{}_{E^\rho_{L_0}}$.
Let $\varepsilon > 0$
be such that $C \Vert u \Vert^{}_{E^\rho_{L_0}} \leq 1/2$ for $u \in {\cal
O}_\infty$ and let $\beta_{n+1} = 2 C_{n, \overline{l}}$ and $\alpha_{n+1} = 2
C_{n, \overline{l}}  \overline{{\cal R}}^{(\overline{l} + 1)}_{L_0,
N_0+3+n+1+\overline{l}+1} (u ; u^{}_1,\ldots,u^{}_{\overline{l}+1})$.
It then follows from inequalities (6.161a) and (6.161b) that
$$\xi^{}_{n+1,i} \leq \alpha_{n+1} + \beta_{n+1}
(\xi^{}_{n+1,0})^{\varepsilon'}  (\xi^{}_{n+1,i+1})^{1 - \varepsilon'},
\quad 0 \leq i \leq n+1. \eqno{(6.162)}$$
Proceeding as in solving system (5.182c) we obtain that $\xi^{}_{n+1,0} \leq
C'_{n+1}  \alpha_{n+1}$, where $C'_{n+1}$ is a polynomial in
$\beta_{n+1}$. This proves together with inequality (5.159), that
$$\Vert (K^{(\overline{l}+1)},	\Phi^{(\overline{l}+1)})
\Vert^{}_{{\cal F}_{n+1}} \leq C_{n+1, \overline{l}+1}
\overline{{\cal R}}^{(\overline{l} + 1)}_{L_0,
N_0+3+n+1+\overline{l}+1} (u ; u^{}_1,\ldots,u^{}_{\overline{l}+1}),
\eqno{(6.163)}$$
which by induction in $n$ proves that inequalities (6.157a) and (6.157b) are
true for each $j \in \Nrm$, since they are true for $j \leq 37$. Inequalities
(6.157a)--(6.158a) are then also true for each $j \in \Nrm$. Inequalities
(6.157a) and (6.157b) for $j \in \Nrm$ and inequality (6.144a), together with
$\Phi^{(\overline{l} + 1)} = \sum_{0 \leq i \leq 4}
\Phi^{(\overline{l}+1)}_i$, give that
$$\Vert (K^{(\overline{l}+1)},	\Phi^{(\overline{l}+1)}) \Vert^{}_{{\cal F}_j}
\leq C_{n+1, \overline{l}+1}	\overline{{\cal R}}^{(\overline{l} + 1)}_{L_0,
N_0+3+j+\overline{l}+1} (u ; u^{}_1,\ldots,u^{}_{\overline{l}+1}),
\quad j \geq 0,\eqno{(6.164{\rm a})}$$
where $C_{n+1, \overline{l}+1}$ depend only on $\rho$ and $\Vert u
\Vert^{}_{E^\rho_{L_0}}$. Inequality (6.158a), with $j \in \Nrm$, and statement
iii) of proposition 6.4 prove that inequality (6.156c) is true for each $j \in
\Nrm$. This shows, together with inequality (6.144b) that
$$H_j (\Phi^{(\overline{l}+1)}, t) \leq (1+t)^{- 3/2 + \rho}  C_{j,
\overline{l}+1}
\overline{{\cal R}}^{(\overline{l} + 1)}_{L_0, N_0+3+j+\overline{l}+14} (u ;
u^{}_1,\ldots,u^{}_{\overline{l}+1}),\quad  j \geq 0, \eqno{(6.164{\rm b})}$$
where $C_{j, \overline{l}+1}$ depends only on $\rho$ and $\Vert u
\Vert^{}_{E^\rho_{L_0}}$. Inequality (6.124) is true, when $\Phi$ is replaced
by
$\Phi^{(\overline{l}+1)}$ and $R^1_{n_2-k,k} (t)$ (resp. $R'_{n_4+7} (t)$,
$R^2_{n_4+9} (t)$,  $R^\infty_{n_4} (t)$) is replaced by $R^1_{n_2-k,k}
(\Phi^{(\overline{l}+1)}, t)$ (resp. $R'_{n_4+7} (\Phi^{(\overline{l}+1)}, t)$,
 $R^2_{n_4+9} (\Phi^{(\overline{l}+1)}, t)$,  $R^\infty_{n_4}
(\Phi^{(\overline{l}+1)}, t)$). This inequality, inequalities (6.112),
(6.157a)--(6.157c), (6.164b) and the definition of $R^1_{n,k}$ in Corollary 5.9
give that
$$\eqalignno{
&\wp^D_j \big((1+\lambda_1 (t))^{k/2}  \Phi^{(\overline{l}+1)}
(t)\big)&(6.164{\rm c})\cr
&\quad{}\leq
(1+t)^{- 3/2 + \rho}  C_{j, \overline{l}+1}
\overline{{\cal R}}^{(\overline{l} + 1)}_{L_0, N_0+3+j+k+\overline{l}+1} (u ;
u^{}_1,\ldots,u^{}_{\overline{l}+1}),\quad  j,k \geq 0,\cr
}$$
where $C_{j, \overline{l}+1}$ depend only on $\rho$ and $\Vert u
\Vert^{}_{E^\rho_{L_0}}$.

Inequalities (6.164a)--(6.164c) prove that the induction hypothesis in $l$ is
true for $l = \overline{l}+1$. Since we already have proved that it is true for
$l=0$ it follows by induction in $l$ that it is true for each $l \in \Nrm$. To
complete the proof we only have to prove the announced decrease properties of
$C_{n,l}$ and $C'_{n,l}$, when $\Vert u \Vert^{}_{E^\rho_{L_0}} \fl 0$.
We observe that since the solution $(\Phi, K) \in {\cal O}_\infty$ of equations
(6.126a)--(6.126c), with $l=0$, is unique, since $(\Phi^{(l)},
K^{(l)}) \in {\cal F}_j$,
$l\geq 1$, is the unique solution of equations (6.126a)--(6.126c) with $l \geq
1$ and since, when $u=0$, $(\Phi^{(l)}, K^{(l)}) = 0$ for $0 \leq l \leq 2$ and
$(\Phi^{(3)}, 0)$ are solutions it follows that $\Phi^{(l)} = 0$ for $0 \leq l
\leq 2$ and $K^{(l)} = 0$ for $0 \leq l \leq 3$, when $u=0$. The use of Taylor
formula now gives the announced properties of $C_{n,l}$ and $C'_{n,l}$ when
$\Vert u \Vert^{}_{E^\rho_{L_0}} \fl 0$. This proves the theorem.

In the situation of Theorem 6.9 let $u \in {\cal O}_\infty$ and let $v(u) = (K,
\Phi)$. According to the variable substitution (6.30) we introduce
$$\eqalignno{
A_\mu (t) &= K_\mu (t) + A^*_\mu (t), \quad 0 \leq \mu \leq 3 &(6.165)\cr
\psi' (t) &= \Phi (t) + \phi^* (t),\cr
}$$
and we introduce
$$\eqalignno{
(A_Y)_\mu &= (\xi^M_Y	A)_\mu,\cr
(\dot{A}_Y)_\mu &=(\xi^M_{P_0 Y}A),\cr
\psi'_Y &= \xi^D_Y  \psi',&(6.166)\cr
}$$
for $Y \in U (\p)$. When we want to indicate the dependence of $A_\mu (t),
(A_Y)_\mu (t)$ etc. on $u$ we shall  write $(A(u))_\mu (t),  (A_Y
(u))_\mu (t)$ etc. The functions $A_\mu$,  $0 \leq \mu \leq 3$, and
$\psi'$ satisfy, as we shall  prove it in next theorem,
the following equations on
$\Rrm^+ \times \Rrm^3$:
$$\eqalignno{
\carre  A_\mu &= (\psi')^+  \gamma^0  \gamma_\mu\psi',&(6.167{\rm a})\cr
(i  \gamma^\mu  \partial_\mu + m)  \psi' &=
\big(A_\mu - (\partial_\mu{\vartheta} (A))\big)  \gamma^\mu
\psi',&(6.167{\rm b})\cr
\noalign{\noindent and}
\partial_\mu A^\mu &= 0.&(6.167{\rm c})\cr
}$$
We recall that $A_{0,Y}, \dot{A}_{0, Y}, \phi'_{0,Y}$ are free solutions given
in (4.137a), (4.137b) and (4.137c).

In the sequel of this chapter $N_0$ will not necessarily denote the same
integer as in Proposition 6.2.
\saut
\noindent{\bf Theorem 6.10.}
{\it
Let $1/2 < \rho < 1$. Then there exists $N_0 \geq 0$ and an open ball ${\cal
O}_{N_0}$ in $E^{\circ\rho}_{N_0}$ with center at the origin such that $A$,
$\dot{A}$, $\psi'$ defined by (6.166) satisfy
$$\eqalignno{
&\vvv(D^l (A - A_0, \psi' - \phi'_0)) (u;u^{}_1,\ldots,u^{}_l)
\vvv^{}_{\rho',\varepsilon, L}& (6.168)\cr
&\quad{}\leq C_{l+L}({\cal R}^l_{N_0, L+l} \big(u^{}_1,\ldots,u^{}_l) + \Vert u
\Vert^{}_{E^\rho_{N_0+L+l}}  \Vert u^{}_1 \Vert^{}_{E^\rho_{N_0}}\cdots\Vert
u^{}_l \Vert^{}_{E^\rho_{N_0}}\big),\cr
}$$
for all $l \geq 0$,  $L \geq 0$,  $1/2 < \rho' \leq 1$,
 $\varepsilon = (\varepsilon (0), \varepsilon (1))$,
$\varepsilon (0) > 0$, $\varepsilon (1) \geq \rho$,
$u \in {\cal O}_\infty = {\cal O}_{N_0} \cap E^\rho_\infty$,
$u^{}_1,\ldots,u^{}_l \in E^{\circ\rho}_\infty$, where
$$\eqalignno{
\vvv (a, \Phi)\vvv^{}_{\rho', \varepsilon,L}
&= \sum_{i = 0,1}
 \sum_{\scr Y \in \ssg^i\atop\scr  \vert Y \vert + k \leq L}
\sup_{t\geq 0}\Big((1+t)^{\rho'-1/2}  \Vert (\xi^M_Y a, \xi^M_{P_0 Y} a) (t)
\Vert^{}_{M^{\rho'}_0}\cr
&\qquad{}+ (1+t)^{2(1-\rho)}  \Vert (1+\lambda_1
(t))^{k/2}  (\xi^D_Y  \Phi) (t) \Vert^{}_{D_0}\cr
&\qquad{}+ \Vert (\delta (t))^{1+i-\varepsilon (i)}(1+t)^{\varepsilon
(i) (1-i)}  (\xi^M_Y a) (t) \Vert^{}_{L^\infty}\cr
&\qquad{}+ \Vert (\delta (t))^{3/2+2(1-\rho)}  (1+ \lambda_1 (t))^{k/2}
 (\xi^D_Y \Phi) (t) \Vert^{}_{L^\infty}\Big)\cr
}$$
and where $C_{l+L}$ depends only on $\rho'$, $\varepsilon$, $\rho$ and $\Vert u
\Vert^{}_{E^\rho_{N_0}}$. Moreover the function $u \mapsto (A - A_0, \psi' -
\phi'_0) (u)$ from ${\cal O}_\infty$ to the Fr\'echet space with seminorms
$\vvv\cdot  \vvv^{}_{\rho', \varepsilon, L}$,  indexed by $L \geq 0$ has a zero
of order two at $u = 0$,
$$(D^l	A_Y, D^l  \dot{A}_Y, D^l  \psi'_Y) (u ;
u^{}_1,\ldots,u^{}_l) \in C^0(\Rrm^+, E^\rho),\quad  Y \in \Pi'$$
and equations (6.167a), (6.167b) and (6.167c) are satisfied.
}\saut
\noindent{\it Proof.}
Since
$$\eqalignno{
&(A_Y - A_{0,Y}, \dot{A}_Y - \dot{A}_{0,Y}, \psi'_Y - \phi'_{0,Y})\cr
&\quad{}=
(K_Y, K_{P_0 Y}, \Phi_Y) + (A^*_Y - A_{0,Y}, A^*_{P_0 Y} - A_{0, P_0 Y},
\phi^*_Y - \phi'_{0,Y})\cr
}$$
according to (6.165) and (6.166), it follows from
Theorem 6.9 that it is sufficient to prove inequality (6.168) with $A_Y$ (resp.
$\dot{A}_Y, \psi'_Y)$ replaced by $A^*_Y$ (resp. $A^*_{P_0 Y}, \phi^*_Y).$
Since
$$\eqalignno{
&(A^*_Y - A_{0,Y}, A^*_{P_0 Y} - A_{0, P_0 Y}, \phi^*_Y - \phi'_{0,Y}) \cr
&\quad{}=
(\Delta^{*M}_Y, \Delta^{*M}_{P_0 Y}, \Delta^{*D}_Y)
+ (A^*_{0,Y} - A_{J+1,Y}, A^*_{0,P_0 Y} - A_{J+1, P_0 Y}, \phi^*_{0,Y} -
\phi'_{J+1,Y})\cr
&\qquad{}+ \sum_{0 \leq n \leq J}  (A_{n+1,Y} - A_{n,Y}, A_{n+1,P_0 Y} -
A_{n, P_0 Y}, \phi'_{n+1,Y} - \phi'_{n,Y}),\cr
}$$
inequality (6.168) now follows from Theorem 4.10, Lemma 6.1 and Proposition 6.2
as does also the statement concerning the second order zero.

It follows from equations (6.1a)--(6.2b), from equations (4.137b) and (4.137c)
for $A_{n+1, \mu}$ and $\phi'_{n+1}$ and from definitions (4.138a) and (4.138b)
of $\Delta^M_n$ and $\Delta^D_n$ that equations (6.167a) and (6.167b) are
satisfied. Equations (6.167a) and (6.167b) give $\carre  \partial_\mu
 A¬\mu = 0$, i.e. with the notation (6.166), $\carre
\partial_\mu (A_{\un})^\mu = 0$. It then follows from inequality (6.168) with
$\rho' = 1$, $l = 1$, $Y = \un$ that
$$\partial_\mu (A_{\un})^\mu = \partial_\mu (A_{0, \un})^\mu = 0,$$
where the last equality follows from the definition of $E^{\circ\rho}$. The
continuity from $\Rrm^+$ to $E^\rho$ follows from the fact that $(K, \Phi) \in
{\cal F}_n$ for $n \geq 0$. This proves the theorem.

Next we introduce the {\it manifold on which the gauge condition}
defined by (1.3a) and (1.3b) is {\it satisfied}. Let $V^\rho_N, N \geq 1$,
be the subset of all elements
$(f, \dot{f}, \alpha) \in E^\rho_N,  1/2 < \rho < 1$, such that
$$\eqalignno{
\Delta  f_0 - \sum_{1 \leq i \leq 3}	\partial_i
 \dot{f}_i + \vert \alpha \vert^2 &= 0,&(6.169)\cr
\dot{f}_0 - \sum_{1\leq i \leq 3}\partial_i  f_i &= 0\cr
}$$
and let $V^\rho_\infty = \displaystyle{\cap_{N \geq 1}}  V^\rho_N.$
We also introduce the map
$$F^{-1}\colon (g, \dot{g}, \beta) \mapsto (f, \dot{f}, \alpha),
f_0 =g^{}_0 - \Delta^{-1}  \vert \beta \vert^2,
f_i = g^{}_i, 1 \leq i \leq 3,
\dot{f} = \dot{g},
\alpha =\beta,\eqno{(6.170)}$$
which maps $E^\rho_N$ onto $E^\rho_N$,  $N \geq 1$ as will be proved.
\saut
\noindent{\bf Theorem 6.11.}
{\it
Let $I$ be the identity map on $E_N$ and let $F^{-1}$ be defined by (6.170).
Then $F = I + F^{(2)}$ and $F^{-1} = I - F^{(2)}$ where $F^{(2)}$ is a (real)
bilinear continuous map from $E_N$ to $E_N,  N \geq 1$. $F^{(2)}$
satisfies
$$\eqalignno{
&\Vert F^{(2)} (u^{}_1,u^{}_2) \Vert^{}_{E^\rho_N}\cr
&\quad{}\leq C_N \big(\Vert \alpha_1 \Vert^{}_{D_N}
 \Vert \alpha_2 \Vert^{}_{D_1} + \Vert \alpha_1 \Vert^{}_{D_1}
\Vert \alpha_2 \Vert^{}_{D_N}\big),
\quad N \geq 1, u^{}_i = (f_i, \dot{f}_i, \alpha_i),\quad i = 1,2,\cr
}$$
where $C_N$ are independent of $u^{}_1,u^{}_2$. Moreover $V^\rho_N,  N \geq
1$, (with its topology inherited from $E^\rho_N$) is a differentiable Hilbert
manifold globally diffeomorphic to $E^{\circ\rho}_N$ by the map
$F\colon V^\rho_N \fl E^{\circ\rho}_N$ and $V^\rho_N = V^\rho_1
\cap E^\rho_N.$
}\saut
\noindent{\it Proof.}
Let $u = (f, \dot{f}, \alpha),	u^{}_i = (f_i, \dot{f}_i, \alpha_i),
 i = 1,2,u,u^{}_1,u^{}_2 \in E_N$. By definition (6.170) of $F$ it follows
that
$$\eqalignno{
F(u) &= u + F^{(2)} (u,u),\cr
F^{-1} (u) &= u - F^{(2)} (u,u),&(6.171)\cr
F^{(2)} (u^{}_1,u^{}_2)
&= \big(({1 \over 2}  \Delta^{-1} (\alpha_1^+  \alpha_2 + \alpha_2^+
\alpha_1), 0,0,0),0,0\big).\cr
}$$
According to Theorem 2.9
$$\eqalignno{
\Vert F^{(2)} (u^{}_1,u^{}_2) \Vert^{}_{E^\rho_N} &= \Vert \big(({1 \over 2}
\Delta^{-1} (\alpha^+_1  \alpha_2 + \alpha_2^+  \alpha_1), 0,0,0),0\big)
\Vert^{}_{M^\rho_N}\cr
&\leq C_N  \sum_{0 \leq \vert \mu \vert \leq \vert \nu \vert \leq n}
 \Vert \vert \nabla \vert^\rho	x^\mu  \partial^\nu
 \Delta^{-1} (\alpha^+_1  \alpha_2 + \alpha_2^+  \alpha_1)
\Vert^{}_{L^2}\cr
}$$
where $\mu, \nu \in \Nrm^3,  x^\mu = x^{\mu_1}_1
x^{\mu_2}_2  x^{\mu_3}_3, \partial^\nu = \partial^{\nu_1}_1
\partial^{\nu_2}_2  \partial^{\nu_3}_3$. Since $\vert \mu \vert \leq
\vert \nu \vert$ it follows that
$$\Vert F^{(2)} (u^{}_1,u^{}_2) \Vert^{}_{E^\rho_N}
\leq C'_N  \sum_{0 \leq \vert \mu \vert \leq \vert \nu \vert \leq n}
 \Vert \vert \nabla \vert^{\rho-2}  x^\mu
\partial^\nu  (\alpha^+_1 \alpha_2 + \alpha_2^+ \alpha_1)
\Vert_{L^2}$$
for some constants $C'_N$ and it follows by continuity, using the fact that
$\Vert \vert \nabla \vert^{\rho-2} h \Vert^{}_{L^2} \leq C_p
\Vert h \Vert^{}_{L^p}$,
 $p = 6(7 - 2 \rho)^{-1}$, for $1/2 < \rho \leq 2$, where $h$ has compact
support and the constant $C_p$ is independent of the support:
$$\Vert F^{(2)} (u^{}_1,u^{}_2) \Vert^{}_{E^\rho_N} \leq C''_N  \sum_{\vert \mu
\vert \leq \vert \nu \vert \leq n}  \Vert x^\mu
\partial^\nu  (\alpha^+_1 \alpha_2 + \alpha_2^+  \alpha_1)
\Vert^{}_{L^p},$$
$1/2 < \rho < 1,  p = 6(7 - 2 \rho)^{-1}$. Using the inequalities
$$\eqalignno{
\Vert h_1  h_2 \Vert^{}_{L^p} &\leq \Vert h_1 \Vert^{}_{L^2}
\Vert h_2 \Vert^{}_{L^{3/(2-\rho)}},\cr
\Vert h \Vert^{}_{L^{3/(2-\rho)}}&\leq C_\rho  \Vert h_1 \Vert^{}_{W^{1,2}},\cr
}$$
for $1/2 < \rho < 1$, we obtain
$$\eqalignno{
&\Vert F^{(2)} (u^{}_1,u^{}_2) \Vert^{}_{E^\rho_N}\cr
&\quad{}\leq C_N  \sum_{0 \leq i \leq N/2}  \big(\Vert \alpha_1
\Vert^{}_{D_{N-i}}  \Vert \alpha_2 \Vert^{}_{D_{i+1}} + \Vert \alpha_1
\Vert^{}_{D_{1+i}}  \Vert \alpha_2 \Vert^{}_{D_{N-i}}\big),\quad  N \geq1.\cr
}$$
Corollary 2.6 now gives (with new constants $C_N$ independent of
$u^{}_1,u^{}_2$)
$$\Vert F^{(2)} (u^{}_1,u^{}_2) \Vert^{}_{E^\rho_N} \leq C_N
\big(\Vert \alpha_1 \Vert^{}_{D_N}
 \Vert \alpha_2 \Vert^{}_{D_1} + \Vert \alpha_1 \Vert^{}_{D_1}
\Vert \alpha_2 \Vert^{}_{D_N}\big),\quad  N \geq 1.$$
This proves the inequality of the theorem.

Let $u = (f, \dot{f}, \alpha) \in V^\rho_N,  N \geq 1$. Then $v = (g,
\dot{g}, \beta) = F(u) \in E^\rho_N$ and according to (6.171):
$g^{}_0 = f_0 + \Delta^{-1}  \vert \alpha \vert^2$, $ g^{}_i = f_i$
for $ 1 \leq i \leq 3$ and  $\dot{g} = \dot{f}$,
$\beta =\alpha$.
It follows from (6.169) that
$$\eqalignno{
\Delta g^{}_0 - \sum_{1 \leq i \leq 3}  \partial_i
\dot{g}_i &= 0,\cr
\dot{g}_0 - \sum_{1 \leq i \leq 3}
\partial_i  g^{}_i &= 0,\cr
}$$
which, according to the definition of $E^{\circ\rho}_N$ proves that $v \in
E^{\circ\rho}_N$. Similarly it follows that $F^{-1} (u) \in V^\rho_N$ if $u
\in E^{\circ\rho}_N$. This proves the theorem since $E^{\circ\rho}_N =
E^\rho_N \cap E^\rho_0$.

We next define a {\it modified wave operator} $\Omega_1$ for the
M-D equations (1.1a), (1.1b) and (1.1c), when $t \fl \infty$.
Let $(A_Y (u)) (t), (\dot{A}_Y (u)) (t), (\psi'_Y (u)) (t), N_0$
and ${\cal O}_{N_0}$ be as in Theorem 6.10 and let
$$\Omega_1 (u) = \big((A (u)) (0), (\dot{A} (u)) (0),
e^{-i {\vartheta}(A,0)} (\psi' (u)) (0)\big),
\quad u \in {\cal O}_\infty = {\cal O}_{N_0} \cap
E^\rho_\infty.\eqno{(6.172)}$$
\saut
\noindent{\bf Theorem 6.12.}
{\it
Let $1/2 < \rho < 1$. There exists $N_0 \geq 0$ and a neighbourhood
${\cal O}_{N_0}$ of zero in $E^{\circ\rho}_{N_0}$ such that $\Omega_1\colon
{\cal
O}_\infty \fl E^\rho_\infty$ is a one to one $C^\infty$ mapping satisfying
$$\eqalignno{
&\Vert (D^l \Omega_1)(u ; u^{}_1,\ldots,u^{}_l) \Vert^{}_{E_L}\cr
&\quad{}\leq F_{L,l} (\Vert u \Vert^{}_{E_{N_0}})  {\cal R}^l_{N_0,l+L}
(u^{}_1,\ldots,u^{}_l)
+ F'_{L,l} (\Vert u \Vert^{}_{E_{N_0}})  \Vert u \Vert^{}_{E_{N_0+l+L}}
 \Vert u^{}_1 \Vert^{}_{E_{N_0}}\cdots \Vert u^{}_l \Vert^{}_{E_{N_0}}\cr
}$$
for each $L \geq 0$, $l \geq 0$, $u^{}_1,\ldots,u^{}_l \in
E^{\circ\rho}_\infty$,
where
$F_{L,l}$ and $F'_{L,l}$ are increasing continuous functions. Moreover
$\Omega_1 (0) = 0$ and $\Omega_1 ({\cal O}_\infty) \subset V^\rho_\infty$.
}\saut
\noindent{\it Proof.}
We first prove that $\Omega_1$ is one to one. Let $u, u' \in {\cal O}_\infty,$
where ${\cal O}_\infty$ is as in Theorem 6.10. If $\Omega_1 (u) = \Omega_1
(u')$, then it follows because of the uniqueness of the local (in time)
solution of the Maxwell-Dirac equations (1.1a), (1.1b) and (1.1c), according to
Theorem 1 of \refGR\ that the solution $(A(u), \psi' (u))$ and $(A(u'), \psi'
(u'))$ of equations (6.167a), (6.167b) and (6.167c) are equal. Theorem 6.10
then give that the free solutions $(A_{0, \un} (u), \psi'_{0, \un} (u))$ and
$(A_{0, \un} (u'), \psi'_{0, \un} (u'))$ are equal, which according to their
definition after formula (4.137c) proves that $u=u'$.

Let
$$a_N = \Big(\sum_{\scr Y \in \Pi'\atop\scr \vert Y \vert \leq N}
\Vert (A_Y (u),
\dot{A}_Y (u), \psi'_Y (u)) (0) \Vert^2_{E^\rho_0}\Big)^{1/2}, \quad  N \geq
0. \eqno{(6.173)}$$
Introduce also
$$\psi_Y (u) = \xi^D_Y	(e^{-i {\vartheta} (A)}  \psi' (u))$$
and
$${\vartheta}'_Y (u) = \xi^{}_Y  {\vartheta} (A),\quad  Y \in U(\p).$$
By Leibniz rule we obtain that
$$\eqalignno{
\Vert (\psi'_Y (u)) (0) \Vert^{}_D &= \Vert (\xi^D_Y  e^{-i {\vartheta}
(A,0)}	\psi (u)) (0) \Vert^{}_D&(6.174)\cr
&\geq \Vert (\xi^D_Y  \psi (u)) (0) \Vert^{}_D
- C_{\vert Y \vert}  \sum \Vert ({\vartheta}'_{Y_1} (u)\cdots
{\vartheta}'_{Y_l} (u)  \psi_Z (u)) (0) \Vert^{}_D,\cr
}$$
$Y \in \Pi'$, $\vert Y \vert \geq 1$, where the sum is taken over
$1 \leq l \leq \vert Y \vert$,  $Y_i \in
\Pi'$,  $\vert Y_1 \vert +\cdots + \vert Y_l \vert + \vert Z \vert \leq
\vert Y \vert$,	$\vert Z \vert \leq \vert Y \vert - 1$,
$\vert Y_i \vert \geq 1$. It follows from Lemma 4.4 and Theorem 6.10 that
$$\sup_{x \in \Rrm^3}  (1+\vert x \vert)^{1/2-\rho}  \vert
({\vartheta}'_{Y_i} (u)) (0,x) \vert \leq F'_{\vert Y_i \vert, 0}
(\Vert u \Vert^{}_{E^\rho_{N_0}})  \Vert u \Vert^{}_{E^\rho_{N_0+\vert Y_i
\vert}},$$
which gives using that $T^D_Z  (\Omega_1 (u)) = (\psi_Z (u)) (0)$:
$$\eqalignno{
&\Vert ({\vartheta}'_{Y_1} (u)\cdots {\vartheta}'_{Y_l} (u)  \psi_Z (u)) (0)
\Vert^{}_D\cr
&\quad{}\leq G_{\vert Y \vert}  (\Vert u \Vert^{}_{E^\rho_{N_0}})
\Vert u \Vert^{}_{E^\rho_{N_0+\vert Y_1 \vert}}\cdots\Vert u
\Vert^{}_{E^\rho_{N_0+\vert Y_l\vert}}
\Vert (1+\vert\cdot \vert)^{l(\rho-1/2)}
 T^D_Z (\Omega_1 (u)) \Vert^{}_D,\cr
}$$
where $G_{\vert Y \vert}$ is a polynomial. Since $\vert Y_i \vert \geq 1$ it
follows from Corollary 2.6 that
$$\Vert u \Vert^{}_{E^\rho_{N_0+\vert Y_1 \vert}}\cdots \Vert u
\Vert^{}_{E^\rho_{N_0+\vert Y_l\vert}}
\leq C_{\vert Y \vert}  \Vert u \Vert^{l-1}_{E^\rho_{N_0+1}}
 \Vert u \Vert^{}_{E^\rho_{N_0+1+\vert Y_1 \vert+\cdots+\vert Y_l
\vert-l}},$$
which after redefinition of the polynomial $G_{\vert Y \vert}$ and by the fact
that $0 \leq \vert Y_1 \vert +\cdots
+ \vert Y_l \vert - l \leq \vert Y \vert -\vert Z \vert - l$ gives:
$$\eqalignno{
&\Vert ({\vartheta}'_{Y_1} (u)\cdots {\vartheta}'_{Y_l} (u)  \psi_Z (u)) (0)
\Vert^{}_D&(6.175)\cr
&\quad{}\leq G_{\vert Y \vert}  (\Vert u \Vert^{}_{E^\rho_{N_0+1}})
\Vert u \Vert^{}_{E_{N_0+1+\vert Y \vert - \vert Z \vert - l}}	\Vert
(1+\vert \cdot \vert)^{l(\rho-1/2)}  T^D_Z (\Omega_1 (u)) \Vert^{}_D.\cr
}$$
The inequality $\Vert (1+\vert\cdot \vert)^b  f \Vert^{}_{L^2} \leq \Vert
(1+\vert \cdot \vert)  f \Vert^b_{L^2}	\Vert f
\Vert^{1-b}_{L^2}$,  $0 \leq b \leq 1$, gives:
$$\eqalignno{
&\Vert (1+\vert\cdot \vert)^{l(\rho-1/2)}  T^D_Z (\Omega_1 (u)) \Vert^{}_D\cr
&\quad{}\leq \Vert (1+\vert \cdot \vert)^{1+(l-1)(\rho-1/2)}  T^D_Z (\Omega_1
(u)) \Vert^{\rho-1/2}_D  \Vert (1+\vert \cdot \vert)^{(l-1)(\rho-1/2)}
 T^D_Z (\Omega_1 (u)) \Vert^{3/2-\rho}_D,\cr
}$$
which together with Theorem 2.9, statement i) of Corollary 2.21 and inequality
(6.175) shows that
$$\eqalignno{
&\Vert ({\vartheta}'_{Y_1} (u)\cdots{\vartheta}'_{Y_l} (u)  \psi_Z (u)) (0)
\Vert^{}_D\cr
&\quad{}\leq G_{\vert Y \vert}  (\Vert u \Vert^{}_{E^\rho_{N_0+1}})
(C_{l,\vert Z \vert} (\Vert \Omega_1 (u) \Vert^{}_{E^\rho_0}))^{\rho-1/2}
 (C_{l-1,Z} (\Vert \Omega_1 (u) \Vert^{}_{E^\rho_0}))^{3/2-\rho}\cr
&\qquad{}\Vert u \Vert^{}_{E_{N_0+1+\vert Y \vert - \vert Z \vert - l}}  \Vert
\Omega_1 (u) \Vert^{\rho-1/2}_{E_{l+\vert Z \vert}}  \Vert \Omega_1
(u) \Vert^{3/2-\rho}_{E_{l-1+\vert Z \vert}}.\cr
}$$
Since
$$\Vert \Omega_1 (u) \Vert^{}_{E^\rho_0} = \Vert (A(u), \dot{A} (u), \psi' (u))
(0) \Vert^{}_{E^\rho_0},$$
it follows from the preceeding inequality and from Theorem 6.10 that
$$\eqalignno{
&\Vert ({\vartheta}'_{Y_1} (u) \cdots {\vartheta}'_{Y_l} (u)  \psi_Z (u)) (0)
\Vert^{}_D &(6.176)\cr
&\qquad{}\leq G'_{\vert Y \vert}  (\Vert u \Vert^{}_{E_{N_0+1}})
\Vert u \Vert^{}_{E_{N_0+1+\vert Y \vert - \vert Z \vert - l}}  \Vert
\Omega_1 (u) \Vert^{\rho-1/2}_{E_{l+\vert Z \vert}}  \Vert \Omega_1
(u) \Vert^{3/2-\rho}_{E_{l-1+\vert Z \vert}},\cr
}$$
where $G'_{\vert Y \vert}$ is a continuous function. Since $(x+y)^a \leq x^a +
y^a$ for $x,y \geq 0$, $0 \leq a \leq 1$, it follows from Corollary 2.6 that
$$\eqalignno{
&\Vert u \Vert^{}_{E_{N_0+1+\vert Y \vert - \vert Z \vert - l}}  \Vert
\Omega_1 (u) \Vert^{\rho-1/2}_{E_{l+\vert Z \vert}}  \Vert \Omega_1
(u) \Vert^{3/2-\rho}_{E_{l-1+\vert Z \vert}}\cr
&\quad{}\leq C_{\vert Y \vert} (\Vert u \Vert^{}_{E_{N_0+1}}	\Vert \Omega_1
(u) \Vert^{\rho-1/2}_{E_{\vert Y \vert}}  \Vert \Omega_1 (u)
\Vert^{3/2-\rho}_{E_{\vert Y \vert-1}}
+ \Vert u \Vert^{}_{E_{N_0+1+\vert Y \vert}}	\Vert \Omega_1 (u)
\Vert^{}_{E_0}),\cr
}$$
which together with Theorem 6.10 and inequality (6.176) give
$$\eqalignno{
&\Vert ({\vartheta}'_{Y_1} (u)\cdots {\vartheta}'_{Y_l} (u)  \psi_Z (u)) (0)
\Vert^{}_D& (6.177)\cr
&\quad{}\leq G_{\vert Y \vert} (\Vert u \Vert^{}_{E_{N_0+1}})  \Vert u
\Vert^{}_{E_{N_0+1+\vert Y \vert}} + G'_{\vert Y \vert}
(\Vert u \Vert^{}_{E_{N_0+1}})
 \Vert \Omega_1 (u) \Vert^{\rho-1/2}_{E_{\vert Y \vert}}
\Vert \Omega_1 (u) \Vert^{3/2-\rho}_{E_{\vert Y \vert-1}},\cr
}$$
where $G_{\vert Y \vert}$ and $G'_{\vert Y \vert}$ are continuous functions.

Definition (6.173) and inequalities (6.174) and (6.177) give after redefining
$N_0+1$ by $N_0$ and noting that
$$\eqalignno{
&\wp^{}_N (T(\Omega_1 (u))) = \Big(\sum_{\scr Y \in \Pi'\atop\scr
\vert Y \vert \leq N}
\Vert (A_Y (u), \dot{A}_Y (u), \psi_Y (u)) (0) \Vert^2_D\Big)^{1/2}:\cr
&\wp^{}_N (T(\Omega_1 (u)))\cr
&\quad{}\leq a_N + G_N (\Vert u \Vert^{}_{E_{N_0}})
\Vert u \Vert^{}_{E_{N_0+N}}
+ G'_N (\Vert u \Vert^{}_{E_{N_0}})  \Vert \Omega_1 (u)
\Vert^{\rho-1/2}_{E_{N}}  \Vert \Omega_1 (u)
\Vert^{3/2-\rho}_{E_{N-1}},\quad  N \geq 1,&(6.178)\cr
}$$
where $G_N$ and $G_{N'}$ are some continuous functions.

Let ${\cal O}_{N_0}$ be sufficiently small so that $\Omega_1 (u) \in {\cal O},$
where ${\cal O}$ is given by Theorem 2.22. This is possible according to
theorem 6.10. It follows from inequality (6.178) statement ii) of Theorem 2.22
and Theorem 6.10 that
$$\eqalignno{
&\wp^{}_1 (T(\Omega_1 (u))) &(6.179)\cr
&\quad{}\leq a_1 + G_1 (\Vert u \Vert^{}_{E_{N_0}})
\Vert u \Vert^{}_{E_{N_0+1}}
+ H (\Vert u \Vert^{}_{E_{N_0}})  \wp^{}_1 (T( \Omega_1
(u)))^{\rho-1/2}  \Vert \Omega_1 (u) \Vert^{3/2-\rho}_{E_0},\cr
}$$
where $H$ is a continuous function. This inequality gives
$$\eqalignno{
&\wp^{}_1 (T(\Omega_1 (u)))&(6.180)\cr
&\quad{} \leq 2 a_1 + 2 G_1 (\Vert u \Vert^{}_{E_{N_0}})
\Vert u \Vert^{}_{E_{N_0+1}}
+ (H(\Vert u \Vert^{}_{E_{N_0}}))^2  \Vert \Omega_1 (u) {\Vert^{}_{E_0}}.\cr
}$$
As a matter of fact if $\Vert \Omega_1 (u) \Vert^{}_{E_0} = 0$
then this is obvious.
Let $\Vert \Omega_1 (u) \Vert^{}_{E_0} > 0$ and let $x = \wp^{}_1
(T (\Omega_1 (u))) /\Vert \Omega_1 (u) \Vert^{}_{E_0}$.
Since $x \geq 1$ and $0 < \rho - 1/2 < 1/2$ we
obtain
$$x \leq (a_1 + G_1 (\Vert u \Vert^{}_{E_{N_0}})  \Vert u
\Vert^{}_{E_{N_0+1}}) / \Vert \Omega_1 (u) \Vert^{}_{E_0} + H (\Vert u
\Vert^{}_{E_{N_0}})  x^{1/2},$$
which gives the result. Inequality (6.180) and Theorem 6.10 (used for $a_1$ and
$\Vert \Omega_1 (u) \Vert^{}_{E_0})$ now show that
$$\wp^{}_1 (T (\Omega_1 (u))) \leq F (\Vert u \Vert^{}_{E_{N_0}})  \Vert u
\Vert^{}_{E_{N_0+1}}, \eqno{(6.181)}$$
where $F$ is a continuous function. Since $\Omega_1 (u) \in {\cal O}$, it
follows from (6.181) and statement~ii) of Theorem 2.22 that
$$\Vert \Omega_1 (u) \Vert^{}_{E_1} \leq F'_{1,0} (\Vert u \Vert^{}_{E_{N_0}})
  \Vert u \Vert^{}_{E_{N_0+1}}, \eqno{(6.182)}$$
where $F'_{1,0}$ is a continuous function.

Inequalities (6.178), (6.181) and (6.182) and statement iii) of
Theorem 2.22 give, after replacing $N_0+1$ by $N_0$, that
$$\eqalignno{
&\wp^{}_N (T (\Omega_1 (u)))&(6.183)\cr
&\quad{} \leq H_N (\Vert u \Vert^{}_{E_{N_0}})	\Vert u
\Vert^{}_{E_{N_0+N}}
+ H'_N (\Vert u \Vert^{}_{E_{N_0}})  \big(\wp^{}_N (T (\Omega_1
(u)))\big)^{\rho-1/2}
 \big(\wp^{}_{N-1} (T (\Omega_1 (u)))\big)^{3/2-\rho},\cr
}$$
$N \geq 1$, where we have estimated $a^{}_N$ by Theorem 6.10 and where
$H_N$ and $H'_N$ are continuous functions.

In the same way as we obtained inequality (6.180) from inequality (6.179) we
obtain from inequality (6.183) that
$$\eqalignno{
&\wp^{}_N (T (\Omega_1 (u)))\cr
&\quad{}\leq 2 H_N (\Vert u \Vert^{}_{E_{N_0}})  \Vert
u \Vert^{}_{E_{N_0+N}}
+ (H'_N (\Vert u \Vert^{}_{E_{N_0}}))^2  \wp^{}_{N-1} (T (\Omega_1
(u))),\quad  N \geq 1.\cr
}$$
Since $\wp^{}_0 (T (\Omega_1 (u))) = \Vert \Omega_1 (u) \Vert^{}_{E_0}
\leq H (\Vert u
\Vert^{}_{E_{N_0}})  \Vert u \Vert^{}_{E_{N_0}}$, according to Theorem
6.10, where $H$ is a continuous function it now follows that
$$\wp^{}_N (T (\Omega_1 (u))) \leq G_N (\Vert u \Vert^{}_{E_{N_0}})	\Vert
u \Vert^{}_{E_{N_0+N}},\quad  N \geq 0, \eqno{(6.184)}$$
where $G_N$ are continuous functions. Since $\Omega_1 (u) \in {\cal O}$ it
follows from (6.184) and statement iii) of Theorem 2.22 (replacing $N_0+1$ by
$N_0$) that
$$\Vert \Omega_1 (u) \Vert^{}_{E_N} \leq F'_{N,0} (\Vert u \Vert^{}_{E_{N_0}})
 \Vert	u \Vert^{}_{E_{N_0+N}},\quad  N \geq 0. \eqno{(6.185)}$$
This proves the theorem in the case of $l=0$. The proof for the case $l>0$ is
so similar that we omit it. That the gauge conditions (1.3a) and (1.3b) are
satisfied follows from the fact that equation (6.167c) is satisfied according
to Theorem 6.10. This shows that $\Omega_1 ({\cal O}_\infty) \subset
V^\rho_\infty$, which proves the theorem.

In order to prove that $\Omega_1$ has a {\it local inverse}, we shall use,
for the space $E^\rho_\infty$, the inverse mapping theorem in the case
of Fr\'echet spaces. To do that we first extend the map
$\Omega_1\colon {\cal O}_\infty \fl V^\rho_\infty$,
where ${\cal O}_\infty$ is an open neighbourhood of zero in
$E^{\circ\rho}_\infty$ to a map from ${\cal O}'_\infty$ to $E^\rho_\infty,$
where ${\cal O}'_\infty$ is a neighbourhood of zero in $E^\rho_\infty$ and
then prove that the derivative of this extended map has a right inverse.
\saut
\noindent{\bf Theorem 6.13.}
{\it
Let $1/2 < \rho < 1$. There exists $N_0 \geq 0$ and $M_0 \geq 0$ and there
exists neighbourhoods ${\cal O}_{N_0}$ and ${\cal U}_{M_0}$ of zero in
$E^{\circ\rho}_{N_0}$ and $V^\rho_{M_0}$ respectively,  such that the map
$\Omega_1 \colon {\cal O}_\infty \fl {\cal U}_\infty$,
${\cal O}_\infty ={\cal O}_{N_0}\cap
E^\rho_\infty$,	${\cal U}_\infty = {\cal U}_{M_0} \cap E^\rho_\infty$,
satisfies the conclusions of Theorem 6.12 and has a $C^\infty$ inverse
$\Omega^{-1}_1\colon {\cal U}_\infty \fl {\cal O}_\infty$ satisfying
$$\eqalignno{
&\Vert (D^l (\Omega^{-1}_1 \circ F^{-1}))  (u ; u^{}_1,\ldots,u^{}_l)
\Vert^{}_{E_L}\cr
&\quad{}\leq C_{L,l}{\cal R}^l_{M_0,l+L} (u^{}_1,\ldots,u^{}_l)
+ C'_{L,l}  \Vert u \Vert^{}_{E_{M_0+l+L}}  \Vert u^{}_1
\Vert^{}_{E_{M_0}}\cdots\Vert u^{}_l \Vert^{}_{E_{M_0}}\cr
}$$
for each $L \geq 0,  l \geq 0$ and each $u \in F ({\cal U}_\infty),
u^{}_1,\ldots,u^{}_l \in E^{\circ\rho}_\infty$, where $F^{-1}$
is defined by (6.170) and where $C_{L,l}$ and $C'_{L,l}$ are constants
depending only on $\rho$ and $\Vert u \Vert^{}_{E^\rho_{M_0}}.$
}\saut
\noindent{\it Proof.}
Let $Q$ be the orthogonal projection on $E^{\circ\rho}_0$ in $E^\rho_0$, let
$N_0 \geq 1$, ${\cal O}_{N_0}$, ${\cal O}_\infty$ be as in Theorem 6.12, let
the map $F$ be as in Theorem 6.11, let ${\cal O}'_{N_0}$ be an open
neighbourhood of zero in $E^\rho_{N_0}$ such that ${\cal O}'_{N_0} \cap
E^{\circ\rho}_{N_0} = {\cal O}_{N_0}$ and let ${\cal O}'_\infty = {\cal
O}'_{N_0} \cap E^\rho_\infty.$
$$G(u) = F (\Omega_1 (Qu)) + (1 - Q)u,	u \in {\cal O}'_\infty,
\eqno{(6.186)}$$
defines a $C^\infty$ function $G\colon {\cal O}'_\infty \fl E^\rho_\infty$
according to Theorem 6.11 and Theorem 6.12 and since $Q$ is a linear continuous
mapping from $E^\rho_N$ to $E^\rho_N,  N \geq 0$. It follows from
Theorem~6.11, Theorem 6.12 and Corollary 2.6 that
$$\eqalignno{
&\Vert (D^l G)  (u ; u^{}_1,\ldots,u^{}_l) \Vert^{}_{E_L}& (6.187)\cr
&\quad{}\leq C_{l,L}	{\cal R}^l_{N_0,l+L} (u^{}_1,\ldots,u^{}_l) + C'_{L,l}
 \Vert u \Vert^{}_{E^\rho_{N_0+L+l}}  \Vert u^{}_1
\Vert^{}_{E^\rho_{N_0}}\cdots \Vert u^{}_l \Vert^{}_{E_{N_0}},\cr
}$$
for all $u \in {\cal O}'_\infty, u^{}_1,\ldots,u^{}_l \in E^\rho_\infty$,
$L \geq 0$, $l \geq 0$, for some constants $C_{l,L}$ and $C'_{l,L}$
depending only on $\Vert u \Vert^{}_{E^\rho_{N_0}}$. Since
$\Omega_1 (Qu) \in V^\rho_\infty$ for $u \in {\cal
O}'_\infty$ according to Theorem 6.12 it follows from Theorem 6.11 that
$$\eqalignno{
Q G(u) &= F (\Omega_1 (Qu)),&(6.188{\rm a}) \cr
(1-Q) G(u) &= (1-Q)u,\quad  u \in {\cal O}'_\infty ,\cr
}$$
which shows that
$$\eqalignno{
Q D G(u).v &= DF (\Omega_1 (Qu)).(D \Omega_1 (Qu).Qv), &(6.188{\rm b})\cr
 (1 - Q)D G(u).v &= (1-Q) v,\cr
}$$
$u \in {\cal O}'_\infty,  v \in E^\rho_\infty.$

We shall prove that $DG(u) \in L(E^\rho_\infty, E^\rho_\infty)$, $u \in {\cal
O}'_\infty$ has a right inverse $w'(u) \in L(E^\rho_\infty, E^\rho_\infty)$,
i.e. $DG (u). w'(u) v = v$ for $v \in E^\rho_\infty$ and $u \in {\cal
O}'_\infty$. For $w(u) \in L(E^{\circ\rho}_\infty, E^{\circ\rho}_\infty)$, $u
\in {\cal O}_\infty$ we define $w'(u) = w(Qu) Q + 1-Q$ for $u \in {\cal
O}'_\infty$. According to (6.188b) $w'(u)$ is then a right inverse of $DG(u)$,
$u \in {\cal O}'_\infty$ if and only if
$$DF (\Omega_1 (Qu)).(D \Omega_1 (Qu).w(Qu) Qv) = Qv, \quad u \in {\cal
O}'_\infty,  v \in E^\rho_\infty,$$
i.e. if
$$DF (\Omega_1 (u)).(D \Omega_1 (u).w(u) v) = v,\quad  u \in {\cal
O}_\infty,  v \in E^{\circ\rho}_\infty.$$
This equation and Theorem 6.11 give
$$D \Omega_1 (u).w(u) v = DF^{-1} (F(\Omega_1 (u))).v,\quad u \in {\cal
O}_\infty,  v \in E^{\circ\rho}_\infty, \eqno{(6.189)}$$
where $D \Omega_1 (u).w(u) v$ belongs to the tangent space of $V^\rho_\infty$
at $\Omega_1 (u)$. Let $H(u;v) =$
\penalty-1000$DF^{-1} (F(\Omega_1 (u))).v$. It follows from
Theorem 6.11 that $H(u;v) \in E^\rho_\infty$ is an element of the tangent space
of $V^\rho_\infty$ at the point $\Omega_1 (u)$. Theorem 6.11, Theorem 6.12 and
Corollary 2.6 give the following result, where $D$ is differentiation only with
respect to $u$:
$$\eqalignno{
&\Vert (D^l H)  (u ;v; u^{}_1,\ldots,u^{}_l) \Vert^{}_{E_L} &(6.190)\cr
&\quad{}\leq C_{l,L}\big({\cal R}^{l+1}_{N_0,L+l} (v,u^{}_1,\ldots,u^{}_l) +
\Vert u \Vert^{}_{E_{N_0+L+l}}	\Vert v \Vert^{}_{E_{N_0}}  \Vert u^{}_1
\Vert^{}_{E_{N_0}}\cdots \Vert u^{}_l \Vert^{}_{E_{N_0}},\big)\cr
}$$
for all $l \geq 0$, $L \geq 0$, where $C_{l,L}$ depends only on $\Vert u
\Vert^{}_{E_{N_0}}$ and where $N_0$, which we have chosen sufficiently large,
is independent of $l,L,u,v,u^{}_1,\ldots,u^{}_l.$

Since $H(u;v)$ belongs to the tangent space of $V^\rho_\infty$ at the point
$\Omega_1 (u)$ we can take $H(u;v) = (a(0), \dot{a} (0), \Psi (0))$ as initial
conditions for the derivative of the Maxwell-Dirac equations in the form
(1.12):
$${d \over dt} (a(t), \dot{a} (t), \Psi (t)) = DT_{P_0} (A(t), \dot{A} (t),
\psi(t)).(a(t), \dot{a} (t), \Psi (t)),\quad t \geq 0, \eqno{(6.191{\rm a})}$$
where
$${d \over dt} (A(t), \dot{A} (t), \psi (t)) = T_{P_0} (A(t), \dot{A} (t),
\psi(t)), \quad t \geq 0, \eqno{(6.191{\rm b})}$$
and $(A(t), \dot{A} (t), \psi (t))$ is given by Theorem 6.10 with $Y = \un$ and
$\psi (t) = e^{- i {\vartheta} (A,t)} \psi' (t)$ and where
$$(A(0), \dot{A} (0), \psi(0)) = \Omega_1 (u),\quad u \in {\cal O}_\infty.
\eqno{(6.191{\rm c})}$$
The variable substitution $\psi' (t) = e^{i {\vartheta} (A,t)} (\psi (t)$ gives
that the derivative $\Psi' (t)$ of $\psi' (t)$ satisfies
$$\Psi' (t) = e^{i {\vartheta} (A,t)}  \Psi (t) + i {\vartheta} (a,t)
  \psi' (t), \quad t \geq 0. \eqno{(6.192)}$$
It follows from (6.191a), (6.191b) and (6.192) that
$$\carre a_\mu = (\psi')^+  \gamma^0  \gamma_\mu
\Psi' + (\Psi')^+  \gamma^0  \gamma_\mu  \psi'
\eqno{(6.193{\rm a})}$$
and
$$(i  \gamma^\mu  \partial_\mu + m)  \Psi' =
(A_\mu + B_\mu)  \gamma^\mu  \Psi' + (a_\mu + b_\mu)
 \gamma^\mu  \psi', \eqno{(6.193{\rm b})}$$
where $B_\mu = - \partial_\mu  {\vartheta} (A)$ and $b_\mu = - {\vartheta}(a)$.
Moreover
$$\partial_\mu	A^\mu = 0\quad \hbox{ and }\quad \partial_\mu  a^\mu= 0.
\eqno{(6.193{\rm c})}$$

We shall solve the system (6.193a), (6.193b) and (6.193c) for initial
conditions \penalty-1000 $(a(0), \dot{a} (0), \Psi' (0))$ at $t=0$, defined by
$H(u;v)=(a(0), \dot{a} (0), \Psi (0))$ and formula (6.192).

Let
$$a_Y = \xi^M_Y a,\quad   \dot{a}_Y = \xi^M_{P_0 Y} a, \quad  \Psi'_Y =
\xi^D_Y  \Psi', \quad  \Psi_Y = \xi^D_Y \Psi \eqno{(6.194{\rm a})}$$
and
$${\vartheta}_Y (a) = \xi^{}_Y	{\vartheta} (a), \quad  b_Y = \xi^M_Y b,
\quad  Y \in U(\p). \eqno{(6.194{\rm b})}$$
Since the initial condition $H(u;v)$, where $u \in {\cal O}_\infty$, $v \in
E^{\circ\rho}_\infty$ for equation (6.193a) and (6.193b) is a function of $u$
and $v$, this is also the case for $a_Y, \dot{a}_Y$, $\Psi_Y$ and $\Psi'_Y$. We
introduce the notation
$$ a^{(l)}_Y\quad \hbox{(resp.} \dot{a}^{(l)}_Y, \Psi^{(l)}_Y,
\Psi'^{(l)}_Y),\quad  l \geq 0, \eqno{(6.195)}$$
for the $l^{\rm th}$ derivative of $a^{}_Y$ (resp.
$\dot{a}_Y, \Psi_Y, \Psi'_Y)$ with respect to $u$, in the directions
$u^{}_1,\ldots,u^{}_l \in E^{\circ\rho}_\infty.$

It follows from (6.192) and from Leibniz rule that
$$\eqalignno{
&\Psi'^{(l)}_Y (0)&(6.196)\cr
&\quad{} = \sum_{l_1+l_2=l}  C_{l_1,l_2}
\suma_{Y_1,Y_2}^Y  ((\xi^{}_{Y_1}  e^{i {\vartheta} (A)})^{(l_1)}
(0)  \Psi^{(l_2)}_{Y_2}
+ i {\vartheta}_{Y_1} (a,0)^{(l)}  \psi'^{(l)}_{Y_2} (0)), \quad
l \geq 0,  Y = \Pi',\cr
}$$
where $f^{(l)}$ is defined as in (6.195). Since $\Psi_Y (0)$ is the derivative
of $T^D_Y (A(0), \dot{A} (0), \psi (0))$ in the direction $(a(0), \dot{a} (0),
\psi (0))$, where $Y \in U(\p)$ and $T^D_Y$ is the Dirac component of $T_Y,$
it follows from i) of Lemma 2.19, Corollary 2.6 and from inequality (6.190)
that
$$\eqalignno{
&\Vert \Psi^{(l)}_Y (0) \Vert^{}_{D_L} &(6.197)\cr
&\ {}\leq C_{l, \vert Y \vert,L}
{\cal R}^{l+1}_{N_0, L+l+\vert Y \vert}  (v,u^{}_1,\ldots,u^{}_l)
+ C'_{l, \vert Y \vert,L}  \Vert u \Vert^{}_{E^\rho_{N_0+L+l+\vert Y
\vert}}  \Vert v \Vert^{}_{E^\rho_{N_0}}  \Vert u^{}_1
\Vert^{}_{E^\rho_{N_0}} \cdots\Vert u^{}_l \Vert^{}_{E^\rho_{N_0}},\cr
}$$
for $Y \in \Pi'$,  $L \geq 0$,  $l \geq 0$,  $u \in
{\cal O}_\infty$, $v, u^{}_1,\ldots,u^{}_l \in E^{\circ\rho}_\infty$.
 Here $C_{l, \vert Y\vert,L}$ and $C'_{l, \vert Y \vert,L}$ are constants
depending only on $\Vert u \Vert^{}_{E^\rho_{N_0}}$ and $N_0$ is
an integer independent of
$L$, $Y$, $l$, $u$, $v,u^{}_1,\ldots,u^{}_l$. It now follows by
Corollary 2.6 and as in the proof of
(6.175) that
$$\eqalignno{
&\Vert (\xi^{}_{Y_1}  e^{i {\vartheta} (A)})^{(l_1)} (0)
\Psi^{(l_2)}_{Y_2} (0) \Vert^{}_{D_L}&(6.198)\cr
&\ {}\leq C_{l, \vert Y \vert,L}  {\cal R}^{l+1}_{N_0, L+l+\vert Y
\vert}	(v,u^{}_1,\ldots,u^{}_l)
+ C'_{l, \vert Y \vert,L}  \Vert u \Vert^{}_{E^\rho_{N_0+L+l+\vert Y
\vert}}  \Vert v \Vert^{}_{E^\rho_{N_0}}  \Vert u^{}_1
\Vert^{}_{E^\rho_{N_0}}\cdots\Vert u^{}_l \Vert^{}_{E^\rho_{N_0}},\cr
}$$
for $Y_1,Y_2 \in \Pi',  \vert Y_1 \vert + \vert Y_2 \vert
\leq \vert Y \vert,  l \geq 0,	l_1+l_2=l,  L
\geq 0$, and some constants $C_{l, \vert Y \vert,L}$ and $C'_{l, \vert Y
\vert,L}$ and integer $N_0$ having the same properties as those in (6.197).
Similarly it is proved that the second term in the sum in (6.196) also satisfy
estimate (6.198). This gives together with (6.190) that
$$\eqalignno{
&\Vert (a^{(l)}_Y (0), \dot{a}^{(l)}_Y (0), \Psi'^{(l)}_Y (0))
\Vert^{}_{E^\rho_L}
&(6.199)\cr
&\ {}\leq C_{l, \vert Y \vert,L}  {\cal R}^{l+1}_{N_0, L+l+\vert Y
\vert}	(v,u^{}_1,\ldots,u^{}_l)
+ C'_{l, \vert Y \vert,L}  \Vert u \Vert^{}_{E^\rho_{N_0+L+l+\vert Y
\vert}}  \Vert v \Vert^{}_{E^\rho_{N_0}}  \Vert u^{}_1
\Vert^{}_{E^\rho_{N_0}}\cdots \Vert u^{}_l \Vert^{}_{E^\rho_{N_0}},\cr
}$$
for $Y \in \Pi',  L \geq 0,  l \geq 0,  u \in
{\cal O}_\infty, v, u^{}_1,\ldots,u^{}_l \in E^{\circ\rho}_\infty$.
 Here $C_{l, \vert Y \vert,L}$ and $C'_{l, \vert Y \vert,L}$ are constants
 depending only on $\Vert
u \Vert^{}_{E^\rho_{N_0}}$ and $N_0$ is an integer independent of $L, Y, l, u,
v,
u^{}_1,\ldots,u^{}_l$.

Next we shall derive estimates for equations (6.193a), (6.193b) and (6.193c).
Application of $\xi^M_Y, Y \in \Pi'$, to equation (6.193a) gives for $1/2 <
\rho' \leq 1$:
$$\eqalignno{
&\Vert (a_Y (t), \dot{a}_Y (t)) \Vert^{}_{M^{\rho'}_0}&(6.200)\cr
& \ {}\leq \Vert (a_Y (0),
\dot{a}_Y (0)) \Vert^{}_{M^{\rho'}_0}
+ \sum_{0 \leq \mu \leq 3}  \int^t_0	\Vert \vert \nabla
\vert^{\rho-1} \big(\xi^M_Y ((\psi')^+  \gamma^0
\gamma_\mu  \Psi' + (\Psi')^+  \gamma^0
\gamma_\mu  \psi')\big) (s) \Vert^{}_{L^2}  ds.\cr
}$$
Since $\Vert \vert \nabla \vert^{\rho'-1} f \Vert^{}_{L^2} \leq C_p
\Vert f \Vert^{}_{L^p}$,  $p = 6 (5-2 \rho')^{-1}$, and since
$$\Vert fg
\Vert^{}_{L^p} \leq \Vert f \Vert^{(1+2 \rho')/3}_{L^\infty}  \Vert f
\Vert^{(2-2 \rho')/3}_{L^2}  \Vert g \Vert^{}_{L^2},$$
we obtain:
$$\eqalignno{
&\Vert \vert \nabla \vert^{\rho'-1} \big(\xi^M_Y \big((\psi')^+  \gamma^0
 \gamma_\mu  \Psi' + \Psi'^+  \gamma^0
\gamma_\mu  \psi')\big) (s) \Vert^{}_{L^2}\cr
&\quad{}\leq C_{\vert Y \vert}  \sum_{\scr \vert Y_1 \vert + \vert Y_2 \vert
\leq \vert Y \vert\atop\scr  \vert Y_2 \vert < \vert Y \vert}  \Vert
\psi'_{Y_1} (s) \Vert^{(1+2\rho')/3}_{L^\infty}  \Vert \psi'_{Y_1}
(s) \Vert^{(2-2 \rho')/3}_{L^2}  \Vert \Psi'_{Y_2} (s) \Vert^{}_{L^2}\cr
&\qquad{}+C_0\Vert \psi'_{\un}(s)\Vert^{(1+2\rho')/3}_{L^\infty}
\Vert \psi'_{\un}(s)\Vert^{(2-2 \rho')/3}_{L^2}
\Vert \Psi'_{Y} (s) \Vert^{}_{L^2}.\cr
}$$
This inequality, inequality (6.200) and Theorem 6.10, give
$$\eqalignno{
&\Vert (a_Y (t), \dot{a}_Y (t)) \Vert^{}_{M^{\rho'}_0}&(6.201)\cr
& \quad{}\leq \Vert (a_Y (0),
\dot{a}_Y (0)) \Vert^{}_{M^{\rho'}_0}\cr
&\qquad{}+ \sum_{\scr L+\vert Y_2\vert\leq\vert Y\vert
\atop\scr \vert Y_2\vert< \vert Y\vert}
F_L(\Vert u\Vert^{}_{E^\rho_{N_0}})
\Vert u\Vert^{}_{E^\rho_{N_0+L}}
\int^t_0 (1+s)^{-(1+2\rho')/2}\Vert\Psi'_{Y_2} (s) \Vert^{}_{L^2}ds\cr
&\qquad{}+
F_0(\Vert u\Vert^{}_{E^\rho_{N_0}})
\Vert u\Vert^{}_{E^\rho_{N_0}}
\int^t_0 (1+s)^{-(1+2\rho')/2}\Vert\Psi'_{Y} (s) \Vert^{}_{L^2}ds,\cr
}$$
where $1/2<\rho'\leq 1$ and $Y\in\Pi'$, $1/2<\rho<1$. It follows from
(6.201) that
$$\eqalignno{
&\wp^{M^{\rho'}}_n (a (t), \dot{a} (t)) &(6.202)\cr
& \quad{}\leq \wp^{M^{\rho'}}_n (a (0), \dot{a} (0))
+C_n \sum_{0\leq i\leq n-1}
 \Vert u\Vert^{}_{E^\rho_{N_0+n-i}}
\int^t_0 (1+s)^{-(1+2\rho')/2}\wp^D_i(\Psi' (s))ds\cr
&\qquad{}+C_0\Vert u\Vert^{}_{E^\rho_{N_0}}
\int^t_0 (1+s)^{-(1+2\rho')/2}\wp^D_n(\Psi' (s))ds,
\quad n\geq 0,t\geq 0,\cr
}$$
where $C_n$, $n\geq 0$, are constants depending only on
$\Vert u\Vert^{}_{E^\rho_{N_0}}$.
Statement i) of Corollary 5.18, with $G_\mu=A_\mu+B_\mu$, $t_0=0$,
$\varepsilon=0$, $\eta=0$, $\rho'=\rho$, and equation (6.193b) give for
$n\geq0$, $1/2<\rho<1$, $t\geq0$:
$$\eqalignno{
&\wp^D_n(\Psi'(t))&(6.203{\rm a})\cr
&\quad{}\leq C_n \Big(\wp^D_n(\Psi'(0))
+\sum_{0\leq i\leq n-1}\Vert u\Vert^{}_{E^\rho_{N_0+n-i}}
\wp^D_i(\Psi' (0))
+ \sum_{0\leq i\leq n}\Vert u\Vert^{}_{E_{N_0+n-i}}\tau^{}_i(t)\big),\cr
}$$
where
$$\eqalignno{
\tau^{}_i(t)&=\wp^{M^{\rho}}_i (a (0), \dot{a} (0))
+\wp^{M^1}_i (a (0), \dot{a} (0))&(6.203{\rm b})\cr
&\quad{}+\int^t_0 (1+s)^{\rho-2}\Big(\sup_{0\leq s'\leq s}
\big(\wp^{M^{\rho}}_i (a (s'), \dot{a} (s'))
+\wp^{M^1}_i (a (s'), \dot{a} (s'))\big)\cr
&\quad{}+\sum_{\scr Y\in \Pi'\atop\scr \vert Y\vert\leq i }
\sup_{0\leq s'\leq s}\big((1+s')^{3/2-\rho}
\Vert \carre a^{}_{Y}(s')\Vert^{}_{L^2(\Rrm^3,\Rrm^4)}\big)\Big)ds\cr
}$$
and where $C_n$, $n\geq0$, depends only on $\Vert u\Vert^{}_{E^\rho_{N_0}}$.
Estimating $\Vert(1+t+\vert\cdot \vert)
\carre a^{}_{Y}(t)\Vert^{}_{L^2(\Rrm^3,\Rrm^4)}$, by using equation (6.193a)
and Theorem 6.10, it follows from (6.202), (6.203a) and (6.203b), using
Corollary 2.6 and denoting
$$e^{}_n(t)=\sup_{0\leq s\leq t}\big(
\wp^{E^{\rho}}_n (a (s), \dot{a} (s),\Psi'(s))\big)
+\sup_{0\leq s\leq t}\big(
\wp^{E^1}_n (a (s), \dot{a} (s),\Psi'(s))\big),$$
that:
$$\eqalignno{
e^{}_n(t)&\leq C_n\Big( e^{}_n(0)+\sum_{0\leq i\leq n-1}
\Vert u\Vert^{}_{E^\rho_{N_0+n-i}}e^{}_i(0)\Big)&(6.204)\cr
&\quad{}+C_n \sum_{0\leq i\leq n-1}\Vert u\Vert^{}_{E^\rho_{N_0+n-i}}
\int_0^t (1+s)^{-\varepsilon}e^{}_i(s)ds\cr
&\quad{}+C_n\Vert u\Vert^{}_{E^\rho_{N_0}}
\int_0^t (1+s)^{-\varepsilon}e^{}_n(s)ds,\quad n\geq0,t\geq0,\cr
}$$
where $\varepsilon=\min(2-\rho,Y_2+\rho)>1$ and where $C_n$ depends only
on $\Vert u\Vert^{}_{E^\rho_{N_0}}$. It follows by
Gr\"onvall inequality that
$$\eqalignno{
e^{}_n(t)&\leq C'_n\Big( e^{}_n(0)+\sum_{0\leq i\leq n-1}
\Vert u\Vert^{}_{E^\rho_{N_0+n-i}}e^{}_i(0)&(6.205)\cr
&\quad{}+\sum_{0\leq i\leq n-1}\Vert u\Vert^{}_{E^\rho_{N_0+n-i}}
\int_0^t (1+s)^{-\varepsilon}e^{}_i(s)ds\Big)\quad, t\geq 0,n\geq 0,\cr
}$$
where $C'_n$, $n\geq0$, are constants depending only on
$\Vert u\Vert^{}_{E^\rho_{N_0}}$. We obtain from (6.205) by induction and
using Corollary 2.6 that
$$e^{}_n(t)\leq C_n\Big( e^{}_n(0)+\sum_{0\leq i\leq n-1}
\Vert u\Vert^{}_{E^\rho_{N_0+n-i}}e^{}_i(0)\Big),\quad t\geq0,n\geq 0,
\eqno{(6.206)}$$
for some constants depending only on $\Vert u\Vert^{}_{E^\rho_{N_0}}$
and for some integer $N_0$ independent of $n$, $t$, $u$.
Combining (6.149) and (6.206) we obtain using Corollary 2.6 that
the solution of (6.193a), (6.193b) and (6.193c) with initial conditions
$(a (0), \dot{a} (0),\Psi'(0))$ sastify
$$\wp^{E^{\rho}}_n (a (t), \dot{a} (t),\Psi'(t))
\leq C_n \Vert v\Vert^{}_{E^\rho_{N_0+n}}
+C'_n \Vert u\Vert^{}_{E^\rho_{N_0+n}}\Vert u\Vert^{}_{E^\rho_{N_0}},
\quad t\geq0,n\geq 0,\eqno{(6.207)}$$
where $C_n$, $C'_n$, are constants depending only on
$\Vert u\Vert^{}_{E^\rho_{N_0}}$ and where $N_0$ is an integer independent of
$n$, $t$, $u$.

We can now use directly Corollary 5.2, with $w(t,s)=e^{{\cal D}(t-s)}$,
on the different terms on the right-hand side of the equation
$$(i\gamma^\mu\partial_\mu +m )\Psi'_Y=
\xi^D_Y\big((A_\mu+B_\mu)\gamma^\mu\Psi'+(a^{}_\mu+b^{}_\mu)\gamma^\mu\psi'\big),
\quad Y\in\Pi',\eqno{(6.208)}$$
which proves that there exists $\beta'\in D^{}_\infty$ such that
$$\lim_{t\fl\infty}\Vert\Psi'_Y(t)-e^{{\cal D}t}T^{D1}_Y(\beta')
\Vert^{}_D=0.\eqno{(6.209)}$$
Since the integral in (6.200) converges as $t\fl\infty$, it follows
using (6.209) that there exists $(g,\dot{g})\in M^\rho_\infty$ such that
$$\lim_{t\fl\infty}\Vert(a (t), \dot{a} (t),\Psi'(t))
-U^1_{\exp(tP_0)}T^{1}_Y(v')\Vert^{}_{M^\rho_0}=0,\eqno{(6.210)}$$
where $v'=(g,\dot{g},\beta)\in E^\rho_\infty$. It follows from (6.207)
and (6.210) that
$$\Vert v'\Vert^{}_{E^\rho_{N_0}}
\leq C_n \Vert v\Vert^{}_{E^\rho_{N_0+n}}
+C'_n \Vert u\Vert^{}_{E^\rho_{N_0+n}}\Vert v\Vert^{}_{E^\rho_{N_0}},
\quad n\geq 0,\eqno{(6.211)}$$
where $C_n$ and $C'_n$ are constants depending only on
$\Vert u\Vert^{}_{E^\rho_{N_0}}$ and where $N_0$ is an integer independent of
$n$ and $u$.

We now define $w(u)v=v'$, $u\in {\cal O}_\infty$, which proves that
equation (6.189) has a solution $w(u)$ with the property (6.209).
Differentiation of equations (6.193a), (6.193b) and (6.193c) with respect
to $u$ in the direction $u^{}_1,\ldots,u^{}_l\in E^{\circ\rho}_\infty$ and
induction give in the same way as (6.211) was obtained that
$$\eqalignno{
&\Vert(D^l (w(u)v))(u^{}_1,\ldots,u^{}_l)\Vert^{}_{E^\rho_{n}}&(6.212)\cr
&\quad{}\leq C_{l,n}{\cal R}^{l+1}_{N_0,l+n}(v,u^{}_1,\ldots,u^{}_l)
+C'_{l,n}\Vert u\Vert^{}_{E^\rho_{N_0+l+n}}
\Vert v\Vert^{}_{E^\rho_{N_0}}
\Vert u^{}_1\Vert^{}_{E^\rho_{N_0}}\cdots
\Vert u^{}_l\Vert^{}_{E^\rho_{N_0}},\cr
}$$
for $l\geq0$, $n\geq0$, $u\in {\cal O}_\infty$,
$v$, $u^{}_1,\ldots,u^{}_l\in E^{\circ\rho}_\infty$.
Here  $C_{l,n}$ and  $C'_{l,n}$ are constants depending only on
$\Vert u\Vert^{}_{E^\rho_{N_0}}$ and $N_0$ is an integer independent of
$l$, $n$, $u$, $v$, $u^{}_1,\ldots,u^{}_l$.
By construction of $w'(u)$, $u\in {\cal O}'_\infty$, it now follows that
$FG(u).w'(u)v=v$ for $v\in E^{\rho}_\infty$ and it follows
from (6.212) that
$$\eqalignno{
&\Vert(D^l (w'(u)v))(u^{}_1,\ldots,u^{}_l)\Vert^{}_{E^\rho_{n}}&(6.213)\cr
&\quad{}\leq C_{l,n}{\cal R}^{l+1}_{N_0,l+n}(v,u^{}_1,\ldots,u^{}_l)
+C'_{l,n}\Vert u\Vert^{}_{E^\rho_{N_0+l+n}}
\Vert v\Vert^{}_{E^\rho_{N_0}}
\Vert u^{}_1\Vert^{}_{E^\rho_{N_0}}\cdots
\Vert u^{}_l\Vert^{}_{E^\rho_{N_0}},\cr
}$$
for $l\geq0$, $n\geq0$, $u\in {\cal O}'_\infty$,
$v$, $u^{}_1,\ldots,u^{}_l\in E^{\rho}_\infty$.
Here  $C_{l,n}$, $C'_{l,n}$ and $N_0$ have the same properties as in (6.212).

Theorem 2.5,  property (6.187) of the map $G\colon{\cal O}'_\infty\fl
E^{\rho}_\infty$ and the existence of a right inverse $w'(u)\in
L(E^{\rho}_\infty,E^{\rho}_\infty)$ of $DG(u)\in
L(E^{\rho}_\infty,E^{\rho}_\infty)$ with  property (6.123) prove,
according to the implicit function theorem for Fr\'echet spaces
(Theorem 4.1.1 of \refSERG), that there exists a positive integer $M_0$,
an open neighbourhood ${\cal U}'_{M_0}$ of zero in $E^\rho_{M_0}$ and
a $C^\infty$ map $H\colon {\cal U}'_{\infty}={\cal U}'_{M_0}\cap
E^{\rho}_\infty\fl {\cal O}'_\infty$ such that $G(H(u))=u$
for $u\in {\cal U}'_{\infty}$. Let ${\cal Q}'_{\infty}=G^{-1}
[{\cal U}'_{\infty}]$ be the inverse image of ${\cal U}'_{\infty}$ by $G$.
Since $G$ is continuous and ${\cal U}'_{\infty}$ is an open neignbourhood of
zero in $E^{\rho}_\infty$ and since $G(0)=0$, it follows that
${\cal Q}'_{\infty}$ is an open neignbourhood of zero in $E^{\rho}_\infty$.
The map $G\colon{\cal Q}'_{\infty}\fl{\cal U}'_{\infty}$ is onto
because $H({\cal U}'_{\infty})\subset {\cal Q}'_{\infty}$ and $G(H(u))=u$
for $u\in{\cal U}'_{\infty}$, and according to Theorem 6.11 and
Theorem 6.12 $G$ is also one-to-one. Since
$G\colon{\cal Q}'_{\infty}\fl{\cal U}'_{\infty}$ is a bijection
and $G(H(u))=u$ for $u\in{\cal U}'_{\infty}$, it follows that
$H\colon {\cal U}'_{\infty}\fl {\cal O}'_\infty$ is a bijection. This
proves that the $C^\infty$ function $G\colon{\cal Q}'_\infty\fl
{\cal U}'_{\infty}$ has a $C^\infty$ inverse
$G^{-1}\colon{\cal U}'_{\infty}\fl{\cal Q}'_{\infty}$.

Similarly, as we obtained (6.213), we now obtain that
$$\eqalignno{
&\Vert(D^lG^{-1}) (u;u^{}_1,\ldots,u^{}_l)\Vert^{}_{E^\rho_{N}}&(6.214)\cr
&\quad{}\leq C_{l,n}{\cal R}^{l}_{M_0,l+n}(u^{}_1,\ldots,u^{}_l)
+C'_{l,n}\Vert u\Vert^{}_{E^\rho_{M_0+l+n}}
\Vert u^{}_1\Vert^{}_{E^\rho_{M_0}}\cdots
\Vert u^{}_l\Vert^{}_{E^\rho_{M_0}},\cr
}$$
for $l\geq0$, $n\geq0$, $u\in {\cal U}'_\infty$,
$u^{}_1,\ldots,u^{}_l\in E^{\rho}_\infty$.
Here  $C_{l,n}$ and  $C'_{l,n}$ are constants depending only on
$\Vert u\Vert^{}_{E^\rho_{M_0}}$ and $M_0$ is an integer independent of
$l$, $n$, $u$, $u^{}_1,\ldots,u^{}_l$.

Let us define ${\cal U}_\infty=F^{-1}({\cal U}'_\infty\cap
E^{\circ\rho}_\infty)$
and let us redefine ${\cal O}_\infty$ by ${\cal O}_\infty=
{\cal Q}'_\infty\cap E^{\circ\rho}_\infty$. According to definition
(6.186) of $G$, Theorem 6.11 and the fact that
$G\colon{\cal Q}'_{\infty}\fl{\cal U}'_{\infty}$ is a
diffeomorphism, it follows that
$\Omega_1\colon{\cal O}_{\infty}\fl{\cal U}_{\infty}$ is a
diffeomorphism, which satisfies the inequality of the theorem
since $G^{-1}$ satisfies inequality (6.214). This proves the theorem.

The construction of a modified wave operator $\Omega_1\colon {\cal O}_\infty
\fl {\cal U}_\infty $, for $t\rightarrow \infty$, being a diffeomorphism
according to Theorem 6.13, could of course as well has been done for
$t\rightarrow -\infty $. To distinguish between the {\it two modified wave
operators} so constructed, we use the notation
$\Omega ^{(\varepsilon )}_1\colon {\cal O}_{1,\infty (\varepsilon )}
\fl {\cal U}_{\infty (\varepsilon )}$ for
the wave operator and $A_{(\varepsilon )} (u), \psi'_{(\varepsilon )}(u)$ for
the functions given by definition (6.165), with $u\in {\cal O}_{1,\infty
(\varepsilon)}$, for the case $t\fl\varepsilon\infty$,
$\varepsilon = \pm $. ${\cal O}_{1,\infty (\varepsilon )}$
and ${\cal U}_{\infty (\varepsilon )}$ are given by Theorem 6.13, with
$N_\varepsilon $ and $M_\varepsilon $ instead of $N_0$ and $M_0$ respectively.
By definition we then have
$$\Omega^{(\varepsilon )}_1 (u)=\big((A_{(\varepsilon )}(u)) (0),
 ( {\dot A}_{(\varepsilon )}(u) )(0), e^{-i\vartheta (A_{(\varepsilon
)}(u),0 )}  (\psi'_{(\varepsilon )} (u) )(0)\big),\quad \varepsilon =\pm, \eqno
{(6.215)}$$
which should be compared with (6.172). Theorem 6.10 is then
true, after the obvious modification that $ (A_{0,Y}(t), {\dot
A}_{0,Y}(t), \phi'_{0,Y}(t) )$ is replaced by $U^1_{{\exp} (tP_0)}
T^1_Y(u)$, $(A_Y, {\dot A}_Y, \psi'_Y)$ is replaced by $(A_{(\varepsilon
)Y}, {\dot A}_{(\varepsilon )Y}, \psi'_{(\varepsilon )Y})$. $N_0$ is replaced
by $N_\varepsilon $ and $t\geq 0$ is replaced by $\varepsilon t \geq 0$. This
gives the following corollary:
\saut
\noindent{\bf Corollary 6.14.}
{\it
The function $u\longmapsto   (A_{(\varepsilon )} (u),{\dot A}_\varepsilon
(u),\psi'_{(\varepsilon )}(u)  )$ satisfies the conclusions of Theorem
6.10 for $t\fl\varepsilon \infty $ and
$\Omega^{(\varepsilon )}_1 \colon
{\cal O}_{1,\infty (\varepsilon )}\fl{\cal U}_{\infty (\varepsilon )}$
satisfies the conclusions of Theorem 6.13, where $\varepsilon =\pm$.
}\saut
Corollary 6.14 permits to {\it solve the Cauchy problem for the
Maxwell-Dirac equations} (1.1.a)--(1.1.c) for times $t\in {\Rrm}$
and intial conditions $v$
belonging to the open neighbourhood ${\cal U}_{\infty (0)} ={\cal U}_{\infty
(+)} \cap {\cal U}_{\infty (-)}$ of zero in $V^\rho _\infty $. We note that it
follows using Corollary 6.14 and the notation proceeding it, that ${\cal
U}_{\infty (0)}={\cal U}_{M_0}\cap V^\rho _\infty$, where $M_0={\max
}(M_+,M_-)$ and ${\cal U}_{M_0}={\cal U}_{M_+}\cap {\cal U}_{M_-}$ is a
neighbourhood of zero in $V^\rho_{M_0}$. Let $\Omega ^{(\varepsilon )}_1\colon
{\cal O}_{1,\infty (\varepsilon )}\fl {\cal U}_{\infty (0)}$,
$\varepsilon = \pm$, be the diffeomorphism, which is the restriction of
the former
$\Omega^{(\varepsilon )}_1$ to ${\cal O}_{1,\infty (\varepsilon )} =
\big(\Omega ^{(\varepsilon )}_1  \big)^{-1}[{\cal U}_{\infty (0)}]$.
We recall that for differentiation of functions defined on
$V^\rho _\infty$, that $V^\rho _\infty$
is diffeomorphic to $E^{\circ\rho}_\infty$, according to Theorem 6.11.
\saut
\noindent{\bf Theorem 6.15.}
{\it
Let $1/2<\rho <1$. If $v=(f,{\dot f},\alpha )\in {\cal U}_{\infty (0)}$, then
there exists a unique solution $h(v)=(A,{\dot A},\psi )\in C^0   ({\Rrm},
(1-\Delta )^{-1/4}E^1_0  )\cap C^1  ({\Rrm},(1-\Delta
)^{1/4}E^1_0  )$ of the M-D equations (1.1a)--(1.1c), with initial data
$v$ at $t=0$. Moreover $h\in C^\infty   ({\cal U}_{\infty (0)},C^\infty_b
({\Rrm},V^\rho_\infty)  )$, where $b$ stands for the topology of
convergence on bounded subsets of ${\Rrm}$, the conclusion of Theorem 6.10,
with $  (A_{0,Y}(t), \dot{A}_{0,Y} (t),\phi'_{0,Y}(t)  )$
replaced by $U^1_{\exp(tP_0)} T^1_Y(u^{}_\varepsilon )$, $u$ replaced by
$u^{}_\varepsilon =  \big(\Omega ^{(\varepsilon )}_1  \big)^{-1}(v)$, $N_0$
replaced by $N_\varepsilon $ and $t\geq 0$ replaced by $\varepsilon t \geq 0$,
is true for $\varepsilon = \pm 1$ and if $h^{(l )}(t)$ is the
$l^{\rm th}$ derivative of the function $v\mapsto (h(v))(t)$ at
$v\in {\cal U}_{\infty (0)}$ in the
directions of the elements $v^{}_1,\cdots ,v^{}_l $ of the tangent space of
$V^\rho_\infty $ at $v$, then there exists $N\geq 0$ such that
$$\Vert h^{(l )}(t) \Vert_{E^\rho _n} \leq C_{n+l ,t}  \big({\cal R}^l_{N,n+l}
(v^{}_1,\cdots,v^{}_l) +\Vert v\Vert_{E^\rho _{N+n+l}}
\Vert v^{}_1 \Vert_{E^\rho_N }\cdots
\Vert v^{}_l \Vert_{E^\rho _N}\big),$$
for $t\in {\Rrm}$, $ n$, $l \in {\Nrm}$, where $C_{n+l ,t}$ depends only on
$\rho $ and $\Vert v \Vert^{}_{E^\rho_N}$.
}\saut
\noindent{\it Proof.}
Let $g^{(l )}_Y$ be the $l^{\rm th}$ derivative of
the function $v\mapsto g^{}_Y (v)=  (A_Y, A_{P_0Y}, \xi^D_Y
(e^{i\vartheta(A)}\psi )  )$, $Y\in \Pi'$. It then follows by definition
(6.215)
of $\Omega_{1,(\varepsilon )}$ and by applying Theorem 6.10 and Corollary 6.14
with $u^{}_\varepsilon = \Omega ^{-1}_{1(\varepsilon )}(v)$ that $g^{(l )}_Y
\in
C^0({\Rrm},E^\rho _0)$, for $Y\in \Pi'$ and $l \geq 0$ and that $g^{}_{\un}(v)$
is a solution of equations (6.167a)--(6.176c).

We prove that $h^{(0)}_Y=T_Y(h(v))\in C^0({\Rrm},E^{\rho}_0)$ for $Y\in \Pi'$.
Let $V_Y=(A_Y, A_{P_0Y},\psi_Y)$, where $\psi_Y=\xi ^D_Y\psi$,
$g^{}_Y (v)=  (A_Y, A_{P_0Y},\psi'_Y)$, $\psi'_Y=\xi^D_Y\psi'$,
$\psi=e^{-i\vartheta(A)}\psi'$. It then follows that
$$\eqalignno{
&{\wp}_n^{E^\rho }(V(t))&(6.216)\cr
&\quad{}\leq {\wp}_n^{E^\rho }\big((g(v))(t)\big)
+C_n\sum \Vert \vartheta_{Y_1} (A,t)\cdots \vartheta_{Y_l}(A,t) \psi'_{Z} (t)
\Vert^{}_D,\quad n\geq 0, t\in {\Rrm},\cr
}$$
where $\vartheta_{Y_i} (A,t)=(\xi^{}_{Y_i}\vartheta(A))(t)$,
$C_n$ is a numerical constant and the sum is taken over $1 \leq l \leq n$,
$Y_i\in \Pi'$, ${Z}\in \Pi'$, $\vert Y_1\vert +\cdots +\vert Y_l \vert +\vert
{Z}\vert \leq n$, $\vert Y_i \vert \geq 1$. It follows from Lemma 4.4 and
Theorem 6.10 that
$$\Vert (\delta (t))^{1/2 -\rho } \vartheta_{Y_i} (A,t)\Vert^{}_{L^\infty }
 \leq C_{\vert Y_i\vert } \Vert
u^{}_\varepsilon \Vert^{}_{E^\rho _{N_0+\vert Y_i\vert }},\eqno{(6.217)}$$
where $u^{}_\varepsilon =\Omega ^{-1}_{1(\varepsilon )}(v)$ and where
$C_{\vert Y_i\vert }$ depends only on $\rho $ and $\Vert u^{}_\varepsilon
\Vert^{}_{E^\rho _{N_0}}$. Inequalities (6.216) and (6.217) give,
similarly as in the proof of Theorem 6.12 (see (6.174)--(6.175)) that
$${\wp}^{E^\rho }_n (V(t)) \leq {\wp}^{E^\rho}_n
((g(v))(t))
+C_n \sum_{{1\leq l \leq n }\atop {j\leq n-l }} \Vert u _\varepsilon
\Vert^{}_{E^\rho _{N_0+1+n-\vert {Z} \vert -l }} {\wp}^D_j \big((\delta(t))^{l
(\rho -1/2)} \psi'(t)\big),\eqno{(6.218)}$$
where $C_n$ depends only on $\rho $ and $\Vert u^{}_\varepsilon
\Vert^{}_{E^\rho _{N_0+1}}$. Since $(\delta (t)) (x) \leq
C(1+(\lambda_1(t))(x)+\vert t\vert)$, where $C$ is independent of $t$
and $x$, and
since $0 < \rho -1/2 < 1/2$, it follows from Theorem 6.10 that
$${\wp}^D_j \big((\delta (t) )^{l (\rho -1/2)}  \psi' (t)  \big)
\leq (1+t)^{l (\rho -1/2)} C_{j+l }  \Vert u^{}_\varepsilon
\Vert^{}_{E^\rho_{N_0+j+l }} ,\quad j,l \geq 0 ,$$
where $C_{j+l }$ depends only on $\rho $ and $\Vert u^{}_\varepsilon
\Vert^{}_{E^\rho _{N_0}}$. This inequality and inequality (6.218) give that
$${\wp}_n^{E^\rho } (V(t))\leq {\wp}_n^{E^\rho }((g(v))(t))
+(1+t)^{n(\rho -1/2)} C_n \Vert u^{}_\varepsilon \Vert^{}_{E^\rho _{N_0+1}}
\Vert u^{}_\varepsilon \Vert^{}_{E^\rho _{N_0+1+n}} ,\quad n\geq 0,$$
where $C_n$ depends only on $\rho $ and $\Vert u^{}_\varepsilon
\Vert^{}_{E^\rho_{N_0+1}}$. Estimating the first term on the right-hand
side of this inequality by Theorem 6.10 then gives that
$${\wp}^{E^\rho}_n   (V(t)  )\leq C_n (1+t)^{n(\rho -1/2)} \Vert
u^{}_\varepsilon \Vert^{}_{E^\rho_{N_0+1+n}} ,\quad n \geq 0, \eqno{(6.219)}$$
where $C_n$ depends only on $\rho $ and $\Vert u \Vert^{}_{E^\rho _{N_0+1}}.$
Choosing $N$ sufficiently large, it follows from Theorem 6.13 that
$${\wp}^{E^\rho}_n  (V(t)  ) \leq C_n (1+t)^{n(\rho -1/2)} \Vert v
\Vert^{}_{E^\rho_{N+n}} ,\quad n\geq 0, t\in {\Rrm}, \eqno{(6.220)}$$
where $C_n$ depends only on $\Vert v \Vert^{}_{E^\rho_N}$ and $\rho $. Let
$Y(t)= {\exp }(t{\rm ad}_{P_0})Y$, as in definition (1.11). Then $V_Y
(t)=T_{Y(t)}  ((h(v))(t) )$, which together with inequalities (6.220)
proves that
$$\eqalignno{
\Vert T_Y  ( (h(v))(t)  ) \Vert^{}_D
&\leq C_{\vert Y\vert } (1+t)^{\vert
Y\vert }\Vert T_{Y(t)}   ((h(v))(t)  )\Vert^{}_D &(6.221)\cr
&\leq C'_{\vert Y\vert} (1+t)^{\vert Y\vert (\rho +1/2)}
\Vert v \Vert^{}_{E^\rho_{N+\vert Y \vert }} ,\quad Y\in \Pi' , t\in {\Rrm},\cr
}$$
where $C'_{\vert Y\vert }$ depends only on $\rho$ and $\Vert u
\Vert^{}_{E^\rho_N}$. This proves that
$${\wp}^{E^\rho }_n   (h^{(0)}(t)  )\leq C_n (1+t)^{n(\rho +1/2)}
\Vert v \Vert^{}_{E^\rho _{N+n}} ,\quad n\geq 0, t\in {\Rrm} ,\eqno{(6.222)}$$
where $C_n$ depends only on $\rho $ and $\Vert v \Vert^{}_{E^\rho_N}.$

According to inequality (6.222), $\Vert h^{(0)}_{\un} (t) \Vert^{}_{E^\rho_0}
\leq C_0 \Vert v \Vert^{}_{E^\rho_N}$, which shows that we can choose ${\cal
U}_{\infty (0)}$ such that the hypothesis of Theorem 2.22 is verified.
Statements ii) and iii) of Theorem 2.22 give that
$$\eqalignno{
\Vert h^{(0)}_{\un}(t)\Vert^{}_{E^\rho _1}&\leq C_1 (1+t)^{\rho +1/2} \Vert
v \Vert^{}_{E^\rho_{N+1}} ,\quad t \in {\Rrm} &(6.223{\rm a})\cr
\noalign{\noindent and then that}
\Vert h^{(0)}_{\un}(t)\Vert^{}_{E^\rho _n}&\leq C_{n,t} \Vert v
\Vert^{}_{E^\rho_{N+n}} ,\quad n \geq 1 , t\in {\Rrm} ,&(6.223{\rm b})\cr
}$$
where $C_1$ and $C_{n,t}$ depend only on $\Vert v \Vert^{}_{E^\rho_{N+1}}$.
It follows from inequality (6.223a) and (6.223b) and from statement i) of
Theorem 2.22 that
$$\Vert h^{(0)}_Y (t)\Vert^{}_{E^\rho_n}\leq C_{n,\vert Y\vert ,t} \Vert v
\Vert^{}_{E^\rho_{N+n+ \vert Y \vert }} ,\quad n \in {\Nrm}, Y \in \Pi',
t \in {\Rrm},
\eqno{(6.224)}$$
where $C_{n, \vert Y \vert ,t}$ depends only on
$\Vert v \Vert^{}_{E^\rho_{N+1}}$.
This proves that $h^{(0)}_Y \in C^0({\Rrm}, E^\rho _\infty )$ for $Y \in \Pi'$.

Since $g^{}_{\un}(v)$ is a solution of equations (6.167a)--(6.167c) it follows
that $h^{(0)}_{\un}$ is a solution of the M-D equations, i.e. $(d/dt)
h^{(0)}_{\un}(t)=T_{P_0}  ( h^{(0)}_{\un}(t)  )$. Equations (1.10)
then gives that $(d/dt)^k  h^{(0)}_{\un}(t)=T_{P_0^{k+1}}   (h^{(0)}_{\un}(t))
=h^{(0)}_{P_0^{k+1}} (t)$, which proves that $h^{(0)}_{\un}=h(v) \in
C^\infty ({\Rrm},V^\rho_\infty )$. This proves the theorem for the case of
$l =0$. Since the case $l \geq 1$ is so similar, we omit it.

According to Theorem 6.15 and Theorem 6.10 the solution $h(v)$, $v \in  {\cal
U}_{\infty (0)}$, satisfies the asymptotic conditions
$$  \Vert   (A(t), {\dot A}(t)  )-U^{M1}_{{\rm exp }(tP_0)}
(f_{1 (\varepsilon )}, {\dot f}_{1(\varepsilon )}   )  \Vert^{}_{M_0^\rho }
+  \Vert \psi (t)-e^{-i\vartheta(A,t)} U^{D1}_{{\exp }(tP_0)}
\alpha_{1(\varepsilon )}  \Vert^{}_{D_0} \fl 0, \eqno{(6.225)}$$
when $\varepsilon t \fl\infty$, $\varepsilon =Ê\pm$, where
$u^{}_{1(\varepsilon )}=  (f_{1(\varepsilon )}, {\dot f}_{1(\varepsilon )},
\alpha_{1(\varepsilon )}   )=  (\Omega ^{(\varepsilon )}_1  )^{-1}
(v).$

To obtain {\it asymptotic representations of the Poincar\'e group,}
which do not require the integration of the M-D equations for their
construction, we shall replace (6.225) by
$$\Vert   ( A(t), {\dot A} (t)  )-U^{M1}_{{\exp  } (tP_0)}
(f_\varepsilon ,{\dot f}_\varepsilon )\Vert^{}_{M^\rho _0} +\Vert \psi
(t)-e^{-i\vartheta(A^{(\varepsilon )},t)}U^{D1}_{{\rm exp }(tP_0)}
\alpha _\varepsilon
\Vert^{}_{D_0}\fl0 , \eqno{(6.226)}$$
when $\varepsilon t\fl\infty $, $\varepsilon = \pm$, where
$A^{(+)}$ is given by (1.22a), with $(f_+, {\dot f}_+,
\alpha_+)$ instead of $(f,{\dot f},\alpha )$, where $f_+=f_{1(+)},  {\dot
f}_+={\dot f}_{1(+)}$ and where we shall determine $\alpha_+$. Similar
relations are valid for the case $\varepsilon = -$, and we shall only state
the following results for $\varepsilon =+$. We recall that
the function $\chi^{}_0$ in formula (1.22a) is given by
$\chi^{}_0 =1$ in this chapter.
To state next proposition let
$$\eqalignno{
E_+  (f_{1(+)}, {\dot f}_{1(+)}, \alpha_{1(+)}  )&=  (f_{1(+)},
{\dot f}_{1(+)}, \alpha _+  ),&(6.227{\rm a})\cr
{\hat \alpha }_+(k)&=\sum_{\varepsilon = \pm } e^{i\vartheta^\infty (A^{(+)}-A,
(\omega (k), -\varepsilon k))} P_\varepsilon (k){\hat \alpha }_{1(+)}(k),
&(6.227{\rm b})\cr
}$$
for $u^{}_{1(+)}=  (f_{1(+)}, {\dot f}_{1(+)}, \alpha _{1(+)}  ) \in
{\cal O}_{1,\infty (+)}$, where $h(v)=(A,{\dot A},\psi )$ is the solution
 of the M-D equations given by Theorem 6.15 with $v=
 (\Omega_1^{(+)} )^{-1}(u^{}_{1(+)})$,
where $\vartheta^\infty $ is given by  (1.23b) and where
$A^{(+)}=A^{(+)}(u^{}_{1(+)})$ is given by (1.22a) with
$u^{}_{1(+)}$ instead of $(f_+, {\dot f}_+, \alpha_+)$.
\saut
\noindent{\bf Proposition 6.16.}
{\it
Let $1/2<\rho <1$. One can choose the open neighbourhood ${\cal O}_{1,\infty
(+)}$ of zero in $E^{\circ\rho}_\infty$ such that $E_+$ is a
diffeomorphism of ${\cal O}_{1,\infty (+)}$ onto an open neighbourhood ${\cal
O}_{\infty (+)}$ of zero in $E^{\circ\rho}_\infty $, such that there
exists $N_+\in {\Nrm}$ and such that:
$$\eqalignno{
\hbox{\rm i)}\
&\Vert (D^l E_+)(u ; u^{}_1,\cdots , u^{}_l )\Vert^{}_{E^\rho_n}\cr
&\quad{}\leq C_{n+l}
\big({\cal R}^l_{N_+,n+l } (u^{}_1,\cdots ,u^{}_l)
+\Vert u\Vert^{}_{E^\rho _{N_++n+l }} \Vert u^{}_1\Vert^{}_{E^\rho_{N_+}}\cdots
\Vert u^{}_l \Vert^{}_{E^\rho_{N_+}}  \big),\hskip29.7mm\cr
}$$
for $n$, $l \in {\Nrm}$, $u\in {\cal O}_{1,\infty (+)},u^{}_1,\cdots, u^{}_l
\in
E^{\circ\rho}_\infty $, where $C_{n+l }$ depends only on $\rho$ and
$\Vert u \Vert^{}_{E^\rho_{N_+}}$,
$$\eqalignno{
\hbox{\rm ii)}\
&\Vert (D^l E^{-1}_+) (u ; u^{}_1,\cdots ,u^{}_l )\Vert^{}_{E^\rho_n}\cr
&\quad{}\leq
C_{n+l } \big({\cal R}^l_{N_+,n+l} (u^{}_1,\cdots ,u^{}_l )
+ \Vert u \Vert^{}_{E^\rho _{N_++n+l }} \Vert u \Vert^{}_{E^\rho _{N_+}}\cdots
\Vert u^{}_l \Vert^{}_{E^\rho_{N_+}}\big),\hskip30.4mm\cr
}$$
for $n$, $l \in {\Nrm}$, $u\in {\cal O}_{\infty (+)}$, $u^{}_1,\ldots ,u^{}_l
\in E^{\circ\rho}_\infty$, where $C_{n+l }$ depends only on $\rho$
and $\Vert u \Vert^{}_{E^\rho_{N_+}}$,
\psaut
\noindent\hbox{\rm iii)} Let $\chi^{}_0 \in C^\infty({\Rrm})$,
$0\leq \chi^{}_0 \leq 1$,
$\chi^{}_0 (s)=0$ for $s\leq 4$, $\chi^{}_0 (s )=1$ for $s \geq 14$,
let $1/2<\kappa<1$,
$\chi^{}_1(t,x)=\chi^{}_0((t^2-\vert x\vert^2)/t^{2\kappa})$ for $t\geq 1$ and
$\vert x \vert <t$, $\chi^{}_1 (t,x)=0$ elsewhere, let $(A^{(+)}_0, {\dot
A}^{(+)}_0, \psi^{(+)}_0)(t)=U^1_{{\exp}(tP_0)} u^{}_+$, $u^{}_+=E_+(u)$, let
$(A,
{\dot A}, \psi )=h  (\Omega _1^{(+)}\circ E^{-1}_+ (u_+))$ be the solution
given by Theorem 6.15 and let $\varphi = \vartheta  (A^{(+)}_0
)+\chi^{}_1\vartheta
(A^{(+)2})$, where $A^{(+)2}$ is given by (1.22a). Then
$$\eqalignno{
&\vvv (D^l (A-A^{(+)}_0 e^{i\varphi } \psi-\psi^{(+)}_0
) (u^{}_+; u^{}_{+1},\ldots ,u^{}_{+ l })\vvv^{}_{\rho', r,L}\cr
&\quad{}\leq C_{L+l } \big({\cal R}^{l}_{N_+,L+l }
(u^{}_{+1},\ldots, u^{}_{+l})
+\Vert u^{}_+\Vert^{}_{E^\rho_{N_++L+l }}
\Vert u^{}_{+1} \Vert^{}_{E^\rho_{N_+}}
\cdots \Vert u^{}_{+l }\Vert^{}_{E^\rho_{N_+}}\big),\cr
}$$
for $L$, $l \in {\Nrm}$, $1/2<\rho'\leq 1$, $r=(r(0),r(1))$,
$r(0)>0$, $r(1)\geq \rho $, $u^{}_+ \in {\cal O}_{\infty (+)}$,
$u^{}_{+1},\cdots,
u^{}_{+l }\in E^{\circ\rho}_\infty $, where $\vvv\cdot   \vvv^{}_{\rho',r,L}$
is given in Theorem 6.10 and where $C_{L+l }$
depends only on $\rho'$, $r$, $\rho $ and $\Vert u^{}_+
\Vert^{}_{E^\rho_{N_+}}$,
\psaut
\noindent\hbox{\rm iv)} Let ${\dot A} =\xi^M_{P_0} A$ and
${\dot A}^{(+)}_0 = \xi^M_{P_0}
A^{(+)}_0$. Then
$$\Vert   (A(t), {\dot A}(t)  )-  (A_0^{(+)}(t), {\dot
A}^{(+)}_0(t)  )\Vert^{}_{M^\rho }
+\Vert \psi (t)-  (e^{-i\varphi }\psi _0^{(+)}  ) (t)\Vert^{}_D
\fl 0,$$
when $t \fl \infty $, for $u \in {\cal O}_{\infty (+)}$.
}\saut
\noindent{\it Proof.}
We shall first estimate norms of $  (D^l (A-A^{(+)})  )(u ;
u^{}_1,\ldots ,u^{}_l )$, for $l \in {\Nrm}$, $u \in {\cal O}_{1,\infty (+)}$,
$u^{}_l \in E^\rho _\infty $, where $A^{(+)}(u)$ is given by (1.22a)
with $\chi^{}_0=1$ and
with $u=(f,{\dot f},\alpha )$ instead of $(f_+,{\dot f}_+,\alpha _+)$
and where $(A(u), {\dot A}(u), \psi (u)  )=h(\Omega_1^{(+)}(u))$ is,
according to Theorem 6.15 the solution of the M-D equations in $V^\rho_\infty
$ with initial conditions $\Omega^{(+)}_1(u)\in {\cal U}_{\infty (0)}$.
Recall that
$A_n$,  $n\geq 0$, and $A^*$ are  the functions of $u\in {\cal
O}_{1,\infty (+)}$ given by (4.137b) and (6.30) respectively and let $A^{(l
)}$ (resp. $A^{(+)(l )}, A^{*(l )}$) be the $l^{\rm th}$
derivative of $A$ (resp. $A^{(+)}, A^*$) at $u$ in the
directions $u^{}_1,\ldots ,u^{}_l $. It follows from Lemma 6.3, statement i) of
Theorem 6.9 and by a suitable definition of $N_+$ that
$$\eqalignno{
&\Vert \delta (t)(1 + \big\vert t - \vert\cdot\vert\big\vert)^{1/2}   (\xi^M_Y
(A^{(l )} -A^{*(l )}   )  )(t)  \Vert^{}_{L^\infty }
+   \Vert   (\xi^M_Y  (A^{(l )}-A^{*(l )}  )  )(t)  \Vert^{}_{L^2}&(6.228)\cr
&\quad{}\leq (1+t)^{-1+\rho } C_{l +\vert Y\vert }
\big({\cal R}^l _{N_+,\vert Y\vert +l } (u^{}_1,\ldots ,u^{}_l )
+\Vert u \Vert^{}_{E^\rho _{N_+ +\vert Y \vert +l }}
\Vert u^{}_1\Vert^{}_{E^\rho
_{N_+}}\cdots \Vert u^{}_l \Vert^{}_{E^\rho _{N_+}}  \big),\cr
}$$
for $t \geq 0$, $l \in {\Nrm}$, $Y\in \Pi'$, $u\in {\cal O}_{1,\infty
(+)}$, $u^{}_1,\ldots ,u^{}_l \in E^\rho_\infty $, where
$C_{l +\vert Y \vert }$
depends only on $\rho $ and $\Vert u\Vert^{}_{E^\rho_{N_+}}$.
Inequality (6.29) and the
inequality following (6.29), with $\rho'=1$ and the analog inequalities for $l
\geq 1$ give that
$$\eqalignno{
&\Vert\delta (t)(1 + \big\vert t - \vert\cdot\vert\big\vert)^{1/2}
 (\xi^M_Y  (A^{*(l )} -A^{(l )}_J))(t)  \Vert^{}_{L^\infty}
+   \Vert   (\xi^M_Y   (A^{*(l)}-A^{(l )}_J ) )(t)  \Vert^{}_{L^2} &(6.229)\cr
&\quad{}\leq (1+t)^{-1+\rho }
C_{l+ \vert Y \vert }
\big({\cal R}^l_{N_+,\vert Y\vert +l } (u^{}_1,\ldots,u^{}_l )
+\Vert u \Vert^{}_{E^\rho _{N_+ +\vert Y\vert +l }} \Vert u^{}_1
\Vert^{}_{E^\rho_{N_+}}\cdots
\Vert u^{}_l \Vert^{}_{E^\rho _{N_+}}  \big),\cr
}$$
for $t\geq 0$, $l \in {\Nrm}$, $Y\in \Pi'$, $u\in {\cal O}_{1,\infty
(+)}$, $u^{}_1,\ldots ,u^{}_l \in E^\rho_\infty$, where $C_{l +\vert Y\vert }$
depends only on $\rho $ and $\Vert u \Vert^{}_{E^\rho_{N_+}}$.
Theorem 4.10 gives that
$$\eqalignno{
& \Vert (\delta (t))^{1+\overline{\chi}-\eta+i}
(\xi^M_Y(A^{(l)}_J-A^{(l )}_1))(t)
\Vert^{}_{L^\infty } (1+t)^n&(6.230)\cr
&\quad{} + \Vert   (\xi^M_Y   (A^{(l )}_J-A^{(l)}_1  ),
\xi^M_{P_0Y}   (A^{(l )}_J-A^{(l )}_1  )  )(t)
\Vert_{M^{\rho'}_0} (1+t)^{\rho'-1/2+\overline{\chi}}\cr
&\qquad{}\leq C_{l +\vert Y\vert }  \big({\cal R}^l _{N_+,\vert Y\vert +l }
(u^{}_1,\ldots ,u^{}_l )+\Vert u\Vert^{}_{E^\rho _{N_++\vert Y\vert +l }} \Vert
u^{}_1\Vert^{}_{E^\rho _{N_+}}\cdots
\Vert u^{}_l \Vert^{}_{E^\rho_{N_+}} \big)\cr
}$$
for $t\geq 0$, $l \in {\Nrm}$, $Y\in {\sg}^i$, $i=0,1$, $0\leq
\rho'\leq 1$, $\rho'-1/2+\overline{\chi}>0$, $\eta>0$,
$u\in {\cal O}_{1,\infty (+)}$,
$u^{}_1,\ldots ,u^{}_l \in E^\rho _\infty$, where
$\overline{\chi}=2(1-\rho )$ and where
$C_{l +\vert Y\vert }$ depends only on $\eta$, $\rho$, $\rho'$ and $\Vert u
\Vert^{}_{E^\rho_{N_+}}$. We note that by (4.137b), with $n=0$,
and by definitions
(1.22a)  of $A^{(+)}$ it follows that $A_1 - A^{(+)}=A_1^2-A^{(+)2}$,
where $A^2_1$  $({\rm resp.} A^{(+)2})$ is the bilinear term of $A_1$ $ ({\rm
resp. }A^{(+)})$ in $u$. Let, for $u^{}_i = (f_i,{\dot f}_i,\alpha _i) \in
E^\rho_\infty $,
$$\eqalignno{
&({ J}_\mu (u^{}_1\otimes u^{}_2)  )(t)& (6.231{\rm a})\cr
&\quad{}= {1\over 2}  \big(
(U^{D1}_{{\exp }(tP_0)}\alpha_1  )^+ \gamma_0 \gamma_\mu
(U^{D1}_{{\exp }(tP_0)}\alpha_2  )
+  (U^{D1}_{{\exp }(tP_0)}\alpha_2  )^+ \gamma _0 \gamma_\mu
(U^{D1}_{{\exp }(tP_0)}\alpha_1  )  \big), \quad t\in {\Rrm},\cr
}$$
and
$$\eqalignno{
&  \big(  ({ J}^{(+)}_\mu (u^{}_1\otimes u^{}_2)  )(t)  \big)(x)&(6.231{\rm
b})\cr
&\quad{}= {1\over 2} \big({m\over t}\big)^3
\big({t\over \sqrt {t^2-\vert x\vert ^2}}  \big)^5
\sum_{\varepsilon = \pm }
\Big(  \big(P_\varepsilon (p(t,x)){\hat \alpha }_1 (p(t,x))  \big)^+
\gamma_0 \gamma_\mu   \big(P_\varepsilon (p(t,x)){\hat \alpha }_2 (p(t,x)\big)
)\cr
&\qquad{}+  \big(P_\varepsilon (p(t,x)){\hat \alpha }_2 (p(t,x))
\big)^+\gamma_0
\gamma_\mu   \big(P_\varepsilon (p(t,x)){\hat \alpha }_1 (p(t,x))
\big) \Big),\cr
}$$
for $t>0$, $t^2 - \vert x \vert ^2 > 0$ and $  \big(  ({ J}^{(+)}_\mu
(u^{}_1 \otimes u^{}_2 )  ) (t)  \big)(x) = 0$ for $t > 0$,
$t^2 - \vert x \vert ^2
\leq 9$. Then $  \big({ J}^{(+)}_\mu (u^{}_1\otimes u^{}_2)  \big)(t)\in
D_\infty$ for $t > 0$. We have $\carre \xi^\mu_{Y_1}A^2_1 (u^{}_1\otimes
u^{}_2)=\xi^M_{Y_1} { J} (u^{}_1\otimes u^{}_2)$ in ${\Rrm}^4$ for
$Y_1\in U(\p)$
and $\carre \xi^M_{Y_1}A^{(+)2}=\xi^M_{Y_1}{ J}^{(+)}(u^{}_1
\otimes u^{}_2)$ in
$\{(t,x)\in {\Rrm}^+\times {\Rrm}^3\big\vert t^2 -
\vert x\vert ^2\geq 4\}$ for
$Y_1\in U(\p )$. This covariance property for $Y_1 \in \Pi' \cap +
U({\frak{sl}}(2,{\Crm}))$, $Y={Z}Y_1$, with ${Z}\in \Pi'\cap U({\Rrm}^4)$ and
Corollary 4.2 and this covariance property for $Y_1\in \Pi'$, $Y=Y_1$ and
statement iii) of Lemma 4.3 give:
$$ \eqalignno{
& \Vert   \big(\xi^M_Y   (A^{(2)}_1(u^{}_1\otimes u^{}_2)-A^{(+)2}
(u^{}_1 \otimes u^{}_2)  )  \big)(t)  \Vert^{}_{L^\infty (\vert x \vert ^2\leq
t^2-4)} (1+t)^2 &(6.232)\cr
&\quad{}+  \Vert \big(\xi^M_Y   (A_1^{(2)}(u^{}_1 \otimes
u^{}_2)-A^{(+)2} (u^{}_1 \otimes u^{}_2)  )  \big) (t)
\Vert^{}_{L^2(\vert x \vert^2\leq t^2-4)} (1+t)^{1/2}\cr
&\qquad{}\leq C_{\vert Y\vert }   \big(\Vert \alpha _1 \Vert^{}_{D_{N_+}} \Vert
\alpha_2
\Vert^{}_{D_{N_++\vert Y\vert }} +\Vert \alpha_1 \Vert^{}_{D_{N_++\vert Y\vert
}}
\Vert \alpha_2 \Vert^{}_{D_{N_+}}  \big),\cr
}$$
for $u^{}_1$, $u^{}_2\in E^\rho_\infty$, $Y\in \Pi'$, $t\geq 2$,
where $C_{\vert Y\vert }$ is independent of $t$, $u^{}_1$, $u^{}_2$.
Since the support of $s\mapsto(d/ds)^n\chi (s)$, $s \in {\Rrm}$,
is a subset of the interval $[1,2]$, it follows
that
$$\big\vert   (\partial_0^{\alpha_0} \partial_1^{\alpha_1}
\partial_2^{\partial_2}
\partial_3^{\alpha_3}\chi (\rho )  )(t,x)\big\vert \leq
C_{\vert \alpha \vert
}(1+t)^{\vert \alpha \vert },\quad  t\geq 0, x\in {\Rrm}^3,
\eqno{(6.233)}$$
where we have defined $(\chi (\rho ))(t,x) = 0$ for $t^2 -
\vert x \vert^2 < 0$.
Since $\xi^{}_Y \rho = 0$ for $Y\in U (sl (2,{\Crm}))$ and since
the volume of
$\{x \in {\Rrm}^3 \big\vert t^2 - 4 \leq \vert x \vert^2 \leq t^2\}$
is bounded by
$C(1 + {t})$, $t \geq 0$, where $C$ is independent of $t$, it follows from
(4.47) of Corollary 4.2 that
$$\eqalignno{
&\Vert   (\xi^M_{{Z}Y} A^{(+)2}(u^{}_1\otimes u^{}_2)  )(t)
\Vert_{L^p (\vert x \vert^2 \geq t^2-4)} &(6.234)\cr
&\qquad{}\leq C_{\vert {Z} \vert + \vert Y \vert}
\sum_{0 \leq n \leq \vert {Z} \vert } (1+t)^{-1+1/p+ \vert {Z} \vert -n}\cr
&\qquad\qquad{}\big(\Vert \alpha_1 \Vert^{}_{D_{N_+}}
\Vert \alpha_2 \Vert^{}_{D_{N_++n+ \vert Y
\vert }} +\Vert \alpha_1 \Vert^{}_{D_{N_++n+ \vert Y \vert }} \Vert \alpha_2
\Vert^{}_{D_{N_+}}  \big),\cr
}$$
for $1 \leq p \leq \infty $, $t \geq 0$, $u^{}_1, u^{}_2 \in E^\rho_\infty$,
$Y \in \Pi'\cap U(sl (2,{\Crm}))$, ${Z}\in \Pi'\cap U ({\Rrm}^4)$, where
$C_{\vert {Z} \vert + \vert Y \vert }$ depends only on $p$. It follows
similarly
from Theorem 4.10, reminding that there is no cut-off function in $A^2_1$, that
$$ \eqalignno{
& \Vert   (\xi^M_Y A^2_1(u^{}_1\otimes u^{}_2)  )(t)
\Vert^{}_{L^p(t^2-4\leq \vert x\vert^2\leq t^2)} &(6.235)\cr
&\quad{}\leq C_{\vert Y \vert } (1+t)^{-1+1/p} \big(\Vert \alpha_1
\Vert^{}_{D_{N_+}} \Vert
\alpha_2 \Vert^{}_{D_{N_++ \vert Y \vert }} +\Vert
\alpha_1Ê\Vert^{}_{D_{N_+ + \vert
Y \vert }} \Vert \alpha_2 \Vert^{}_{D_{N_+}}\big),\cr
}$$
for $1 \leq p \leq \infty$, $t \geq 0$, $u^{}_1, u^{}_2 \in
E^\rho_\infty$, $Y \in
\Pi'$, where $C_{\vert Y \vert }$ depends only on $p.$

According to (6.227a) and (6.227b) let $E_+(u) = (f,{\dot f},\alpha_+(u))$
for $u = (f,{\dot f},\alpha )\in {\cal O}_{1,\infty (+)}$, where
$$(\alpha_+ (u))^\wedge (k)= \sum_{\varepsilon = \pm } e^{i\vartheta^\infty (A
^{(+)}-A,(\omega (k),-\varepsilon k))} P_\varepsilon (k){\hat \alpha} (k) .
\eqno{(6.236)}$$
To prove that $E_+$ is a $C^\infty $ function from ${\cal O}_{1,\infty
(+)}$ to $E^\rho_\infty $ satisfying the inequality of statement i), it is
sufficient to prove this for the function $\alpha_+$ from
${\cal O}_{1,\infty (+)}$
to $D_\infty$. If ${\hat \beta }(k)=\sum_\varepsilon F_\varepsilon (k)
P_\varepsilon (k){\hat \alpha }(k)$, then if follows using
definition (1.5d) of $T^{D1}_{M_{0i}}$ that
$$
(T^{D1}_{M_{0i}}\beta)^\wedge (k)=\sum_\varepsilon
\big((-\varepsilon \omega (k){\partial \over \partial k_i}F_\varepsilon (k)
)P_\varepsilon (k){\hat \alpha }(k)
+ F_\varepsilon (k) P_\varepsilon (k)(T^{D1}_{M_{0i}} \alpha )^\wedge
(k)  \big),\eqno{(6.237{\rm a})}$$
and if $F_\varepsilon (k)=f(\omega (k),-\varepsilon k)$ then
$$-\varepsilon \omega (k){\partial \over \partial k_i}F_\varepsilon (k)=
(\xi^{}_{M_{0i}}f  )(\omega (k),-\varepsilon k).\eqno{(6.237{\rm b})}$$
A similar relation is valid for the rotations which gives that
$$\eqalignno{
&(T^{D1}_X \beta )^\wedge (k) &(6.237{\rm c})\cr
&\quad{}= \sum_\varepsilon   ((\xi^{}_X f)(\omega
(k), - \varepsilon k)P_\varepsilon (k){\hat \alpha }(k)
+ f(\omega (k), - \varepsilon k) P_\varepsilon (k)(T^{D1}_X \alpha )^\wedge
(k), \quad X\in sl (2,{\Crm}).\cr
}$$
Using the SL$(2,{\Crm})$ covariance of the function $\vartheta^\infty $
it follows from
(6.237c), with $f(y) = \vartheta^\infty (H,y)$ that
$$(T^{D1}_{{Z}Y} \beta )(k) = \suma_{Y_1,Y_2}^Y\sum_\varepsilon
\vartheta^\infty   (\xi^M_{Y_1}H,(\omega (k),-\varepsilon k)
)P_\varepsilon (k)
(T^{D1}_{{Z}Y_2}\alpha )^\wedge (k),\eqno{(6.238)}$$
for $Y\in \Pi'\cap U(sl (2,{\Crm}))$ and ${Z} \in \Pi' \cap U({\Rrm}^4)$,
where ${\hat \beta }(k)=\sum_\varepsilon \vartheta^\infty (H,(\omega
(k),-\varepsilon k))$ and $H\colon {\Rrm}^+\times {\Rrm}^3\rightarrow {\Rrm}^4$
is sufficiently smooth and decreasing. It follows from equalities (6.236) and
(6.238) that  $  \big(T^{D1}_{{Z}Y}((D^l \alpha_+) u;
u^{}_1,\ldots ,u^{}_l))\big)^\wedge (k)$ is a sum of terms
$$\eqalignno{
&C_{Y_1,\ldots ,Y_{j+1} ; i_1,\ldots ,i_{j+1}} \vartheta^\infty
(\xi^M_{Y_1}H^{(i_1)},l_\varepsilon (k)  )\cdots \vartheta^\infty
(\xi^M_{Y_j}H^{(i_j)},l_\varepsilon (k))&(6.238)\cr
&\qquad{}e^{i\vartheta^\infty (H,l_\varepsilon (k))} P_\varepsilon (k)
(T^{D1}_{{Z}Y_{j+1}} \alpha^{(i_{j+1})})^\wedge (k),\cr
}$$
$ {Z}\in \Pi' \cap U({\Rrm}^4)$, $Y \in \Pi' \cap U(sl (2,{\Crm}))$,
where $H = A^{(+)}-A$, $l_\varepsilon (k)=(\omega (k),-\varepsilon k)$,
where $H^{(i_1)},\ldots ,H^{(i_j)}$, $\alpha ^{(i_{j+1})}$ are the
derivatives of order $i_1, \ldots ,i_j$ respectively $i_{j+1}$ at $u$, each
depending respectively on $i_1,\cdots ,i_j$, $i_{j+1}$
distinct arguments among
$u^{}_1,\ldots ,u^{}_l$ and where $i_1+\cdots +i_{j+1}=l$, $i_1,\ldots
,i_{j+1}\geq 0$, $Y_1,\ldots ,Y_{j+1}\in \Pi' \cap U(sl (2,{\Crm}))$, $\vert
Y_1\vert +\cdots +\vert Y_{j+1}\vert =\vert Y\vert$,
$\vert Y_q\vert +i_q\geq 1$ for
$q\leq j$. Since the function $(t,x)\mapsto \vartheta^\infty (H,(t,x))$ is
homogeneous of degree zero, we can use statement ii) of Lemma 3.1 to the
function $(\omega (k),-\varepsilon k)\mapsto \vartheta^\infty (H,m^{-1}(\omega
(k),-\varepsilon k))=\vartheta^\infty (H,(\omega (k),-\varepsilon k))$.
This gives
$$\eqalignno{
&\vert \vartheta^\infty   (\xi^M_Y H^{(l )},
(\omega (k),-\varepsilon k ))\vert &(6.240)\cr
&\quad{}\leq
2m  \big(\ln \big(1+{\omega (k)\over m}\big) +{1\over 2\tau }  \big)
\sup_{\scr  t \geq 1\atop\scr \vert x\vert <t}
\big((1+t)(1+t- \vert x \vert )^\tau \vert   (\xi^M_Y H^{(l
)}  )(t,x)\vert   \big),\quad \tau >0,\cr
}$$
for $Y \in \Pi' \cap U(sl (2,{\Crm}))$, $k \in {\Rrm}^3$, $\varepsilon
=\pm $. Since $t - \vert x \vert \leq 4(1+t)^{-1}$, for $t\geq 1$,
$t^2 - \vert x \vert^2 \leq 2$ and $\vert x \vert \leq t$ it follows
from inequalities (6.228), (6.229), (6.230), (6.232), (6.234) and (6.235),
choosing $\tau = 1-\rho $ in (6.240)
that
$$\eqalignno{
&\big\vert \vartheta^\infty   (\xi^M_Y (A^{(+)(l )}-A^{(l )}),(\omega (k),
-\varepsilon k)  \big)\big\vert& (6.241)\cr
&\quad{}\leq C_{\vert Y \vert + l }\ln \big(1+{\omega
(k) \over m}  \big)
\big({\cal R}^l _{N_+, \vert Y \vert + l } (u^{}_1,\ldots ,u^{}_l ) +
\Vert u \Vert^{}_{E^\rho_{N_+ + \vert Y \vert + l }} \Vert u^{}_1
\Vert^{}_{E^\rho_{N_+}} \cdots \Vert u^{}_l \Vert^{}_{E^\rho_{N_+}}\big),\cr
}$$
for $l \geq 0$, $Y \in \Pi'\cap U(sl (2,{\Crm}))$, $k \in {\Rrm}^3$,
$\varepsilon = \pm$, $u \in {\cal O}_{1, \infty (+)}$, $u^{}_1, \ldots
,u^{}_l \in E^\rho_\infty $, where $C_{\vert Y \vert + l }$
depends only on $\rho $ and
$\Vert u \Vert^{}_{E^\rho_{N_+}}$. Reindexing for the moment the sequence
$u^{}_1,\ldots ,u^{}_l $ we can suppose that $H^{(i_1)} ;\ldots ; H^{(i_j)}$
depend on the arguments $u^{}_1,\ldots ,u^{}_{i_1}$ ; $\ldots ;u^{}_{i_1+\cdots
+i_{j-1}+1},\ldots $, $u^{}_{i_1+\cdots +i_j}$ respectively and that $\alpha
^{(i_{j+1})}$ depends on $u^{}_{l -i_{j+1}+1},\ldots ,u^{}_l $. Using the
notation
$$\overline{{\cal R}}^{(l )}_{N_+,N_+ + \vert Y \vert + l } (u ; u^{}_1,\ldots
,u^{}_l )={\cal R}^l _{N_+,\vert Y \vert + l } (u^{}_1, \ldots ,u^{}_l )
+ \Vert u \Vert^{}_{E^\rho_{N_++\vert Y\vert +l }} \Vert u^{}_1
\Vert^{}_{E^\rho_{N_+}} \cdots \Vert u^{}_l \Vert^{}_{E^\rho_{N_+}}$$
using the fact that $\ln(1+\omega (k)/m) \leq \omega (k)/m$ and that
$$\overline{{\cal R}}^{(l )}_{N_+,N_+ + \vert Y \vert + l } (u ; u^{}_1,
\ldots, u^{}_l )
\leq \overline{{\cal R}}^{(l )}_{N_+ + 1,N_+ + 1+ \vert Y \vert + l -1} (u
; u^{}_1,\ldots ,u^{}_l ),$$
it follows from Corollary 2.6 and inequality (6.241) that the expression
(6.239) is majorized by
$$C_{\vert Y\vert +l } \overline{{\cal R}}^{(l )}_{N_++1,  N_++1+\vert {Z}\vert
+\vert	Y\vert +l } (u ; u^{}_1,\ldots ,u^{}_l ).\eqno{(6.242)}$$
Here we have  also used the fact that
$$  \Vert \omega (-i \partial)^{j} T^{D1}_{{Z}Y_{j+1}}\
\alpha^{(i_{j+1})}  \Vert^{}_D \leq C_{\vert {Z}\vert +\vert Y_{j+1} \vert +j}\
\overline{{\cal R}}^{(i_{j+1})}_{N_+,N_+ + \vert {Z} \vert + \vert Y_{j+1}\vert
+i_{j+1}} (u ; u^{}_{l -i_{j+1}+1},\ldots ,u^{}_l )$$
and that $\vert Y_q \vert + i_q \geq 1$ for $q \leq j$. Since the expression
(6.242) is independent of permutations of $u^{}_1, \ldots , u^{}_l $, this
expression majorizes all the expressions (6.239). This proves, after
redefinition of $N_+$, that $E_+ \colon {\cal O}_{1,\infty (+)} \fl
E^\rho_\infty $ is $C^\infty $ and that statement i) of the proposition is
true.

We shall next study the asymptotic behavior of $e^{i\varphi (t)}\psi (t)$ in
statement iii) of the proposition for $u^{}_+=(f_+, {\dot f}_+, \alpha_+ )
=E_+(u)$, $u=(f,\dot{f},\alpha)\in {\cal
O}_{1,\infty (+)}$ and $v=\Omega ^{(+)}_1(u)$. It follows from (4.47)
of Corollary 4.2 that
$$\eqalignno{
& \Vert   (\xi^M_Y A^{(+)2} (u^{}_1\otimes u^{}_2)  )(t)
\Vert_{L^P(\vert x \vert ^2 \leq t^2 - 4)} &(6.243)\cr
&\quad{}\leq (1+t)^{-1 - i + 3/p} C_{\vert Y \vert}
\big(\Vert \alpha _1 \Vert^{}_{D_{N_+}} \Vert \alpha_2 \Vert^{}_{D_{N_+ + \vert
Y \vert }} + \Vert \alpha _1 \Vert^{}_{D_{N_+ + \vert Y \vert}} \Vert \alpha_2
\Vert^{}_{D_{N_+}}\big),\cr
}$$
for $1 \leq p \leq \infty $, $t \geq 0$,  $u^{}_j = (f_j, {\dot f}_j, \alpha_j)
\in E^{\circ\rho}_\infty$, $j = 0$, $1$, $Y \in \sg ^i$, $i = 0,1$, where
$N_+$ and $C_{\vert Y \vert }$ are independent of $p$, $t$, $u^{}_j$, $Y$.
Let $\chi^{}_1 (t,x) =
\chi^{}_0   ((t^2 - \vert x \vert ^2)/t^{2\kappa}  )$ for $t \geq 1$ and
$\chi^{}_1 (t,x) = 0$ elsewhere for $t \geq 0$. Since $\chi^{}_1 (t,x) = 0$
for $t^2 - \vert
x \vert^2 \leq 4t^{2\kappa}$, it follows that $t^2 - \vert x \vert^2 \geq
4t^{2\kappa}>16$ in
the support of $\varphi$. Here we have  used the fact that $1/2 < \kappa < 1$.
Inequalities
(6.234) and (6.243) then give for $Y \in \Pi'$ and for $(t,x) \in {\rm supp}\
 {\chi^{}_1}$:
$$\eqalignno{
&\big\vert \xi^{}_Y \vartheta  (A^{(+)2} (u^{}_1 \otimes u^{}_2),(t,x)   )
\big\vert&(6.244)\cr
&\quad{} \leq
(1+t)^{-i+\kappa'} C_{\vert Y \vert }
\big(\Vert \alpha_1 \Vert^{}_{D_{N_+}} \Vert \alpha_2 \Vert^{}_{D_{N_+ + \vert
Y \vert }} + \Vert \alpha_1 \Vert^{}_{D_{N_++\vert Y\vert }} \Vert \alpha_2
\Vert^{}_{D_{N_+}}  \big), \cr
}$$
for $t \geq 0$, $x \in {\Rrm}^3$, $Y \in \sg^i$, $i = 0,1$, where
$\kappa' > 0$ can be made arbitrary small by choosing $\kappa \in ]1/2,1[$
sufficiently close to 1 and where $C_{\vert Y \vert }$ depends only on
$\kappa$. According to the definition of $\chi^{}_1$ we obtain that
$$\vert \xi^{}_Y  \chi^{}_1 \vert (t,x) \leq C_{\vert Y \vert }
(1+t)^{-i(2\kappa-1)} \quad {\rm for } Y \in \sg^i,  i \in {\Nrm}.
\eqno{(6.245)}$$
Hence, redefining $\kappa'$ and identifying the second order part of the map
$\varphi$ with a bilinear symmetric map $\varphi^2$, it follows from inequality
(6.244) that
$$\eqalignno{
&\vert \xi^{}_Y  \varphi^2(u^{}_1 \otimes u^{}_2) \vert (t,x) &(6.246)\cr
&\quad{}\leq (1+t)^{-i+\kappa'} C_{\vert Y \vert }
\big(\Vert \alpha_1 \Vert^{}_{D_{N_+}} \Vert \alpha_2 \Vert^{}_{D_{N_+ + \vert
Y \vert }} + \Vert \alpha _1 \Vert^{}_{D_{N_+ + \vert Y \vert }} \Vert \alpha_2
\Vert^{}_{D_{N_+}}   \big),\cr
}$$
for $t \geq 0$, $x \in {\Rrm}^3$, $Y \in \sg^i$, $i \in \{0,1\}$,
$u^{}_j = (f_j,{\dot f}_j,\alpha_j) \in E^\rho_\infty $, where $\kappa'$ can be
made arbitrary small by choosing $\kappa \in ]1/2,1[$ sufficiently close to
1 and where $C_{\vert Y \vert }$ depends only on $\kappa$.
Let $(t,x) \in {\rm
supp}\
(1-{\chi^{}_1} )$ and $\vert x \vert \leq t$. Then $0 \leq t - \vert x \vert
\leq 16 t^{2 \kappa -1}$, which shows that $(1 + \lambda_1 (t))(x)
\geq C(1+t)^{2(1-\kappa )}$ for such $t$ and $x$. Hence by the definition
of $\lambda_1 $, it follows that $(1 + t + \vert x \vert )^{2(1 - \kappa)}
\leq C(1 + \lambda_1 ,(t)) (x)$ for $(t,x) \in {\rm  supp}\ (1-\chi^{}_1)$
and $t \geq 0$, where $C$ is independent of $t$ and $x$. This gives together
with Theorem 6.10 and Lemma 4.4 that
$$\eqalignno{
&  \Vert (1 + \lambda_1 (t))^{-\tau /(2(1-\kappa))}
\big(\xi^{}_Y (1-\chi^{}_1)\
\vartheta(A^{(l)}-A_0^{(l)}) \big) (t) \Vert^{}_{L^\infty}&(6.247)\cr
&\quad{}+   \Vert (\delta (t))^{2-\rho }   \big(\xi^{}_{Z}(1-\chi^{}_1)
\vartheta
(A^{(l )}-A_0^{(l)})\big) (t)   \Vert^{}_{L^\infty }\cr
&\qquad{}\leq C_L   \big({\cal R}^l_{N_+,L} (u^{}_1,\ldots ,u^{}_l ) + \Vert
u \Vert^{}_{E^\rho_{N_++L}} \Vert u^{}_1 \Vert^{}_{E^\rho_{N_+}} \cdots
\Vert u^{}_l
\Vert^{}_{E^\rho_{N_+}} \big),\cr
}$$
for $t \geq 0$, $\tau > 0$, $Y \in \Pi'$, ${Z} \in \sg^1$,
$u^{}_1,\ldots,u^{}_l \in E^{\circ\rho}_\infty $,
$\max (\vert Y \vert,\vert {Z} \vert ) + l \leq L$,
where $A^{(l)} - A_0^{(l)} = (D^l (A-A_0)) (u; u_{1},\ldots ,u^{}_l)$
and where $C_L$ depends only on $\tau, \rho , \kappa$ and $\Vert
u \Vert^{}_{E^{\rho }_{{N_+} + L}}$. It follows from inequalities
(6.228), (6.229),
(6.230) and (6.232) that if $H^{(l)} = (D^l (A^{(+)} - A)) (u ; u^{}_1,\ldots ,
u^{}_l)$ then
$$\eqalignno{
\vert \xi_Y^M H^{(l)} \vert (t,x) &\leq (1 + t)^{-3+2 \rho - j (\rho -1/2)}
(1+ (\lambda_1 (t)) (x))^{1 - \rho + j (\rho - 1/2)} &(6.248{\rm a})\cr
&\quad{} C_{\vert Y \vert, l }
\big({\cal R}^l_{N_+, \vert Y \vert + l } (u^{}_1,\ldots , u^{}_l) + \Vert u
\Vert^{}_{E^\rho_{N_+ + \vert Y \vert + l }}
\Vert u^{}_1 \Vert^{}_{E^\rho_{N_+}}
\cdots \Vert u^{}_l \Vert^{}_{E^\rho _{N_+}}  \big),\cr
}$$
for $t^2 - \vert x \vert^2 \geq 4$, $t \geq 0$,  $l \geq 0$,  $Y \in
\sg^j$,  $j \in \{ 0, 1 \}$, $u \in {\cal O}_{1, \infty (+)}$, $u^{}_1,\ldots ,
u^{}_l \in E^{\circ\rho}_\infty $, where $C_{\vert Y \vert, l}$
depends only on $\rho $ and $\Vert u \Vert^{}_{E^\rho _{N_+}}$.
It follows from Theorem
6.10 and inequality (6.234) that
$$\eqalignno{
\vert \xi ^M_{{Z} Y} H^{(l )} \vert (t,x)
&\leq C_{\vert {Z}\vert + \vert
Y \vert +l } \sum_{0\leq n\leq \vert {Z} \vert } (1+t)^{-1+ \vert {Z}
\vert -n}&(6.248 {\rm b})\cr
&\quad{} \big({\cal R}^l_{N_+,n + \vert Y \vert + l }(u^{}_1,\ldots ,u^{}_l ) +
\Vert u \Vert^{}_{E^\rho_{N_+ +n+ \vert Y \vert + l }} \Vert u^{}_1
\Vert^{}_{E^\rho_{N_+}} \cdots \Vert u^{}_l \Vert^{}_{E^\rho _{N_+}}\big),\cr
}$$
for $t^2 - \vert x \vert^2 \leq 4$, $t \geq 0$, $l \geq 0$, ${Z} \in
\Pi' \cap U ({\Rrm}^4)$, $Y \in \Pi' \cap U(sl (2,{\Crm}))$, $u \in {\cal
O}_{1,\infty (+)}$, $u^{}_1,\ldots ,u^{}_l \in E^{\circ\rho}_\infty $, where
$C_{\vert {Z} \vert + \vert Y \vert + l }$ depends only on $\rho$ and $\Vert
u \Vert^{}_{E^\rho_{N_+}}$. If $(t,x), t > 0$, is inside the light cone then it
follows from inequality (6.248a) and (6.248b) that the line integral
$\vartheta^\infty  (\xi ^M_Y H^{(l )}, (t,x)   )$ converges absolutely for
$Y \in \Pi' \cap U(sl (2,{\Crm}))$. The function $(t,x)\mapsto \vartheta^\infty
(\xi^M_Y H^{(l )}, (t,x)  )$ from the interior of the forward light cone
into ${\Rrm}$ is homogeneous of degree zero and $\xi^{}_Y \vartheta^\infty
(H^{(l)})=\vartheta^\infty   (\xi^M_Y H^{(l )}  )$ for $Y\in \Pi' \cap U(sl
(2,{\Crm}))$. Expressing the derivatives $\partial_\mu $
as linear functions of
$\xi^{}_{M_{0i}}$ and the dilatation operator as in equality (5.67),
it follows from
inequality (6.248a) with $Y\in \Pi' \cap U(sl (2,{\Crm}))$, inequality
(6.248b) with ${Z}={\un}$ and by using $t^2-\vert x\vert^2\geq 4t^{2\kappa}$,
$t\geq 0$,
in the support of $\chi^{}_1$, that
$$  \big(1+\ln(1+t/(t-\vert x\vert )) \big)^{-1}
t^{(2\kappa-1)\vert {Z}\vert } \vert \xi^{}_{{Z}Y} \vartheta^\infty (H^{(l
)})\vert (t,x),$$
${Z}\in \Pi'\cap U({\Rrm}^4)$, $Y\in \Pi'\cap U({\frak{sl}}(2,{\Crm}))$
is uniformly bounded in $(t,x)$  on the support of $\chi^{}_1$. According to
inequality (6.245) and using the bounds in inequalities (6.248a) and (6.248b)
we obtain that
$$\eqalignno{
&\big\vert \xi^{}_{{Z}Y} \chi^{}_1\vartheta^\infty
(H^{(l)})\big\vert(t,x)&(6.249)\cr
&\quad{}\leq\big(1 + \ln(1+t / (t - \vert x \vert))\big)(1+t)^{- \vert {Z}
\vert (2 \kappa-1)}\cr
&\qquad{}  C_{\vert {Z}Y \vert + l }
 \big({\cal R}^l _{N_+,\vert {Z} \vert + \vert Y \vert + l }\
(u^{}_1,\ldots ,u^{}_l ) + \Vert u \Vert^{}_{E^\rho_{N_+ + \vert {Z}
\vert + \vert
Y \vert + l }} \Vert u^{}_1 \Vert^{}_{E^\rho_{N_+}} \cdots \Vert u^{}_l
\Vert^{}_{E^\rho_{N_+}}\big),\cr
}$$
for $(t,x)\in {\rm supp}\ (\chi^{}_1)$, $l \geq 0$, ${Z} \in \Pi' \cap
U({\Rrm}^4)$, $Y \in \Pi'\cap U(sl (2,{\Crm}))$, $u \in {\cal O}_{1,\infty
(+)}, u^{}_1,\ldots ,u^{}_l \in E^{\circ\rho}_\infty $, where $C_{\vert
{Z}Y \vert + l }$ depends only on $\rho$ and $\Vert u \Vert^{}_{E^\rho_{N_+}}$.
Similarly as we obtained (6.246) from inequality (6.244), it follows from
inequalities (6.248a) and (6.248b) and from $1+\lambda_1(sy) \leq
C(1+\lambda_1(y))$ for $y$ inside the light cone and for $0 \leq s \leq 1$,
that
$$\eqalignno{
&\big\vert \xi^{}_Y  \chi^{}_1\vartheta (H^{(l)}) \big\vert (t,x)&(6.250)\cr
&\quad{} \leq (1+t)^{-2  (1- \rho)}(1 + \lambda_1 (t,x))^{1/2} \cr
&\qquad{} C_{\vert Y \vert + l } \big({\cal R}^l_{N_+, \vert Y \vert + l }
(u^{}_1, \ldots, u^{}_l) + \Vert
u \Vert^{}_{E^\rho_{N_+ + \vert  Y \vert + l}}
\Vert u^{}_1 \Vert^{}_{E^\rho_{N_+}}
\cdots \Vert u^{}_l \Vert^{}_{E^\rho_{N_+}}\big),\cr
}$$
for $t \geq 0$, $x \in {\Rrm}^3$, $Y \in \sg^1$, $u \in {\cal O}_{1,\infty
(+)}$, $u^{}_1, \ldots , u^{}_l  \in E^{\circ\rho}_\infty$, where $\kappa$ is
chosen sufficiently close to one and where $C_{\vert Y \vert + l }$
depends only on $\kappa$, $\rho$ and $\Vert u \Vert^{}_{E^\rho_{N_+}}$.
Since $\varphi
-\vartheta(A) = \chi^{}_1  \vartheta(A^{(+)}-A) + (1-\chi^{}_1)
\vartheta(A_0-A)$, it follows from
inequalities (6.247) and (6.250) that
$$\eqalignno{
&  \Vert (\delta (t))^{2(1-\rho )}(1+\lambda _1(t))^{-1/2}   (\xi^{}_Y
  \big(D^l (\varphi -\vartheta(A))   (u ; u^{}_1,\ldots ,u^{}_l )\big)(t)
\Vert^{}_{L^\infty }& (6.251)\cr
&\quad{}\leq C_{\vert Y \vert +l }   \big({\cal R}^l _{N_+,\vert Y \vert + l}
(u^{}_1,\ldots ,u^{}_l ) +
\Vert u \Vert^{}_{E^\rho_{N_+ + \vert Y \vert + l }} \Vert
u^{}_1 \Vert^{}_{E^\rho_{N_+}} \cdots \Vert u^{}_l
\Vert^{}_{E^\rho_{N_+}}\big),\cr
}$$
for $t \geq 0$, $Y \in \sg^1$, $u \in {\cal O}_{1,\infty (+)},$
$u^{}_1,\ldots ,u^{}_l \in E^{\circ\rho}_\infty$, where $C_{\vert Y \vert
+ l }$ depends only on $\rho$ and $\Vert u \Vert^{}_{E^\rho_{N_+}}$. It also
follows from inequality (6.247), with $\tau /(2(1-\kappa ))$
replaced by $\tau$, and from inequalities (6.248a) and (6.248b), that
$$ \eqalignno{
& \Vert (1 + \lambda_1 (t))^{- \tau}   \big(\xi^{}_Y (D^l (\varphi
- \vartheta(A)))(u ;  u^{}_1,\cdots , u^{}_l)\big) (t)   \Vert^{}_{L^\infty }
&(6.252)\cr
&\quad {}\leq C_{\vert Y \vert + l }
\big({\cal R}^l _{N_+,\vert Y \vert + l }
(u^{}_1, \ldots ,u^{}_l) +
\Vert u\Vert^{}_{E^\rho_{N_+ + \vert Y \vert + l }}\Vert
u^{}_1 \Vert^{}_{E^\rho_{N_+}} \cdots
\Vert u^{}_l \Vert^{}_{E^\rho_{N_+}}\big),\cr
}$$
for $t \geq 0$, $Y \in \Pi' \cap U(sl (2,{\Crm}))$, $u \in
{\cal O}_{1,\infty (+)}$, $u^{}_1,\ldots ,u^{}_l  \in E^{\circ\rho}_\infty$,
$\tau > 0$, where $C_{\vert Y \vert + l }$ depends only on $\tau$, $\rho$ and
$\Vert u \Vert^{}_{E^\rho_{N_+}}$. Let $\psi_0 (t) = U^{D1}_{\exp (tP_0)}
\alpha$. Using Theorem 6.10, Corollary 2.6, inequalities (6.251) and (6.252),
with $\tau = 1/2$ and redefining $N_+$ we obtain, with the notation
$$q^{(l )}_Y(t) =   \big(\xi^D_Y   \big(D^l (e^{i(\varphi - \vartheta(A))}
(e^{i\vartheta(A)} \psi - \psi_0))\big)(u ; u^{}_1,\ldots ,u^{}_l )  \big)(t),
\eqno{(6.253)}$$
that
$$\eqalignno{
&(1+t)^{2(1-\rho )}\Vert (1+\lambda_1(t))^{k/2} q^{(l)}_Y (t)
  \Vert^{}_D
+ \Vert (\delta (t))^{3/2 + 2(1-\rho)} (1 + \lambda_1 (t))^{k/2}
q^{(l)}_Y  (t)   \Vert^{}_{L^\infty }&(6.254)\cr
&\qquad{}\leq C_{\vert Y \vert + l + k}   \big({\cal R}^l_{N_+,\vert Y \vert +
l
+ k} (u^{}_1,\ldots ,u^{}_l )+ \Vert u \Vert^{}_{E^\rho_{N_+ + \vert Y \vert +
l
+ k}} \Vert u^{}_1 \Vert^{}_{E^\rho_{N_+}}\cdots
\Vert u^{}_l \Vert^{}_{E^\rho_{N_+}}\big),\hskip5mm\cr
}$$
for $t \geq 0$, $l \in {\Nrm}$, $k \in {\Nrm}$, $Y \in \Pi'$, $u \in
{\cal O}_{1,\infty (+)}$, $u^{}_1,\ldots ,u^{}_l \in E^{\circ\rho}_\infty $,
where $C_{\vert Y \vert + l + k}$ depends only on $\rho$ and $\Vert u
\Vert^{}_{E^\rho_{N_+}}$. Similarly, if
$$r^{(l )}_{Y,{Z}} (t) =   \big(\xi^D_{Y{Z}}   (D^l
(e^{i(\varphi -\vartheta(A))} \psi_0 )  )
- \xi^D_Y   (D^l (e^{i(\varphi - \vartheta(A))} \xi^D_{Z} \psi_0))\big)(t),
\eqno{(6.255)}$$
then
$$\eqalignno{
&(1+t)^{2(1-\rho )}\Vert (1 + \lambda_1 (t))^{k/2} r^{(l)}_{Y,{Z}} (t)
\Vert^{}_D
+   \Vert (\delta (t))^{3/2 + 2(1 - \rho)} (1 + \lambda_1 (t))^{k/2}
r^{(l )}_{Y,{Z}} (t)   \Vert^{}_{L^\infty}&(6.256)\cr
&\ {}\leq \vert {Z} \vert C_{\vert Y \vert + \vert {Z} \vert + l + k}
\big({\cal R}^l_{N_+,\vert Y \vert + \vert {Z} \vert +l +k} (u^{}_1,\ldots,
u^{}_l )
+ \Vert u \Vert^{}_{E^\rho_{N_+ +\vert Y \vert + \vert {Z} \vert + l + k}}
\Vert u^{}_1 \Vert^{}_{E^\rho_{N_+}}\cdots
\Vert u^{}_l \Vert^{}_{E^\rho_{N_+}}\big),\cr
}$$
for $t \geq 0$, $l \in {\Nrm}$, $k \in {\Nrm}$, $Y \in \Pi'\cap U(sl
(2,{\Crm}))$, ${Z} \in \Pi'\cap U({\Rrm}^4)$, $u \in {\cal O}_{1,\infty
(+)}$, $u^{}_1,\ldots , u^{}_l \in E^{\circ\rho}_\infty $, where $C_{\vert
Y \vert + \vert {Z} \vert + l + k}$ depends only on $\rho$ and $\Vert u
\Vert^{}_{E^\rho_{N_+}}$. Development of the expression
$$\xi^D_Y   \big(  (D^l (e^{i(\varphi - \vartheta(A))} \xi^D_{Z}
\psi_0))(u ; u^{}_1,\ldots , u^{}_l)\big),\quad  Y \in \Pi' \cap
U({\frak{sl}}(2,{\Crm})),$$
shows that it is a sum of expressions
$$\eqalignno{
&C_{Y_1, \ldots , Y_{j+1} ; i_1, \ldots ,i_{j+1}}
\big(\xi^{}_{Y_1}(\varphi^{(i_1)} - \vartheta(A^{(i_1)}))\big)\cdots
\big(\xi^{}_{Y_j}  (\varphi^{(i_j)} - \vartheta(A^{(i_j)}))\big)&(6.257)\cr
&\quad{} e^{i(\varphi - \vartheta(A))}   (\xi^D_{Y_{j+1}{Z}}
\psi_0^{(i_{j+1})}),
\quad Y , Y_1\ldots ,Y_{j+1} \in \Pi' \cap U({\frak{sl}}(2,{\Crm})),\cr
}$$
where the coefficients and the dependence of the arguments are as in
expression (6.239). According to Theorem A.1 there exists two functions
$\beta^{(i_{j+1})}_{(\varepsilon )X} \in C^\infty   ({\Rrm}^+ \times
{\Rrm}^3-\{0\})$, $\varepsilon =\pm $, given by formula (A.1), homogeneous of
degree $-3/2$ and with support in the forward light cone such that, if
$$(\phi^{(i_{j+1})}_{(\varepsilon )X} (t)) (x) = e^{i\varepsilon m
(t^2 - \vert x \vert^2)^{1/2}} \beta^{(i_{j+1})}_{(\varepsilon )X} (t,x), $$
then there exists $N\in {\Nrm}$ such that
$$\Vert (1 + \lambda_1(t))^{k/2}   \big(P_\varepsilon (-i \partial)
U^{D1}_{{\rm exp } (tP_0)} T^{D1}_X \alpha^{(i_{j+1})} -
\phi^{(i_{j+1})}_{(\varepsilon )X} (t)   \big)   \Vert^{}_D
\leq t^{-1} C_k \Vert \alpha^{(i_{j+1})}\Vert^{}_{D_{N + k + \vert X \vert }}$$
for $t > 0$, $k \in {\Nrm}$, $X \in \Pi'$, where $C_k$ is independent of
$t$ and $X$. This gives together with inequality (6.252), taking $0 < \tau \leq
1/2$ and after redefining $N_+$ by $\max(N_+ + 1,N)$ that, if
$$\eqalignno{
&Q_j \big(Y_1,\ldots ,Y_j ; i_1,\ldots ,i_j ; (t,x)\big)& (6.258{\rm a})\cr
&\quad{}= \big((\xi^{}_{Y_1}(\varphi ^{(i_1)} - \vartheta(A^{(i_1)})))\cdots
  (\xi^{}_{Y_j}   (\varphi^{(i_j)}-\vartheta(A^{(i_j)})))\big) (t,x),\cr
}$$
then
$$\eqalignno{
&\Vert (1+\lambda_1(t))^{k/2} Q_j \big(Y_1,\ldots ,Y_j ; i_1,\ldots ,i_j
; (t,\cdot)\big)
\big(P_\varepsilon (-i\partial )   (\xi^D_{Y_{j+1} {Z}}\
\psi_0^{(i_{j+1})}   ) (t) - \phi^{(i_{j+1})}_{(\varepsilon ) Y_{j+1}{
Z}} (t)\big)  \Vert^{}_D \cr
&\quad{}\leq t^{-1} C_{\vert Y \vert + \vert {Z} \vert + k + l }
\overline{{\cal R}}^{(l)}_{N_+,N_+ + \vert Y \vert +
\vert {Z} \vert + k + l } (u ;
u^{}_1,\ldots ,u^{}_l ),\quad  t>0,&(6.258{\rm b})\cr
}$$
where $Y_1,\ldots, Y_{j+1}$, $i_1, \ldots , i_{j+1}$ are as in expression
(6.239), where we have used the notation $ \overline{{\cal R}}^{(l)}$
as in expression (6.242) and where we have used that
$\vert Y_q \vert + i_q \geq 1$ for $q \leq j$.
Here $C_{\vert Y \vert + \vert {Z} \vert + k + l }$ depends only on $\rho$
and $\Vert u \Vert^{}_{E^\rho _{N_+}}$ for $\tau \in ]0,1/2]$ fixed.
Similarly it follows, using Theorem A.1 with $L^\infty$-norms, that
$$\eqalignno{
&\Vert(\delta (t))^{3/2 + 2 (1-\rho )}  (1 + \lambda_1 (t))^{k/2} Q_j
\big(Y_1,\ldots ,Y_j  ;  i_1,\ldots , i_j ; (t,\cdot)\big)&(6.258{\rm c})\cr
&\quad{}  \big(P_\varepsilon (-i \partial)   (\xi^D_{Y_{j+1} {Z}}
\psi_0^{(i_{j+1})}  ) (t) - \phi^{(i_{j+1})}_{(\varepsilon) Y_{j+1}{Z}}
(t)\big)\Vert^{}_{L^\infty} \cr
&\qquad{}\leq t^{- (2\rho -1)}  C_{\vert Y \vert + \vert {Z } \vert + k
+ l}  \overline{{\cal R} }^{(l)}_{N_+, N_+ + \vert Y \vert + \vert {Z } \vert
+ k + l}  (u; u^{}_1,\ldots , u^{}_l), \quad t > 0.\cr
}$$
Since
$$  \big({\vartheta} (H^{(l )}) - {\vartheta}^\infty (H^{(l )})   \big)(y) = -
\int^\infty_1  y^\mu  H^{(l )}_\mu (sy) ds,$$
for $y^\mu  y_\mu > 0$, $ y_0 > 0$, since $\xi^{}_Y $ $\chi^{}_1$ is uniformly
bounded in the half space $t \geq 0$ for $Y \in \Pi'$ and since
$(1 + \lambda_1 (t)) (x) \leq C (1+t / (t - \vert x \vert))$ inside the
forward light cone, it follows, using
the $sl (2, {\Crm})$ covariance of ${\vartheta}$ and ${\vartheta}^\infty$
and using inequality (6.248a) with $j=0$, that
$$\eqalignno{
&\big\vert   \big(\xi^{}_Y  \chi^{}_1   ({\vartheta} (H^{(l )}) -
{\vartheta}^\infty
(H^{(l)})  )  \big) (t,x) \big\vert &(6.259)\cr
&\quad{} \leq (1 + t)^{-1+\rho -\tau}  (1+t/(t-\vert x
\vert))^\tau
C_{\vert Y \vert + l}   \overline{{\cal R}}^{(l)}_{N_+, N_+ + \vert Y \vert +
l}  (u  ;  u^{}_1,\ldots , u^{}_l),\cr
}$$
for $(t,x) \in {\rm supp}\ \chi^{}_1$, $ l \in {\Nrm}$, $ Y \in \Pi' \cap
\cup (sl (2, {\Crm}))$, $ 0 \leq \tau \leq 1 - \rho$,
$ u \in {\cal O}_{1, \infty (+)}$, $ u^{}_1,\ldots , u^{}_l \in
E^{\circ\rho}_\infty$, where $C_{\vert Y
\vert + l}$ depends only on $\Vert u \Vert^{}_{E^\rho_{N_+}}$ and $\rho $.
It follows from inequality (6.247), from the fact that ${\vartheta}^\infty
(H^{(l )})$ is homogeneous of degree zero inside the forward light cone and
from inequality (6.249) that
$$\eqalignno{
&\big\vert   \big(\xi^{}_Y (1-\chi^{}_1) {\vartheta}^\infty (H^{(l )})
\big)(t,x) \big\vert
+ \big\vert   \big(\xi^{}_Y (1-\chi^{}_1) {\vartheta}(A^{(l )}-A_0^{(l )})
\big) (t,x)
\big\vert&(6.260)\cr
&\quad{}\leq (1+t/(t - \vert x \vert ))^\tau  C_{\vert Y \vert + l}
\overline{{\cal
R}}^{(l)}_{N_+, N_+ + \vert Y \vert + l}  (u  ;  u^{}_1,\ldots , u^{}_l),\cr
}$$
for $(t,x) \in {\rm supp}\ (1-\chi^{}_1)\cap \{(t,x) \big\vert t > 0$ and
$t > \vert x
\vert \}$, $Y\in \Pi'\cap U(sl (2,{\Crm}))$, $l \in {\Nrm}$, $u \in {\cal
O}_{1,\infty (+)}$, $u^{}_1, \ldots , u^{}_l \in E^{\circ\rho}_\infty$,
$\tau > 0$, where $C_{\vert Y \vert + l } $ depends only on $\tau$, $\rho $ and
$\Vert u \Vert^{}_{E^\rho_{N_+}}$. Since, inside the forward light cone
$$ \eqalignno{
& \big(D^l (\varphi - \vartheta(A)-\vartheta^\infty (A^{(+)}-A))\big) (u ;
u^{}_1,\cdots, u^{}_l )&(6.261)\cr
&\quad{}= \chi^{}_1 \big(\vartheta(H^{(l )}) - \vartheta^\infty (H^{(l )})
\big) + (1-\chi^{}_1)   \big(\vartheta^\infty (H^{(l )}) +
\vartheta(A^{(l )}_0 - A^{(l )})
\big )\cr
}$$
it follows from inequalities (6.259) and (6.260) that if $e$ is the
characteristic function of the support of $\chi^{}_1$ and $ \overline{e}$
the characteristic function of ${\rm supp}\ (1 - \chi^{}_1) \cap \{(t,x)
\big\vert t>0, \vert x \vert <t\}$, then
inequalities (6.259) and (6.260) give that
$$\eqalignno{
&\big\vert \big(\xi^{}_Y   (D^l (\varphi - \vartheta(A) -
\vartheta^\infty (A^{(+)}-A))
  ) (u  ;  u^{}_1,\ldots , u^{}_l)\big) (t,x) \big\vert&(6.262)\cr
&\quad{}\leq \big(e(t,x) (1+t)^{-1+\rho -\tau^{}_1} (1+t/(t - \vert x \vert
))^{\tau^{}_1}+ \overline{e} (t,x) (1+t/(t -
\vert x \vert ))^{\tau^{}_2}\big)\cr
&\qquad{}C_{\vert Y \vert + l}   \overline{{\cal R}}^{(l)}_{N_+, N_+ +
\vert Y \vert +
l}  (u  ;  u^{}_1,\ldots , u^{}_l),\cr
}$$
for $0\leq \vert x \vert <t$, $t>0$, $l \in {\Nrm}$, $Y \in \Pi' \cap
U({\frak{sl}}(2,{\Crm}))$, $u \in {\cal O}_{1,\infty (+)}$, $u^{}_1,\ldots,
u^{}_l \in
E^{\circ\rho}_\infty$, $0\leq \tau^{}_1\leq 1-\rho $, $\tau^{}_2>0$,
where $C_{\vert Y \vert +l}$ depends only on $\tau^{}_2$, $\rho $
and $\Vert u \Vert^{}_{E^\rho_{N_+}}$. Let
$$Q_j^\infty (Y_1,\ldots ,Y_j ; i_1,\ldots ,i_j ; (t,x))
=\big((\xi^{}_{Y_1} \vartheta^\infty (H^{(i_1)}))\cdots (\xi^{}_{Y_j}
\vartheta^\infty (H^{(i_j)}))\big)(t,x),\eqno{(6.263)}$$
where $Y_1,\ldots ,Y_j$, $i_1,\ldots ,i_j$ are as in expression (6.239).
Let $a^{}_q$ (resp. $b^{}_q)$ be the $q^{\rm th}$ factor in the product
defining $Q_j
(Y_1,\ldots ,Y_j ; i_1,\ldots ,i_j)$  (resp. $Q_j^\infty (Y_1,\ldots,Y_j ;
i_1,\ldots ,i_j)$) in expression (6.258a) (resp. (6.263)). Then
$$\vert (Q_j-Q_j^\infty) (Y_1,\ldots ,Y_j ; i_1,\ldots ,i_j) \vert
\leq C_j \sum_{1 \leq q \leq j} \vert a^{}_q-b^{}_q \vert M_q ,$$
where $M_q$ is a monomial of degree $j-1$ given by
$$M_q = \sum  c^{}_1\cdots c^{}_{q-1}  c^{}_{q+1} \cdots c^{}_q ,$$
where the domain of summation is defined by $c^{}_p \in \{\vert a^{}_p\vert,
\vert b^{}_p \vert \}$ for $p \in \{1,\ldots ,q-1\}\cup \{q+1,\ldots ,j\}$.
We estimate
$M_q(t,x)$ for $t>0$, $\vert x \vert <t$ by using inequality (6.252),
with $0<\tau \leq 1/2$, and by using inequality (6.249), with ${Z}={\un}$,
together with the fact that $\xi^{}_Y \vartheta^\infty (H^{(l)})$ is
homogeneous of degree zero. We estimate $\vert a^{}_q-b^{}_q \vert (t,x)$,
for $t>0$, $\vert x \vert <t$, by using inequality (6.262) with
$\tau^{}_1=1-\rho $ and $0<\tau^{}_2\leq 1/2$ and by observing
that $t^{2(1-\kappa)}\leq 4t/(t-\vert x \vert )$ in the support of
$ \overline{e}$ and that
$$\eqalignno{
&e(t,x) (1+t)^{-2(1-\rho )}(1+t/(t-\vert x \vert ))^{1-\rho }
+\overline{e}(t,x) (1+t/(t-\vert x \vert ))^{\tau^{}_2}\cr
&\quad{}\leq C(1+t)^{-2(1-\rho )}(1+t/(t-\vert
x \vert ))^{\tau^{}_2 + (1-\rho )/(1-\kappa)},\cr
}$$
for $t>0$, $\vert x \vert <t$, where $C$ is
independent of $(t,x)$. Since $t/(t-\vert x\vert)
\leq 2t^2/(t^2-\vert x \vert^2)$
for $t>0$, $\vert x \vert <t$, it then follows from Theorem A.1,
after redefinition of $N_+$, that
$$\eqalignno{
&\Vert(1+\lambda_1(t))^{k/2} (Q_j-Q_j^\infty ) (Y_1,\ldots ,Y_j ;
i_1,\ldots ,i_j ; (t,\cdot) ) \phi^{(i_{j+1})}_{(\varepsilon )Y_{j+1}
{Z}}(t)   \Vert^{}_D&(6.264{\rm a})\cr
&\quad{}\leq t^{-2(1-\rho )} C_{\vert Y \vert + \vert {Z} \vert + k+ l}
{\overline{\cal R}}^{(l)}_{N_+, N_+ + \vert Y
 \vert + \vert {Z} \vert + k + l }
(u; u^{}_1,\ldots , u^{}_l) \cr
}$$
and that
$$\eqalignno{
&\Vert (1 + \lambda_1(t))^{k/2} (Q_j - Q^\infty_j) (Y_1, \ldots ,Y_j
;  i_1, \ldots ,i_j ; (t,\cdot)) \phi^{(i_{j+1})}_{(\varepsilon ) Y_{j+1}
{Z}} (t)   \Vert^{}_{L^\infty }&(6.264{\rm b})\cr
&\quad{}\leq t^{-2(1-\rho ) - 3/2}  C_{\vert Y \vert + \vert {Z} \vert + k + l}
\
 \overline{{\cal R}}^{(l)}_{N_+, N_+ + \vert Y \vert + \vert {Z} \vert + k + l
}
 (u  ;  u^{}_1, \ldots , u^{}_l)\cr
}$$
$t > 0$, where  $Y_1,\ldots ,Y_{j+1} ; i_1,\ldots ,i_{j+1}$ are as in
expression (6.239), where $C_{\vert Y \vert + \vert {Z} \vert + k + l}$
depends only on $\rho $ and $\Vert u \Vert^{}_{E^\rho_{N_+}}$ for $\tau $ and
$\tau^{}_2$
fixed and where we have used that $\vert Y_q \vert + i_q \geq 1$ for $q \leq
j$.
$\xi^D_{Y{Z}}   ((D^l  \psi^{(+)}_0) (u ; u^{}_1, \ldots , u^{}_l
)  )$, $Y\in \Pi'\cap U(sl (2,{\Crm}))$, ${Z}\in \Pi'\cap U({\Rrm}^4)$
is a solution of the Dirac equation with initial condition
$T^{D1}_{Y{Z}}((D^{l }\alpha_+)(u ; u^{}_1,\ldots ,u^{}_l))$. Since we have
already proved statement i) of the proposition we can apply Theorem A.1 to
this solution. Hence there exists two functions $\beta^{(+)
(l)}_{(\varepsilon ) Y{Z}}\in C^\infty ({\Rrm}^+ \times {\Rrm}^3-\{0\})$,
$\varepsilon = \pm $, given by formula (A.1), homogeneous of
degree $-3/2$ and with support contained in the forward light cone such that,
if
$$\eqalignno{
  (\phi^{(+)(l )}_{(\varepsilon ) Y{Z}}(t)  ) (x) &=
e^{i\varepsilon m (t^2-\vert x \vert ^2)^{1/2}} \beta^{(+)
(l)}_{(\varepsilon) Y{Z}} (t,x),\cr
\alpha^{(+)(l )}&=(D^l \alpha_+)(u ; u^{}_1,\ldots ,u^{}_l ),\cr
}$$
then there exists $N \in {\bf N}$ such that
$$\eqalignno{
&  \Vert (1 + \lambda_1(t))^{k/2}    \big(P_\varepsilon (-i \partial )
U^{D1}_{{\rm exp }(tP_0)} T^{D1}_{Y{Z}} \alpha^{(+) (l)} -
\phi^{(+)(l )}_{(\varepsilon ) Y{Z}} (t)\big)\Vert^{}_D\cr
&\quad{}\leq t^{-1} C_k \Vert T^{D1}_{Y{Z}} \alpha^{(+) (l)}
\Vert^{}_{D_{N+k}},\cr
}$$
for $t > 0$, $k \in {\Nrm}$, where $C_k$ is independent of $t$. Theorem A.1
also gives a similar statement for the $L^\infty $-norm. Hence according to
statement i) of this proposition
$$\eqalignno{
&t \Vert (1 + \lambda_1(t))^{k/2}   \big(P_\varepsilon (-i \partial )
U^{D1}_{\exp(tP_0)} T^{D1}_{Y{Z}} \alpha^{(+)(l)} -
\phi^{(+)(l)}_{(\varepsilon ) Y{Z}} (t)\big) \Vert^{}_D&(6.265)\cr
&\quad{}+ t^{-1+2\rho } \Vert (\delta (t))^{3/2+2(1-\rho)}
(1+\lambda_1(t))^{k/2}   \big(P_\varepsilon (-i\partial ) U^{D1}_{\exp(tP_0)}
T^{D1}_{Y{Z}} \alpha^{(+)(l )}-\phi^{(+)(l)}_{(\varepsilon)Y{Z}}(t)\big)
\Vert^{}_{L^\infty }\cr
&\qquad{}\leq C_{\vert Y \vert +\vert {Z}\vert +k+ l}   \overline{{\cal
R}}^{(l)}_{N_+, N_+ + \vert Y \vert + \vert {Z}\vert +l +k}  (u  ;
u^{}_1,\ldots , u^{}_l),\cr
}$$
for $t \geq 1$, $\varepsilon = \pm$, $Y \in \Pi' \cap
U({\frak{sl}}(2,{\Crm}))$,
${Z} \in \Pi' \cap U({\Rrm}^4)$, $k \in \Nrm$, $l \in \Nrm$, $u
\in {\cal O}_{1,\infty (+)}$, $u^{}_1, \ldots ,u^{}_l \in
E^{\circ\rho}_\infty $, where $C_{\vert Y \vert + \vert {Z} \vert + k + l }$
depends only on $\rho $ and $\Vert u \Vert^{}_{E^\rho _{N_+}}$. We have here
suitably redefined $N_+$. It follows from (A.1a) that
$\beta ^{(+)(0)}_{(\varepsilon ){Z}}=e^{i\vartheta^\infty (A^{(+)}-A)}
\beta ^{(0)}_{(\varepsilon ){Z}}$. The
construction (A.1a)--(A.3) is sl$(2,{\Crm})$ covariant, which shows that
$$\beta ^{(+)(0)}_{(\varepsilon ) Y{Z}} = \xi^D_Y \big((D^l \beta
^{(+)(0)}_{(\varepsilon ){Z}})(u ;u^{}_1,\ldots ,u^{}_l)\big),\quad
Y \in \Pi' \cap U (sl (2,{\Crm})).$$
Since $\xi^{}_X \Lambda = 0$ for $\Lambda (t,x) = \sqrt {t^2 - \vert x
\vert^2}$, $t > 0$, $\vert x \vert < t$, $X \in {_frak{sl}}(2,{\Crm})$, it
follows that
$\sum_\varepsilon \phi^{(+)(l)}_{(\varepsilon) Y{Z}}$ is the sum of the
expressions (6.257) with $\varphi^{(i_q)}-\vartheta(A^{(i_q)})$
replaced by $\vartheta^\infty (H^{(i_q)})$ and $\xi^D_{Y_{j+1} {Z}}
\psi_0^{(i_{j+1})}$ replaced by $\sum_\varepsilon
\phi_{(\varepsilon ) Y_{j+1}{Z}}^{(i_{j+1})}$. Hence, inequality (6.256),
development (6.257) and inequalities (6.258b), (6.258c), (6.264a), (6.264b) and
(6.265) give that
$$\eqalignno{
&(1+t)^{2(1-\rho )}\Vert (1 +\lambda_1(t))^{k/2}
\big(\xi^D_Y((D^l (e^{i(\varphi -\vartheta(A))} \psi_0 - \psi^{(+)}_0) )
(u ; u^{}_1,\ldots, u^{}_l))\big) (t) \Vert^{}_D&(6.266)\cr
&\quad{}
+ \Vert (\delta (t))^{3/2+2 (1-\rho)} (1+\lambda_1(t))^{k/2}
\big(\xi^D_Y ((D^l (e^{i (\varphi-\vartheta (A))} \psi_0 - \psi_0^{(+)}))
(u ; u^{}_1,\ldots, u^{}_l))\big) (t) \Vert^{}_{L^\infty}\cr
&\qquad{}\leq C_{ \vert Y \vert + k + l }
\overline{{\cal R}}^{(l)}_{N_+, N_+ +
\vert Y \vert + k + l} (u ; u^{}_1,\ldots, u^{}_l),\cr
}$$
for $t \geq 0$, $Y \in \Pi'$, $k$, $l \in \Nrm$,
$u \in {\cal O}_{1,\infty (+)}$,
$u^{}_1,\ldots, u^{}_l \in E^{\circ\rho}_\infty$,
where $C_{\vert Y \vert + k + l}$ depends only on $\rho$ and $\Vert u
\Vert^{}_{E^\rho_{N_+}}$. Since
$$e^{i\varphi} \psi - \psi_0^{(+)} = e^{i(\varphi - \vartheta (A))}
(e^{i\vartheta (A)} \psi - \psi_0)
+ e^{i (\varphi - \vartheta (A))} \psi_0 - \psi_0^{(+)},$$
inequalities (6.254) and (6.266) and theorem 6.10 for $A$ give that
$$\eqalignno{
&\vvv (D^l (A - A_0^{(+)}, e^{i\varphi} \psi - \psi_0^{(+)})) (u ;
u^{}_1,\ldots, u^{}_l) \vvv_{\rho', \varepsilon, L}&(6.267)\cr
&\qquad{}\leq C_{L + l} \overline{{\cal R}}^{(l)}_{N_+, N_+ + L + l}
(u ; u^{}_1,\ldots, u^{}_l)  , \cr
}$$
for $L$, $l \in \Nrm$, $1/2 < \rho' \leq 1$, $\varepsilon (0) > 0$,
$\varepsilon (1) \geq \rho$, $u \in {\cal O}_{1, \infty (+)}$,
$u^{}_1,\ldots, u^{}_l \in
E^{\circ\rho}_\infty$, where $C_{L + l}$ depends only on $\rho'$,
$\varepsilon$,
$\rho$ and $\Vert u \Vert^{}_{E^\rho_{N_+}}$, where
$u^{}_+ = (f_+, {\dot f}_+, \alpha_+) = E_+ (u)$, $(A_0^{(+)},
{\dot A}^{(+)}_0, \psi_0^{(+)}) (t) =
U^1_{\exp (t P_0)} u^{}_+$ and where $(A, {\dot A}, \psi) = h (v)$ is the
solution of the M-D equations given by theorem 6.15 with $v =
\Omega_1^{(+)}(u)$.

To prove that $E_+\colon {\cal O}_{1,\infty(+)}
\fl E^{\circ\rho}_\infty$ has a differentiable
inverse $E_+^{-1}\colon {\cal O}_{\infty (+)} \fl {\cal O}_{1, \infty (+)}$,
where
${\cal O}_{\infty (+)}$ is an open neighbourhood of zero in
$E^{\circ\rho}_\infty$, we shall use the implicit function theorem
for Fr\'echet spaces. We therefore have to prove that the linear continuous
operator $DE_+ (u) \in L (E^{\circ\rho}_\infty, E^{\circ\rho}_\infty)$
has a right inverse $w(u) \in L (E^{\circ\rho}_\infty, E^{\circ\rho}_\infty)$,
i.e. $(DE_+)  (u ; w(u) V_+) =
V_+$ for $u \in {\cal O}_{1, \infty (+)}$ and $V_+ \in E^{\circ\rho}_\infty$.
Since $\Omega^{(+)}_1\colon {\cal O}_{1, \infty (+)} \fl {\cal U}_{\infty (0)}$
is a diffeormorphism and if $V' (u)$ is a tangent vector of ${\cal U}_{\infty
(0)}$ at the point $\Omega^{(+)}_1 (u)$ satisfying the equation $(DG) (u ; V'
(u)) = V_+$ for $V_+ \in E^{\circ\rho}_\infty$, where $G = E_+ \circ
(\Omega^{(+)}_1)^{-1}$ then it follows by the chain rule that $V(u) =
(D(\Omega^{(+)}_1)^{-1} (\Omega^{(+)}_1 (u) ; V' (u))$
is a solution of the equation $(DE_+) (u ;
V(u)) = V_+$. Let $h (\Omega^{(+)}_1 (u)) = (A, \dot{A}, \psi)$ be the solution
of the M-D equations, with initial condition $\Omega^{(+)}_1 (u) = (A(0),
\dot{A} (0), \psi (0))$ at $t = 0$, given by Theorem 6.15. If $V(u) \in
E^{\circ\rho}_\infty$ and $(D(h \circ \Omega^{(+)}_1)) (u ; V(u)) = (a,
\dot{a},
\Psi)$, then $(a, \dot{a}, \Psi)$ is the solution of the equations given by the
derivative of the M-D equations, i.e.
$${d \over dt}	(a(t), \dot{a} (t),\Psi (t)) = (D T_{P_0})
 \big((A(t), \dot{A} (t), \psi (t));(a(t), \dot{a} (t), \Psi (t))\big),
\eqno{(6.268{\rm a})}$$
with initial data at $t = 0$ given by
$$(a(0), \dot{a} (0), \Psi (0)) = V' (u) = (D \Omega^{(+)}_1) (u ; V(u)).
\eqno{(6.268{\rm b})}$$
Let $\tilde{h} (u) = (A, \dot{A}, e^{i \varphi} \psi),	\tilde{\psi}
= e^{i \varphi} \psi$ and $(D \tilde{h}) (u ; V(u)) = (a, \dot{a},
\tilde{\Psi})$. Since $\varphi (t) = 0$ for $0 \leq t \leq 2$ it follows from
(6.268a) and (6.268b) that
$$\eqalignno{
\big(i \gamma^\mu  \partial_\mu + m - \gamma^\mu (A_\mu -
\partial_\mu \varphi)\big)
 \tilde{\Psi} &= \gamma^\mu \big(a^{}_\mu -
 \partial_\mu (D \varphi) (u ; V(u))\big)
\tilde{\psi},&(6.269{\rm a})\cr
\carre a^{}_\mu &= {\tilde{\Psi}}^+ {\gamma_0}  {\gamma_\mu}  {\tilde{\psi}}
+ {\tilde{\psi}^+}  {\gamma_0}  {\gamma_\mu}  {\tilde{\Psi}},&(6.269{\rm b})\cr
\noalign{\noindent and}
(a(0), \dot{a} (0), \tilde{\Psi} (0)) &= V' (u). &(6.269{\rm c})\cr
}$$
Moreover defining $V_+ (u) = (DG) (u ; V' (u))$ then $V_+ (u) = (DE_+ (u ;
V(u))$ by the definition of $G$ and so it follows from inequality (6.267) that
$$\Vert (a(t), \dot{a}(t), \tilde{\Psi} (t)) - U^1_{\exp (t P_0)}
V_+ (u) \Vert^{}_D
\fl 0, \eqno{(6.269{\rm d})}$$ when $t \fl \infty$. Therefore,
if for fixed $V_+ \in E^{\circ\rho}_\infty$ equations (6.269a) and
(6.269b) have, for each $u \in
{\cal O}_{1, \infty (+)}$, a solution $(a, \dot{a}, \tilde{\Psi})$, in a
suitable space, satisfying the asymptotic condition (6.269d), then $V'(u)$
defined by (6.269c) is a solution of the equation $(DG) (u ; V'(u)) = V_+$. Let
$V_+ = (g^{}_+, \dot{g}_+, \beta_+) \in E^{\circ\rho}_\infty$, $\phi' (t) =
e^{i
(\varphi - {\vartheta} (A))}  U^{D1}_{\exp (tP_0)}  \beta_+$,  $\psi' =
e^{i {\vartheta} (A)} \psi$,  $\Psi' = e^{i ({\vartheta} (A) - \varphi)}
(\tilde{\Psi} - U^{D1}_{\exp (tP_0)}  \beta_+)$ and let $b^{}_\mu =
\partial_\mu ((D \varphi) (u ; V(u)))$. Equations (6.269a) and
(6.269b) give that
$$\eqalignno{
\big(i \gamma^\mu  \partial_\mu + m - \gamma^\mu (A_\mu -
\partial_\mu  {\vartheta} (A))\big) \Psi'
&= \gamma^\mu \big(a^{}_\mu - b^{}_\mu)  \psi' +
\gamma^\mu (A_\mu - \partial_\mu \varphi)  \phi',&(6.270{\rm a})\cr
\carre a^{}_\mu &= (\Psi' + \phi')^+  \gamma_0  \gamma_\mu  \psi' + (\psi')^+
 \gamma_0  \gamma_\mu (\Psi' + \phi')\hskip10mm&(6.270{\rm b})\cr
}$$
and the asymptotic condition (6.269d) gives that
$$\Vert \Psi' (t) \Vert^{}_D + \Vert (a(t), \dot{a} (t)) - U^{M1}_{\exp (tP_0)}
(g^{}_+, \dot{g}_+) \Vert^{}_{M^\rho_0} \fl 0, \eqno{(6.270{\rm c})}$$
when $t \fl \infty$. Denoting for the moment by $\alpha \mapsto J^{(+)}
(\alpha)$
the function defined in (1.21). Since $\hat{\alpha} (k) = c(k)  \hat{\alpha}_+
(k)$, where $c(k) \in \Crm$ and $\vert c(k) \vert = 1$, it follows that
$J^{(+)} (\alpha) = J^{(+)} (\alpha_+)$. According to the definition of
$\varphi$, we therefore obtain that
$$b^{}_\mu = \partial_\mu ((D \varphi) (u^{}_+ ; V_+)), \eqno{(6.271)}$$
where $u^{}_+ = E_+ (u)$. We shall prove that system (6.270a)--(6.270b)
have a solution $(a, \Psi')$ with finite norms $\vvv (a - a_0, \Psi')
\vvv^{}_{\rho',r,L}$ for $L \in \Nrm$,  $1/2 < \rho' \leq 1$ and $r = (r(0),
r(1))$,  $r(0) > 0$, $r(1) \geq \rho$. This is done by using Theorem 5.14 with
$t_0 = \infty$ and by using inequalities (6.176) and (6.177). Since the proof
is
standard at this point, though tedious, we only state the result, which after
taking $N_+$ sufficiently large and ${\cal O}_{1, \infty (+)}$ sufficiently
small, reads:
$$\vvv (D^l (a - a_0, \Psi'))  (u ; u^{}_1,\ldots,u^{}_l) \vvv^{}_{\rho',r,L}
\leq C_{L+l}  \overline{{\cal R}}^{(l+1)}_{N_+,N_++L+l}
(u ; u^{}_1,\ldots,u^{}_l,V_+),
\eqno{(6.273)}$$
for $L$, $l \in \Nrm$,  $1/2 < \rho' < 1$,  $r = (r(0), r(1))$,
$r(0) > 0$,  $r(1) \geq \rho$, $ u \in {\cal O}_{1, \infty (+)}$,
$u^{}_1,\ldots,u^{}_l \in
E^{\circ\rho}_\infty$, where $C_{L+l}$ depends only on $\rho'$, $r$,
$\rho$ and $\Vert u
\Vert^{}_{E^\rho_{N_+}}$ and where $(a_0(t), \dot{a}_0 (t)) =$
\penalty-1000 $ U^{M1}_{\exp (tP_0)}
(g, \dot{g})$. It then follows using Lemma 2.19 that, if $V' (u)$ is defined by
(6.269), then
$$\Vert (D^l V')  (u ; u^{}_1,\ldots,u^{}_l) \Vert^{}_{D_n}
\leq C_{n+l}  \overline{{\cal R}}^{(l+1)}_{N_+,N_++n+l}  (u ; u^{}_1,
\ldots,u^{}_l,V_+), \eqno{(6.274)}$$
for $n$, $l \in \Nrm$,  $V_+ \in E^{\circ\rho}_\infty$,
$u \in {\cal O}_{1, \infty (+)}$, $u^{}_1,\ldots,u^{}_l \in
E^{\circ\rho}_\infty$, where $C_{n+l}$ depends only on
$\rho$ and $\Vert u \Vert^{}_{E^\rho_{N_+}}$. Defining $V(u) = (D
(\Omega^{(+)}_1)^{-1})  (\Omega^{(+)}_1 (u) ; V' (u))$, Theorem 6.13 gives,
after redefining $N_+$, that
$$\Vert (D^l V)  (u ; u^{}_1,\ldots,u^{}_l) \Vert^{}_{D_n}
\leq C_{n+l}  \overline{{\cal R}}^{(l+1)}_{N_+,N_++l+n}
(u ; u^{}_1,\ldots,u^{}_l,V_+),
\eqno{(6.275)}$$
for $n$, $l \in \Nrm$,  $V_+ \in E^{\circ\rho}_\infty$,
$u \in {\cal O}_{1, \infty (+)}$, $u^{}_1,\ldots,u^{}_l \in
E^{\circ\rho}_\infty$, where $C_{n+l}$ depends only on
$\rho$ and $\Vert u \Vert^{}_{E^\rho_{N_+}}$. It follows from inequality
(6.275)
that $w(u)$, defined by $V(u) = w(u)  V_+$, is a right inverse of $DE_+(u)$,
i.e.
$(DE_+) (u ; V(u)) = V_+$, satisfying the hypotheses of the implicit function
theorem in Fr\'echet spaces (Theorem 4.1.1. of \refSERG), so there exists an
integer $M_+$, an open neighbourhood ${\cal O}_{M_+(+)}$ of zero in
$E^{\circ\rho}$ and a $C^\infty$ map $H\colon {\cal O}_{\infty (+)} =
{\cal O}_{M_+ (+)} \cap
E^{\circ\rho}_\infty \fl {\cal O}_{1, \infty (+)}$ such that $E_+(H(u)) = u$
for
$u \in {\cal O}_{\infty (+)}$. A similar discussion as in the end of the proof
of Theorem 6.13 shows that ${\cal O}_{\infty (+)}$ and ${\cal O}_{1, \infty
(+)}$ can be chosen such that $E_+\colon {\cal O}_{1, \infty (+)} \fl {\cal
O}_{\infty (+)}$ is a diffeomorphism and such that the inequality of statement
ii) of the proposition is satisfied. This proves statement ii).

Statement iii) follows from statement ii) and inequality (6.267). Statement
iv) is a particular case of statement iii). This proves the proposition.

We could, of course, have defined a map $E_-$ which satisfies the suitably
modified statements in Proposition 6.16, when $t \fl - \infty$. We can
therefore define two new wave operators $\Omega^{(\varepsilon)}\colon {\cal
O}_{\infty (\varepsilon)} \fl {\cal U}_{\infty (0)}$, where
$$\Omega^{(+)} = \Omega^{(+)}_1 \circ E^{-1}_+, \quad \Omega^{(-)} =
\Omega^{(-)}_1 \circ E^{-1}_-. \eqno{(6.276)}$$

We shall extend the diffeomorphisms
$\Omega^{(\varepsilon)}\colon {\cal O}_{\infty
(\varepsilon)} \fl {\cal U}_{\infty (0)}$,  $\varepsilon = \pm$,
to a diffeomorphism
$\Omega_\varepsilon\colon {\cal O}^{(\varepsilon)}_\infty \fl {\cal U}_\infty$,
where ${\cal U}_{\infty (0)} \subset {\cal U}_\infty$ (resp. ${\cal O}_{\infty
(\varepsilon)} \subset {\cal O}^{(\varepsilon)}_\infty$) is an open
neighbourhood of zero in $V^\rho_\infty$ (resp. $E^{\circ\rho}_\infty$), such
that the nonlinear representation $X \mapsto T_X$ of the Poincar\'e Lie algebra
is integrable to a {\it nonlinear representation $g \mapsto U_g$ of the
Poincar\'e group} ${\cal P}_0$ on ${\cal U}_\infty$ and such that
$\Omega_\varepsilon$, $\varepsilon = \pm$, are modified wave operators.
We shall do this in two steps.
First we construct, using the explicit covariance of the M-D equations, an
invariant set $S$ of solutions of the M-D equations and prove that each
solution in $S$ has an initial condition at $t = 0$ in the manifold
$V^\rho_\infty$  and satisfies
asymptotic conditions analog to those in statement iv) of Proposition 6.16
when $\varepsilon t \fl \infty,  \varepsilon = \pm$. Second $U_g$ is defined
by the action of ${\cal P}_0$ on $S$, the extension $\Omega_\varepsilon$ is
defined by considering $U_g \circ \Omega^{(\varepsilon)} \circ
U^{(\varepsilon)}_{g-1}$ and $\Omega_\varepsilon$ is proved to be a
diffeomorphism by using the implicit function theorem in Fr\'echet spaces. With
suitably chosen finite-dimensional matrix representations $\Lambda \mapsto
V(\Lambda)$,  $\Lambda \mapsto V'(\Lambda)$ of $SL (2, \Crm)$,
solutions $(A, \psi)$ of the M-D equations transform under
$g = (a, \Lambda) \in {\cal P}_0$ as
$$\eqalignno{
A_g (y) &= V(\Lambda)  A(V(\Lambda)^{-1}  (y-a))& (6.277{\rm a})\cr
\noalign{\noindent and}
\psi_g (y) &= V'(\Lambda)  \psi (V(\Lambda)^{-1}  (y-a)),\quad  y \in \Rrm^4.
&(6.277{\rm b})\cr
}$$
For $v \in {\cal U}_{\infty (0)}$ and $h(v) = (A, {\dot A}, \psi)$ as in
Theorem
6.15, we introduce the notation, $h_g (v) = (A_g, {\dot A}_g, \psi_g),$
$$S_0 = \{ h_e (v) \big\vert v \in {\cal U}_{\infty (0)} \},\quad
S = \{ h_g (v) \big\vert v \in {\cal U}_{\infty (0)},  g \in {\cal P}_0 \},
\eqno{(6.277{\rm c})}$$
$${\cal U}_\infty = \{ (h_g (v)) (t, .) \big\vert h_g (v) \in S, t = 0 \},
\eqno{(6.277{\rm d})}$$
where $e$ is the identity element in ${\cal P}_0$. It follows from Theorem 6.15
and the definitions of $S$ and ${\cal U}_\infty$ that $S \subset C^\infty
(\Rrm^4, \Rrm^4 \oplus \Rrm^4 \oplus \Crm^4)$ and ${\cal U}_\infty \subset
C^\infty(\Rrm^3, \Rrm^4 \oplus \Rrm^4 \oplus \Crm^4)$ respectively.
\saut
\noindent{\bf Lemma 6.17.}
{\it
Let $\Gamma_g (A, \psi) = (A_g, \psi_g)$ be defined by (6.277a) and (6.277b).
In
the situation of statement iii) of proposition 6.16 for $t \fl \infty$ and its
analog for $t \fl - \infty$, there exists $N_+$ such that
$$\eqalignno{
&\vvv \Gamma_g \big((D^l (A - A^{(+)}_0,  e^{i \varphi} \psi -
\psi^{(+)}_0))  (u^{}_+; u^{}_{+1},\ldots,u^{}_{+l})\big) \vvv^0_{\rho',r,L}\cr
&\quad{}\leq C_{L+l}  \big({\cal R}^l_{N_+,L+l}  (u^{}_{+1},\ldots,u^{}_{+l})
+ \Vert u^{}_+ \Vert^{}_{E^\rho_{N_++L+l}}
\Vert u^{}_{+1} \Vert^{}_{E^\rho_{N_+}}\cdots
\Vert u^{}_{+l} \Vert^{}_{E^\rho_{N_+}}\big),\cr
}$$
for $L$, $l \in \Nrm$,  $1/2 < \rho' \leq 1$,  $r = (r(0), r(1))$,
$r(0) > 0$, $r(1) \geq \rho$,  $u^{}_+ \in {\cal O}_{\infty (+)}$,
$u^{}_{+1},\ldots, u^{}_{+l} \in
E^{\circ\rho}_\infty$, where $C_{L+l}$ depends only on $\rho'$, $r$, $\rho$ and
$\Vert u^{}_+\Vert^{}_{E^\rho_{N_+}}$ and where
$$\eqalignno{
\vvv (a, \Phi) \vvv^0_{\rho',r,L} &= \sum_{i \in \{ 0,1 \}}
\sum_{\scr Y \in \ssg^i\atop\scr  k + \vert Y \vert \leq L}  \sup_{t \geq 0}
\Big((1+t)^{\rho'-1/2}  \Vert (\xi^M_Y a, \xi^M_{P_0 Y} a) (t)
\Vert^{}_{M^{\rho'}_0}\cr
&\qquad{}+ (1+t)^{2(1-\rho)}  \Vert (1+\lambda_1 (t))^{k/2}
(\xi^D_Y  \Phi) (t) \Vert^{}_{D_0}\cr
&\qquad{}+ \Vert (\delta (t))^{1+i-r(i)}  (\xi^M_Y a) (t)
\Vert^{}_{L^\infty}\cr
&\qquad{}+ \Vert (\delta (t))^{3/2+2(1-\rho)}  (1+\lambda_1 (t))^{k/2}
(\xi^D_Y
\Phi) (t) \Vert^{}_{L^\infty}\Big).\cr
}$$
}\saut
\noindent{\it Proof.}
For a moment we write $\lambda_1 (y)$ (resp. $\delta (y)$) instead of
$(\lambda_1 (t)) (x)$ (resp. $(\delta (t)) (x)$) for $y = (t,x) \in \Rrm^4$. It
follows from the definition of $\lambda_1$ and $\delta$ that there exists a
constant $C_g > 0$ such that if $g = (a, \Lambda) \in {\cal P}_0$ then
$$\eqalignno{
C^{-1}_g (1+\lambda_1 (y)) &\leq 1 + \lambda_1 (\Lambda^{-1} (y-a)) \leq C_g (1
+ \lambda_1 (y))& (6.278{\rm a})\cr
\noalign{\noindent and}
C^{-1}_g \delta (y) &\leq \delta (\Lambda^{-1} (y-a)) \leq C_g
\delta (y),&(6.278{\rm b})\cr
}$$
for each $y \in \Rrm^4$. It follows from these two inequalities that, if
$$\Delta^{D(l)}_Y = \xi^D_Y (D^l(e^{i \varphi} \psi - \psi^{(+)}_0))  (u^{}_+ ;
u^{}_{+1},\ldots, u^{}_{+l}),\quad  Y \in \Pi', \eqno{(6.279{\rm a})}$$
then
$$\eqalignno{
&\sup_{y \in \Rrm^+ \times \Rrm^3}  \Big((\delta (y))^{3/2 + 2(1-\rho)}  (1 +
\lambda_1 (y))^{k/2}  \big\vert V' (\Lambda)  \Delta^{D(l)}_Y
(V(\Lambda)^{-1} (y-a)) \big\vert\Big)&(6.279{\rm b})\cr
&\quad{}\leq C_g \Big(\sup_{y \in H_1}  \big((\delta (y))^{3/2 + 2(1-\rho)}
(1 + \lambda_1 (y))^{k/2}  \big\vert \Delta^{D(l)}_Y (y) \big\vert\big)\cr
&\qquad{}+ \sup_{y \in H_2}  \big((\delta (y))^{3/2 + 2(1-\rho)}  (1 +
\lambda_1
(y))^{k/2}\big\vert  \Delta^{D(l)}_Y (y) \big\vert\big)\Big),
\quad g \in {\cal P}_0,\cr
}$$
where $H_1 = \{ y \in \Rrm^4 \big\vert y_0 \geq 0$ and $V(\Lambda) y+a
\in \Rrm^+ \times \Rrm^3 \}$ and $H_2 = \{ y \in \Rrm^4 \big\vert y_0
\leq 0$ and $V(\Lambda)
y+a \in \Rrm^+ \times \Rrm^3 \}$. Since $H_2 \cap \{ y \in \Rrm^4
\big\vert y^\mu y_\mu \geq 0 \}$ is a bounded region of $\Rrm^4$,
it follows that $\delta (y) \leq C_g (1 + \lambda_1 (y))$ for $y \in H_2$.
Hence the second term on the right-hand side is, according to the analog
of Proposition 6.16 for $t \fl -
\infty$ and Theorem 5.7 applied to a free field, majorized by
$$\eqalignno{
&C_g  \sup_{y \in H_2}  \big((\delta (y))^{3/2}  (1 + \lambda_1
(y))^{k/2+2(1-\rho)}  \big\vert \Delta^{D(l)}_Y (y) \big\vert\big)\cr
&\quad{}\leq C_{g,\vert Y \vert+k+l}  \big({\cal R}^l_{N_+,\vert Y \vert+k+l}
(u^{}_{+1},\ldots,u^{}_{+l})
+ \Vert u^{}_+ \Vert^{}_{E^\rho_{N_++\vert Y \vert+k+l}}  \Vert u^{}_{+1}
\Vert^{}_{E^\rho_{N_+}}\cdots\Vert u^{}_{+l} \Vert^{}_{E^\rho_{N_+}}\big),\cr
}$$
where $N_+$ has been suitably chosen. Application of Theorem 6.16 to the first
term on the right-hand side of inequality (6.279b) then gives that
$$\eqalignno{
&\sup_{y \in \Rrm^+ \times \Rrm^3}  \big((\delta (y))^{3/2 + 2(1-\rho)}  (1 +
\lambda_1 (y))^{k/2}  \big\vert V' (\Lambda)  \Delta^{D(l)}_Y
(V(\Lambda)^{-1} (y-a))
\big\vert\big)&(6.280)\cr
&\quad{}\leq C_{g,\vert Y \vert+k+l}  \big({\cal R}^l_{N_+,\vert Y \vert+k+l}
(u^{}_{+1},\ldots,u^{}_{+l})
+ \Vert u^{}_+ \Vert^{}_{E^\rho_{N_++\vert Y \vert+k+l}}  \Vert u^{}_{+1}
\Vert^{}_{E^\rho_{N_+}} \cdots\Vert u^{}_{+l} \Vert^{}_{E^\rho_{N_+}}\big),\cr
}$$
where $C_{g,\vert Y \vert+k+l}$ depends only on $\rho$ and $\Vert u^{}_+
\Vert^{}_{E^\rho_{N_+}}$. Let
$$\Delta^{M(l)}_Y = \xi^M_Y \big((D^l (A - A^{(+)}_0))  (u^{}_+ ;
u^{}_{+1},\ldots,u^{}_{+l})\big). \eqno{(6.281{\rm a})}$$
Then it follows from inequalities (6.278a) and (6.278b) that
$$\eqalignno{
&\sup_{y \in \Rrm^+ \times \Rrm^3}  \big((\delta (y))^{1+i-r(i)}  \big\vert V
(\Lambda)  \Delta^{M(l)}_Y (V(\Lambda)^{-1} (y-a)) \big\vert\big)&(6.281{\rm
b})\cr
&\quad{}\leq C_g \Big(\sup_{y \in H_1}  \big((\delta (y))^{1+i-r(i)}
\big\vert \Delta^{M(l)}_Y (y)\big\vert\big)
+ \sup_{y \in H_2}  \big((\delta (y))^{1+i-r(i)}  \big\vert \Delta^{M(l)}_Y (y)
\big\vert\big)\Big),\cr
}$$
where $Y \in \sg^i$,  $i \in \{ 0,1 \}$. Let $F(t) = A^{(+)}_0 (t) -
A^{(-)}_0 (t)$,  $t \in \Rrm$, where $A^{(+)}_0$ is given by Proposition 6.16
and
$A^{(-)}_0$ by the analog of Proposition 6.16 for $t \fl - \infty$. By the
definition of $A^{(\varepsilon)}_0$,  $\carre A^{(\varepsilon)}_0 = 0$ for
$\varepsilon = \pm$, so $\carre  \xi^M_Y  F = 0$ for $Y \in \Pi'$. Since
$(\xi^M_Y F) (0) = (\xi^M_Y (A^{(+)}_0 - A)) (0) - (\xi^M_Y (A^{(-)}_0 - A))
(0)$, it follows from Proposition 6.16 and its analog for $t \fl - \infty$ that
$((\xi^M_Y F^{(l)}) (0),  (\xi^M_{P_0 Y} F^{(l)}) (0)) = T^{M1}_Y ((F^{(l)}
(0), \dot{F}^{(l)} (0)) \in M^{\rho'}_0$ for $Y \in \Pi'$,  $1/2 < \rho' \leq
1$, where $\dot{F}^{(l)} (0) = (\xi^M_{P_0} F^{(l)}) (0)$ and $F^{(l)}$ is the
$l^{\rm th}$ derivative of $F$. Hence, by the definition of the spaces
$M^{\rho'}_n$,
$n\geq 0$, $(F^{(l)} (0), \dot{F}^{(l)} (0)) \in M^{\rho'}_n$ for
$n \geq 0$ and
$1/2 < \rho' \leq 1$. Proposition 2.15 then gives that
$$\eqalignno{
&(1+\vert t \vert+\vert x \vert)^{3/2 - \rho'}  \vert (\xi^M_Y F^{(l)}) (t,x)
\vert + (1+\vert t \vert+\vert x \vert)(1+\big\vert \vert t \vert-\vert x
\vert \big\vert)^{3/2 - \rho'}  \vert (\xi^M_{P_\mu Y} F^{(l)}) (t,x) \vert\cr
&\qquad{}\leq C_{\vert Y \vert, \rho'}  \Vert (F^{(l)} (0), \dot{F}^{(l)} (0))
\Vert^{}_{M^{\rho'}_{\vert Y \vert + 2}},\cr
}$$
for $Y \in \Pi'$,  $0 \leq \mu \leq 3$ and $1/2 < \rho' < 1$. Since $\delta
(y) \leq C_g (1+\big\vert \vert y_0 \vert - \vert \vec{y} \vert \big\vert)$
for $y = (y_0, \vec{y}) \in H_2$ it follows, using the bounds in
Proposition 6.16 and its analog when $t \fl - \infty$ that
$$\eqalignno{
&\sup_{y \in H_2}  \big((\delta (y))^{1+i-r(i)}  \vert (\xi^M_Y F^{(l)}) (y)
\vert\big)&(6.282)\cr
&\quad{}\leq C_{g,\vert Y \vert+l}  \big({\cal R}^l_{N_+,\vert Y \vert+l}
(u^{}_{+1},\ldots,u^{}_{+l})
+ \Vert u^{}_+ \Vert^{}_{E^\rho_{N_++\vert Y \vert+l}}  \Vert u^{}_{+1}
\Vert^{}_{E^\rho_{N_+}} \cdots \Vert u^{}_{+l} \Vert^{}_{E^\rho_{N_+}}\big),\cr
}$$
where $Y \in \sg^i$,  $i \in \{ 0,1 \}$. The definition of $F$ gives that $A
- A^{(+)}_0 = A - A^{(-)}_0 - F$. Inequality (6.282) and the analog of
Proposition 6.16 for $t \fl - \infty$ give that
$$\eqalignno{
&\sup_{y \in H_2}  \big((\delta (y))^{1+i-r(i)}  \vert \Delta^{M(l)}_Y (y)
\vert\big) &(6.283{\rm a})\cr
&\quad{}\leq C_{g,\vert Y \vert+l}  \big({\cal R}^l_{N_+,\vert Y \vert+l}
(u^{}_{+1},\ldots,u^{}_{+l})
+ \Vert u^{}_+ \Vert^{}_{E^\rho_{N_++\vert Y \vert+l}}  \Vert u^{}_{+1}
\Vert^{}_{E^\rho_{N_+}} \cdots \Vert u^{}_{+l} \Vert^{}_{E^\rho_{N_+}}\big),\cr
}$$
where $Y \in \sg^i$,  $i \in \{ 0,1 \}$. This inequality and Proposition 6.16
or the first term on the right-hand side of inequality (6.181b) give that
$$\eqalignno{
&\sup_{y \in \Rrm^+ \times \Rrm^3}  \big((\delta (y))^{1+i-r(i)}  \big\vert V
(\Lambda)  \Delta^{M(l)}_Y (V(\Lambda)^{-1} (y-a)) \big\vert\big)& (6.283{\rm
b})\cr
&\quad{}\leq C_{g,\vert Y \vert+l}  \big({\cal R}^l_{N_+,\vert Y \vert+l}
(u^{}_{+1},\ldots,u^{}_{+l})
+ \Vert u^{}_+ \Vert^{}_{E^\rho_{N_++\vert Y \vert+l}}  \Vert u^{}_{+1}
\Vert^{}_{E^\rho_{N_+}} \cdots u^{}_{+l} \Vert^{}_{E^\rho_{N_+}}\big),\cr
}$$
for $Y \in \sg^i$,  $i \in \{ 0,1 \}$,  $l \in \Nrm$, where $C_{g, \vert Y
\vert+l}$ depends only on $r$, $\rho$ and $\Vert u^{}_+
\Vert^{}_{E^\rho_{N_+}}$.

We now go back to the usual notations $(\lambda_1 (t)) (x)$ and $(\delta (t))
(x)$ for $(t,x) \in \Rrm \times \Rrm^3$. Since $(\lambda_1 (t)) (x) \geq \vert
x \vert$ for $\vert x \vert \geq \vert t \vert$, it follows from inequality
(6.280) that
$$\eqalignno{
&(1+t)^{2(1-\rho)}  \Vert (1+\lambda_1 (t))^{k/2}  V' (\Lambda)
\Delta^{D(l)}_Y  (V(\Lambda)^{-1} ((t,\cdot)-a)) \Vert^{}_{D_0}&(6.284)\cr
&\quad{}\leq C_{g,\vert Y \vert+k+l}  \big({\cal R}^l_{N_+,\vert Y \vert+k+l}
(u^{}_{+1},\ldots,u^{}_{+l})
+ \Vert u^{}_+ \Vert^{}_{E^\rho_{N_++\vert Y \vert+k+l}}  \Vert u^{}_{+1}
\Vert^{}_{E^\rho_{N_+}} \cdots \Vert u^{}_{+l} \Vert^{}_{E^\rho_{N_+}}\big),\cr
}$$
for $Y \in \Pi'$,  $k$, $l \in \Nrm$, where $C_{g, \vert Y \vert+k+l}$ depends
only
on $\rho$ and $\Vert u^{}_+ \Vert^{}_{E^\rho_{N_+}}$ and where we have
redefined
$N_+$. By the invariance of the current under local phase transformations, it
follows that $\carre A_\mu = (e^{i \varphi} \psi)^+  \gamma_0  \gamma_\mu
e^{i \varphi} \psi$. Inequality (2.67), then gives that
$$\eqalignno{
&\Vert \big(V(\Lambda)  \Delta^{M(l)}_Y  (V(\Lambda)^{-1} ((t,\cdot)-a)),
V(\Lambda)  \Delta^{M(l)}_{P_\nu Y}  (V(\Lambda)^{-1} ((t,\cdot)-a))\big)
\Vert^{}_{M^{\rho'}}\cr
&\quad{}\leq C_{\rho'}  \sum_{0 \leq \mu \leq 3}  \int^\infty_t
\suma_{Y_1,Y_2}^Y  \Vert (V' (\Lambda)  (\xi^D_{Y_1} (e^{i \varphi} \psi))
(V(\Lambda)^{-1}
((s,\cdot)-a)))^+\cr
&\qquad{}\qquad{}\gamma_0  \gamma_\mu  V' (\Lambda)  (\xi^D_{Y_2}
(e^{i \varphi} \psi))
(V(\Lambda)^{-1} ((s,\cdot)-a)) \Vert^{}_{L^p}  ds,\cr
}$$
where $p = 6/(5-2 \rho)$,  $0 \leq \nu \leq 3$,  $Y \in \Pi'$,
$l \in \Nrm$. The inequality
$$\Vert fg \Vert^{}_{L^p} \leq \Vert f \Vert^{}_{L^2}  \Vert g
\Vert^{2(1-\rho')/3}_{L^2}  \Vert g \Vert^{(1+2 \rho')/3}_{L^\infty},$$
inequalities (6.281b) and (6.284), the use of Theorem 5.5 and Theorem 5.7 for
the free field $\psi^{(+)}_0$ show together with the above inequality that
$$\eqalignno{
&\Vert \big(V(\Lambda)  \Delta^{M(l)}_Y  (V(\Lambda)^{-1} ((t,\cdot)-a)),
V(\Lambda)  \Delta^{M(l)}_{P_\nu Y}  (V(\Lambda)^{-1} ((t,\cdot)-a))\big)
\Vert^{}_{M^{\rho'}}&(6.285)\cr
&\quad{}\leq C_{\rho',\vert Y \vert+l}  (1+t)^{- \rho'+1/2}\cr
&\qquad{}\big({\cal R}^l_{N_+,
\vert Y \vert+l}  (u^{}_{+1},\ldots,u^{}_{+l})
+ \Vert u^{}_+ \Vert^{}_{E^\rho_{N_++\vert Y \vert+l}}  \Vert u^{}_{+1}
\Vert^{}_{E^\rho_{N_+}}\cdots \Vert u^{}_{+l} \Vert^{}_{E^\rho_{N_+}}\big),\cr
}$$
for $Y \in \Pi'$,  $l \in \Nrm$,  $1/2 < \rho' \leq 1$, where
$C_{\rho', \vert Y \vert+l}$ depends only on $\rho$ and
$\Vert u^{}_+ \Vert^{}_{E^\rho_{N_+}}$.

Inequalities (6.280), (6.283b), (6.284) and (6.285) together with the fact that
natural action of ${\cal P}_0$ on $U (\p)$ leaves invariant the ideal spanned
by $\sg^1$, prove the lemma.
\saut
\noindent{\bf Proposition 6.18.}
{\it
Let the group action $\Gamma$, the set $S$ and the set ${\cal U}_\infty$ be
defined  by (6.277a) and (6.277b), (6.277c) and (6.277d). Then
${\cal U}_\infty \subset V^\rho_\infty$, the group action
$\Gamma\colon {\cal P}_0 \times S \fl S$ defines a group action
$U\colon {\cal P}_0 \times {\cal U}_\infty \fl {\cal U}_\infty$ and
$U\colon {\cal P}_0 \times {\cal U}_{\infty (0)} \fl {\cal
U}_\infty$ is $C^\infty$. One can choose ${\cal O}_{\infty (+)}$
such that if $u^{}_+ \in {\cal O}_{\infty (+)}$ and $g \in {\cal P}_0$
are such that $U^{(+)}_g (u^{}_+) \in {\cal O}_{\infty (+)}$, then
$U_g (\Omega^{(+)} (u^{}_+)) = \Omega^{(+)} (U^{(+)}_g (u^{}_+))$.
Moreover there exists $N_0\in\Nrm$ such that if $F\colon V^\rho_\infty
\fl E^{\circ\rho}_\infty$ is as in Theorem 6.11 then
$$\eqalignno{
&\Vert (D^l (F \circ U_g \circ F^{-1})) (u ; u^{}_1,\ldots,u^{}_l)
\Vert^{}_{E^\rho_n}\cr
&\quad{}\leq C_{g,n+l}  \big({\cal R}^l_{N_0,n+l}  (u^{}_1,\ldots,u^{}_l)
+ \Vert u \Vert^{}_{E^\rho_{N_0+n+l}}  \Vert u^{}_1
\Vert^{}_{E^\rho_{N_0}} \cdots \Vert
u^{}_l \Vert^{}_{E^\rho_{N_0}}\big),\cr
}$$
for $g \in {\cal P}_0$,  $n$, $l \in \Nrm$,  $u \in F({\cal U}_{\infty (0)})$,
$u^{}_1,\ldots,u^{}_l \in E^{\circ\rho}_\infty$, where
$C_{g, n+l}$ depends only on $\rho$ and $\Vert u \Vert^{}_{E^\rho_{N_0}}.$
}\saut
\noindent{\it Proof.}
The proof that ${\cal U}_\infty \subset V^\rho_\infty$ which is based on
Lemma 6.17 is so similar to the proof of Theorem 6.12 based on Theorem 6.10,
that we omit it.

It follows by the definition of $\Gamma$ that if $t \mapsto s(t)$ is an element
of $S$ and $a^{}_0 \in \Rrm$, then $t \mapsto s(t-a^{}_0)$ is also an element
of
$S$. Hence $s(-a^{}_0) \in {\cal U}_\infty \subset V^\rho_\infty$. Moreover
$(d/dt)^n  s(t) = T_{P^n_0} (s(t)) \in E^\rho_\infty$ according to statement
i) of Corollary 2.21, so $s \in C^\infty (\Rrm, V^\rho_\infty)$. The
uniqueness of the local solutions in a larger topological space than
$V^\rho_\infty$, given by \refGR\ then proves that the map $p\colon s
\mapsto s(0)$ of
$S$ onto ${\cal U}_\infty$ is a bijection. Since $U_g = p \circ \Gamma_g \circ
p^{-1}$, it follows that $U\colon {\cal P}_0 \times {\cal U}_\infty \fl {\cal
U}_\infty$ is a group action. The fact that $U\colon {\cal P}_0 \times
{\cal U}_{\infty (0)}
\fl {\cal U}_\infty$ is $C^\infty$ and the inequality of the proposition
follow from Lemma 2.19 and Lemma 6.17. We omit the details.

To prove that $\Omega^{(+)}$ intertwines $U_g$ and $U^{(+)}_g$, let $u^{}_+ =
(f,
\dot{f}, \alpha) \in {\cal O}_{\infty (+)}$. Let $(A, \psi)$ be the solution in
$S$ with initial condition $\Omega^{(+)} (u^{}_+) = v$, let $(A_g, \psi_g) =
\Gamma_g (A, \psi)$ be the transformed solution in $S$ and let $\dot{A} (t) =
{d \over dt}  A(t), \dot{A}_g (t) = {d \over dt}  A_g (t)$ and $((\varphi_g
(u^{}_+)) (t)) (x) = ((\varphi (u^{}_+)) (y_0)) (\vec{y})$, where
$(y_0, \vec{y}) = y = V(\Lambda)^{-1} ((t,x) - a),  g = (a, \Lambda)$
and where $\varphi$ is given
by statement iii) of Proposition 6.16. Lemma 6.17 then gives that
$$\Vert (A_g, \dot{A}_g, e^{i \varphi_g (u^{}_+)} \psi_g) (t) -
U^1_{\exp(t P_0)} U^1_g u^{}_+ \Vert^{}_{E^\rho_0} \fl 0,
\eqno{(6.286)}$$
when $t \fl \infty$. Let us study the limit of
$$I (t) = \Vert e^{- i (\varphi_g (u^{}_+))(t)}  U^{D1}_{\exp(t P_0)}  U^{D1}_g
\alpha - e^{- i \big(\varphi(U^{(+)}_g (u^{}_+))\big)(t)}  U^{D1}_{\exp(t P_0)}
\alpha_g
\Vert^{}_D,$$
when $t \fl \infty$. Let $\theta_g (t,x) = \big((\varphi_g (u^{}_+) - \varphi
(U^{(+)}_g (u^{}_+))) (t)\big) (x)$. Since $P_\varepsilon (-i \partial)
U^{D1}_g \alpha \in D_\infty$ (resp. $P_\varepsilon (-i \partial)
\alpha_g \in D_\infty)$ we can apply Theorem A.1 to $e^{i \varepsilon
\omega(-i \partial)t} P_\varepsilon (-i \partial)  U^{D1}_g
\alpha$ (resp. $e^{i \varepsilon \omega(-i \partial)}  P_\varepsilon
(-i \partial) \alpha_g)$ and let $\beta_\varepsilon$
(resp. $\beta'_\varepsilon$)$ \in C^\infty (\Rrm^+ \times \Rrm^3 -\{ 0 \})$ be
the corresponding function homogeneous of degree $- 3/2$ defined by (A.1) to
(A.3). This gives that
$$I_\infty = \lim_{t \fl \infty}  I(t) = \sum_{\varepsilon = \pm}  \lim_{t
\fl \infty}  \Vert \beta_\varepsilon (t,\cdot) - e^{i \theta_g (t,\cdot)}
\beta'_\varepsilon (t,\cdot) \Vert^{}_D.$$
The variable transformation $x \mapsto p^{}_\varepsilon (t,x)$ gives,
since $x/t = - \varepsilon  p^{}_\varepsilon (t,x) / \omega(p^{}_\varepsilon
(t,x))$ and since the support of $\beta_\varepsilon$ and $\beta'_\varepsilon$
is contained in the forward light cone:
$$I_\infty = \sum_\varepsilon  \lim_{t \fl \infty}  \Vert P_\varepsilon
\hat{\alpha}_g - e^{-i \theta_g (t,t K_\varepsilon)}  P_\varepsilon (U^{D1}_g
\alpha)^\wedge \Vert^{}_D, \eqno{(6.287)}$$
where $K_\varepsilon (k) = - \varepsilon k / \omega(k)$,  $k \in \Rrm^3$.
It follows from the definition of $\chi^{}_1$ and $\varphi$ in statement
iii) of Proposition
6.16, from the definition of $A^{(+)}$ in (1.22a) (with $\chi_0 = 1$)
and from the fact that the
function $\alpha \mapsto J^{(+)} (\alpha)$ defined in (1.21) satisfies $J^{(+)}
(U^{D1}_{(a,I)} \alpha) = J^{(+)} (\alpha)$ that $I_\infty = 0$. It then
follows from (6.286) that
$$\Vert (A_g (t), \dot{A}_g (t), \psi_g (t))
- (I \oplus I \oplus e^{-i \varphi (U^{(+)}_g (u^{}_+),t)})  U^1_{\exp (t P_0)}
U^{(+)}_g (u^{}_+) \Vert^{}_{E^\rho_0} \fl 0,\eqno{(6.288)}$$
when $t \fl \infty$. We now choose ${\cal O}_{\infty (+)}$ such that if $u^{}_+
\in {\cal O}_{\infty (+)}$ and $U^{(+)}_g (u^{}_+) \in {\cal O}_{\infty (+)}$
then there is a product of one parameter subgroups $g^{}_i (s)$
such that $U^{(+)}_{g^{}_i
(s)} (u^{}_+) \in {\cal O}_{\infty (+)}$ for $0 \leq s \leq 1$ and $g =
\displaystyle{\Pi_i}  g^{}_i (1)$, where the product is finite. Replacing $g$
by
$g^{}_i (s)$ in (6.288), it follows from statement iv) of Proposition 6.16 that
$U_{g^{}_i (s)} (\Omega^{(+)} (u^{}_+)) = \Omega^{(+)} (U^{(+)}_{g^{}_i (s)}
(u^{}_+)),$
for $0 \leq s \leq 1$ since $U_{g^{}_i (s)}  (\Omega^{(+)} (u^{}_+)) \in {\cal
U}_{\infty (0)}$ for $s$ sufficiently small by continuity. This shows that $U_g
(\Omega^{(+)} (u^{}_+)) = \Omega^{(+)} (U^{(+)}_g (u^{}_+))$, which proves the
proposition.

We shall now extend $\Omega^{(+)}\colon {\cal O}_{\infty (+)}
\fl {\cal U}_{\infty (0)}$ to a Poincar\'e invariant domain.
Let ${\cal O}^{(+)}_\infty = \cup_{g \in
{\cal P}_0}  U^{(+)}_g ({\cal O}_{\infty (+)})$. ${\cal O}^{(+)}_\infty$ is an
open neighbourhood of zero in $E^{\circ\rho}_\infty$, since this is the case
for ${\cal O}_{\infty (+)}$ and since $U^{(+)}_g = (U^{(+)}_{g^{-1}})^{-1}$ and
$u \mapsto U^{(+)}_{g^{-1}} (u)$ is continuous on $E^{\circ\rho}_\infty$. If $u
\in  {\cal O}^{(+)}_\infty$, then there exists $g$ such that $U^{(+)}_{g^{-1}}
(u) \in  {\cal O}_{\infty (+)}$ and we define
$$\Omega_+ (u) = U_g \big(\Omega^{(+)} (U^{(+)}_{g-1} (u))\big),
\eqno{(6.289)}$$
which is an element of ${\cal U}_\infty$ since $U^{(+)}_{g-1} (u) \in {\cal
O}_{\infty (+)}$. Next theorem shows that $\Omega_+$ is a map from ${\cal
O}^{(+)}_\infty$ to ${\cal U}_\infty$.
\saut
\noindent{\bf Theorem 6.19.}
{\it
If $u^{}_+ \in {\cal O}^{(+)}_\infty$,  $g^{}_1$, $g^{}_2 \in {\cal P}_0$,
$U^{(+)}_{g^{-1}_1} (u^{}_+) \in {\cal O}_{\infty (+)}$ and $U^{(+)}_{g^{-1}_2}
(u^{}_+) \in {\cal O}_{\infty (+)}$, then $U_{g^{}_1} (\Omega^{(+)}
(U^{(+)}_{g^{-1}_1} (u^{}_+))) = U_{g^{}_2} (\Omega^{(+)} (U^{(+)}_{g^{-1}_2}
(u^{}_+)))$. The set ${\cal U}_\infty$ is an open neighbourhood of zero in
$V^\rho_\infty$,  $\Omega_+\colon {\cal O}^{(+)}_\infty \fl {\cal U}_\infty$
is a diffeomorphism, the nonlinear group representation $U\colon {\cal P}_0
\times {\cal U}_\infty \fl {\cal U}_\infty$ is $C^\infty$, $\Omega_+
\circ U^{(+)}_g = U_g \circ \Omega_+$ for each $g \in {\cal P}_0$ and
$$\eqalignno{
&\Vert U^M_{\exp (t P_0)} (\Omega_+ (u)) - U^{M1}_{\exp (t P_0)} (f, \dot{f})
\Vert^{}_{M^\rho_0}\cr
&\quad{}+ \big\Vert U^D_{\exp (t P_0)} (\Omega_+ (u)) -
\sum_{\varepsilon = \pm}  e^{i{s}^{(+)}_\varepsilon (u,t,-i \partial)}
P_\varepsilon (-i \partial)
U^{D1}_{\exp (t P_0)} \alpha \big\Vert^{}_D \fl 0,\cr
}$$
when $t \fl \infty$, for $u = (f,
\dot{f}, \alpha) \in {\cal O}^{(+)}_\infty$, where ${s}^{(+)}_\varepsilon$ is
defined by formula (1.18).
}\saut
\noindent{\it Proof.}
Since $U^{(+)}_{g^{-1}_i} (u^{}_+) \in {\cal O}_{\infty (+)}$ for $i \in \{ 0,1
\}$, it follows from Proposition 6.18 that
$$\eqalignno{
U_{g^{}_1} (\Omega^{(+)} (U^{(+)}_{g^{-1}_1} (u^{}_+))) &= U_{g^{}_2}
\big(U_{(g^{-1}_1 g^{}_2)^{-1}}  \big(\Omega^{(+)} (U^{(+)}_{g^{-1}_1
g^{}_2} (U^{(+)}_{g^{-1}_2} (u^{}_+)))\big)\big)\cr
&= U_{g^{}_2} (\Omega^{(+)} (U^{(+)}_{g^{-1}_2} (u^{}_+)))\cr
}$$
and it follows that $\Omega^{(+)} (U^{(+)}_{g^{-1}_1} (u^{}_+)) \in {\cal
U}_{\infty (0)}$. This proves that $\Omega_+\colon {\cal O}^{(+)}_
\infty \fl {\cal U}_\infty$ is a map.

If $u^{}_+ \in {\cal O}^{(+)}_\infty$, then there exists
$h \in {\cal P}_0$ such
that $U^{(+)}_{h^{-1}} (u^{}_+) \in {\cal O}_{\infty (+)}$
so\penalty-10000
 $U^{(+)}_{(gh)^{-1}}
(U^{(+)}_g (u^{}_+)) \in {\cal O}_{\infty (0)}$ also.
Definition (6.289) then
gives that
$$\eqalignno{
\Omega^{(+)} (U^{(+)}_{g^{}_1} (u^{}_+))&= U_{gh}
\big(\Omega^{(+)} (U^{(+)}_{(gh)^{-1}}
(U^{(+)}_g (u^{}_+)))\big)\cr
&= U_g \big(U_h (\Omega^{(+)} (U^{(+)}_{h^{-1}} (u^{}_+)))\big) =
U_g (\Omega_+ (u^{}_+)),\cr
}$$
which proves the intertwining property. If $u \in {\cal U}_{\infty (0)}$, then
$u^{}_+ = (\Omega^{(+)})^{-1} (u) \in {\cal O}_{\infty (+)}$ and $U_g (u) =
\Omega_+ (U^{(+)}_g (u^{}_+))$, which shows that $\cup_{g \in {\cal P}_0}
U_g ({\cal U}_{\infty (0)})$ is a subset of the image of $\Omega_+$. Hence
according to the definition ${\cal U}_\infty$, the map $\Omega_+$ is
onto. If $u \in {\cal U}_\infty$ and $u = \Omega_+ (u^{}_+) = U_g
(\Omega^{(+)} (U^{(+)}_{g^{-1}} (u^{}_+)))$, where
$U^{(+)}_{g^{-1}} (u^{}_+) \in {\cal O}_{\infty (+)}$, then
$u^{}_+ = U^{(+)}_g  ((\Omega^{(+)})^{-1} (U_{g^{-1}}
(u)))$ is independent of the choice of $g$ and shows, since $\Omega^{(+)}\colon
{\cal O}_{\infty (+)} \fl {\cal U}_{\infty (0)}$ is a bijection, that
$\Omega_+$ is one-to-one. Hence $\Omega_+\colon {\cal O}^{(+)}_\infty \fl {\cal
U}_\infty$ is a bijection.

It follows from Theorem 3.12, Theorem 6.12, statement ii) of Proposition 6.16
and Proposition 6.18 that $\Omega_+\colon {\cal O}^{(+)}_\infty \fl {\cal
U}_\infty$ is $C^\infty$ and that there exists an integer $N_+$ such that
$$\eqalignno{
&\Vert (D^l  \Omega_+)  (u ; u^{}_1,\ldots,u^{}_l) \Vert^{}_{E^\rho_n}
&(6.290)\cr
&\quad{}\leq C_{n+l}  \big({\cal R}^l_{N_+,n+l}  (u^{}_1,\ldots,u^{}_l)
+ \Vert u \Vert^{}_{E^\rho_{N_++n+l}}  \Vert u^{}_1 \Vert^{}_{E^\rho_{N_+}}
\cdots \Vert
u^{}_l \Vert^{}_{E^\rho_{N_+}}\big),\cr
}$$
for $u \in {\cal O}^{(+)}_\infty$,  $u^{}_1,\ldots,u^{}_l \in
E^{\circ\rho}_\infty$,
$n$, $k \in \Nrm$, where $C_{n+l}$ depends only on $\rho$ and $\Vert u
\Vert^{}_{E^\rho_{N_+}}$. Let $\overline{u} \in {\cal O}^{(+)}_\infty$.
Since ${\cal
O}^{(+)}_\infty \subset E^{\circ\rho}_\infty$ is an open set there exists a
open neighbourhood ${\cal O}$ of zero in $E^{\circ\rho}_\infty$ such that
$\overline{u} + {\cal O} \subset {\cal O}^{(+)}_\infty$. The function
$F\colon v \mapsto
\Omega_+ (\overline{u} + v)$ from ${\cal O}$ to ${\cal U}_\infty$ satisfies
estimate (6.290) with $\overline{u} + v$ instead of $u$ and with $C_{n+l}$
depending only on $v$, for fixed $\overline{u}$. To prove that $F$ has a
$C^\infty$ local
inverse it is therefore sufficient to prove that $DF(v)$ has a right inverse,
for $v$ in a neighbourhood of zero, satisfying the hypotheses of the implicit
function theorem in Fr\'echet spaces. The existence of such a right inverse is
proved using Lemma 6.17 and following the proof of Theorem 6.13. We leave out
the details since the proofs are so similar. Since $\Omega_+$ is a bijection
this proves that $\Omega_+\colon {\cal O}^{(+)}_\infty \fl {\cal U}_\infty$
is a diffeomorphism and that ${\cal U}_\infty$ is open.

Since $\Omega_+$ is a diffeomorphism and $U_g = \Omega_+ \circ U^{(+)}_g \circ
\Omega^{-1}_+$, it follows that $U\colon {\cal P}_0 \times {\cal U}_\infty
\fl {\cal U}_\infty$ is $C^\infty$.

To prove the statement concerning the limit let $u^{}_+ = (f, \dot{f},
\alpha) \in {\cal O}^{(+)}_\infty$, let $g \in {\cal P}_0$ be such that
$U^{(+)}_{g^{-1}} (u^{}_+) \in {\cal O}_{\infty (+)}$, let $(A, \psi)$
be the solution in $S$ with initial condition $\Omega_+ (u^{}_+)$ and let
$(A', \psi')$ be the solution in $S$ with initial condition $U_{g^{-1}}
(\Omega_+ (u^{}_+))$. According to the already proved intertwining property
and the definition of $\Omega_+$, it follows that $U_{g^{-1}} (\Omega_+
(u^{}_+)) = \Omega^{(+)} (U^{(+)}_{g^{-1}} (u^{}_+)) \in {\cal
U}_{\infty (0)}$. Defining $\Gamma_g (A', \psi') = (A'_g, \psi'_g)$ it follows
from the definition of $U_g$ that $(A'_g, \psi'_g) = (A, \psi)$. Therefore
applying (6.288) with $(A'_g, \dot{A}'_g, \psi_g)$ instead of $(A_g, \dot{A}_g,
\psi_g)$ and $U^{(+)}_{g^{-1}} (u^{}_+)$ instead of $u^{}_+$, we obtain that
$$\Vert (A(t), \dot{A} (t), \psi (t)) - (I \otimes I \otimes e^{-i \varphi
(u^{}_+,t)})  U^1_{\exp (t P_0)} u^{}_+ \Vert^{}_D \fl 0,$$
when $t \fl \infty$. This shows that we only have to prove that
$$\Vert e^{-i \varphi (u^{}_+,t)}  U^{D1}_{\exp (t P_0)} \alpha -
\sum_{\varepsilon = \pm}  e^{i s_\varepsilon (u^{}_+,t,-i \partial)}
P_\varepsilon (-i \partial)  U^{D1}_{\exp (t P_0)} \alpha \Vert^{}_D \fl 0,
\eqno{(6.291)}$$
when $t \fl \infty$. Using that
$$\eqalignno{
\omega(k)  {\partial \over \partial k_i}
h(t,tk/\omega(k)) &= h^{}_{M_{0i}} (t,tk/\omega(k)) - (k_i/\omega(k))
h^{}_D (t,tk/\omega(k)),\cr
\noalign{\noindent where}
h^{}_{M_{0i}} (t,x) &= x_i (\partial / \partial t)  h(t,x)+
t (\partial / \partial x_i)  h(t,x)\cr
\noalign{\noindent and}
h^{}_{D} (t,x) &= t (\partial / \partial t)  h(t,x)+
\sum_{1\leq i\leq 3}x_i (\partial / \partial x_i)  h(t,x),\cr
}$$
using Lemma 4.4 and
Corollary 4.2 and using formula (1.20b) it follows that there exists $N \in
\Nrm$ such that
$$\big\vert {\partial^n \over \partial t^n}  \partial^\alpha_k  s_\varepsilon
(u^{}_+,t,k) \big\vert\leq C_{\vert \alpha \vert + n}
\omega(k)^{- \vert \alpha \vert}
(1+t)^{\rho - 1/2}  (1+\Vert u^{}_+ \Vert^{}_{E^\rho_N})  \Vert u^{}_+
\Vert^{}_{E^\rho_{N+\vert \alpha \vert+n}}, \eqno{(6.292)}$$
for some constants $C_{\vert \alpha \vert + n}$ depending only on $\rho$. Let
$B(r)$, $r>0$, be the open ball of radius $r$ in $\Rrm^3$. If $0<R<R'$,
then it follows from inequality (6.292), formula
(A.2) and from the inverse  function theorem in $\Rrm^3$ there exists
$T\geq0$ such
that the equation
$$\varepsilon  {q^{}_\varepsilon (t,x) \over \omega(q^{}_\varepsilon
(t,x))} + t^{-1}
F_\varepsilon (u^{}_+, t, q_\varepsilon (t,x)) + {x \over t}=0,
\eqno{(6.293)}$$
where $F_\varepsilon (u^{}_+,t,k) = \nabla_k  s_\varepsilon (u^{}_+,t,k)$,
has a
unique solution $q_\varepsilon (t,x) \in B(mR')$ for $(t,x) \in \{ (s,y)
\big\vert s > T, y/s <R(1+R^2)^{1/2} \} = Q(T,R)$.
The function $q^{}_\varepsilon \in
C^\infty (Q(T,R), \Rrm^3)$ and the inverse $y_\varepsilon
(t, \cdot)$ of $x \mapsto
q^{}_\varepsilon (t,x)$ is an element of $C^\infty (]T, \infty[ \times B(mR'),
\Rrm^3)$. Moreover if $p^{}_\varepsilon (t,x)$ is given by formula (A.2) then
$$\vert q^{}_\varepsilon (t,x) - p^{}_\varepsilon (t,x) \vert \leq C(1+t)^{-
3/2+\rho}, \quad (t,x)\in Q(T,R), \eqno{(6.294)}$$
where $C$ is independent of $(t,x)$. Since $\vert (\varphi (u^{}_+,t)) (x) +
{\vartheta} (A^{(+)}, (t,x)) \vert$ converges to zero uniformly
 inside every conic
neighbourhood which is included in the interior of the forward light cone
when $t \fl \infty$, it
follows from (6.294) and Theorem 7.7.5 of \refHA\ that if ${\rm supp}\
\hat \alpha_1
\subset B(mR)$, then
$$\Vert e^{-i \varphi (u^{}_+,t)}  U^{D1}_{\exp (t P_0)} \alpha_1 -
\sum_{\varepsilon = \pm}  e^{i s_\varepsilon (u^{}_+,t,-i \partial)}
P_\varepsilon (-i \partial)  U^{D1}_{\exp (t P_0)} \alpha_1 \Vert^{}_{D} \fl
0.\eqno{(6.295)}$$
Let $\alpha\in D^{}_\infty$ and $\nu>0$. Then there exists $R>0$ and
$\alpha_1\in D^{}_\infty$ such that ${\rm supp}\ \hat \alpha_1
\subset B(mR)$ and $\Vert\alpha-\alpha_1 \Vert^{}_{D}\leq \nu/4$. According to
inequality (6.294) there exists $T_0\geq 0$ such that for this $ \alpha_1$
the norm in (6.295) is majorized by $\nu/2$ for $t\geq T_0$.
The norm in expression (6.291) is therefore majorized by $\nu$
for $t\geq T_0$. This proves the theorem.
\vfill\eject
\noindent{\titre Appendix}
\saut
We shall prove here auxiliary $L^2$- and $L^\infty$-estimates for approximate
solutions of the linear homogeneous Klein-Gordon equation. This can be done by
a direct application of the method of stationary phase. However here we shall
combine it with Theorem 5.5 for $G=0$, Theorem 5.7 and with the
method of symbolic calculus developed in \refHKG\ which is
easier  and leads to the same result.

For $f \in D_\infty$  $(\simeq S(\Rrm^3,\Crm^4)$) and $\varepsilon = \pm 1$, we
introduce the sequence
 $g^{}_l \in\penalty-10000 C^\infty ((\Rrm^+ \times \Rrm^3)- \{ 0 \})$
defined by $g^{}_l(t,x)=0$ for $0\leq t \leq \vert x\vert$, and
$$\eqalignno{
g^{}_0 (t,x) &= e^{i 3\varepsilon \pi /4}  (D(t,x))^{- 1/2}  \hat{f}
(p^{}_\varepsilon (t,x)), \quad \varepsilon = \pm, &({\rm A}.1\hbox{a})\cr
g^{}_l &= {\rho \over 2i \varepsilon l m}  \carre g^{}_{l-1},
\quad l \geq 1,&({\rm A}.1\hbox{b})\cr
}$$
(i.e. $g = e^{(2i \varepsilon m )^{-1}\rho \carrre} g^{}_0$ as formal power
 series in the invariable $m^{-1}$), for
$0\leq \vert x\vert< t$, where
$$\rho (t,x) = (t^2 - \vert x \vert^2)^{1/2},\quad  p^{}_\varepsilon (t,x) = -
\varepsilon m x / \rho (t,x), \eqno{({\rm A}.2)}$$
and
$$D(t,x) =   m^2 (\omega(p^{}_\varepsilon (t,x)))^{-5} = (t/m)^3
(\rho(t,x)/t)^5 \eqno{({\rm A}.3)}$$
is the Jacobian of the transformation $x \mapsto p^{}_\varepsilon (t,x)$
for fixed
$t$. The support of $g^{}_l$ is contained in the set $\{y\in \Rrm^4\big\vert
y_\mu y^\mu \geq 0,y^0\geq 0 \}$
and $g^{}_l$ is homogeneous of degree $- 3/2 - l$.

We introduce the representation $X\mapsto \xi^{}_X$ of the Poincar\'e Lie
algebra $\p$ by:
$$\eqalignno{
\xi^{}_{P_0}&={\partial\over\partial t},\cr
\xi^{}_{P_i}&={\partial\over\partial x^{}_i},\quad 1\leq i\leq 3,\cr
\xi^{}_{M_{0i}}&=x^{}_i{\partial\over\partial t}
+t{\partial\over\partial x^{}_i},\quad 1\leq i\leq 3,\cr
\xi^{}_{M_{ij}}&=-x^{}_i{\partial\over\partial x^{}_j}
+x^{}_j{\partial\over\partial x^{}_i},\quad 1\leq i<j\leq 3.\cr
}$$
We recall that $\Pi$ is an ordered basis of $\p$ and that $\Pi'$
is the corresponding standard basis of the enveloping algebra
$U(\p)$ of $\p$. Given an ordering on the basis $Q=\{M_{\mu\nu}\big\vert
0\leq \mu<\nu\leq 3\}$ of ${\frak {so}}(3,1)$, let $Q'$ be the corresponding
standard basis of the enveloping algebra
$U({\frak {so}}(3,1))$ of ${\frak {so}}(3,1)$.

To state the results on decrease properties of the Klein-Gordon equation,
we introduce
$$\eqalignno{
\varphi_0 &= e^{i\varepsilon \omega(-i\partial)t}f,\cr
\varphi_n &=\varphi_0-e^{i\varepsilon m\rho(t)}\sum_{0\leq l\leq n-1}g^{}_l
(t),
\quad n\geq 1,\cr
}$$
and if $X\mapsto a^{}_X$ is a map from $\Pi'$ to $D^{}_\infty$,
then we introduce
$$E^{(p)}_j(a)=\sum_{\scr Y\in \Pi'\atop\scr \vert Y\vert \leq j}
\Big( m\Vert a^{}_Y\Vert^{}_{L^p}+\sum_{0\leq \mu\leq 3}\Vert a^{}_{P_\mu Y}
\Vert^{}_{L^p}\Big),$$
for $j\in\Nrm$ and $1\leq p\leq \infty$. We also introduce $\lambda(t)$ and
$\delta(t)$ for $t\geq0$ by
$$\eqalignno{
(\lambda(t))(x)&= t/(1+t-\vert x\vert)\quad \hbox{for}\ 0\leq
\vert x\vert \leq t,\cr
(\lambda(t))(x)&=\vert x\vert \quad \hbox{for}\ 0\leq t \leq
\vert x\vert,\cr
(\delta(t))(x)&=1+t+\vert x\vert.\cr
}$$
If $\mu\colon\Rrm_+\fl D$, we define
$$(\tilde{\mu}(t))_Y=(\xi^{}_Y\mu)(t), Y \in \Pi', t \in \Rrm_+$$
\saut
\noindent{\bf Theorem A.1.}
{\it
There exist positive numbers $C_i\in\Rrm^+$, $i\geq0$, such that
$$\eqalignno{
E^{(2)}_j\big((1+\lambda(t))^{k/2}\tilde{\varphi}^{}_0(t)\big)
&\leq C_{j+k}\Big(\Vert f\Vert^{}_{D_{j+k}}+\sum_{1\leq i\leq 3}
\Vert \partial_i f\Vert^{}_{D_{j+k}}\Big),\cr
\noalign{\noindent for $j,k\in\Nrm$, $t\geq0$, $f\in D_\infty$,}
E^{(2)}_j\big((1+\lambda(t))^{k/2}\tilde{\varphi}^{}_{n+1}(t)\big)
&\leq C_{j+k+n}\ t^{-n-1}\Vert f\Vert^{}_{D_{3(j+k+n)+4}},&({\rm A}.4)\cr
\noalign{\noindent for $j,k,n\in\Nrm$, $t\geq1$, $f\in D_\infty$,}
E^{(\infty)}_j\big(\delta(t)^{3/2}(1+\lambda(t))^{k/2}\tilde{\varphi}^{}_0(t)\big)
&\leq C_{j+k}\Big(\Vert f\Vert^{}_{D_{j+k+8}}+\sum_{1\leq i\leq 3}
\Vert \partial_i f\Vert^{}_{D_{j+k+8}}\Big),\cr
\noalign{\noindent for $j,k\in\Nrm$, $t\geq0$, $f\in D_\infty$ and}
E^{(\infty)}_j\big(\delta(t)^{3/2}(1+\lambda(t))^{k/2}\tilde{\varphi}^{}_{n+1}(t)\big)
&\leq C_{j+k+n}\ t^{-n-1}\Vert f\Vert^{}_{D_{3(j+k+n)+28}},&({\rm A}.5)\cr
}$$
for $j,k,n\in\Nrm$, $t\geq1$, $f\in D_\infty$. Moreover ${\rm supp}\ g^{}_l(t)
\subset \{x\in \Rrm\big\vert \vert x\vert \leq t\}$ for $t>0$,
$$\eqalignno{
\Vert (\rho^{-j}\xi^{}_{XY}g^{}_l)(t)\Vert^{}_{L^2}
&\leq C_{j+\vert X\vert +l}\ t^{-j-l-\vert X\vert}\Vert
f\Vert^{}_{D_{3\vert X\vert
+3l+\vert Y\vert+j}},&({\rm A}.6)\cr
\noalign{\noindent and}
\Vert (\rho^{-j}\xi^{}_{XY}g^{}_l)(t)\Vert^{}_{L^\infty}
&\leq C_{j+\vert X\vert+ l}\ t^{-3/2-j-l-\vert X\vert}\Vert
f\Vert^{}_{D_{3\vert X\vert
+3l+\vert Y\vert+j+5}},&({\rm A}.7)\cr
}$$
for $t>0$, $j,l\in\Nrm$, $X\in\Pi'\cap U(\Rrm^4)$ and $Y\in Q'$.
}\saut
\noindent This theorem will be proved at the end of the appendix.

The development in Theorem A.1 can be inverted. Given a homogeneous function
$g \in C^\infty ((\Rrm^+ \times \Rrm^3) - \{0\})$ of degree $- 3/2$ such that
${\rm supp}\ g(1, \cdot) \subset \{ x \in \Rrm^3 \big\vert
\vert x \vert \leq 1 \}$, we
construct by iteration $f_0,\ldots,f_n \in D_\infty$:
$$\eqalignno{
\hat{f}_l (k) &= e^{-i 3\varepsilon \pi/4}  (m^2 /
\omega(k)^5)^{1/2}  g^{}_{l,0}
(1, {- \varepsilon k /\omega(k)}), \quad g^{}_{0,0} =
g,\quad 0 \leq l \leq n,
&({\rm A}.8{\rm a})\cr
g^{}_{l,0} (t,x) &= - \sum_{1 \leq j \leq l}  t^j
g^{}_{l-j,j} (t,x),\quad  1 \leq l
\leq n,&({\rm A}.8{\rm b})\cr
g^{}_{l,j} &= {\rho \over 2ij \varepsilon m}  \carre
g^{}_{l,j-1},\quad  1 \leq j
\leq n-l. &({\rm A}.8{\rm c})\cr
}$$
By this construction $g^{}_{l,j} \in C^\infty ((\Rrm^+
\times \Rrm^3) -\{ 0 \})$ is
homogenous of degree $- 3/2-j$ with support in the
forward light cone.
\saut
\noindent{\bf Theorem A.2.}
{\it
Let $g \in C^\infty ((\Rrm^+ \times \Rrm^3) -\{ 0 \})$ be a
homogeneous function of degree $- 3/2$ with ${\rm supp}\
g(1,\cdot) \subset \{ x \in \Rrm^3 \big\vert \vert x \vert \leq 1 \}$.
If $f_0,\ldots,f_n$ is given by the construction (A.8a)--(A.8c) and
$$u^{}_n(t)=e^{i\varepsilon m\rho(t)}g(t)
-\sum_{0\leq l\leq n} t^{-l}e^{i\varepsilon\omega(-i\partial)t}f^{}_l,$$
then
$$\eqalignno{
&E^{(2)}_j\big((1+\lambda(t))^{k/2}\tilde{u}^{}_n(t)\big)&({\rm A}.9)\cr
&\qquad{}\leq C_{j+k+n}\sum_{\scr Y\in Q'\atop\scr q+\vert Y \vert
\leq 3(j+k+n)+4}
\Vert ({m / \rho(1,\cdot)})^q  (\xi^{}_Y g) (1,\cdot) \Vert^{}_{L^2}t^{-n-1},
\quad t\geq 1,\cr
\noalign{\noindent and}
&E^{(\infty)}_j\big(\delta(t)^{3/2}(1+\lambda(t))^{k/2}
\tilde{u}^{}_n(t)\big)&({\rm A}.10)\cr
&\qquad{}\leq C_{j+k+n}\sum_{\scr Y\in Q'\atop\scr q+\vert Y \vert
\leq 3(j+k+n)+28}
\Vert ({m / \rho(1,\cdot)})^q  (\xi^{}_Y g) (1,\cdot) \Vert^{}_{L^2}t^{-n-1},
\quad t\geq 1,\cr
\noalign{\noindent for $j,k,n\in\Nrm$. Moreover}
&\Vert f^{}_l \Vert^{}_{D_j} \leq C_{l+j}  \sum_{\scr Y\in Q'\atop\scr
q+\vert Y \vert \leq j+2l}
\Vert ({m / \rho(1,\cdot)})^{q+l}  (\xi^{}_Y g) (1,\cdot) \Vert^{}_{L^2},
\quad j,l\in\Nrm.&({\rm A}.11)\cr
}$$
}\saut
\noindent{\it Proof.}
Since the proofs of the statements are so similar for the $L^2$ case and the
$L^\infty$ case, we only prove the $L^2$ case.

Application of Theorem A.1 to the functions $f^{}_l$, $0 \leq l \leq n$,
with a development up to order $n-l$ gives, if
$v^{}_n(t)=e^{i\varepsilon m\rho(t)}\big(
g(t)-\sum_{0\leq l\leq n} t^{-l}\sum_{0\leq j\leq n-l}g^{}_{l,j}(t)\big)$:
$$\eqalignno{
E^{(2)}_j\big((1+\lambda(t))^{k/2}\tilde{u}^{}_n(t)\big)
&\leq E^{(2)}_j\big((1+\lambda(t))^{k/2}\tilde{v}^{}_n(t)\big)&({\rm A}.12)\cr
&\leq C_{j+k+n}\ t^{-n-1}\sum_{0 \leq l \leq n}\Vert
f^{}_l\Vert^{}_{D_{3(j+k+n-l)+4}},
\quad t\geq 1,\cr
}$$
where $g^{}_{l,j}$ is given by (A.1) with $f^{}_l$ instead of $f$ i.e.:
$$\eqalignno{
g^{}_{l,0} (t,x) &= e^{i 3\varepsilon\pi/4} (D(t,x))^{- 1/2}
\hat{f}_l (p^{}_\varepsilon
(t,x)),\quad  0 \leq l \leq n, &({\rm A}.13{\rm a})\cr
g^{}_{l,j} &= {\rho \over 2i \varepsilon jm}
\carre g^{}_{l,j-1},\quad  1 \leq j \leq
n-l.& ({\rm A}.13{\rm b})\cr
}$$
According to  definition (A.8) and formulas (A.13a),
(A.13b) we have $g^{}_{0,0}= g$ and
$$\sum_{0 \leq j \leq r}  t^j  g^{}_{r-j,j} = 0,\quad  1 \leq r \leq n,$$
which proves that
$$\sum_{0 \leq l \leq n}  \sum_{0 \leq j \leq n-l}  t^{-l}  g^{}_{l,j} (t) =
\sum_{0 \leq r \leq n}  t^{-r}  \sum_{0 \leq j \leq r}  t^j  g^{}_{r-j,j} = g,
\eqno{({\rm A}.14)}$$
so $v^{}_n=0$. By (A.12) we get
$$E^{(2)}_j\big((1+\lambda(t))^{k/2}\tilde{u}^{}_n(t)\big)\leq
C_{j+k+n}\ t^{-n-1}\sum_{0 \leq l \leq n}\Vert f^{}_l
\Vert^{}_{D_{3(j+k+n-l)+4}},
\quad t \geq 1. \eqno{({\rm A}.15)}$$
Inequality (A.9) follows from (A.15) and
inequality (A.11), which we now prove.

Introduce the linear representation $X \mapsto \eta^{}_X$ of the Poincar\'e Lie
algebra ${\p}$ by
$$\eqalignno{
(\eta^{}_{P_0} h) (k) &= - \varepsilon i \omega(k)  h(k),
\quad (\eta^{}_{P_j} h) (k) = i
k_j h(k),\quad  j = 1,2,3,& ({\rm A}.16{\rm a})\cr
(\eta^{}_{M_{ij}} h) (k) &= - (k_i  {\partial \over \partial k_j} - k_j
{\partial \over \partial k_i}) h(k), \quad 1 \leq i < j \leq 3,& ({\rm
A}.16{\rm b})\cr
(\eta^{}_{M_{0j}} h) (k) &= - \varepsilon
{\partial \over \partial k_j} (\omega(k)
h(k)), \quad 1 \leq j \leq 3,&({\rm A}.16{\rm c})\cr
}$$
where $h \in D_\infty$ and $\varepsilon = 1$ or $\varepsilon = - 1$.

The norms of $f$:
$$\Vert f \Vert^{}_{D_{\scr s}},\quad  \Vert \hat{f} \Vert^{}_{D_{\scr s}},
\quad
\Big(\sum_{\scr Y \in \Pi'\atop\scr \vert Y \vert \leq s}
\Vert \eta^{}_Y  \hat{f} \Vert^2_{L^2}\Big)^{1/2}, \eqno{({\rm A}.17)}$$
are then equivalent. Let $X \mapsto \xi^{}_X$ be the representation
of the Lorentz Lie algebra ${\frak {so}}(3,1)$ on functions on $\Rrm^4$ given
by
$$\eqalignno{
(\xi^{}_{M_{ij}} h) (t,x) &= -\big(x_i  {\partial \over \partial x_j} - x_j
{\partial
\over \partial x_i}\big) h(t,x),\quad  1 \leq i < j \leq 3,&({\rm A}.18{\rm
a})\cr
(\xi^{}_{M_{0i}} h) (t,x) &= \big(x_i  {\partial \over \partial t} + t
{\partial  \over \partial x_i}\big) h(t,x),\quad  1 \leq i \leq 3.&({\rm
A}.18{\rm b})\cr
}$$

Since $g^{}_{l,0}$ is homogeneous of degree $- 3/2$ it follows from (A.8a) that
$$\hat{f}_l (k) = e^{- i 3\varepsilon\pi/4}  {(m /\omega(k))}  g^{}_{l,0}
(\omega(k), -\varepsilon k). \eqno{({\rm A}.19)}$$
Since $(\eta^{}_X \hat{f}^{}_l) (k) = e^{-i3\varepsilon \pi/4}  {(m
/\omega(k))}
(\xi^{}_X
g^{}_{l,0}) (\omega(k), - \varepsilon k)$ for $X \in {\frak {so}}(3,1)$, we get
$$(\eta^{}_Y \hat{f}^{}_l)(k) = e^{-i 3\varepsilon\pi/4}  {(m / \omega(k))}
(\xi^{}_Y  g^{}_{l,0})(\omega(k), - \varepsilon k),$$
for $Y \in U({\frak {so}}(3,1)),$ the enveloping algebra of ${\frak
{so}}(3,1)$.
$\xi^{}_Yg^{}_{l,0}$ and $g^{}_{l,0}$ have the same degree of homogeneity, so
$$(\eta^{}_Y \hat{f}^{}_l) (k) = e^{-i 3\varepsilon\pi/4}  (m /
(\omega(k))^{5/2})
(\xi^{}_Y g^{}_{l,0}) (1, - \varepsilon k / \omega(k)),\quad Y \in
U({\frak {so}}(3,1)). \eqno{({\rm A}.20)}$$
Choosing an ordering on $\Pi$ such that
$$P_0 < P_1 < P_2 < P_3 < M_{\mu \nu},\quad  0 \leq \mu < \nu \leq 3,$$
we obtain, because of the equivalence of the norms in (A.17) and because of the
explicit expression (A.16a) of $\eta^{}_{P^{}_\mu}$, that
$$\Vert f \Vert^{}_{D_{\scr s}} \leq C_s
\Big(\sum_{q+\vert Z \vert \leq s}  \sum_{Z \in Q'}
\Vert \omega^q  \eta^{}_Z  \hat{f} \Vert^2_{L^2}\Big)^{1/2}, \eqno{({\rm
A}.21)}$$
where $Q'$ is a basis of $U({\frak {so}}(3,1))$. Since $m (\omega(k))^{- 5/2}$
is
the Jacobian of the transformation $k \mapsto {- \varepsilon
k / \omega(k)}$, it follows from formula (A.20) that
$$\Vert \omega^q  \eta^{}_Z  \hat{f}^{}_l \Vert^{}_{L^2} =
\Vert (m  \rho (1, \cdot))^q (\xi^{}_Z  g^{}_{l,0}) (1, \cdot)
\Vert^{}_{L^2}, \quad  Z \in U({\frak {so}}(3,1)), \eqno{({\rm A}.22)}$$
and then from (A.21) that
$$\Vert f^{}_l \Vert^{}_{D_{\scr s}} \leq C'_s
\sum_{q + \vert Z \vert \leq s}  \sum_{Z
\in Q'}  {  }Ê\Vert ({m / \rho (1, \cdot)})^q
(\xi^{}_Z  g^{}_{l,0}) (1, \cdot)
\Vert^{}_{L^2}, \eqno{({\rm A}.23)}$$
where $0 \leq l \leq n$,  $s \geq 0$.

It follows from definition (A.8c) of $g^{}_{l,j}$,
$1 \leq j \leq n-l$, that
$$g^{}_{l,j} = (j!(2i \varepsilon m)^j)^{-1} (\rho \carre)^j
g^{}_{l,0}. \eqno{({\rm A}.24)}$$
Since $\xi^{}_X$,  $X \in {\frak {so}}(3,1)$, commutes with $\carre$
and with the multiplication by  $\rho$, we get
$$\xi^{}_Z  g^{}_{l,j} = (j!(2i \varepsilon m)^j)^{-1}  (\rho
\carre)^j  \xi^{}_Z  g^{}_{l,0}, \quad Z \in U({\frak {so}}(3,1)).
\eqno{({\rm A}.25)}$$
Using that inside the light cone
$$\carre = \rho^{-2} \Big(L^2 + 2L - \sum_{0 \leq \mu < \nu \leq 3}
\xi^{}_{M_{\mu
\nu} M^{\mu \nu}}\Big), \eqno{({\rm A}.26)}$$
where $L = t {\partial \over \partial t} + \sum^3_{i = 1}  x_i  {\partial
\over \partial x_i}$ and $M^{0i} = - M_{0i}$, $1 \leq i \leq 3$, and $M^{ij} =
M_{ij}$, $1 \leq i < j \leq 3$, we obtain for $h$ being a homogeneous function
of degree $\chi$ on $\Rrm^4$:
$$\carre h = \rho^{-2} \Big((\chi+1)^2 - 1 - \sum_{0 \leq \mu < \nu \leq 3}
\xi^{}_{M_{\mu \nu} M^{\mu \nu}}\Big), \eqno{({\rm A}.27)}$$
which is homogeneous of degree $\chi-2.$ It follows from (A.24) and (A.27) that
$$\xi^{}_Z  g^{}_{l,j} = (j! (2i \varepsilon m)^j)^{-1}
\rho^{-j} \Big( \prod^j_{q=1} \big((q-1/2)^2 - 1 -
\sum_{0 \leq \mu < \nu \leq 3}
\xi^{}_{M_{\mu \nu} M^{\mu \nu}}\big)\Big) \xi^{}_Z  g^{}_{l,0},
\eqno{({\rm A}.28)}$$
$Z \in U({\frak {so}}(3,1))$,  $1 \leq j \leq n-l$,  $0 \leq l \leq n$,
since $\xi^{}_Z  g^{}_{l,0}$ is homogeneous of degree $- 3/2$
and since $\xi^{}_X$,  $X \in {\frak {so}}(3,1)$, commutes with the
multiplication by $\rho$. We get from (A.28) that
for $Z \in U({\frak {so}}(3,1))$:
$$\eqalignno{
&\Vert ({m / \rho (1, \cdot)})^q  (\xi^{}_Z  g^{}_{l,j})
(1, \cdot) \Vert^{}_{L^2}
&({\rm A}.29)\cr
&\qquad{}\leq C_j  \sum_{\scr Y \in Q'\atop\scr
\vert Y \vert \leq 2j}  \Vert ({m/ \rho
(1, \cdot)})^{q+j}  (\xi^{}_Y  \xi^{}_Z
g^{}_{l,0}) (1, \cdot)  \Vert^{}_{L^2},
\quad 1 \leq j \leq n-l.\cr
}$$
Definition (A.8b) shows that
$$\eqalignno{
&\Vert ({m / \rho (1, \cdot)})^q  (\xi^{}_Z  g^{}_{l,0})
(1, \cdot) \Vert^{}_{L^2}
&({\rm A}.30)\cr
&\qquad{}\leq \sum_{1 \leq i \leq l}  C_{\vert Z \vert}
\sum_{\scr  Y \in Q'\atop\scr
\vert Y \vert \leq \vert Z \vert}  \Vert ({m/ \rho  (1,
\cdot)})^q  (\xi^{}_Y
g^{}_{l-j,j}) (1, \cdot)  \Vert^{}_{L^2},\quad Z \in Q',
1 \leq l \leq n,\cr
}$$
since $\vert \xi^{}_{M_{0j}} t \vert \leq t$ inside
the light cone.
Let $1 \leq l \leq n$,  it then follows from (A.29) and (A.30) that
$$\eqalignno{
&\Vert ({m/ \rho (1, \cdot)})^q  (\xi^{}_Z  g^{}_{l,0})
(1, \cdot) \Vert^{}_{L^2}
&({\rm A}.31)\cr
&\qquad{}\leq C_{\vert Z \vert,l}  \sum_{\scr Y \in Q'\atop
\scr \vert Y \vert \leq \vert Z \vert}
 \sum_{1 \leq j \leq l}  \sum_{\scr Y \in Q'\atop\scr\vert X \vert \leq 2j}
\Vert ({m/ \rho  (1, \cdot)})^{q+j}
(\xi^{}_X  \xi^{}_Y  g^{}_{l-j,0}) (1, \cdot)
\Vert^{}_{L^2}.\cr
}$$
Using that $\xi^{}_X  \xi^{}_Y = \xi^{}_{XY}$ and inequality (A.31) we get
$$\eqalignno{
&\Vert ({m / \rho (1, \cdot)})^q  (\xi^{}_Z  g^{}_{l,0})
(1, \cdot) \Vert^{}_{L^2}
&({\rm A}.32)\cr
&\qquad{}\leq C_{\vert Z \vert,l}  \sum_{1 \leq j \leq l}  \sum_{\scr Y \in Q'
\atop\scr 0 \leq \vert Y \vert \leq \vert Z \vert+2j}
\Vert ({m / \rho  (1, \cdot)})^{q+j}
(\xi^{}_Y  g^{}_{l-j,0}) (1, \cdot) \Vert^{}_{L^2},\quad l \geq 1.\cr
}$$
Inequality (A.32) shows that, (the case $l=0$ being trivial):
$$\eqalignno{
&\Vert ({m / \rho (1, \cdot)})^q  (\xi^{}_Z  g^{}_{l,0})
(1, \cdot) \Vert^{}_{L^2}
&({\rm A}.33)\cr
&\qquad{}\leq C_{\vert Z \vert,l}  \sum_{\scr Y \in Q'\atop
\scr \vert Y \vert \leq \vert Z \vert+2l}
\Vert ({m /\rho  (1, \cdot)})^{q+l}  (\xi^{}_Y  g^{}_{0,0}) (1, \cdot)
\Vert^{}_{L^2}, \quad  0 \leq l \leq n.\cr
}$$
Since $g^{}_{0,0} = g$ we obtain according to inequalities (A.23) and (A.33)
$$\Vert f^{}_l \Vert^{}_{D_{\scr s}} \leq C'_{l,s}
\sum_{\scr Y \in Q'\atop\scr
q+\vert Y \vert \leq s+2l}  \Vert ({m/ \rho  (1, \cdot)})^{q+l}
(\xi^{}_Y  g) (1,\cdot)
\Vert^{}_{L^2},\quad  0 \leq l \leq n. \eqno{({\rm A}.34)}$$
This proves the theorem.
\saut
\noindent{\bf Theorem A.3.}
{\it
If $F \in C^\infty ((\Rrm^+ - \{ 0 \}) \times \Rrm^3)$ is homogeneous
of degree
$0$ and if $f \in D_\infty,$ then there exists a unique sequence of functions
$g^{}_l \in D_\infty$,   $l \geq 0$, such that for every $n \geq 0$,
$j\geq 0$ and $L \geq 0$
$$\eqalignno{
&\Vert \big({d \over dt}\big)^j  \big(e^{-i \varepsilon
\omega(-i \partial)t}  F(t)  e^{i
\varepsilon \omega(-i \partial)t}  f -
\sum_{0 \leq l \leq n}  t^{-l}  g^{}_l\big)
\Vert^{}_{D_L}& ({\rm A}.35)\cr
&\ {}\leq C_{n+L+j}\ t^{-n-1-j}
\Big(\sum_{\scr X\in\Pi'\cap U(\Rrm^4)\atop\scr
\vert X\vert\leq L+j} \Vert
\big(\xi^{}_X F(t) \Vert^{}_{L^\infty (\Rrm^3)} +
\sum_{\scr Y\in Q'\atop\scr \vert Y\vert\leq N}
\Vert (\xi^{}_Y F) (1, \cdot) \Vert^{}_{L^\infty (B)}\Big)  \Vert f
\Vert^{}_{D_N},\cr
}$$
$ t \geq 1$, where $B$ is the unit ball in $\Rrm^3,$ $N$ depends on
$n,j$ and $L$. Moreover
$$\hat{g}_0 (k) = F(1, - {\varepsilon k/ \omega(k)})  \hat{f} (k)
\eqno{({\rm A}.36)}$$
and $g^{}_l$ is a bilinear function of $F$ and $f$ satisfying
$$\Vert g^{}_l \Vert^{}_{D_i} \leq C_{i,l}  \sum_{\scr Y \in Q'\atop
\scr \vert Y\vert\leq M} \Vert (\xi^{}_Y F)
(1,\cdot) \Vert^{}_{L^\infty (B)}  \Vert f \Vert^{}_{D_M},\quad l,i\geq0,
\eqno{({\rm A}.37)}$$
where $M$ is an integer depending on $l$ and $i$.
}\saut
Before proving the theorem we remark that formula (A.36) can be generalized to
give explicit expressions for  $g^{}_n$,  $n \geq 0$. As a matter of fact
$$\hat{g}^{}_n (k) = {i^n \over n!}  \sum_j  {\partial \over \partial k_{j_1}}
\cdots {\partial \over \partial k_{j_n}}
\big((\partial_{j_1}\cdots\partial_{j_n}  F)
(1, - {\varepsilon k / \omega(k)})  \hat{f} (k)\big). \eqno{({\rm A}.38)}$$
Since we shall  not use this result for $n \geq 1$, we shall only prove the
particular case $n=0$ in (A.36).
\saut
\noindent{\it Proof.}
To prove the the uniqueness of the sequence $g^{}_l$, $l \geq 0$, let
$g'_l$, $l\geq 0$ be  another sequence satisfying (A.35)
and let $n\in\Nrm$ be such that $g^{}_l=g'_l$ for $0\leq l\leq n-1$. Then
$$\eqalignno{
\Vert g^{}_n - g'_n \Vert^{}_{D_L} &\leq t^n \Vert \sum_{0 \leq l \leq n}
t^{-l}
(g^{}_l - g'_l) \Vert^{}_{D_L}\cr
&\leq t^n \Big(\Vert e^{-i \varepsilon \omega(-i \partial) t}  F(t)  e^{i
\varepsilon \omega(-i \partial) t}  f - \sum_{0 \leq l \leq n}  t^{-l}  g^{}_l
\Vert^{}_{D_L}\cr
&\quad{}+ \Vert e^{-i \varepsilon \omega(-i \partial) t}  F(t)
e^{i \varepsilon \omega(-i\partial) t}  f - \sum_{0 \leq l \leq n}
t^{-l}  g'_l \Vert^{}_{D_L}\Big)\cr
&\leq t^{-1}  2 C_{N,L} \Big(\Vert F(1, \cdot)
\Vert^{}_{L^\infty (\Rrm^3)} + \sum_{Y
\in D(N)}  \Vert (\xi^{}_Y F) (1, \cdot) \Vert^{}_{L^\infty (B)}\Big)  \Vert f
\Vert^{}_{D_N}.\cr
}$$
Taking the limit $t \fl \infty$ now proves that $g^{}_n = g'_n$,  $n \geq 0$.
Hence by induction $g^{}_l = g'_l$ for $l\geq 0$.

To construct the sequence $g^{}_n$,  $n \geq 0$, we define
$r \in C^\infty ((\Rrm^+ -\{ 0 \}) \times \Rrm^3)$, homogeneous of
degree $- 3/2$ by formula (A.1a),
$$r^{}_0 (t,x) = e^{i3 \varepsilon \pi/4}  \big({m / t}\big)^{3/2}  \big({t /
\rho(t,x)}\big)^{5/2}  \hat{f} \big({- \varepsilon mx / \rho(t,x)}\big),
\eqno{({\rm A}.39{\rm a})}$$
we define the sequence $r^{}_n \in C^\infty ((\Rrm^+ -\{ 0 \})
\times \Rrm^3)$,  $n \geq 0$, of homogeneous functions of degree
$-n-3/2$ by formula (A.1b), i.e.
$$r^{}_n = {1 \over n!}  \Big({\rho \carre \over 2i \varepsilon m}\Big)^n
r^{}_0,
\quad  n \geq0. \eqno{({\rm A}.39{\rm b})}$$
It now follows from Theorem A.1 and Leibniz rule that
if $X\in\Pi'\cap U(\Rrm^4)$, then
$$\eqalignno{
&\Vert(1+\lambda(t))^{k/2}\xi^{}_X \Big(F(t)  e^{i \varepsilon
\omega(-i \partial) t}
f - e^{im \varepsilon \rho (t)}  F(t)  \sum_{0 \leq l \leq K}  r^{}_l (t)\Big)
\Vert^{}_{L^2}&({\rm A}.40)\cr
&\quad{}\leq C_{\vert X \vert}  \sum_{\vert X_1\vert +\vert X_2\vert =
\vert X\vert}
\Vert (\xi^{}_{X_1}F)(t)\xi^{}_{X_2}\Big(
e^{i \varepsilon\omega(-i \partial) t}  f
- e^{im \varepsilon \rho (t)}  \sum_{0 \leq l \leq K}
r^{}_l (t)\Big) \Vert^{}_{L^2}\cr
&\quad{}\leq C_{K, \vert X \vert,k} \Big(\sum_{\vert X_1\vert \leq
\vert X\vert}
 \Vert \xi^{}_{X_1}F(t) \Vert^{}_{L^\infty}\Big)  \Vert f
\Vert^{}_{D_{N'}}  t^{-K-1},\quad  t \geq1,\cr
}$$
where $X_1,X_2\in\Pi'\cap U(\Rrm^4)$ in the summation domains,
where $N'$ depends on $K, X,k$  and where
$$\Vert \rho (t)^{-j}\xi^{}_X  r^{}_l (t) \Vert^{}_{L^2} \leq C_{l, \vert
X \vert,j}  \Vert f \Vert^{}_{D_{N'}}  t^{-(l+\vert X \vert+j)}, \quad
t >0. \eqno{({\rm A}.41)}$$
The support of $r^{}_l$ is contained in the light cone.
Since the function $(t,x)\mapsto F(t,x)  r^{}_l (t,x)$ is homogeneous
of degree $-l-3/2,$ we can apply
Theorem A.2 to the function $(t,x) \mapsto t^l  F(t,x)  r^{}_l (t,x) = q^{}_l
(t,x)$, which shows that there are functions $g^{}_{l,j} \in D_\infty$,
$l \geq 0$, $j \geq 0$, given by the construction (A.8) satisfying
$$\eqalignno{
&\Vert (1+\lambda(t))^{k/2}\xi^{}_X \Big(e^{im \varepsilon
\rho (t)}  q^{}_l (t,x) - \sum_{0 \leq j \leq L}  t^{-j}
e^{im \varepsilon \rho (t)}  g^{}_{l,j}\Big)
\Vert^{}_{L^2}\cr
&\qquad{}\leq C_{L,\vert X\vert,k}  \sum_{\scr Y \in Q'\atop\scr
s+\vert Y \vert \leq \vert Z \vert\leq N'}
\Vert ({m /\rho  (1, \cdot)})^s  (\xi^{}_Y  g^{}_l) (1,\cdot)
\Vert^{}_{L^2}\  t^{-L-1},\quad  t \geq1,\cr
}$$
and
$$\Vert g^{}_{l,j} \Vert^{}_{D_i} \leq C_{j,i}  \sum_{\scr Y \in Q'\atop\scr
s+\vert Y \vert \leq i+2j}
\Vert ({m/ \rho  (1, \cdot)})^{s+j}  (\xi^{}_Y  q^{}_l) (1,\cdot)
\Vert^{}_{L^2}.$$
These two inequalities, the definition
of $q^{}_l$ and inequality (A.41) give
$$\eqalignno{
&\Vert(1+\lambda(t))^{k/2}\xi^{}_X  \Big(
e^{im \varepsilon \rho (t)}  F(t)  r^{}_l (t)
- \sum_{0 \leq j \leq L}  t^{-j-l}  e^{i \varepsilon \omega(-i \partial)t}
g^{}_{l,j}\Big) \Vert^{}_{L^2}&({\rm A}.42{\rm a})\cr
&\quad{}\leq C_{L,\vert X\vert,k}  \sum_{\scr Y\in Q'
\atop\scr \vert Y\vert \leq N''} \Vert (\xi^{}_Y  F)
(1,\cdot) \Vert^{}_{L^\infty (B)}  \Vert f \Vert^{}_{D_{N''}}  t^{-L-1-l},
\quad t > 0,\cr
}$$
and
$$\Vert g^{}_{l,j} \Vert^{}_{D_i} \leq C_{l,j,i}
\sum_{\scr Y \in Q'\atop\scr \vert Y\vert \leq M}  \Vert (\xi^{}_Y
F) (1,\cdot) \Vert^{}_{L^\infty (B)}  \Vert f \Vert^{}_{D_M},
\eqno{({\rm A}.42{\rm b})}$$
where $N''$ depends on $L$, $\vert X \vert$, $k$ and $M$ on $l$, $j$, $i$.
We have used repeatedly here that
$$\eqalignno{
&\xi^{}_{M_{\mu \nu}}  t^l  F(t,x)  r^{}_l (t,x) \cr
&\quad{}= (\xi^{}_{M_{\mu \nu}} F) (t,x)
t^l  r^{}_l (t,x) + F(t,x)  [\xi^{}_{M_{\mu \nu}}, t^l]
r^{}_l (t,x) + F(t,x)  t^l
\xi^{}_{M_{\mu \nu}}  r(t,x)\cr
}$$
and that the commutator of a monomial of degree $l$ in $(t,x)$ with
$\xi^{}_{M_{\mu \nu}}$ is a monomial of degree $l$.

We now define
$$g^{}_n = \sum_{l+j=n}  g^{}_{l,j},\quad  n \geq 0, \eqno{({\rm A}.43)}$$
which, according to (A.42b), proves the inequality (A.37) and which, together
with (A.40), (A.42a) and (A.42b) gives, choosing $L$ and $K$ sufficiently
large,
$$\eqalignno{
&\Vert (1+\lambda(t))^{k/2}\xi^{}_X\Big(F(t)
e^{i \varepsilon \omega(-i \partial)t}  f
- \sum_{0 \leq l \leq n}  t^{-l}
e^{i \varepsilon \omega(-i \partial)t}  g^{}_l\Big)
\Vert^{}_{L^2}&({\rm A}.44)\cr
&\quad{}\leq C_{n,\vert X \vert,k} \Big(
\sum_{\scr X_1\in\Pi'\cap U(\Rrm^4)\atop\scr\vert X_1 \vert \leq \vert X\vert}
\Vert \xi^{}_{X_1}  F(t) \Vert^{}_{L^\infty (\Rrm^3)} + \sum_{\scr Y \in
Q'\atop\scr \vert Y\vert \leq N}  \Vert (\xi^{}_Y F) (1, \cdot)
\Vert^{}_{L^\infty (B)}\Big) \Vert f \Vert^{}_{D_N}
t^{-n-1},\cr
}$$
$t \geq 1$, where $N$ depends on $n$, $\vert X\vert$, $k$.
If
$$h^{}_n (t) = e^{-i \varepsilon \omega(-i \partial)t}  F(t)  e^{i \varepsilon
\omega(-i\partial)t} - \sum_{0 \leq l \leq n}  t^{-l}  g^{}_l,$$
then
$$\Vert \big({d\over dt}\big)^l h^{}_n (t) \Vert^{}_{D_L} \leq \sum_{\scr \vert
\alpha \vert \leq L\atop\scr  \vert \beta \vert \leq L}  \Vert x^\beta
\partial^\alpha \big({d\over dt}\big)^l h^{}_n (t) \Vert^{}_{L^2},$$
according to Theorem 2.9. Since $(i \varepsilon x_j \omega(-i \partial)-t
\partial_j)  e^{-i \varepsilon \omega(-i \partial)t} =
e^{-i \varepsilon \omega(-i
\partial)t} x_j$, it follows that
$$
\Vert \big({d\over dt}\big)^l h^{}_n (t) \Vert^{}_{D_L} \leq C_{L,l}
\sum_{\scr s+\vert \beta \vert \leq L\atop \scr j+ \vert \alpha \vert \leq L+l}
t^s \Vert
x^\beta  \partial^\alpha\big({d\over dt}\big)^j
e^{i \varepsilon \omega(-i \partial)t} h^{}_n (t) \Vert^{}_{L^2}.$$
This shows together with (A.44) that
$$\eqalignno{
&\Vert \big({d\over dt}\big)^l h^{}_n (t) \Vert^{}_{D_L} &({\rm A}.45)\cr
&\quad{}\leq C_{n,L,l}  \Big(\sum_{\scr X\in \Pi'\cap U(\Rrm^4)\atop\scr
\vert X\vert \leq L+l}
\Vert \xi^{}_X F(t) \Vert^{}_{L^\infty (\Rrm^3)} + \sum_{\scr Y \in Q'
\atop\scr \vert Y\vert\leq N}
\Vert (\xi^{}_Y F) (1, \cdot)
\Vert^{}_{L^\infty (B)}\Big)
\Vert f \Vert^{}_{D_N}t^{-n-1+L}, \cr
}$$
$t \geq 1,  n \geq L$, where $N$ is redefined. Let $n' = n+l+L$.
Then, it follows from (A.37) and (A.45) that
$$\eqalignno{
&\Vert\big({d\over dt}\big)^l h^{}_n (t) \Vert^{}_{D_L}&({\rm A}.46)\cr
&\quad{}\leq \Vert \big({d\over dt}\big)^l h^{}_{n'} (t) \Vert^{}_{D_L} +C_l
\sum_{0 \leq j \leq L-1}  t^{-(n-1+l+j)}  \Vert g^{}_{n+1+j} \Vert^{}_{D_L}\cr
&\quad{}\leq C_{n,L,l}  \Big(\sum_{\scr X\in \Pi'\cap U(\Rrm^4)
\atop\scr \vert X \vert \leq L+l}  \Vert \xi^{}_X
F(1) \Vert^{}_{L^\infty (\Rrm^3)} +
\sum_{\scr Y \in Q'\atop\scr\vert Y\vert \leq N }
\Vert (\xi^{}_Y F) (1, \cdot)
\Vert^{}_{L^\infty (B)}\Big)  \Vert f \Vert^{}_{D_N}  t^{-n-1-l},\cr
}$$
$t \geq 1$, $L,l\geq 0$,  where we have redefined $N$. This proves (A.35).

To prove (A.36) we note that $g^{}_0 = g^{}_{0,0}$ according to (A.43) and that
$\hat{g}_{0,0}$ is obtained from $q^{}_0$ by formula (A.8a)
$$\hat{g}_{0,0} (k) = e^{- i \varepsilon \pi/4}
(m^2 / \omega(k)^5)^{1/2}  q^{}_0
(1, {- \varepsilon k / \omega(k)}).$$
Since $q^{}_0 (t,x) = F(t,x)  r^{}_0 (t,x)$ formula (A.39a) now gives
$$\hat{g}_0 (k) = F(1, {- \varepsilon k / \omega(k)}) \hat{f} (k),$$
which proves the theorem.
\saut
\noindent{\it Proof of Theorem A.1}
We recall that
$$\eqalignno{
\varphi_0(t) &= e^{i\varepsilon \omega(-i\partial)t}f,&({\rm A}.47{\rm a})\cr
\noalign{\noindent and}
\varphi_n(t) &=\varphi_0(t)-e^{i\varepsilon m\rho(t)}
\sum_{0\leq l\leq n-1}g^{}_l (t),
\quad n\geq 1.&({\rm A}.47{\rm b})\cr
}$$
The construction of $g^{}_l$ by (A.1a) and (A.1b) then gives that
$$(\carre +m^2)\varphi_0=0,\quad (\carre +m^2)\varphi_{n+1}
=-e^{i\varepsilon m\rho}\carre g^{}_n,\quad n\geq0.\eqno{({\rm A}.48)}$$
Let $\Gamma^A$, $A\in \{1,\ldots,16\}$ be the matrices  $I$, $\gamma^\mu$,
$0\leq \mu\leq3$,
$\gamma^\mu\gamma^\nu$, $0\leq \mu<\nu\leq3$,
$\gamma^\mu\gamma^\nu\gamma^\tau$, $0\leq \mu<\nu<\tau\leq3$,
$\gamma^0\gamma^1\gamma^2\gamma^3$, in a given order. The
set $\{\Gamma^A\big\vert 1\leq A\leq 16\}$ is then a basis of the
complex vector space of $4\times4$ complex matrices, and
$\Gamma^A$ is invertible for $1\leq A\leq 16$.
If $a^{}_\mu\in\Crm$ for $0\leq \mu\leq3$ and $b\in\Crm$, and if
$$
M=a^{}_0 I - \sum_{1\leq j\leq 3}a^{}_j\gamma^0\gamma^j+ib\gamma^0,
\eqno{({\rm A}.49{\rm a})}$$
then there exist complex numbers $c^{}_\mu(A)$ and $d^{}(A)$, $0\leq \mu\leq3$,
$1\leq A\leq 16$, independent of $a^{}_\mu$ and $b$, such that
$$a^{}_\mu I=\gamma_\mu\gamma^0\sum_A c^{}_\mu(A)(\Gamma^A)^{-1}M\Gamma^A,
\eqno{({\rm A}.49{\rm b})}$$
(where there is no summation over $\mu$ on the right-hand side) and
$$b I=\gamma^0\sum_A d(A)(\Gamma^A)^{-1}M\Gamma^A.
\eqno{({\rm A}.49{\rm c})}$$

As a matter of fact, conjugation of $M$ with $\gamma^1\gamma^2\gamma^3$
(resp. $\gamma^0\gamma^1\gamma^2\gamma^3$, $\gamma^0\gamma^i\gamma^j$)
corresponds to replace   $(a^{}_1,a^{}_2,a^{}_3,b)$ (resp. $b,a^{}_k$) by
$(-a^{}_1,-a^{}_2,-a^{}_3,-b)$ (resp. $-b,-a^{}_k$), in (A.49a),
where $(i,j,k)$ is a permutation of $(1,2,3)$. Let
$$h^{(A)}_n =\big({\partial\over \partial t}+{\cal D}\big)\Gamma^A
\varphi^{}_n,
\eqno{({\rm A}.50{\rm a})}$$
where ${\cal D}$ is as in equation (1.2b), and let
$$r^{(A)}_0=0,\quad r^{(A)}_n =-i \gamma^0\Gamma^Ae^{i\varepsilon m\rho}
\carre g^{}_{n-1},\quad n\geq 1.
\eqno{({\rm A}.50{\rm b})}$$
Using that $\carre=({d/dt} -{\cal D})({d/dt} +{\cal D})$, it then follows
from (A.48) that
$$(i\gamma^\mu\partial_\mu + m)h^{(A)}_n=r^{(A)}_n,\quad n\geq0,1\leq A\leq 16.
\eqno{({\rm A}.51)}$$
The representation $\xi$ of the Poincar\'e Lie algebra $\p$ satisfies
$$\eqalignno{
\xi^{}_{P_0}&=t\rho^{-2}\Big(L-\sum_{j=1}^{3}(x_j/t)
\xi^{}_{M_{0j}}\Big)&({\rm A}.52{\rm a})\cr
\xi^{}_{P_i}&=t^{-1}\xi^{}_{M_{0i}}-x_i\rho^{-2}
\Big(L-\sum_{j=1}^{3}(x_j/t)\xi^{}_{M_{0j}}\Big),\quad 1\leq i\leq 3,
&({\rm A}.52{\rm b})\cr
}$$
where $\rho^2=t^2 - \vert x\vert^2 \neq 0$,
$$L= t {\partial\over\partial t}+\sum_{i=1}^{3}x_i
{\partial\over\partial x_i}.
\eqno{({\rm A}.52{\rm c})}$$
Let $H\in C^\infty ((\Rrm^+\times\Rrm^3) - \{0\})$ be a homogeneous function
of degree $\chi\in\Rrm$ and let ${\rm supp}\ H\subset \{y\in\Rrm^+\times\Rrm^3
\big\vert y^\mu y_\mu\geq 0\}$. It follows from (A.52a)--(A.52c) that
$$\vert (\xi^{}_{P_\mu}H)(t,x)\vert
\leq (2+\vert\chi\vert)t\rho^{-2}\sum_{\scr Y\in Q'\atop\scr
\vert  Y\vert \leq 1}
\vert (\xi^{}_{Y}H)(t,x)\vert, \quad t - \vert x\vert>0.$$
Using that $\Rrm^4$ is an ideal of $\p$, we obtain by induction that
$$\vert (\xi^{}_{Y}H)(t,x)\vert\leq C_{\vert Y\vert,\vert \chi\vert}
(t\rho^{-2})^{\vert Y\vert}
\sum_{\scr Z\in Q'\atop\scr \vert  Z\vert \leq \vert  Y\vert }
\vert (\xi^{}_{Z}H)(t,x)\vert,
\eqno{({\rm A}.53)}$$
where $Y\in\Pi'\cap U(\Rrm^4)$ and $ t - \vert x\vert>0$.

The operator $\carre$ and the multiplication by $\rho$ commute with
$\xi^{}_{Z}$
when $Z\in Q'$.
Since $(\rho\carre)^j\xi^{}_{X}g^{}_0$, $X\in Q'$, is homogeneous of degree
$-j-3/2$, it follows from (A.53) that
$$\vert (\xi^{}_{YX}(\rho\carre)^j g^{}_0)(t,x)\vert
\leq C_{\vert Y\vert,j+3/2}(t\rho^{-2})^{\vert Y\vert}
\sum_{\scr Z\in Q'\atop\scr \vert  Z\vert \leq \vert  Y\vert }
\vert ((\rho\carre)^j \xi^{}_{ZX}g^{}_0)(t,x)\vert,
\eqno{({\rm A}.54)}$$
where $j\geq0$, $X\in Q'$, $Y\in\Pi'\cap U(\Rrm^4)$, $ t - \vert x\vert>0$.
Using, as in (A.28), expression (A.27) for the operator $\carre$,
we obtain from (A.54) that
$$\vert (\xi^{}_{YX}(\rho\carre)^j g^{}_0)(t,x)\vert
\leq C_{\vert Y\vert,j}(t\rho^{-2})^{\vert Y\vert}\rho^{-j}
\sum_{\scr Z\in Q'\atop\scr \vert  Z\vert \leq \vert  Y\vert+2j}
\vert (\xi^{}_{ZX} g^{}_0)(t,x)\vert,
\eqno{({\rm A}.55{\rm a})}$$
where $j\geq0$, $X\in Q'$, $Y\in\Pi'\cap U(\Rrm^4)$, $ t - \vert x\vert>0$ and
where $C_{\vert Y\vert,j}$ is a new constant. Similarly it follows that
$$\vert (\xi^{}_{YX}\carre(\rho\carre)^j g^{}_0)(t,x)\vert
\leq C_{\vert Y\vert,j}(t\rho^{-2})^{\vert Y\vert}\rho^{-2-j}
\sum_{\scr Z\in Q'\atop\scr \vert  Z\vert \leq \vert  Y\vert+2j+2}
\vert (\xi^{}_{ZX} g^{}_0)(t,x)\vert,
\eqno{({\rm A}.55{\rm b})}$$
where $j\geq0$, $X\in Q'$, $Y\in\Pi'\cap U(\Rrm^4)$, $ t - \vert x\vert>0$.
Formulas (A.55a) and (A.55b), and definition (A.1b) of $g^{}_l$ give that
$$\vert (\xi^{}_{YX}g^{}_l)(t,x)\vert
\leq C_{\vert Y\vert,l}(t\rho^{-2})^{\vert Y\vert}\rho^{-l}
\sum_{\scr Z\in Q'\atop\scr \vert  Z\vert \leq \vert  Y\vert+2l}
\vert (\xi^{}_{ZX} g^{}_0)(t,x)\vert
\eqno{({\rm A}.56{\rm a})}$$
and
$$\vert (\xi^{}_{YX}\carre g^{}_l)(t,x)\vert
\leq C_{\vert Y\vert,l}(t\rho^{-2})^{\vert Y\vert}\rho^{-l-2}
\sum_{\scr Z\in Q'\atop\scr \vert  Z\vert \leq \vert  Y\vert+2l+2}
\vert (\xi^{}_{ZX} g^{}_0)(t,x)\vert,
\eqno{({\rm A}.56{\rm b})}$$
where $l\in\Nrm$, $X\in Q'$, $Y\in\Pi'\cap U(\Rrm^4)$, $ t - \vert x\vert>0$.
Similarly as we obtained equality (A.20), it follows from definition (A.1a)
of $g^{}_0$ that
$$(\eta^{}_Z\hat f)(k)
=e^{-3\varepsilon\pi/4}m(\omega(k))^{-5/2}t^{3/2}
(\xi^{}_Zg^{}_0)(t,-t\varepsilon k/\omega(k)),
\eqno{({\rm A}.57)}$$
where $t>0$, $k\in\Rrm^3$ and $Z\in U({\frak {so}}(3,1))$.
After multiplication with $\omega(k)^j$ and integration, we obtain
using the variable substitution (A.2) that
$$\Vert\omega^j\eta^{}_Z\hat f\Vert^2_{L^2}
=\Vert(mt/\rho(t))^j(\xi^{}_Z g^{}_0)(t)\Vert^2_{L^2},
\eqno{({\rm A}.58)}$$
where $t>0$ and $Z\in U({\frak {so}}(3,1))$.
Inequalities (A.56a) and (A.56b), and inequality (A.58) give that
$$\Vert\rho(t)^{-j}(\xi^{}_{XY} g^{}_l)(t)\Vert^{}_{L^2}
\leq C_{\vert X\vert+j+l}\ t^{-j-l-\vert X\vert}
\sum_{\scr Z\in Q'\atop\scr \vert  Z\vert \leq \vert  X\vert+2l}
\Vert\omega^{2\vert  X\vert+l+j}\eta^{}_{ZY}\hat f\Vert^{}_{L^2}
\eqno{({\rm A}.59{\rm a})}$$
and
$$\eqalignno{
&\Vert\rho(t)^{-j}(\xi^{}_{XY} \carre g^{}_l)(t)\Vert^{}_{L^2}
&({\rm A}.59{\rm b})\cr
&\qquad{}\leq C_{\vert X\vert+j+l}\ t^{-j-l-\vert X\vert-2}
\sum_{\scr Z\in Q'\atop\scr \vert  Z\vert \leq \vert  X\vert+2l+2}
\Vert\omega^{2\vert  X\vert+l+j+2}\eta^{}_{ZY}\hat f\Vert^{}_{L^2},\cr
}$$
where $t>0$, $l,j\in\Nrm$, $X\in\Pi'\cap U(\Rrm^4)$ and $Y\in Q'$.
The equivalence of the  norms in (A.17) and inequalities (A.59a) and (A.59b)
give
$$\Vert\rho(t)^{-j}(\xi^{}_{XY}  g^{}_l)(t)\Vert^{}_{L^2}
\leq C_{\vert X\vert+j+l}\ t^{-j-l-\vert X\vert}
\Vert f\Vert^{}_{D_{3\vert X\vert+3l+\vert Y\vert+j}}
\eqno{({\rm A}.60{\rm a})}$$
and
$$\Vert\rho(t)^{-j}(\xi^{}_{XY}\carre  g^{}_l)(t)\Vert^{}_{L^2}
\leq C_{\vert X\vert+j+l}\ t^{-j-l-\vert X\vert-2}
\Vert f\Vert^{}_{D_{3\vert X\vert+3l+\vert Y\vert+j+4}},
\eqno{({\rm A}.60{\rm b})}$$
where $t>0$, $l,j\in\Nrm$, $X\in\Pi'\cap U(\Rrm^4)$ and $Y\in Q'$.
Inequality (A.60a) proves inequality (A.6) of the theorem.
It follows from inequality (A.56a) and equality (A.57) that
$$\eqalignno{
&\Vert\rho(t)^{-j}(\xi^{}_{XY} g^{}_l)(t)\Vert^{}_{L^\infty}&({\rm A}.61)\cr
&\qquad{}\leq C_{\vert X\vert+j+l}\ t^{-3/2-\vert X\vert-j-l}
\sum_{\scr Z\in Q'\atop\scr \vert  Z\vert \leq \vert  X\vert+2l}
\Vert\omega^{5/2+2\vert  X\vert+j+l}\eta^{}_{ZY}\hat f\Vert^{}_{L^\infty},\cr
}$$
where $t>0$, $l,j\in\Nrm$, $X\in\Pi'\cap U(\Rrm^4)$ and $Y\in Q'$.
If $h\in S(\Rrm^3,\Crm^4)$, then $\Vert \hat h\Vert^{}_{L^\infty}
\leq\Vert h\Vert^{}_{L^1}\leq C \Vert h\Vert^{}_{D_2}
\leq C'\Vert\hat h\Vert^{}_{D_2}$, so it follows from (A.61) and from
the equivalence of norms in (A.17) that
$$\Vert\rho(t)^{-j}(\xi^{}_{XY} g^{}_l)(t)\Vert^{}_{L^\infty}
\leq C_{\vert X\vert+j+l}\ t^{-3/2-\vert X\vert-j-l}
\Vert f\Vert^{}_{D_{3\vert X\vert+3l+\vert Y\vert+j+5}},
\eqno{({\rm A}.62)}$$
where $t>0$, $l,j\in\Nrm$, $X\in\Pi'\cap U(\Rrm^4)$ and $Y\in Q'$.
This proves inequality (A.7) of the theorem.

It follows by induction that there exist polynomials $R^{(k)}_{X,Y}$ of degree
$k$ such that
$$
e^{-i\varepsilon m\rho}\xi^{}_X e^{i\varepsilon m\rho}
=\xi^{}_X +\sum_{\scr Y\in \Pi'\cap U(\Rrm^4)\atop\scr
\vert Y\vert\leq\vert X\vert-1}
\sum_{0\leq i\leq \vert X\vert-\vert Y\vert-1}
\rho^{-i}R^{(\vert X\vert-\vert Y\vert+i)}_{X,Y}(y/\rho) \xi^{}_Y,
\eqno{({\rm A}.63)}$$
for $X\in\Pi'\cap U(\Rrm^4)$, $y\in\Rrm^+\times\Rrm^3$, $y^\mu y_\mu > 0$.
This equality and inequalities (A.60a) and (A.60b) give that
$$
\Vert(\rho(t)^{-j}\xi^{}_{XY}e^{i\varepsilon m\rho} g^{}_l)(t)\Vert^{}_{L^2}
\leq C_{\vert X\vert+j+l}\ t^{-j-l}(1+t^{-\vert X\vert})
\Vert f\Vert^{}_{D_{3\vert X\vert+3l+\vert Y\vert+j}},
\eqno{({\rm A}.64{\rm a})}$$
and
$$
\Vert(\rho(t)^{-j}\xi^{}_{XY}e^{im\rho}\carre  g^{}_l)(t)\Vert^{}_{L^2}
\leq C_{\vert X\vert+j+l}\ t^{-j-l-2}(1+t^{-\vert X\vert})
\Vert f\Vert^{}_{D_{3\vert X\vert+3l+\vert Y\vert+j+4}},
\eqno{({\rm A}.64{\rm b})}$$
where $t>0$, $l,j\in\Nrm$, $X\in\Pi'\cap U(\Rrm^4)$ and $Y\in Q'$.

Since $\Vert(\xi^{D}_{X}h)(t)\Vert^{}_{D}\leq\Vert(\xi^{}_{X}h)(t)\Vert^{}_{D}
+ C_X\Vert h(t)\Vert^{}_{D}$, for $X\in\p$, it follows that
$$\eqalignno{
\wp^D_n(h(t))
&\leq C_n \Big( \sum_{\scr Y\in \Pi'\atop\scr \vert Y\vert\leq n}
\Vert(\xi^{}_{Y}h)(t)\Vert^{2}_{D}\Big)^{1/2}&({\rm A}.65)\cr
&\leq C'_n \wp^D_n(h(t)),\quad n\geq 0,\cr
}$$
for some constants $C_n$ and $C'_n$. If $\lambda_1$ is defined as in Theorem
5.5, then $(\lambda_1(t))(x)=(1+t)(1+t-\vert x\vert)^{-1}$ for
$0\leq \vert x\vert< t$, so $(\lambda_1(t))(x)\leq C t^2 \rho^{-2}$
for $0\leq \vert x\vert< t$. Therefore it follows from (A.50b), (A.64b)
and (A.65) that
$$
\wp^D_j\big((1+\lambda_1(t))^{k/2}r^{(A)}_n(t)\big)
\leq n C_{j+k+n}\ t^{-n-1}(1+t^{-j})\Vert f\Vert^{}_{D_{3j+3n+k+1}},
\eqno{({\rm A}.66)}$$
for $t>0$, $j,k,n\in\Nrm$.

Let $Y\in U(\p)$. It then follows from definition (A.47b) of $\varphi^{}_n$
and from theorem 1 of \refHKG\ (and \refHKGD\ for the details)
that there exists an integer $N\geq1$ such that
$\Vert \delta(t)^2(\xi^{}_Y \varphi^{}_N)(t)\Vert^{}_{L^\infty}$
$\leq C$ for $t\geq1$, where $C\in \Rrm^+$ is independent of $t$
and where $(\delta(t)(x)=1+t+\vert x\vert$. Therefore
$\lim_{t\fl\infty}\Vert (\xi^{}_Y \varphi^{}_N)(t)\Vert^{}_{L^2}=0$.
It then follows from (A.47b) and (A.64a) with $j=0$ that
$\lim_{t\fl\infty}\Vert (\xi^{}_Y \varphi^{}_n)(t)\Vert^{}_{L^2}=0$
for  $Y\in U(\p)$ and $n\geq 1$. Definition (A.50a) and inequality
(A.65) now give that
$$
\lim_{t\fl\infty}\wp^D_j(h^{(A)}_n(t))=0\quad \hbox{for}\ j\geq 0,n\geq 1.
\eqno{({\rm A}.67)}$$
It follows from equality (A.51), inequality (A.66) with $k=0$ and limit
(A.67) that
$$
\wp^D_j(h^{(A)}_n(t))
\leq  C_{j+n}\ t^{-n}\Vert f\Vert^{}_{D_{3j+3n+1}},\quad t\geq 1,
j\geq 0,n\geq 1.
\eqno{({\rm A}.68{\rm a})}$$
We also note that, since $r^{}_0=0$, definition (A.50a) give that
$$
\wp^D_j(h^{(A)}_0(t))
=\Vert h^{(A)}_0(0)\Vert^{}_{D_{j}}
\leq C\Big(\Vert f\Vert^{}_{D_{j}}+\sum_{1\leq i\leq 3}
\Vert \partial_j f\Vert^{}_{D_{j}}\Big),\quad j\geq 0.
\eqno{({\rm A}.68{\rm b})}$$

Theorem 5.5, with $G=0$, $g=0$, and inequality (A.68b) give that
$$
\wp^D_j\big((1+\lambda_1(t))^{k/2}h^{(A)}_0(t)\big)
\leq C_{j+k}\Big(\Vert f\Vert^{}_{D_{j+k}}+\sum_{1\leq i\leq 3}
\Vert \partial_j f\Vert^{}_{D_{j+k}}\Big),
\eqno{({\rm A}.69)}$$
fot $t\geq0$ and $j,k\in\Nrm$. Since $(1+\lambda_1(t))(x)\leq C(1+t)$
in the support of $r^{(A)}_n$, it follows from Theorem 5.5, with $G=0$, that
$$\eqalignno{
&\wp^D_j\big((1+\lambda_1(t))^{k/2}h^{(A)}_n(t)\big)\cr
&\qquad{}\leq C_{j+k}\Big(\wp^D_{j+k}(h^{(A)}_n(t))
+t\sum_{0\leq i\leq k-1}
\wp^D_{j+i}\big((1+\lambda_1(t))^{(k-1-i)/2}r^{(A)}_n(t)\big)\Big),\quad
t>0,\cr
}$$
Inequalities (A.66) and (A.68a) then give that
$$\wp^D_j\big((1+\lambda_1(t))^{k/2}h^{(A)}_n(t)\big)
\leq C_{j+k+n}\ t^{-n}\Vert f\Vert^{}_{D_{3(j+k+n)+1}},
\eqno{({\rm A}.70)}$$
for $t\geq 1$, $j,k\in\Nrm$, $n\geq1$.

Applying the operator $\xi^D_Y$, $Y\in\Pi'$, on both sides of equation (A.51),
it follows from Theorem 5.7 and inequality (A.68b) that
$$
\Vert(\delta(t))^{3/2}
(1+\lambda_1(t))^{k/2}(\xi^D_Yh^{(A)}_0)(t)\Vert^{}_{L^\infty}
\leq C\Big(\Vert f\Vert^{}_{D_{\vert Y\vert+k+8}}
+\sum_{1\leq i\leq 3}
\Vert \partial_j f\Vert^{}_{D_{\vert Y\vert+k+8}}\Big),
\eqno{({\rm A}.71)}$$
for $t\geq0$, $k\in\Nrm$ and $Y\in \Pi'$.
Similarly, it follows from Theorem 5.7 and inequalities (A.66) and (A.68a) that
$$
\Vert(\delta(t))^{3/2}
(1+\lambda_1(t))^{k/2}(\xi^D_Yh^{(A)}_n)(t)\Vert^{}_{L^\infty}
\leq C_{k+\vert Y\vert+n}\ t^{-n}
\Vert f\Vert^{}_{D_{3(k+\vert Y\vert+n)+25}},
\eqno{({\rm A}.72)}$$
for $t\geq1$, $k,n\in\Nrm$, $n\geq1$ and $Y\in \Pi'$.

According to equalities (A.49b), (A.49c) and (A.50a),
$\partial_\mu \varphi^{}_n$, $0\leq \mu\leq3$, and $m\varphi^{}_n$ are
linear combinations of $h^{(A)}_n$, $1\leq A\leq 16$.
This fact and inequalities (A.70) and (A.72), prove inequalities (A.4) and
(A.5).
This proves the theorem.
\vfill\eject
\noindent{\titre References}
\gsaut
\item{\refBFFLSI} Bayen, F., Flato, M., Fronsdal, C., Lichnerowicz, A.
and Sternheimer, D.:
Deformation theory and quantization I:
Deformation of symplectic structures. Ann. Phys. {\bf 111}, 61--110 (1978).
\saut
\item{\refBFFLSII} Bayen, F., Flato, M., Fronsdal, C., Lichnerowicz, A.
and Sternheimer, D.:
Deformation theory and quantization II: Physical applications. Ann. Phys.
{\bf 111}, 111--151 (1978).
\saut
\item{\refCAL} Calderon, A.P.: Lebesgue spaces of differentiable functions
and distributions. Proc. Symp. Pure Math. IV, AMS 1961, 33--49.
\saut
\item{\refD} Dito, J.: Star-products and nonstandard quantization
for Klein-Gordon equation. J. Math. Phys. {\bf 33}, 791--801 (1992).
\saut
\item{\refFPS} Flato, M., Pinczon, G. and Simon, J.C.H.: Non-linear
representations of Lie groups. Ann. Sci. Ec. Norm. Sup\'er., 4$^{\rm e}$
s\'erie, {\bf 10}, 405--418 (1977).
\saut
\item{\refFS} Flato, M. and Simon, J.C.H. : Non-linear equations and
covariance. Lett. Math. Phys. {\bf 2}, 155--160 (1979).
\saut
\item{\refFSYM} Flato, M. and Simon, J.C.H.: Yang-Mills equations are
formally linearizable. Lett. Math. Phys. {\bf 3}, 279-283 (1979).
\saut
\item{\refFST} Flato, M., Simon, J.C.H. and Taflin, E.:
On global solutions of the Maxwell-Dirac Equations. Commun. Math. Phys.
{\bf 112}, 21--49 (1987).
\saut
\item{\refFSTA} Flato, M., Simon, J.C.H. and Taflin, E.:
The Maxwell-Dirac equations: Asymptotic completeness and infrared problem.
Reviews in Math. Phys. {\bf 6}, N$^{\rm o}$ 5a, 1071--1083 (1994).
\saut
\item{\refGR} Gross, L.: The Cauchy problem for the coupled Maxwell and
Dirac equations. Commun. Pure Appl. Math {\bf 19}, 1--15 (1966).
\saut
\item{\refHA} H\"ormander, L.: The analysis of linear partial differential
operators. Vol. I, Berlin, Heidelberg, New-York: Springer 1985.
\saut
\item{\refHKG} H\"ormander, L.: Remarks on the Klein-Gordon equation.
Journ\'ee \'Equations aux d\'eriv\'ees partielles, Saint-Jean de Monts
1987, Soc. Math. France (1987).
\saut
\item{\refHKGD} H\"ormander, L.: Non-Linear Hyperbolic Differential
Equations, Lectures 1986--1987, N$^{\rm o}$ 1988:2 /ISSN 0327--8475,
Dept. Mathematics, Lund.
\saut
\item{\refHSL} H\"ormander, L.: On Sobolev spaces associated with some Lie
algebras, in ``Current topics in partial differential equations'',
Tokyo: Kinokuniya Company Ltd 1985, pp. 261--287.
\saut
\item{\refKI} Kato, K.: Linear evolution equations of ``hyperbolic" type,
II. J. Math. Soc. Japan {\bf 25}, 648--666 (1973).
\saut
\item{\refKII} Kato, T.: Quasilinear equations of evolution,
with application to partial differential equations. Lecture Notes
in Mathematics, Vol. 448. Berlin, Heidelberg, New-York: Springer 1975,
pp. 25--70.
\saut
\item{\refSERG} Sergeraert, F.: Un th\'eor\`eme de fonctions
implicites sur certains espaces de Fr\'echet et quelques applications.
Ann. Sci. Ec. Norm. Sup\'er. 4$^{\rm e}$ s\'erie, {\bf 5}, 599--660 (1972).
\saut
\item{\refSI} Simon, J.C.H.: A wave operator for a non-linear
Klein-Gordon equation. Lett. Math. Phys. {\bf 7}, 387--398 (1983).
\saut
\item{\refSTI} Simon, J.C.H. and Taflin, E.: Wave operators and
analytic solutions for systems of non-linear Klein-Gordon equations
and non-linear Schr\"odinger equations. Commun. Math. Phys. {\bf 99},
541--562 (1985).
\saut
\item{\refST} Simon, J.C.H. and Taflin, E.: The Cauchy problem for
non-linear Klein-Gordon equations. Commun. Math. Phys. {\bf 152},
433--478 (1993).
\saut
\item{\refVW} Von Wahl, W.: $L^p$ decay rates for homogeneous wave
equations. Math. Z. {\bf 120}, 93--106 (1971).
\saut
\item{\refW} Warner, G.: Harmonic analysis on semi-simple Lie group I,
Die Grundlehren der mathematischen Wissenschaften vol. 188,
Berlin, Heidelberg, New-York: Springer 1972.
\vfill \eject
\centerline{\bf Contents}

\bigskip

\def\toto{\leaders \hbox to 5mm{\hfil.\hfil}\hfill}
\noindent \line{1. Introduction \toto 2}\par
\vskip4mm
\noindent \line{2. The nonlinear representation $T$ and spaces of
differentiable vectors \toto 17}\par
\vskip4mm
\noindent \line{3. The asymptotic nonlinear representation \toto 48}\par
\vskip4mm
\noindent \line{4. Construction of the approximate solution \toto 75}\par
\vskip4mm
\noindent \line{5. Energy estimates and $L^2 - L^\infty$ estimates for
the Dirac field \toto 117}
\vskip4mm
\noindent \line{6. Construction of the modified wave operator and its inverse
\toto 192}\par
\vskip4mm
\noindent \line{   Appendix \toto 291}\par
\vskip4mm
\noindent \line{   References \toto 306}\par
\end